\documentclass[ngerman,UKenglish,epj,nopacs,numbook]{svjour}
\pdfoutput=1
\newcommand*{\texlive}{2011}

\newcommand*{\epjcreview}{2015}

\usepackage{ubonn-thesis}
\usepackage[switch]{lineno}
\usepackage{listings}
\usepackage{multirow}
\usepackage{multicol}
\usepackage{pdflscape}
\usepackage{tabu}
\usepackage{braket}
\usepackage{afterpage}
\usepackage{adjustbox}
\usepackage{ifthen}
\usepackage{xspace}
\usepackage[title]{appendix}
\usepackage[colorlinks=true,linkcolor=blue,citecolor=blue,urlcolor=blue,bookmarksopen]{hyperref}
\usepackage{rotating}
\usepackage{morefloats}
\usepackage{microtype}
\usepackage{tabularx}
\usepackage{calc}
\usepackage{fixltx2e}
\usepackage{thesis_defs}

\graphicspath{%
  {figs/},%
  {figs/cover/},%
  {figs/graphics/},%
  {feynmf/}%
}

\begin{document}
\onecolumn
\title{Charm Production and QCD Analysis at HERA and LHC}
\titlerunning{Charm Production and QCD Analysis at HERA and LHC}

\author{O.~Zenaiev
}                     
%
%
\institute{DESY}
%
%
\abstract{
This review is devoted to the study of charm production in $ep$ and $pp$ collisions. 
The total set of measurements obtained by the two collaborations H1 and ZEUS from HERA and 
their combination is outlined, as well as complementary data obtained by the LHCb collaboration at the LHC. 
After fitting the parton distribution functions the charm production cross sections are predicted 
within perturbative QCD at next-to-leading order using the fixed-flavour-number scheme.
Agreement with the data is found. 
The combined  HERA charm data  are sensitive to the $c$-quark mass and enabled its accurate determination.
The predictions crucially depend upon the knowledge of the gluon distribution function. 
It is shown that the shape of the gluon distribution based on the HERA data is considerably improved 
by adding the measurements from LHCb and applicable down to 
values $x$ of about $10^{-6}$, where $x$ is the proton momentum fraction carried by a parton.
\PACS{
      {PACS-key}{discribing text of that key}   \and
      {PACS-key}{discribing text of that key}
     } 
} 
\maketitle

\twocolumn

\tableofcontents





\renewcommand\ozmod[1]{{#1}}
\renewcommand\ozmodN[1]{{#1}}
\renewcommand\ozmodNN[1]{{#1}}
\renewcommand\ozmodNNN[1]{{#1}}
\renewcommand\ozmodNNNN[1]{{#1}}

\clearpage
\section{Introduction}
\label{sec:intro}

The HERA collider was the first and unique machine in which electrons and protons were collided. 
It emerged from a series of earlier lepton--nucleon accelerator studies as the highest energy electron--proton collider 
to investigate simultaneously neutral and charged current reactions and their electroweak unification. 
The pointlike electron serves as probe to study the internal structure of the proton governed by strong interactions, 
i.e.\ Quantum Chromodynamics (QCD). 
The generic electron--proton scattering process occurs via the exchange of an 
electroweak boson. 
The uniqueness of HERA consists in the clean distinction between electroweak and strong processes. 
The precise knowledge of electroweak interactions makes HERA ideal for investigation of QCD. 

Measurements of deep inelastic scattering (DIS) at HERA have been the central topic in the investigation of the proton structure 
for the two collider experiments, H1 and ZEUS~\cite{Abramowicz:2015mha,DIScomb}. 
Such measurements are the core data to determine the proton structure in terms of parton distribution functions (PDFs). 
The inclusive cross section at HERA contains contributions from all active quark and antiquark flavours. 
It is remarkable that a large contribution, up to one third, is coming from events with charm. 
This necessitates the understanding of heavy-flavour production for global QCD analyses of HERA data and is the main subject of the present review. 

The tests of perturbative QCD depend on phenomenological input, in particular on the knowledge of the gluon distribution function. 
For this reason an additional piece has been included in the analysis coming from charm production in the LHCb experiment at the Large Hadron Collider (LHC). 

This review describes various aspects of heavy-flavour physics at HERA and LHC. 
It presents one new measurement of charm production at HERA, 
which is further combined with other precise H1 and ZEUS charm measurements in order to obtain the most precise charm dataset from HERA. 
These combined data are extensively used in a comparison of data and theory and in a QCD analysis to extract the $c$-quark mass. 
Another combination is performed at the more exclusive level of \Dstar visible cross sections. 
In contrast to the inclusive one, it does not include theory-related uncertainties.
Furthermore, charm and beauty measurements from LHCb are considered and included in a QCD analysis. 
They provide sensitivity to the gluon distribution at low values of fractions of the proton momenta carried by a parton. 
This is a kinematic range that is currently not covered by other input data, and therefore improves the PDF fits. 

The review is organised in the following way. 
Section~\ref{sec:th} introduces the theoretical concepts, relevant for the subsequent contents. 
Section~\ref{sec:exp:hera} gives a description of the HERA experimental set-up, 
while Section~\ref{sec:exp:hera:hfmeas} describes tagging techniques used to measure charm production at HERA and presents existing measurements. 
Section~\ref{sec:dch} deals with the new physics results, the measurement of \Dch-meson production performed with the ZEUS detector at HERA. 
Section~\ref{sec:comb} describes a combination of charm measurements from H1 and ZEUS, performed at the two levels: 
for \Dstar visible cross sections and for inclusive reduced charm cross sections. 
Section~\ref{sec:hflhcb} switches to the LHC: it introduces measurements of heavy-flavour productions at the LHCb experiment 
and discusses the impact of the phase-space coverage which is comprementary to the one from HERA. 
Section~\ref{sec:pdffit} presents a QCD analysis including the LHCb heavy-flavour data.
Finally, Section~\ref{sec:concl} summarises the results.

This review is based on the PhD thesis of the author~\cite{Zenaiev:2015qea}. 
The physics results presented in detail in Sections~\ref{sec:dch}--\ref{sec:pdffit} were part of the work for the thesis, 
and later on most of them were published~\cite{zeusdch_hera2,dstarcombpaper,Zenaiev:2015rfa}. 

%
\clearpage
\section{Theoretical overview of heavy-flavour production in QCD}
\label{sec:th}

Section~\ref{sec:th:qcd:pqcd} gives a short introduction to perturbative calculations in QCD.
Section~\ref{sec:th:hq:treat} discusses \ozmod{ways of treating} of heavy-quark production and 
\ozmod{focuses on the fixed-flavour-number scheme, \ozmodN{the preferred scheme in this review}}.
In Section~\ref{sec:th:qcd:pqcd:mass} \ozmodN{various} defintions of the heavy-quark mass are given.
Sections~\ref{sec:th:hq:ep} and \ref{sec:th:hq:pp} are the central part of this theoretical overview: 
they provide information on the current status of 
the calculations for heavy-quark production in different schemes in \ep and \pp collisions, respectively. 
Section~\ref{sec:th:hq:frag} reviews an important non-perturbative aspect of heavy-flavour production:
the fragmentation process of partons into hadrons.
Finally, Section~\ref{sec:th:hq:outlook} gives a summary.

\subsection{Perturbative calculations}
\label{sec:th:qcd:pqcd}

\begin{figure}[htbp]
  \centering
  \includegraphics[width=1.00\figwidth,trim=0 0 0 0mm,clip=true]{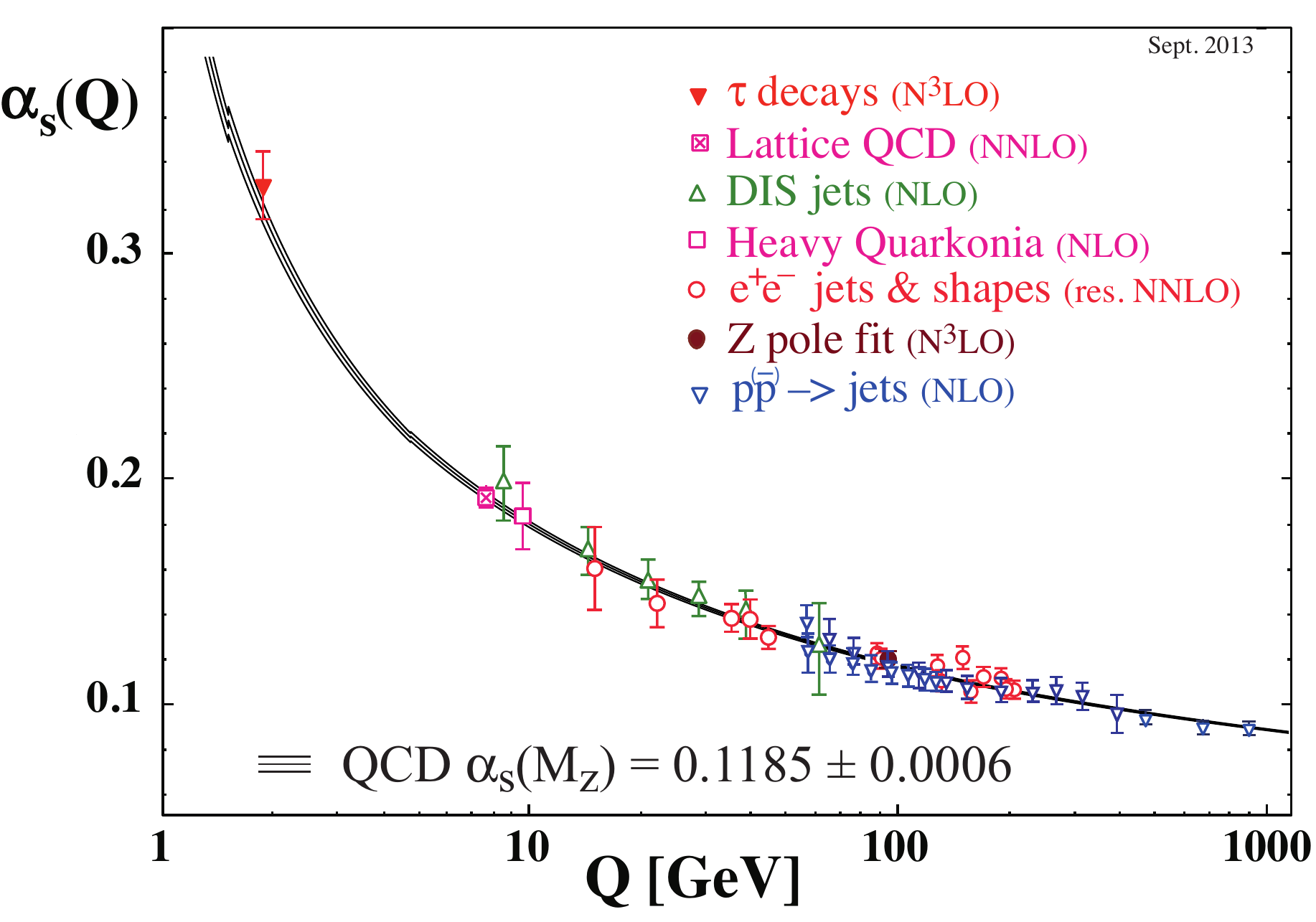}
  \caption[Summary of measurements of $\alpha_s$ as function of energy scale]
  {Summary of measurements of $\alpha_s$ as a function of the energy scale $Q$. 
	The respective order of pQCD used in the extraction of $\alpha_s$ is
	indicated in brackets. The plot was taken from~\cite{pdg2014}.}
	\label{fig:th:as}
\end{figure}

In the approach of perturbative QCD (pQCD), any physical quantity, $\Gamma$, is given as a power series in the strong coupling constant, 
$\alpha_s=g^2/4\pi$, where $g$ is the constant representing the coupling strength in the QCD Lagrangian:
\begin{equation}
	\Gamma = \sum_{i=0}^n c_i \alpha_s^i,
\label{th:pert}
\end{equation}
where $n$ is the order of the calculation and the coefficients $c_i$ are determined using the Feynman rules. 
Contributions to the perturbative expansion of scattering amplitudes beyond the leading order (LO) are usually formally divergent. 
In order to \ozmod{regularise} these divergences, different \emph{renormalisation schemes} exist. 
Moreover, in subtracting the divergences in any renormalisation scheme, an arbitrary mass scale is introduced, known as the \emph{renormalisation scale}, $\mu_r$.
Most commonly the modified minimal subtraction scheme, \msbar, is used~\cite{Bardeen:1978yd}. 
The renormalised coupling, $g_r$, turns out to be scale dependent; 
keeping only the one-loop order, the running coupling is given by 
\begin{equation}
	g_r^2=\frac{1}{\beta_0 \ln(\frac{\mu_r^2}{\Lambda_{\rm QCD}^2})},
\label{th:run}
\end{equation}
where \ozmod{the constant of integration $\Lambda_{\rm QCD}$ is a dimensionful quantity, known as the \emph{QCD scale}, 
$\beta_0=(33-2n_f)/(48\pi^2)$ is the one-loop beta-function coefficient with $n_f$ being the number of massless quark flavours.}
\ozmod{The strong coupling can be determined \ozmodN{through} experimental observables, 
e.g.\ jet production cross sections, event shapes, $\tau$ decay width etc.} 
The measurements of $\alpha_s$ as a function of the energy scale 
are shown in Fig.~\ref{fig:th:as}~\cite{pdg2014}. 
The running of $\alpha_s$ \ozmod{agrees with} the expectation from pQCD. 
The renormalised coupling \ozmod{decreases} as the relevant momentum scale grows. 
This behaviour is known as \emph{asymptotic freedom}; it \ozmodN{enables} perturbative calculations \ozmodN{at} large momentum scales (short distances). 
On the other hand, \ozmod{the perturbative approach breaks down at $\Lambda_{\rm QCD}$ (long distances) as the coupling \ozmodN{gets too large}.} 
\ozmodN{This phenomenon is known as \emph{confinement}.} 
Quarks and gluons are not observed as free particles, because, with increasing distance between them, 
\ozmodN{the production of a new quark-antiquark pair instead is energetically preferred.}

Because of confinement hadrons are considered to be made up of massless constituents, known as \emph{partons}, held together by their mutual interactions. 
Application of perturbative calculations to any process involving hadrons requires \emph{factorisation} of short- and long-distance effects~\cite{Collins:1989gx}. 
To define the separation, an arbitrary mass scale appears, known as the \emph{factorisation scale}, $\mu_f$. 
It is introduced in a \ozmod{way} similar to the way the renormalisation scale $\mu_r$ appears in renormalisation, although serves different purposes. 
In the factorisation approach hadrons are described by \emph{parton distribution functions} (PDFs), which are not perturbatively calculable and must be extracted from data, 
however pQCD predicts their evolution with $\mu_f$ (see Section~\ref{sec:th:qcd:factoris} for more details).

Thus the region where pQCD calculations are reliable is given by $\mu_r,\mu_f \gg \Lambda_{\rm QCD}$. 
One usually chooses the two scales to be of the order of the energy involved in the hard process; 
\ozmod{e.g.\ for the inclusive production of heavy quarks $\mu_r^2=\mu_f^2= m_Q^2$ is a possible choice, 
where $m_Q$ denotes the heavy-quark mass (see e.g.~\cite{Behnke:2015qja} for an exhaustive discussion).}

\subsection{Treatment of heavy flavours in pQCD}
\label{sec:th:hq:treat}

The masses of the heavy quarks \ozmodNN{satisfy} $m_Q\gg\Lambda_{\rm QCD}$ 
($m_c \approx \SI{1.5}{GeV}$, $m_b \approx \SI{4.5}{GeV}$, $m_t \approx \SI{170}{GeV}$) \ozmodNN{and then} provide a hard scale 
for pQCD calculations. \ozmodNN{At the same time} they complicate calculations, since \ozmodNN{the} new hard scale leads to the appearance of terms proportional to $\ln(\frac{p_T^2}{m_Q^2})$, 
\ozmodN{where $p_T$ is the transverse momentum of the produced heavy quark}, 
known as the \emph{multi-scale problem}. 
\ozmodNN{One has a freedom to treat the heavy quarks \ozmodN{either} as massive or massless in perturbative calculations; 
both choices have their advantages and disadvantages at different phase-space regions\ozmod{, as discussed below}.
The PDF evolution and $\alpha_s$ running depend on the number of quark flavours assumed to be massless and appearing in loops and legs.
Several schemes exist for the treatment of heavy flavours in pQCD.} 

In the present review in most cases the \emph{fixed-flavour-number scheme} (FFNS) 
is used in comparisons of theory to the data, 
since it provides most reliable predictions in the phase space of existing experimental data. 
In this scheme, heavy quarks are treated as massive at all energy scales, 
\ozmod{thus they do not enter the PDF evolution of massless quarks and gluons and $\alpha_s$ running}.%
\footnote{Note that in some variants of the FFNS, heavy quarks contribute to the loops in the PDF evolution and $\alpha_s$ running (see, e.g.~\cite{Gluck:1993dpa,Kuprash:2014nfa}); 
sometimes these variants are called the \emph{mixed-flavour-number scheme}~\cite{qcdnum}.}
One has to specify which particular quark \ozmod{flavours} are treated as massless, 
\ozmodNN{e.g.\ the number of flavours $n_f = 3$ for massless $u$, $d$ and $s$.}
The FFNS is expected to be most precise in the threshold region 
$p_T^2 \sim m_Q^2$, 
while at high 
$p_T$ 
terms proportional to 
$\ln(\frac{p_T^2}{m_Q^2})$ may spoil the convergence of the perturbative series.

Other schemes are known as variants of the \emph{variable-flavour-number scheme} (VFNS), 
in which heavy quarks are treated as massive or massless depending on the energy scale:
\begin{itemize}
	\item in the \emph{zero-mass variable-flavour-number scheme} (ZM-VFNS)~\cite{Collins:1986mp}, heavy flavours are treated as infinitely massive 
	(and thus \ozmodNN{completely} vanishing) below a certain threshold and as massless above it. 
	This scheme is expected to be appropriate at high energy scales, since the PDF evolution of the heavy quarks 
	and the renormalisation of collinear and infrared singularities	provides 
	a resummation of terms proportional to $\ln \frac{p_T^2}{m_Q^2}$;
	\item in the \emph{general-mass variable-flavour-number scheme} (GM-VFNS), an interpolation is made between the FFNS and the ZM-VFNS, avoiding double counting of common terms 
	in the PDF evolution and coefficient functions. This scheme is expected to combine the advantages of the FFNS and ZM-VFNS, although some level of arbitrariness 
	is unavoidably introduced in the treatment of the interpolation. Therefore, different variants of the GM-VFNS are 
	available~\cite{Aivazis:1993kh,Aivazis:1993pi,Buza:1995ie,Collins:1998rz,Kramer:2000hn,Tung:2001mv,Thorne:2006qt,Forte:2010ta}.
	Moreover, this arbitrariness prevents a clear interpretation of the heavy-quark masses in terms of a specific scheme; 
	therefore the heavy-quark masses in GM-VFNS \ozmod{must} be treated as effective mass parameters.\ozmodN{%
  \footnote{Although at a certain order of pQCD a VFNS can be converted to use other heavy-quark mass definition, 
  for example see~\cite{Bertone:2016pbr} for FONLL structure functions with \msbar running masses.}}
\end{itemize}
In the context of VFNS many non-perturbative models, particularly those based on the light-cone wave-function picture, expect
an ``intrinsic charm'' component of the nucleon at an energy scale comparable to the $c$-quark mass. 
This intrinsic-charm component, if present at a low-energy scale, will participate
fully in QCD dynamics and evolve along with the other partons as the energy scale
increases; for more details see, e.g.~\cite{Pumplin:2007wg} and references therein. 
Such models predict a sizeable intrinsic-charm contribution to heavy-flavour production, 
\ozmod{but} in the phase-space regions which are difficult to be probed with currently available experimental data~\cite{Lykasov:2012hf} 
(see Section~\ref{sec:exp:lhcb:charm}). 
\ozmod{In the recent analysis~\cite{Ball:2016neh} some evidence was found that the \ozmodNN{intrinsic} charm PDF at large parton momentum and low-energy scale 
carries about 1\% of the total momentum of the proton. 
Future LHC data are expected to further constrain \ozmodNN{a} possible intrinsic-charm component of the proton.}

\subsection{Quark masses}
\label{sec:th:qcd:pqcd:mass}

Since free quarks are unobservable, one can \ozmod{consider} different definitions of the quark mass $m_Q$. 
One of the most popular choices is the pole quark mass, $m_Q^{\rm pole}$, defined as \ozmodNN{the mass at} the position of the pole in the quark
propagator in perturbation theory. This quantity is introduced in a gauge invariant way and 
is well defined in each finite order of perturbation theory. 
This convenient feature has made it very popular and widely used in perturbative calculations, 
although it has an important drawback: any definition of this quantity suffers from an 
intrinsic uncertainty of order $\frac{\Lambda_{\rm QCD}}{m_Q}$.
The problem arises for the reason, that the pole mass is sensitive to large-distance dynamics (infrared contributions).%
\footnote{In other words, the pole mass is unobservable, because of confinement no free colored quarks exist. 
Perturbation theory itself produces clear evidence for this non-perturbative correction to $m_Q^{\rm pole}$: 
the signal is the peculiar factorial growth of the high-order terms in the $\alpha_s$ expansion corresponding to a renormalon; 
for more details see, e.g.~\cite{Bigi:1994em} and references therein.}

Alternative mass definitions \ozmod{avoid} this problem. The most prominent example is
the \msbar mass, $m_Q(\mu_r)$, which is to be evaluated at the renormalisation scale $\mu_r$, where $\mu_r\gg\Lambda_{\rm QCD}$,
and which is free of ambiguities of order $\Lambda_{\rm QCD}$. 
One benefit of theoretical predictions using the \msbar mass is improved stability of
the perturbative series with respect to scale variations as compared to the result in the pole-mass scheme~\cite{Alekhin:2010sv}. 
The scale dependence of the running mass at LO is given by
\begin{equation}
	m_Q(\mu_r)=m_Q(m_Q)\left(1-\frac{\alpha_s(\mu_r)}{\pi}\ln \frac{\mu_r^2}{m_Q^2}\right).
\end{equation}
Here $m_Q(m_Q)$ denotes the \msbar running mass evaluated at the scale $\mu_r=m_Q$.

\begin{figure*}[htbp]
  \sidecaption
  \centering
  \includegraphics[width=0.80\figwidth,trim=0 0 20mm 0mm,clip=true]{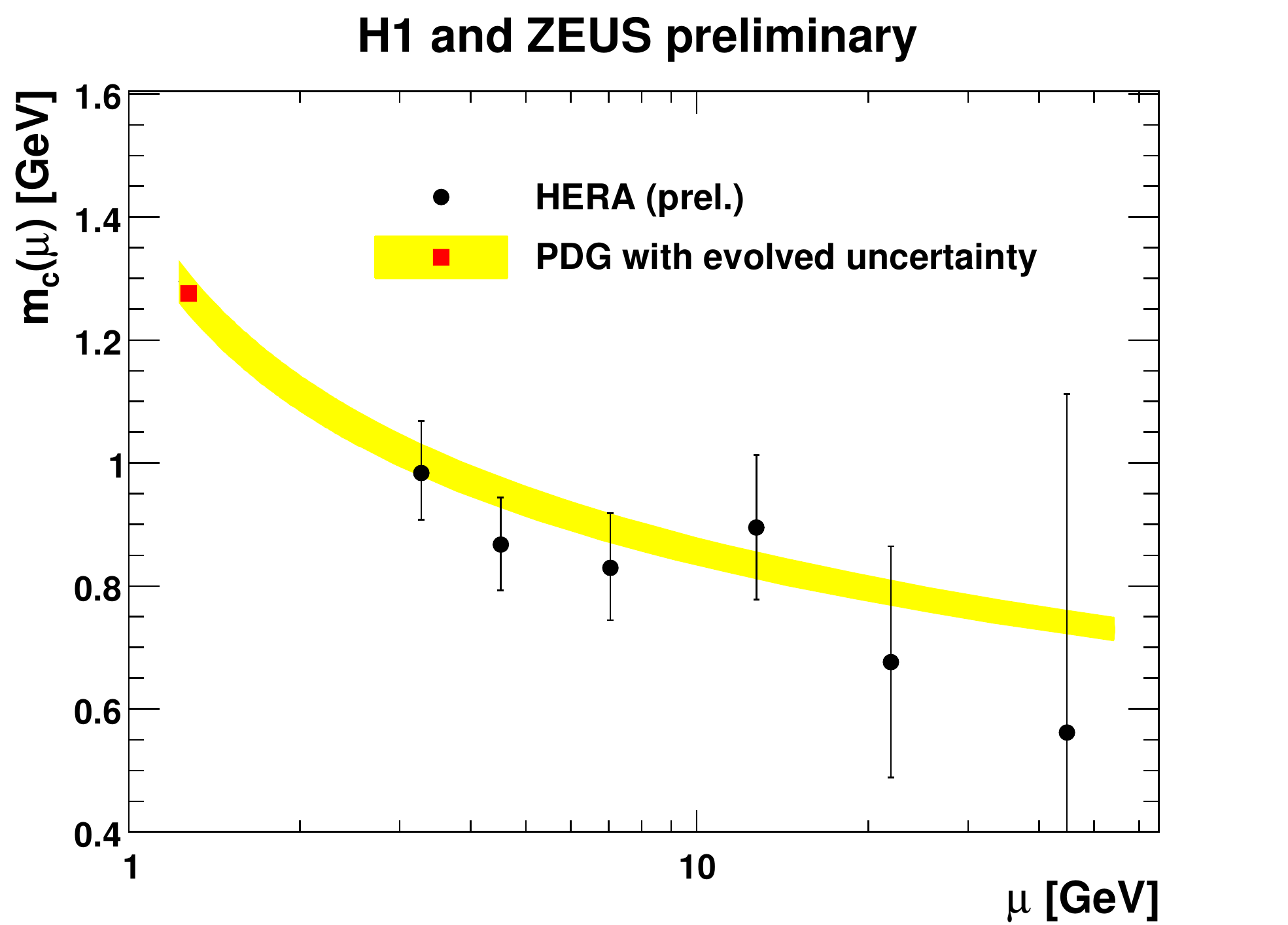}
  \includegraphics[width=0.80\figwidth,trim=0 0 20mm 0mm,clip=true]{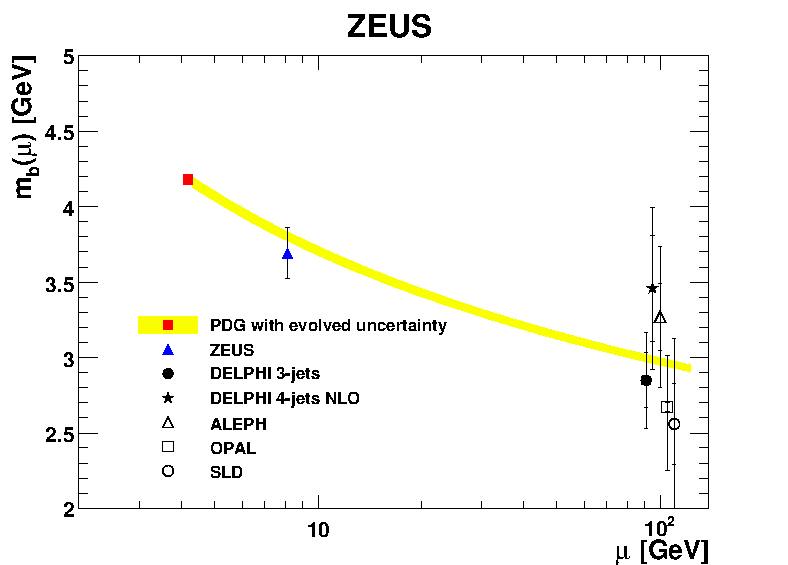}
  \caption[Measurements of charm and beauty running masses]
  {Measurements of the charm (left) and beauty (right) \msbar running masses as a function of the energy scale $\mu$~\cite{Abdallah:2008ac,zeussecvtx_hera2,mcrunprel}.}
	\label{fig:th:mqrun}
\end{figure*}

The scale dependence of the charm and beauty running masses has been measured 
at LEP and HERA%
\footnote{\ozmod{As of November 2016} only \emph{preliminary} data from HERA on charm-mass running is available, 
i.e.\ only the most important figures have been publicly released by the H1 and ZEUS Collaborations, 
while the \ozmodN{complete} publication of the results is expected soon.}~\cite{Abdallah:2008ac,zeussecvtx_hera2,mcrunprel} 
and is shown in Fig.~\ref{fig:th:mqrun}. 
It is found to be consistent with the QCD expectation.

The relation between the pole mass $m_Q^{\rm pole}$ and the \msbar running mass $m_Q(m_Q)$ 
is known to three loops~\cite{Gray:1990yh,Broadhurst:1991fy,Chetyrkin:1999ys,Melnikov:2000qh}; 
at one-loop order it is given by
\begin{equation}
	m_Q^{\rm pole}=m_Q(m_Q)\left(1+\frac{4\alpha_s(m_Q)}{3\pi}\right).
\label{eq:th:runpolmass}
\end{equation}

\subsection{Heavy-quark production in \ep collisions}
\label{sec:th:hq:ep}

Heavy-quark production in deep inelastic \ep scattering collisions serves as a test of pQCD \ozmod{(see Sections~\ref{sec:dch} and~\ref{sec:comb})}; 
moreover, it is directly sensitive to the gluon density of the proton and to the heavy-quark masses \ozmod{(see Section~\ref{sec:comb:red:ffns})}. 
The charm contribution to the inclusive cross section at HERA \ozmod{reaches $30\%$}~\cite{zd9697}, 
thus necessitating its understanding for any global QCD analysis based on HERA data. 
\ozmod{The contents of this Section is partially based on~\cite{vanNeerven:2001tb}, where more details can be found.}

\subsubsection{Kinematics of \ep collisions and heavy-flavour structure functions}
\label{sec:th:dis:kin}

\begin{figure}[htbp]
\centering
\begin{fmffile}{ep}
\begin{fmfgraph*}(180,120)
	\fmfpen{thin}
	\fmfleft{i1,i2}
	\fmfright{o1,o2}
	\fmf{fermion,label=$e({k})$,label.side=left}{i2,v2}
	\fmf{fermion,label=$l^{\prime}({k^{\prime}})$,label.side=left}{v2,o2}
	\fmf{photon,label=$V^{*}({q})$,label.side=right}{v1,v2}
	\fmf{quark,label=$P({p})$,label.side=left,label.dist=0.075h}{i1,v1}
	\fmf{quark,label=$X({p^{\prime}})$,label.side=left,label.dist=0.075h}{v1,o1}
	\fmfblob{.15w}{v1}
	\fmfdot{v2}
	\fmffreeze
	\fmfshift{(0,-0.10w)}{v1}
	\fmfshift{(0,0.07w)}{i1}
	\fmfshift{(0,0.07w)}{o1}
	\fmfshift{(-0.02,0)}{o1}
	\fmfi{plain}{vpath (__v1,__o1) shifted(thin*(0,4))}
	\fmfi{plain}{vpath (__v1,__o1) shifted(thin*(1,-4))}
\end{fmfgraph*}
\end{fmffile}
\caption[Schematic diagram of \ep scattering]
{Schematic diagram of \ep scattering.}
\label{fig:th:ep}
\end{figure}
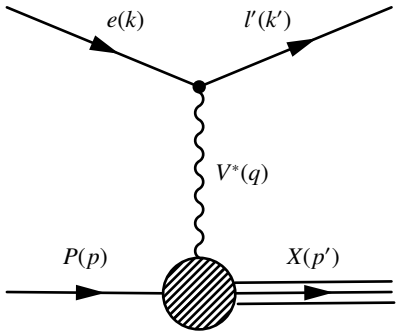

The generic electron--proton%
\footnote{Both electrons and positrons are referred to as electrons, unless explicitly stated otherwise.} 
scattering process, $ep \to l^{\prime}X$, where $l^{\prime}$ is the scattered lepton and $X$ is the hadronic final state, is shown in Fig.~\ref{fig:th:ep}. 
It occurs via the exchange of an electroweak boson $V^{*}$ 
(the superscript ${}^{*}$ denotes \ozmod{an intermediate vector boson}.) 
of two types:
\begin{itemize}
	\item a neutral $\gamma$ or $Z^{0}$ boson; these reactions are called \emph{neutral current} (NC);
	\item a charged $W^{\pm}$ boson; these reactions are called \emph{charged current} (CC).
\end{itemize}

Denoting the incoming electron and proton four-momenta with ${k}$ and ${p}$, respectively, and the 
scattered-lepton four-momentum with ${k^{\prime}}$, the event kinematics can be described by the following Lorentz invariant variables:
\begin{equation}
\begin{aligned}
	Q^2&=-q^2=(k-k^{\prime})^2,\\
	W^2&=(p+q)^2,\\
	y&=\frac{p\cdot q}{p\cdot k},\\
	x&=\frac{Q^2}{2p\cdot q}.\\
\end{aligned}
\end{equation}
\ozmodN{$Q^2$ is the virtuality of the exchanged boson, $W^2$ is the boson--proton energy squared, $x$ and $y$ are Bjorken scaling variables. 
The $y$ variable is also referred to as inelasticity. 
The variables $x$, $y$ and $Q^2$ are related by 
\begin{equation}
	Q^2 = s x y,
\end{equation}
with $s = 2 k \cdot p \approx s_{\rm tot}$ approximately equals to the centre-of-mass energy $s_{\rm tot}$  of the experiment.}

The virtuality $Q^2$ can be interpreted as the power with which the exchanged boson can resolve the proton structure. 
Depending on $Q^2$, the \ep scattering phase space is divided into two regions:
\begin{itemize}
	\item \emph{deep inelastic scattering} (DIS), if $Q^2 \gtrsim \SI{1}{GeV^2}$;
	\item \emph{photoproduction} (PHP), if $Q^2 \approx \SI{0}{GeV^2}$.
\end{itemize}

The inelasticity $y$ defines the relative fraction of the electron energy transferred to the hadronic system in the proton rest frame, 
while the Bjorken variable $x$ determines the relative fraction of the proton energy involved in the \ozmodNN{partonic subprocess}. 
More details on \ep scattering physics, 
including description of the quark-parton model (QPM), can be found \ozmodN{for instance} in~\cite{sf}. 

\subsubsection{Factorisation approach}
\label{sec:th:qcd:factoris}
\ozmod{The inclusive differential cross section of heavy-flavour production in DIS, $\frac{\dif^2\sigma^{Q\bar{Q}}}{\dif x \dif Q^2}$, 
where $Q\bar{Q}$ stands for $c$ or $b$ quark-antiquark pairs (top production is not accessible at HERA), 
is expressed in terms of the dimensionless \emph{reduced cross sections}:
\begin{equation}
\sigma_{\rm red}^{c\bar{c}}=\frac{{\rm d}^2\sigma^{c\bar{c}}}{{\rm d}x {\rm d}Q^2} \cdot \frac{xQ^4}{2\pi\alpha^2\,(1+(1-y)^2)},
\label{eq:th:hqcsred}
\end{equation}
where $\alpha$ is the running electromagnetic coupling. 
The reduced cross sections can be \ozmodNN{expressed} in terms of the heavy-flavour \emph{structure functions} $F_2^{Q\bar{Q}}$, $F_L^{Q\bar{Q}}$:
\begin{equation}
\sigma_{\rm red}^{c\bar{c}}=F_2^{c\bar{c}}-\frac{y^2}{1+(1-y)^2}F^{c\bar{c}}_L.
\label{eq:th:hqcssf}
\end{equation}
Here the term proportional to the parity-violating structure function, $xF_3$, is neglected since $Q^2 \ll M^2_Z$, 
where $M_Z$ is the $Z^0$-boson mass. Conventionally the structure functions $F_2^{Q\bar{Q}}$, $F_L^{Q\bar{Q}}$ are defined at the 
Born level without QED and electroweak radiative corrections, except for the running electromagnetic coupling $\alpha=\alpha(Q^2)$.}
The heavy-flavour structure functions are \ozmodN{predicted in the FFNS using light-flavour PDFs as input and} 
the factorisation approach, 
which gives the field-theory realisation of the parton model in the form of the theorem of the separation of the 
long-distance from the short-distance dependence for DIS~\cite{Collins:1989gx}. 
This theorem states that the sum of all the diagrammatic contributions to the 
structure functions is a direct generalisation of the parton-model results, 
given by
\begin{equation}
	\ozmod{F^{Q\bar{Q}}_{2,L}(x,Q^2)=\sum_{i} \int_{x/x_{\rm max}}^{1} {\dif \xi} C^{Q\bar{Q}~i}_{2,L} \left(\frac{x}{\xi},Q^2, \mu_r^2,\mu_f^2\right) f_i({\xi},\mu_f^2),}
\label{eq:th:f2factoris}
\end{equation}
where $i$ denotes the sum over all partons (massless quarks, antiquarks and gluons), $\xi$ is the momentum fraction of the parton $i$, 
which goes from $x / x_{\rm max}$ to $1$, $x_{\rm max} = \frac{1}{1 + 4M^2/Q^2}$, 
$M = m_Q^{\rm pole}$ is the heavy-quark pole mass, 
$C^{Q\bar{Q}~i}_{2,L}$ are the heavy-flavour coefficient functions (known also as the hard-scattering functions, or Wilson coefficients, or matrix elements) and 
$f_i$ are the massless PDFs.
The factorisation scale $\mu_f$ serves to define the separation of short-distance from long-distance effects: 
any propagator that is off-shell by $\mu_f^2$ or more will contribute to $C^{Q\bar{Q}~i}_{2,L}$, while below this scale it will be \ozmodNN{absorbed} into $f_i$. 
\ozmod{Note that the left-hand side of Eq.~\ref{eq:th:f2factoris}, which is an observable quantity, 
does not depend on arbitrary scales $\mu_r$ and $\mu_f$ by definition. 
This is a requirement of the factorisation theorem. 
However, if the right-hand side of Eq.~\ref{eq:th:f2factoris} is expressed 
as a perturbative series truncated at a certain order, the calculated value for the observable turns out to be scale-dependent \ozmodNN{due to} the neglected orders. 
For the \ozmodNN{actual} calculation of $F^{Q\bar{Q}}_{2,L}(x,Q^2)$ one sets the two scales to some fixed values 
and varies \ozmodNN{them} within a certain range to estimate \ozmodNN{the effect of} the missing higher-order corrections.}
The $C^{Q\bar{Q}~i}_{2,L}$ are calculated in perturbation theory (see Section~\ref{sec:th:hq:ep:ffns}) 
but the $f_i$ must be extracted by comparing Eq.~\ref{eq:th:f2factoris} to 
some standard set of cross sections. 

\ozmodNN{The factorisation prescription} \ozmod{is not unique and allows different choices}. A set of rules that \ozmodNN{defines} these choices
is called a \emph{factorisation scheme}. 
\ozmodN{Common factorisation schemes are \msbar~\cite{Bardeen:1978yd} or DIS~\cite{Diemoz:1987xu,Allaby:1987vs}.} 
Within such a scheme the PDFs have no \ozmod{physical} meaning, 
since they are dominated by infrared effects and thus by infrared parameters that cannot be measured, 
although they can be extracted from data by comparing the theoretical calculation~\ref{eq:th:f2factoris} with measured cross sections. 
The factorisation theorem ensures that the hard-scattering functions determined in this calculation
are insensitive to infrared scales and parameters, and are applicable to cross sections 
calculated with phenomenologically determined PDFs.

A remarkable consequence of factorisation is that measuring PDFs for one scale ${\mu_f}_1$ 
allows their prediction for any other scale ${\mu_f}_2$, as long as both ${\mu_f}_1$ and ${\mu_f}_2$ 
are large enough \ozmod{which means} both $\alpha_s({\mu_f}_1)$ and $\alpha_s({\mu_f}_2)$ are small. 
The evolution of PDFs in $\mu_f$ is most often, and most conveniently, \ozmodN{described in
terms of the integro-differential \emph{Dokshitzer-Gribov-Lipatov-Altarelli-Parisi (DGLAP) 
equations}~\cite{dglap_Gribov:1972ri,Dokshitzer:1977sg,dglap_Altarelli:1977zs,dglap_Curci:1980uw,dglap_Furmanski:1980cm,dglap_Moch:2004pa,dglap_Vogt:2004mw}:} 

\begin{equation}
	\mu_f^2 \frac{\dif {}}{\dif \mu_f^2} f_i(x,\mu_f^2) = \sum_{j} \int_x^1 \frac{\dif \xi}{\xi} P_{ij}\left(\frac{x}{\xi},\alpha_s\right) f_i(\xi,\mu_f^2).
\label{eq:th:evol}
\end{equation}
The \emph{evolution kernels} $P_{ij}(x)$, or the \emph{splitting functions}, are given by perturbative expansions, beginning with $O(\alpha_s)$; 
they represent the probability of a parton $i$ to emit a parton $j$ carrying a fraction $z=\frac{x}{\xi}$ of the momentum of the parton $i$. 

Note that the integral on the right-hand side of Eq.~\ref{eq:th:evol} begins at $x$. 
\ozmod{Thus, it is only necessary to know $f_i(\xi,{\mu_f}_1^2)$ for $\xi > x$ at some starting value of the scale ${\mu_f}_1$, 
in order to derive $f_i(x,{\mu_f}_2^2)$ at a higher value ${\mu_f}_2 > {\mu_f}_1$.} 
This is a great simplification, since data at small $x$ are hard to \ozmod{obtain} at moderate energies.

At very low values of $x$, terms proportional to $\alpha_s \ln (\frac{1}{x})$ may spoil the accuracy of the DGLAP approach; 
there other evolution schemes, e.g.\ BFKL~\cite{Kuraev:1976ge,Kuraev:1977fs,Balitsky:1978ic} or CCFM~\cite{Ciafaloni:1987ur,Catani:1989yc,Catani:1989sg,Marchesini:1994wr}, might be more appropriate to use. 
The difference between the schemes comes from the ordering of the emitted partons before entering the hard-scattering process.

\ozmodN{Since} the perturbative series is truncated at a certain order, the approximation is $\mu_f$ dependent \ozmodNN{due to} neglected orders. 
In practice the two scales are often set to be equal, although it is not a requirement. 
To estimate the perturbative uncertainties of the neglected higher orders, the $\mu_r$ and $\mu_f$ scales are varied 
around the central values, simultaneously or independently.

\subsubsection{Calculations in FFNS}
\label{sec:th:hq:ep:ffns}

\begin{figure}[htbp]
\centering
\begin{fmffile}{bgf}
\begin{fmfgraph*}(160,150)
	\fmfpen{thin}
	\fmfleft{i1,i2}
	\fmfright{o1,o2,o3,o4}
	\fmf{fermion,label=$e$,label.side=left}{i2,v4}
	\fmf{fermion,label=$e^{\prime}$,label.side=left}{v4,o4}
	\fmf{photon,label=$\gamma^{*}$}{v4,v3}
	\fmf{quark,label=$\bar{Q}$,label.side=left}{o2,v2}
	\fmf{quark,label=$Q$}{v2,v3}
	\fmf{quark,label=$Q$,label.side=left}{v3,o3}
	\fmf{gluon,label=$g$,label.dist=0.06h}{v1,v2}
	\fmf{fermion,label=$P$,label.side=right}{i1,v1}
	\fmf{fermion,label=$ $}{v1,o1}
	\fmfblob{.15w}{v1}
	\fmfdot{v4,v3,v2}
	\fmffreeze
	\fmfshift{(-0.15w,0)}{v1}
	\fmfshift{(-0.15w,0)}{v2}
	\fmfshift{(-0.15w,0)}{v3}
	\fmfshift{(-0.15w,0)}{v4}
	\fmfi{plain}{vpath (__v1,__o1) shifted(thin*(0,4))}
\end{fmfgraph*}
\end{fmffile}
\caption[BGF diagram]
{The BGF diagram.}
\label{fig:th:hq:ep:bgf}
\end{figure}
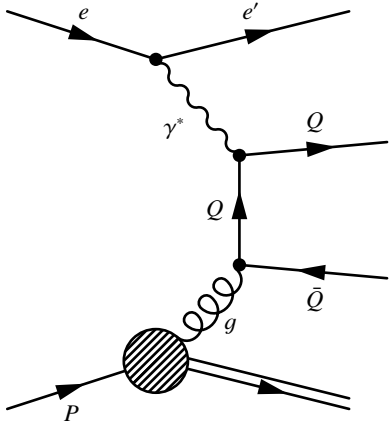

In the FFNS \ozmod{with $n_f=3$} there are no $c$ and $b$ quarks in the proton at any scale, 
therefore the leading-order (LO) process (${\cal O}(\alpha_s)$) for heavy-flavour production in DIS is the boson--gluon-fusion (BGF) process~\cite{Witten:1975bh,Babcock:1977fi,Novikov:1977yc,Leveille:1978px,Gluck:1979aw}, $g\gamma^{*} \to Q\bar{Q}$, 
shown in Fig.~\ref{fig:th:hq:ep:bgf}. The corresponding hard-scattering functions $C^{g\gamma^{*} \to Q\bar{Q}~(0)}_{2,L}$ are given by:
\begin{equation}
\begin{split}
	C^{g\gamma^{*} \to Q\bar{Q}~(0)}_{2} &= e^2_c \frac{\alpha_s(\mu_r^2)}{\pi} \{ \nu [4z^2(1-z)-z/2-2\epsilon z^2(1-z)] \\
	                     &+ [z/2-z^2(1-z)+2\epsilon z^2(1-3z) -4\epsilon^2 z^3){\rm ln}\frac{1+\nu}{1-\nu} \},\\
	C^{g\gamma^{*} \to Q\bar{Q}~(0)}_{L} &= e^2_c \frac{\alpha_s(\mu_r^2)}{\pi} \left[ 2z^2(1-z)\nu - 4\epsilon^2 z^3{\rm ln}\frac{1+\nu}{1-\nu} \right],
\end{split}
\label{eq:th:c20}
\end{equation}
\ozmod{where $e_c = +2/3$ is the $c$-quark charge in units of the proton charge,} 
$\epsilon = \frac{{M}^2}{Q^2}$, $M = m_Q^{\rm pole}$ is the heavy-quark pole mass, $\nu = \sqrt{1 - 4\epsilon \frac{z} {1-z}}$ 
\ozmodNN{and $z \equiv \frac{x}{\xi}$ running from $x$ to $\frac{1}{1+4\epsilon}$}. 
\ozmod{Note that the LO hard-scattering functions in Eq.~\ref{eq:th:c20} do not depend on the factorisation and renormalisation scales 
(the dependence on the latter appears only via the strong coupling), therefore the structure functions in Eq.~\ref{eq:th:f2factoris}, 
calculated at LO, necessarily turn out to be dependent on these arbitrary scales. 
For higher-order calculations this dependence partially cancels in the convolution of the hard-scattering functions, PDFs and running strong coupling. 
The two scales are chosen to be of the order of $Q^2$ or $M^2$; 
\ozmodNN{a typical choice is} $\mu_r^2=\mu_f^2= Q^2 + 4M^2$~\cite{heracharmcomb}.
}

Next-to-leading-order (NLO) corrections ($O(\alpha_s^2)$) were calculated in~\cite{Laenen:1992zk,Laenen:1992xs}.
They can be classified into three groups:
\begin{enumerate}
	\item real corrections to the BGF process, i.e.\ all processes containing an extra gluon in the final state $g\gamma^{*} \to Q\bar{Q}g$;
	\item virtual corrections to the BGF process coming from the interference of $O(\alpha_s)$ and $O(\alpha_s^3)$ terms;
	\item a new process, when the virtual photon interacts with a light quark $q$ in the proton: $\gamma^{*}q(\bar{q}) \to Q\bar{Q}q(\bar{q})$.
\end{enumerate}
\ozmodNN{The NLO predictions~\cite{Laenen:1992zk,Laenen:1992xs} are available in the HVQDIS program~\cite{hvqdis}, 
which calculates fully differential double-particle inclusive cross sections.} 
The pole-mass definition is used in these calculations. 
\ozmod{The NLO corrections for charm production are important as they change both the shape 
and normalisation of the transverse momentum, pseudorapidity, and $x$ distributions, 
while the $Q^2$ distribution only receives a shift in normalisation~\cite{hvqdis}. 
In the kinematic region of HERA, the scale dependence of the NLO calculations for charm production is moderate:
it varies from $10\%$ at high $Q^2$ to $30\%$ at low $Q^2$ (see Sections~\ref{sec:dch:res}, \ref{sec:comb:dstar:single} and~\ref{sec:comb:red:ffns}).}

In a recent variant of the FFNS from the ABM group the running-mass definition in the \msbar scheme is used~\cite{Alekhin:2010sv}. 
This scheme has the advantage of improving the convergence of the perturbative series (see Section~\ref{sec:th:qcd:pqcd:mass}). 
These predictions are provided for inclusive quantities only, i.e.\ at the $F_{2,L}^{Q\bar{Q}}$ level.

At next-to-next-to-leading order (NNLO) ($O(\alpha_s^3)$) only approximate calculations are available. 
For $F_2$, four out of the five massive Wilson coefficients are known 
at large scales $Q^2$~\cite{Ablinger:2014nga,Behring:2014eya,Ablinger:2010ty,Blumlein:2012vq,Ablinger:2014vwa}, 
and an estimate has been made for the remaining coefficient~\cite{Kawamura:2012cr} based on the anticipated small-$x$ behavior, 
a series of moments~\cite{Bierenbaum:2009mv}, and two-loop operator matrix elements~\cite{Bierenbaum:2007qe,Bierenbaum:2008yu}. 
E.g.\ in Ref.~\cite{Kawamura:2012cr} 
combined approximate expressions for 
three kinematic limits are given: in the limit of high partonic centre-of-mass energy squared, $\hat{s} \gg m_Q^2$, 
in the threshold region, $\hat{s} \gtrsim 4m_Q^2$, 
and in the high-scale region $Q^2 \gg m_Q^2$.

\subsubsection{Calculations in VFNS}
\label{sec:th:hq:ep:vfns}

In the VFNS, the LO process for heavy-flavour production in \ep collisions is the QPM scattering. 
At NLO, fully differential calculations exist only in the ZM-VFNS\ozmodN{~\cite{Heinrich:2004kj,Sandoval:2009dw,Sandoval:2009zz}}. 

The main difference between the FFNS and ZM-VFNS mechanisms can be attributed to the fact that 
for heavy-quark production in the FFNS, two heavy particles appear in the final state instead
of one as in the case of the intrinsic heavy-quark approach%
\footnote{In this case the other heavy quark belongs to the proton remnant and thus is effectively integrated over.}. 
This reveals itself in the $p_T$-distribution where for the FFNS the quark and antiquark appear back to
back in the Breit frame. 
The heavy-flavour data from HERA~\cite{h1dstarhighQ2,h1dstar_hera2} clearly \ozmod{prefer} the $p_T$-spectrum predicted by the FFNS production mechanism 
\ozmodN{(see Fig.~\ref{fig:hera:hfmeas:dstar:h1cs} in Section~\ref{sec:exp:hera:hfmeas:dstar})}.

Calculations in the GM-VFNS for heavy-flavour production in DIS exist only at the inclusive $F_2^{Q\bar{Q}}$ level. 
Some of the most popular GM-VFNS are the Thorne-Roberts (RT)~\cite{Thorne:2006qt,Thorne:1997ga,Thorne:2012az}, 
Aivazis-Collins-Olness-Tung (ACOT)~\cite{Aivazis:1993pi} and FONLL~\cite{Forte:2010ta} schemes.
The calculations are available at NLO and (approximate) NNLO orders. 
Predictions from various variants of GM-VFNS were compared to the combined HERA charm data in~\cite{heracharmcomb};
they are generally found to describe the data well in the region $Q^2 \gtrsim \SI{5}{GeV^2}$.

\subsection{Heavy-quark production in \pp collisions}
\label{sec:th:hq:pp}

Similar to the case of \ep collisions, 
heavy-quark production in hadronic collisions is interesting either as a benchmark process for the study of pQCD 
or as a probe of the nucleon structure~\cite{Mangano:1997ri}. Most important examples of the latter are:
\begin{itemize}
	\item inclusive heavy-flavour production at high energy mostly probes the gluon density of the proton, since the leading process is $gg \to Q\bar{Q}$~\cite{Guzzi:2014wia,Zenaiev:2015rfa,Gauld:2015yia}. 
		This covers a wide kinematic range, because a hard scale provided by the mass of heavy quarks allows applicability of pQCD 
		even at low transverse momentum $p_T \sim \Lambda_{\rm QCD}$;
	\item $W^{\pm}+c$ final states probe the strange-quark content of the proton, 
	since the LO production mechanism is $gs \to W^{\pm}c$~\cite{Alekhin:2014sya}.
\end{itemize}
The understanding of heavy-quark production is also important \ozmodN{in searches for} possible new physics, where QCD-initiated heavy-quark final states \ozmodN{cause} large backgrounds for such analyses.
\ozmod{The contents of this Section is partially based on~\cite{Mangano:1997ri,Klasen:2014dba}, where more details can be found.}

The cross sections for heavy-flavour production in \pp collisions are calculated in pQCD using the factorisation approach, 
similar to Eq.~\ref{eq:th:f2factoris}:
\begin{equation}
\begin{split}
	\sigma^{pp \to Q\bar{Q}} =& \sum_{i,j} \iint_{0}^{1} {\dif x_1} {\dif x_2} {f_i(x_1,\mu_f^2) f_j(x_2,\mu_f^2)}\\
  \times& \hat{\sigma}_{ij \to Q\bar{Q}}(x_1,x_2,\mu_f^2\ozmod{,\mu_r^2},\dots).
	\label{eq:th:ppfactoris}
\end{split}
\end{equation}
Here the sum in $i$, $j$ goes over all \ozmodNN{relevant} partons, $\hat{\sigma}_{ij \to Q\bar{Q}}$ is the perturbatively calculated partonic cross section, 
$x_1$, $x_2$ are momentum fractions carried by the two incoming partons, 
and $f_i$, $f_j$ are the PDFs, introduced in Section~\ref{sec:th:qcd:factoris}, for the two incoming protons $p_1$ and $p_2$. 
\ozmod{Note that similarly to Eq.~\ref{eq:th:f2factoris} the left-hand side of Eq.~\ref{eq:th:ppfactoris} is an observable quantity and does not depend 
on scales $\mu_r$ and $\mu_f$ by definition.}

\begin{figure}[htbp]
\centering
\vspace{0.25cm}
\begin{fmffile}{hqpplo1}
\begin{fmfgraph*}(70,45)
	\fmfpen{thin}
	\fmfleft{i1,i2}
	\fmfright{o1,o2}
	\fmf{quark}{i1,v1}
	\fmflabel{$q$}{i1}
	\fmf{quark}{v1,i2}
	\fmflabel{$\bar{q}$}{i2}
	\fmf{gluon}{v1,v2}
	\fmf{quark}{o1,v2}
	\fmflabel{$Q$}{o1}
	\fmf{quark}{v2,o2}
	\fmflabel{$\bar{Q}$}{o2}
	\fmfdot{v1}
	\fmfdot{v2}
\end{fmfgraph*}
\end{fmffile}
\\ \vspace{1.0cm}
\begin{fmffile}{hqpplo2}
\begin{fmfgraph*}(70,45)
	\fmfpen{thin}
	\fmfleft{i1,i2}
	\fmfright{o1,o2}
	\fmf{gluon}{i1,v1}
	\fmflabel{$g$}{i1}
	\fmf{gluon}{i2,v2}
	\fmflabel{$g$}{i2}
	\fmf{quark}{v1,v2}
	\fmf{quark}{o1,v1}
	\fmflabel{$Q$}{o1}
	\fmf{quark}{v2,o2}
	\fmflabel{$\bar{Q}$}{o2}
	\fmfdot{v1}
	\fmfdot{v2}
	\fmffreeze
	\fmfshift{(0,-0.14w)}{v1}
	\fmfshift{(0,0.14w)}{v2}
\end{fmfgraph*}
\end{fmffile}
\hspace{0.25cm}
\begin{fmffile}{hqpplo3}
\begin{fmfgraph*}(70,45)
	\fmfpen{thin}
	\fmfleft{i1,i2}
	\fmfright{o1,o2}
	\fmf{gluon}{i1,v2}
	\fmflabel{$g$}{i1}
	\fmf{gluon}{i2,v1}
	\fmflabel{$g$}{i2}
	\fmf{quark}{v1,v2}
	\fmf{quark}{o1,v1}
	\fmflabel{$Q$}{o1}
	\fmf{quark}{v2,o2}
	\fmflabel{$\bar{Q}$}{o2}
	\fmfdot{v1}
	\fmfdot{v2}
	\fmffreeze
	\fmfshift{(0,-0.275w)}{v1}
	\fmfshift{(0,0.275w)}{v2}
\end{fmfgraph*}
\end{fmffile}
\hspace{0.25cm}
\begin{fmffile}{hqpplo4}
\begin{fmfgraph*}(70,45)
	\fmfpen{thin}
	\fmfleft{i1,i2}
	\fmfright{o1,o2}
	\fmf{gluon}{i1,v1}
	\fmflabel{$g$}{i1}
	\fmf{gluon}{v1,i2}
	\fmflabel{$g$}{i2}
	\fmf{gluon}{v1,v2}
	\fmf{quark}{o1,v2}
	\fmflabel{$Q$}{o1}
	\fmf{quark}{v2,o2}
	\fmflabel{$\bar{Q}$}{o2}
	\fmfdot{v1}
	\fmfdot{v2}
\end{fmfgraph*}
\end{fmffile}
\vspace{0.25cm}
\caption[LO diagrams for heavy-quark production in \pp collisions]
{LO diagrams for heavy-quark production in \pp collisions.}
\label{fig:th:hq:pp:lo}
\end{figure}
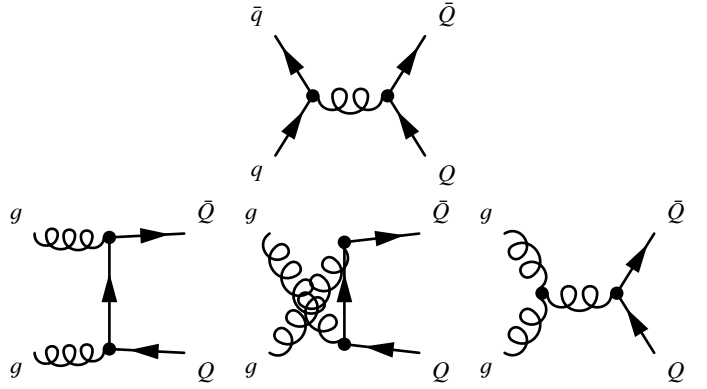

In the FFNS at LO ($O(\alpha_s^2)$) two processes are responsible for heavy-quark production:
\begin{equation}
	q\bar{q} \to Q\bar{Q}~~~~~{\rm and}~~~~~gg \to Q\bar{Q}.
\end{equation}
The diagrams are shown in Fig.~\ref{fig:th:hq:pp:lo} and the corresponding differential partonic cross sections 
are~\cite{Gluck:1977zm,Combridge:1978kx,Hagiwara:1978hw,Jones:1977di,Georgi:1978kx,Babcock:1977fi}:
\begin{equation}
\begin{split}
	\frac{{\rm d}\hat{\sigma}_{ij \to Q\bar{Q}}(\hat{s}, \theta)}{{\rm d}\phi_{(2)}} =& \left|M(ij \to Q\bar{Q})\right|^2,\\
	\left|M(q\bar{q} \to Q\bar{Q})\right|^2 =& (4\pi\alpha_s)^2\frac{V}{2N^2}\left(\tau_1^2+\tau_2^2+\frac{\rho}{2}\right),\\
	\left|M(gg \to Q\bar{Q})\right|^2 =& (4\pi\alpha_s)^2\frac{1}{2VN}\left(\frac{V}{\tau_1\tau_2}-2N^2\right)\\
	&\times \left(\tau_1^2+\tau_2^2+\rho-\frac{\rho^2}{4\tau_1\tau_2}\right),\\
\end{split}
\label{eq:th:pp0}
\end{equation}
where $N=3$ is the number of colors, $V=N^2-1=8$ is the dimension of the ${\rm SU}(3)$ gauge group, i.e.\ the number of gluons, 
$\hat{s} = (x_1 p_1 + x_2 p_2)^2$ is the squared partonic centre-of-mass energy, $\tau_{1,2} = (1 \mp \beta{\rm cos}\theta)/2$, 
$\theta$ is the partonic scattering angle, $\rho = 4M^2/\hat{s}$, $\beta = \sqrt{1-\rho}$, 
$M = m_Q^{\rm pole}$ is the heavy-quark pole mass, and ${\rm d}\phi_{(2)}$ is \ozmodNN{the two-body phase-space element} given by
\begin{equation}
\begin{split}
	{\rm d}\phi_{(2)} &\equiv \frac{1}{2\hat{s}} \frac{{\rm d}^3Q}{(2\pi)^32Q^0}\frac{{\rm d}^3\bar{Q}}{(2\pi)^32\bar{Q}^0}(2\pi)^4\delta^4(i+j-Q-\bar{Q}) \\
	&= 
	\frac{\pi}{2\hat{s}} \left(\frac{1}{4\pi}\right)^2\beta{\rm d}{\rm cos}\theta.
\end{split}
\end{equation}
The total production cross section for heavy quarks is finite at LO, owing to the fact that 
$M$ is the minimum virtuality exchanged in the $t$-channel, therefore no poles can develop
in the intermediate propagators. This is not the case for light quarks: the total production cross section for $u$
or $d$ quarks is not calculable in pQCD~\cite{Mangano:1997ri}. 
\ozmod{Note that the LO partonic cross sections in Eq.~\ref{eq:th:pp0}, similarly to the LO hard-scattering functions in Eq.~\ref{eq:th:c20}, 
do not depend on the factorisation and renormalisation scales, except for the $\alpha_s^2(\mu_r^2)$ dependence. 
The scales are chosen to be of the order of the energy involved in the hard process. 
\ozmodNN{A typical choice} is \ozmodN{$\mu_r^2=\mu_f^2= M^2$ or $\mu_r^2=\mu_f^2= M^2 + \langle p_T^2 \rangle$}, 
where $\langle p_T^2  \rangle$ is the average squared transverse momentum of the produced heavy quark and antiquark. 
Other possible choices are the off-shell of the internal lines in the diagrams in Fig.~\ref{fig:th:hq:pp:lo} 
$\mu_r^2=\mu_f^2=\hat{s}$, $\mu_r^2=\mu_f^2=M^2-(g-c)^2$ or $\mu_r^2=\mu_f^2=M^2-(g-\bar{c})^2$ (see e.g.~\cite{Combridge:1978kx}). 
} 

The total partonic cross section can be obtained by integrating over the partonic scattering angle:
\begin{equation}
\begin{split}
	\hat{\sigma}_{q\bar{q} \to Q\bar{Q}}(\hat{s}) =& \frac{\alpha_s^2}{M^2} \left( \frac{V}{N^2} \right) \frac{\pi \beta}{24} \rho (2+\rho),\\
	\hat{\sigma}_{gg \to Q\bar{Q}}(\hat{s})       =& \frac{\alpha_s^2}{M^2} \left( \frac{1}{NV} \right) \frac{\pi \beta}{24} \rho \{3L(\beta)(\rho^2+2V(\rho+1))\\
	                                                      &+ 2(V-2)(1+\rho)+\rho(6\rho-N^2)\},
\end{split}
\label{eq:th:pp0totalpartonic}
\end{equation}
where $L(\beta)=\frac{1}{\beta}{\rm log}\left( \frac{1+\beta}{1-\beta}\right) - 2$. 

At large $\hat{s}$ the $q\bar{q}$ rate \ozmodNN{drops} more quickly than $gg$, 
\ozmod{as can be seen from Eq.~\ref{eq:th:pp0totalpartonic} (this remains true also when NLO effects are considered)}. 
\ozmod{In addition}, threshold effects for the $q\bar{q}$ channel vanish very quickly as soon as $\hat{s} > 4m_Q^2$; 
this is related to the spin 1/2 of quarks~\cite{Mangano:1997ri}. 

To calculate the differential hadronic cross section of Eq.~\ref{eq:th:ppfactoris}, the partonic cross section in Eq.~\ref{eq:th:pp0} needs 
to be convoluted with the PDFs in the hadrons. The kinematics of the final state can be parametrised in terms of the transverse momenta ${p_T}_1$, ${p_T}_2$ 
and rapidities $y_1$, $y_2$ of the produced quark and antiquark, which are related at LO to the parton momentum fractions $x_1$, $x_2$:
\begin{equation}
\begin{split}
	x_1 &= ({\rm e}^{y_1}+{\rm e}^{y_2})/\epsilon,\\
	x_2 &= ({\rm e}^{-y_1}+{\rm e}^{-y_2})/\epsilon,\\
\end{split}
\label{eq:th:pp0x1x2}
\end{equation}
where $\epsilon = \sqrt{s}/M_{T}$, $M_{T} = \sqrt{M^2+p_T^2}$, $p_T = {p_T}_1 = {p_T}_2$ and $s$ is the squared hadron centre-of-mass energy. 
\ozmodNN{The resulting phase-space element is}
\begin{equation}
\begin{split}
	{\rm d}\phi_{p\bar{p}} \equiv& \frac{1}{2\hat{s}} {\rm d}x_1{\rm d}x_2 \frac{{\rm d}^3Q}{(2\pi)^32Q^0}\frac{{\rm d}^3\bar{Q}}{(2\pi)^32\bar{Q}^0}(2\pi)^4\delta^4(x_1p_1+x_2p_2 \\
	-&Q-\bar{Q}) = \frac{\pi x_1 x_2}{4M_T^4[1+{\rm cosh}(y_1-y_2)]^2} \left(\frac{1}{4\pi}\right)^2 {\rm d}y_1{\rm d}y_2{\rm d}p_T^2,
\end{split}
\label{eq:th:pp0x1x2ps}
\end{equation}
and the differential cross section at LO is:
\begin{equation}
\begin{split}
	\frac{{\rm d}\sigma^{pp \to Q\bar{Q}}}{{\rm d}y_1{\rm d}y_2{\rm d}p_T^2} = \frac{\pi}{{4M_T^4}}\frac{\sum_{i,j}x_1 f_i(x_1) x_2 f_{j}(x_2)\left|M(ij \to Q\bar{Q})\right|^2}{(4\pi)^2[1+{\rm cosh}(y_1-y_2)]^2}.
\end{split}
\label{eq:th:pp0x1x2cs}
\end{equation}
As can be seen from Eq.~\ref{eq:th:pp0x1x2cs}, for a fixed value of $p_T$ the rate is suppressed when $|y_1-y_2|$ becomes large, 
therefore the quark and antiquark tend to be produced with the same rapidity. 
At the LHC the bulk of the contribution to heavy-flavour production directly probes the gluon content 
of the proton \ozmodN{and serves to improve its knowledge over the HERA determination \ozmodNN{(see Section~\ref{sec:pdffit})}.} 

\subsubsection{MNR calculations}
\label{sec:th:hq:pp:nlo}

NLO corrections come from two sources of $O(\alpha_s^3)$ diagrams:
real- and virtual-emission diagrams. In the first case, the corrections come from the square of the real-emission matrix elements; 
in the second case, from the interference of the virtual matrix elements (of $O(\alpha_s^4)$) with the tree-level ones (of $O(\alpha_s^2)$). 
Ultraviolet divergences in the virtual diagrams are removed by the renormalisation
process. Infrared and collinear divergences, which appear both in the virtual diagrams
and in the integration over the emitted parton in the real-emission processes, cancel each other or
are absorbed in the PDFs. The complete calculations of NLO corrections
to the production of heavy-quark pairs in hadro- and in photoproduction were done in~\cite{mnrtotal,Beenakker:1988bq} 
(total hadroproduction cross sections), \cite{mnrsingle,Beenakker:1990maa} (one-particle inclusive distributions in hadroproduction),
\cite{Ellis:1988sb,Smith:1991pw} (total and one-particle inclusive distributions in \ep photoproduction), ~\cite{mnrdouble}
(two-particle inclusive distributions in hadroproduction) and \cite{Frixione:1993dg} (two-particle inclusive distributions
in \ep photoproduction). They are known as \ozmod{Mangano-Nason-Ridolfi} (MNR) calculations and are available in the MNR program\ozmod{~\cite{mnrplace}}, 
which calculates double- or single-particle inclusive or total cross sections.
The pole-mass definition is used in the calculations.

\begin{figure}[htbp]
  \centering
  \includegraphics[width=1.0\figwidth,trim=0 0 0 0mm,clip=true]{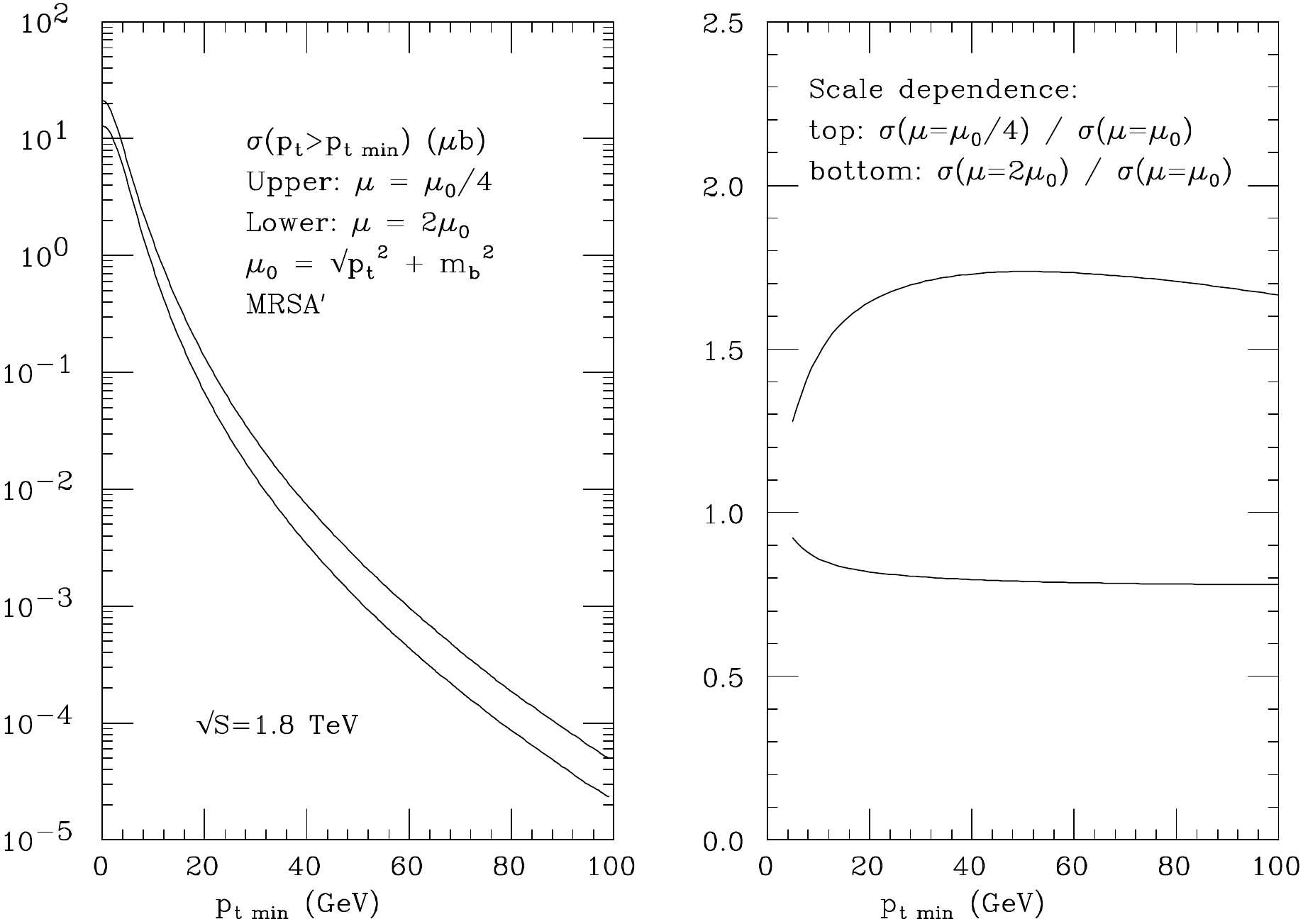}
  \caption[Scale dependence of inclusive $p_T$ distribution for $b$ quarks at Tevatron]
  {Scale dependence of the inclusive $p_T$ distribution for $b$ quarks, in $p\bar{p}$
		collisions at $\sqrt{s} = \SI{1.8}{TeV}$. 
		The plot was taken from~\cite{Mangano:1997ri}.}
	\label{fig:th:hq:pp:scale}
\end{figure}

There are a few important remarks concerning the NLO calculations:
\begin{itemize}
	\item no collinear singularities appear when gluons are emitted from the final-state heavy quarks,
		since they are screened by the heavy-quark mass. Therefore, contrary to the case of a light parton,
		the $p_T$ distribution for a heavy quark is a well-defined quantity in NLO. For light partons, 
		a collinear singularity would be encountered that requires the introduction of a fragmentation function,
		not calculable from first principles (see also Section~\ref{sec:th:hq:frag});
	\item at large $p_T$, nevertheless, large $\ln(p_T/m_Q)$ factors appear, signalling the increased probability
		of collinear gluon emission. At large $p_T$, the massive quark \ozmod{behaves similar to} 
		a massless particle. These logarithms can be resummed using the fragmentation-function
		formalism (see next Section~\ref{sec:th:hq:pp:fonll});
	\item new processes appear at NLO which drastically change the $\hat{s}$ dependence of the
		cross sections and/or the kinematic distributions;
	\item there is evidence however that NLO is not sufficient to get accurate estimates,
		since a large scale dependence is still present. This is demonstrated in Fig.~\ref{fig:th:hq:pp:scale}, which shows \ozmod{as an example} the scale
		dependence of the inclusive $p_T$ distribution of $b$ quarks at the Tevatron. 
		\ozmod{At the LHC, the estimated uncertainty is of the order of $50\%$ for $b$-quark production, and it is even larger 
		for $c$-quark production, owing to the smaller value of the $c$-quark mass.}
		Large scale dependence is a symptom of large NNLO corrections.%
		\footnote{\ozmod{The complete NNLO calculations for differential distributions in the top-quark pair production process in hadronic collisions have recently appeared in the literature~\cite{Czakon:2015owf,Czakon:2016dgf}.}}
\end{itemize}
\ozmod{For an extensive and a more quantitative analysis of the NLO corrections, 
as well as a general discussion of the corrections beyond the Born level, see Ref.~\cite{Mangano:1997ri}.}

\subsubsection{FONLL calculations}
\label{sec:th:hq:pp:fonll}

The fixed-order plus next-to-leading-logarithms (FONLL) calculations~\cite{Cacciari:1998it} 
were developed for improving the large-$p_T$ differential cross section for heavy-quark production in hadron--hadron collisions 
and then were extended to photoproduction in \ep collisions~\cite{Cacciari:2001td}. 
This approach is a variant of GM-VFNS, based on the matching of NLO massive and massless calculations according to the prescription~\cite{Klasen:2014dba}:
\begin{equation}
	\dif \sigma_{\rm FONLL}=\dif \sigma_{\rm FO} +(\dif \sigma_{\rm RS}- \dif \sigma_{\rm FOM0}) \times G(m_Q,p_T).
\end{equation}
Here FO denotes the massive NLO cross section, where a heavy quark enters
only in the partonic scattering through the flavour-creation processes, but not in the
PDFs, and its mass is kept as a non-vanishing parameter. 
This part, which is singular in the massless limit, and the finite parts related
to its different definition in dimensional and mass regularisation are denoted FOM0 and
therefore resummed to next-to-leading-logarithm order in the contribution denoted RS. 
The RS contribution is then added to the FO calculation, while the overlap FOM0 is subtracted to avoid double counting. 
This is controlled by the matching function $G(m_Q,p_T)$, 
which must tend to unity in the massless limit $p_T \gg m_Q$,
where FO approaches FOM0 and the mass logarithms must be resummed. 
In FONLL its functional form is
\begin{equation}
	G(m_Q,p_T) = \frac{p_T^2}{p_T^2+a^2 m_Q^2},
\end{equation}
with an ad-hoc constant $a = 5$. 

\begin{figure}[htbp]
  \centering
  \includegraphics[width=1.0\figwidth,trim=0 0 0 0mm,clip=true]{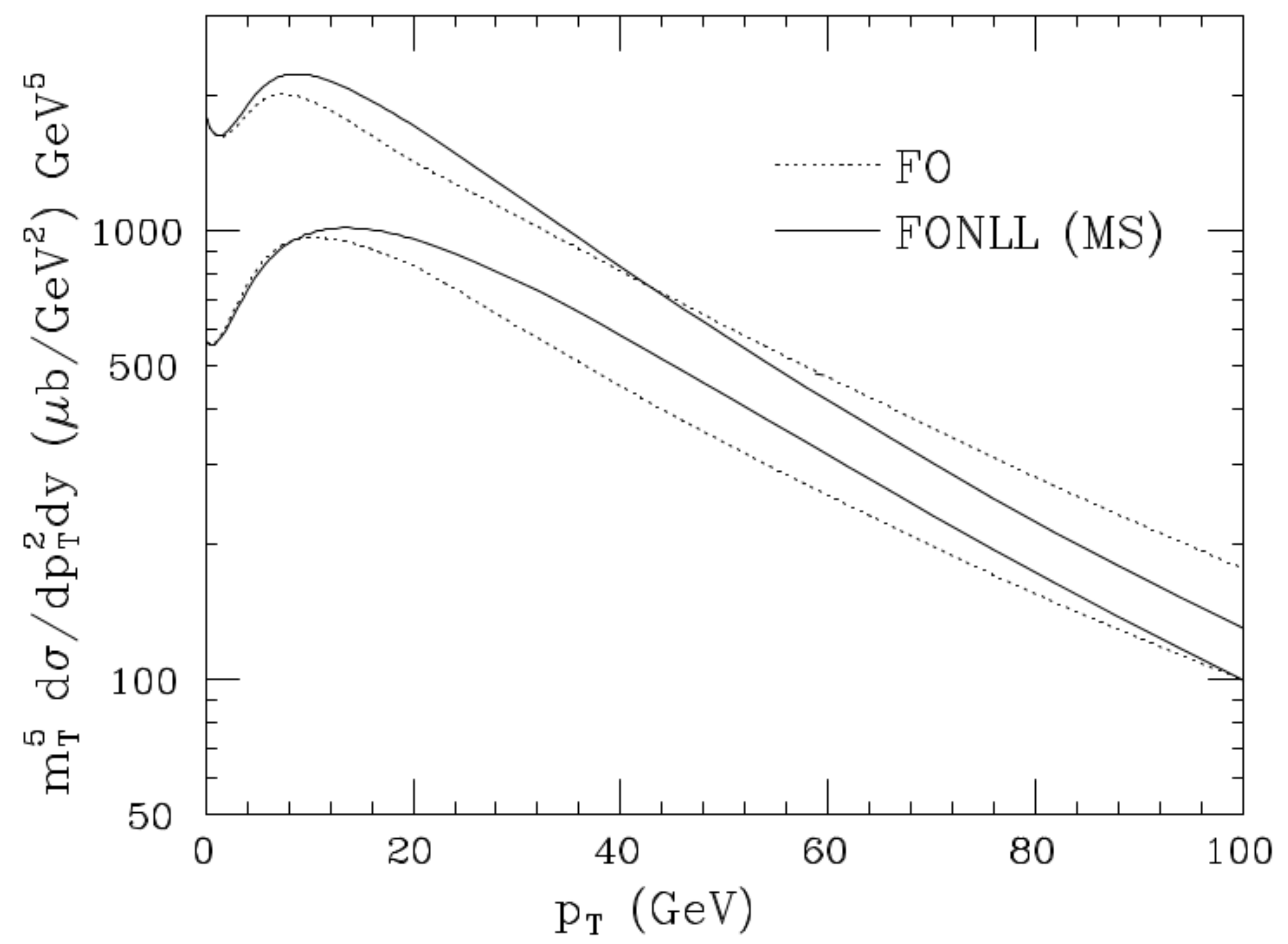}
  \caption[Comparison of NLO and FONLL uncertainty bands from scale variations]
  {Comparison of the uncertainty bands from the scale variations of the NLO and FONLL calculations for beauty production at the Tevatron. 
  The plot was taken from~\cite{Cacciari:1998it}.}
	\label{fig:th:hq:pp:fonll:nlovsfonllorig}
\end{figure}

Comparison of the NLO and FONLL calculations for beauty production at the Tevatron is shown in Fig.~\ref{fig:th:hq:pp:fonll:nlovsfonllorig}, 
where uncertainty bands obtained from the scale variations are shown. The resummation procedure indicates the presence 
of a small enhancement in the intermediate-$p_T$ region, followed by a reduction of the cross section (and of the uncertainty band) at larger $p_T$~\cite{Cacciari:1998it}. 
Both uncertainty bands fully overlap in a wide $p_T$ range.
FONLL predictions for LHC data are given in~\cite{Cacciari:2012ny}; 
they can be also obtained using the public web interface~\cite{FONLLWeb}.

\subsubsection{Other GM-VFNS calculations}
\label{sec:th:hq:pp:gmvfns}

Other GM-VFNS calculations~\cite{Kniehl:2005mk} were originally performed in the
massless limit, valid at high $p_T$, and therefore include flavour-creation, gluon-splitting and
flavour-excitation processes~\cite{Klasen:2014dba}. Subsequently the calculations were improved by identifying the
previously omitted finite-mass terms through a comparison with the massive NLO calculation,
where together with the mass logarithms, finite terms were also subtracted in such
a way that in the limit $m_Q \to 0$ the correct massless \msbar result was recovered, 
since the PDFs and perturbative fragmentation functions that are convoluted with the partonic cross sections
are defined in the ZM-VFNS. 

\subsection{Fragmentation of heavy quarks}
\label{sec:th:hq:frag}

The production of hadrons in QCD can only be described by taking into account a 
non-perturbative hadronisation phase, i.e.\ the processes which transform objects amenable to a perturbative description 
(quarks and gluons) into real particles \ozmodNN{(see Fig.~\ref{fig:dch:mc:bgf} in Section~\ref{sec:dch})}. 
\ozmod{The contents of this Section is partially based on~\cite{Baines:2006uw}, where more details can be found.} 

In the case of light hadrons, the QCD factorisation 
theorem~\cite{Mueller:1978xu,Collins:1981ta,Collins:1989gx,Collins:1985ue,Collins:1988ig,Bodwin:1984hc} 
allows for factorisation of these non-perturbative effects into universal (but factorisation-scheme dependent)
fragmentation functions:
\begin{equation}
	\frac{\dif \sigma_h}{\dif p_T}(p_T)=\sum_i \int_0^1\frac{\dif z}{z}\frac{\dif \sigma_i}{\dif p_T}\left( \frac{p_T}{z},\mu\right) D_{i \to h}(z,\mu)+O\left(\frac{\Lambda_{\rm QCD}}{p_T}\right).
\end{equation}
In this equation, valid up to higher-twist corrections of order $\frac{\Lambda_{\rm QCD}}{p_T}$, the partonic cross sections $\frac{\dif \sigma_h}{\dif p_T}$ for
production of the \ozmod{hadron $h$} are calculated in pQCD, while the fragmentation functions, $D_{i \to h}(z,\mu)$, 
are usually extracted from fits to experimental data 
(not to be confused with the heavy-quark perturbative fragmentation functions, introduced for the GM-VFNS calculations, 
which initial values at the starting scale are calculable perturbatively~\cite{Mele:1990cw}). 
The fragmentation function $D_{i \to h}(z,\mu)$ describes the probability that a parton $i$ fragments into a hadron $h$ 
carrying a fraction $z$ of the momentum of the parton $i$. 
Due to their universality they can be used to make predictions 
for different processes. The factorisation scale $\mu$ is a reminder of the non-physical character of both
the partonic cross sections and the fragmentation functions: it is usually taken of the order of the hard scale 
of the process ($p_T$), and $D_{i \to h}(z,\mu)$ are evolved from a low scale up to $\mu$ by means of the DGLAP
evolution equations.%
\footnote{Note, that this scale $\mu$ it is not the $\mu_f$, which was introduced in Section~\ref{sec:th:qcd:factoris}, although the argument for its 
introduction is the same: to separate long-distance effects. This scale may be called the fragmentation scale.}

This general picture becomes different for the production of heavy-flavoured 
hadrons. 
NLO QCD calculations describe the production of an on-shell heavy quark. 
Still, mimicking the factorisation theorem given above, the quark to hadron transition can be described by convoluting the heavy-quark production cross section 
with a suitable scale-independent non-perturbative fragmentation function, $D^{\rm np}_{Q \to H}(z)$, describing the hadronisation of the heavy quark: 
\begin{equation}
	\frac{\dif \sigma_H}{\dif p_T}(p_T)=\int_0^1 \frac{\dif z}{z} \frac{\dif \sigma_Q^{\rm pert}}{\dif p_T} \left( \frac{p_T}{z},m_Q\right) D^{\rm np}_{Q \to H}(z).
\label{eq:th:fragnp}
\end{equation}
It is worth noting that at this stage this formula is not the \ozmod{result} of a rigorous theorem, 
but is used on a purely phenomenological basis. 
Moreover, it will in general fail (or at least \ozmodNN{be subjected} to large uncertainties) in the region where
the mass $m_Q$ of the heavy quark is not much smaller than its transverse momentum $p_T$, since the choice of the
scaling variable, $z$, is no longer unique, and $O(m_Q/p_T)$ corrections cannot be neglected. 
This leads to a modelling uncertainty which is, however, small compared to the perturbative uncertainty at NLO.

An important characteristic of the non-perturbative fragmentation function is 
that the average fraction of momentum lost by the heavy quark when hadronising into a heavy-flavoured hadron, 
${\langle z \rangle}^{\rm np}$, is given by~\cite{Bjorken:1977md,Suzuki:1977km}
\begin{equation}
	{\langle z \rangle}^{\rm np} \simeq 1-\frac{\Lambda_{\rm QCD}}{m_Q}.
\label{eq:th:hq:pp:frag:mean}
\end{equation}
Since (by definition) the mass of a heavy quark is much larger than the scale $\Lambda_{\rm QCD}$, this amounts to saying
that the non-perturbative fragmentation function for a heavy quark from Eq.~\ref{eq:th:fragnp} is very hard, i.e.\ the quark loses very little momentum when
hadronising. This can also be seen \ozmod{by noting that} a fast massive quark will lose
very little speed (and hence momentum) when picking up a light quark of mass $\Lambda_{\rm QCD}$ from the vacuum to form a
heavy meson.%
\footnote{More modern and more rigorous derivations of this result can be found in~\cite{Jaffe:1993ie,Nason:1996pk,Cacciari:2002xb}.}

This basic behaviour is to be found as a common \ozmod{feature} in all the non-perturbative heavy-quark fragmentation functions, derived
from various phenomenological models. Among the most commonly used are the 
Kartvelishvili-Likhoded-Petrov~\cite{Kartvelishvili:1977pi}, Bowler~\cite{Bowler:1981sb}, Peterson-Schlatter-Schmitt-Zerwas~\cite{Peterson:1982ak} and Collins-Spiller~\cite{Collins:1984ms} 
functions. These models all provide some functional form for the $D^{\rm np}_{Q \to H}(z)$ function and one or more
free parameters that control its hardness. Such parameters are usually not predicted by the models (or only very roughly), and must be fitted to experimental data.

There are two important aspects concerning the fragmentation of heavy quarks:
\begin{enumerate}
	\item a non-perturbative fragmentation function is designed to describe the transition from the heavy quark to the hadron, 
involving many soft gluons \ozmodNN{with} energies of the order of $\Lambda_{\rm QCD}$. However, if a heavy quark is produced
in a high-energy event, it will initially be far off-shell: hard gluons will be emitted to bring it
on-shell, reducing the heavy-quark momentum and yielding in the process large collinear logarithms. 
The amount of gluon radiation is related to the distance between the heavy-quark mass scale and the hard scale
of the interaction, and is therefore process dependent. 
To account for this dependence, different free parameters of the non-perturbative fragmentation function 
\ozmodNN{are} used at different centre-of-mass energies or transverse momenta (see, e.g.~\cite{h1frag});
	\item since only the final heavy-flavoured hadron is observed, both the non-perturbative fragmentation function and the perturbative cross
section for producing heavy quarks must be regarded as non-physical objects. The details of the fitted
non-perturbative fragmentation function (e.g.\ the precise value(s) of its free parameter(s)) depend on those of the perturbative
cross sections: different perturbative calculations (LO, NLO, FONLL etc.) 
and different perturbative parameters (heavy-quark masses, strong coupling etc.) lead to
different non-perturbative fragmentation functions. These in turn will have to be used only with a perturbative description 
similar to the one within which they have been determined~\cite{tevatronbexcess}.
\end{enumerate}

\subsection{\ozmod{Concluding remarks}}
\label{sec:th:hq:outlook}
QCD provides robust predictions for heavy-flavour production, owing to the presence of the finite heavy-quark mass 
which provides a hard scale for perturbative calculations. However application of perturbative calculations 
to any process involving hadrons requires \textit{a priori} knowledge of proton PDFs which are not calculable perturbatively, but 
must be extracted from data. In addition to describe the transistion of heavy quarks into colorless heavy-flavoured hadrons 
phenomenological fragmentation functions have to be used. 
\ozmodNN{Alternative treatments of the heavy-quark mass effects in perturbative calculations lead to several schemes.} 
In the phase space of currently available experimental data 
the most rigorous calculations are performed in the FFNS, when mass effects of heavy quarks are fully taken into 
account in all parts of calculations at the price that this potentially \ozmod{may} spoil the convergence of the perturbative series 
at high energy scales.

The dominant heavy-flavour production process at HERA is the boson--gluon-fusion process and at the LHC it is the gluon--gluon fusion. 
Therefore at both colliders the gluon distribution is an essential ingredient \ozmod{to predict} the production rate. 
\ozmodN{In other words existing precise heavy-flavour data help to \ozmod{pin down} the gluon distribution 
\ozmod{(see Section~\ref{sec:pdffit})}.}

Currently exact pQCD calculations exist at NLO for heavy-flavour production both in $ep$ and $pp$ collisions; 
\ozmodNN{but} only approximate NNLO calculations are \ozmod{available}. 
Since the calculations depend on non-perturbative input (PDFs and fragmentation), it is important to remember that 
a careful treatment of the latter is crucial for \ozmod{a meaningful comparison of data \ozmodNN{and} theory}. 
Uncertainties \ozmod{in} the predictions come from missing higher orders 
(known as scale uncertainties), QCD parameters (the heavy-quark masses and strong coupling constant), input PDFs 
and phenomenological fragmentation functions; in the bulk of the available phase space, even at NLO, they are 
dominated by scale uncertainties (especially in the case of \pp collisions\ozmod{, where these uncertainty are of the order of factor $2$ for charm production at the LHC}). 
Therefore progress in theoretical calculations is crucial for performing \ozmodNN{strong} tests of QCD. \ozmodNN{Nevertheless already} with currently available 
calculations one can not only test QCD but also use experimental data for significant improvement 
in the precision of parameters of QCD (mainly the heavy-quark masses) and \ozmodNN{the gluon content of the proton}, 
and \ozmodNN{leading to an improved} predictive power of the Standard Model.

%
\clearpage
\section{HERA collider, H1 and ZEUS experiments}
\label{sec:exp:hera}


\subsection{HERA collider}
\label{sec:exp:hera:coll}

\ozmod{HERA played a prominent role in the exploration of the proton structure.
It emerged from a series of electron--proton accelerator
studies in the 70's as the highest energy \ep collider possible, 
which made it possible to produce both NC and CC reactions \ozmodNN{simulataneously} 
and study electroweak unification.
The description below is partially based on~\cite{Klein:2008di}.}

\begin{figure}[htbp]
  \centering
  \includegraphics[width=1.0\figwidth,trim = 0 0 0 86mm,clip=true]{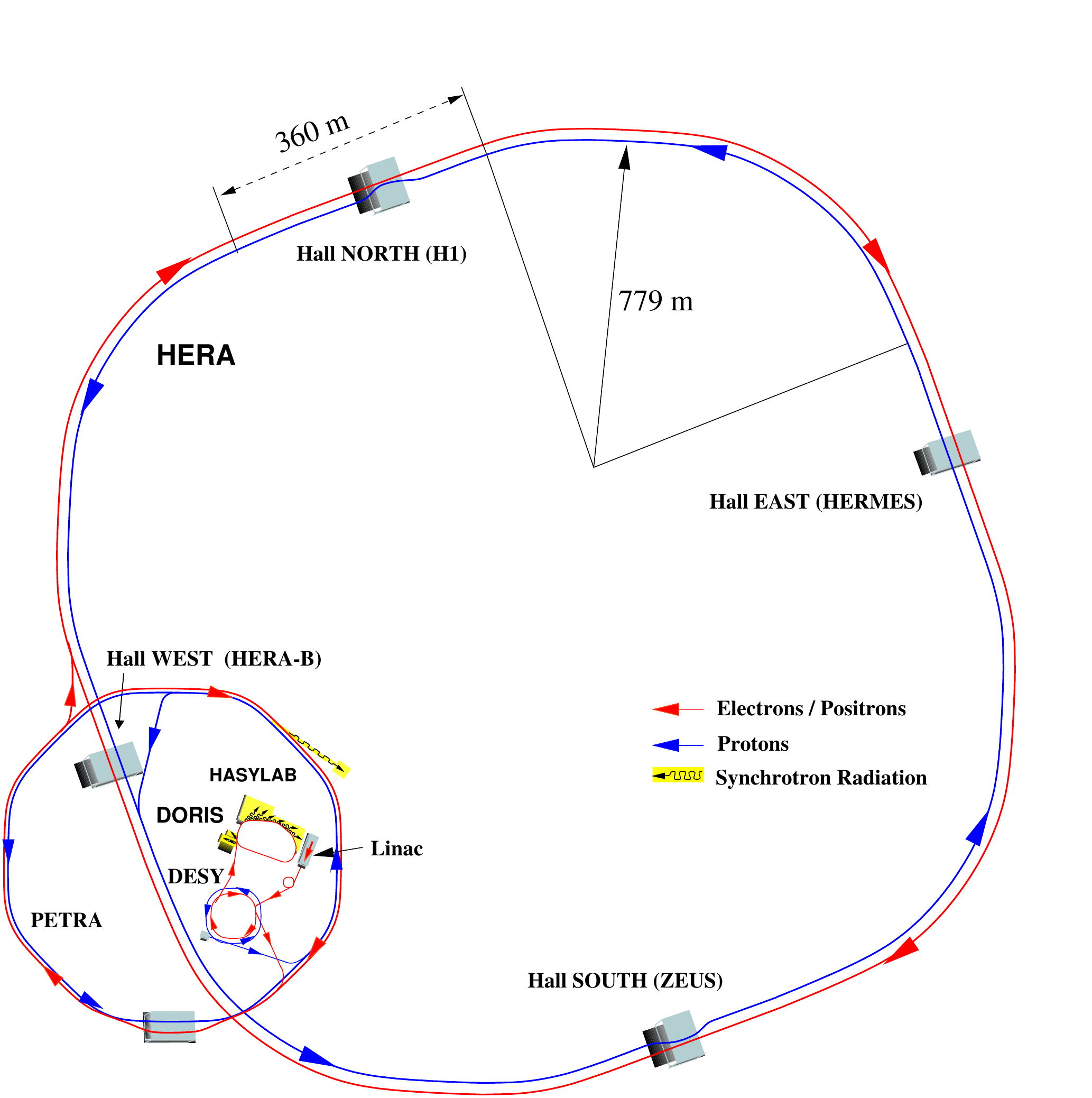}
  \caption[Schematic view of HERA accelerator ring and preaccelerators]
	{A schematic view of the HERA accelerator ring and preaccelerators. 
	The plot was taken from~\cite{verenapictures}.}
  \label{fig:hera}
\end{figure}

HERA (German: Hadron-Elektron Ring Anlage), 
at DESY, Hamburg, was the first, and so far the only, accelerator complex in which electrons and protons were collided~\cite{hera}.
It was built in the 80’s with the capability to scatter polarised electrons and positrons off protons, 
at an energy of the proton beam of initially $\SI{820}{GeV}$ until it was increased to $\SI{920}{GeV}$, in 1998. 
Together with an electron energy of $\SI{27.5}{GeV}$, this resulted in a centre-of-mass energy, $\sqrt{s}$, of about 320 GeV. 
The energy was high enough to probe the phase space in $x$ down to $10^{-6}$ and $Q^2$ up to $30000$ GeV$^2$. 
The protons were accelerated and stored in a ring of superconducting magnets. 
The electron ring was normal conducting.
A schematic view of the HERA accelerator ring and preaccelerators is shown in Fig.~\ref{fig:hera}.

Two general-purpose detectors with nearly $4\pi$ acceptance were proposed in 1985, H1~\cite{Abt:1996hi} 
and ZEUS~\cite{zeusbb}. They were operated over the 16 years of HERA operation.
Two further experiments at HERA were built and run in the fixed-target mode. 
The HERMES experiment~\cite{HERMES:1993aa} (1994--2007) used the polarised electron beam 
to study spin effects in lepton--nucleon interactions using a polarised nuclear target. 
The HERA-B experiment~\cite{Hartouni:1995cf} (1998--2003) was designed to investigate
$B$-meson physics and nuclear effects in the interactions of the proton-beam halo with a nuclear wire target.

The first HERA data were taken in summer 1992. 
HERA had its first phase of operation (referred to as HERA-I) from 1992 through 2000. 
In this period, the collider experiments H1 and ZEUS each
recorded data corresponding to integrated luminosities of approximately $\SI{120}{pb^{-1}}$ of $e^{+}p$ 
and $\SI{15}{pb^{-1}}$ of $e^{-}p$ collisions. 
The HERA collider was then upgraded to increase the specific luminosity by a factor of about four,
as well as to provide longitudinally polarised lepton beams to the collider experiments~\cite{HERAUpgrade}. 
The second data-taking phase (referred to as HERA-II) began in 2003, after completion of the machine and detector upgrades, 
and ended in 2007. 
The H1 and ZEUS experiments each recorded approximately $\SI{200}{pb^{-1}}$ of $e^{+}p$ and 
$\SI{200}{pb^{-1}}$ of $e^{-}p$ data with
electron (positron) energy of approximately $\SI{27.5}{GeV}$ and proton energy of $\SI{920}{GeV}$. 
The lepton beams had an average polarisation of approximately $\pm 30\%$ 
with roughly equal samples of opposite polarities recorded.
In the last three months of HERA operation, 
data with lowered proton-beam energies of $\SI{460}{GeV}$ (referred to as LER, Low Energy Run)
and $\SI{575}{GeV}$ (referred to as MER, Middle Energy Run) were taken; 
each experiment recorded approximately $\SI{13}{pb^{-1}}$ and $\SI{7}{pb^{-1}}$ of the LER and MER data, respectively.
The primary purpose of the LER and MER data was the measurement of the longitudinal proton structure function $F_L$.

HERA ceased operations in June 2007 after a long, successful data-taking period of 16 years. 
\ozmodN{A wealth of results have been published. The studies have considerably enlaged 
the knowledge on the proton structure and provided tests of the Standard Model. 
There are still ongoing analyses.}

\subsection{H1 and ZEUS experiments}
\label{sec:exp:hera:h1zeus}

The collider detectors H1~\cite{Abt:1996hi,Abt:1996xv,Appuhn:1996na} and ZEUS~\cite{zeusbb} were designed primarily for 
\ozmod{deep inelastic scattering (DIS)} \ozmod{at large virtuality $Q^2$ and large} final-state energies. 
Thus, much attention was paid to the
electromagnetic and hadron calorimeters. The H1 Collaboration chose liquid argon as active material
for their main calorimeter to maximise long-term reliability. The ZEUS Collaboration chose scintillator
active media and \ozmodN{depleted} uranium as the absorber material \ozmodN{with the property of equal 
``$e\pi$'' response} to electrons and hadrons. The calorimeters were complemented by large-area wire
chamber systems to measure muon momentum and the tail of hadron-shower energy. Because the electron-
and proton-beam energies were very different, the detectors were asymmetric, with extended coverage of the
forward (proton-beam) direction. Drift chambers inside the calorimeters, both in H1 and in ZEUS,
were segmented into a forward and a central part. Later, in H1 starting in 1996 and in ZEUS from 2003
onwards, silicon detectors near the beampipe were installed for precision vertexing and tracking. Both
apparatus were complemented with detector systems positioned near the beam axis in the accelerator
tunnel, to measure backward photons and electrons, mainly for the determination of the \ozmod{beam} interaction
luminosity, and to tag leading protons and neutrons in the forward direction. 
Both experiments took data \ozmod{over} the entire time of HERA’s operation with efficiency of 70--80\%.
The main components of the H1 and ZEUS detectors are briefly described below with 
the main emphasis on the components most relevant for studies of heavy-flavour production: the tracking systems 
\ozmod{for the precise reconstruction of tracks and vertices, which is crucial for the identification of heavy-flavoured hadrons,} 
and the calorimeters\ozmod{, needed for the identification of the scattered electron and for the reconstruction of event kinematical variables}.

\subsubsection{H1 detector}
\label{sec:exp:hera:h1zeus:h1det}

\begin{figure}[htbp]
  \centering
  \includegraphics[width=1.0\figwidth,trim = 4mm 2mm 0mm 8mm,clip=true]{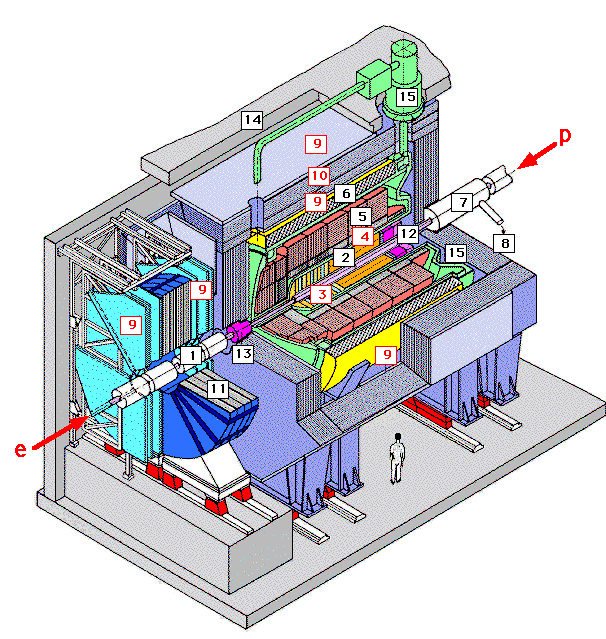}
  \begin{tabu} to 1.0\figwidth {lX[l]lX[l]}
	$1$ & Beam pipe                      & $9$ & Muon chambers                                       \\
	$2$ & Central tracking detector                       & $10$& Return yoke                                        \\
	$3$ & Forward tracking detector       & $11$& Myon-Toroid-Magnet                                 \\
	$4$ & Electromagnetic LAr & $12$& SpaCal                          \\
	$5$ & Hadronic LAr           & $13$& PLUG calorimeter                                   \\
	$6$ & Superconducting coil                     & $14$& Concrete screen                                    \\
	$7$ & Compensating magnet                             & $15$& Liquid Argon cryostat                              \\
	$8$ & Helium cryogenics                               &     &                                                    \\
  \end{tabu}
  \caption[Three-dimensional view of H1 Detector]
	{A three-dimensional view showing the layout of the H1 detector. 
	The components are indicated by numbers on the figure.}
  \label{fig:hera:h1}
\end{figure}

A schematic view of the H1 detector~\cite{Abt:1996hi,Abt:1996xv,Appuhn:1996na} 
along the beampipe and the main detector components are shown in Fig.~\ref{fig:hera:h1}.
The coordinate system is a right-handed Cartesian system, with the $\text{Z}$
axis in the proton-beam direction, referred to as the ``forward
direction'', and the $\text{X}$ axis pointing towards the centre of HERA.
The coordinate origin is at the nominal interaction point.
The pseudorapidity is defined as $\eta=-\ln\left(\tan\frac{\theta}{2}\right)$,
where the polar angle, $\theta$, is measured with respect to the
$\text{Z}$ axis. The azimuthal angle, $\phi$, is
measured with respect to the $X$ axis.

Charged particles were measured within the central tracking detector (CTD) in the pseudorapidity
range $-1.85 < \eta < 1.85$. The CTD consisted of two large cylindrical jet chambers
(CJCs), surrounding a system of three silicon detectors consisting of 
the Central Silicon Tracker (CST)~\cite{Pitzl:2000wz}, and the Forward and Backward Silicon Trackers~\cite{Nozicka:2003ru}. 
The CJCs were separated by a drift
chamber which improves the $z$-coordinate reconstruction. A multiwire proportional chamber~\cite{Becker:2007ms}, 
which was mainly used in the trigger, is situated inside the inner CJC. These detectors
are arranged concentrically around the interaction region in a magnetic field of $\SI{1.16}{T}$. 
The trajectories of charged particles were measured with a transverse momentum resolution of 
$\sigma(p_T)/p_T = 0.005 \cdot p_T/\SI{}{GeV} \oplus \SI{0.015}{}$~\cite{h1ctd_res}.%
\footnote{The $\oplus$ sign indicates that the terms are added in quadrature.} 
The interaction vertex was reconstructed from CTD tracks. 
The CTD also provided triggering information based on track segments measured in 
the CJCs~\cite{Wolff:1992rj,Baird:2001xc,Meer:2001im} and a measurement of the specific ionisation energy loss, $dE/dx$, of charged
particles. The Forward Silicon Tracker measured tracks of charged particles at smaller polar 
angles ($1.5 < \eta < 2.8$) than the central tracker.

Charged and neutral particles were measured in the liquid argon (LAr) calorimeter, which
surrounded the tracking chambers and covers the range $-1.5 < \eta < 3.4$ with full azimuthal
acceptance~\cite{Andrieu:1993kh}. Electromagnetic shower energies were measured with a precision of 
$\sigma(E)/E = 12\%/\sqrt{E}/$\SI{}{GeV} $\oplus 1\%$ and hadronic energies with $\sigma(E)/E = 50\%\sqrt{E}/$\SI{}{GeV} $\oplus 2\%$, 
as determined in test beam measurements~\cite{Andrieu:1993xn,Andrieu:1994yn}. 
A lead-scintillating fibre calorimeter, 
also referred to as the Spaghetti Calorimeter (SpaCal),~\cite{Appuhn:1996na} 
covered the backward region $-4.0 < \eta < -1.4$ (the region of high $Q^2$) completed the measurement of charged and
neutral particles. For electrons the SpaCal had a relative energy resolution of 
$\sigma(E)/E = 7\%/\sqrt{E}/$\SI{}{GeV} $\oplus 1\%$, as determined in test beam measurements~\cite{Nicholls:1995di}. The SpaCal
provided energy and time-of-flight information used for triggering purposes. 
Because the LAr calorimeter was non-compensating and had on average a larger response to electromagnetic compared to 
hadron energy depositions, a software weighting method had to be applied for the energy reconstruction. 
The hadronic final state was reconstructed using an energy flow algorithm which combines
charged particles measured in the CTD and the forward tracking detector with information
from the SpaCal and LAr calorimeters.

The luminosity determination was based on the measurement of the Bethe-Heitler process~\cite{Bethe:1934za} 
where the photon was detected in a calorimeter located at $Z = \SI{-103}{m}$ downstream
of the interaction region in the electron beam direction. 
Additionally, the overall integrated luminosity normalisation was determined using a precision measurement of the QED Compton process~\cite{Aaron:2012kn} 
\ozmodNN{with the best achieved relative uncertainty on the measured luminosity was 2.3\%.}

To reduce the event rate to technickally acceptable $\approx \SI{10}{Hz}$ 
a sophisticated multilevel trigger system was used at H1~\cite{h1trigger_Eichler,h1trigger_Elsen}. 
The first trigger level was supposed to stop the pipeline. 
The decision was based on special trigger signals from various detector 
components. The second trigger level started the readout and used neural networks and topological triggers. 
The third trigger level was placed into operation in 2005 and was mainly used for heavy-quark decays identification. 
It used time-optimised routines for the reconstruction of decay resonances and event properties, 
therefore event building was started on this level. 
On the fourth trigger level an on-line event reconstruction was performed.

\subsubsection{ZEUS detector}
\label{sec:exp:hera:h1zeus:zeusdet}

\begin{figure}[htbp]
  \centering
  \includegraphics[width=1.0\figwidth]{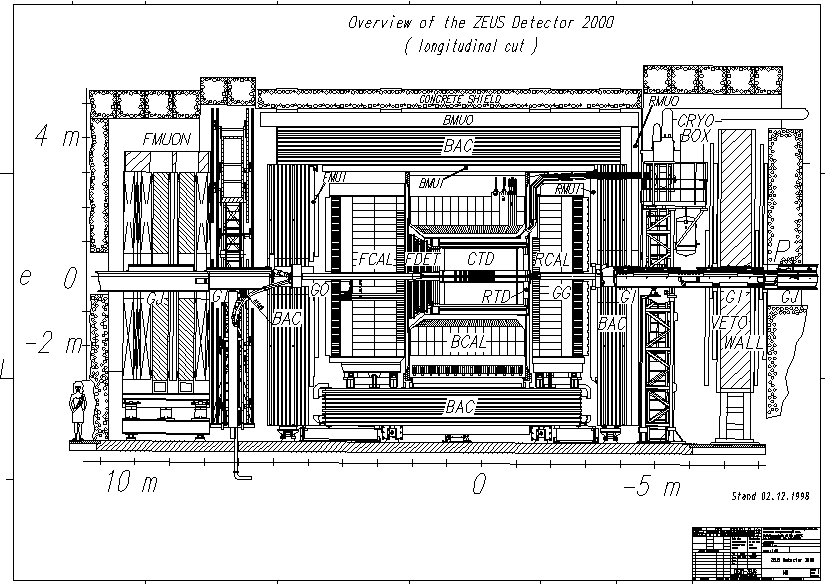}
  \caption[Schematic view of ZEUS detector along beampipe]
	{A schematic view of the ZEUS detector along the beampipe.}
  \label{fig:hera:zeus}
\end{figure}

A schematic view of the ZEUS detector~\cite{zeusbb} along the beampipe and the main detector components 
are shown in Fig.~\ref{fig:hera:zeus}.
The ZEUS coordinate system is the same as for the H1 detector. 
The main detector components are briefly described below.

The momenta of charged particles were measured by the Central Tracking Detector (CTD)~\cite{Harnew:1988ye,Foster:1992mj,Foster:1993ja} 
in the $\SI{1.43}{T}$ magnetic field of the solenoid~\cite{Acerbi:1987cw}. 
The CTD was a cylindrical drift chamber measuring the direction, momentum and energy loss ($dE/dx$). 
It was filled with a gas mixture of argon, carbon dioxide and ethane. 
The CTD was made of 72 layers of wires, which were grouped in 9 superlayers.
The angular coverage of the CTD was $\SI{15}{\degree}<\theta<\SI{164}{\degree}$ and 
the momentum resolution for the full-length tracks in the HERA-I period was determined to be 
$\sigma(p_T)/p_T = \SI{0.0058}{} \cdot p_T/\SI{}{GeV} \oplus \SI{0.0065}{} \oplus \SI{0.0014}{GeV}/p_T$.

\begin{figure}[htbp]
  \centering
  \includegraphics[width=1.0\figwidth,trim = 15mm 0mm 5mm 0mm,clip=true]{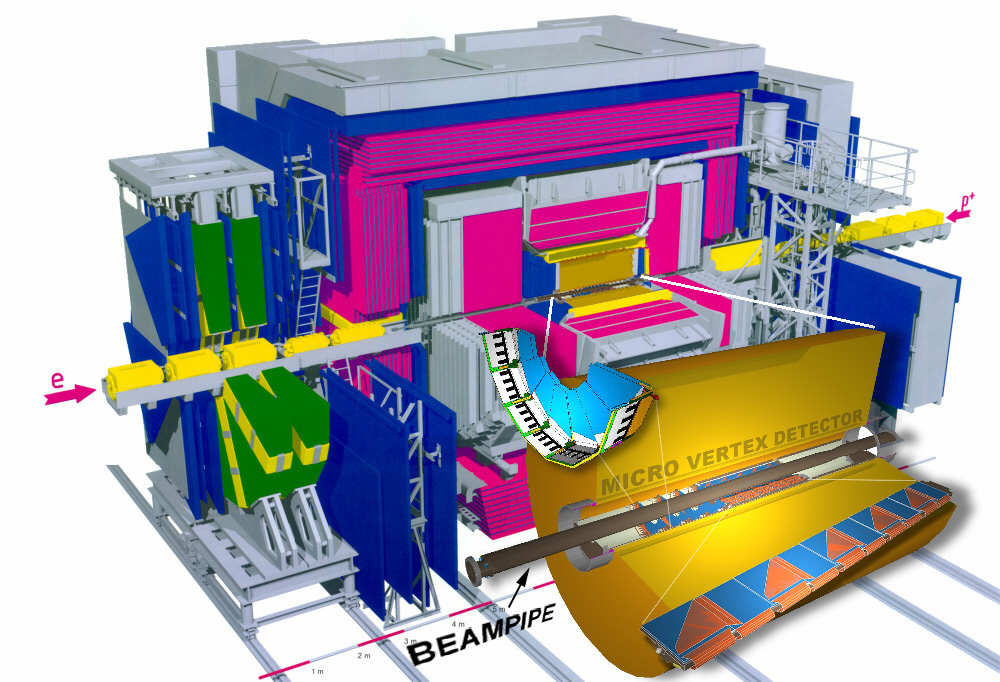}
  \caption[Schematic view of ZEUS detector with installed MVD]
	{A schematic view of the ZEUS detector with installed MVD.}
  \label{fig:hera:zeusmvd}
\end{figure}

At the time of the HERA luminosity upgrade during the shutdown period 2000--2001, 
the tracking system of the ZEUS detector was upgraded with the Microvertex Detector (MVD)~\cite{Polini:2007sw} (Fig.~\ref{fig:hera:zeusmvd}).
The MVD was a silicon-strip vertex detector, mainly supposed to allow reconstruction of secondary vertices 
and track impact parameters from heavy-quark decays.
The MVD consisted of two sections: barrel (BMVD) with an angular coverage $\SI{30}{\degree}<\theta<\SI{150}{\degree}$ 
and forward (FMVD), which extended the coverage to $\SI{7}{\degree}$.
The momentum resolution of the combined tracking system MVD+CTD for full-length tracks in the HERA-II period 
was determined to be $\sigma(p_T)/p_T = \SI{0.0029}{} \cdot p_T/\SI{}{GeV} \oplus \SI{0.0081}{} \oplus \SI{0.0012}{GeV}/p_T$, 
indicating an improved transverse momentum resolution, although the MVD material between the interaction point 
and the CTD increases the probability for multiple scattering.

The forward region of the ZEUS detector required enhanced tracking and 
particle identification capabilities due to the asymmetric beam energies. 
It consisted of the Forward Tracking Detector (FTD) and the Transition Radiation Detector (TRD). 
The purpose of the FTD was to reconstruct low-angle tracks of ionising particles whereas the TRD separated electrons from hadrons. 
During the HERA luminosity upgrade programme the TRD was replaced by the Straw Tube Tracker (STT)~\cite{Fourletov:2004iu}, 
which improved the tracking efficiency in events with high multiplicities.
In the rear direction the Rear Tracking Device (RTD) was located.
To determine the position of the scattered electron near the beampipe, 
the small-angle rear tracking detector (SRTD)~\cite{Bamberger:1997fg} was used.

The most important sub-detector that measured energies was 
the calorimeter (CAL)~\cite{Derrick:1991tq,Andresen:1991ph,Caldwell:1992wc,Bernstein:1993kj}. 
The CAL was mechanically subdivided in three parts:
\begin{itemize}
	\item the Barrel Calorimeter (BCAL) covering polar angles from $\SI{36.7}{\degree}<\theta<\SI{129.1}{\degree}$;
	\item the Forward Calorimeter (FCAL) covering polar angles from $\SI{2.2}{\degree}<\theta<\SI{39.9}{\degree}$;
	\item the Rear Calorimeter (RCAL) covering polar angles from $\SI{128.1}{\degree}<\theta<\SI{176.5}{\degree}$.
\end{itemize}
The CAL was a sampling calorimeter consisting of plates of depleted uranium interleaved 
with plastic scintillator as active material. 
The ratio of absorber and scintillator thickness had been chosen to achieve equal signals from hadrons 
and electromagnetic showers, thereby producing the best possible resolution for hadrons. 
The CAL provided precise energy measurements for hadrons and jets, 
an angular resolution for jets better than $\SI{10}{mrad}$, 
the ability to discriminate between hadrons and electrons using their different energy depositions, 
and a time resolution of $\SI{1}{ns}$. 
The energy resolution for electrons and hadrons as determined under test-beam conditions was 
$18\%/\sqrt{E/\SI{}{GeV}}$ and $35\%/\sqrt{E/\SI{}{GeV}}$, respectively.

The Backing Calorimeter (BAC) was built to fulfill two tasks: 
to achieve a hermetic hadron jet-energy measurement and 
to aid the tracking of muons passing through the iron yoke of the detector. 
To measure the energy of hadron-shower leakages out of the CAL and to correct jet-energy measurements, 
the BAC was equipped with an analog readout, 
giving precise information on the deposited energy but only approximate information on the deposit position. 
To enable muon tracking in the iron yoke, a complementary digital readout 
was designed, giving basically no information about the deposited energy, but exact position in two dimensions. 
This information was used for better positioning of shower leakages 
and for discrimination between leaking hadron cascades and muons. 
To identify muons, the forward muon detector (FMUON) was located in front of the magnet yoke and 
the barrel and rear muon detectors (BMUON, RMUON)~\cite{Abbiendi:1993mi} 
inside and outside the iron yoke. Note that one of techniques used at HERA to measure the production of 
charmed and beauty hadrons is to identify their decays into muons (see Section~\ref{sec:exp:hera:hfmeas}).

The luminosity was measured at ZEUS 
using the bremsstrahlung reaction $ep \to e^{\prime}\gamma p$ by a lead-scintillator calorimeter (PCAL)~\cite{Andruszkow:2001jy}, 
located at $Z=\SI{-107}{m}$, 
and (after the HERA upgrade) an independent magnetic spectrometer (SPEC)~\cite{Helbich:2005qf}, located at $Z=\SI{-104}{m}$. 
The best achieved relative uncertainty on the measured luminosity was 1.8\%.

To reduce the event rate from the highest collision rate $\approx \SI{10}{MHz}$ 
to technically acceptable $\approx \SI{10}{Hz}$, a three-level trigger system was used at ZEUS. 
The First Level Trigger (FLT)~\cite{Smith:1994nx,Heath:1991rk} consisted of hardware trigger systems in individual sub-detectors, 
which sent the information to the Global First Level Trigger (GFLT) to perform the decision. 
Events that passed the GFLT were processed further by the Second Level Trigger (SLT), 
based on software triggers, which used information on charged-particle tracks, the interaction vertex, 
calorimeter timing and global energy sums~\cite{Quadt:1999hr}. Events that passed the SLT were 
processed by the Third Level Trigger (TLT)~\cite{Bhadra:1989kz}, 
which took the decision based on the global information from an event.
Finally, events that passed the TLT were written to tape to be fully reconstructed offline.

\clearpage
\section{\ozmodN{Overview of existing measurements of charm production at HERA}}
\label{sec:exp:hera:hfmeas}

This Section describes tagging techniques used to measure open%
\footnote{When the measured final state contains \ozmodNN{a charmed hadron}.} 
charm production at HERA and
gives an overview of the measurements in deep inelastic scattering (DIS) done by the H1 and ZEUS \ozmodNN{Collaborations}. 
This overview is restricted \ozmodN{to those measurements which are later (see Section~\ref{sec:comb}) used for their combination;} 
the measuremerents are listed in Table~\ref{tab:comb:red:input_ext}. 
\ozmodN{A \ozmodNN{detailed} description of event reconstruction and inclusive DIS selection can be found in next Section~\ref{sec:dch}, 
where the ZEUS measurement of $D^{+}$ production~\cite{zeusdch_hera2} is \ozmodNN{outlined}, 
whereas the present Section \ozmodNN{merely} discusses techniques used to identify charm production.} 

\begin{table*}[tbp]
\caption[Datasets used in combination of reduced charm cross sections]
{Datasets used in the combination of the reduced charm cross sections. For each dataset the charm tagging method, 
the $Q^2$, $p_T$ ($E_T$) and $\eta$ range, the number of cross-section measurements, $N$, the integrated luminosity, 
${\cal L}$, and the centre-of-mass energy, $\sqrt{s}$, are given. The dataset with the $D^{0,\text{ no }D^{*+}}$ tagging method
is based on an analysis of $D^{0}$ mesons not originating from detectable $D^{*+}$
decays. 
}
\label{tab:comb:red:input_ext}
\begin{center}
\tabcolsep1.2mm
\renewcommand*{\arraystretch}{1.2}
\begin{tabu} to \textwidth [t] {|l|l|l|l|l|l|r|r|r|}\hline 
       \multicolumn{2}{|l|}{Dataset} & Tagging method & $Q^2$ range & $p_T$ ($E_T$) range & $\eta$ range       & $N$ & ${\cal L}$& ${\sqrt{s}}$  \\ 
     & & & [GeV$^2$] & [GeV] & &  & [pb$^{-1}$] & [$\SI{}{GeV}$] \\ \hline
1&H1 VTX~\cite{h1ltt_hera2}    &Inclusive&  $5<Q^2<2000$ & not restricted & not restricted              & $29$ & $245$ & $318$ \\
2&H1 $\Dstar$ HERA-I~\cite{h1dstar_hera1}      &$D^{*+}$    &  $2<Q^2<100$ & $\SI{1.5}{}<p_T(\Dstar)<\SI{15}{}$ & $|\eta(\Dstar)|<1.5$                  & $17$ & $47$& $318$\\
3&H1 $\Dstar$ HERA-II (med. $Q^2$)~\cite{h1dstar_hera2}     &$D^{*+} $     &  $5<Q^2<100$ & $\SI{1.25}{}<p_T(\Dstar)<\SI{20}{}$ & $|\eta(\Dstar)|<1.8$                 &  $25$ & $348$& $318$\\
4&H1 $\Dstar$ HERA-II (high $Q^2$)~\cite{h1dstarhighQ2}     &$D^{*+}$     &  $100<Q^2<1000$ & $\SI{1.5}{}<p_T(\Dstar)<\SI{20}{}$ & $|\eta(\Dstar)|<1.8$           &  $6$ & $351$& $318$ \\
5&ZEUS $\Dstar$ 96-97~\cite{zd9697}           &$D^{*+}$      & $1<Q^2<200$ & $\SI{1.5}{}<p_T(\Dstar)<\SI{15}{}$ & $|\eta(\Dstar)|<1.5$                 & $2$1 & $37$& $300$ \\
6&ZEUS $\Dstar$ 98-00~\cite{zd00}           &$D^{*+}$      &  $1.5<Q^2<1000$ & $\SI{1.5}{}<p_T(\Dstar)<\SI{15}{}$ & $|\eta(\Dstar)|<1.5$           & $31$ & $82$& $318$ \\
7&ZEUS $D^0$ 2005~\cite{zd0dp}                  &$D^{0,{\rm no}D^{*+}}$ &$ 5<Q^2<1000$ & $\SI{1.5}{}<p_T(D^{0})<\SI{15}{}$ & $|\eta(\Dstar)|<1.6$   & $9$  & $134$& $318$\\
8&ZEUS $\mu$ 2005~\cite{zeus_muon}                    &Semi-leptonic         & $20<Q^2<10000$ & $p_T(\mu)>\SI{1.5}{}$ & $-1.6<\eta(\mu)<2.2$         & $8$  &  $126$& $318$  \\
9&ZEUS $D^+$ HERA-II~\cite{zeusdch_hera2}&$D^+$         & $5<Q^2<1000$ & $\SI{1.5}{}<p_T(\Dch)<\SI{15}{}$ & $|\eta(\Dstar)|<1.6$              & $14$   & $354$& $318$\\
10&ZEUS $\Dstar$ HERA-II~\cite{zeusdstar_hera2}                  &$\Dstar$         & $5<Q^2<1000$ & $p_T(\Dstar)>\SI{1.5}{}$ & $|\eta(\Dstar)|<1.5$              & $31$   & $363$& $318$\\
11&ZEUS VTX HERA-II~\cite{zeussecvtx_hera2}                  &Inclusive         & $5<Q^2<1000$ & $E_T^{\rm jet}>\SI{4.2}{}$ & $-1.6<\eta^{\rm jet}<2.2$              & $18$   & $354$& $318$\\ \hline
\end{tabu}
\end{center}
\end{table*}


\subsection{Reconstruction of \Dstar mesons in the ``golden'' decay channel}
\label{sec:exp:hera:hfmeas:dstar}

\Dstar mesons are reconstructed in the ``golden'' decay channel ${\Dstar \to D^{0} \pi_s^{+}}$ and subsequently ${D^{0} \to K^{-}\pi^{+}}$. 
The $\pi_s^{+}$ denotes a ``slow'' pion with a low momentum in the \Dstar centre-of-mass frame, 
since the mass of \Dstar is only slightly above the sum of the masses of $D^{0}$ and $\pi^{+}$. 
This results in a narrow peak for the mass difference $\Delta M = M(K^{-}\pi^{+}\pi_s^{+})- M(K^{-}\pi^{+})$ near the threshold, 
accompanied with a not too large combinatorial background and hence the best signal-to-background ratio.
\ozmodN{The smallness of the mass difference in a charmed hadron decay with emission of a low momentum pion 
was first observed in a bubble chamber event at BNL~\cite{Andrews:1975xt}. 
It was proposed for charmed mesons in~\cite{Nussinov:1975ay} and widely used in various 
experiments (e.g.~\cite{Feldman:1977ir,Gibbons:1997ag,zd9697,h1dstar_hera1,LHCbCharm}).}
The main shortcoming is that in practice \Dstar mesons can be measured in the limited 
kinematic space $p_T(\Dstar) \gtrsim \SI{1.25}{GeV}$ only, 
otherwise the transverse momentum of the slow pion is too small and its track cannot be reconstructed.
Another limitation comes from the fact that all decay products have to be reconstructed in the tracking system, 
thus the production of \Dstar mesons can be measured in the central region only, typically $|\eta(\Dstar)| \lesssim 1.8$.
Also, the product of the branching ratios for the decay channels $\Dstar \to D^{0} \pi_s^{+}$ and $D^{0} \to K^{-}\pi^{+}$ is about $3\%$ only~\cite{pdg2014}.
However still the most precise measurements of open charm production at HERA were obtained using this technique.

Both the H1 and ZEUS \ozmodNN{Collaborations} have measured the production of \Dstar mesons using the ``golden'' decay channel using the HERA-I and HERA-II 
data~\cite{h1dstar_hera1,zd9697,zd00,zeusdstar_hera2,h1dstarhighQ2,h1dstar_hera2} 
(see Table~\ref{tab:comb:red:input_ext}). 
The best phase-space coverage was achieved in the HERA-II H1 measurement~\cite{h1dstar_hera2}: $p_T(\Dstar)>\SI{1.25}{GeV}$, $|\eta(\Dstar)|<1.8$.

\begin{figure*}[htbp]
  \sidecaption
  \centering
  \includegraphics[width=0.80\figwidth,trim = 3mm 0 3mm 0,clip=true]{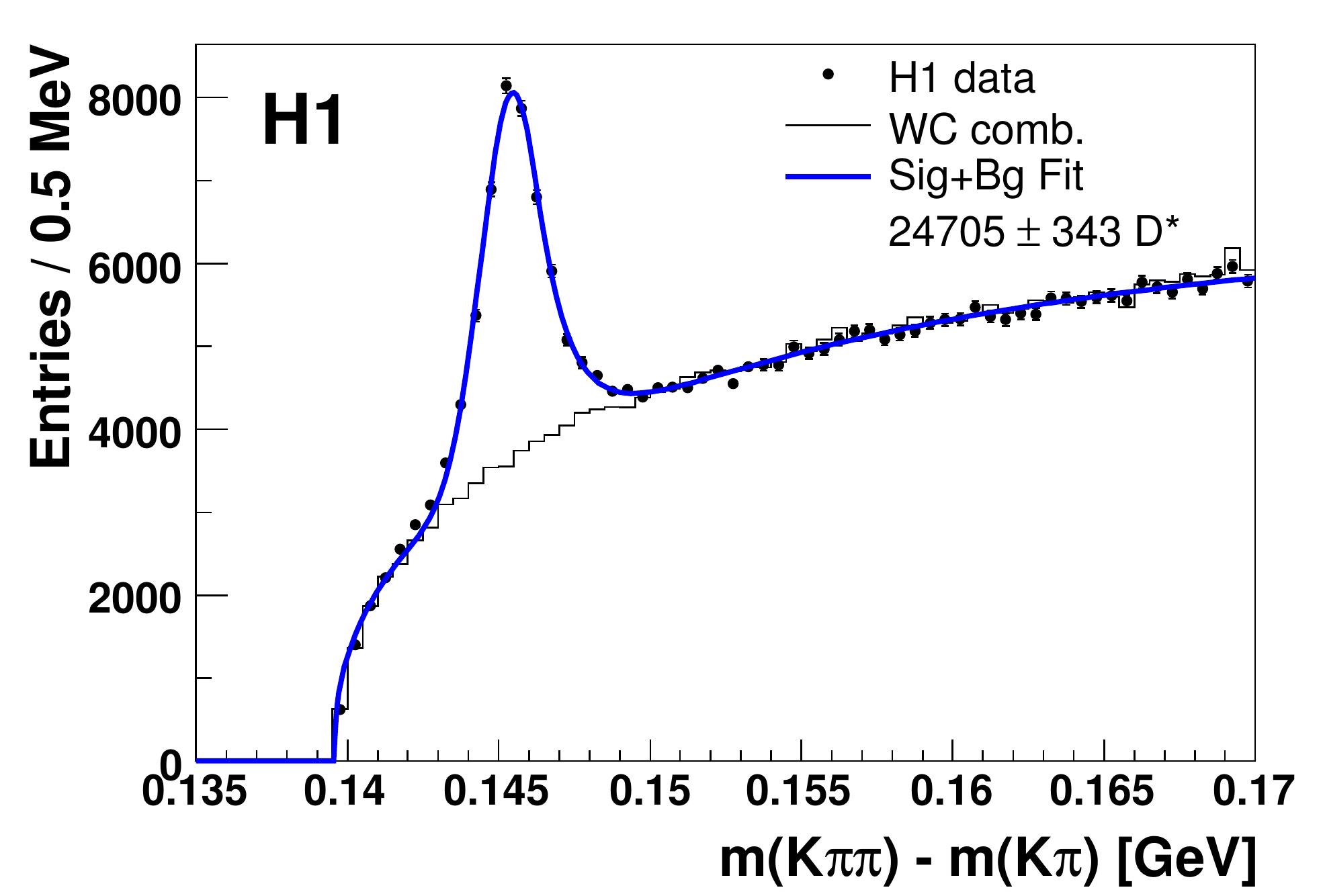}
  \includegraphics[width=0.80\figwidth,trim = 7mm 0 28mm 8mm,clip=true]{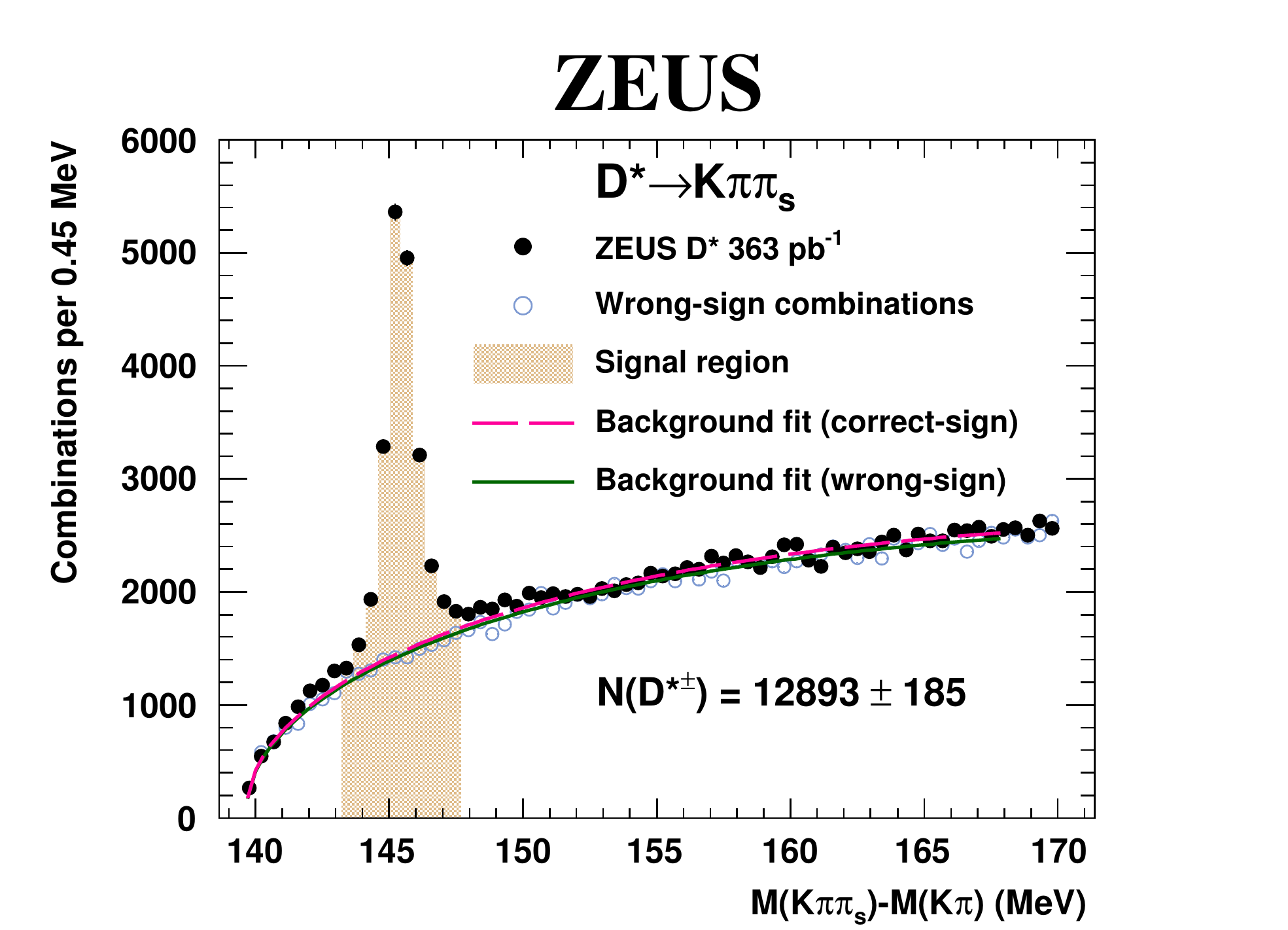}
  \caption[Distributions of $\Delta M$ from H1 and ZEUS \Dstar measurements]
	{Distributions of the reconstructed mass difference $\Delta M$ from the H1~\cite{h1dstar_hera2} (left) and ZEUS~\cite{zeusdstar_hera2} (right) \Dstar measurements, respectively.}
  \label{fig:hera:hfmeas:dstar:mass}
\end{figure*}

Distributions of the reconstructed mass difference $\Delta M$ for the most precise H1 and ZEUS HERA-II measurements~\cite{zeusdstar_hera2,h1dstar_hera2} are shown in Fig.~\ref{fig:hera:hfmeas:dstar:mass}. 
Note that these measurements are performed in slightly different ranges of $p_T(\Dstar)$ and $\eta(\Dstar)$, therefore 
the ZEUS measurement has a better signal-to-background ratio and narrower peak at the cost of two times smaller statistics. 
Both experiments performed a subtraction of the background using the wrong-sign combinations, obtained by forming ``$D^0$ candidates'' by combining two tracks with the same sign.

\begin{figure}[htbp]
  \centering
  \includegraphics[width=0.495\figwidth,trim = 0 30mm 6mm 10mm,clip=true]{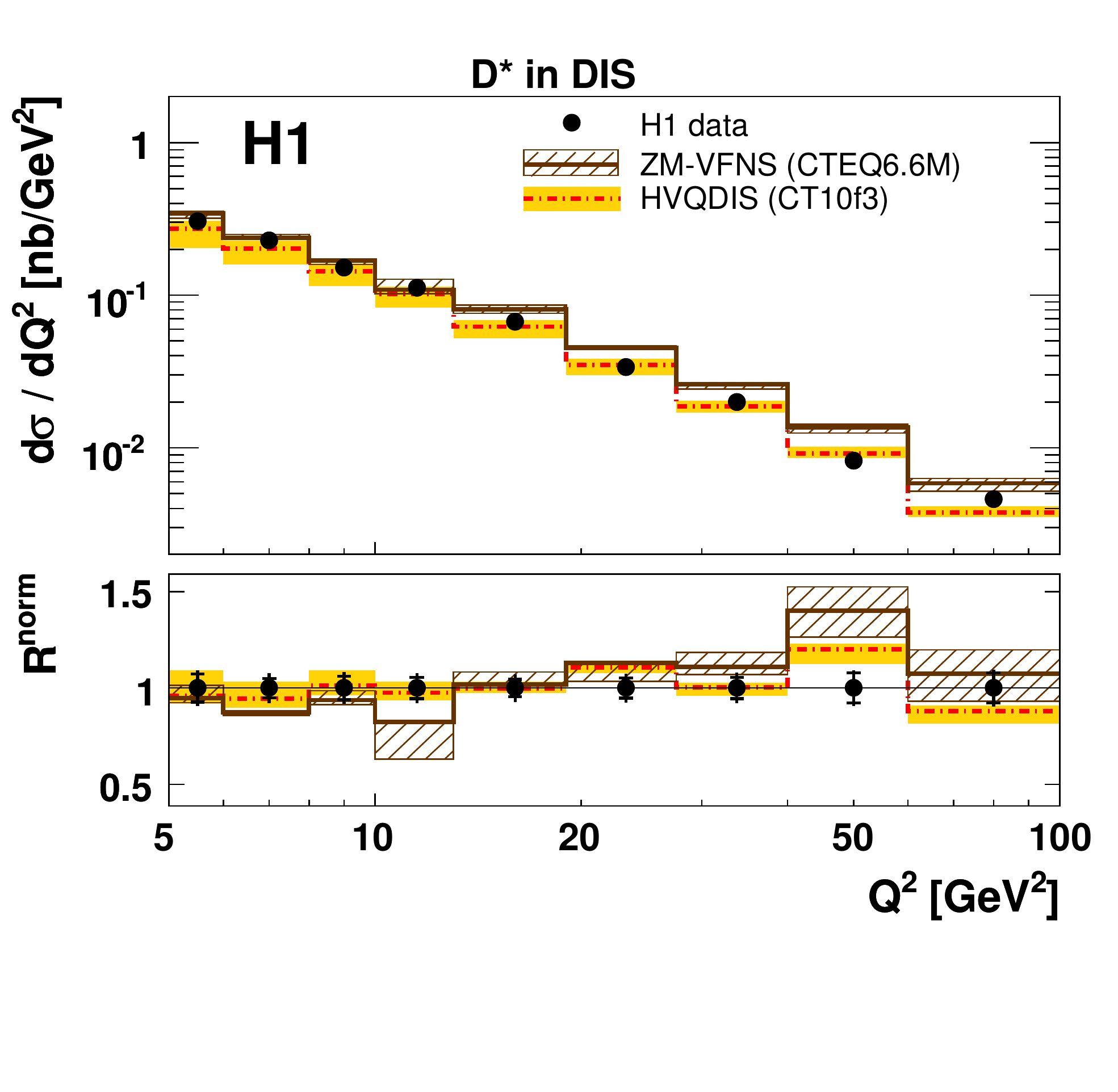}
  \put(-17,57){(a)}
  \includegraphics[width=0.495\figwidth,trim = 0 30mm 6mm 10mm,clip=true]{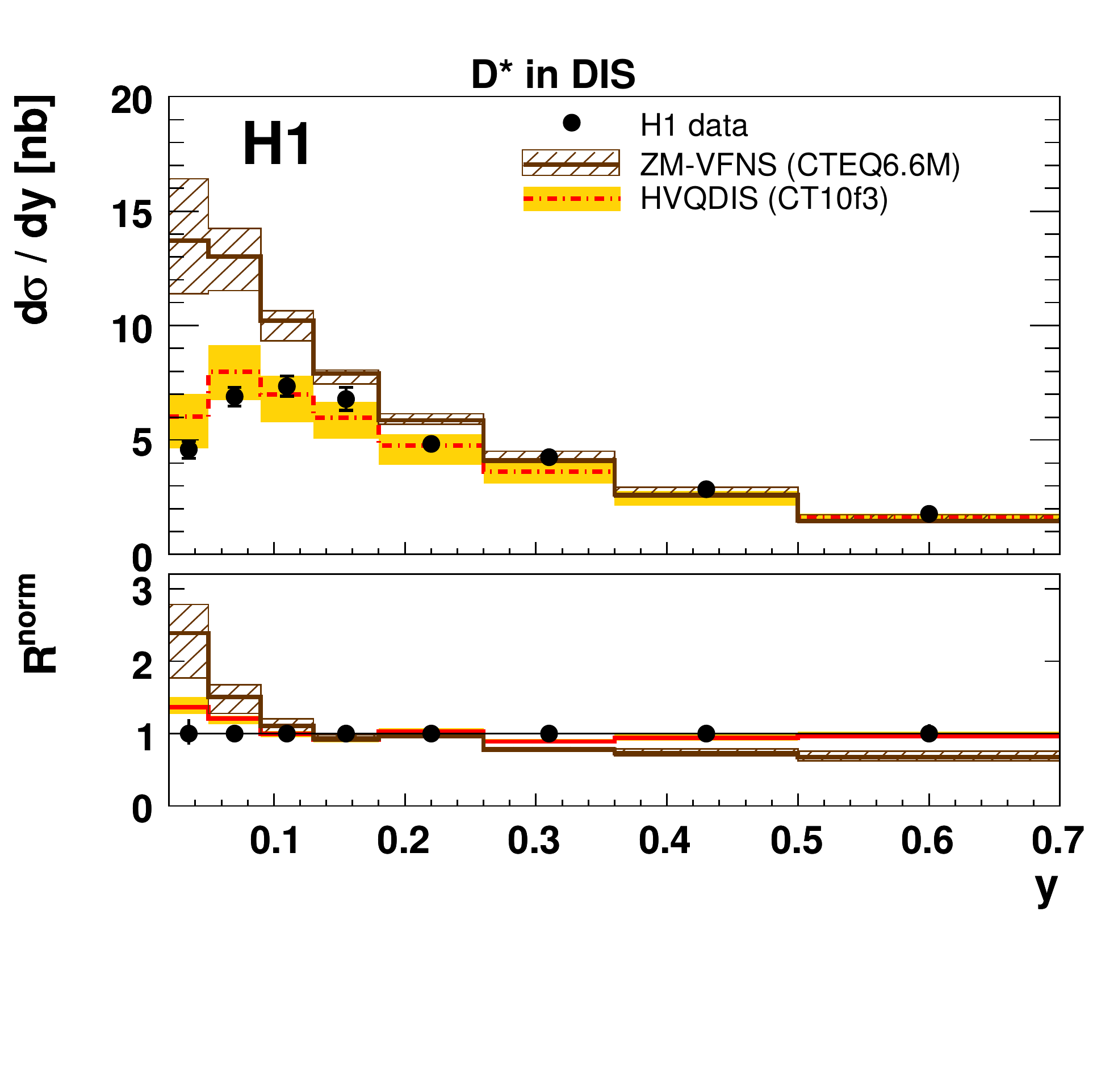}
  \put(-17,57){(b)}
  \newline
  \includegraphics[width=0.495\figwidth,trim = 0 30mm 6mm 10mm,clip=true]{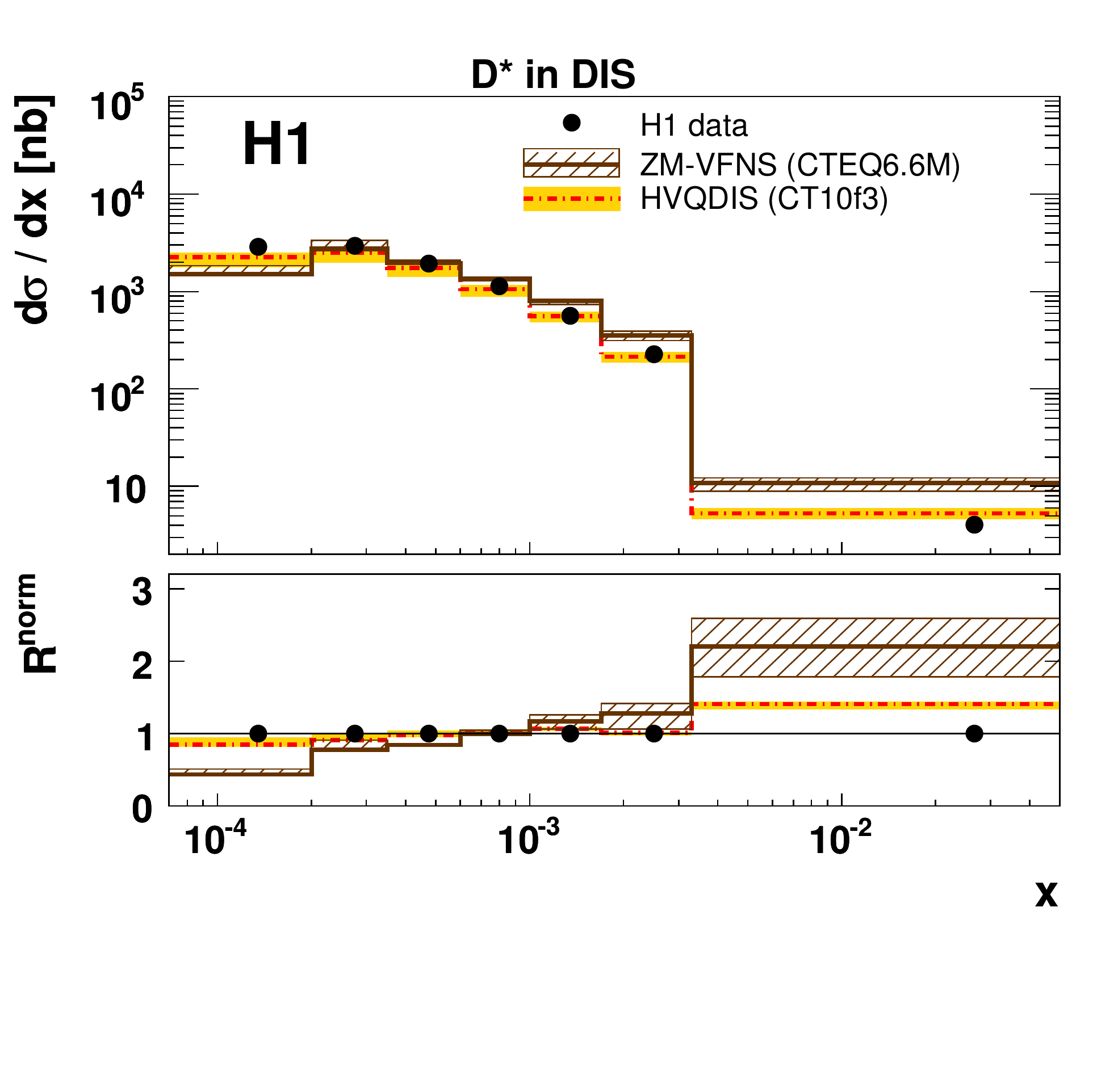}
  \put(-17,57){(c)}
  \includegraphics[width=0.495\figwidth,trim = 0 30mm 6mm 10mm,clip=true]{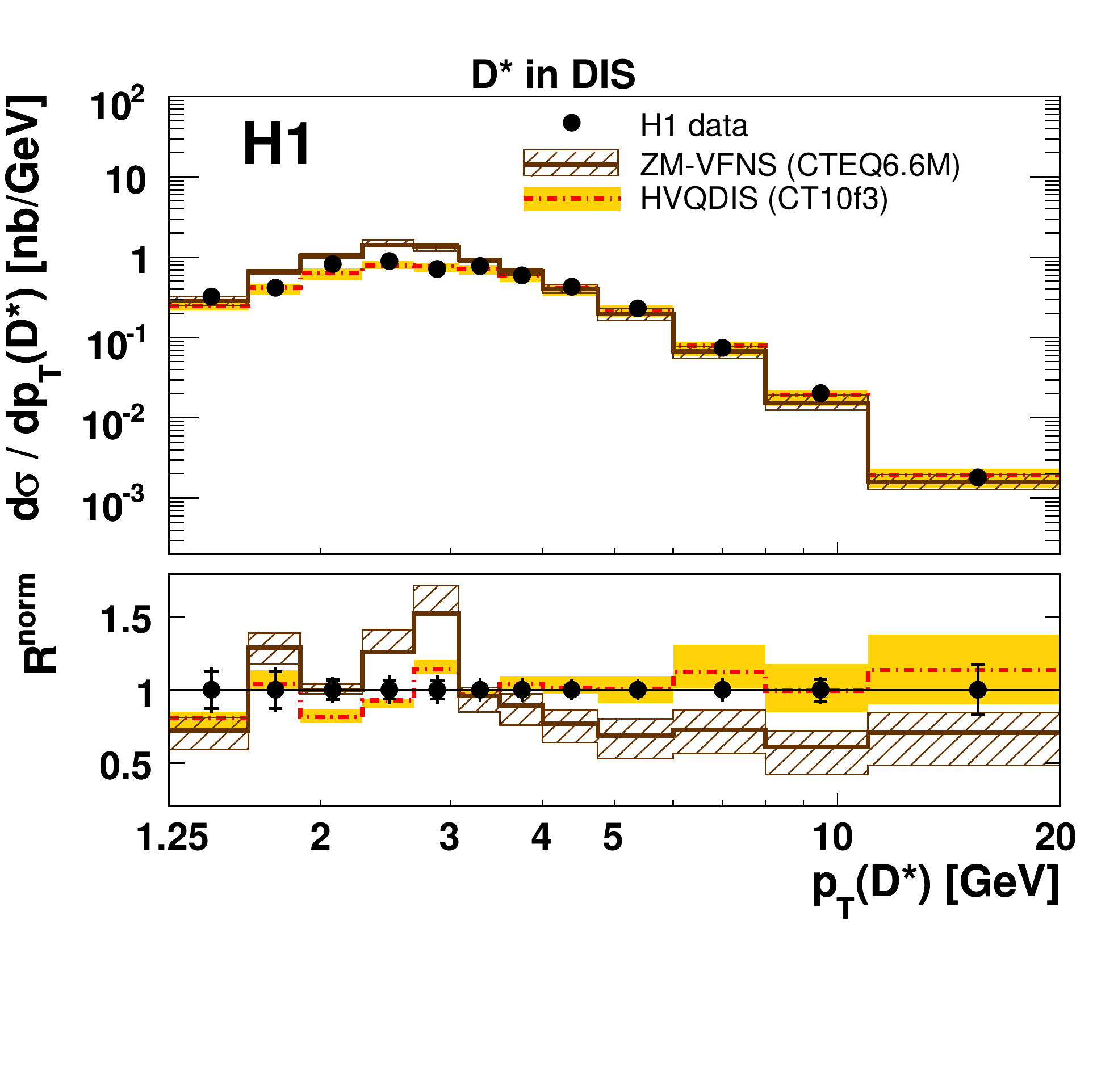}
  \put(-17,57){(d)}
  \newline
  \includegraphics[width=0.495\figwidth,trim = 0 30mm 6mm 10mm,clip=true]{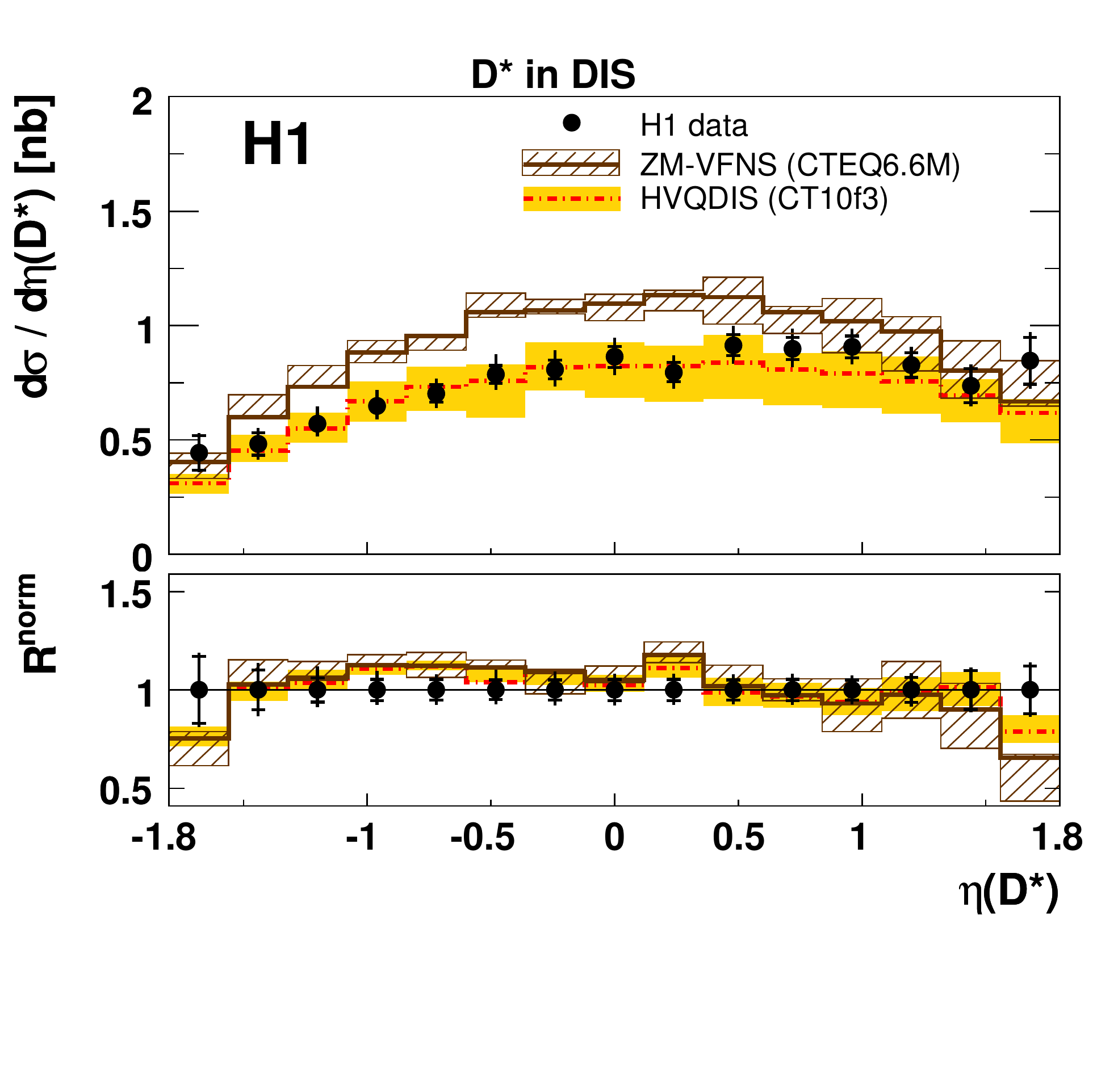}
  \put(-17,45){(e)}
  \includegraphics[width=0.495\figwidth,trim = 0 30mm 6mm 10mm,clip=true]{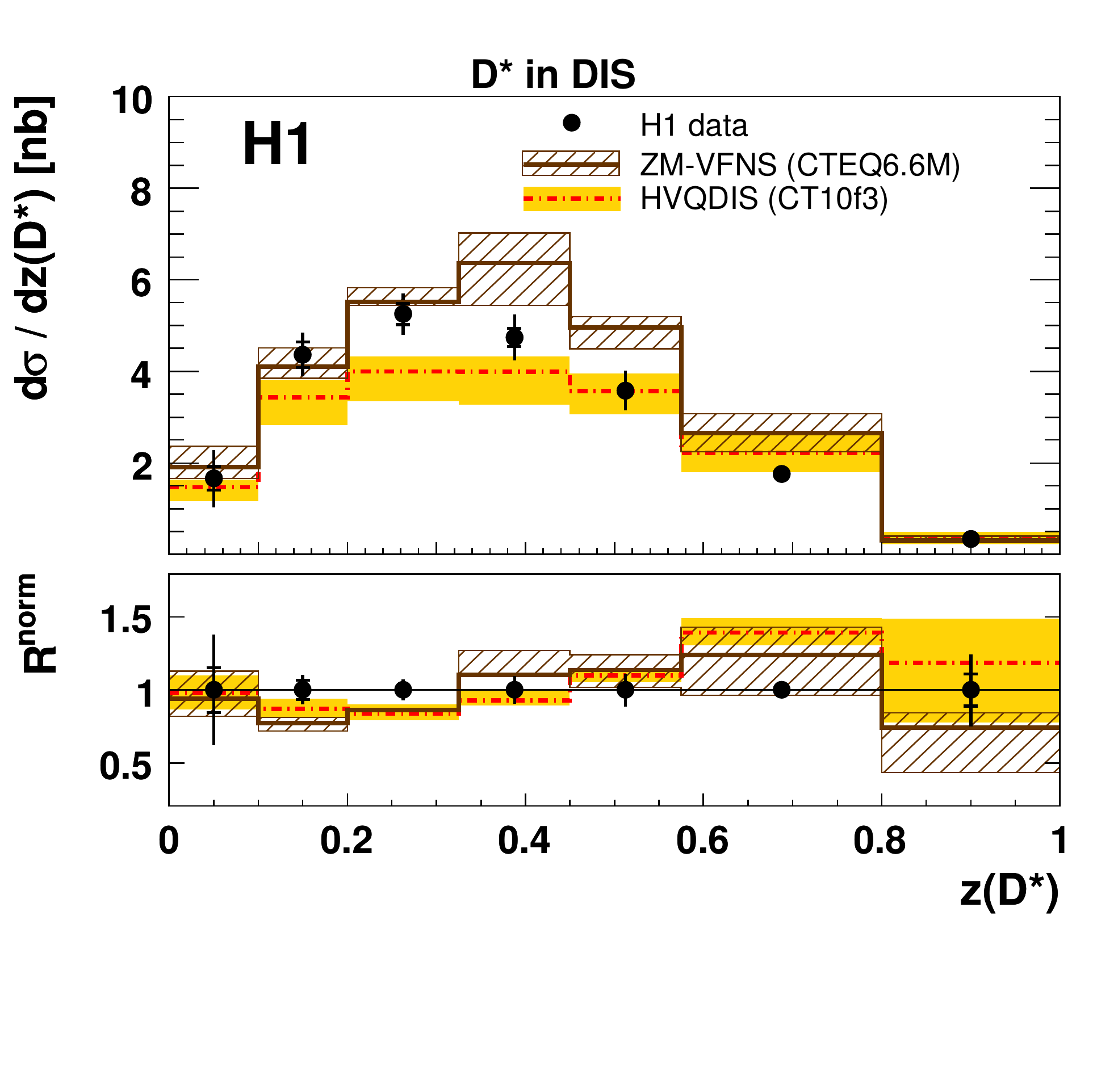}
  \put(-17,47){(f)}
  \caption[Differential \Dstar cross sections from H1 HERA-II measurement]
	{Differential \Dstar cross sections as a function of $Q^2$ (a), $y$ (b), $x$ (c), 
	$p_T(\Dstar)$ (d), $\eta(\Dstar)$ (e) and $z(\Dstar)$ (f), measured in~\cite{h1dstar_hera2}. 
	The data are compared to NLO predictions obtained in the ZM-VFNS and FFNS (HVQDIS). 
	In the lower part of the figures the normalised ratio, $R^{\rm norm}$, of theory to data is shown, 
	defined in Eq.~{3} of~\cite{h1dstar_hera2}, 
	which has reduced normalisation uncertainties.}
  \label{fig:hera:hfmeas:dstar:h1cs}
\end{figure}

The measured cross sections of \Dstar production as a function of $Q^2$, $y$, $x$, $p_T(\Dstar)$, $\eta(\Dstar)$ and $z(\Dstar)= (E(\Dstar)-p_Z(\Dstar))/(2E_e y)$, 
with $E_e$ being the incoming electron energy, $E(\Dstar)$ and $p_Z(\Dstar)$ the energy and longitudinal momentum of \Dstar, respectively, 
are shown in Fig.~\ref{fig:hera:hfmeas:dstar:h1cs} 
and compared to the NLO predictions, obtained in the ZM-VFNS and FFNS (see Sections~\ref{sec:th:hq:treat} and~\ref{sec:th:hq:ep}). 
The dominant experimental uncertainty is the systematic uncertainty on the tracking efficiency ($\approx 4\%$); 
in most of the bins the statistical uncertainty is smaller than the total systematical one.
The FFNS predictions describe the data reasonably well within uncertainties, with a possible exception for the shape of the $z(\Dstar)$ distribution.
The ZM-VFNS predictions describe the data significantly less well; in particular, they fail to describe the shape of $p_T(\Dstar)$, $y$ and $x$ distributions.

\subsection{Reconstruction of weakly decaying $D$ mesons}
\label{sec:exp:hera:hfmeas:dweak}

\ozmodN{The exploitation of the long lifetime of weakly decaying charmed hadrons allows their identification.} 
All final decay products must be charged particles reconstructed in the tracking system. 
Examples of such decay channels are $D^{+} \to K^{-}\pi^{+}\pi^{+}$ and $D^{0} \to K^{-}\pi^{+}$.
Large combinatorial background can be significantly suppressed by applying a cut on lifetime information 
(e.g.\ track impact parameters or decay-length significance), although since the background rises steeply 
towards lower values of $p_T(D)$, a lower cut on $p_T(D)$ has to be applied; a cut on 
transverse momentum also improves the \ozmodN{efficiency} of the lifetime information. 
\ozmodN{It should be noted that there are limitations of this technique, which are similar to those of the previous one:} 
a measurement can be performed only in a fiducial transverse-momentum and pseudorapidity phase-space region 
and the branching ratios are small. 

ZEUS measured the production of $D^{0}$~\cite{zd0dp} and $D^{+}$~\cite{zeusdch_hera2} mesons using the weak decays $D^{0} \to K^{-}\pi^{+}$ and $D^{+} \to K^{-}\pi^{+}\pi^{+}$, respectively. 
The measurement of $D^0$ production was based on the $\SI{134}{pb^{-1}}$ of data from 2005 only,
while the measurement of $D^{+}$ production used the full HERA-II data of $\SI{354}{pb^{-1}}$.%
\footnote{The \Dch measurement~\cite{zeusdch_hera2} superseded the previous measurement of $D^{+}$ production in~\cite{zd0dp}, based on data from 2005.}
Both measurements were performed in the phase-space region $p_T(D^{+},D^0)>\SI{1.5}{GeV}$, $|\eta(D^{+},D^0)|<1.6$, $5<Q^2<\SI{1000}{GeV^2}$, $0.02<y<0.7$.
Lifetime information was used to reduce combinatorial background substantially, applying a cut on the decay-length significance of the secondary vertex.
This technique benefits from the MVD tracking and vertexing, \ozmodN{which is not} feasible using the HERA-I data.
\ozmodN{The measurement of $D^{+}$ production is one of the important results further described in detail in Section~\ref{sec:dch}, where also 
an example of an event with a selected \Dch candidate can be seen in Fig.~\ref{fig:zevis_zeviscn_analysistrkvtx}.}

\subsection{Usage of semi-leptonic decays}
\label{sec:exp:hera:hfmeas:slep}

\ozmodNN{Charmed particles with semi-leptonic decays can be identified} 
using discriminating variables, e.g.\ the missing transverse momentum caused by a neutrino or the impact parameter of the lepton track.
The measurements benefit from large branching ratios and a better pseudorapidity coverage at the cost of a worse signal-to-background ratio. 
		
ZEUS measured charm and beauty production \ozmodNN{exploiting} their decays into muons~\cite{zeus_muon}.
The measurement was based on the $\SI{134}{pb^{-1}}$ of data from 2005.
The measured observables were cross sections of muons originating from charm and beauty \ozmodNN{decays}. 
The fractions of muons originating from charm, beauty and light flavours were extracted by \ozmodNN{using} three discriminating variables: 
the muon impact parameter, the muon momentum component transverse to the associated jet axis, and the missing transverse momentum, 
which is sensitive to the neutrino from semi-leptonic decays.
The kinematic \ozmodNN{region} of the measurement was $p_T(\mu)>\SI{1.5}{GeV}$, $-1.6<\eta(\mu)<2.3$, 
$Q^2>\SI{20}{GeV^2}$ and $0.01<y<0.7$ (note \ozmodNN{the} extended coverage of the forward region compared to $D$ measurements).

\begin{figure}[htbp]
  \centering
  \includegraphics[width=1.0\figwidth,trim = 6mm 7mm 17mm 9mm,clip=true]{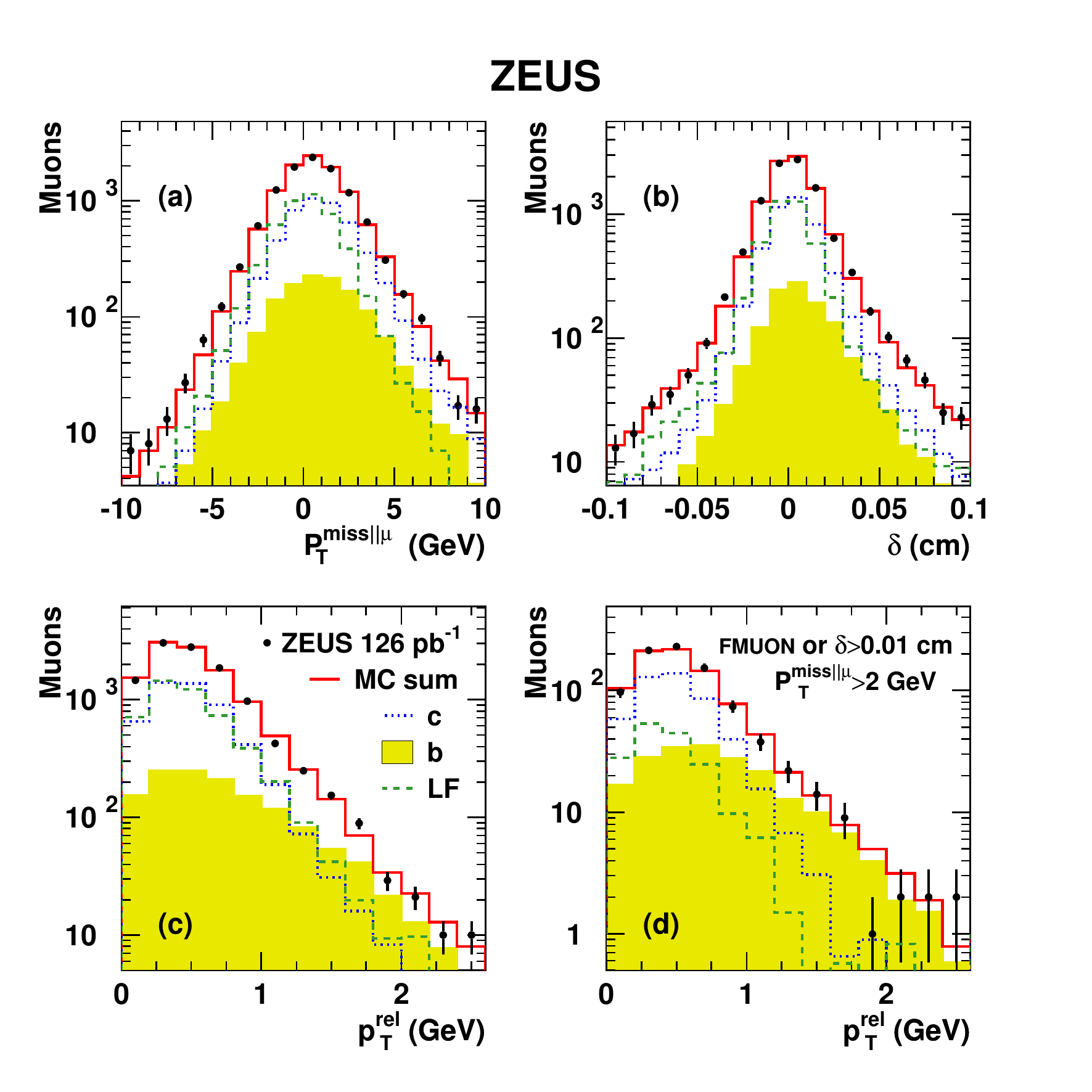}
  \caption[Distributions of discriminating variables from ZEUS muon measurement]
  {Distributions of the discriminating variables from ZEUS muon measurement~\cite{zeus_muon}: the missing transverse momentum, $p_T^{{\rm miss}||\mu}$, (a), 
  muon impact parameter, $\delta$, (b), muon momentum component transverse to the axis of the associated jet\ozmod{, 
  $p_T^{\rm rel} = |\mathbf{p}^{\mu} \times \mathbf{p}^{\rm jet}| / |\mathbf{p}^{\rm jet}|$}, (c) and 
  \ozmodN{$p_T^{\rm rel}$ for a heavy-flavour-enriched sample with $p_T^{{\rm miss}||\mu} > 2$~GeV 
  and either a muon in the forward tracking detector of the muon system (FMUON) or $\delta > 0.01$~cm (d).} 
  The data are compared to the MC expectation 
  with the normalisation of the charm, beauty and light-flavour, LF, components obtained from the global fit. 
  The charm, beauty and light-flavour contributions are shown separately. 
  \label{fig:hera:hfmeas:slep:zeusmuon}}
\end{figure}

\begin{figure}[htbp]
  \centering
  \includegraphics[width=1.0\figwidth,trim = 6mm 8mm 16mm 9mm,clip=true]{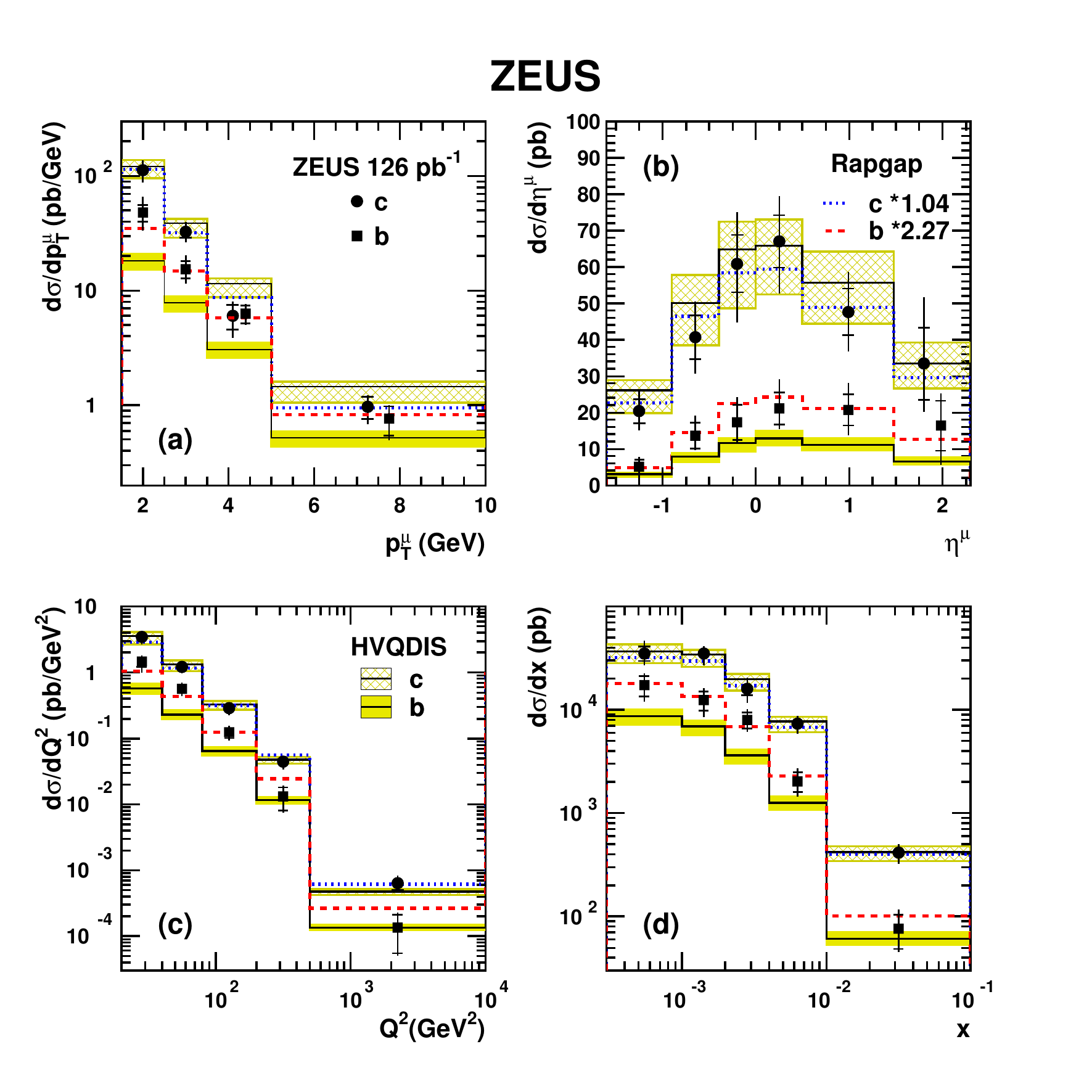}
  \caption[Differential muon cross section for charm and beauty production from ZEUS]
	{Differential muon cross section for charm and beauty production as a function of $p_T(\mu)$ (a), $\eta(\mu)$ (b),	$Q^2$ (c) and $y$ (d), measured in~\cite{zeus_muon}. 
	The data are compared to the NLO predictions obtained in the FFNS (HVQDIS) and to the predictions from MC RAPGAP.}
  \label{fig:hera:hfmeas:slep:zeusmuoncs}
\end{figure}

Distributions of the discriminating variables are shown in Fig.~\ref{fig:hera:hfmeas:slep:zeusmuon}. 
Contributions from charm and beauty production are separated from light flavours and from each other by using a global template fit to the Monte Carlo (MC) expectation.
The measured muon differential cross sections as a function of $p_T(\mu)$, $\eta(\mu)$, $Q^2$ and $x$ are shown in Fig.~\ref{fig:hera:hfmeas:slep:zeusmuoncs} 
and compared to the NLO predictions, obtained in the FFNS, and RAPGAP MC, normalised according to the result of the global fit. 
The NLO FFNS predictions describe the data well.
The RAPGAP MC gives a good description of the shape of all the differential cross sections. 
\ozmodN{Since MC was normalised according to the result of the fit, this can be \ozmodNN{considered} as a verification of the validity of the fit results.}

\subsection{Fully inclusive analyses based on lifetime information}
\label{sec:exp:hera:hfmeas:secvtx}

\ozmodNN{Events with charmed particles} are identified by reconstruction of displaced secondary vertices based on the lifetime information.
\ozmodN{The measurement results benefit from the larger phase-space coverage} and largest statistics, since they are not limited by any particular branching ratio, 
although the signal-to-background ratio is usually worst.

\begin{figure}[htbp]
  \centering
  \includegraphics[width=0.495\figwidth,trim = 9mm 0mm 0mm 2mm,clip=true]{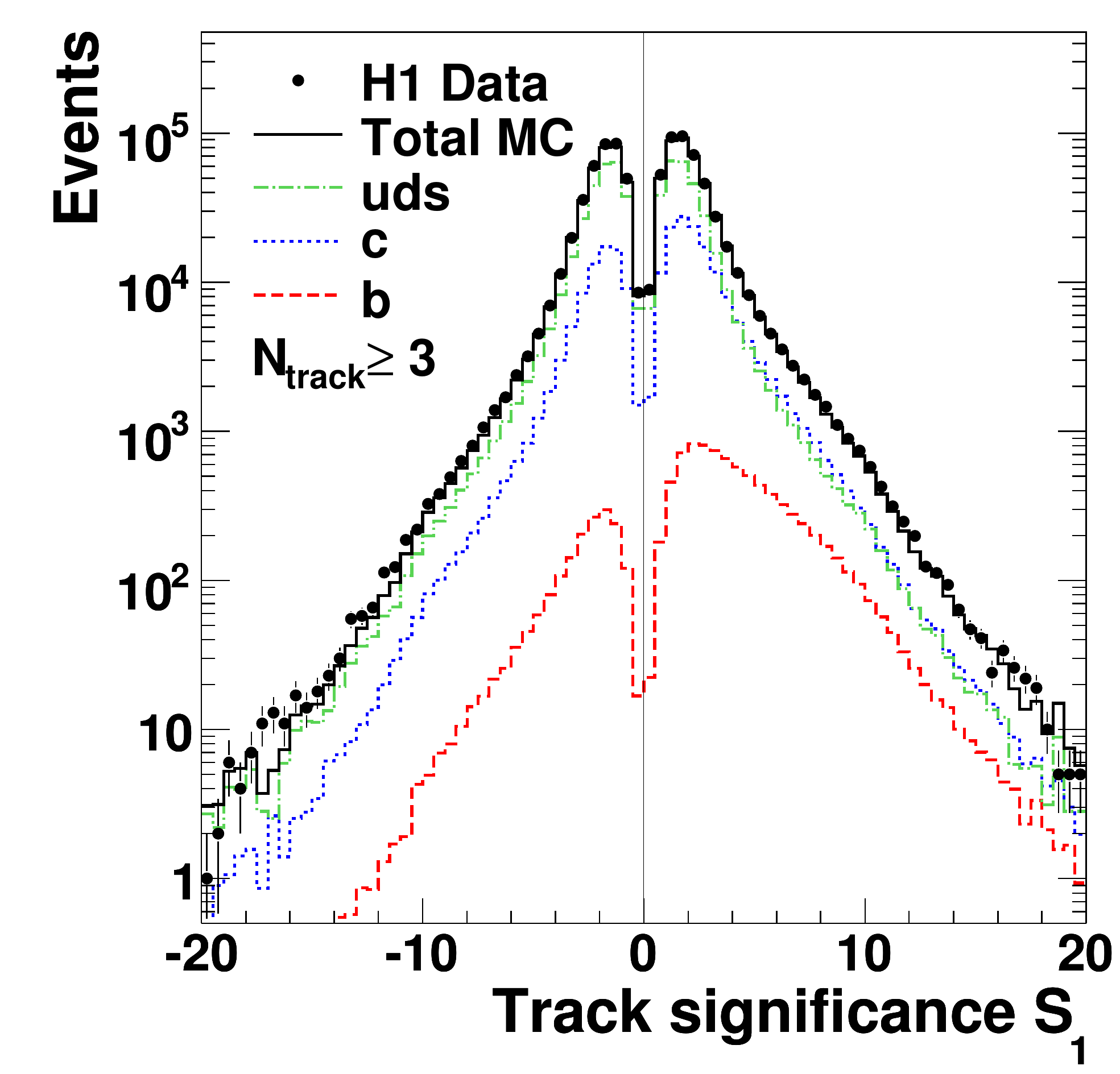}
  \includegraphics[width=0.495\figwidth,trim = 9mm 0mm 0mm 2mm,clip=true]{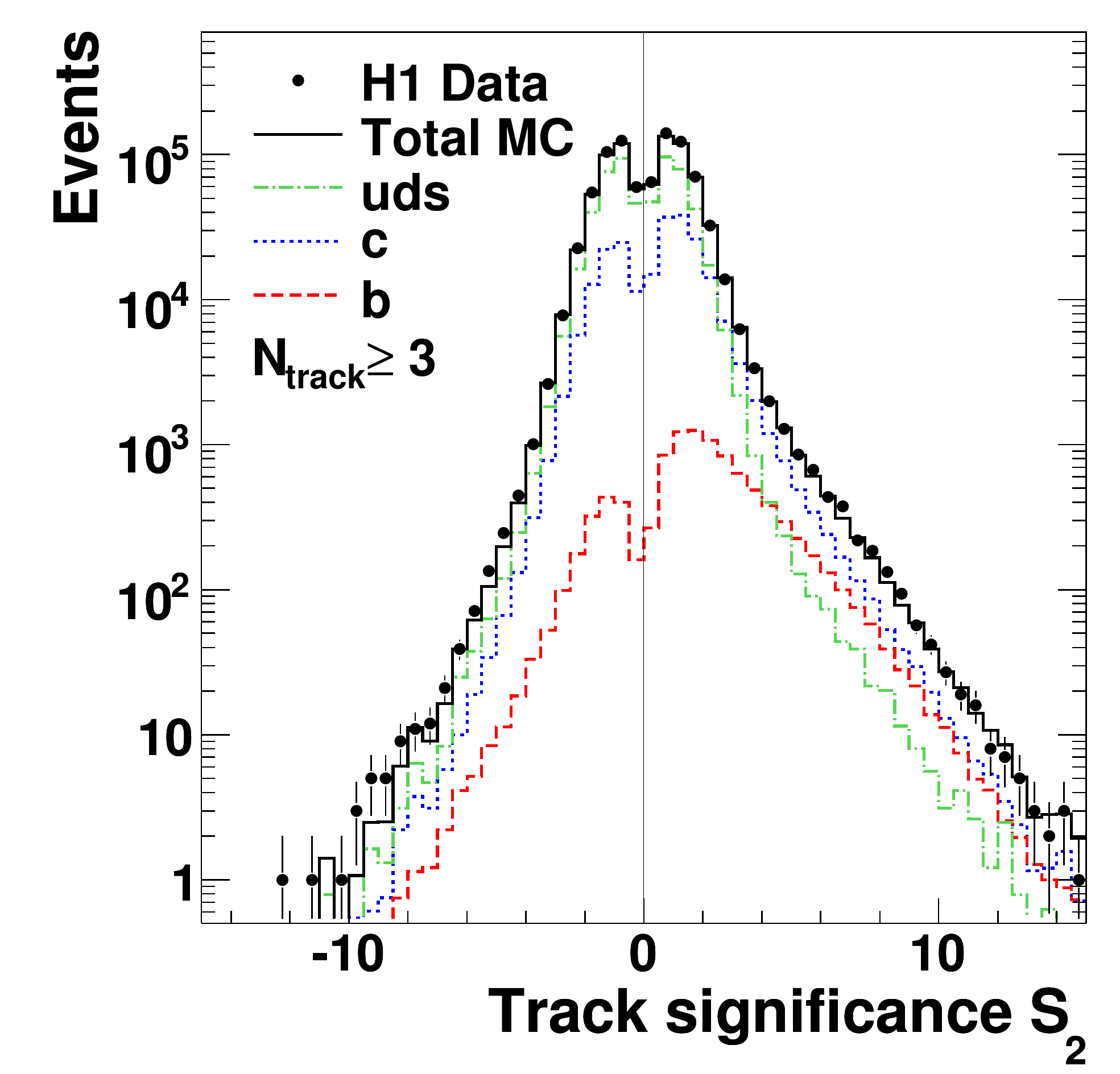}
  \includegraphics[width=0.495\figwidth,trim = 9mm 0mm 0mm 2mm,clip=true]{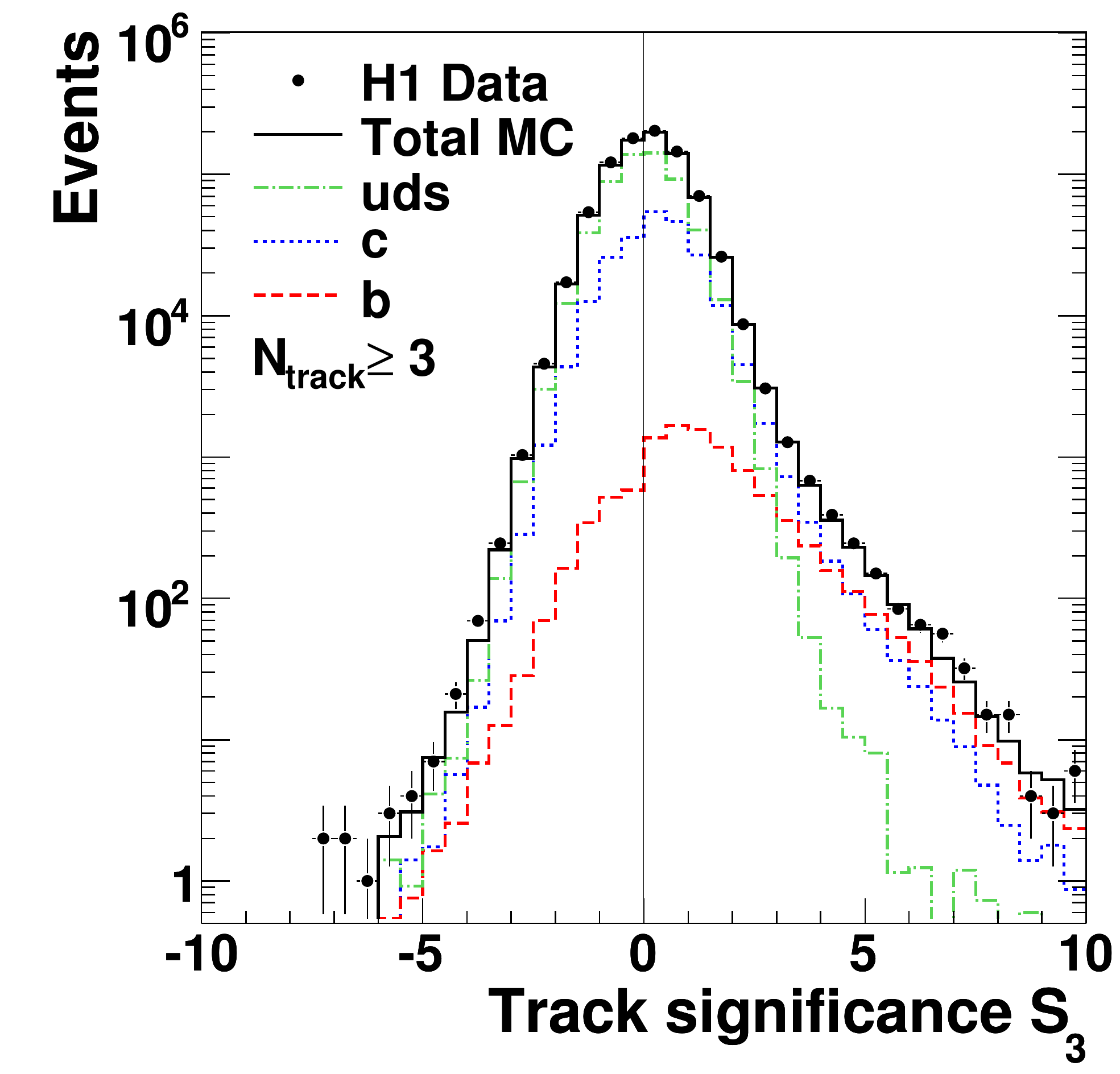}
  \includegraphics[width=0.495\figwidth,trim = 9mm 0mm 0mm 2mm,clip=true]{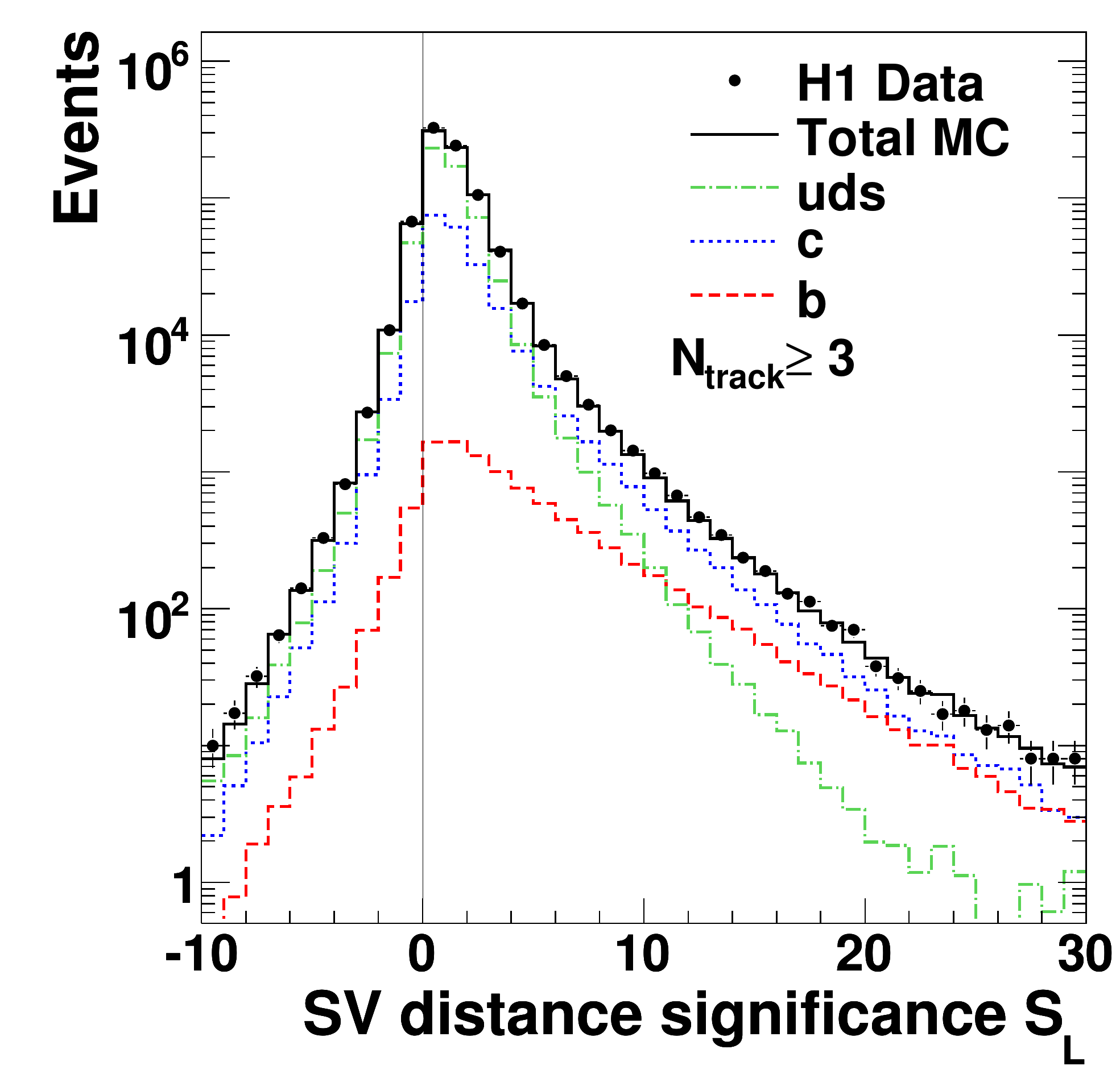}
  \caption[Distributions of discriminating variables from H1 vertex measurement]
  {Distributions of discriminating variables from the H1 vertex measurement~\cite{h1ltt_hera2}: 
  the impact-parameter significances, defined as the significance of the track with the highest, $S_1$, (top left), 
  second highest, $S_2$, (top right) and third highest, $S_3$, (bottom left) absolute significances, respectively, 
  and the secondary-vertex significance, $S_L$, (bottom right). The data are compared to the MC expectation, 
  obtained after applying the scale factors from the fit to the complete data sample. 
  The charm, beauty and light-flavour contributions are shown separately.}
  \label{fig:hera:hfmeas:secvtx:h1}
\end{figure}

H1 measured inclusive charm and beauty cross sections using variables reconstructed by the vertex detector, 
including the impact parameter of tracks to the primary vertex and the position of the secondary vertex~\cite{h1ltt_hera2}. 
The measurement was based on the $\SI{189}{pb^{-1}}$ of data from 2006--2007. 
The phase space of the measurement was $5<Q^2<\SI{2000}{GeV^2}$ and $0.0002<x<0.05$.
Similar to the technique used for measurements with semi-leptonic decays, described in Section~\ref{sec:exp:hera:hfmeas:slep}, 
this measurement was based on the discrimination of charm and beauty contributions, performed with a neural network, using long-lifetime discriminating variables.
Fig.~\ref{fig:hera:hfmeas:secvtx:h1} shows the distributions of the discriminating variables, used as input for the neural network. 
The measured quantities were the charm and beauty reduced cross sections as a function of $Q^2$ and $x$ in the full $p_T$ and $\eta$ range.
The measurement~\cite{h1ltt_hera2} was then combined with previous H1 measurements~\cite{Aktas:2004az,Aktas:2005iw} based on HERA-I data.

\begin{figure}[htbp]
  \centering
  \includegraphics[width=1.0\figwidth,trim = 1mm 1mm 9mm 6mm,clip=true]{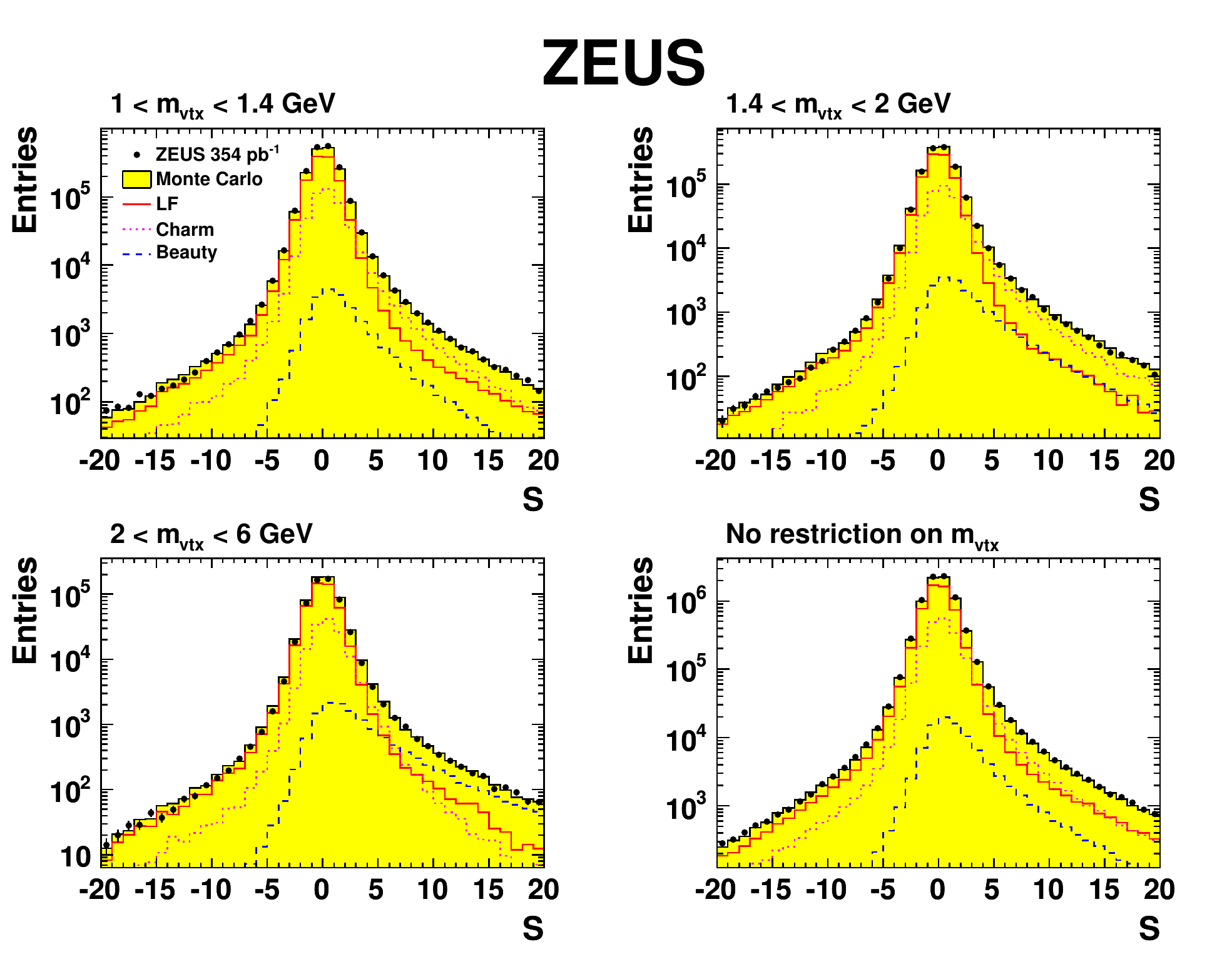}
  \caption[Distributions of discriminating variables from ZEUS vertex measurement]
  {Distributions of the decay-length significance, $S$, for different bins of the secondary-vertex mass, $m_{\rm vtx}$: 
  $1<m_{\rm vtx}<\SI{1.4}{GeV}$ (top left), $1.4<m_{\rm vtx}<\SI{2}{GeV}$ (top right), $2<m_{\rm vtx}<\SI{6}{GeV}$ (bottom left) and no restriction on $m_{\rm vtx}$ (bottom right) from the ZEUS vertex measurement~\cite{zeussecvtx_hera2}. 
  The data are compared to the sum of all MC distributions. 
  The individual contributions from the beauty, charm and light-flavour MC subsamples are shown separately.}
  \label{fig:hera:hfmeas:secvtx:zeus}
\end{figure}

\begin{figure}[htbp]
  \centering
  \includegraphics[width=0.495\figwidth,trim = 0 0 0 0,clip=true]{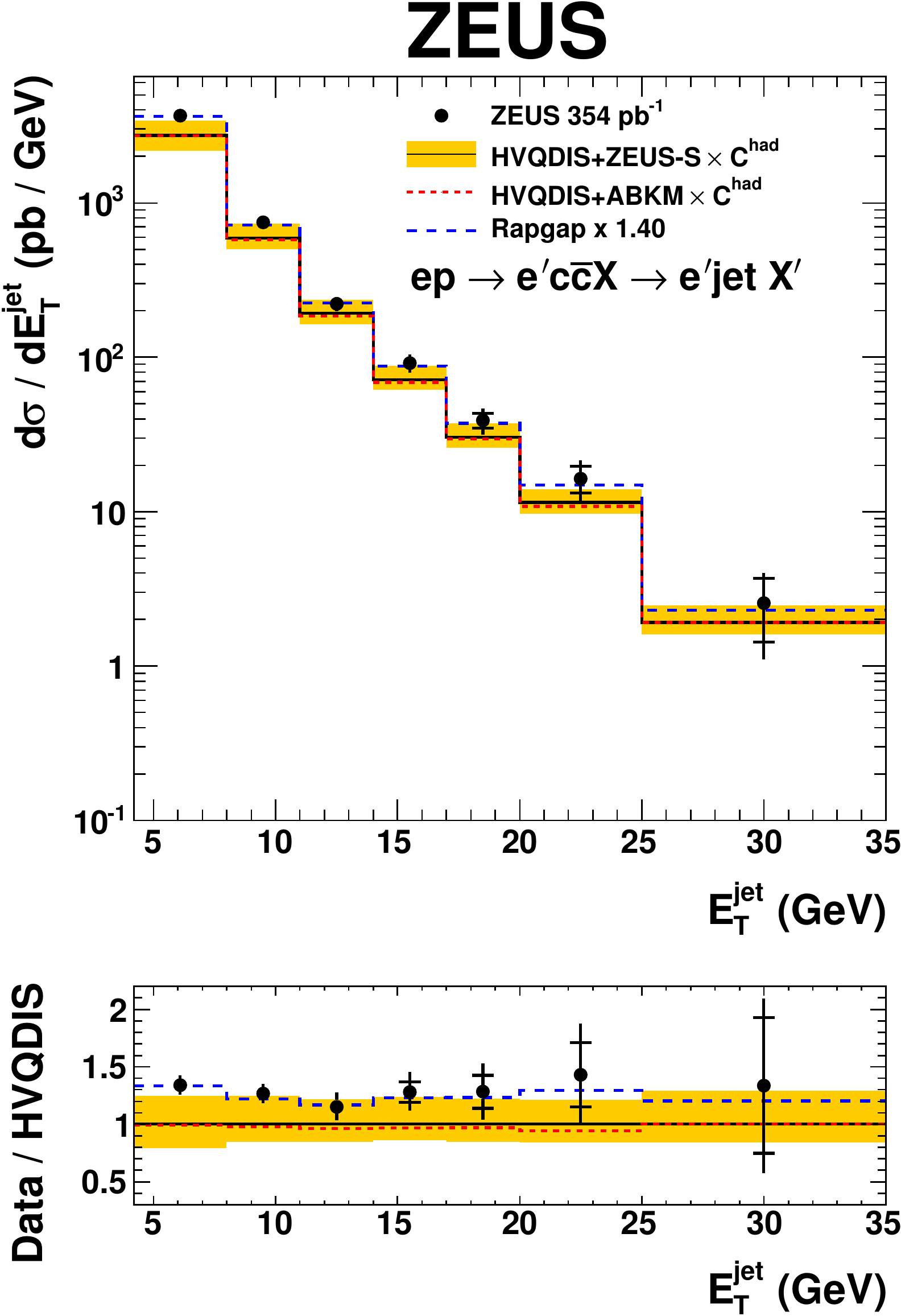}
  \includegraphics[width=0.495\figwidth,trim = 0 0 0 0,clip=true]{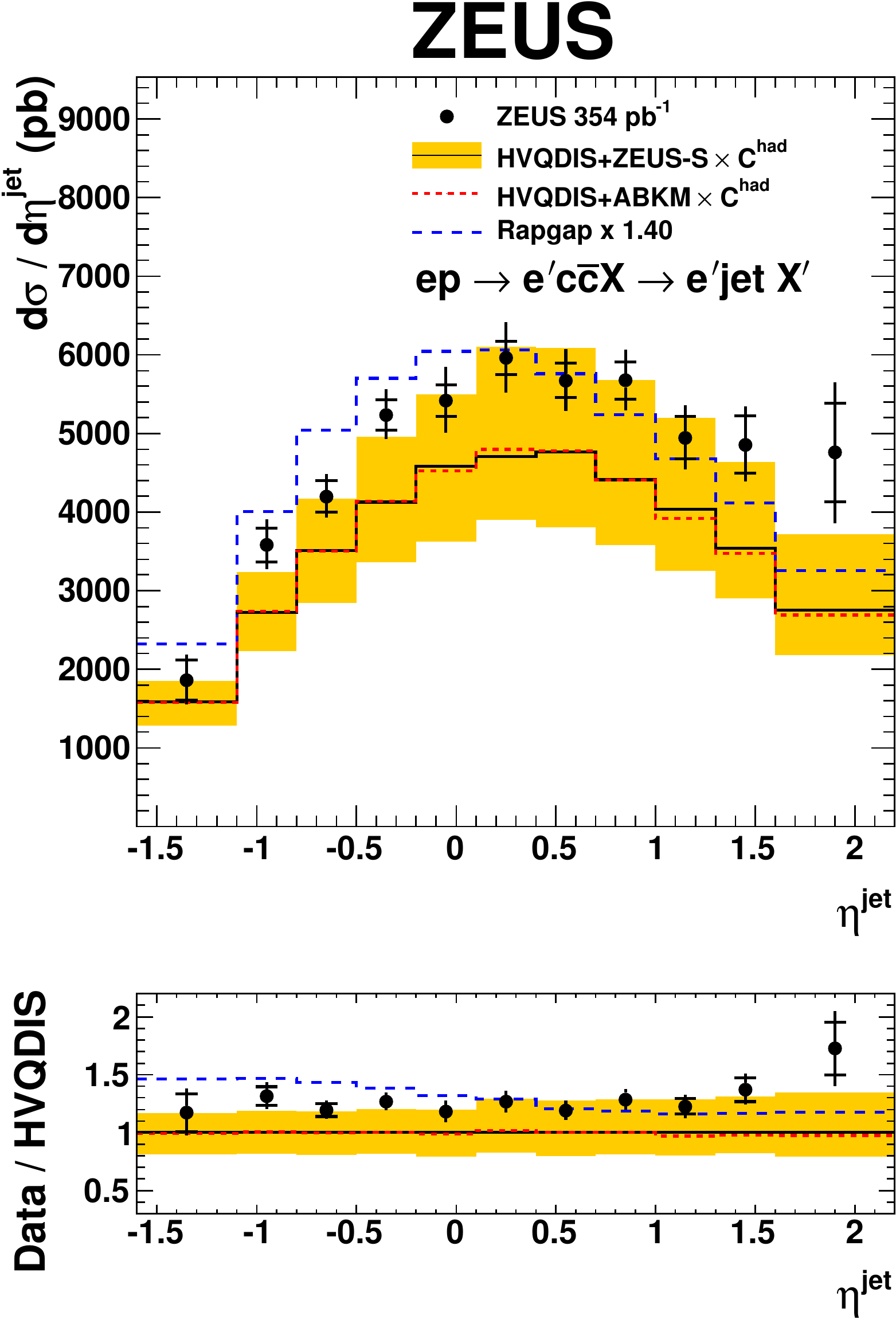}
  \includegraphics[width=0.495\figwidth,trim = 0 0 0 0,clip=true]{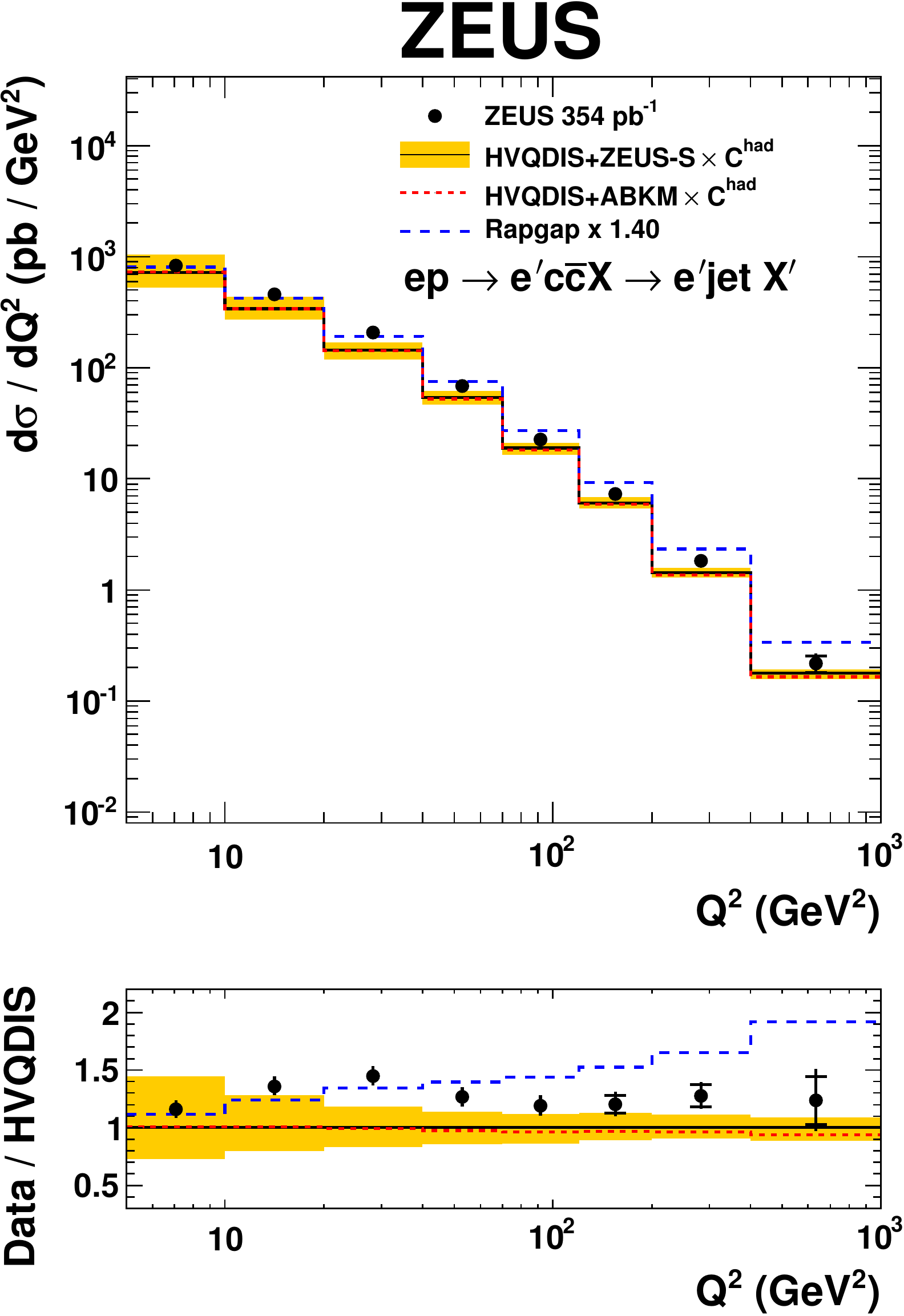}
  \includegraphics[width=0.495\figwidth,trim = 0 0 0 0,clip=true]{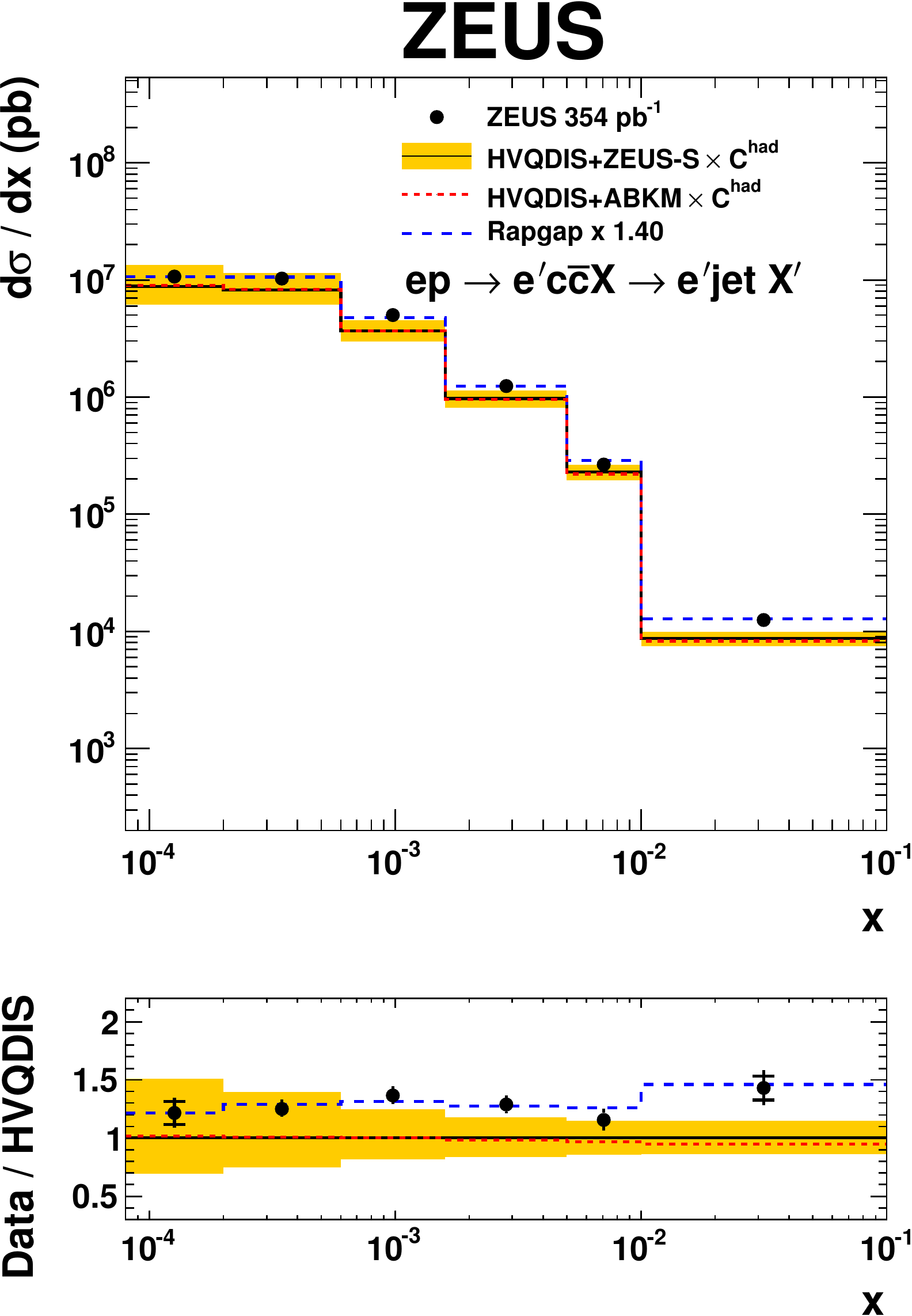}
  \caption[Differential cross section for inclusive jet production in charm events from ZEUS]
	{Differential cross section for inclusive jet production in charm events as a function of 
	$E_T^{\rm jet}$ (top left), $\eta^{\rm jet}$ (top right), $Q^2$ (bottom left) and $x$ (bottom right) measured in~\cite{zeussecvtx_hera2}. 
	The data are compared to the NLO predictions obtained in the FFNS (HVQDIS) with different input PDFs and to the predictions from MC RAPGAP.}
  \label{fig:hera:hfmeas:secvtx:zeuscs}
\end{figure}

ZEUS measured the production of charm and beauty with at least one jet using the invariant mass of the charged tracks associated with secondary vertices 
and the decay-length significance of these vertices~\cite{zeussecvtx_hera2}.
The measurement was based on the full HERA-II data of $\SI{354}{pb^{-1}}$.
The kinematic phase \ozmodNN{space} of the charm measurement was $E_T^{\rm jet}>\SI{4.2}{GeV}$, $-1.6<\eta^{\rm jet}<2.2$, $5<Q^2<\SI{1000}{GeV^2}$ and $0.02<y<0.7$, 
where $E_T^{\rm jet}$ is the transverse energy of the jet. 
Contributions from charm and beauty production were separated from light flavours and from each other by using a global template fit to the MC expectation.
Fig.~\ref{fig:hera:hfmeas:secvtx:zeus} shows the distributions of the decay-length significance for different bins of the secondary-vertex mass, $m_{\rm vtx}$. 
All MC samples were normalised according to the scaling factors obtained from the fit.
A good agreement between data and MC is observed. 
The first two mass bins corresponding to the region $1 < m_{\rm vtx} < \SI{2}{GeV}$ are dominated by charm events. 
In the third mass bin, $2 < m_{\rm vtx} < \SI{6}{GeV}$, beauty events are dominant at high values of the decay-length significance. 
The measured differential cross sections for inclusive jet production in charm events as a function of $E_T^{\rm jet}$, $\eta^{\rm jet}$, $Q^2$ and $x$ are shown in Fig.~\ref{fig:hera:hfmeas:secvtx:zeuscs} 
and compared to the NLO predictions obtained in the FFNS with different proton PDFs and to the predictions from the RAPGAP MC, 
scaled to the ratio of the measured visible cross section to the RAPGAP prediction. 
\ozmod{All measured cross sections are better described
by the NLO FFNS, while RAPGAP provides a worse description of the shape of the
charm cross sections than the NLO FFNS calculations. The data are typically $20\text{--}30\%$ above the
NLO predictions, but in agreement within uncertainties.} The differences
between the NLO predictions using different proton PDFs are mostly very small.

\subsection{\ozmod{Concluding remarks}}
\label{sec:exp:hera:hfmeas:summary}

Different \ozmodN{tagging} techniques have been used to measure open charm production in DIS at HERA. 
The most precise results were obtained in measurements of \Dstar production using the ``golden'' decay channel.
In all cases (except for the H1 vertex measurement~\cite{h1ltt_hera2}) the measured quantities were visible cross sections in a limited $p_T$ ($E_T^{\rm jet}$) and $\eta$ phase-space region.
The largest phase-space coverage was obtained in fully inclusive analyses based on lifetime information.
Techniques based on the usage of semi-leptonic decays and fully inclusive analyses are often used for a simultaneous measurement of charm and beauty production, 
while in measurements using the full reconstruction of $D$ mesons usually the sum of heavy-flavoured hadron production \ozmodN{of} charm and beauty processes are measured (dominated by charm). 
Techniques that rely on the usage of lifetime information require precise tracking and vertexing, 
thus can be fully exploited only with the data taken with the silicon detectors near the beampipe.

Results of precise charm measurements, which provide a double-differential cross section, are \ozmod{used} to extract the inclusive cross section, 
i.e.\ the charm structure function $F_2^{c\bar{c}}$ or the reduced cross section \red. 
The extraction is based on the extrapolation procedure, which used the shape of theoretical 
predictions, thus measurements with larger transverse-momentum and pseudorapidity phase-space coverage are preferable 
(more details on the extrapolation procedure are provided in Section~\ref{sec:comb:proc:pscorr} in the context of charm data combination).

All measured cross sections were compared to the theoretical predictions obtained in different schemes. 
The NLO FFNS predictions provide a good description of the data within uncertainties in all cases, 
while the NLO ZM-VFNS predictions do not describe the shape of some kinematic variables well.
A direct comparison of measured visible cross sections to the GM-VFNS predictions is not possible, 
since the GM-VFNS calculations were done only for inclusive cross sections.
Comparisons to MC predictions did not aim to check theory, since in these cases MC simulations were LO and parton showers, 
re-normalised to the data (more details on the technique of MC simulations are provided in Section~\ref{sec:dch:mc}). 
These comparisons mainly aimed \ozmodNN{at justifying} the validity of the template fit procedure or the acceptance corrections, 
which exploited MC.

In general, measurements performed using different methods are complementary to each other 
and thus can be combined to achieve the best precision. 
This is presented in Section~\ref{sec:comb:red}. 
Such combination requires an extrapolation of the visible cross sections to the full phase space using the shape of some theoretical calculations. 
Since the FFNS predictions are consistent with the data in all kinematic regions (including high $Q^2$), 
the NLO FFNS is considered \ozmod{to be} the best \ozmodNN{and presently the only practically available} theoretical calculation for this extrapolation. 


\clearpage
\section{Measurement of \Dch production}
\label{sec:dch}

\ozmod{Among the charm-tagging techniques, described in Section~\ref{sec:exp:hera:hfmeas}, 
measurements of \Dch-meson production are based on the full reconstruction of final-state charmed hadrons 
and crucially depend on the precise tracking and vertexing near the beampipe. 
This Section describes the measurement of \Dch-meson production with the ZEUS detector, using the full HERA-II data set.} 
Combinatorial background can be significantly suppressed by applying a cut on lifetime information.
The \ozmodN{earlier} ZEUS measurement of \Dch production~\cite{zd0dp} was performed using $\SI{134}{pb^{-1}}$ of data from 2005. 
It has demonstrated the high potential of this tagging method, although was not really competitive in precision to other ZEUS charm measurements, e.g.~\cite{zd00}.
The present measurement \ozmodN{is based on} \ozmodNN{about} $2.5$ times larger data sample and improved tracking alignment.
The results were published by the ZEUS Collaboration~\cite{zeusdch_hera2}. 
\ozmodN{The H1 measurement of \Dch-meson production~\cite{Aktas:2004ka} is based on the HERA-I data.}

Section~\ref{sec:dch:evrec} explains general aspects of the event reconstruction in ZEUS, relevant for the present analysis; 
it is partially based on~\cite{mishath}. 
Section~\ref{sec:dch:evsel} describes the selection of DIS events.
Section~\ref{sec:dch:mc} introduces the technique of MC simulations and provides information on MC samples, used in the analysis.
Section~\ref{sec:dch:dch} describes the reconstruction and selection of \Dch candidates, while 
Section~\ref{sec:dch:signal} explains the procedure of \ozmodNN{attributing these candidates to \Dch and background.} 
Section~\ref{sec:dch:cs} describes the cross-section determination procedure and \ozmodNN{corrections applied}. 
Details of the theoretical calculations are given in Section~\ref{sec:dch:cs:th}.
Finally, results are reported \ozmodN{and summarised} in Section~\ref{sec:dch:res}.

\subsection{Event reconstruction}
\label{sec:dch:evrec}

Each single \ep collision is referred to as an \emph{event}. 
Events \ozmod{selected by the Third Level Trigger (TLT)} were written to tape as raw data in the form of signals from all sub-detectors 
(see Section~\ref{sec:exp:hera:h1zeus:zeusdet} for the description of the ZEUS detector).
These data were used offline to reconstruct general characteristics of events which correspond to signatures of physical objects 
(particles, jets etc.). 
\ozmod{Subsequently,} the reconstruction of tracks (Section~\ref{sec:dch:evrec:trk}), vertices (Section~\ref{sec:dch:evrec:vtx}), 
hadronic final-state system (Section~\ref{sec:dch:evrec:zufo}) and DIS kinematic variables (Section~\ref{sec:dch:evrec:dis}) is \ozmod{briefly} described.

\subsubsection{Tracking}
\label{sec:dch:evrec:trk}

\ozmodN{A charged particle is identified 
through \ozmodNN{the} trajectory it leaves in the detector. 
This trajectory is referred to as a \emph{track}.} 
It depends not only on the inhomogeneous magnetic field, 
but also on energy loss and multiple scattering in the material; thus the reconstruction of tracks is a complicated task. 
The approach adopted in ZEUS made use of the Kalman filter~\cite{kalman} and \ozmod{ensured} a rigorous treatment of all factors which affect 
particle trajectories. 
\ozmod{For the most precise reconstruction of tracks the information from the CTD and MVD was used.}

The Kalman filter algorithm~\cite{kalman} is an iterative procedure for the reconstruction of tracks from the measured hits. %
It reconstructs tracks from the outermost point of the tracking system to the origin. 
Unlike other global methods which fit all the measurements to a single set
of track parameters, the Kalman filter causes the track to ``follow the measurements'' through the detector~\cite{trk_avery}.
A detailed description of the procedure can be found in~\cite{kalman} and an extended review of its properties and advantages can be found in~\cite{trk_avery}.

Tracks were reconstructed in two stages:
\begin{itemize}
	\item {\bf pattern recognition.} The first stage was performed in multiple steps
		by the VCTRACK package. 
    It started from the outermost tracking
		detector layer, which was the 9th CTD superlayer for the central region, where
		the track density was lower than close to the interaction point. Combinations
		of three CTD hits from axial
		CTD superlayers
		formed the tracking \emph{seeds}. A track seed was extrapolated inward, gathering
		additional hits with increasing precision as the trajectory parameters
		were updated. A very broad \ozmod{``fictitious''} hit was added at the beam line to
		guide the trajectory. After a ``road'' of hits from the CTD through the
		MVD to the interaction point has been created, a least-squares fit of
		the track was performed using the selected hits on the road in order to determine
		the helix parameters at the beginning of the helix. In general the
		tracking reconstruction was not restricted to tracks with hits in all tracking
		devices; the so-called \emph{CTD-only} and \emph{MVD-only} 
		tracks have hits in only one sub-detector;
	\item {\bf trajectory refinement.} A track fit was performed with the Kalman
		filter to improve the precision of the helix parameters in the vicinity of
		the interaction point. As input it took the fit output from the
		pattern recognition stage. The track fit was applied recursively in three
		steps: \emph{prediction}, \emph{filtering} and \emph{smoothing}. At the prediction step, the
		present state $i$ hits (i.e.\ hits that have already been used for the trajectory
		estimation) was used to predict the position of the next $(i+1)^{\rm th}$ hit on
		the next detector sensor (which could be a CTD wire or an MVD sensor). At
		the following filtering step the predicted and the measured values for the
		$(i+1)^{\rm th}$ hit positions were combined. At the last step a smoothing of
		the whole trajectory was performed and the covariance matrix was updated.
\end{itemize}

\subsubsection{Vertexing}
\label{sec:dch:evrec:vtx}

\ozmod{A vertex is the position where an interaction or decay happened.}
The evaluation of vertices serves two purposes~\cite{verbth}. 
The first is to evaluate the position 
of the primary \ep interaction point and to calculate the appropriate track momenta at that
point with improved precision due to the vertex constraint. The second purpose
of using vertices is to estimate the probability that the tracks originate from
a certain vertex. This probability might be estimated from the vertex fit
quality (e.g.\ the \chisq of the vertex fit) and used for the event selection. 
The essential information that is used in the fit consists of track
parameters and their covariance matrices.

Proper identification of both the primary point of interaction and the \Dch
decay vertex in an event was of particular importance for this analysis. 
Their position was reconstructed first with the VCTRACK package and further refinement was applied later~\cite{mishath}.

The vertex pattern recognition started with a loose constraint that the primary
vertex \ozmod{must} be found along the proton-beam line. Track pairs, that
were compatible with this soft constraint as well as with a common vertex, 
were combined with other track pairs. The final choice of the primary-vertex
position after the pattern recognition stage was the vertex with the best
overall \chisq. To improve the precision of the vertex-position measurement, the
Deterministic Annealing Filter (DAF)~\cite{Fruhwirth:1999tya} was used. 
The main feature of the DAF algorithm is that 
tracks with the best quality get the largest weight in the fit, 
while tracks that are far from the vertex get the smallest weight in the fit~\cite{verbth}.
In the chosen approach the vertex position was measured iteratively by calculating
a weighted sum of the \chisq contributions from individual tracks to the vertex~\cite{mishath}.

For the primary-vertex fit, \ozmod{an important} further improvement in precision was possible
by the introduction of a constraint on the vertex position to be close to the averaged
interaction point, the \emph{beamspot}. The beamspot was defined as
the overlap region of the colliding beams. It had a width of roughly $80 \times \SI{20}{\micro\metre}$ 
in the $\text{XY}$ plane, 
but it was too large in the $\text{Z}$ direction to use this information as a constraint~\cite{mishath}. 

In the case of secondary vertices, e.g.\ the \Dch-decay vertex, the fit was
made with the same algorithm skipping the step of the pattern recognition,
since the combination of tracks was chosen based on its compatibility with
the \Dch mass. For each secondary vertex, the corresponding reduced primary
vertex was recalculated removing the secondary-vertex tracks and repeating
the standard primary-vertex fit~\cite{mishath}.

\subsubsection{Hadronic final states}
\label{sec:dch:evrec:zufo}

To get the most precise hadron energy measurement, information from
the calorimeter and the tracking detectors was combined into the so-called
\emph{ZEUS Unidentified Flying Objects} (ZUFOs)~\cite{Briskin:1998sv}
\footnote{ZUFOs are also referred to as Energy Flow Objects (EFOs) in ZEUS publications.} 
Ideally each ZUFO was supposed to represent one final-state particle.
The energy resolution of the CAL developed for higher particle energies as $\sigma(E)/E \sim 1/E$, 
while the tracking momentum resolution, parametrised by $\sigma(p_T)/p_T = a p_T \oplus b \oplus c/p_T$, 
gave a better energy estimate for lower particle momenta (see Section~\ref{sec:exp:hera:h1zeus:zeusdet}). 
For neutral particles, only CAL information could be used, whereas for charged
particles the tracking information was mainly used below $\SI{10}{GeV}$ while calorimeter energy was used for higher energies.

ZUFOs were constructed in the following steps:
\begin{itemize}
	\item CAL cells were clustered into two-dimensional \emph{cell islands};  
	\item the cell islands from the previous stage were used as input to clustering in ($\theta$, $\phi$) space 
		to form three-dimensional energy clusters 
		called \emph{cone islands};
	\item tracks, that have been fitted to a vertex and passed certain
		requirements, were extrapolated to the surface of the CAL taking into
		account the magnetic field; 
		as a result of this procedure, groups of cone islands and tracks ~--- ZUFOs ~--- were formed;
	\item the combination of the information from the CAL and the tracking system was carried out in the following way:
		\begin{itemize}
			\item[$\circ$] if one track has been matched to one cone island, the ZUFO energy was
				taken either from the CAL cluster or from the matched track momentum, 
				depending on which measurement had better resolution;
			\item[$\circ$] for tracks that have not been associated to islands, the energy was
				derived from the momentum measurement with the assumption
				that the particle was a charged pion;
			\item[$\circ$] cone islands that have not been matched to any track were treated as
				neutral particles and the CAL energy was used;
			\item[$\circ$] cone islands with more than three associated tracks were treated as
				jets and the energy was taken from the CAL;
			\item[$\circ$] if a track has been matched to multiple islands or two tracks have been matched
				to one or two islands, the algorithm was similar to the one-to-one
				matching, but using the sum of energies or momenta instead.
		\end{itemize}
\end{itemize}

\ozmod{Additional} corrections were applied to account for the material 
of the detector, the inefficiency in the regions of cracks between the CAL
sections, the presence of muons%
\footnote{Muons did not release all their energy in the CAL, 
thus if the CAL information was used the energy would be underestimated.} 
and the imbalance in the compensation effect for low momentum ($\sim \SI{1}{GeV}$) hadrons. 
In this analysis the reconstructed ZUFOs have been used to determine
the kinematics of the hadronic system as well as DIS kinematic variables 
(see Section~\ref{sec:dch:evrec:dis}).

\subsubsection{Scattered-electron identification and reconstruction of kinematic variables}
\label{sec:dch:evrec:dis}

The identification of the scattered electron is essential for the NC DIS event selection. 
The scattered electron leaves \ozmod{a signature in the detector} which differentiates the NC DIS events 
from the CC DIS, where the neutrino escapes undetected, and photoproduction (PHP), 
where the scattered electron escapes through the beam hole. 
There have been two main electron finders developed in ZEUS: 
the neural-network-based SINISTRA95 (also referred to just as SINISTRA)~\cite{Abramowicz:1995zi} 
and the probabilistic EM~\cite{Kappes:2001jy}. 
The former was tuned for the kinematic region of the \Dch measurement, 
whereas the latter was better for the high-$Q^2$ region, where the electron was reconstructed in the BCAL.

A scattered electron passing through the CAL created an electromagnetic shower, 
therefore most of its energy was measured in the EMC with a small leakage in the HAC.
SINISTRA started from the search of the cells with maximum energy deposits to form candidate clusters.
These clusters were formed using the next-to-nearest neighbour algorithm on CAL towers 
to produce islands and then merging the islands from different CAL sections. 
This information was passed to the neural network, which had 
been trained using MC simulated hadronic and electromagnetic clusters in the RCAL. 
As an output SINISTRA returned a number between 0 and 1, 
which represents the probability of the cluster to be the scattered electron.
In the following only the candidate with the highest probability was considered. 
The identified electron was assigned the energy of the reconstructed CAL cluster.

After the reconstruction of the scattered electron and the hadronic system in an event, 
the kinematic variables $Q^2$, $x$ and $y$, introduced in Section~\ref{sec:th:dis:kin}, can be calculated.
There were several methods:
%
\begin{itemize}
\item {\bf the electron method} 
used only the electron energy and scattering angle.
The kinematic variables were calculated as follows:
\begin{equation}
\begin{split}
	Q^2_{\rm el}&=2E_eE_e^{\prime}(1+\cos \theta_e),\\
	y_{\rm el}&=1-\frac{E_e^{\prime}}{2E_e}(1-\cos \theta_e),\\
	x_{\rm el}&=\frac{Q^2_{\rm el}}{sy_{\rm el}},
\end{split}
\end{equation}
where $E_e$ is the incoming electron energy (which is known \textit{a priori}), $E_e^{\prime}$ and $\theta_e$ 
are the scattered-electron energy and angle, respectively\ozmod{, and $s$ is the centre-of-mass~energy squared.}
This method relies strongly on the measurement of the electron energy and position. 
Because of the characteristics of the ZEUS detector it is more precise in the rear region, 
therefore it is optimal at low $Q^2$.
In addition this method is strongly affected by initial- and final-state photon radiation, 
which spoils the measurement of the lepton energy and leads to deterioration of the results; 
\item {\bf the Jacquet-Blondel method (JB)} 
relied exclusively on the reconstruction of the hadronic final state~\cite{Amaldi:1979qp}. 
The kinematic variables were calculated as follows:
\begin{equation}
\begin{split}
	y_{\rm JB}&=\frac{\delta_{\rm had}}{2E_e},\\
	Q^2_{\rm JB}&=\frac{P^2_{T~{\rm had}}}{1-y_{\rm JB}},\\
	x_{\rm JB}&=\frac{Q^2_{\rm JB}}{sy_{\rm JB}},
\end{split}
\end{equation}
where $P_{T~{\rm had}}$ and $\delta_{\rm had}$ are given by
\begin{equation}
\begin{split}
	P_{T~{\rm had}}&=\sqrt{\sum_i (P^i_{x~{\rm had}})^2+(P^i_{y~{\rm had}})^2},\\
	\delta_{\rm had}&=\sum_i (E^i_{\rm had}-P_{z~{\rm had}}^i),
\end{split}
\end{equation}
where ($E^i_{\rm had}$, $P^i_{x~{\rm had}}$, $P^i_{y~{\rm had}}$, $P^i_{z~{\rm had}}$) 
is the four-momentum of each hadron final state and 
the sum goes over all hadronic energy, excluding the scattered electron, if any. 
The advantage of this method is that it does not require the scattered electron to be detected and 
thus can be used in PHP or CC events, although it has poor $Q^2$ resolution in DIS events;
\item {\bf The double-angle method (DA)}
combined information from the scattered electron and the 
hadronic system~\cite{Buchmuller:1992rq1,Buchmuller:1992rq2}.
The kinematic variables were calculated as follows:
\begin{equation}
\begin{split}
	Q^2_{\rm DA}&=4E_e^2\frac{\cot(\theta_e/2)}{\tan(\theta_e/2)+\tan(\theta_{\rm had}/2)},\\
	y_{\rm DA}&=\frac{\tan(\theta_{\rm had}/2)}{\tan(\theta_e/2)+\tan(\theta_{\rm had}/2)},\\
	x_{\rm DA}&=\frac{Q^2_{\rm DA}}{sy_{\rm DA}},
\end{split}
\end{equation}
where $\theta_{\rm had}$ is the hadronic angle, defined as
\begin{equation}
\begin{split}
	{\rm tan} (\theta_{\rm had}/2)=\frac{\delta_{\rm had}}{P_{T~{\rm had}}}.
\end{split}
\end{equation}
This method exploits the fact that the angular resolution for the hadronic system is usually better 
than the angular resolution for the scattered electron, while for energy it is \textit{vice versa}. 
Thus the DA method leads to a more precise measurement of the kinematic variables in a large part of the 
phase space and was chosen as the main one for the present analysis~\cite{mishath}. 
\end{itemize}

\subsection{DIS event selection}
\label{sec:dch:evsel}

The \ozmod{analysis} used the full HERA-II data with an integrated luminosity $\SI{354}{pb^{-1}}$. 
Both electron--proton and positron--proton \ozmodN{events} were used, because the charm NC DIS
cross sections at not too high $Q^2$ are invariant with respect to the lepton charge.
The DIS kinematic region of the measurement was restricted to $5<Q^2<\SI{1000}{GeV^2}$ and $0.02<y<0.7$, 
where reliable reconstruction of the scattered electron was possible with the ZEUS detector 
after the HERA-II high-luminosity upgrade~\cite{HERAUpgrade}.

The selected events had to be triggered online by one of the inclusive DIS Third Level Trigger slots 
(see~\cite{mishath} for the description of these slots):
\begin{itemize}
	\item SPP02 for the 2004--2005 data period, or
	\item SPP09, or HFL17, or HPP31 for the 2006--2007 data period.
\end{itemize}
Furthermore to ensure selection of good DIS events, the following cuts were applied offline:
\begin{itemize}
	\item $5<Q^2_{\rm DA}<\SI{1000}{GeV^2}$, $0.02<y_{\rm DA}<0.7$. 
		These criteria selected the considered DIS phase-space region;
	\item $E_e^{\prime}>\SI{10}{GeV}$. The requirement ensured high efficiency of SINISTRA 
		and rejected possible background PHP events with ``fake'' scattered electrons;
	\item $E^{\rm cone}_{{\rm non}~e}<\SI{5}{GeV}$, where $E^{\rm cone}_{{\rm non}~e}$ is the energy deposit 
		in the CAL in the cone centered around the scattered electron 
		with a radius of 0.8 in the ($\eta$, $\phi$) plane, not originating from it. 
		This cut is known as the \emph{electron isolation} and was supposed 
		to improve further the quality of the scattered-electron reconstruction;
	\item ${\rm prob}_{\rm SINISTRA}>0.9$, where ${\rm prob}_{\rm SINISTRA}$ is the output of the SINISTRA 
		neural network.%
		\footnote{Despite the notation, it is not the probability in its mathematical meaning.}
		This selection further ensured high efficiency in SINISTRA;
	\item $y_{\rm JB}>0.02$. This requirement rejected events with the poorly reconstructed hadronic system, 
		for which the DA method was not precise;
	\item $40<\delta_{\rm had}<\SI{65}{GeV}$. The lower cut reduced the PHP contamination (when the scattered electron was not detected) 
		and the upper cut rejected events initiated by cosmic-ray particles;%
		\footnote{For a fully contained NC event, $\delta_{\rm had}=2E_e=\SI{55}{GeV}$.}
	\item $-30<Z_{\rm vtx}<\SI{30}{cm}$, where $Z_{\rm vtx}$ is the ${\text Z}$ coordinate of the primary vertex. 
		This requirement rejected events initiated by beam-gas and satellite-bunch interactions;
	\item a set of cuts on the geometric position of the scattered electron in the CAL ($x_{e^{\prime}}$, $y_{e^{\prime}}$, $z_{e^{\prime}}$), 
		to remove events, in which the scattered electron passed through the regions of the CAL poorly simulated in Monte Carlo; note that these cuts are quoted as exclusion cuts, 
		i.e.\ the events were removed if they satisfied any of the criteria:
	\begin{itemize}
		\item[$\circ$] $|x_{e^{\prime}}|<\SI{13}{cm}$ and $|y_{e^{\prime}}|<\SI{13}{cm}$. 
			This requirement is known as the \emph{box cut} and removed the edges of the CAL;
		\item[$\circ$] $\sqrt{x_{e^{\prime}}^2+y_{e^{\prime}}^2}>\SI{175}{cm}$. This cut rejected the region between 
			the RCAL and BCAL;
		\item[$\circ$] $-104<z_{e^{\prime}}<\SI{-98.5}{cm}$ or $164<z_{e^{\prime}}<\SI{174}{cm}$. 
			This requirement is known as the \emph{super-crack cut} and removed the regions of cracks between the 
			RCAL, BCAL and FCAL;
		\item[$\circ$] $6.5<x_{e^{\prime}}<\SI{12}{cm}$ and $y_{e^{\prime}}>0$, or 
			$-14<x_{e^{\prime}}<\SI{-8.5}{cm}$ and $y_{e^{\prime}}<0$. 
			This requirement is known as the \emph{module-gap cut} and removed the region of gaps between halves 
			of the RCAL;
		\item[$\circ$] $|x_{e^{\prime}}|<\SI{12}{cm}$ and $y_{e^{\prime}}>\SI{80}{cm}$. This requirement is known as the 
			\emph{chimney cut} and removed the region of the RCAL where cooling tubes and supply cables for the solenoid 
			were mounted;
		\item[$\circ$] in addition, for a subset of the data with the run ranges $59600\text{--}60780$, $61350\text{--}61580$, 
			$61800\text{--}63000$ the region $11<x_{e^{\prime}}<\SI{27}{cm}$ and $10.5<y_{e^{\prime}}<\SI{27}{cm}$ 
			was removed, which was not described by the Monte Carlo simulations.
	\end{itemize}
\end{itemize}

\subsection{Monte Carlo simulations}
\label{sec:dch:mc}

\ozmodN{
The complexity of the HERA experiments makes it necessary to apply Monte Carlo (MC) methodes for their evaluation. 
The two tasks are:
\begin{itemize}
	\item the descrption of all relevant physics processes with their complete final state using available MC generators, and
	\item the simulation of the detector response, i.e. the account of the effects as the final state particles pass through the various detector components.
\end{itemize}
}

\subsubsection{Simulation of \ozmodN{physics processes}}
\label{sec:dch:dch:mc:tech:ph}

\begin{figure}[htbp]
  \centering
  \includegraphics[width=1.0\figwidth,trim=0 2mm 2mm 0,clip=true]{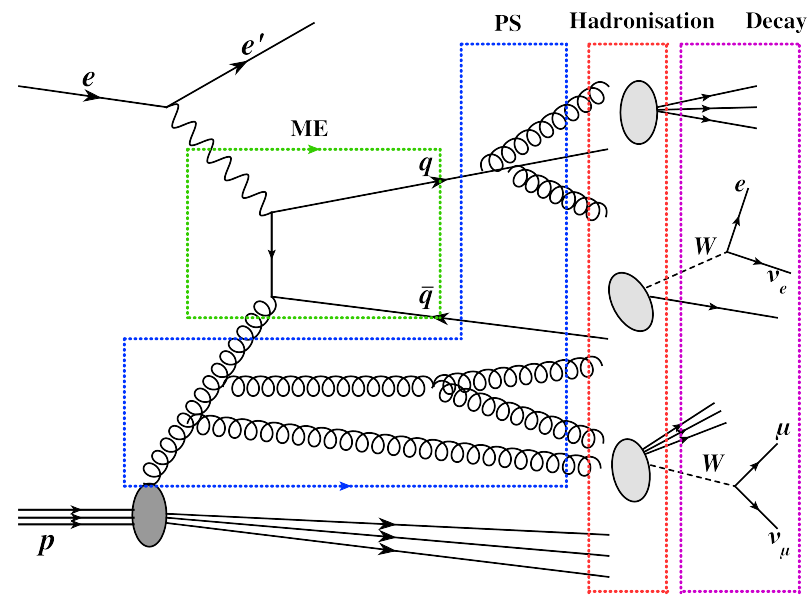}
  \caption[Physics simulation in MC for BGF process]
	{Stages of physics simulation in MC for the BGF process. The figure was taken from~\cite{Shehzadi:2011sxa}.}
  \label{fig:dch:mc:bgf}
\end{figure}

In the generation of MC events the QCD factorisation theorem~\cite{Mueller:1978xu,Collins:1981ta,Collins:1989gx,Collins:1985ue,Collins:1988ig,Bodwin:1984hc} 
is exploited to separate short- and long-distance effects. 
Fig.~\ref{fig:dch:mc:bgf} illustrates the case for 
\ozmodN{different phases in the boson-gluon fusion process (see also Fig.~\ref{fig:th:hq:ep:bgf} for the corresponding diagram):}
\begin{itemize}	
	\item {\bf the hard scattering process} is usually calculated at LO;
	\item {\bf radiation corrections} (referred to also as \emph{parton showers}) are modelled using some phenomenological models. 
		The difference between the fixed-order NLO calculation and LO accompanied by parton showers is that the latter 
		\ozmod{is better reproducing} the whole final state (the \emph{event shape}), 
		which is important for the correct simulation of the detector response, while the former gives a better description of inclusive quantities;%
		\footnote{NLO calculations are much more complicated to be matched with parton showers.}
	\item {\bf hadronisation} is the non-perturbative QCD process of the formation of colourless hadrons from coloured partons. 
		It is performed by using some phenomenological models;
	\item {\bf decays} of unstable particles are simulated accordingly to \ozmod{available} decay tables.%
		\footnote{Some relatively long-lived particles (typically pions, kaons, muons) are usually considered as stable in an MC generator, 
		since they interact with a detector directly.}
\end{itemize}
Examples of event generators commonly used in ZEUS are PYTHIA~\cite{pythia}, ARIADNE~\cite{ariadne}, RAPGAP~\cite{rapgap} etc.

\subsubsection{Simulation of detector response}
\label{sec:dch:mc:tech:dr}

After the simulation of underlying physics processes, final-state particles are passed through a simulated detector. 
Simulation of the ZEUS detector was performed in the MOZART program, which is based on GEANT 3.21~\cite{geant}. 
Furthermore, generated events were passed through the simulated ZEUS trigger system and the reconstruction program ZEPHYR. 
More details on the ZEUS MC production system can be found in~\cite{mishath}. 
Finally, MC events were written to tape as regular data and processed by the same reconstruction and selection algorithms, 
although they contain additional information on generated particles, referred to as \emph{generated}, or \emph{true} information. 
However, the procedure of matching between generated particles and reconstructed ones has some complications (see Section~\ref{sec:dch:cs:acc:mat}).

\subsubsection{MC samples}
\label{sec:dch:mc:sample}

In the present analysis the following MC samples have been used:
\begin{itemize}
	\item the RAPGAP charm DIS MC sample, \ozmodN{$ep \to e' c\bar{c}X$}, was the main sample used to determine acceptance corrections. 
		MC events were simulated with the RAPGAP 3.00~\cite{rapgap} program, interfaced with HERACLES 4.6.1~\cite{heracles} 
		to incorporate first-order electroweak corrections. The CTEQ5L~\cite{cteq5l} PDFs were used for the proton;
	\item the RAPGAP beauty DIS MC sample, \ozmodN{$ep \to e' b\bar{b}X$}, similar to the previous one, was used to estimate the contribution to \Dch production from decays of beauty hadrons;
	\item the RAPGAP charm DIS MC sample without QED radiation, \ozmodN{$ep \to e' c\bar{c}X$}, was used to correct the measured cross sections to the QED Born level;
	\item the ARIADNE inclusive MC sample, \ozmodN{$ep \to e' X$}, was used for simulation of combinatorial background and optimisation of selection cuts;
	\item the PYTHIA PHP charm MC sample, \ozmodN{$ep \to e' c\bar{c}X$}, was used to estimate the contribution from PHP events.
\end{itemize}

\subsection{Reconstruction and selection of \Dch candidates}
\label{sec:dch:dch}

The \Dch mesons were reconstructed in the decay channel $\Dch \to K^{-}\pi^{+}\pi^{+}$. 
\ozmodN{The full final-state particle reconstruction consists in making combinations} 
of all tracks with proper charges, if possible followed by the reconstruction of the secondary vertex (the place where 
the decay happened) and re-fitting of the considered tracks to this vertex, thus improving their reconstruction. 
The invariant mass, $M(K\pi\pi)$, is calculated using the energy and momentum conservation rules, 
and the masses of the daughter particles. 
If \ozmodNN{the reconstructed invariant mass} is found to be close to the mass of \ozmod{the hadron under consideration}, 
the combination is considered as a \emph{candidate}. The tracks from the selected combinations are 
referred to as \emph{daughter} tracks.

Inherently such a method leads to a large combinatorial background, which is \ozmod{due to} combinations of tracks not originating 
from the analysed hadron decay channel or from wrongly combined daughter tracks. In order to suppress this background, 
additional cuts on the parameters of the daughter tracks and the quality of the secondary-vertex reconstruction 
\ozmodN{have been} applied.

The measurement was performed in the \Dch phase-space region $1.5<p_T(\Dch)<\SI{15}{GeV}$, $|\eta(\Dch)|<1.6$. 
At lower \ozmodN{values} of $p_T(\Dch)$ the combinatorial background increases drastically, making the \ozmod{measurement} impossible, 
while at higher values of $p_T(\Dch)$ the production cross section becomes too small to be measured with the available integrated luminosity. 
The $\eta(\Dch)$ range is determined by the coverage of the tracking system, since all daughter tracks have to be detected and well reconstructed.

\subsubsection{Selection of secondary vertices}
\label{sec:dch:dch:secvtx}

\begin{figure}[htbp]
  \centering
  \includegraphics[width=0.8\figwidth]{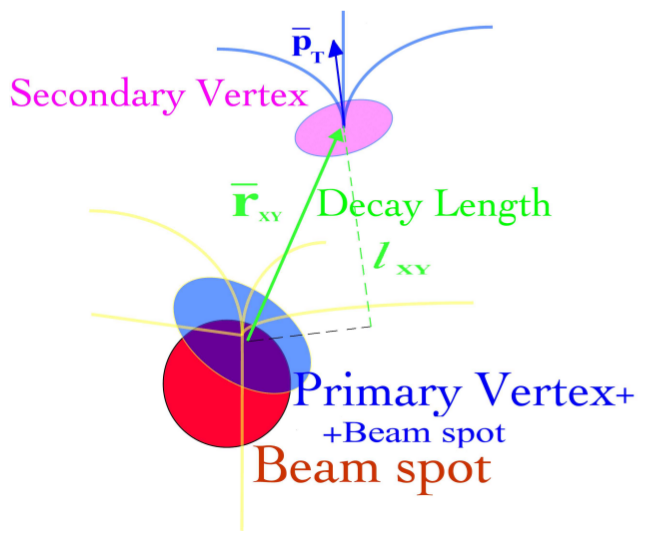}
  \caption[Production and decay of \Dch meson]
	{Production and decay of a \Dch meson. The figure was taken from~\cite{mishath}.}
  \label{fig:dch:dchsignif}
\end{figure}

The \ozmodN{large} lifetime of \Dch mesons, $c\tau(\Dch)=311.8\pm \SI{2.1}{\micro\metre}$~\cite{pdg2012}, 
makes it possible to reconstruct their secondary vertices with \ozmod{the Microvertex Detector (MVD)}. Important characteristics of 
the reconstructed secondary vertices (Fig.~\ref{fig:dch:dchsignif}) include:
\begin{itemize}
	\item \chisq of the secondary-vertex fit, $\chisq_{\rm sec~vtx}$;
	\item the decay length, defined as the distance between the primary and secondary vertices;
	\item the uncertainty on the decay length;
	\item the collinearity of the directions from the primary to the secondary vertex and the \Dch momentum.
\end{itemize}
The most efficient way of using the lifetime information is to combine the last three quantities into the \emph{projected decay-length significance} 
(\ozmod{for simplicity} referred to as just the \emph{decay-length significance}), $S_l$, 
defined as the ratio of the decay length, projected on the $\text{XY}$ plane and on the \Dch momentum, to the uncertainty on this quantity:
\begin{equation}
\begin{split}
	S_l=\frac{l_{\text{XY}}}{\sigma_{l_{\text{XY}}}},
\end{split}
\end{equation}
	where $l_{\text{XY}}$ is the projected decay length, defined as
\begin{equation}
\begin{split}
	l_{\text{XY}}=\frac{(\vec{S}_{\text{XY}}-\vec{P}_{\text{XY}})\cdot \vec{p}(\Dch)}{p_T(\Dch)}
\end{split}
\end{equation}
and $\sigma_{l_{\text{XY}}}$ is the uncertainty on $l_{\text{XY}}$. 
Here $\vec{P}_{\text{XY}}$ and $\vec{S}_{\text{XY}}$ are the vectors pointing to the primary and secondary vertices, respectively, 
and the $\cdot$ sign denotes a scalar product. 
The projection on the $\text{XY}$ plane was used because the resolution of the vertex position was most precise in the transverse plane. 

\begin{figure}[htbp]
  \centering
  \includegraphics[width=0.495\figwidth,trim=0 2mm 2mm 0,clip=true]{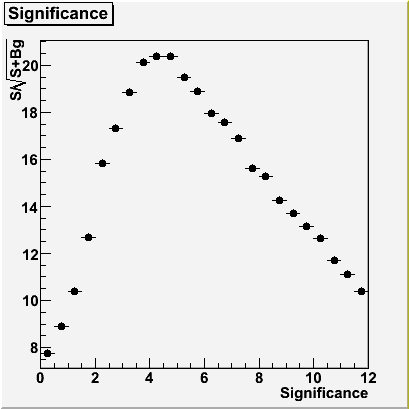}
  \includegraphics[width=0.495\figwidth,trim=0 2mm 2mm 0,clip=true]{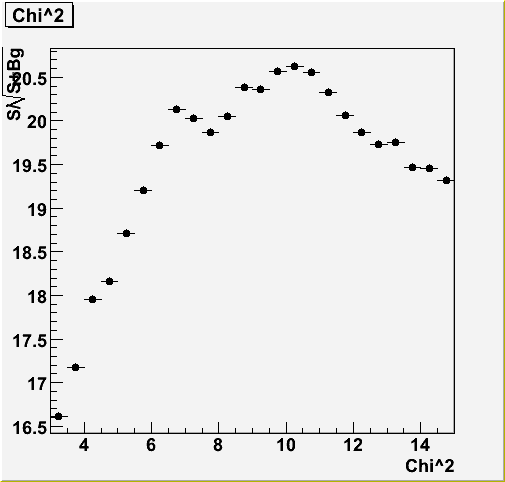}
  \caption[Statistical significance of mass peak as function of lower cut on $S_l$ and $\chisq_{\rm sec~vtx}$]
	{The statistical significance of the mass peak as a function of a lower cut 
	on the decay-length significance (left) and an upper cut on \chisq of the secondary vertex (right).}
  \label{fig:dch:dchoptsecvtx}
\end{figure}

The optimal cuts on $S_l$ and $\chisq_{\rm sec~vtx}$ were determined by maximising the statistical significance of the mass peak, 
$S_P$, defined as the ratio of the signal%
\footnote{Genuine $\Dch \to K^{-}\pi^{+}\pi^{+}$ decays are referred to as \emph{signal}.} 
to its statistical uncertainty, assuming a Poisson distribution:
\begin{equation}
	S_P=\frac{S}{\sqrt{S+Bg}},
\end{equation} 
where $S$ is the number of candidates in the signal peak and $Bg$ is the number of candidates in the background, 
where the region of the signal peak 
is defined within three standard deviations. 
The study was performed on the inclusive ARIADNE MC sample. 
The dependence of $S_P$ on a lower cut on $S_l$ and an upper 
cut on $\chisq_{\rm sec~vtx}$ is shown in Fig.~\ref{fig:dch:dchoptsecvtx}. 
The optimal cuts are:
\begin{itemize}
	\item $S_l>4$,
	\item $\chisq_{\rm sec~vtx}<10$.
\end{itemize}

\subsubsection{Selection of \Dch candidates}
\label{sec:dch:dch:cand}

To ensure the selection of well reconstructed \Dch candidates and to improve the signal-to-background ratio, the following cuts were applied:
\begin{itemize}	
	\item $1.5<p_T(\Dch)<\SI{15}{GeV}$, $|\eta(\Dch)|<1.6$, to select the \Dch phase-space region;
	\item $S_l>4$, $\chisq_{\rm sec~vtx}<10$, to reduce the combinatorial background, as explained in Section~\ref{sec:dch:dch:secvtx};
	\item $l_{\text{XY}}<\SI{1.5}{cm}$, to ensure that selected secondary vertices were inside the beampipe, thus 
		did not originate from interactions with the beampipe or detector material;
	\item $p_T(K)>\SI{0.5}{GeV}$, $p_T(\pi)>\SI{0.35}{GeV}$, to further reduce combinatorial background while still keeping the detector acceptance at a reasonable level at low $p_T(\Dch)$;
	\item $|\eta(K,\pi)|<1.75$, to ensure the selection of well reconstructed daughter tracks; 
	\item each track should have at least two MVD hits in both the $\text{Z}$ and $\phi$ directions and pass through at least three CTD superlayers, 
		to improve further the quality of the daughter tracks;
	\item the mass difference $\Delta M =M(K\pi\pi)-M(K\pi)$ should not be within $0.143 < \Delta M <\SI{0.148}{GeV}$\ozmod{, 
		which is the difference between the \Dstar and $D^{0}$ masses,} 
		to reduce background from \Dstar mesons decaying in the ``golden'' channel 
		\ozmod{${\Dstar \to D^{0} \pi_s^{+}}$, ${D^{0} \to K^{-}\pi^{+}}$} (see Section~\ref{sec:exp:hera:hfmeas:dstar}), 
		which result in identical final states;
	\item the invariant mass of a combination of the kaon and any of two pion daughter tracks assuming that they are kaons, $M(KK)$, should not be within $1.0115 < M(KK) <\SI{1.0275}{GeV}$.
		This cut reduced background from $D_s^{+}$ mesons decaying in the channel $D_S^{+}\to \phi \pi^{+}$ with subsequent $\phi \to K^{-}K^{+}$, 
		which result in similar final states with an asymmetric mass peak (a so-called \emph{reflection}).
\end{itemize}

\begin{figure}[htbp]
  \centering
  \includegraphics[width=1.00\figwidth]{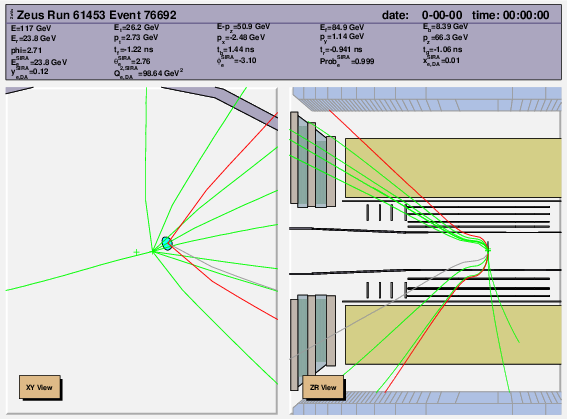}
  \caption[Event with $D^{+} \to K^{-} \pi^{+} \pi^{+}$ candidate]
	{Event with a $D^{+} \to K^{-} \pi^{+} \pi^{+}$ candidate.}
  \label{fig:zevis_zeviscn_analysistrkvtx}
\end{figure}

An example of an event with a selected \Dch candidate, displayed in the ZEUS Event Display program, is shown in Fig.~\ref{fig:zevis_zeviscn_analysistrkvtx}.

\subsection{Extraction of \Dch signal}
\label{sec:dch:signal}

\begin{figure*}[htbp]
  \sidecaption
  \centering
  \includegraphics[width=0.432\figwidth*\real{1.65},trim=1mm 3mm 5mm 0,clip=true]{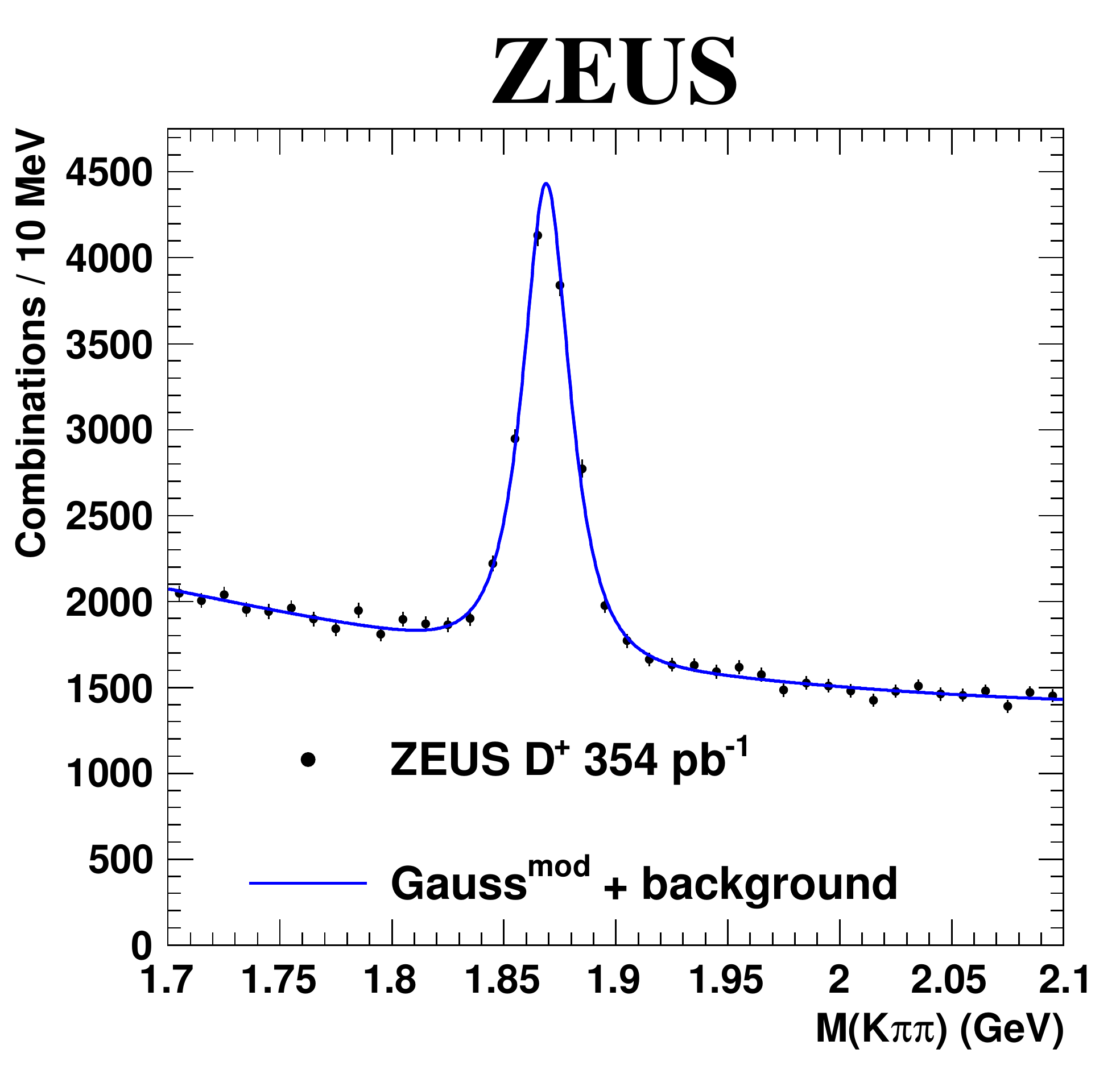}
  \includegraphics[width=0.56\figwidth*\real{1.65},trim=1mm 1mm 13mm 0,clip=true]{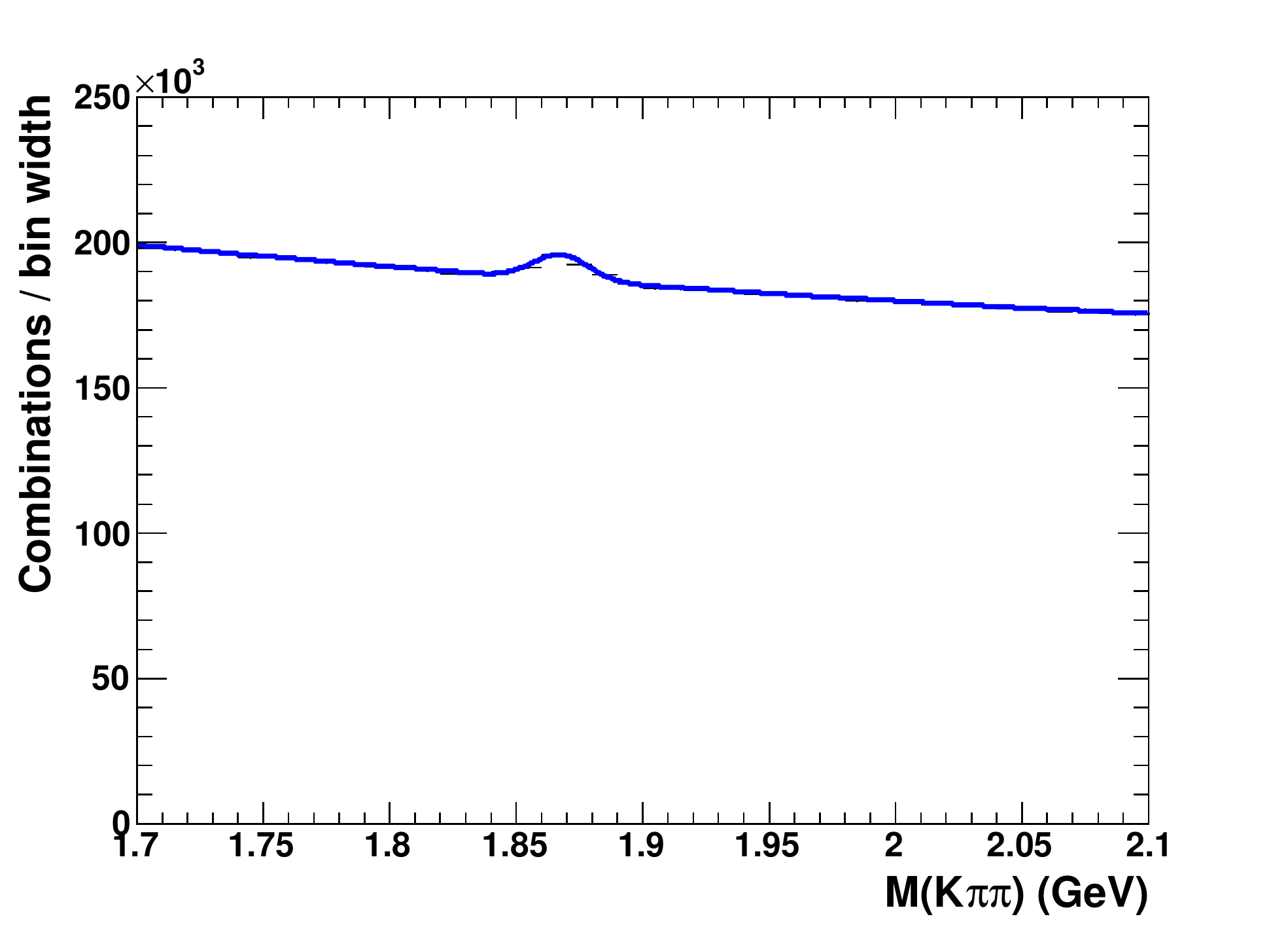}
  \caption[Mass distribution of reconstructed \Dch candidates]
	{Mass distribution of the reconstructed \Dch candidates after final selection (left), and without the cuts on the decay-length significance and \chisq of the secondary vertex (right). 
	The solid curve represents a fit to the sum of a modified Gaussian for the signal and a second-order polynomial for the background.}
  \label{fig:dch:mass}
\end{figure*}

Fig.~\ref{fig:dch:mass} shows the invariant mass distribution $M(K\pi\pi)$ of the selected \Dch candidates. 
For comparison, the same distribution selected without the cuts on the decay-length significance and \chisq of the secondary vertex is also shown. 
The \ozmod{cuts applied} on $S_l$ and $\chisq_{\rm sec~vtx}$ improved the statistical significance by a factor of 3 
(a similar conclusion can also be drawn from Fig.~\ref{fig:dch:dchoptsecvtx}).
To extract the number of reconstructed \Dch mesons, the mass distribution was fitted \ozmodN{to} a function
\begin{equation}
	F(M)=F_{\rm signal}(M)+F_{\rm background}(M),
\end{equation}
where the signal component, $F_{\rm signal}(M)$, is given by a modified Gaussian function:
\begin{equation}
	F_{\rm signal}(M)=C \exp[-0.5X^{1+1/(1+\beta X)}],~~X=\frac{|M-M_0|}{\sigma_M}
	\label{eq:modgaus}
\end{equation}
and the background component, $F_{\rm background}(M)$, is given by a second-order polynomial. 
The signal position, $M_0$, the peak width, $\sigma_M$, as well as the signal normalisation parameter, $C$, and parameters 
of the background component were free parameters in the fit. 
The parameter $\beta$ of the modified Gaussian function controls the deviation of its tails from the normal distribution 
($\beta=0$); the central value $\beta=0.5$ was chosen to get the best description of the peak, 
while it has been varied in order to estimate the systematic uncertainty (see Section~\ref{sec:dch:cs:syst}).
The fit was performed using the least-squares method as implemented in the MINUIT package~\cite{minuit}. 
As the expectation values in the \chisq-function, the integrals of the fit function within each bin of $M$ were used. 
To account for possible non-linearities, the fit uncertainty was calculated as the average of the positive and negative fit uncertainty, 
obtained with the MINOS algorithm~\cite{minuitmanual}.

The number of \Dch mesons yielded by the fit is $N(\Dch)=\SI{8356\pm 198}{}$. 
The fitted position of the peak is $M_0=1868.97\pm \SI{0.26}{MeV}$, where only the statistical uncertainty is quoted, 
\ozmodN{and is consistent} with the PDG value of $1869.62\pm \SI{0.15}{MeV}$~\cite{pdg2012}. 
The peak width is $\sigma=12.2\pm \SI{0.3}{MeV}$, driven by the momentum resolution of the detector.

\subsection{Cross-section determination}
\label{sec:dch:cs}

\ozmodN{The differential cross section as a function of a given observable $Y$ in the $i^{\rm th}$ bin was determined as
\begin{equation}
	\frac{\dif \sigma}{\dif Y}=\frac{N_{\rm DATA}^{}-N_{{\rm MC}~b}^{\rm reco}}{\mathcal{A}\mathcal{L}\mathcal{B}\Delta Y_i}\cdot \mathcal{C^{\rm rad}},
\label{eq:dch:cs}
\end{equation}
where $\Delta Y_i$ is the width of the $i^{\rm th}$ bin, 
$N_{\rm DATA}^{}$ is the number of the reconstructed \Dch mesons in the data and $N_{{\rm MC}~b}^{\rm reco}$ is the number of \Dch mesons from beauty-hadron decays, as predicted by RAPGAP. 
The latter was additionally scaled by 1.6 \ozmodN{according to} previous ZEUS measurements~\cite{Chekanov:2009kj,Abramowicz:2010zq,Abramowicz:2011kj} of beauty production in DIS. 
The determination of the cross section accounts for} the branching ratio, $\mathcal{B}(\Dch\to K^{-}\pi^{+}\pi^{+})=9.13 \pm 0.19\%$~\cite{pdg2012}, 
the acceptance correction, $\mathcal{A}$, the radiative correction, $\mathcal{C^{\rm rad}}$, and the contribution from beauty-hadron decays. 
The radiative corrections were applied to correct the measured cross sections \ozmodNN{corresponding} to the QED Born level; more details on their calculations can be found in~\cite{mishath}. 
The determination of the acceptance corrections is described in Section~\ref{sec:dch:cs:acc}.

\subsubsection{Acceptance correction}
\label{sec:dch:cs:acc}

MC simulations were used to determine efficiency, $\mathcal{E}$, purity, $\mathcal{P}$, and acceptance $\mathcal{A}$. 
For the $i^{\rm th}$ bin these quantities are defined as
\begin{equation}
\begin{split}
	\mathcal{E}_i&=\frac{N_i^{\rm gen}\bigcap N_i^{\rm rec}}{N_i^{\rm gen}},\\
	\mathcal{P}_i&=\frac{N_i^{\rm gen}\bigcap N_i^{\rm rec}}{N_i^{\rm rec}},\\
	\mathcal{A}_i&=\frac{\mathcal{E}_i}{\mathcal{P}_i}=\frac{N_i^{\rm rec}}{N_i^{\rm gen}},
\label{eq:dch:acc}
\end{split}
\end{equation}
where $N_i^{\rm gen}$ and $N_i^{\rm rec}$ are the numbers of the signal events, generated and reconstructed in the $i^{\rm th}$ bin, respectively. 
The notation $N_i^{\rm gen}\bigcap N_i^{\rm rec}$ in the numerators means that events must be generated and reconstructed in the same bin. 
Therefore, the efficiency is the portion of events generated in a given bin, that were also reconstructed in the same bin; 
it determines the dependence of the measurement on the MC simulations. 
The purity is the portion of events reconstructed in a given bin, that were also generated in the same bin; 
it determines the level of migrations of events to different bins.
\ozmodN{The purity plots for $p_T(\Dch)$, $\eta(\Dch)$, $Q^2$ and $y$ are provided in Appendix~\ref{sec:app:dch}; 
the purity values are typically above 80\%.} 
Finally, the acceptance determines the correction from detector to generator level required to calculate the cross section. 

\subsubsection{Comparison of data and MC}
\label{sec:dch:cs:acc:cp}

To get the correct acceptance, the MC simulations must describe the shapes of all kinematic variables in the data. 
The acceptance determined from the MC and integrated over some variable, $x$, is given by:
\begin{equation}
\begin{split}
	\mathcal{A}&=\frac{1}{\sigma^{\rm tot}}\int \mathcal{A}(x) \frac{\dif \sigma}{\dif x} \dif x,\\
	\sigma^{\rm tot}&=\ozmod{\int \frac{\dif \sigma}{\dif x} \dif x},
\end{split}
\end{equation}
where the integration is performed over the full range of the variable $x$, $\mathcal{A}(x)$ is the acceptance at a fixed value of $x$ and 
$\frac{\dif \sigma}{\dif x}$ is the \ozmodNN{differential cross section as a function of $x$, used in the MC simulations.} The correct detector simulation guarantees the correct 
value of $\mathcal{A}(x)$, although $\frac{\dif \sigma}{\dif x}$ is the generator-level cross section, thus even for correct $\mathcal{A}(x)$ 
at all $x$, incorrect $\frac{\dif \sigma}{\dif x}$ will lead to an incorrect total acceptance $\mathcal{A}$.

Therefore the differential distributions of kinematic variables from the MC simulations and from the data were compared to each other; these comparison plots 
are referred to as \emph{control plots}. 
Since the MC simulations usually describe the shapes of kinematic distributions, but not the normalisation, and moreover 
acceptance does not depend on the MC normalisation, the MC distributions are re-normalised to the data.
To estimate the goodness of the description, for each control plot the \chisqndof were calculated as follows:
\begin{equation}
	\chisqndof=\frac{1}{\ndof} \sum_i \frac{(N_i^{\rm DATA}-N_i^{\rm MC})^2}{{\sigma_i^{\rm DATA}}^2+{\sigma_i^{\rm MC}}^2},
\end{equation}
where the sum goes over all bins, $N_i^{\rm DATA}$ and $N_i^{\rm MC}$ are the number of signal events in the $i^{\rm th}$ bin in the data and MC, respectively, 
$\sigma_i^{\rm DATA}$ and $\sigma_i^{\rm MC}$ are the corresponding statistical uncertainties on $N_i^{\rm DATA}$ and $N_i^{\rm MC}$, respectively, and
$\ndof$ is the number of bins minus one%
\footnote{Because of re-normalisation of MC to the data.}.

\begin{figure}[htbp]
  \centering
  \includegraphics[width=0.495\figwidth,trim=1mm 0mm 12mm 6mm,clip=true]{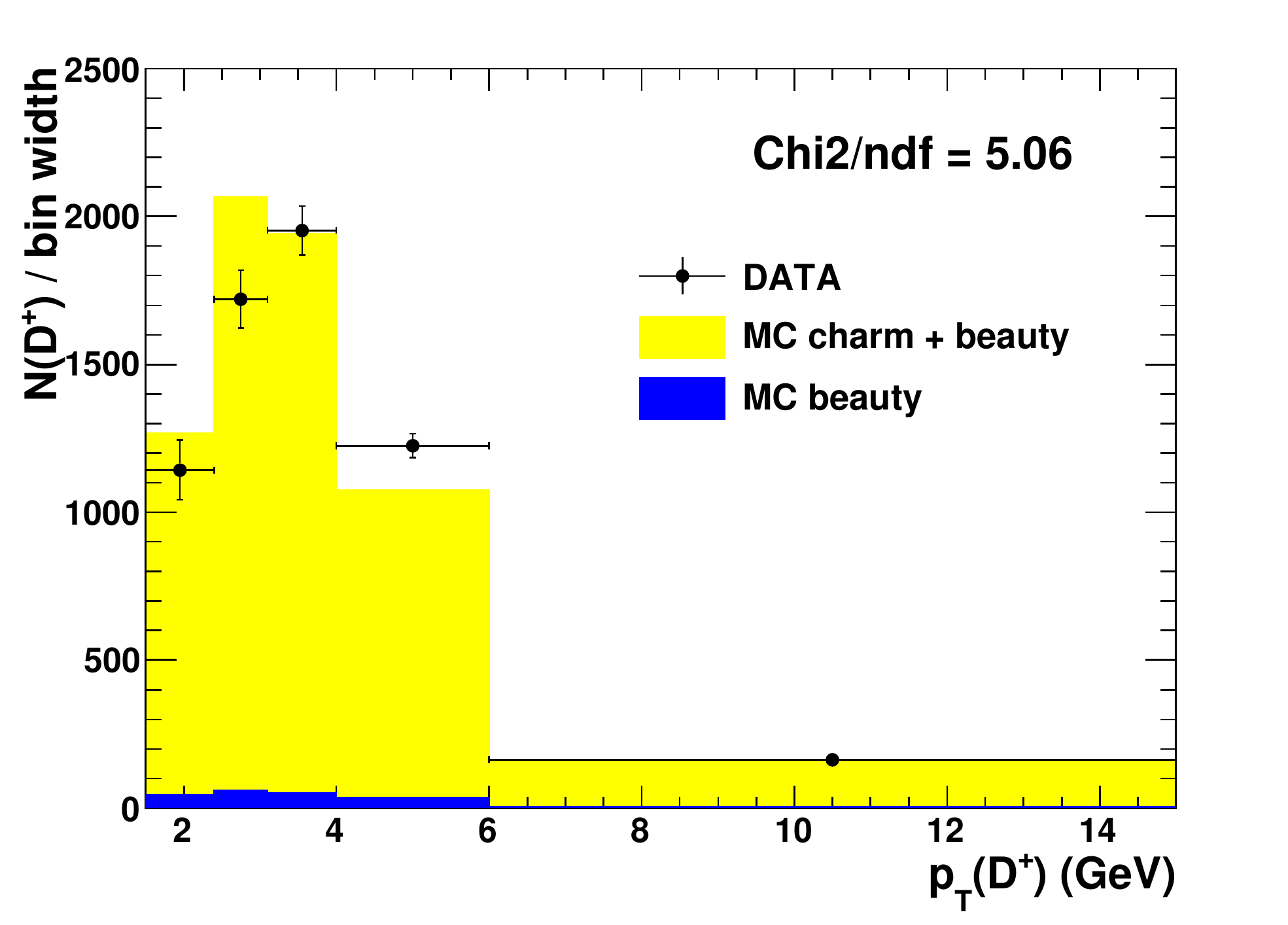}
  \includegraphics[width=0.495\figwidth,trim=1mm 0mm 12mm 6mm,clip=true]{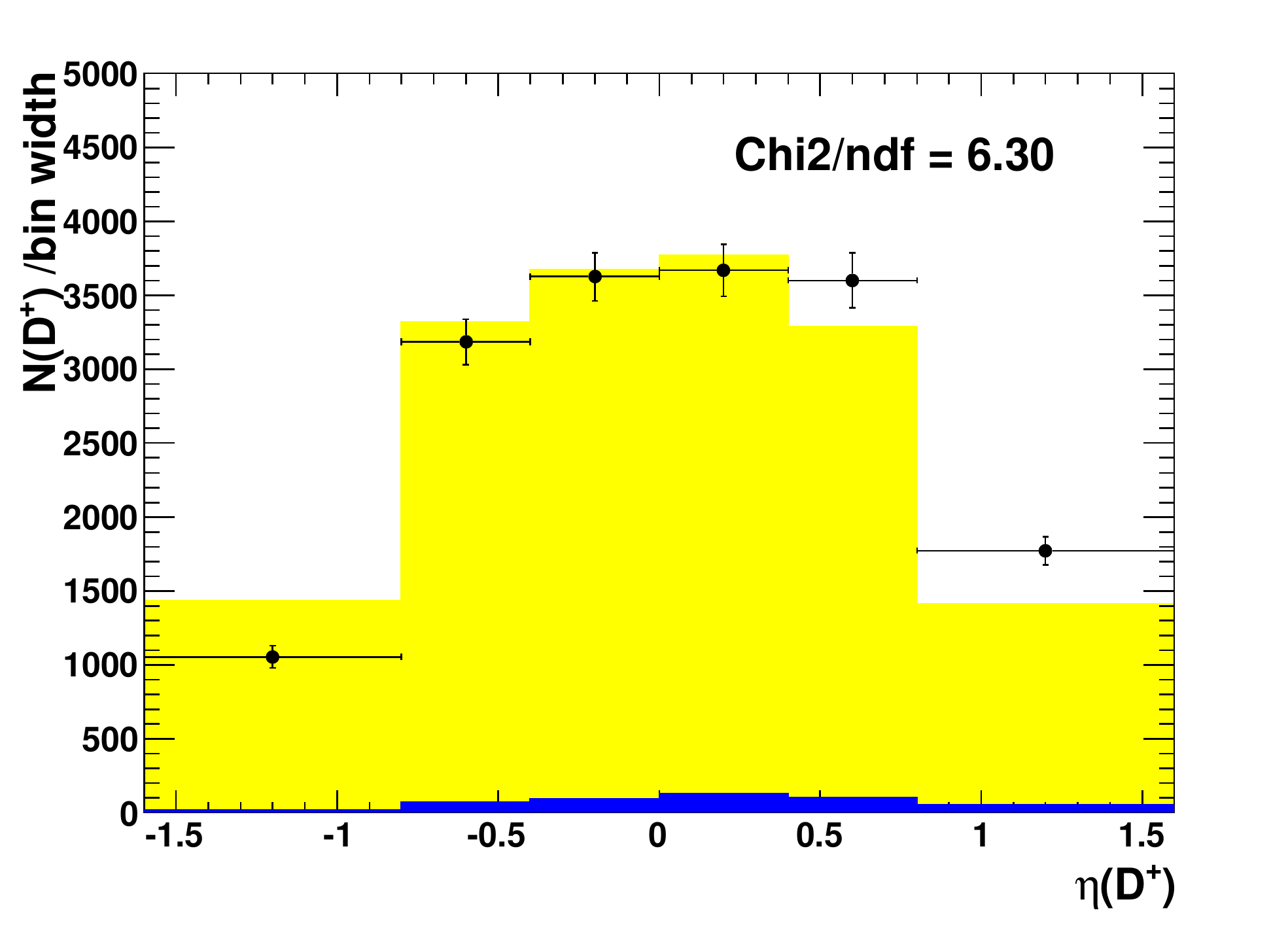}
  \includegraphics[width=0.495\figwidth,trim=1mm 0mm 12mm 6mm,clip=true]{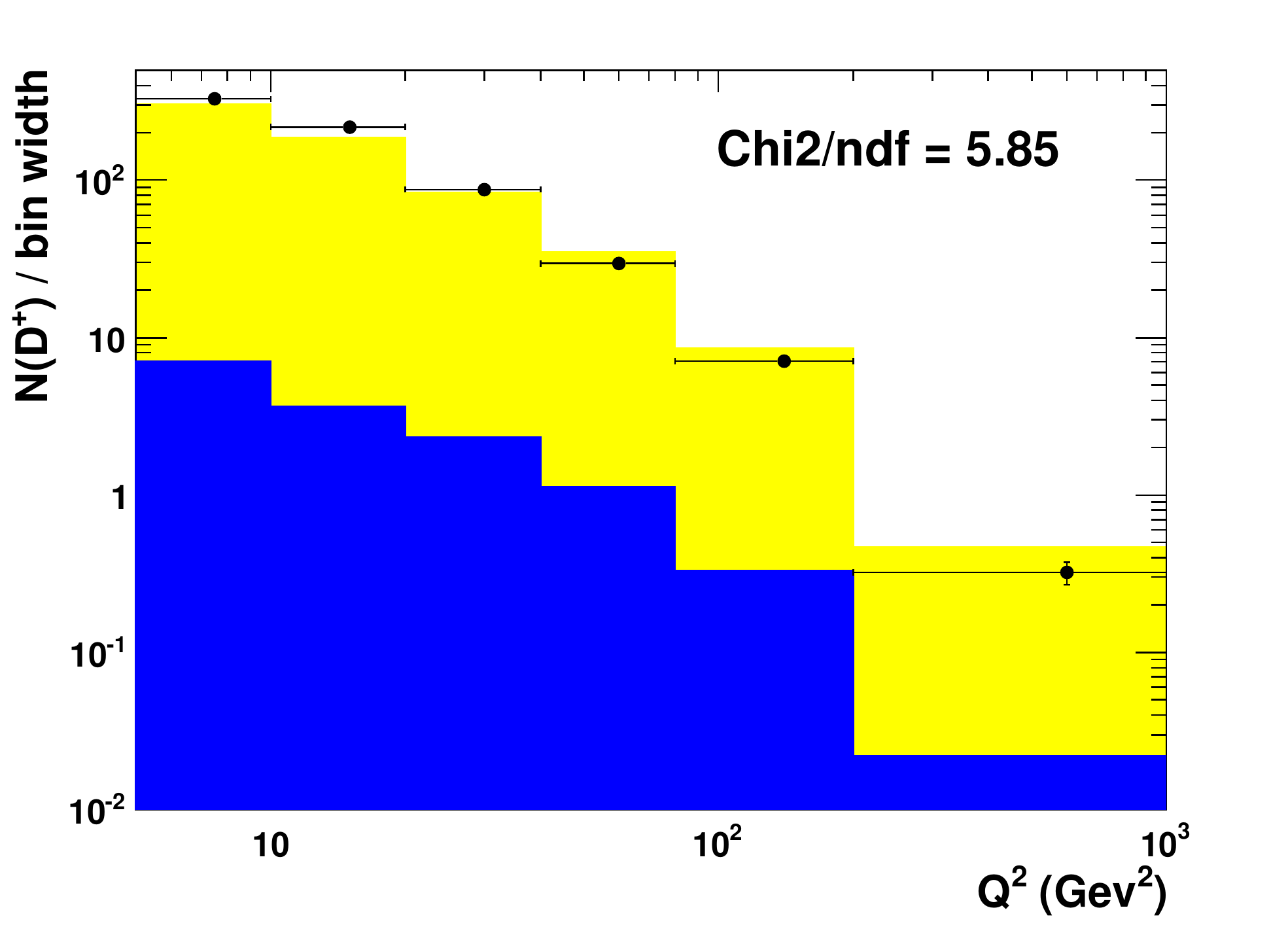}
  \includegraphics[width=0.495\figwidth,trim=1mm 0mm 12mm 6mm,clip=true]{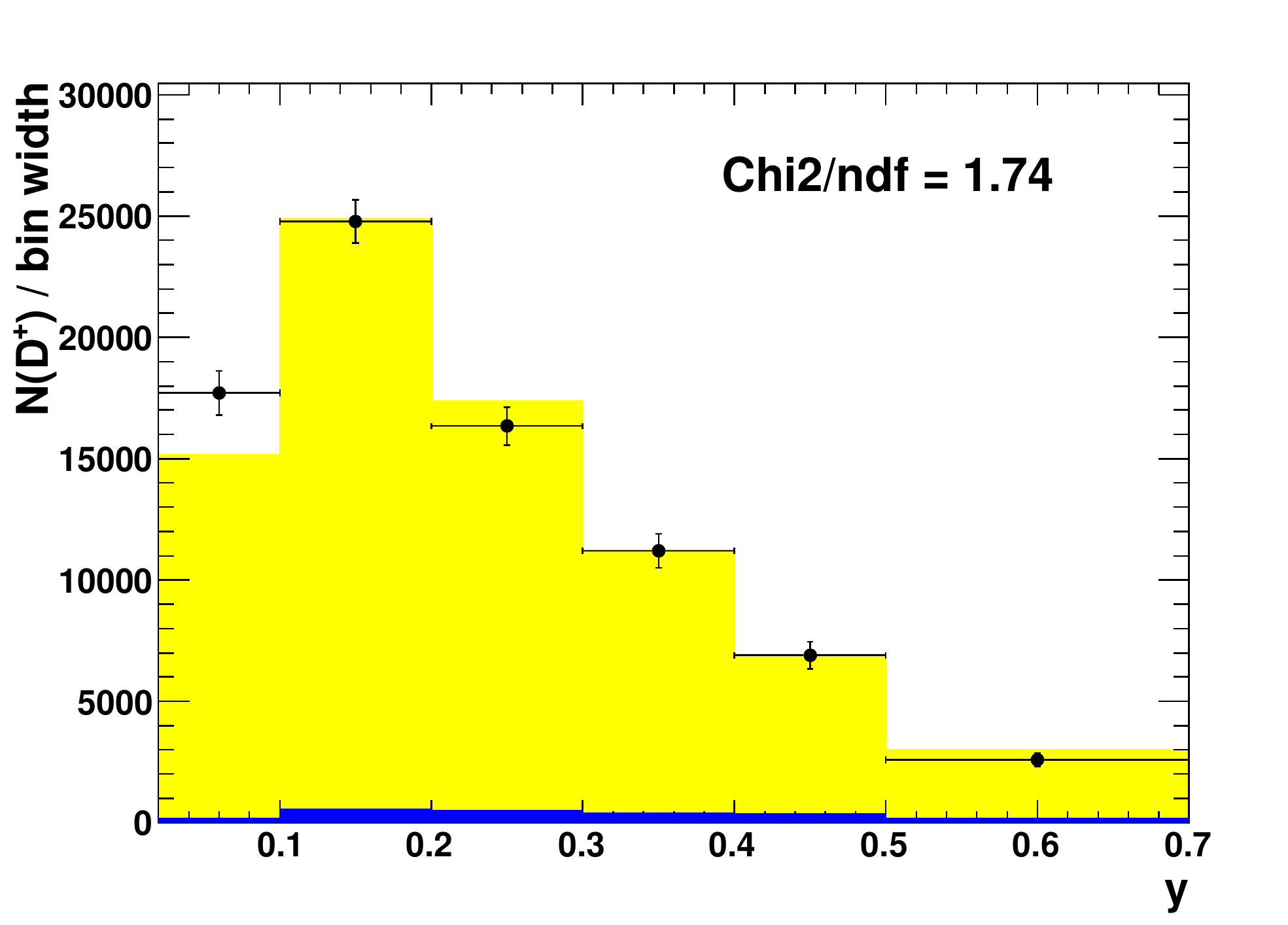}
  \caption[Control plots for $p_T(\Dch)$, $\eta(\Dch)$, $Q^2$ and $y$ before reweighting]
	{Control plots for $p_T(\Dch)$ (top left), $\eta(\Dch)$ (top right), $Q^2$ (bottom left) and $y$ (bottom right). The data are shown as points, with bars representing the statistical uncertainty. 
	The sum of charm and beauty MC is shown as the light shaded area; the beauty contribution is shown separately as the dark shaded area.}
  \label{fig:dch:cpbefore}
\end{figure}

Fig.~\ref{fig:dch:cpbefore} shows the control plots for $p_T(\Dch)$, $\eta(\Dch)$, $Q^2$ and $y$. 
The data are compared to the sum of charm and beauty MC; the beauty contribution was scaled by 1.6~\cite{Chekanov:2009kj,Abramowicz:2010zq,Abramowicz:2011kj}, 
while the charm contribution was re-normalised to the difference between the data and re-scaled MC beauty.%
\footnote{This procedure corresponds to the measurement of charm production, when the beauty contribution is assumed to be known \textit{a priori}.} 
The beauty contribution is shown separately; typically it is below $5\%$. More control plots can be found in Appendix~\ref{sec:app:dch} (Fig.~\ref{fig:dch:cpadd}). 
The MC does not describe well the shapes of $p_T(\Dch)$, $\eta(\Dch)$ and $Q^2$, thus the generator-level MC cross sections had to be reweighted.%
\footnote{For a single correction one should say rather ``weighting'', 
but the term ``reweighting'' is much more convenient and will be used in this work. 
Moreover, this is not a single correction applied to MC in the analysis.} 
The reweighting procedure is described later in this Section.

\begin{figure}[htbp]
  \centering
  \includegraphics[width=0.495\figwidth,trim=1mm 0mm 12mm 6mm,clip=true]{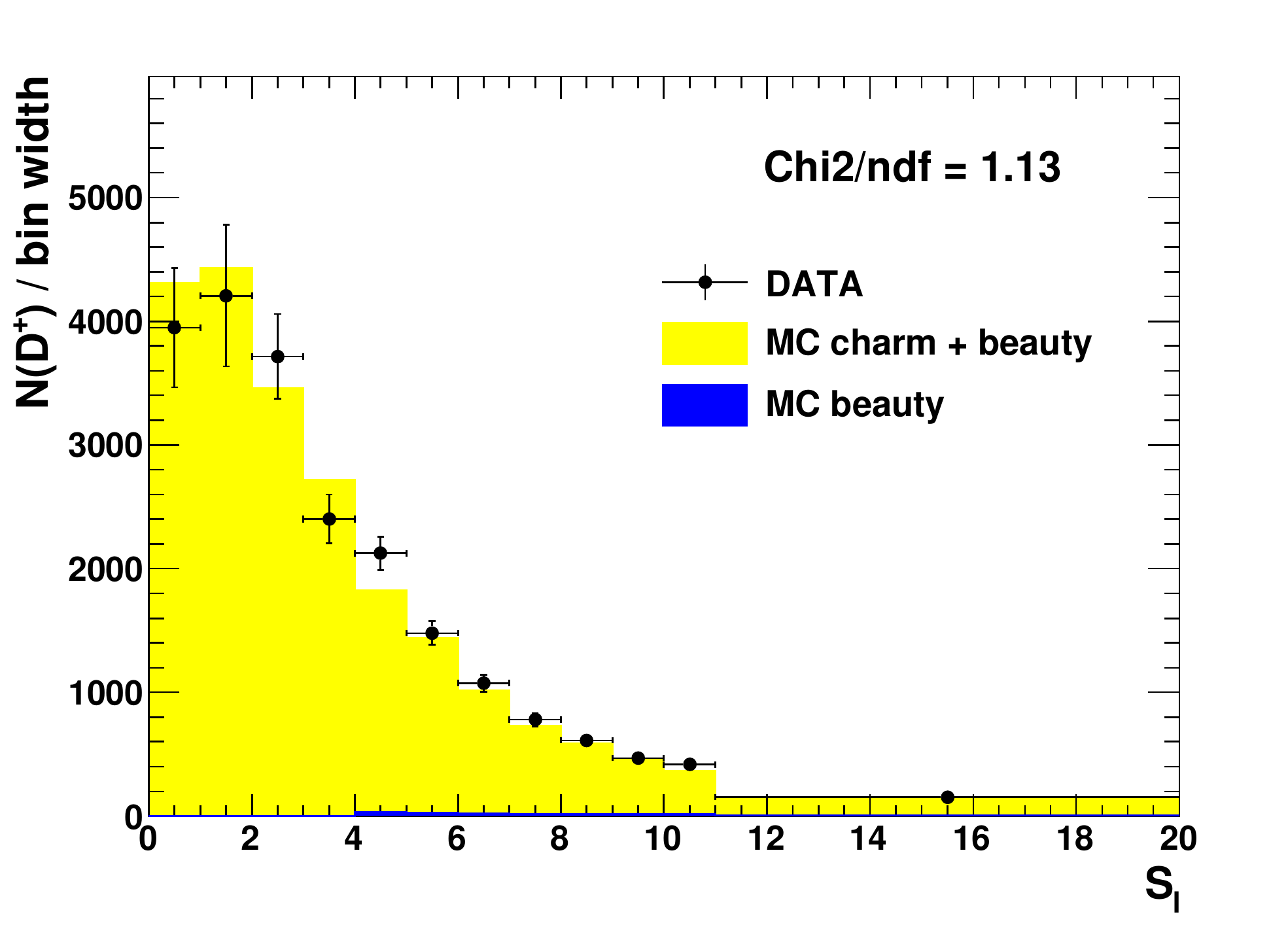}
  \includegraphics[width=0.495\figwidth,trim=1mm 0mm 12mm 6mm,clip=true]{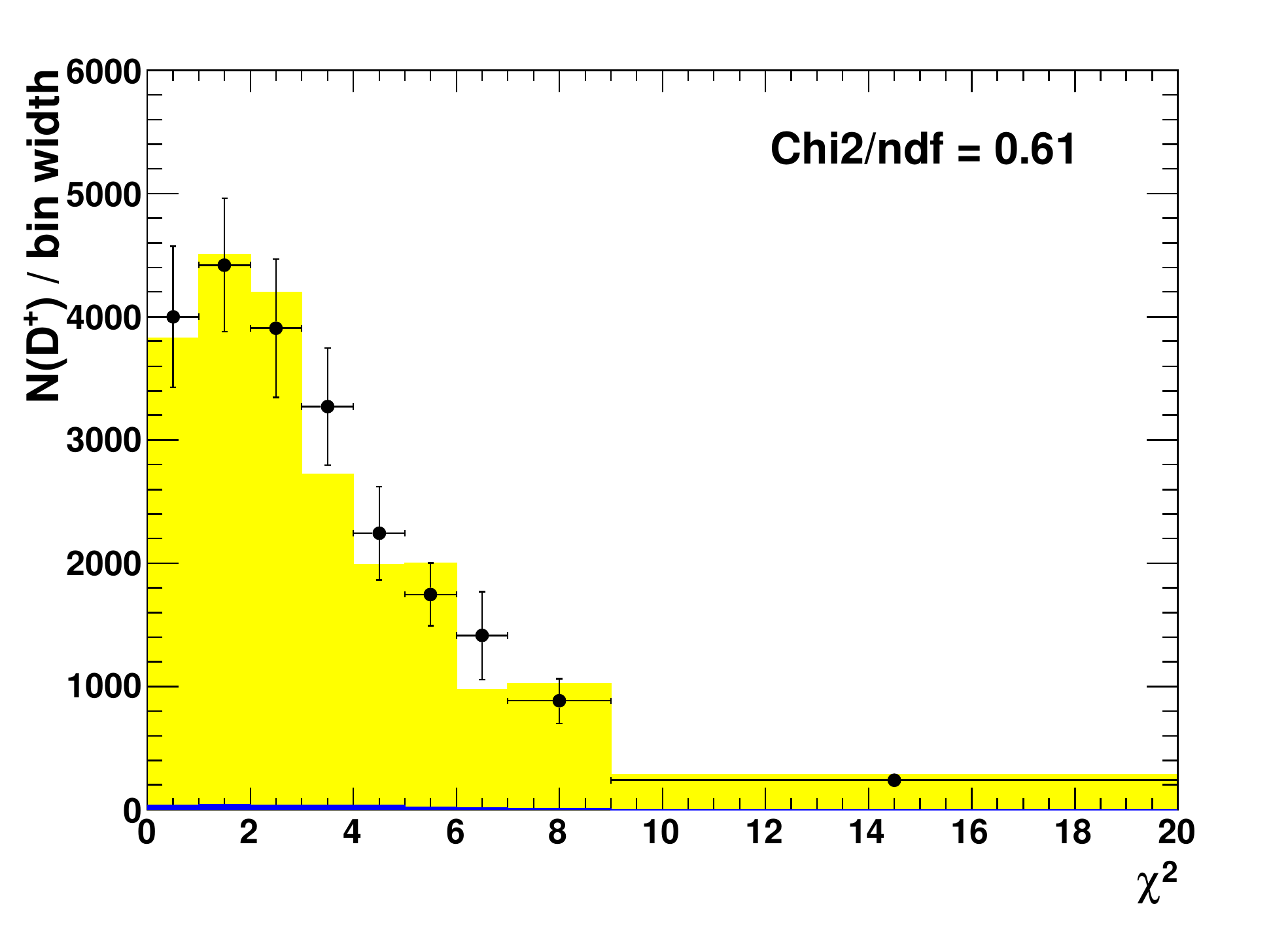}
  \caption[Control plots for $S_l$ and $\chisq_{\rm sec~vtx}$]
	{Control plots for $S_l$ (left) and $\chisq_{\rm sec~vtx}$ (right). The data are shown as points, with bars representing the statistical uncertainty. 
	The sum of charm and beauty MC is shown as the light shaded area; the beauty contribution is shown separately as the dark shaded area.}
  \label{fig:dch:cpsecvtxbefore}
\end{figure}

Fig.~\ref{fig:dch:cpsecvtxbefore} shows control plots for $S_l$ and $\chisq_{\rm sec~vtx}$, obtained before applying the cuts on these quantities. 
MC simulations describe these distributions well. This fact is of crucial importance, because the detector acceptance \ozmod{strongly} depends 
on the cuts applied on $S_l$ and $\chisq_{\rm sec~vtx}$, so that an incorrect simulation of their shape would lead to large systematic uncertainties;
this was the dominant systematic uncertainty in the previous analysis~\cite{zd0dp}, performed with an inferior tracking alignment and calibration.

\subsubsection{Data to MC matching}
\label{sec:dch:cs:acc:mat}

A general rule of the MC reweighting approach is that the kinematic weights must be applied at the generator level only. 
This is straightforward for reweighting in inclusive event quantities, e.g.\ $Q^2$, although it 
becomes complicated if the shapes of \Dch kinematic variables should be corrected 
(namely $p_T(\Dch)$ and $\eta(\Dch)$), because:
\begin{itemize}
	\item in each MC event there may be more than one generated \Dch mesons;
	\item the efficiency of the \Dch reconstruction in the present analysis is not very high 
		($\mathcal{E}=1.5\text{--}15\%$ depending on $p_T(\Dch)$; see Fig.~\ref{fig:dch:eff} in Appendix~\ref{sec:app:dch}), 
		thus for a large fraction of events, the reweighting of all generated \Dch mesons will result 
		in a reweighting of the combinatorial background, 
		which does not make sense and potentially may introduce an additional systematic uncertainty.
\end{itemize}
These complications arise from the fact that according to the general rule weights must be applied for \emph{events}, 
while the control plots allow their determination for \emph{candidates} \ozmod{for \Dch} only. If applied for events, weights are unique 
for both generator and reconstructed level, while if applied for candidates, the uniqueness is lost.

Therefore in the present analysis a procedure of \emph{matching} between true and reconstructed 
\Dch candidates was developed. It contained two steps:
\begin{enumerate}
	\item for each daughter track the corresponding generator-level particle was matched, if the following 
	criteria (motivated by the resolution of the tracking system) were fulfilled:
	\begin{itemize}
	\item $|\Delta p_T|=|p_T^{\rm gen}-p_T^{\rm rec}|<\SI{0.2}{GeV}$, where $p_T^{\rm gen}$ and $p_T^{\rm rec}$ 
		are the transverse momenta of the generated and reconstructed particles, respectively, and
	\item $\Delta R=\sqrt{(\phi^{\rm gen}-\phi^{\rm rec})^2+(\eta^{\rm gen}-\eta^{\rm rec})^2}<0.035$, 
		where $\phi^{\rm gen}$ and $\phi^{\rm rec}$ are the azimuthal angles of the generated and reconstructed 
		particles, respectively, and $\eta^{\rm gen}$ and $\eta^{\rm rec}$ are the pseudorapidities of the 
		generated and reconstructed particles, respectively;
	\end{itemize}
	\item if all daughter tracks were successfully matched to generator-level particles and if the generator-level 
	particles originated from a \Dch meson in the considered decay channel%
	\footnote{The indirect decay channel $\Dch \to \tilde{K}^{*0}(892)\pi^{+}$ with subsequent $\tilde{K}^{*0}(892) \to K^{-}\pi^{+}$ was simulated in the MC and considered in the matching procedure.}, 
	the reconstructed \Dch candidate 
	was considered to be successfully matched to the generator-level one.
\end{enumerate}
The efficiency of this matching procedure was found to be very close to 100\%~\cite{ozmaster}.%
\footnote{The efficiency of the matching procedure is defined as the ratio of the number of matched particles 
to the number of candidates in the fitted signal. This quantity is not to be confused with the efficiency defined in Eq.~\ref{eq:dch:acc}.}
Note that the matching procedure is needed also to determine purity and efficiency (the numerators in~\ref{eq:dch:acc}), 
although it is not needed for the accpetance determination. 

\subsubsection{MC reweighting}
\label{sec:dch:cs:acc:rew}

\begin{figure*}[htbp]
  \sidecaption
  \centering
  \includegraphics[width=0.528\figwidth*\real{1.65},trim=0 0mm 0mm 13mm,clip=true]{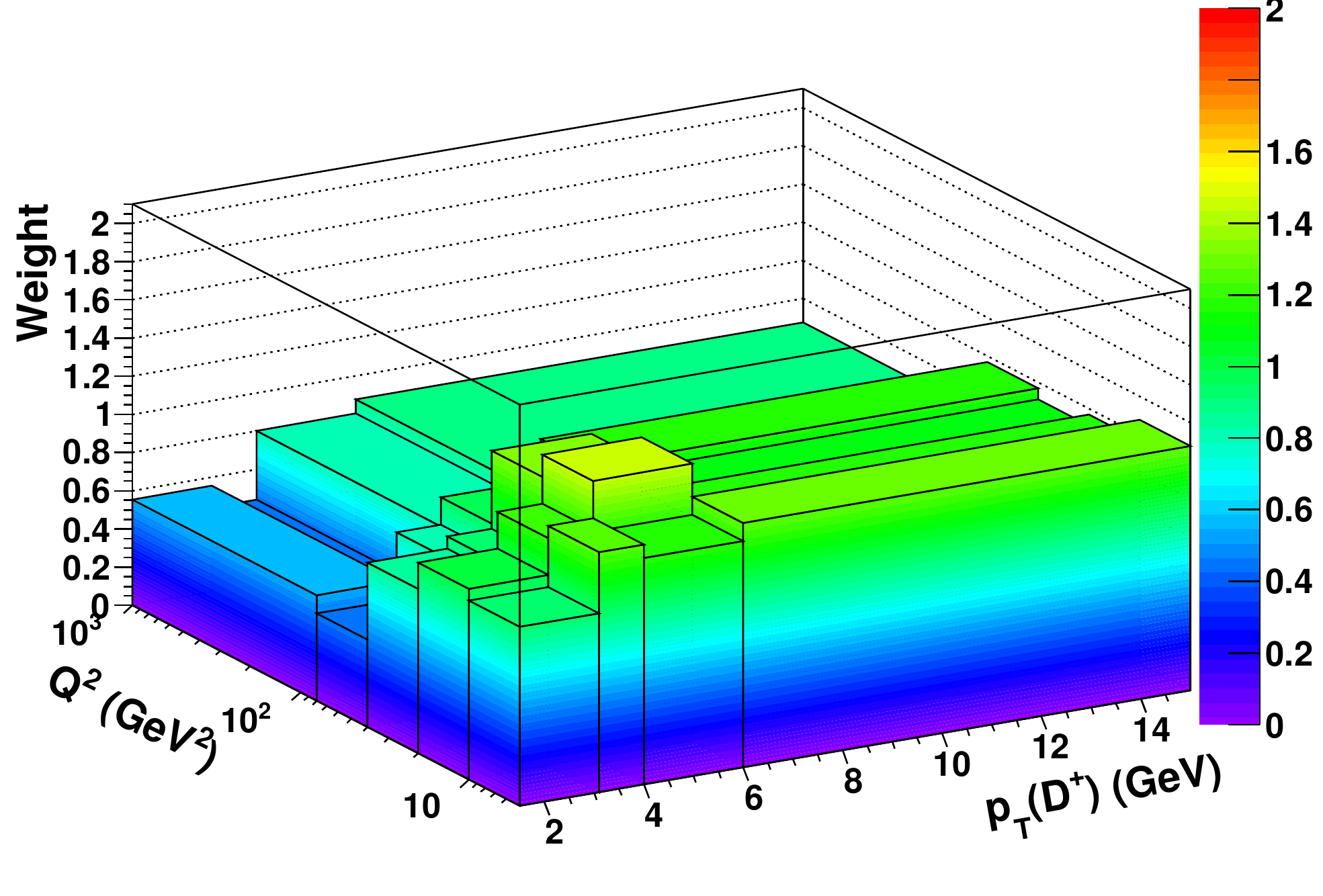}
  \includegraphics[width=0.464\figwidth*\real{1.65},trim=1mm 0mm 19mm 0,clip=true]{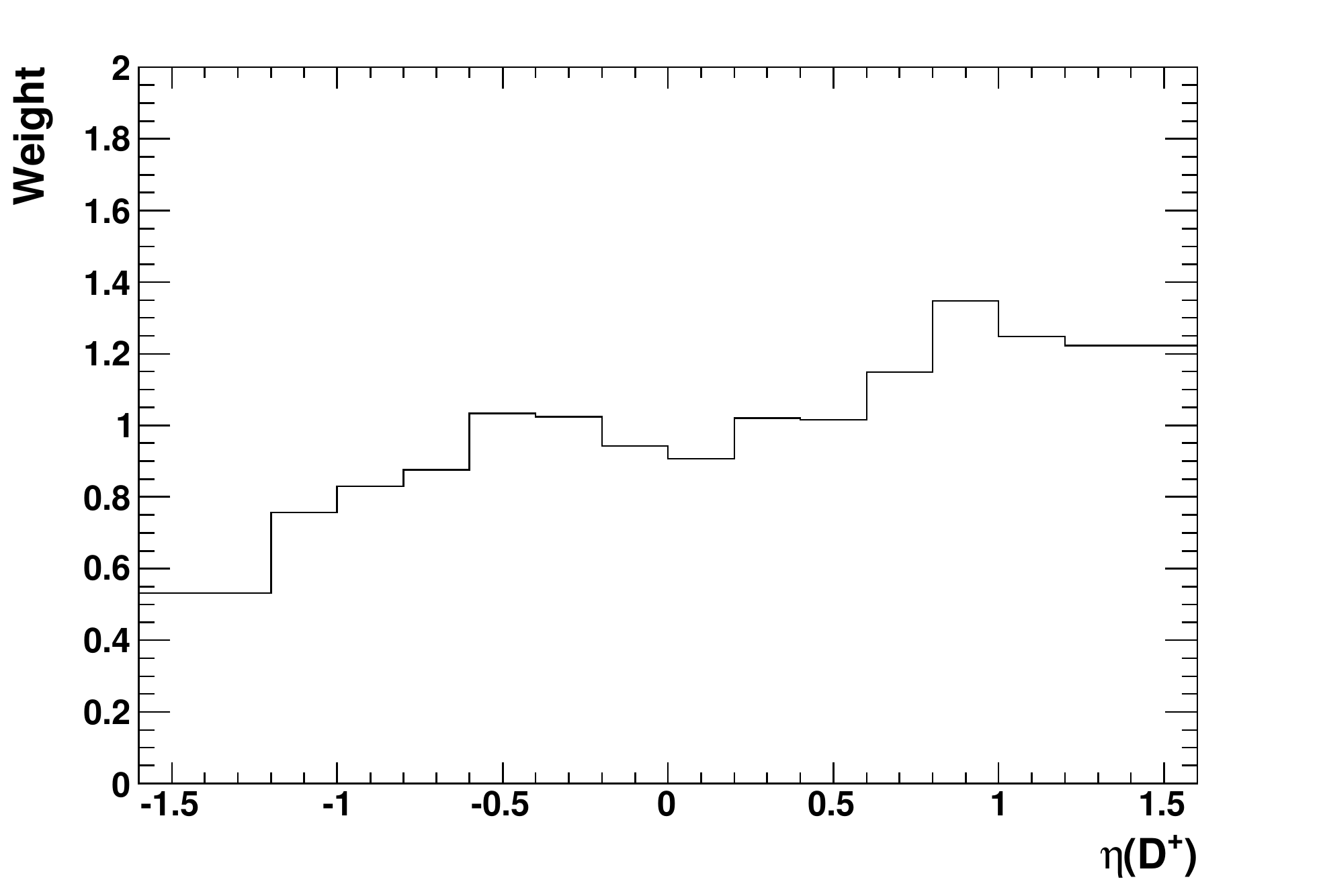}
  \caption[Step functions used for MC reweighting in $p_T(\Dch)\text{--}Q^2$ and $\eta(\Dch)$]
	{Step functions used for MC reweighting in $p_T(\Dch)\text{--}Q^2$ (left) and $\eta(\Dch)$ (right).}
  \label{fig:dch:rewf}
\end{figure*}

Since transverse momentum $p_T(\Dch)$ and virtuality $Q^2$ are significantly correlated, reweighting in these two variables was performed simultaneously, 
while the cross section in pseudorapidity $\eta(\Dch)$ was reweighted independently.
In both cases step functions determined from the control plots as the ratios of the number of signal events in the data 
to the number of signal events in the charm MC were used as reweighting functions. 
The beauty MC contribution was subtracted from the data and reweighting was applied only to the charm MC. 
The reweighting functions are shown in Fig.~\ref{fig:dch:rewf}.
For the reweighting in $p_T(\Dch)$ and $\eta(\Dch)$ only the matched \Dch candidates, as explained in Section~\ref{sec:dch:cs:acc:mat}, 
were reweighted at the reconstruction level, because the reweighting of non-reconstructed \Dch effectively 
would result in a meaningless reweighting of combinatorial background, thus producing additional statistical fluctuations.
For the acceptance calculation according to Eq.~\ref{eq:dch:acc}, all \Dch were reweighted at the generator level (for the $N_i^{\rm gen}$ calculation).%
\footnote{Note that this procedure does not guarantee that the same weights have been applied at both levels (generator and reconstruction), 
and therefore cannot \textit{a priori} guarantee consistency for the determined acceptance. 
In order to check it, the $p_T(\Dch)\text{--}Q^2$ reweighting was performed by applying the same weight, 
derived from the ``best'' \Dch (with highest $p_T(\Dch)$), on both levels. 
The difference between the two procedures was found to be less than $0.5\%$.}

\begin{figure}[htbp]
  \centering
  \begin{minipage}[t]{0.49\textwidth}
  \includegraphics[width=0.5\figwidth,trim=1mm 0mm 12mm 6mm,clip=true]{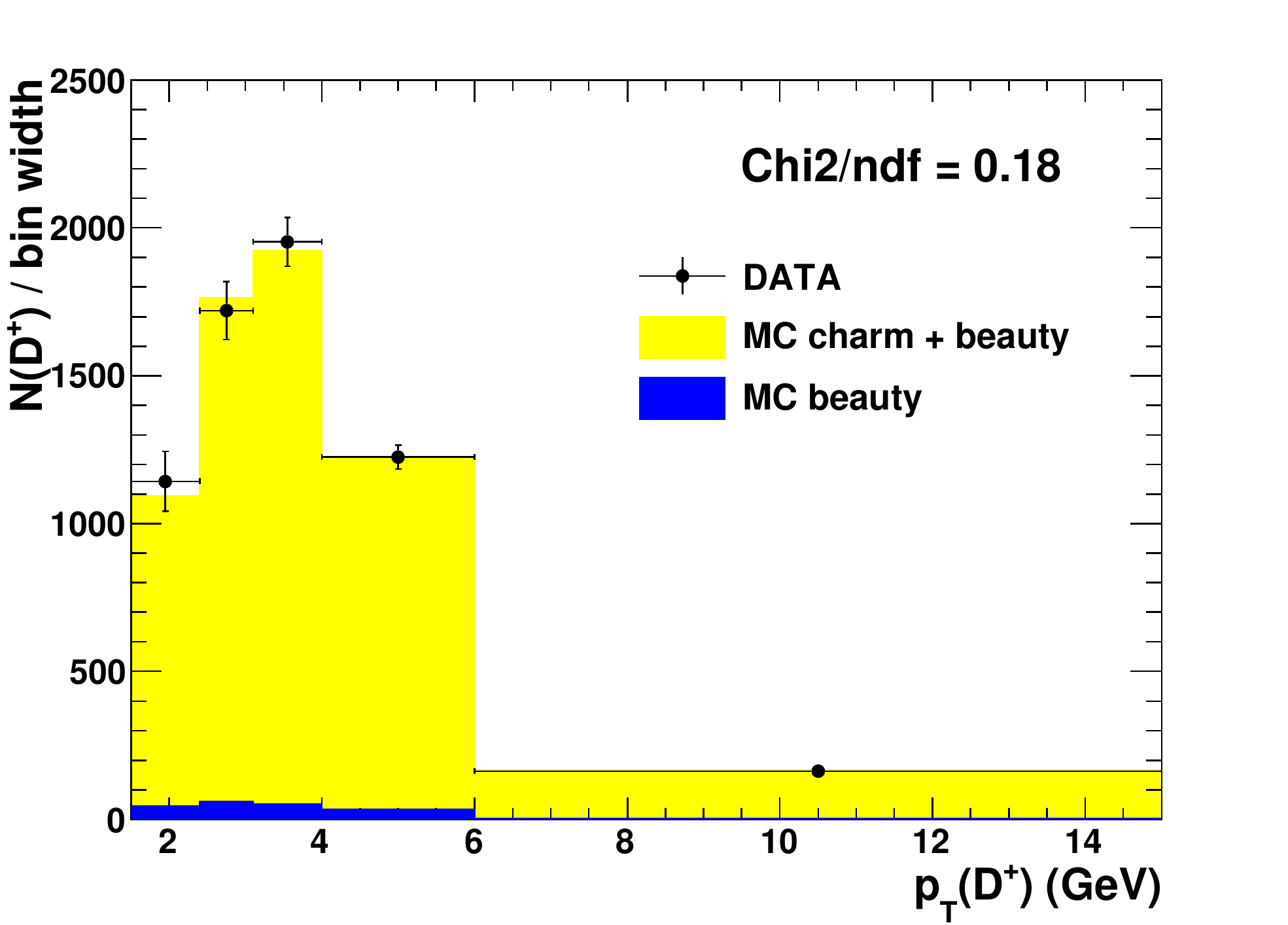}
  \includegraphics[width=0.5\figwidth,trim=1mm 0mm 12mm 6mm,clip=true]{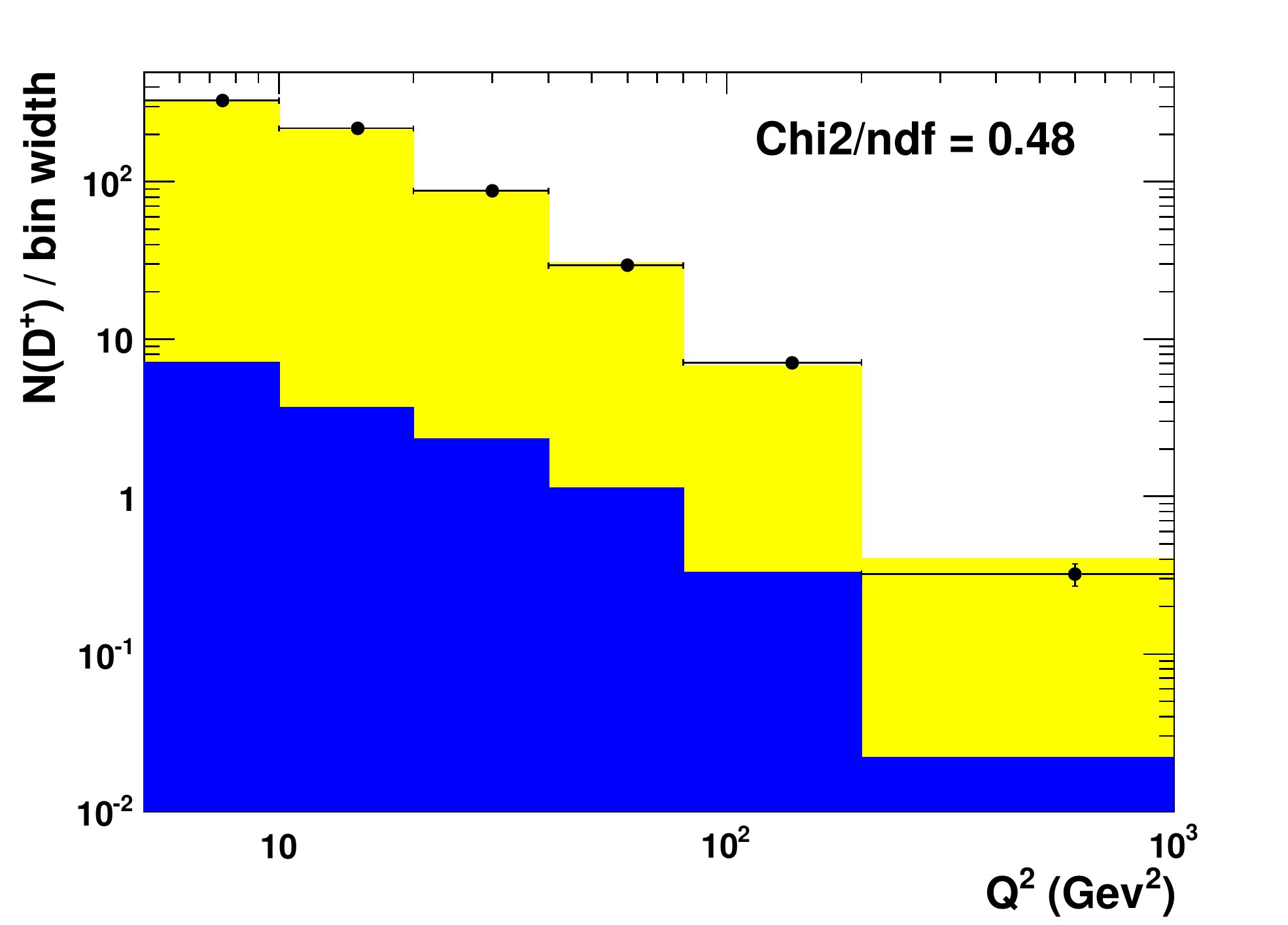}
  \end{minipage}
  \begin{minipage}[t]{0.49\textwidth}
  \includegraphics[width=0.5\figwidth,trim=1mm 0mm 12mm 6mm,clip=true]{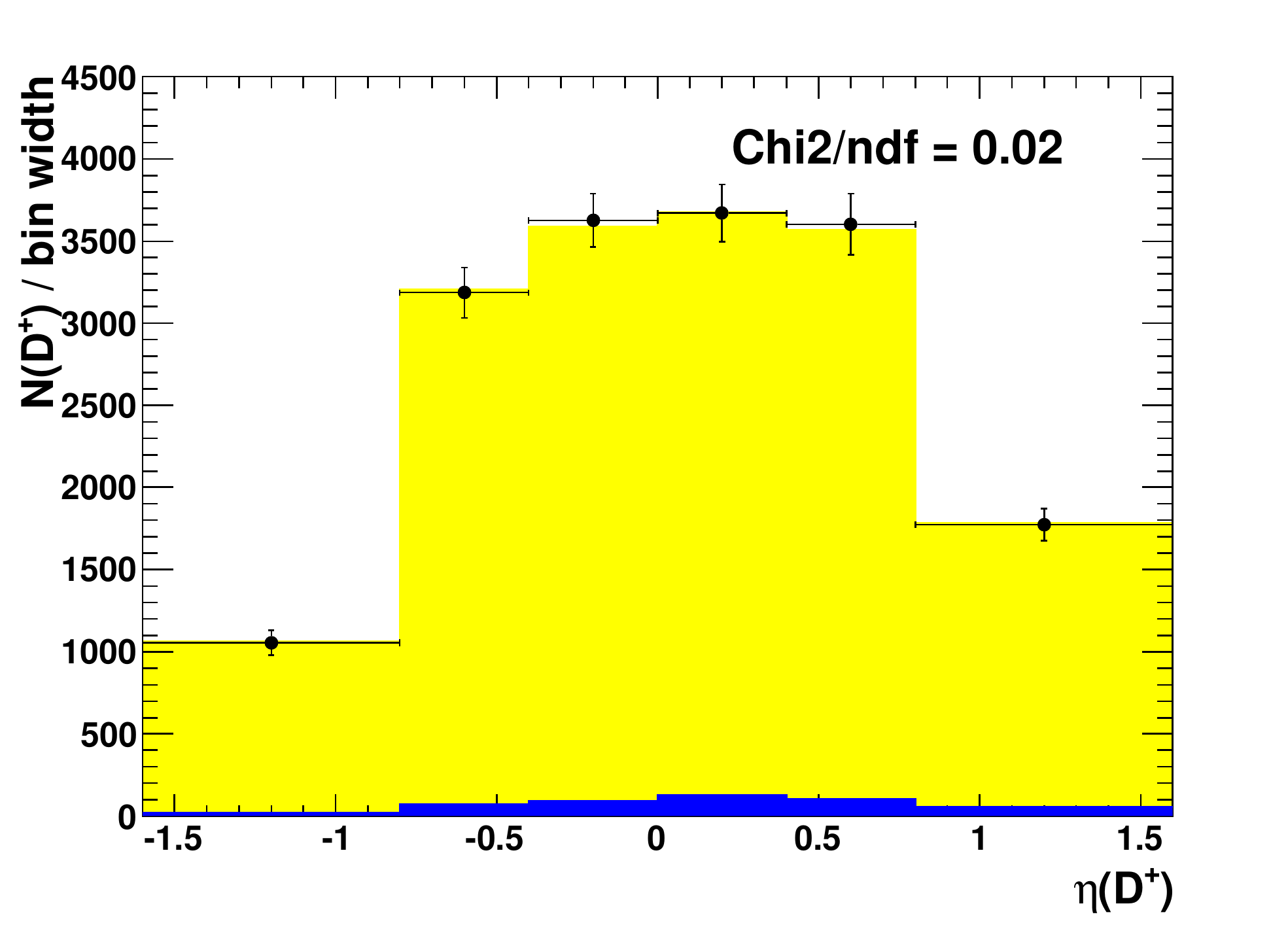}
  \caption[Control plots for $p_T(\Dch)$, $\eta(\Dch)$ and $Q^2$ after reweighting]
	{Control plots for $p_T(\Dch)$ (top left), $\eta(\Dch)$ (top right) and $Q^2$ (bottom) after reweighting. The data are shown as points, with bars representing the statistical uncertainty. 
	The sum of charm and beauty MC is shown as the light shaded area; the beauty contribution is shown separately as the dark shaded area.}
  \label{fig:dch:cpafter}
  \end{minipage}
\end{figure}

\begin{figure}[htbp]
  \centering
  \includegraphics[width=0.495\figwidth,trim=1mm 0mm 16mm 7mm,clip=true]{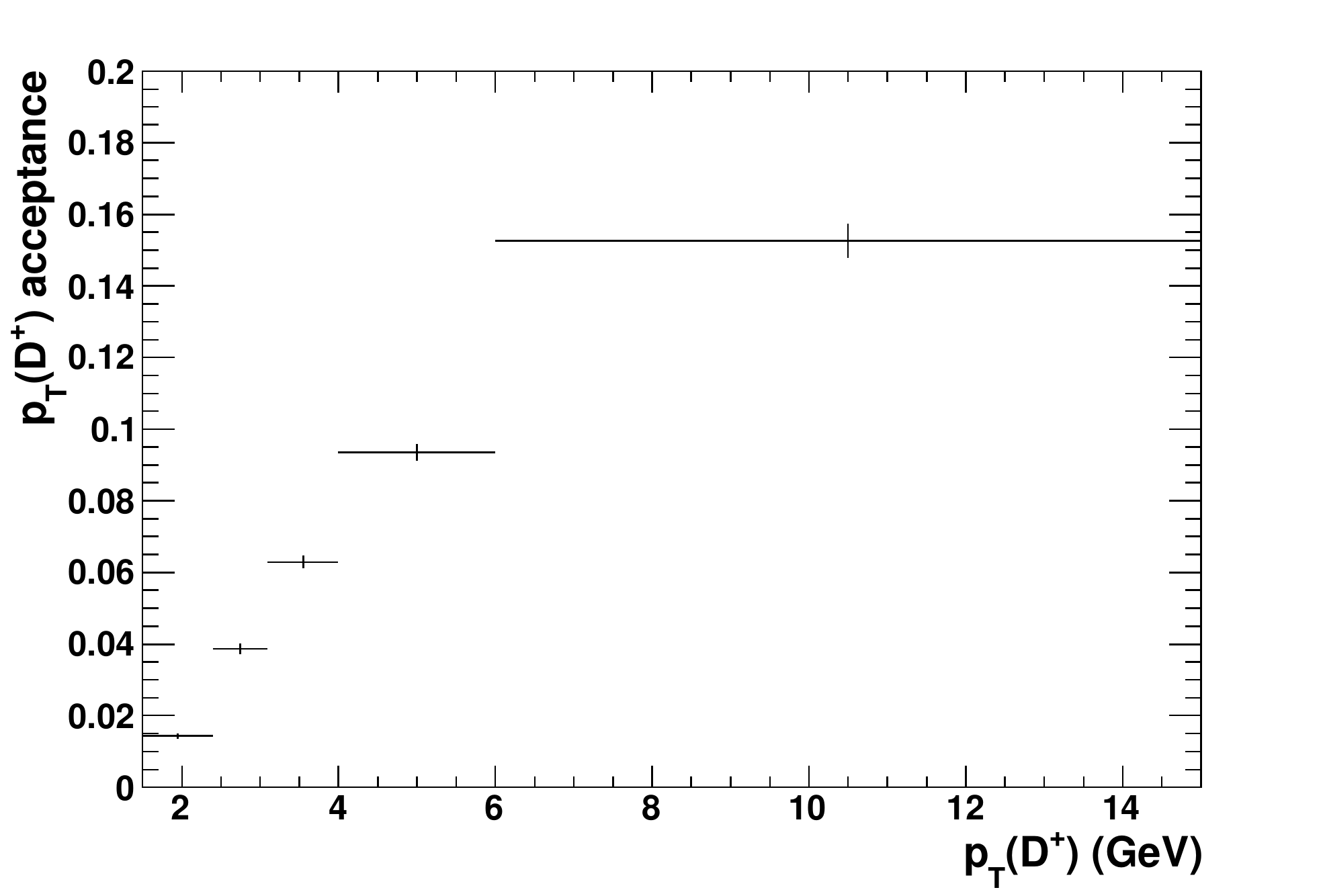}
  \includegraphics[width=0.495\figwidth,trim=1mm 0mm 16mm 7mm,clip=true]{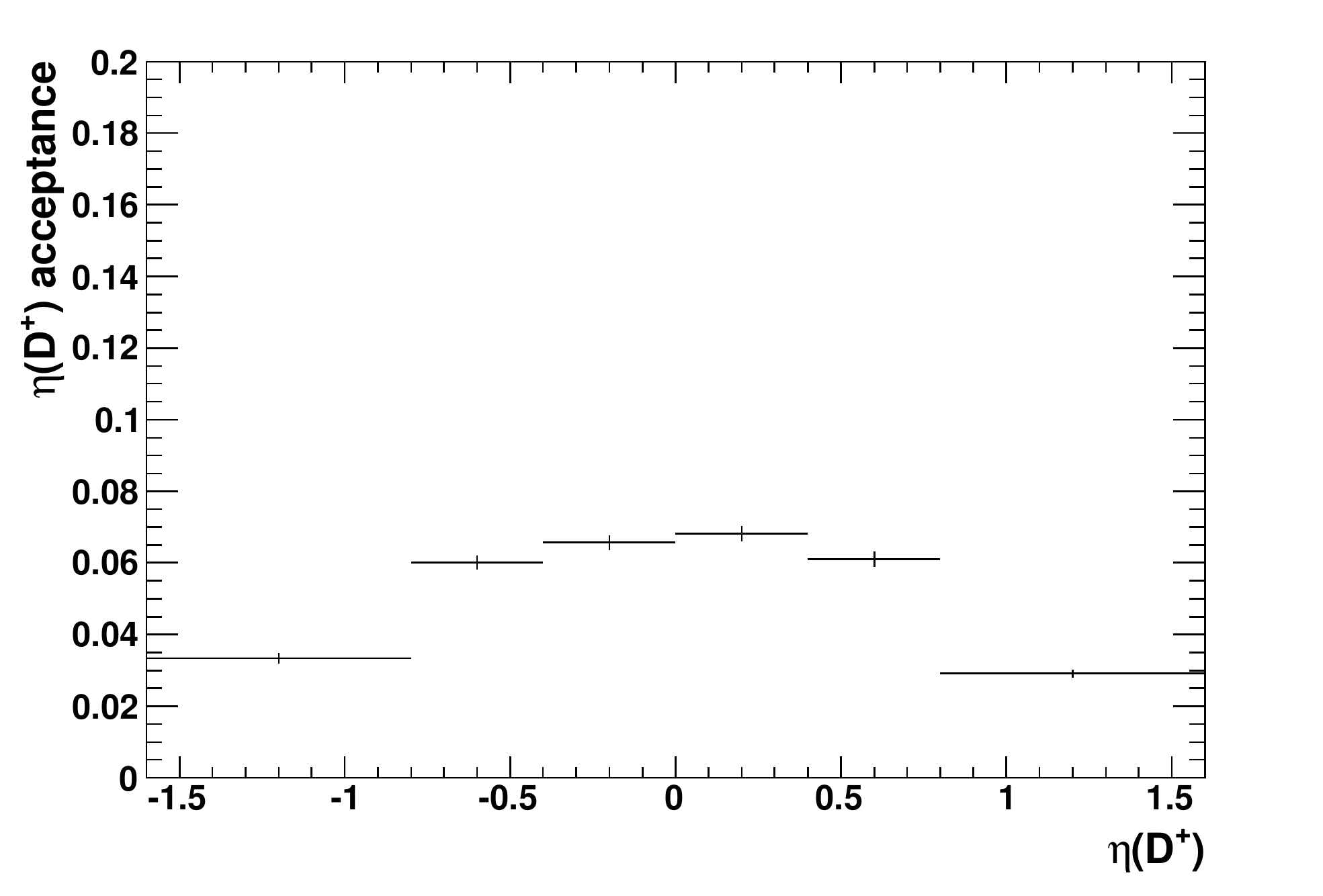}
  \includegraphics[width=0.495\figwidth,trim=1mm 0mm 16mm 7mm,clip=true]{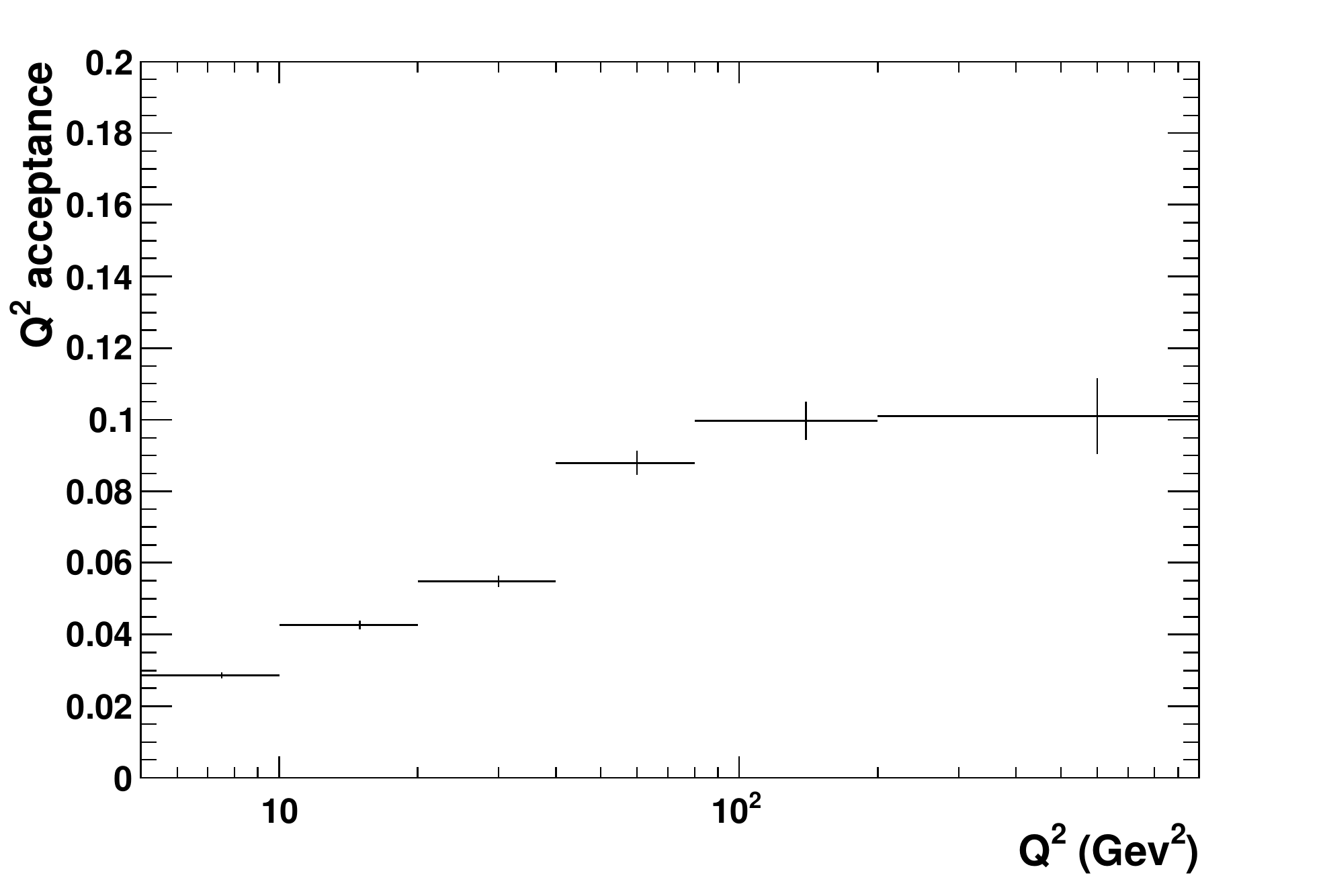}
  \includegraphics[width=0.495\figwidth,trim=1mm 0mm 16mm 7mm,clip=true]{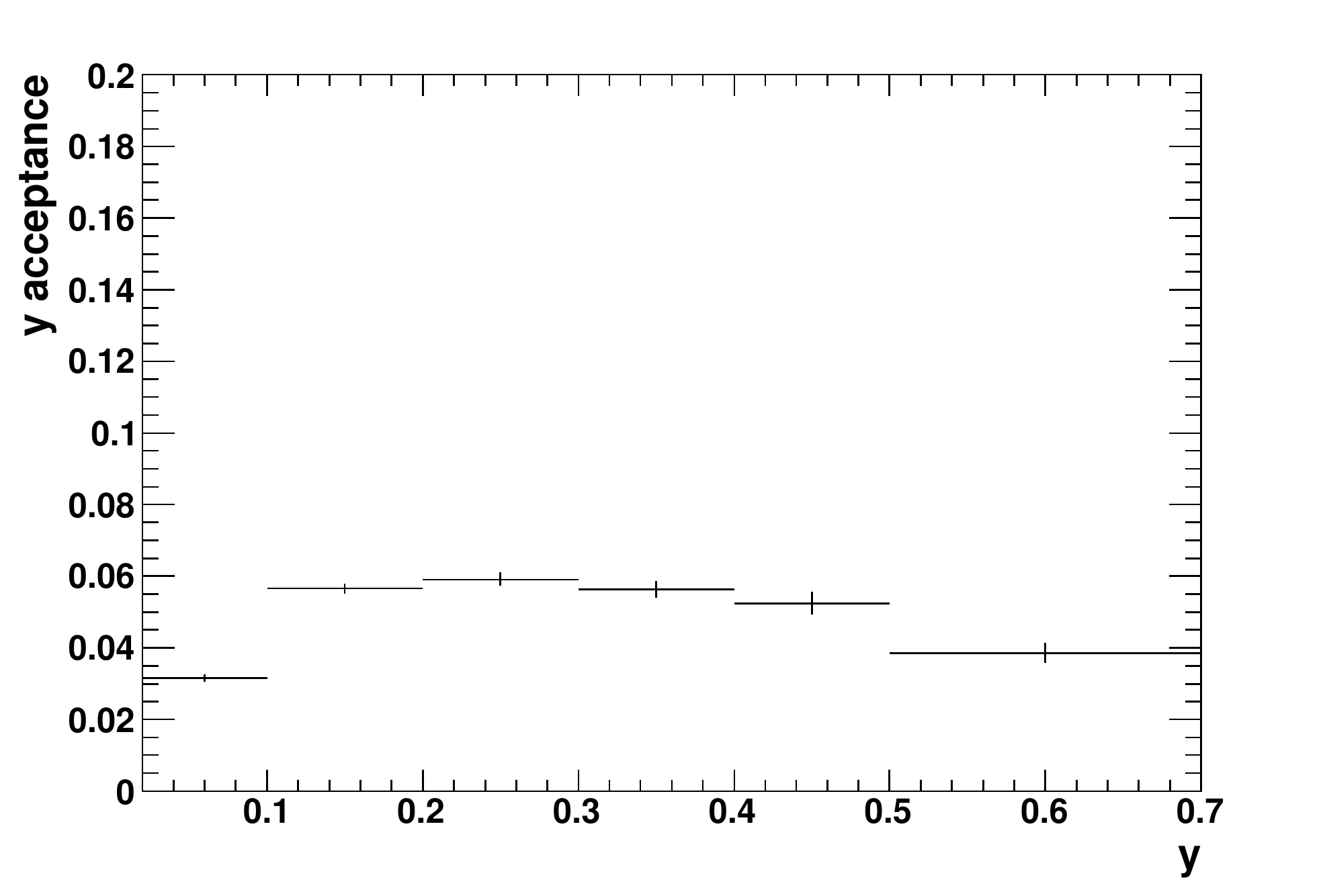}
  \caption[Acceptance as function of $p_T(\Dch)$, $\eta(\Dch)$, $Q^2$ and $y$]
	{Acceptance as a function of $p_T(\Dch)$ (top left), $\eta(\Dch)$ (top right), $Q^2$ (bottom left) and $y$ (bottom right). 
	Error bars represent the statistical uncertainty.}
  \label{fig:dch:acc}
\end{figure}

The control plots for $p_T(\Dch)$, $\eta(\Dch)$ and $Q^2$ after reweighting are shown in Fig.~\ref{fig:dch:cpafter}. 
The reweighted MC simulations describe the data well and were used to determine acceptance corrections.
The acceptance as a function of $p_T(\Dch)$, $\eta(\Dch)$, $Q^2$ and $y$ is shown in Fig.~\ref{fig:dch:acc}. 
It is not high, mainly because of the strong cut applied on the decay-length significance in order to reduce combinatorial background, 
varying from $1.5\%$ at low $p_T(\Dch)$ to $15\%$ at high $p_T(\Dch)$. 
The same plots for purity and efficiency are provided in Appendix~\ref{sec:app:dch} (Fig.~\ref{fig:dch:pur} and~\ref{fig:dch:eff}).

\subsubsection{Additional corrections}
\label{sec:dch:cs:cor}

Additional corrections were applied in the MC simulations:
%
\begin{itemize}
\item {\bf trigger-inefficiency correction.} 
Most of the First Level Trigger bits (see Section~\ref{sec:exp:hera:h1zeus:zeusdet}) used in this analysis had some requirements on the track multiplicity in the events. 
The efficiency of these criteria was measured~\cite{mishath} using a trigger without track requirements and the detector 
simulation was tuned to match the data. The trigger-inefficiency corrections for the MC simulations 
were between $1\text{--}10\%$ for different tracking requirements. 
The corrections changed the overall efficiency of the triggers used in the analysis by a negligible amount 
for medium-$Q^2$ values and up to $\sim 2\%$ for the low- and high-$Q^2$ regions. 
More details on this study can be found in~\cite{mishath}; 
%
\item {\bf tracking-inefficiency correction.} 
A special study~\cite{libovth} was performed to assess the tracking inefficiency for charged pions
due to hadronic interactions in the detector material and how well the MC simulations reproduce
these interactions. The MC simulations were found to underestimate the interaction rate by
about 40\% for $p_T < \SI{1.5}{GeV}$ and to agree with the data for $p_T > \SI{1.5}{GeV}$; 
more details can be found in~\cite{libovth} and references therein. 
A corresponding correction was applied to the MC simulations. 
The effect of the correction on the \Dch-production cross section was found to be about 3\%. 
The effect of the correction on the \Dch differential cross sections is provided in Appendix~\ref{sec:app:dch} 
(Fig.~\ref{fig:dch:trackeff_cs});
%
\item{\bf decay-length smearing.} 
The $S_l$ distribution was found to be asymmetric~\cite{mishath} with respect to zero, 
with charmed mesons dominating in the positive tail. 
Detector resolution effects cause the negative tail, which is dominated by light-flavour events. A
smearing was applied to the decay length of a small fraction of the MC events
in order to reproduce the negative decay-length data. 
The parameters of the smearing had to be tuned to describe the data. 
The effect of the smearing is typically below 3\%.
More details on this correction can be found in~\cite{mishath}. 
\end{itemize}

\subsubsection{Systematic uncertainties}
\label{sec:dch:cs:syst}

The systematic uncertainties were determined by changing the analysis procedure or varying parameter values 
within their estimated uncertainties and repeating the extraction of the signals and the cross-section calculations. 
The following sources of systematic uncertainties were considered with \ozmod{the impact} on the cross sections 
given in parentheses:
\begin{itemize}
	\item the cut on the positions $|x_{e^{\prime}}|$ and $|y_{e^{\prime}}|$ of the scattered electron in the RCAL was
		varied by $\SI{\pm 1}{cm}$ in both the data and the MC simulations, to account for potential
		imperfections of the detector simulation near the inner edge of the CAL (±1\%);
	\item the reconstructed electron energy was varied by $\pm 2\%$ in the MC only, to account
		for the uncertainty in the electromagnetic energy scale (< 1\%);
	\item the energy of the hadronic system was varied by $\pm$3\% in the MC only, to account
		for the uncertainty in the hadronic energy scale (< 1\%);
	\item the FLT tracking-efficiency corrections for the MC were varied
		within their estimated uncertainties (< 1\%)~\cite{mishath};
	\item uncertainties due to the signal-extraction procedure were estimated by repeating the fit
		in both the data and the MC using:
		\begin{itemize}
			\item an exponential function for the background parametrisation (< 1\%);
			\item a signal parametrisation changed by simultaneously varying the $\beta$ parameter
				of the modified Gaussian function\ozmod{(see Eq.~\ref{eq:modgaus})} in the data and MC by ${}^{+0.1}_{-0.2}$ from the nominal 
				value $0.5$. The range was chosen to cover the values which give the best description
				of the mass peaks in the data and MC simulations in bins of the differential cross sections (${}^{+0.7\%}_{-1.5\%}$);
		\end{itemize}
	\item uncertainties due to the decay-length smearing procedure were estimated by varying its parameters by $\pm 50\%$ ($\pm 1\%$)~\cite{mishath}.
		As a further cross check, the cut on the decay-length significance was varied between 3 and 5. The
		resulting variations of the cross sections were compatible with the variation of the
		decay-length smearing and were therefore omitted to avoid double counting;
	\item the scaling factor for the MC beauty-production cross sections was varied by $\pm 0.6$
		from the nominal value $1.6$. This was done to account for the range of the RAPGAP
		beauty-prediction normalisation factors extracted in various 
		analyses~\cite{Chekanov:2009kj,Abramowicz:2010zq,Abramowicz:2011kj} ($\pm 2\%$);
	\item the uncertainties due to the model dependence of the acceptance corrections were
		estimated by varying the shapes of the kinematic distributions in the charm MC
		sample in a range of good description of the data:
		\begin{itemize}
			\item the shape of the $\eta(\Dch)$ reweighting function ($\pm 2\%$);
			\item the shape of the $p_T(\Dch)\text{--}Q^2$ reweighting function ($\pm 4\%$);
		\end{itemize}
	\item the uncertainty of the pion track inefficiency due to nuclear interactions was evaluated by varying 
		the correction applied to the MC by its estimated
		uncertainty of $\pm 50\%$ of its nominal size ($\pm 1.5\%$);
	\item the contribution from the PHP processes was estimated using the PYTHIA MC sample and found to be $<{0.5\%}$, 
		therefore it was neglected;
	\item overall normalisation uncertainties:
	\begin{itemize}
		\item the simulation of the MVD hit efficiency ($\pm 0.9\%$)~\cite{mishath}; %
		\item the effect of the description of $\chisq_{\rm sec~vtx}<10$ was checked by multiplying
			$\chisq_{\rm sec~vtx}$ for \Dch candidates in the MC simulations by a factor $1.1$ to match the
			distribution in the data ($+2\%$)~\cite{mishath}; %
		\item the branching-ratio uncertainty ($\pm 2.1\%$);
		\item the measurement of the luminosity ($\pm 1.9\%$).
	\end{itemize}
The size of each systematic effect was estimated bin-by-bin except for the overall normalisation uncertainties. 
The overall systematic uncertainty was determined by adding the above uncertainties in quadrature. 
The normalisation uncertainties due to the luminosity measurement and that of the branching ratio were not 
included in the systematic uncertainties on the differential cross sections.
\end{itemize}

\subsection{Theoretical calculations}
\label{sec:dch:cs:th}

NLO QCD predictions were obtained in the FFNS with the HVQDIS program~\cite{hvqdis} (see Section~\ref{sec:th:hq:ep:ffns}).%
The renormalisation and factorisation scales were set to $\mu_r = \mu_f = \sqrt{Q^2+4 m_c^2}$ 
and the $c$-quark pole mass to $m_c = \SI{1.5}{GeV}$. 
The FFNS variant of the ZEUS-S NLO QCD PDF fit~\cite{zeuss} to inclusive DIS data was used as the parametrisation of the proton PDFs. 
The same charm mass and choice of scales were used in the fit as in the HVQDIS calculation. 
The strong coupling constant was set to $\alpha_s^{n_f=3}(M_Z) = 0.105$, corresponding to $\alpha_s^{n_f=5}(M_Z) = 0.116$. 

To calculate \Dch observables, events at the parton level were interfaced with a fragmentation model based on the Kartvelishvili function \cite{Kartvelishvili:1977pi}.
The fragmentation was performed in the $\gamma^{*}p$ centre-of-mass frame.
The Kartvelishvili parameter, $\alpha_K$, was parametrised~\cite{mishath} as a smooth function of the invariant mass 
of the $c \bar{c}$ system, $M_{c \bar{c}}$, to fit the measurements of the \Dstar fragmentation function 
by ZEUS \cite{zeusfrag} and H1 \cite{h1frag}: $\alpha_K (M_{c \bar{c}}) = 2.1 + 127/(M_{c \bar{c}}^2 - 4 m_c^2)$, 
with $m_c$ and $M_{c \bar{c}}$ in $\SI{}{GeV}$. 
In addition, the mean value of the fragmentation function was scaled down \ozmod{to} 0.95 since kinematic 
considerations \cite{frag06} and direct measurements~\cite{Seuster:2005tr} show that, on average, 
the momentum of $D^{+}$ mesons is 5\% lower than that of \Dstar mesons; 
this is due to some of the $D^{+}$ mesons originating from \Dstar decays. 
For the fragmentation fraction, $f(c \rightarrow D^{+})$, the value $0.2297 \pm 0.0078$ was used \cite{Lohrmann:2011np}.

The uncertainties on the theoretical predictions were estimated as follows:
 \begin{itemize}
	\item the renormalisation and factorisation scales were independently varied up and down by a factor of $2$;
	\item the $c$-quark mass was consistently changed in the PDF fits and in the HVQDIS calculations by $\pm \SI{0.15}{GeV}$;
	\item the proton PDFs were varied within the total uncertainties of the ZEUS-S PDF fit;
	\item the fragmentation function was varied by changing the functional dependence of the parametrisation function 
		$\alpha (M_{c \bar{c}})$ within uncertainties \cite{mishath};
	\item the fragmentation fraction was varied within its uncertainties.
\end{itemize}
The total theoretical uncertainty was obtained by summing in quadrature the effects of the individual variations. 
The dominant contributions originate from the variations of the $c$-quark mass and the scales. 
In previous studies~\cite{heracharmcomb} the uncertainty due to the variation of $\alpha_s^{n_f=3}(M_Z)$ was found 
to be insignificant and neglected here.

\subsection{Results}
\label{sec:dch:res}

\begin{figure*}[htbp]
  \centering
  \includegraphics[width=0.85\figwidth,trim=3mm 0mm 7mm 0,clip=true]{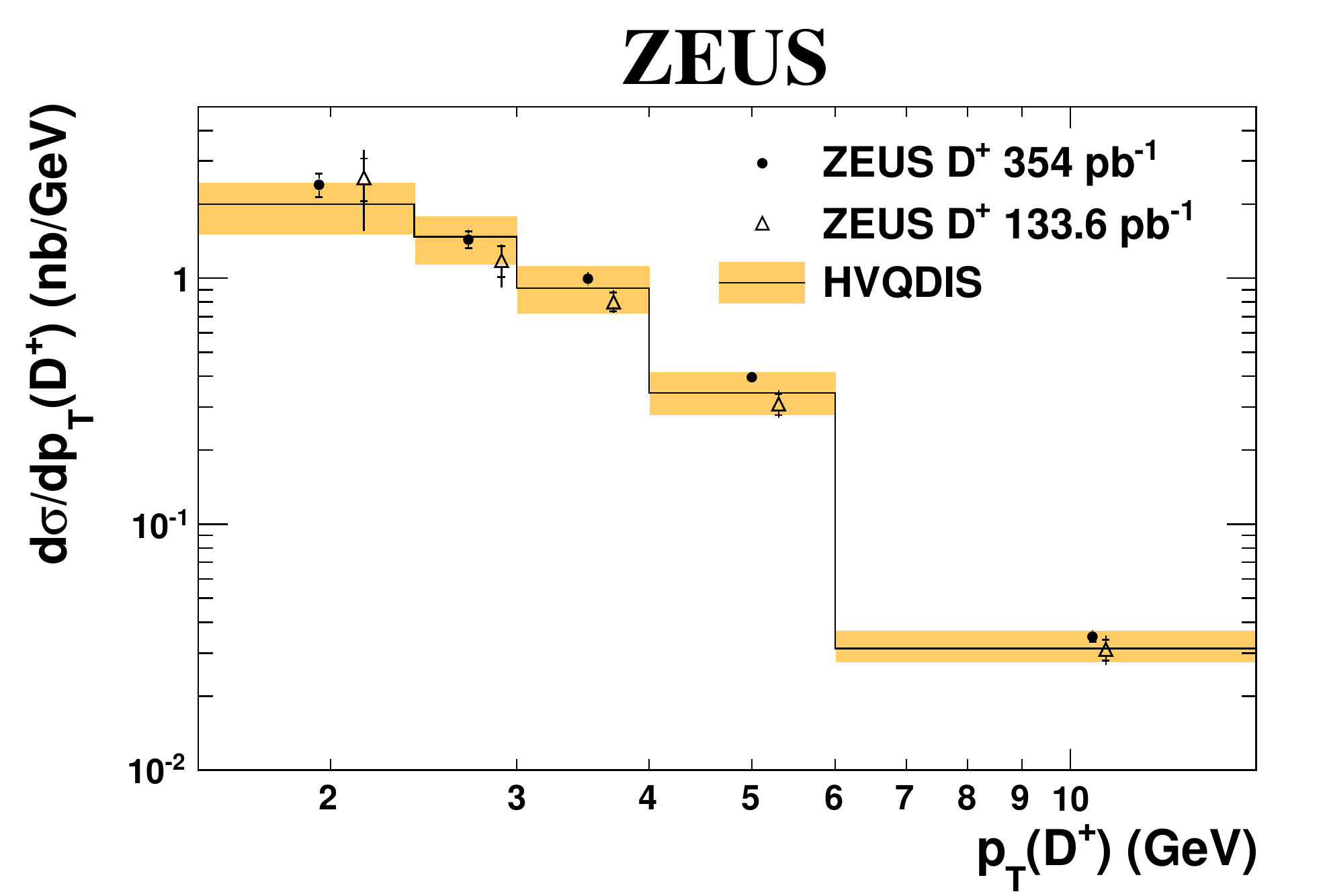}
  \includegraphics[width=0.85\figwidth,trim=3mm 0mm 7mm 0,clip=true]{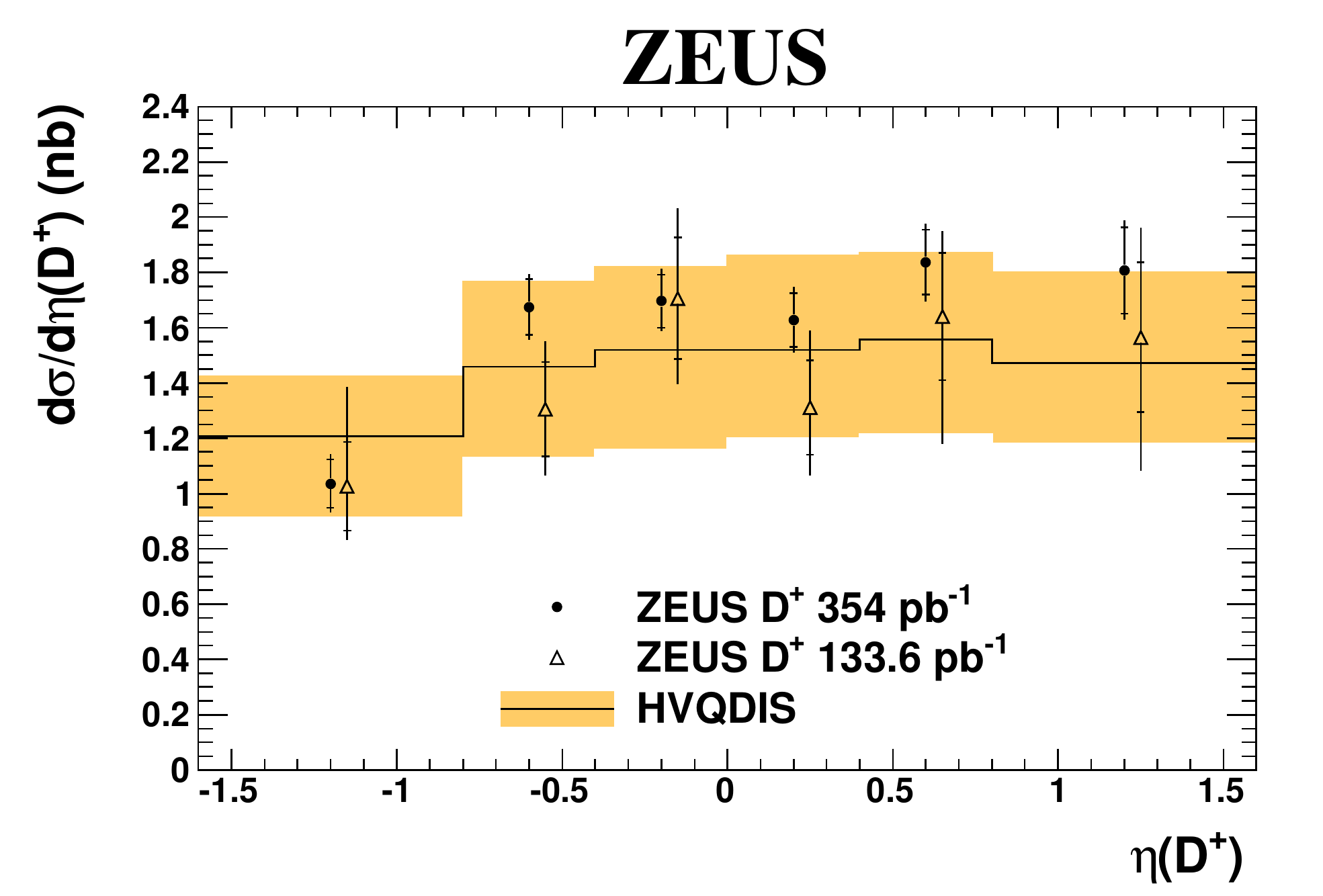}
  \includegraphics[width=0.85\figwidth,trim=3mm 0mm 7mm 0,clip=true]{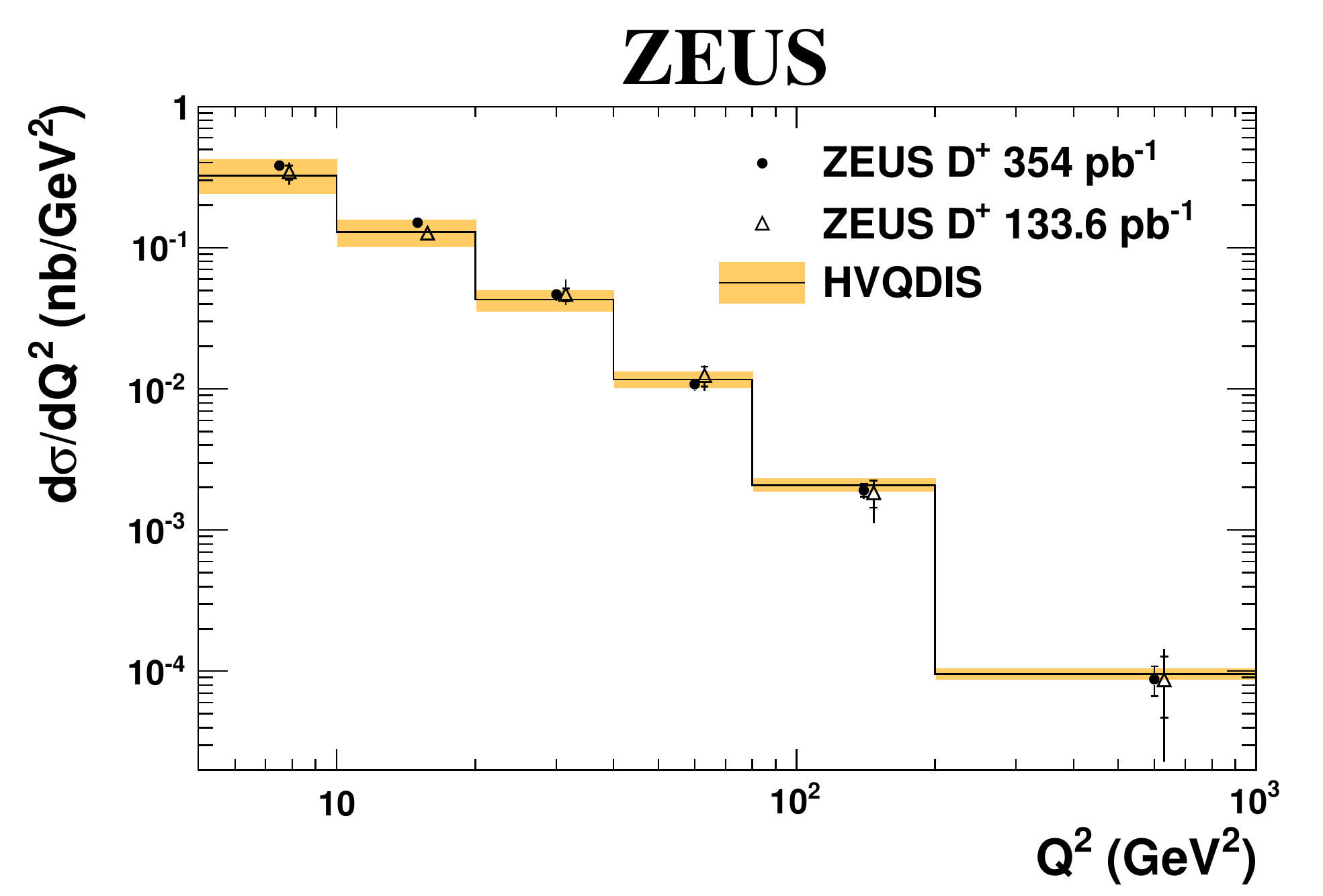}
  \includegraphics[width=0.85\figwidth,trim=3mm 0mm 7mm 0,clip=true]{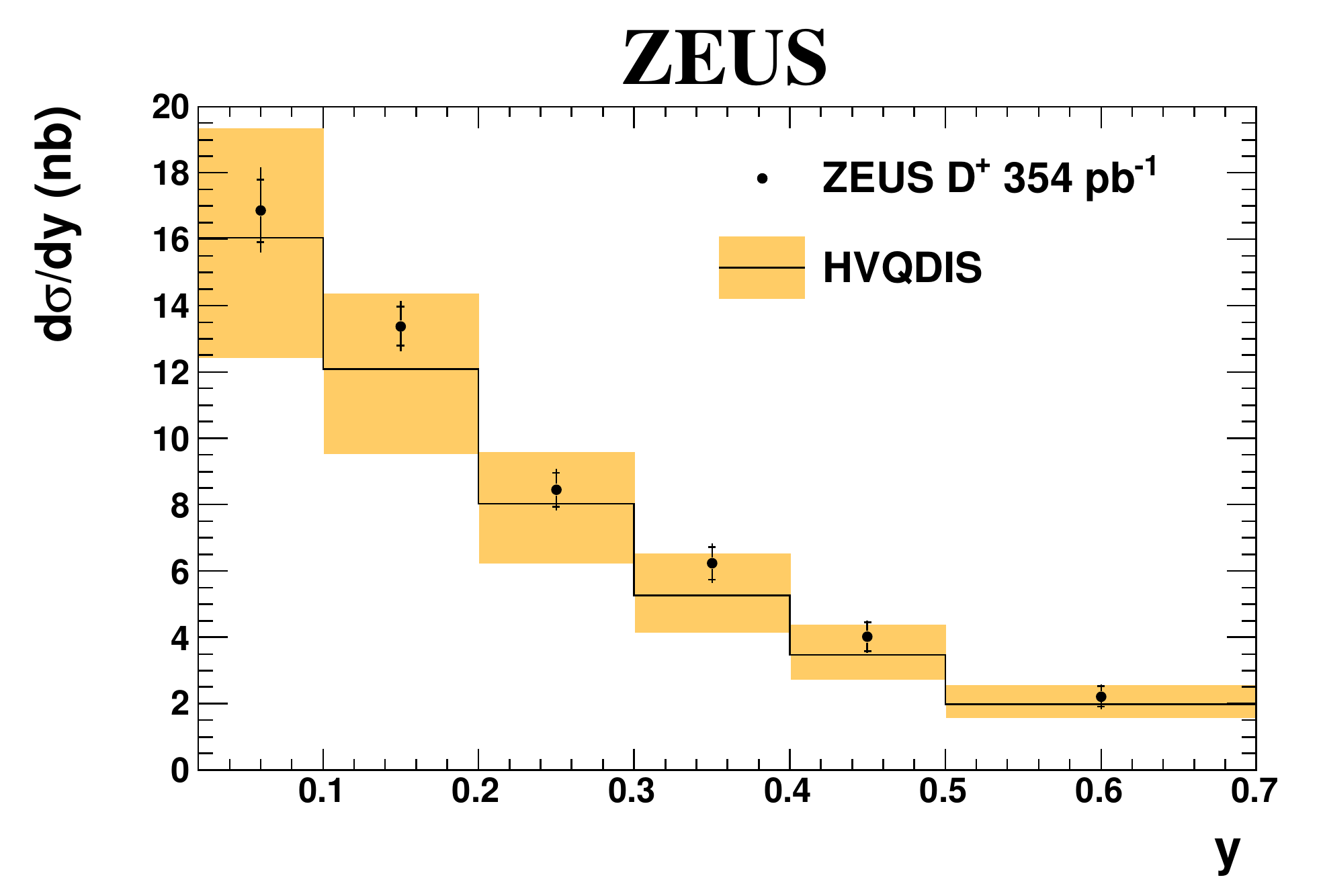}
  \caption[Differential \Dch cross sections as function of $p_T(\Dch)$, $\eta(\Dch)$, $Q^{2}$ and $y$]
	{Differential cross sections for \Dch production as a function of $p_T(\Dch)$ (top left), $\eta(\Dch)$ (top right), $Q^{2}$ (bottom left) and $y$ (bottom right). 
	The cross sections are given in the kinematic region~\ref{eq:dch:ps}. 
	The results obtained in this analysis are shown as filled circles. 
	The inner error bars correspond to the statistical uncertainty, while the outer error bars represent the statistical and systematic uncertainties added in quadrature. 
	For the cross section as a function of $p_T(\Dch)$, $\eta(\Dch)$ and $Q^{2}$, the results of the previous ZEUS measurement are also shown (open triangles). 
	The solid lines and the shaded bands represent the NLO QCD predictions in the FFNS with estimated uncertainties.
	}
  \label{fig:dch:scs}
\end{figure*}

The production of \Dch mesons in the process $\ep \to e' c \bar{c} X \to e' \Dch X$ (i.e.\ not including \Dch mesons from beauty decays) was measured in the kinematic range:
\begin{equation}
\label{eq:dch:ps}
\begin{split}
        &5 < Q^{2} < \SI{1000}{GeV^2},\\
        &0.02 < y < 0.7,\\
        &1.5 < p_{T}(\Dch) < \SI{15}{GeV},\\
        &|\eta(\Dch)| < 1.6.
\end{split}
\end{equation}
The differential cross sections are defined according to Eq.~\ref{eq:dch:cs}. 
The measured cross sections in bins of $p_{T}(\Dch)$, $\eta(\Dch)$, $Q^{2}$ and $y$ are listed in Table~\ref{tab:dch:scs} and shown in Fig.~\ref{fig:dch:scs}. 
The cross section falls by about three orders of magnitude over the measured $Q^2$ range and one order of magnitude in $y$; 
it also falls with the transverse momentum $p_{T}(D^{+})$, but is only mildly dependent on the pseudorapidity $\eta(D^{+})$.
The measured cross sections are compared to the results of the previous ZEUS $D^+$ measurement~\cite{zd0dp}%
\footnote{The contribution of $D^{+}$ mesons from beauty decays was subtracted using the scaled RAPGAP MC predictions~\cite{Chekanov:2009kj,Abramowicz:2010zq,Abramowicz:2011kj}.}, 
based on a subset of the HERA-II data.
The present measurement has significantly smaller uncertainties and supersedes the previous results.
The NLO QCD predictions, calculated in the FFNS, provide a good description of the data. 
The experimental uncertainties are smaller than the theoretical uncertainties, apart from the high-$Q^2$ region, where statistics is limited.

\begin{table}[htbp]
\caption[Differential \Dch cross sections as function of $p_T(\Dch)$, $\eta(\Dch)$, $Q^{2}$ and $y$]
{Differential cross sections for $D^{+}$ production in bins of $p_T(\Dch)$, $\eta(\Dch)$, $Q^{2}$ and $y$. 
The cross sections are given in the kinematic region~\ref{eq:dch:ps}. 
The statistical and systematic uncertainties, $\Delta_{\mathrm{stat}}$ and $\Delta_{\mathrm{syst}}$, are presented separately. 
Normalisation uncertainties of $1.9\%$ and $2.1\%$ due to the luminosity and the branching-ratio measurements, respectively, were not included in $\Delta_{\mathrm{syst}}$. 
The correction factors to the QED Born level, $\mathcal{C}_{\rm rad}$, are also listed. 
For reference, the beauty cross sections predicted by RAPGAP and scaled as described in the text, $\sigma_b$, are also shown.} 
\label{tab:dch:scs}
\begin{center}
\tabcolsep1.22mm
\renewcommand*{\arraystretch}{1.25}
	\begin{tabu} to \columnwidth {|X[c] @{:} X[c]|c c c|c c|}
	\hline
	\multicolumn{2}{|c|}{$p_{T}(D^{+})$} & $\dif \sigma / \dif p_{T}(D^{+})$ & $\Delta_{\mathrm{stat}}$ & $\Delta_{\mathrm{syst}}$& $\mathcal{C}_{\rm rad}$ & $\dif \sigma_b / \dif p_{T}(D^{+})$\\	
	\multicolumn{2}{|c|}{$[\SI{}{GeV}]$} & \multicolumn{3}{c|}{$[\SI{}{nb/GeV}]$} & &$[\SI{}{nb/GeV}]$\\
	\hline
         1.5 & 2.4 & 2.40 & $\pm0.26$ & ${}^{+0.14}_{-0.12}$ & 1.016 & 0.07\\
         2.4 & 3 & 1.44 & $\pm0.12$ & ${}^{+0.07}_{-0.05}$ & 1.020 & 0.05 \\
         3 & 4 & 1.00 & $\pm0.05$ & ${}^{+0.04}_{-0.04}$ & 1.023 & 0.03 \\
         4 & 6 & 0.396 & $\pm0.017$ & ${}^{+0.014}_{-0.013}$ & 1.029 & 0.011 \\
         6 & 15 & 0.0349 & $\pm0.0018$ & ${}^{+0.0011}_{-0.0010}$ & 1.054 & 0.0011\\
	\hline	
	\hline
	\multicolumn{2}{|c|}{$\eta(D^{+})$} & $\dif \sigma / \dif \eta(D^{+})$ & $\Delta_{\mathrm{stat}}$ & $\Delta_{\mathrm{syst}}$& $\mathcal{C}_{\rm rad}$ & $\dif \sigma_b / \dif \eta(D^{+})$\\	
	\multicolumn{2}{|c|}{} & \multicolumn{3}{c|}{$[\SI{}{nb}]$} & & $[\SI{}{nb}]$\\
	\hline
         $-1.6$ & $-0.8$ & 1.04 & $\pm0.09$ & ${}^{+0.06}_{-0.06}$ & 1.034 & 0.02 \\
         $-0.8$ & $-0.4$ & 1.67 & $\pm0.10$ & ${}^{+0.06}_{-0.06}$ & 1.025 & 0.05 \\
         $-0.4$ & $0.0$ & 1.70 & $\pm0.10$ & ${}^{+0.07}_{-0.05}$ & 1.023 & 0.05 \\
         $0.0$ & $0.4$ & 1.63 & $\pm0.10$ & ${}^{+0.07}_{-0.07}$ & 1.017 & 0.06 \\
         $0.4$ & $0.8$ & 1.84 & $\pm0.12$ & ${}^{+0.07}_{-0.08}$ & 1.013 & 0.06 \\
         $0.8$ & $1.6$ & 1.81 & $\pm0.16$ & ${}^{+0.09}_{-0.09}$ & 1.016 & 0.05 \\
	\hline
	\end{tabu}
	\begin{tabu} to \columnwidth {|X[c] @{:} X[c]|c c c|c c|}
	\hline
	\multicolumn{2}{|c|}{$Q^{2}$} & $\dif \sigma / \dif Q^{2}$ & $\Delta_{\mathrm{stat}}$ & $\Delta_{\mathrm{syst}}$ & $\mathcal{C}_{\rm rad}$ & $\dif \sigma_b / \dif Q^{2}$\\	
	\multicolumn{2}{|c|}{$[\SI{}{GeV^2}]$} & \multicolumn{3}{c|}{$[\SI{}{nb/GeV^2}]$} & & $[\SI{}{nb/GeV^2}]$ \\
	\hline
         $5$ & $10$ & 0.382 & $\pm0.022$ & ${}^{+0.027}_{-0.017}$ & 1.018 & 0.007 \\
         $10$ & $20$ & 0.150 & $\pm0.007$ & ${}^{+0.008}_{-0.010}$ & 1.016 & 0.003 \\
         $20$ & $40$ & 0.047 & $\pm0.003$ & ${}^{+0.003}_{-0.004}$ & 1.020 & 0.002 \\
         $40$ & $80$ & 0.0108 & $\pm0.0008$ & ${}^{+0.0008}_{-0.0009}$ & 1.025 & 0.0006 \\
         $80$ & $200$ & 0.00192 & $\pm0.00020$ & ${}^{+0.00014}_{-0.00016}$ & 1.042 & 0.00016 \\
         $200$ & $1000$ & 0.000088 & $\pm0.000021$ & ${}^{+0.000006}_{-0.000007}$ & 1.113 & 0.000013 \\
        \hline	
	\hline
	\multicolumn{2}{|c|}{$y$} & $\dif \sigma / \dif y$ & $\Delta_{\mathrm{stat}}$ & $\Delta_{\mathrm{syst}}$& $\mathcal{C}_{\rm rad}$ & $\dif \sigma_b / \dif y$\\	
	\multicolumn{2}{|c|}{} & \multicolumn{3}{c|}{$[\SI{}{nb}]$} & & $[\SI{}{nb}]$ \\
	\hline
         0.02 & 0.1 & 16.9 & $\pm0.9$ & ${}^{+0.9}_{-0.8}$ & 1.038 & 0.1 \\
         0.1 & 0.2 & 13.4 & $\pm0.6$ & ${}^{+0.5}_{-0.5}$ & 1.022 & 0.3 \\
         0.2 & 0.3 & 8.5 & $\pm0.5$ & ${}^{+0.4}_{-0.4}$ & 1.025 & 0.3 \\
         0.3 & 0.4 & 6.2 & $\pm0.5$ & ${}^{+0.3}_{-0.3}$ & 1.016 & 0.3 \\
         0.4 & 0.5 & 4.0 & $\pm0.4$ & ${}^{+0.3}_{-0.2}$ & 1.008 & 0.2 \\
         0.5 & 0.7 & 2.2 & $\pm0.3$ & ${}^{+0.2}_{-0.2}$ & 0.999 & 0.2 \\
	\hline
	\end{tabu}
\end{center}
\end{table}

\begin{figure*}[htbp]
  \centering
  \begin{minipage}[t]{0.33\textwidth}
  \includegraphics[width=0.995\linewidth,trim=3mm 0mm 3mm 0,clip=true]{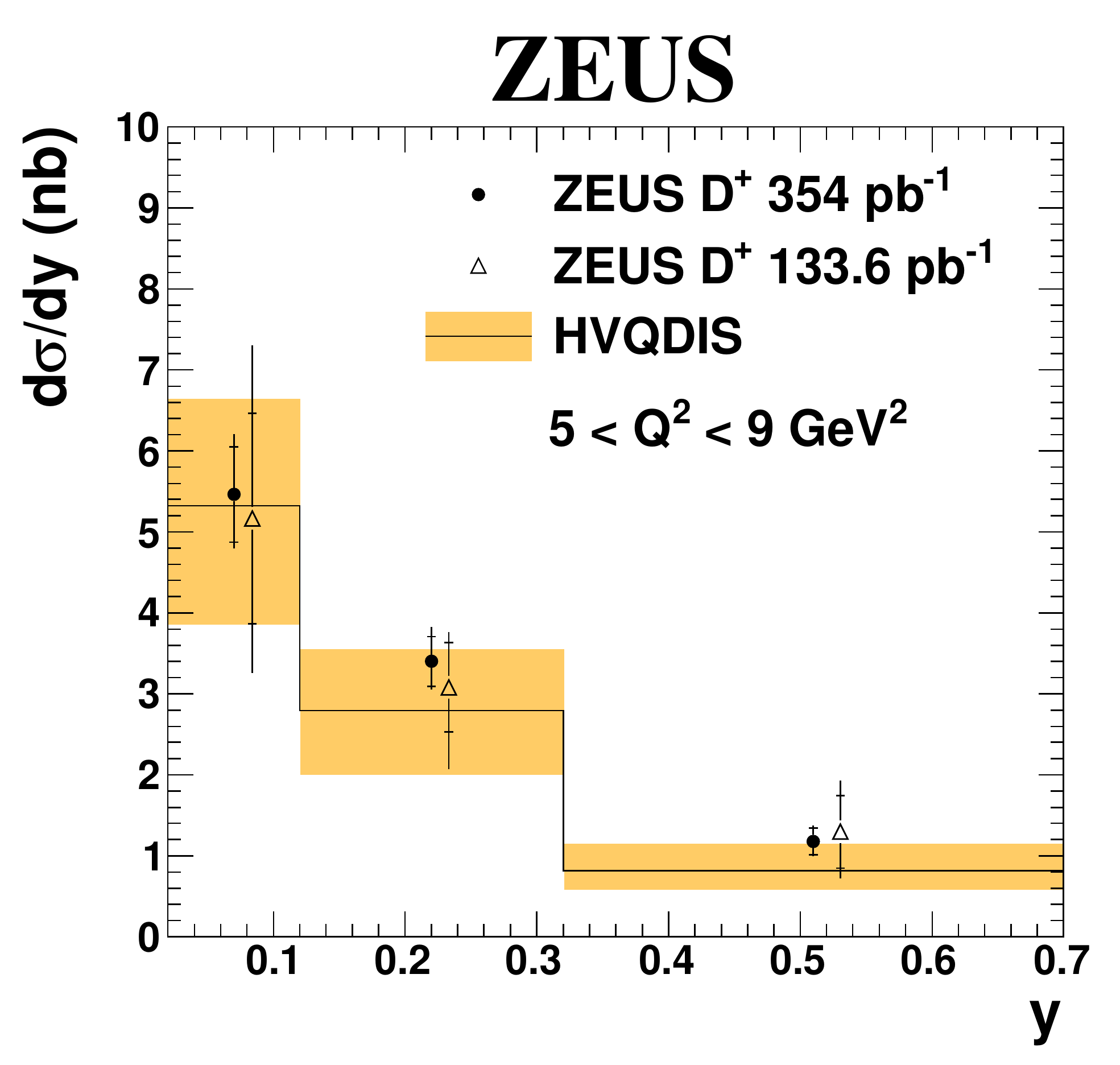}
  \put(-25,60){(a)}\\
  \includegraphics[width=0.995\linewidth,trim=3mm 0mm 3mm 0,clip=true]{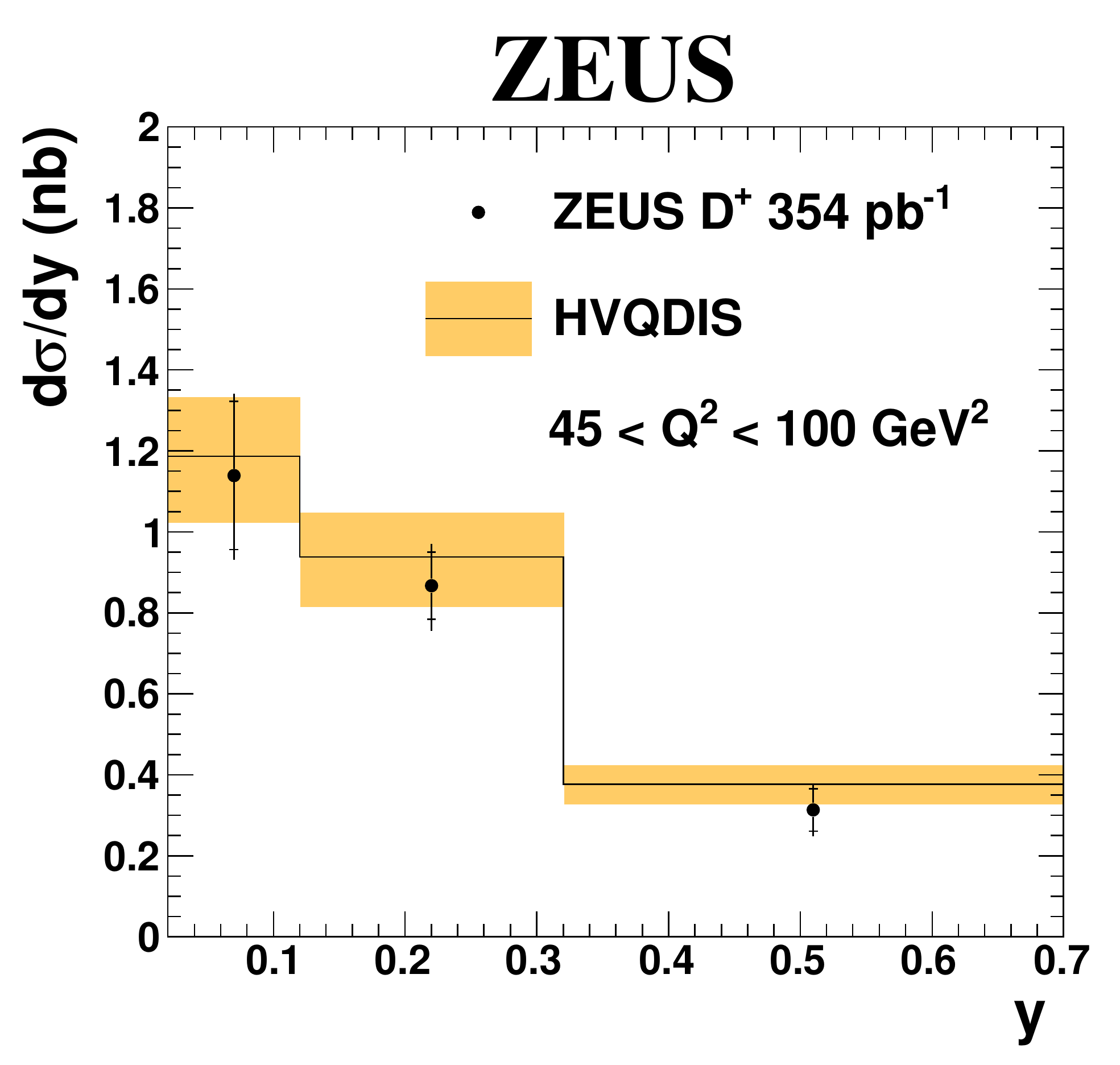}
  \put(-25,60){(d)}
  \end{minipage}
  \begin{minipage}[t]{0.33\textwidth}
  \includegraphics[width=0.995\linewidth,trim=3mm 0mm 3mm 0,clip=true]{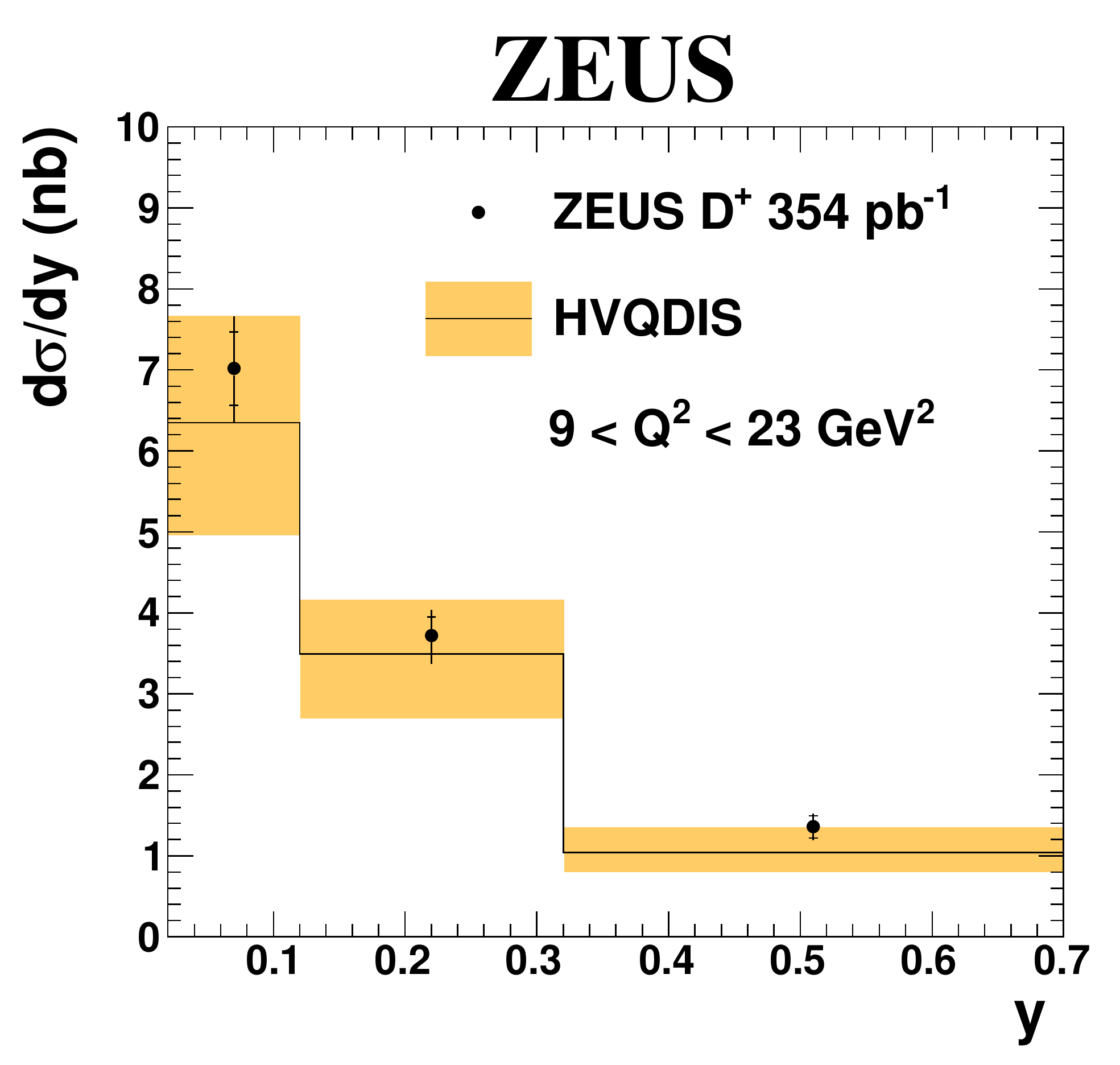}
  \put(-25,60){(b)}\\
  \includegraphics[width=0.995\linewidth,trim=3mm 0mm 3mm 0,clip=true]{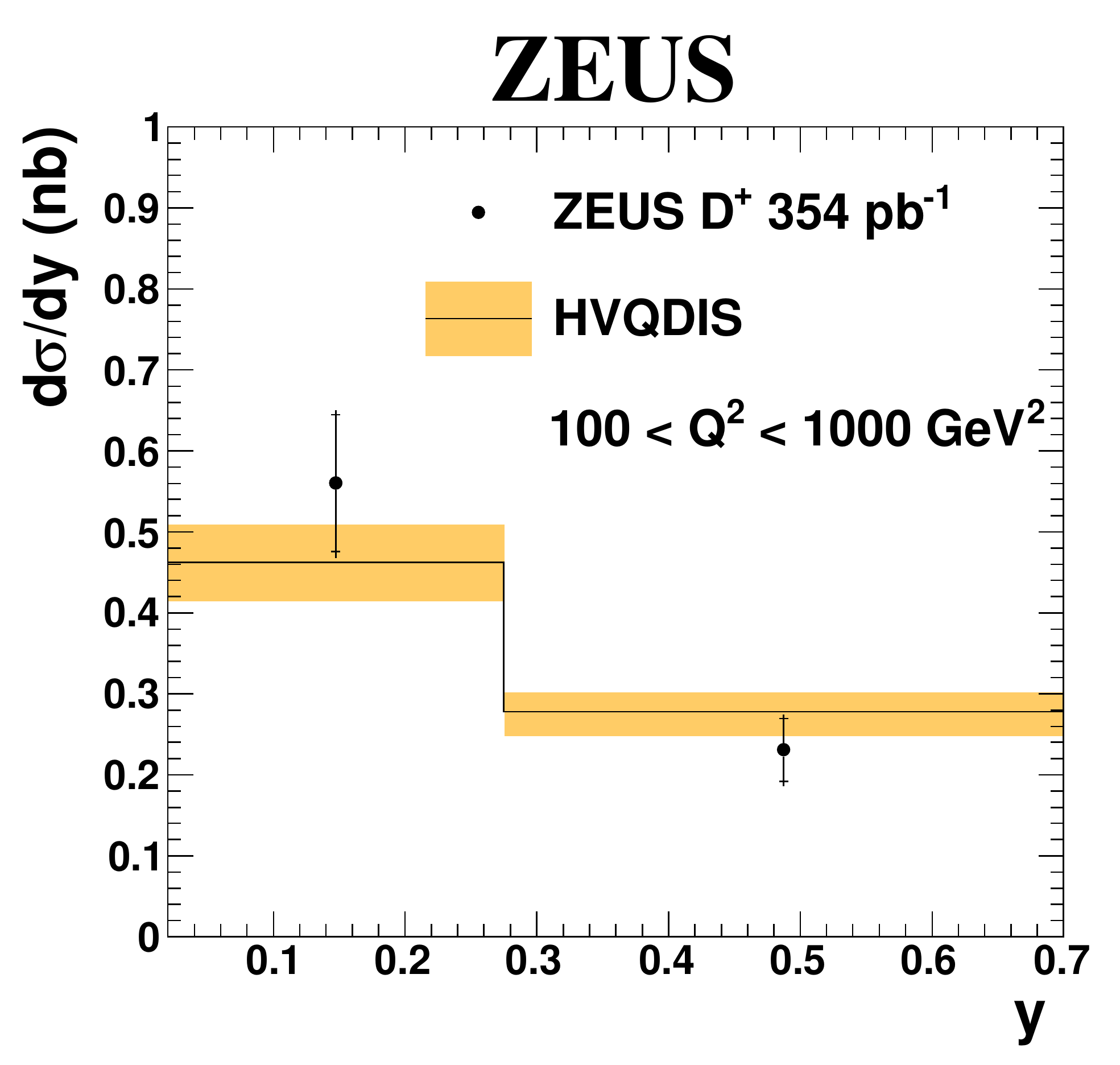}
  \put(-25,60){(e)}
  \end{minipage}
  \begin{minipage}[t]{0.33\textwidth}
  \includegraphics[width=0.995\linewidth,trim=3mm 0mm 3mm 0,clip=true]{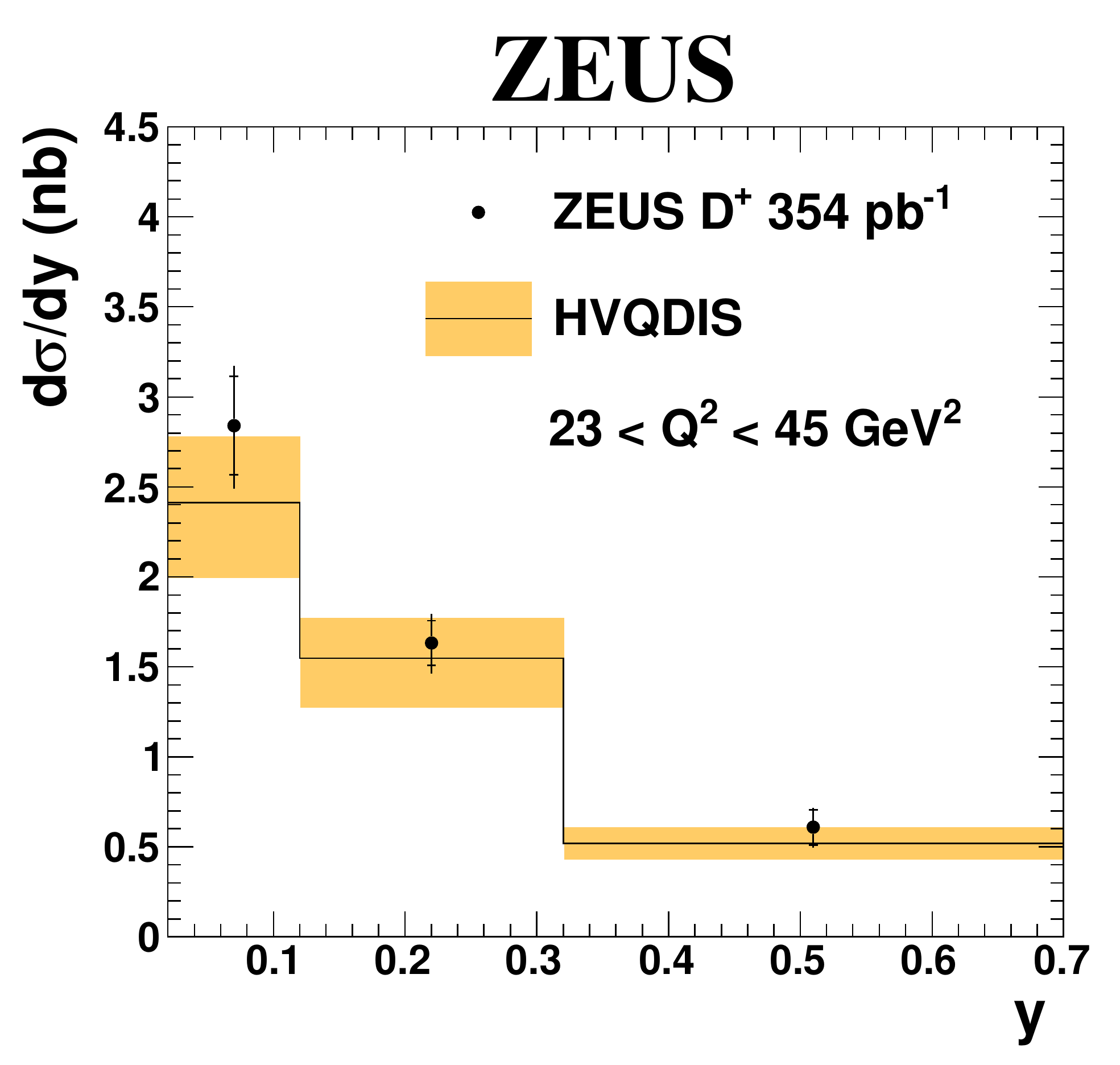}
  \put(-25,60){(c)}
  \end{minipage}
  \caption[Differential \Dch cross sections as function of $y$ in different $Q^2$ ranges]
	{Differential cross sections for \Dch production as a function of $y$ in different $Q^2$ ranges: 
	$5 < Q^{2} < \SI{9}{GeV^2}$ (a), $9 < Q^{2} < \SI{23}{GeV^2}$ (b), $23 < Q^{2} < \SI{45}{GeV^2}$ (c), $45 < Q^{2} < \SI{100}{GeV^2}$ (d) and $100 < Q^{2} < \SI{1000}{GeV^2}$ (e).
	The cross sections are given in the kinematic region~\ref{eq:dch:ps}. 
	The results obtained in this analysis are shown as filled circles. 
	The inner error bars correspond to the statistical uncertainty, while the outer error bars represent the statistical and systematic uncertainties added in quadrature. 
	For the $Q^2$ range $5 < Q^{2} < \SI{9}{GeV^2}$, the results of the previous ZEUS measurement are also shown (open triangles). 
	The solid lines and the shaded bands represent the NLO QCD predictions in the FFNS with estimated uncertainties.}
  \label{fig:dch:dcs}
\end{figure*}

The measured cross sections as a function of $y$ in five $Q^2$ \ozmodNN{bins} are listed in Table~\ref{tab:dch:dcs} and shown in Fig.~\ref{fig:dch:dcs}. 
The data are well reproduced by the HVQDIS calculation. 
The effects of individual sources of systematic uncertainties (described in Section~\ref{sec:dch:cs:syst}) on the cross sections in bins of $Q^2$ and $y$ can be found in~\cite{zeusdch_hera2}. 
The measured double-differential cross section as a function of $Q^2$ and $y$ has been used to extract the charm contribution $F_2^{c\bar{c}}$ to the proton structure function; 
the results can be found in~\cite{zeusdch_hera2}. 

\begin{table}[htbp]
\caption[Differential \Dch cross sections as function of $y$ in different $Q^2$ ranges]
{Differential cross sections for $D^{+}$ production as a function of $y$ in five regions of $Q^{2}$. 
The cross sections are given in the kinematic region~\ref{eq:dch:ps}. 
The statistical and systematic uncertainties, $\Delta_{\mathrm{stat}}$ and $\Delta_{\mathrm{syst}}$, are presented separately. 
Normalisation uncertainties of $1.9\%$ and $2.1\%$ due to the luminosity and the branching-ratio measurements, respectively, were not included in $\Delta_{\mathrm{syst}}$. 
The correction factors to the QED Born level, $\mathcal{C}_{\rm rad}$, are also listed. 
For reference, the beauty cross sections predicted by RAPGAP and scaled as described in the text, $\sigma_b$, are also shown.} 
\label{tab:dch:dcs}
\begin{center}
\tabcolsep0.30mm
\renewcommand*{\arraystretch}{1.35}
	\begin{tabu} to \columnwidth {|c|c |X[c] @{:} X[c]|c c c|c c|}
	\hline
	Bin &$Q^{2}$ & \multicolumn{2}{c|}{$y$} & $\dif \sigma / \dif y$ & $\Delta_{\mathrm{stat}}$ & $\Delta_{\mathrm{syst}}$ & $\mathcal{C}_{\rm rad}$ & $\dif \sigma_b / \dif y$\\	
	& $[\SI{}{GeV^2}]$ & \multicolumn{2}{c|}{} & \multicolumn{3}{c|}{$[\SI{}{nb}]$}& & $[\SI{}{nb}]$\\
	\hline	
         1 & \multirow{3}{*}{5 : 9} & 0.02 & 0.12 & 5.46 & $\pm0.59$ & ${}^{+0.46}_{-0.30}$ &1.026 & 0.04\\
         2 & & 0.12 & 0.32 & 3.40 & $\pm0.31$ & ${}^{+0.29}_{-0.16}$& 1.022 & 0.06 \\
         3 & & 0.32 & 0.7 & 1.18 & $\pm0.17$ & ${}^{+0.10}_{-0.08}$& 1.006 & 0.04 \\
	\hline
         4 & \multirow{3}{*}{9 : 23} & 0.02 & 0.12 & 7.02 & $\pm0.45$ & ${}^{+0.46}_{-0.49}$ &1.028 & 0.05\\
         5 & & 0.12 & 0.32 & 3.72 & $\pm0.23$ & ${}^{+0.21}_{-0.26}$& 1.017 & 0.09 \\
         6 & & 0.32 & 0.7 & 1.36 & $\pm0.14$ & ${}^{+0.09}_{-0.10}$& 0.998 & 0.06 \\
	\hline
         7 & \multirow{3}{*}{23 : 45} & 0.02 & 0.12 & 2.84 & $\pm0.27$ & ${}^{+0.19}_{-0.22}$ &1.040 & 0.03\\
         8 & & 0.12 & 0.32 & 1.63 & $\pm0.12$ & ${}^{+0.10}_{-0.12}$& 1.020 & 0.05 \\
         9 & & 0.32 & 0.7 & 0.609 & $\pm0.097$ & ${}^{+0.047}_{-0.053}$& 1.009 & 0.035 \\
	\hline
         10 & \multirow{3}{*}{45 : 100} & 0.02 & 0.12 & 1.14 & $\pm0.18$ & ${}^{+0.09}_{-0.10}$ &1.046 & 0.03\\
         11 & & 0.12 & 0.32 & 0.867 & $\pm0.083$ & ${}^{+0.063}_{-0.074}$& 1.024 & 0.050 \\
         12 & & 0.32 & 0.7 & 0.313 & $\pm0.052$ & ${}^{+0.032}_{-0.037}$& 1.012 & 0.033 \\
	\hline	
         13 & \multirow{2}{*}{100 : 1000} & 0.02 & 0.275 & 0.560 & $\pm0.085$ & ${}^{+0.031}_{-0.038}$ &1.117 & 0.033\\
         14 & & 0.275 & 0.7 & 0.231 & $\pm0.039$ & ${}^{+0.020}_{-0.022}$& 1.030 & 0.035 \\
	\hline
	\end{tabu}
\end{center}
\end{table}

\ozmodN{In summary,} 
the present results supersede the previous ZEUS $D^{+}$ measurement, based on a subset of the data, and exhibit significantly better precision. 
The improvement in precision comes from the larger data sample, used in the present analysis and from a better control of experimental systematic uncertainties, 
owing to improved tracking alignment and calibration. 
Predictions from NLO QCD in the FFNS describe the measured cross sections well. 
The results presented here are of similar or higher precision than measurements of charm production, previously published by ZEUS%
\footnote{At the moment when the results were being published (February 2013)\ozmod{; 
later on the most precise ZEUS charm measurement became the measurement of \Dstar production using the full HERA-II data set~\cite{zeusdstar_hera2}}.}.
The new precise data provide an improved check of pQCD and \ozmodN{provide further constrains on} the PDFs in the proton.
Section~\ref{sec:comb:red} uses these data for the HERA charm combination.

\clearpage
\section{Combination of the HERA charm measurements}
\label{sec:comb}

This Section is devoted to the combination of the open charm measurements at HERA in DIS. 
Measurements of open charm production at HERA provide an important input for tests of QCD. 
As outlined in Section~\ref{sec:th:hq:ep}, $c$ quarks in \ep collisions are predominantly produced by the boson--gluon-fusion process, 
\ggcc, thus charm production is sensitive to the gluon distribution in the proton, and charm measurements are a valuable input \ozmodNN{for studies of the proton structure} and 
for the extraction of the $c$-quark mass.

Section~\ref{sec:comb:intro} explains the motivation and gives an overview 
of general aspects of the procedure.
Section~\ref{sec:comb:proc} describes the procedure 
of cross-section averaging, the extrapolation to a common phase-space region and the treatment of experimental uncertainties. 
Details of the theoretical calculations in the FFNS, which were used in the combination procedure for phase-space corrections 
and for the comparison with the combined data, are given in Section~\ref{sec:comb:th}. 
In Sections~\ref{sec:comb:dstar} and \ref{sec:comb:red} the main results are presented: 
a combination of visible \Dstar cross sections and reduced charm cross sections, respectively. 
Finally, Section~\ref{sec:comb:summary} gives a summary of the results.

\subsection{Introduction}
\label{sec:comb:intro}

\ozmodN{The main goal of a data combination is to obtain a single consistent dataset for a given physical process. 
State-of-the-art QCD analysis procedures 
use data from a number of individual experiments. 
The data points are correlated through common systematic uncertainties, within and also across the publications. 
\ozmod{The procedure can be significantly simplified} by averaging the input data 
in a model-independent way before performing a QCD analysis of that 
data~\cite{Glazov:2005rn}: e.g. combined into a single dataset DIS charm cross-section measurements 
are much easier to handle compared to a scattered set of individual experimental 
measurements, overviewed in Section~\ref{sec:exp:hera:hfmeas}, while retaining the full correlations between data points. 
Also, a combination serves as a consistency cross check of the input data: a study of the global \chisqndof of the average and the distribution of the 
pulls allows a model-independent consistency examination between the measurements. 
Although a combination is not supposed to provide new information, 
it is possible that a combination will give an extra reduction of correlated uncertainties due to usage of information from 
the phase-space corners which normally would not be used in analyses or theory fits.}

A combination requires input data in the same bins covering the same phase-space region. 
Considering the existing charm measurements at HERA, there are two strategies for the combination: 
\begin{itemize}
	\item to combine a limited number of measurements that closely fulfill the above requirement;
	\item to combine all relevant measurements extrapolated to a common phase-space region and common bins.
\end{itemize}
The former provides a model-independent combination (or with minimised model dependency) which \ozmodNN{retains} most 
of original information; this strategy is followed in the combination of the visible \Dstar cross 
sections (Section~\ref{sec:comb:dstar}). The latter gains from a big number of input measurements, 
\ozmodNN{thus it has ultimate accuracy at the cost of some model dependence and a sizeable theoretical uncertainty from the extrapolation procedure;} 
this strategy is followed in the combination of the reduced charm cross sections (Section~\ref{sec:comb:red}).

\subsection{Combination procedure}
\label{sec:comb:proc}

In this Section common aspects of the combination procedure are described: 
the combination method, needed to average quantities given in a common phase-space region (Section~\ref{sec:comb:proc:method}), 
the treatment of uncertainties of input quantities in the combination method (Section~\ref{sec:comb:proc:unc}), 
and phase-space corrections, 
needed to translate the input quantities to a common phase-space region (Section~\ref{sec:comb:proc:pscorr}).

\subsubsection{Combination method}
\label{sec:comb:proc:method}

The HERAverager package~\cite{heraverager}%
\footnote{HERAverager is based on the earlier program F2averager introduced in~\cite{Glazov:2005rn} 
and used, e.g.\ for the previous HERA charm combination~\cite{heracharmcomb}.}
is used for the combination of the charm data. 
It is an averaging tool developed for the H1 and ZEUS data combination. 
The combination method is based on the minimisation of the \chisq-function which includes 
correlated systematic uncertainties using the nuisance-parameter technique, also known as the Hessian method~\cite{hessian}. 

\paragraph*{\chisq definition\\}
\label{sec:comb:proc:def}

Consider $N_e$ sets of measurements of $N_m$ quantities $\mu_i$ (e.g.\ from different experiments or from one experiment, but obtained in different analises), 
$\mu_i^e$. \ozmodN{Each measurement has one uncorrelated uncertainty, $\sigma_i^e$, and $N_s$ correlated, $\Gamma_i^{e,j}$: 
$\mu_i^e \pm \sigma_i^e \pm \sum_{j=1}^{N_s}\Gamma_i^{e,j}$ 
($1 \le i \le N_m$, $1 \le e \le N_e$, $1 \le j \le N_s$).} 

All correlated uncertainties are assumed to be Gaussian. 
Thus each measurement can be written as
\begin{equation}
	\label{eq:comb:proc:init}
	\mu_i^e=m_i+\sigma_i^e a_i^e-\sum_{j=1}^{N_s}\Gamma_i^{e,j}b^{e,j},
\end{equation}
where $a_i^e$ and $b^{e,j}$ are independent variables distributed according to the unit Gaussian distribution around zero. 
Note that $b^{e,j}$ are independent of $i$; that is, the uncertainties $\Gamma_i^{e,j}$ are $100\%$ correlated 
for all data points denoted with the same $e$. The $b^{e,j}$ are called nuisance parameters of correlated uncertainties. 
Then the generalised \chisq can be written as
\begin{equation}
	\label{eq:comb:proc:extchi2}
	\chisq(\mathbf{m},\mathbf{b})=\sum_{e=1}^{N_e}\sum_{i=1}^{N_m}\frac{\left(m_i-\sum_{j=1}^{N_s}\Gamma_i^{e,j}b^{e,j}-\mu_i^e\right)^2}{{\sigma_i^e}^2}+\sum_{j=1}^{N_s}{b^{e,j}}^2,
\end{equation}
where the vectors $\mathbf{m}$ and $\mathbf{b}$ denote the true parameters $m_i$ and nuisance parameters $b^{e,j}$, respectively. 
Here the first term takes into account the effects of the shifts of the correlated uncertainties, and 
the second term is a penalty for the correlated uncertainty shifts from their nominal (zero) values. 
The uncorrelated uncertainties $\sigma_i^e$ are the total uncorrelated uncertainties which may consist of several independent components 
(e.g.\ a statistical uncertainty and several different systematic ones, assumed to be uncorrelated between the data points) added in quadrature, 
according to the law of combination of errors~\cite{barlow,blobel_lohrmann}. 
Note, that some of $\Gamma_i^{e,j}$ may 
be equal to $0$ if the measurement $\mu_i^e$ is insensitive to the systematic source $j$. 
A formal derivation of the \chisq expression~\ref{eq:comb:proc:extchi2} from the assumption~\ref{eq:comb:proc:init} can be found in~\cite{Stump:2001gu}. 
The averaging problem is solved 
by minimising $\chisq(\mathbf{m},\mathbf{b})$ w.r.t $\mathbf{m}$ and $\mathbf{b}$, providing the average values $\overline{\mathbf{\mathbf{m}}}$ and 
the fitted nuisance parameters $\overline{\mathbf{b}}$; a variation of $\chisq(\mathbf{m},\mathbf{b})$ by $1$ provides the uncertainties on these values. 
The formulas for these quantities are provided in Appendix~\ref{sec:comb:proc:min}.

\ozmodN{The Hessian method usually leads to a reduction of correlated uncertainties in the combination procedure. 
This is a considerable advantage compared to the more conservative offset method~\cite{offset}, 
when error propagation is based on shifting the data by the systematic errors and adding 
the deviations in quadrature, therefore the size of the correlated uncertainties remains unchanged.} 

So far the form of the correlated uncertainties $\Gamma_i^{e,j}$ was not specified. It is useful to define 
the relative correlated systematic uncertainties by the ratio 
\begin{equation}
	\label{eq:comb:proc:rel}
	\gamma_i^{e,j}=\frac{\Gamma_i^{e,j}}{\mu_i^e}.
\end{equation}
Usually the relative, not absolute, systematic uncertainties are provided by the measurements. 
\ozmodN{Several} types of uncertainty treatment can be considered:
\begin{itemize}
	\item the \emph{multiplicative} treatment, when the systematic uncertainties are proportional to the \emph{true} values: 
		\begin{equation}
			\Gamma_i^{e,j}=m_i \gamma_i^{e,j};
		\end{equation}
	\item the \emph{additive} treatment, when the systematic uncertainties are independent of the true value; 
		then they are considered to be proportional to the \emph{measured} values, i.e.\ by the definition~\ref{eq:comb:proc:rel}, 
		or independent of either:
		\begin{equation}
			\Gamma_i^{e,j}=\mu_i \gamma_i^{e,j}
		\end{equation}
		(in other words they are constant and not changed in the combination procedure);
  \item \ozmod{a \ozmodN{mixed} case is the signal-dominated statistical uncertainties, which obey the Poisson statistics; 
    their values are scaled with the square root of $m_i$:
    \begin{equation*}
      \Gamma_i^{e,j}=\sqrt{\mu_i m_i} \gamma_i^{e,j}.
    \end{equation*}}
\end{itemize}
The same options exist for the treatment of the uncorrelated uncertainties in the denominator of~\ref{eq:comb:proc:extchi2}.
The additive treatment is appropriate for background-dominated uncertainties, which do not depend on the true value $\mathbf{m}$, while 
the multiplicative treatment is appropriate for all others.

For the charm measurements at HERA the statistical uncertainties are mainly dominated by background, so in the combination they were treated additively. 
The systematic uncertainties are predominantly proportional to the central values and thus treated multiplicatively. 
So the \chisq-function, used in the present combination, is given by
\begin{equation}
	\label{eq:comb:proc:finalchi2}
	\chisq(\mathbf{m},\mathbf{b})=\sum_{e=1}^{N_e}\sum_{i=1}^{N_m}\frac{\left(m_i-\sum_{j=1}^{N_S}\gamma_i^{e,j} m_i b^{e,j}-\mu_i^e\right)^2}{{\delta_{stat,i}^e \mu_i^e}^2+{\delta_{uncor,i}^e m_i}^2}+\sum_{j=1}^{N_s}{b^{e,j}}^2,
\end{equation}
where in the denominator the statistical and uncorrelated systematic uncertainties are added in quadrature; 
$\delta_{stat,i}^e$ and $\delta_{uncor,i}^e$ are the relative statistical and uncorrelated systematic uncertainties, respectively, 
defined similar to~\ref{eq:comb:proc:rel}:
\begin{equation}
	\begin{split}
	\delta_{stat,i}^e & =\frac{\sigma_{stat,i}^{e}}{\mu_i^e},\\
	\delta_{uncor,i}^e & =\frac{\sigma_{uncor,i}^{e}}{\mu_i^e},
	\end{split}
\end{equation}
where $\sigma_{stat,i}^{e}$ and $\sigma_{uncor,i}^{e}$ are the absolute statistical and uncorrelated systematic uncertainties, respectively. 
In the previous combination of the reduced charm cross sections~\cite{heracharmcomb} the sensitivity of the result 
to the treatment of the uncertainties was studied and procedural uncertainties were assigned; however they turned out to be much smaller 
than the other ones (on average below $0\text{--}10$\% of the total uncertainty, reaching up to 40\% only at few combined points~\cite{heracharmcomb}) and are neglected in the present combination.

\subsubsection{Treatment of systematic uncertainties}
\label{sec:comb:proc:unc}

As explained in Section~\ref{sec:comb:proc:def}, in the combination procedure uncertainties are treated either 
as fully uncorrelated or fully correlated between the data points of certain measurements. 
Neither of these is conservative in general. Experimental uncertainties of the input measurements 
consist of statistical and systematic uncertainties. The statistical component of uncertainties was always treated as uncorrelated 
between all data points.%
\footnote{In fact small correlations exist between inclusive measurements and those where full final states were reconstructed 
(e.g.\ between the measurement~\cite{zeussecvtx_hera2}, where information from secondary vertices from all charm-hadron decays was used, 
and \cite{zeusdstar_hera2}, where \Dstar mesons were reconstructed in the $\Dstar \to D^{0}(K^{-}\pi^{+})\pi^{+}_{s}$ decay channel), 
but since the corresponding branching ratios are much smaller than $\SI{1}{}$, 
phase-space cuts differ and statistical uncertainties in heavy-flavour measurements are usually dominated by background, such correlations have been neglected.}

The systematic component of uncertainties, in general, may have mixed nature: it may be partially correlated between the data points; 
moreover, the level of correlation may differ \ozmod{in different phase-space regions}. In the current combination procedure the following 
``common sense'' strategy was applied:
\begin{itemize}
	\item normalisation uncertainties (reported as a single number) were treated as correlated (e.g.\ luminosity and branching ratios); they are marked as `N';
	\item those uncertainties that have smooth behaviour in the phase space \ozmodN{of the measurement} were also treated as correlated 
		(typically these are different kinds of corrections, reweightings, inefficiencies etc., evaluated using studies based on MC); they are marked as `S';
	\item theory-related uncertainties that arose from the phase-space corrections (see Sections~\ref{sec:comb:proc:pscorr}) were treated as correlated; they are marked as `T';
	\item all other uncertainties were treated as uncorrelated (typically these are uncertainties estimated using cut variations in data, 
		which are subject to statistical fluctuations).
\end{itemize}
Explicit information on the sources that were treated as correlated is given in Sections~\ref{sec:comb:dstar:single:det},~\ref{sec:comb:dstar:double:det} and ~\ref{sec:comb:red:det}.

Many of the experimental systematic uncertainties are quoted \ozmodNN{as} asymmetric and \ozmod{have been symmetrised} before performing a combination. 
\ozmodN{For the measurements~\cite{zeusdch_hera2,zeusdstar_hera2,zeussecvtx_hera2}, which have not been included in the previous charm combination~\cite{heracharmcomb},
symmetrisation \ozmodN{consisted in} taking the largest deviation; no corrections to the central values were applied. 
For those measurements that have been included in the previous charm combination~\cite{heracharmcomb}, the symmetrisation remains the same as in~\cite{heracharmcomb}.}%
\footnote{It was found in~\cite{heracharmcomb} that the results are insensitive to the details of the symmetrisation procedure.}

\subsubsection{Phase-space correction}
\label{sec:comb:proc:pscorr}

Whenever the quantities to be averaged are measured in different \ozmod{phase-space regions}, they have to be corrected before 
performing a combination. Assume that there is a measured quantity (e.g.\ a cross section) in the phase-space region $1$, $\sigma^{meas}_1$, which needs to be \ozmod{shifted} into the phase-space region $2$.
The correction procedure is called \emph{extrapolation} and \ozmod{relies} on usage of theoretical calculations:
\begin{equation}
	\sigma^{extr}_2 = \sigma^{meas}_1 \frac{\sigma^{th}_2}{\sigma^{th}_1}.
\end{equation}
Here $\sigma^{th}_1$ and $\sigma^{th}_2$ are the 
predicted quantities in the phase-space regions $1$ and $2$, respectively. The closer the phase-space regions $1$ and $2$ 
(in particular, the more they overlap), the less model dependency the extrapolated quantity, $\sigma^{extr}_2$, has. In order 
to estimate the remaining model dependence, the parameters of the theoretical calculations are varied; the resulting 
uncertainty is called the \emph{extrapolation uncertainty}. In addition to the extrapolation 
uncertainty the extrapolated quantity $\sigma^{extr}_2$ has an original uncertainty of $\sigma^{meas}_1$ 
(which, for instance, may consist of statistical and systematic uncertainties of the experimental measurement).

It is important to distinguish between ``small'' extrapolations to another region of a measured phase-space region (these can be thought of rather as interpolations), 
referred to in the future as \emph{swimming}, e.g.\ when a quantity 
is translated into a different binning scheme, and actual extrapolation to an unmeasured phase-space region, referred to in the future just as \emph{extrapolation}. 
In the first case it is important that the original measurement in general covers all the phase-space region where the swimming is performed, 
so the predictions can be compared to the measurements in order to check the adequacy of the swimming. 
In contrast, in the second case the results of the extrapolation 
depend on theoretical predictions in unmeasured phase-space corners; thus in general the adequacy of the results cannot be verified unless 
there are other measurements in the uncovered regions. 
Note that in both cases the corrections do not depend on common normalisation factors. 
For the combinations presented in this review, the combination of \Dstar cross sections 
requires only swimming, while the combination of reduced charm cross sections requires \ozmodN{genuine} extrapolation.

For the charm combination presented here, phase-space corrections were always done using the theory \ozmodN{outlined} in Section~\ref{sec:th:hq:ep:ffns}: the NLO QCD calculations ($O(\alpha_s^2)$) 
in the 3-flavour FFNS. 
Details of the theoretical calculations (including the variations which are used to estimate the extrapolation uncertainties) are given in Section~\ref{sec:comb:th}.

\subsection{Theoretical calculations in FFNS}
\label{sec:comb:th}

The FFNS theoretical calculations were used for two purposes:
\begin{itemize}
	\item for the extrapolation and swimming corrections (Sections~\ref{sec:comb:dstar:single:det}, \ref{sec:comb:dstar:double:det}, \ref{sec:comb:red:det:extrcorr});
	\item for the comparison of \ozmodNN{theory with the combined data} and QCD analyses (Sections~\ref{sec:comb:dstar:single:th}, \ref{sec:comb:dstar:double}, \ref{sec:comb:red:ffns}).
\end{itemize}

NLO QCD predictions in the FFNS were obtained with the HVQDIS program~\cite{hvqdis}. 
The parameters used in the calculations, together with the corresponding variations which were used to estimate the uncertainties, are described below.%
\footnote{The settings were mainly \ozmodNN{taken over from~\cite{heracharmcomb}, 
albeit with some modifications} (see also Section~\ref{sec:comb:red:det:extrcorr}).}
In the combination procedure each extrapolation uncertainty was treated as correlated between all points and all measurements. 
Most of the extrapolation uncertainties 
were originally asymmetric and \ozmod{have been} symmetrised before performing \ozmodNN{the} combination. 
Symmetrisation was performed by taking the largest deviation; no corrections to the central values were applied. 
For the data to theory comparison, to obtain total theoretical uncertainties, all the variations were added in quadrature \ozmod{and} the summation was performed separately for positive and negative variations.

\subsubsection{Parton-level cross sections}
The parton-level cross sections were calculated using the following settings:
\begin{itemize}
 \item {\bf the renormalisation and factorisation scales} were set to 
	  $\mu_r=\mu_f=\sqrt{Q^2+4m_c^2}$ and varied up and down by a factor of two. 
	  The variations were performed independently if the theoretical predictions were used for comparison with data, 
	  or simultaneously if they were used for extrapolation or swimming corrections (which are sensitive only to the shape of the predictions);
 \item {\bf the pole mass of the $c$ quark} $m_c=1.50 \pm 0.15$ GeV~\cite{heracharmcomb}; since the renormalisation and 
		factorisation scale definitions include the $c$-quark mass, varying 
		this also slightly affected the two scales;
 \item {\bf the strong coupling constant} $\alpha_s^{n_f=3}(M_Z) = 0.105 \pm 0.002$,
	 corresponding to the value $\alpha_s^{n_f=5}(M_Z) = 0.116 \pm 0.002$;
	\item {\bf the PDFs} were described by a 
		series of 3-flavour FFNS variants of the HERAPDF1.0 set~\cite{DIScomb} 
		at NLO, similar to those used for the cross-section extrapolations in the 
		previous charm combination~\cite{heracharmcomb}, evaluated for $m_c=1.5 \pm 0.15$~GeV, 
		for  $\alpha_s^{n_f=3}(M_Z) = 0.105 \pm 0.002$, and for different scales.
		Charm measurements were not included in the determination of these PDF sets.
		For each of the parameter variations \ozmodNN{(the scales, mass and $\alpha_s$)}, a different respective PDF set 
		was used. By default, the scales for the charm contribution to the inclusive 
		data in the PDF determination were chosen to be consistent with the factorisation scale 
		used in HVQDIS, while the renormalisation scale in HVQDIS was decoupled from the PDF scales, except
		in the cases where the factorisation and renormalisation scales were varied simultaneously.
		As a cross check, instead of fitting the PDFs from inclusive data, 
		3-flavour NLO variants of the ABM \cite{abm11} and MSTW \cite{Martin:2010db} PDFs 
		were also used to evaluate the cross sections. 
		For MSTW, the variant with $m_c=1.5$~GeV was chosen.
		The differences were found to be much smaller compared to those from 
		other parameter variations, therefore the PDF uncertainties are neglected; the plots are provided in Appendix~\ref{sec:app:dstar} (Fig.~\ref{fig:comb:dstar:single:pdfvar}).
\end{itemize}

\subsubsection{Fragmentation}
The fragmentation model described in the previous publication~\cite{heracharmcomb} was used to provide 
hadron-level cross sections, if needed. 
It is based on the measurements by H1~\cite{h1frag} and ZEUS~\cite{zeusfrag} 
using the production of \Dstar mesons, with and without associated jets, in DIS and PHP. 
This model uses the fragmentation 
function of Kartvelishvili et al.~\cite{Kartvelishvili:1977pi}, controlled by the parameter
$\alpha_K$, to describe the longitudinal fraction of the \ozmodNN{$c$-quark} momentum transferred
to the \Dstar mesons. The fragmentation was performed in the photon-proton centre-of-mass frame 
by rescaling the quark three-momentum, then the energy of the produced hadron was calculated and 
the hadron was boosted to the lab frame. The calculation of the hadron energy and the Lorentz boost were done by using 
the hadron mass.%
\footnote{As explained in Section~\ref{sec:th:hq:frag}, a phenomenological fragmentation model \ozmod{must} be applied exactly 
in the same way as it was measured. Here the fragmentation model follows the original H1 and ZEUS measurements~\cite{h1frag,zeusfrag}.}
Different values of $\alpha_K$~\cite{heracharmcomb} 
were used for different bins in the photon-parton centre-of-mass-frame squared energy, $\hat s$, and for different hadrons.
Since ground-state $D$ mesons partly originate from decays of \Dstar and other excited
mesons, the corresponding \ozmodNN{$c$-quark} fragmentation function is softer than
that measured using \Dstar mesons.
From kinematic considerations~\cite{frag06}, supported 
by experimental measurements~\cite{Seuster:2005tr}, the expectation value for
the fragmentation function of $c$ quarks into $D^{0,\text{ not }D^{*+}}$%
\footnote{$D^{0,\text{ not }D^{*+}}$ refers to $D^{0}$ that do not originate from decays of $\Dstar$.}, 
$D^+$ and in the mix of charmed hadrons decaying into muons, has to be
reduced by $\approx 5\%$ with respect to that for \Dstar mesons. The
values of $\alpha_K$ for the fragmentation into ground-state hadrons, used for the $D^{0,\text{ not }D^{*+}}$, $D^+$ and $\mu$ measurements, have been 
re-evaluated accordingly~\cite{heracharmcomb}  and are reported in Table~\ref{tab:comb:proc:pscorr:fragkart}. 
The model also implements a transverse fragmentation
component by assigning to the charmed hadron a transverse momentum, $k_T$,
with respect to the $c$-quark direction, with $\langle k_T \rangle=0.35 \pm 0.15$ GeV. 
If needed (for the phase-space corrections for the ZEUS muon measurement~\cite{zeus_muon}), 
the charm-hadron cross sections were accompanied by the semi-leptonic decays from~\cite{Adam:2006nu}.
Fragmentation fractions were taken from~\cite{pdg2012,Lohrmann:2011np} and are listed in Table~\ref{tab:comb:th:ff}.

\begin{table}[h]
\caption[$\alpha_K$ parameters used for longitudinal fragmentation]
{The $\alpha_K$ parameters used for the longitudinal fragmentation into
   $\Dstar$ mesons and in ground-state (g.s.) charmed hadrons.  The first column shows the $\hat{s}$ range in which a particular value of  $\alpha_K$  is
  used, with $\hat{s}_1=70\pm40$~GeV$^2$ and $\hat{s}_2=324$~GeV$^2$.
   The variations of $\alpha_K$ are given in the second and third column. 
The parameter $\hat{s}_2$ was not varied, since the 
corresponding uncertainty is already covered by the $\alpha_K$ variations.
\label{tab:comb:proc:pscorr:fragkart}}
\tabcolsep0.9mm
\begin{center}
\begin{tabular}{|c|c|c|l|} \hline
  $\hat{s}$ range       &  $\alpha_K (\Dstar)$  &$\alpha_K (\rm g.s.)$   &  Measurement \\ \hline
  $\hat{s}\leq\hat{s}_1$            &  $6.1 \pm 0.9$     & $4.6\pm 0.7$          &\cite{h1frag} \Dstar, DIS, no-jet sample \\ \hline
  $\hat{s}_1<\hat{s} \leq\hat{s}_2$ &  $3.3 \pm 0.4$     & $2.7\pm 0.3$          &\cite{h1frag} \Dstar, DIS, jet sample   \\ \hline
  $\hat{s}>\hat{s}_2$                        &  $2.67 \pm 0.31$ &  $2.19\pm  0.24$    &\cite{zeusfrag} \Dstar jet PHP       \\ \hline
\end{tabular}
\end{center}
\end{table}

\begin{table}[htbp]
\caption[Charm and beauty fragmentation fractions]
{\ozmodNN{$c$-quark} fragmentation fractions to charmed mesons and the charm branching fraction to muons (top), and 
\ozmodNN{$b$-quark} branching fractions to charmed mesons (bottom).}
\label{tab:comb:th:ff}
\tabcolsep5.0mm
\renewcommand*{\arraystretch}{1.25}
\begin{center}
\begin{tabularx}{\columnwidth}[t]{|X|c|}  \hline 
$f(c\rightarrow D^{*+})$ & $0.2287 \pm 0.0056$  \\ \hline
$f(c\rightarrow D^+)$   &  $0.2256 \pm 0.0077$\\    \hline
$f(c\rightarrow D^{0, {\rm not }D^{*+}})$  &$0.409 \pm 0.014$\\ \hline
$B(c\rightarrow \mu)$ &  $0.096 \pm 0.004$\\ \hline
\hline
$f(b\rightarrow D^{*+},D^{*-})$ & $0.173 \pm 0.020$ \\ \hline
$f(b\rightarrow D^{+},D^{-})$ & $0.233 \pm 0.017$ \\ \hline
$f(b\rightarrow D^{0},\bar{D^{0}})$ & $0.598 \pm 0.029$ \\ \hline
\end{tabularx}
\end{center}
\end{table}

In total, the following uncertainties were assigned to the fragmentation:
\begin{itemize}
	\item the variation of $\alpha_K$ (the upward and downward variations were performed simultaneously for all 
		$\hat s$ bins and for all hadrons%
		\footnote{The values of $\alpha_K$, determined in~\cite{h1frag,zeusfrag}, are only partially correlated 
			(the two values from ~\cite{h1frag}) or rather fully uncorrelated (the values from~\cite{h1frag} and~\cite{zeusfrag}), 
			nevertheless their simultaneous variation is the most conservative way to estimate the uncertainty.});
	\item the variation of $\hat{s}_1$%
	\footnote{In the case of extrapolation uncertainties, $\alpha_K$ and $\hat{s}_1$ variations were added in quadrature and treated as one source, referred to as `longitudinal fragmentation'.};
	\item the variation of $\langle k_T \rangle$%
	\footnote{In the case of extrapolation uncertainties this source is referred to as `transverse fragmentation'.};
	\item the uncertainties on the fragmentation fractions. 
\end{itemize}

\subsubsection{Beauty contribution}
\ozmodNN{In most of the \ozmod{analyses} the cross sections of charmed hadrons 
either produced directly or in decays of beauty hadrons were measured.}
For the combination of reduced charm cross sections the beauty contribution needed to be subtracted, 
while for the data to theory comparison for \Dstar-production cross sections it must be added to the charm theoretical predictions. 
In previous H1 and ZEUS charm analyses the beauty contribution had been obtained using the RAPGAP MC~\cite{rapgap}, 
with the normalisation rescaled to dedicated beauty measurements. 
Typical normalisation factors vary from 1.0 to 2.0~\cite{Chekanov:2004tk,Chekanov:2009kj,Abramowicz:2010zq,Abramowicz:2011kj,zeussecvtx_hera2}, thus an uncertainty $\sim 50\%$ has to be 
assigned to the beauty contribution. Propagated to the uncertainty on charm and beauty production, 
this results in an uncertainty up to $\sim 5\%$ and thus becomes a dominant uncertainty at high $Q^2$, where the perturbative calculations are quite accurate. 
Moreover, this \ozmodNN{approach} provides predictions at LO accompanied by parton showers, re-normalised to measured data. 

\begin{figure}[tbp]
\centering
  \begin{minipage}[t]{0.495\columnwidth}
  \includegraphics[width=0.5\figwidth,trim=2mm 0mm 9mm 3mm,clip=true]{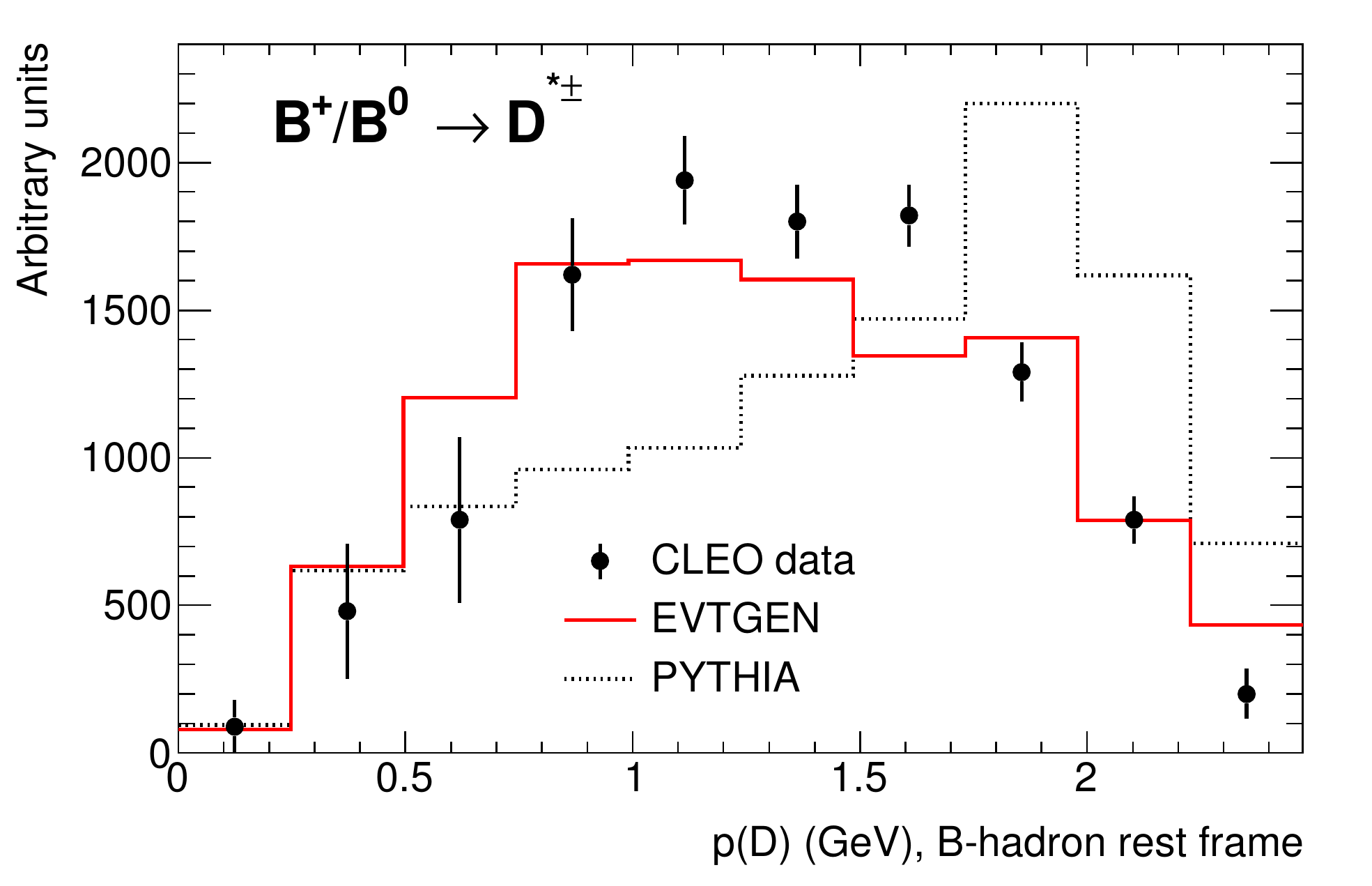}
  \includegraphics[width=0.5\figwidth,trim=2mm 0mm 9mm 3mm,clip=true]{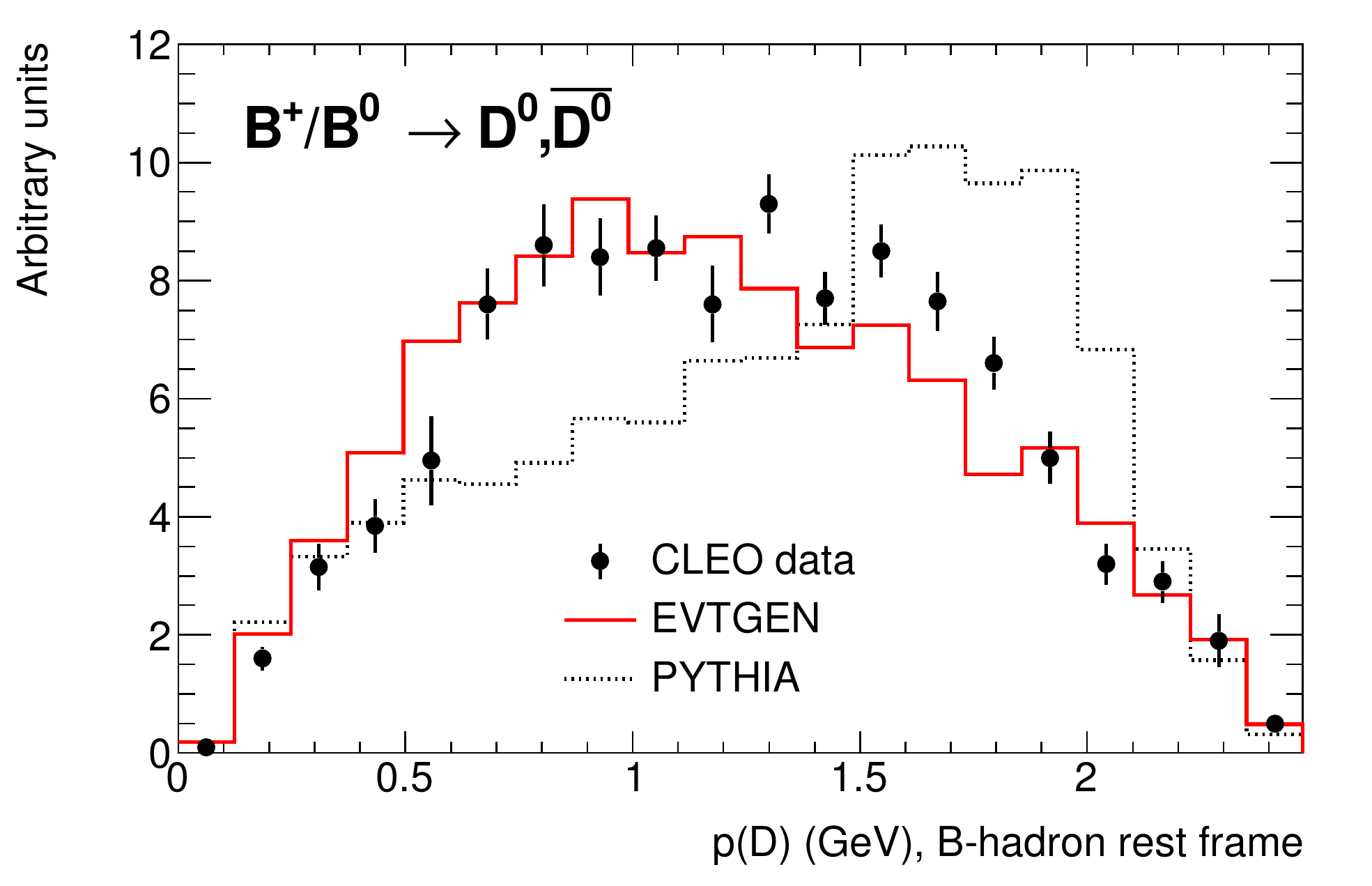}
  \end{minipage}
  \begin{minipage}[t]{0.495\columnwidth}
  \includegraphics[width=0.5\figwidth,trim=2mm 0mm 9mm 3mm,clip=true]{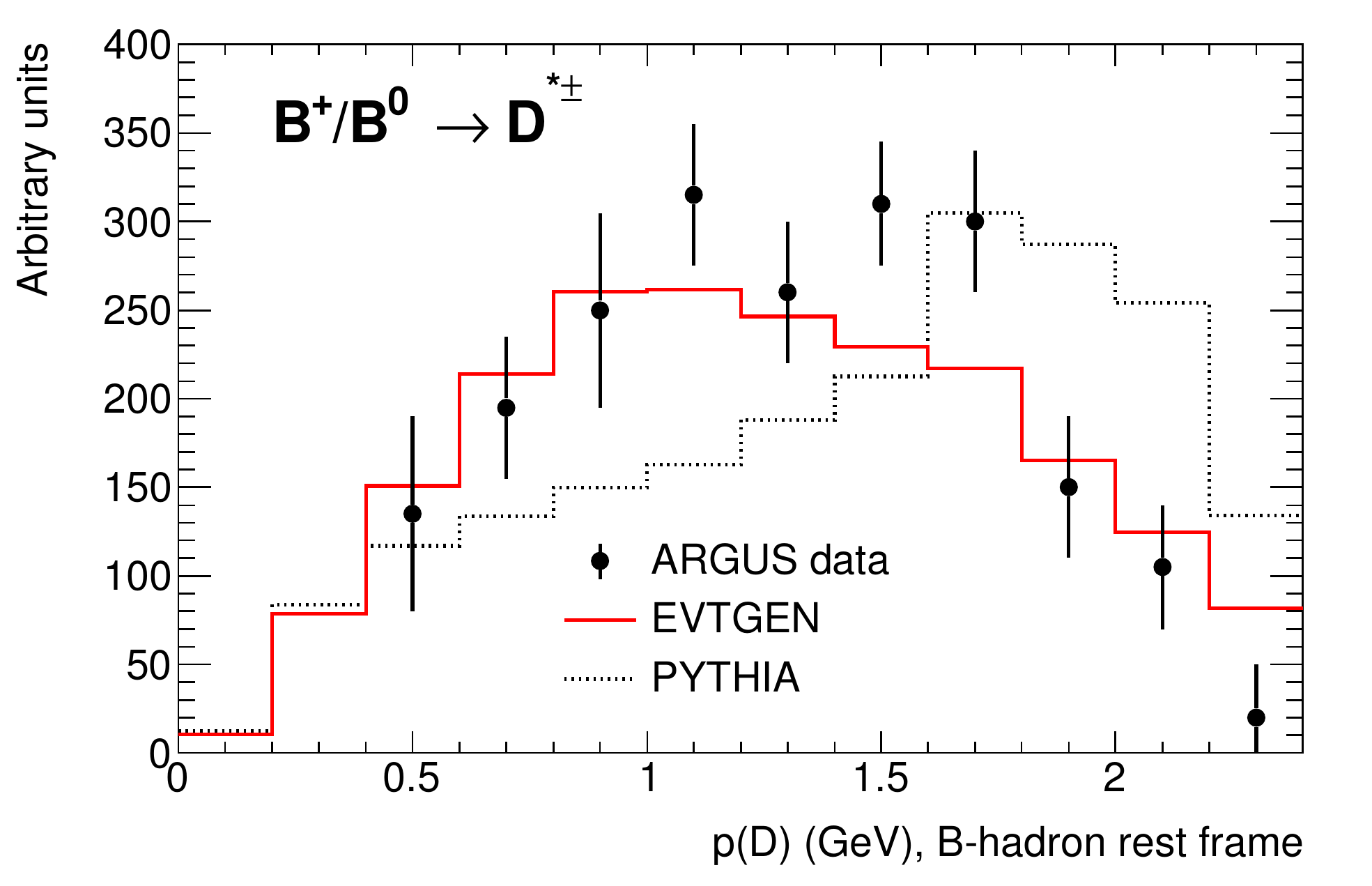}
  \includegraphics[width=0.5\figwidth,trim=2mm 0mm 9mm 3mm,clip=true]{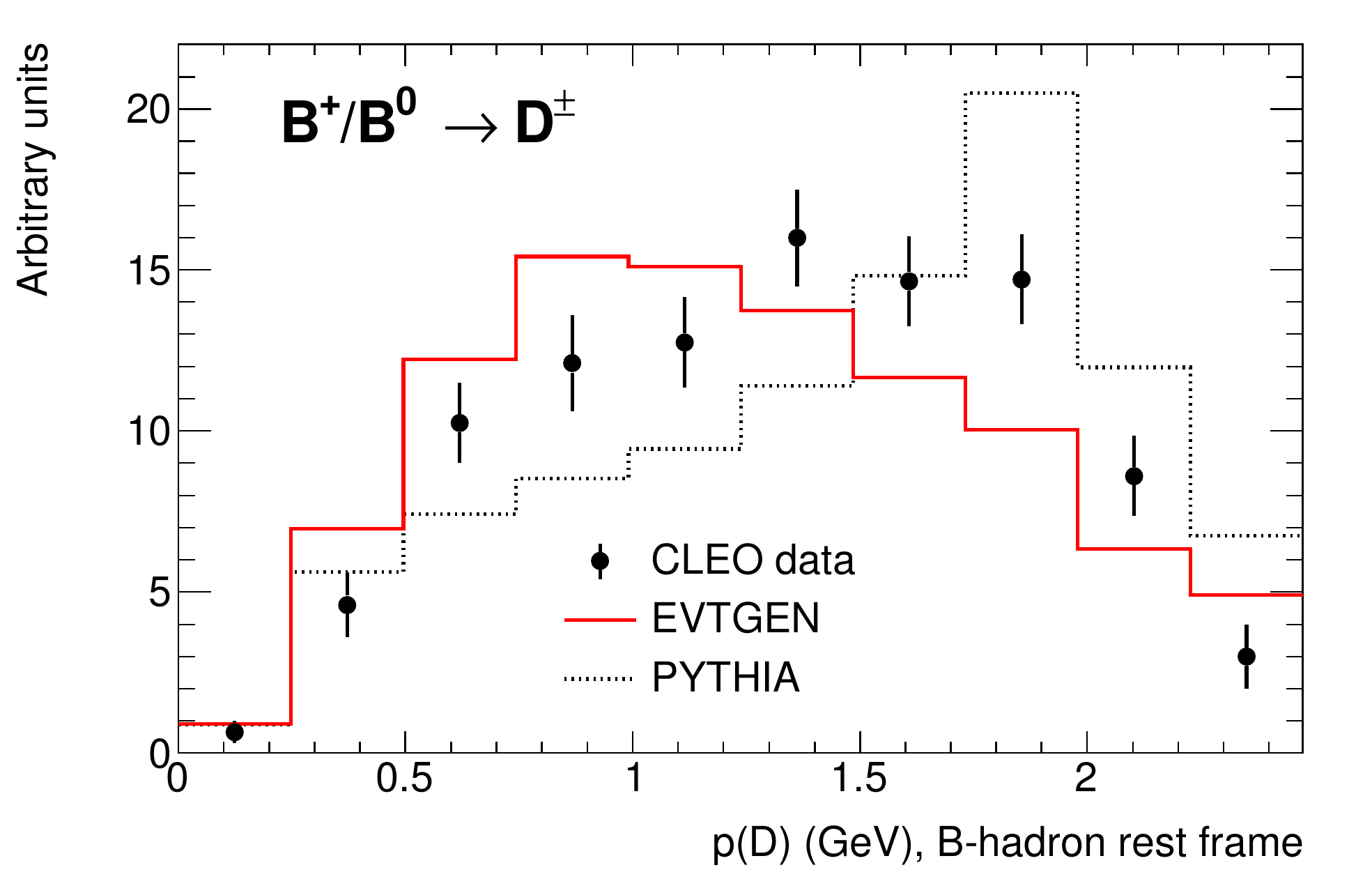}
  \end{minipage}
	\caption[Distributions of $D$-meson momenta in $B$-hadron rest frame]
	{Distributions of $\Dstar$ (top left and top right), $D^{0}$ (bottom left) and $D^{+}$ (bottom right) momenta in the $B$-hadron rest frame 
	as obtained from the PYTHIA~\cite{pythia} and EvtGen~\cite{evtgen} MC generators compared with 
	the data from CLEO~\cite{Gibbons:1997ag} (top left, bottom left and bottom right) and ARGUS~\cite{Albrecht:1995yb} (top right). 
	The distributions from the event generators are normalised to the data.}
	\label{fig:comb:th:ctobdecays}
\end{figure}

In the present study the beauty contribution was obtained at NLO: 
from the NLO QCD predictions for beauty hadrons with subsequent decays into charmed hadrons. 
A non-trivial ingredient of these calculations is the decay kinematics of \ozmodNN{beauty to charmed} hadrons, 
which, since many individual decay channels are involved, has to be obtained from some MC generator. 
In Fig.~\ref{fig:comb:th:ctobdecays} the distributions of $D$-meson momenta in the $B$-hadron rest frame 
as obtained from the PYTHIA~\cite{pythia} and EvtGen~\cite{evtgen} MC generators are compared with 
the data from CLEO~\cite{Gibbons:1997ag} and ARGUS~\cite{Albrecht:1995yb}. 
The shape from EvtGen describes the data reasonably well, therefore it was used for the predictions.

The parameters for the beauty contribution calculations and uncertainties were:
\begin{itemize}
\item {\bf the renormalisation and factorisation scales} 
$\mu_r=\mu_f=\sqrt{Q^2+4m_b^2}$,
varied as for charm. The variations for charm and beauty were applied 
simultaneously;
\item {\bf the pole mass of the $b$ quark} $m_b=4.75 \pm 0.25$ GeV;
\item {\bf the fragmentation model for $b$ quarks} based on 
the Peterson et al. \cite{Peterson:1982ak} parametrisation 
with $\epsilon_{b}=0.0035 \pm 0.0020$~\cite{frag00};
\item {\bf the fraction of beauty hadrons decaying into charmed hadrons} was taken from~\cite{pdg2012} and listed in Table~\ref{tab:comb:th:ff};
\item {\bf the PDFs}, described by the same set 
(the 3-flavour FFNS) as the one used for the corresponding charm prediction.
\end{itemize}

The dominant uncertainty comes from the variation of the fraction of beauty hadrons decaying into $D^{*+}$ mesons; 
although it reaches only $\approx 2\%$ in the highest $Q^2$ bins. 
Since the beauty contribution itself is small (varies from $1\%$ at low $Q^2$ to $7\%$ at high $Q^2$), 
all other uncertainties are negligible.

\subsection{Combination of visible \Dstar cross sections}
\label{sec:comb:dstar}

Among all techniques used at HERA to measure open charm production (see Section~\ref{sec:exp:hera:hfmeas}), 
measurements of \Dstar production have the best signal-to-background ratio and are the most precise. 
ZEUS and H1 have recently published single- and double-differential \Dstar cross sections for inclusive \Dstar-meson 
production in DIS from their respective final HERA-II datasets~\cite{h1dstarhighQ2,h1dstar_hera2,zeusdstar_hera2}. The measurements have been performed in 
similar \ozmod{phase-space regions} and used similar binning schemes%
\footnote{An agreement on the phase-space region and binning schemes was achieved between the H1 and ZEUS Collaborations before performing the measurements.}
thus fulfilling the requirement for a combination with minimised theory dependence.

The phase-space region of HERA-II measurements in DIS is restricted compared to that of HERA-I measurements. 
Due to beam-line modifications related to HERA-II high-luminosity running~\cite{HERAUpgrade} the visible phase 
space of these \Dstar cross sections at HERA-II is restricted to virtualities $Q^2 > \SI{5}{GeV}^2$. 
This fact prevents straightforward combination with HERA-I measurements for most of the single-differential \Dstar cross sections, although 
in the case of the single- or double-differential \Dstar cross sections as a function of $Q^2$, the above 
restriction does not apply and the kinematic range can be extended to lower $Q^2$ using earlier HERA-I 
measurements. In fact only the double-differential \Dstar cross sections as a function of $Q^2$ and $y$ 
can be combined with HERA-I measurements without applying extensive swimming corrections.%
\footnote{Although the single-differential \Dstar cross sections as a function of $Q^2$, in principle, also can be combined with HERA-I measurements without applying swimming corrections, 
this combination is not provided, because 
\begin{itemize}
\item information on the single-differential \Dstar cross sections as a function of $Q^2$ can be obtained from the double-differential \Dstar cross sections as a function of $Q^2$ and $y$, 
	provided in Section~\ref{sec:comb:dstar:double};
\item to keep consistency between all the combined single-differential \Dstar cross sections, which requires the same input data.
\end{itemize}}
For this reason the \ozmodN{treatment} of the visible \Dstar cross section combination consists of two parts: 
a combination of single-differential \Dstar cross sections, 
described in Section~\ref{sec:comb:dstar:single}, and a combination of the double-differential cross section, 
described in Section~\ref{sec:comb:dstar:double}. 
While the common method and strategy for both parts remain 
the same, the input measurements and phase-space regions differ.
All measurements to be combined for the single- and double-differential \Dstar cross sections are already corrected 
to the QED Born level with a running fine-structure constant and 
include both the charm and beauty contributions to \Dstar production. 
The results reported in this Section have been published by the H1 and ZEUS Collaborations~\cite{dstarcombpaper}.

\subsubsection{Combination of single-differential cross sections}
\label{sec:comb:dstar:single}

First the input measurements, combination phase-space region and 
details of the combination procedure are given, which includes all necessary corrections needed to transform the input data to the common phase-space region. 
Then the results of the combination are presented and discussed. 
Afterwards the combined data are compared to NLO QCD predictions. 
As a result of a detailed comparison of data and theory a `customised' theoretical calculation is introduced.

\paragraph*{Input measurements, phase-space region and combination details\\}
\label{sec:comb:dstar:single:det}

Table~\ref{tab:comb:dstar:single:input} presents the datasets used for the combination together with their visible phase-space 
regions and integrated luminosities. Note that the H1 Collaboration has published \Dstar cross-section measurements separately 
for $5<Q^2<100$~GeV$^2$ (dataset I)~\cite{h1dstar_hera2}%
\footnote{From the two sets of measurements in \cite{h1dstar_hera2}, the one 
compatible with the cuts on $p_T(\Dstar)$ and $\eta(\Dstar)$ quoted in Table~\ref{tab:comb:dstar:single:input}, which are compatible with 
the phase-space region of the ZEUS measurement~\cite{zeusdstar_hera2}, was chosen and referred to as dataset I.} 
and for $100<Q^2<1000$~GeV$^2$ (dataset II)~\cite{h1dstarhighQ2} because different 
sub-detectors had to be employed  for the detection and  measurement of the scattered electron in these two regions.
Thus the overall phase-space region for the combined \Dstar cross sections is given by
\begin{equation}
	\begin{split}
		&5< Q^2 < \SI{1000}{GeV}^2,\\
		&0.02<y<0.7,\\
		&p_T(\Dstar)>\SI{1.5}{GeV},\\
		&|\eta(\Dstar)|<1.5.
	\end{split}
\label{eq:comb:dstar:single:phasespace}
\end{equation}

\begin{table*}[tbp]
\caption[Datasets used for combination of \Dstar single-differential cross sections]
{Datasets used in the combination of the visible \Dstar single-differential cross sections. For each dataset the respective kinematic region 
and the integrated luminosity, ${\cal L}$, are given.}
\label{tab:comb:dstar:single:input}
\begin{center}
\tabcolsep2.0mm
\renewcommand*{\arraystretch}{1.2}
\begin{tabularx}{\textwidth}{|X|l|l|l|l|l|}
\hline
\multirow{3}{*}{Dataset} & \multicolumn{4}{c|}{Kinematic range}      & ${\cal L}$  \\ \cline{2-5}
                          & $Q^2$        & $y$ & $p_T(\Dstar)$ & $\eta(\Dstar)$ & \\ 
                          & [$\SI{}{GeV}^{2}$] &     & [$\SI{}{GeV}$]     &             & [$\text{pb}^{\text{-1}}$]            \\ \hline
I:\ \ \  H1 $\Dstar$ HERA-II (medium $Q^2$) \cite{h1dstar_hera2} & $~~~~5:100$ & $0.02:0.70$ & $1.5:\infty$ & $-1.5:1.5$ & $348$ \\ \hline
II:\ \  H1 $\Dstar$ HERA-II (high $Q^2$) \cite{h1dstarhighQ2} & $100:1000$ & $0.02:0.70$ & $1.5:\infty$ & $-1.5:1.5$ & $351$ \\ \hline
III: ZEUS $\Dstar$ HERA-II \cite{zeusdstar_hera2} & $~~~~5:1000$ & $0.02:0.70$ & $1.5:20.0$ & $-1.5:1.5$ & $363$ \\ \hline
\end{tabularx}
\end{center}
\end{table*}

The combination was done for single-differential \Dstar cross sections as a function of the 
\Dstar transverse momentum, $p_T(\Dstar)$,  pseudorapidity, $\eta(\Dstar)$, and inelasticity, 
$z(\Dstar)= (E(\Dstar)-p_Z(\Dstar))/(2E_e y)$, with $E_e$ being the incoming electron energy, 
$E(\Dstar)$ and $p_Z(\Dstar)$ the energy and longitudinal momentum of \Dstar, respectively, 
as well as of the DIS kinematic variables  $Q^2$ and $y$.%
\footnote{Although all input measurements from Table~\ref{tab:comb:dstar:single:input} \ozmod{give} also the single-differential cross section as 
a function of the Bjorken variable $x$, the binning differs significantly, preventing a combination without large swimming corrections.}

Since the H1 datasets I and II are complementary to each other and give the phase-space region of the combination~\ref{eq:comb:dstar:single:phasespace}, their differential \Dstar cross sections 
\ozmod{are} summed up on a bin-by-bin basis and enter the combination as a single dataset. 
However, due to the limited statistics at high $Q^2$ a coarser binning scheme in $p_T(\Dstar)$, $\eta(\Dstar)$, $z(\Dstar)$ and $y$ had to be used in dataset II compared to dataset I. 
This made a straightforward summation of the differential \Dstar cross sections from the two measurements \ozmod{complicated}. 
Therefore the cross section in a bin $i$ of a given observable integrated in the range $5<Q^2<1000$~GeV$^2$ was calculated according to
\begin{eqnarray}
\lefteqn{\sigma_i(5<\frac{Q^2}{\text{GeV}^2}<1000)=\sigma_i(5<\frac{Q^2}{\text{GeV}^2}<100)}\\\nonumber
&&{}+\sigma_i^{NLO}(100<\frac{Q^2}{\text{GeV}^2}<1000)\cdot\frac{\sigma_{int}(100<\frac{Q^2}{\text{GeV}^2}<1000)}{\sigma_{int}^{NLO}(100<\frac{Q^2}{\text{GeV}^2}<1000)}.
\end{eqnarray} 
Here $\sigma_{int}$ denotes the integrated visible cross section and $NLO$ stands for the NLO predictions 
obtained from HVQDIS.%
\footnote{Since the normalisation was taken from another measurement, not from theory predictions, 
this is swimming, as explained in Section~\ref{sec:comb:proc:pscorr}.}
In this calculation both the experimental uncertainties of the visible cross section at high $Q^2$ and the theoretical uncertainties (described in Section~\ref{sec:comb:proc:pscorr}) were included. 
The contribution from the region $100<Q^2<1000$~GeV$^2$ to the full $Q^2$ range amounts to $4$\% on average and reaches up to $50$\% at the highest $p_T(\Dstar)$; 
the extrapolation uncertainty is negligible in most of the bins compared to the corresponding experimental uncertainty; only at the two highest $p_T(\Dstar)$ bins it approaches $35$\% of the experimental uncertainty. 
Thus in the combination procedure the extrapolation uncertainties from all theoretical parameter variations were added in quadrature and treated as an uncorrelated uncertainty. 
The sensitivity of the shape to the beauty contribution was found to be negligible and therefore was ignored.

For the single-differential \Dstar cross sections as a function of $Q^2$, the procedure described above was not needed. 
However the binning schemes used for these \Dstar cross sections differ between datasets I--II and dataset III. At low $Q^2$ this was solved by combining the cross-section measurements of the first two bins 
of dataset I into a single bin. For $Q^2>100$~GeV$^2$ no consistent binning scheme could be defined from the single-differential cross-section measurements ${\rm d}\sigma/{\rm d}Q^2$ itself. 
However, the measurements of the double-differential \Dstar cross section ${\rm d}^2\sigma/{\rm d}y{\rm d}Q^2$ have been performed in a common binning scheme. 
By integrating these \Dstar cross sections in $y$, single-differential \Dstar cross sections in $Q^2$ were obtained at $Q^2>100$~GeV$^2$ 
from datasets II, III which were used directly in the combination.
The contribution to dataset III from the range $p_T(\Dstar)>\SI{20}{GeV}$ was found to be negligible ($\ll 1\%$). 

Applying the procedure described above provided exactly two input measurements for each combined bin: one from H1 (datasets I--II) and one from ZEUS (dataset III). 
Thus \ndof is equal to the number of combined bins. Since the data are statistically correlated between the different distributions, each distribution was combined separately.

The branching ratios for datasets I, II were updated to the PDG value~\cite{pdg2012}.
A full list of considered correlated sources is provided in Appendix~\ref{sec:app:dstar} (Table~\ref{tab:comb:dstar:syst}). 
All systematic uncertainties were treated as uncorrelated between the H1 and ZEUS measurements, except for the branching-ratio uncertainty; 
since the latter is fully correlated between all datasets, it is not changed in the combination and was not included in the combination 
but applied as an external uncertainty on the results.

\paragraph*{Combined \Dstar cross sections\\}
\label{sec:comb:dstar:single:res}

The results of combining the HERA-II measurements 
\cite{h1dstar_hera2,h1dstarhighQ2,zeusdstar_hera2}
as a function of $p_T(\Dstar)$, $\eta(\Dstar)$, $z(\Dstar)$, $Q^2$ and $y$  are given 
in Table~\ref{tab:comb:dstar:single:combined}, together with their uncorrelated and correlated uncertainties.
The total uncertainties were obtained by adding the uncorrelated and correlated uncertainties in quadrature. 
A detailed breakdown of the correlated uncertainties is provided in Appendix~\ref{sec:app:dstar} (Table~\ref{tab:comb:dstar:single:combinedfull12}).

\begin{table*}[tbp]
\caption[Combined single-differential \Dstar cross section]
{The combined single-differential \Dstar cross sections as a function of $p_T(\Dstar)$, $\eta(\Dstar)$, $z(\Dstar)$, $Q^2$ and $y$, with their 
uncorrelated ($\delta_{unc}$), correlated ($\delta_{cor}$) and total ($\delta_{tot}$) uncertainties.
The cross sections are given in the kinematic region~\ref{eq:comb:dstar:single:phasespace}. 
} 
\label{tab:comb:dstar:single:combined}
\tabcolsep4.1mm
\renewcommand*{\arraystretch}{1.1}
\begin{minipage}[t]{0.5\textwidth}
\begin{tabu} to \columnwidth [t]{|X[c]|c|c|c|c|}
\hline
$p_T(\Dstar)$ &$ \frac{d\sigma}{dp_T(\Dstar)}$ &$\delta_{unc}$  &$\delta_{cor}$&$\delta_{tot}$ \\
  ($\SI{}{GeV}$)       & ($\SI{}{nb/GeV}$)                         &$(\%)$            &$(\%)$ &$(\%)$ \\ 
\hline
1.50 : 1.88 & 2.35 & 6.4 & 4.7 & 8.0 \\ \hline
1.88 : 2.28 & 2.22 & 4.9 & 4.2 & 6.4 \\ \hline
2.28 : 2.68 & 1.98 & 3.7 & 4.0 & 5.5 \\ \hline
2.68 : 3.08 & 1.55 & 3.5 & 3.7 & 5.1 \\ \hline
3.08 : 3.50 & 1.20 & 3.7 & 3.5 & 5.1 \\ \hline
3.50 : 4.00 & 9.29 $\times 10^{-1}$ & 3.2 & 3.4 & 4.7 \\ \hline
4.00 : 4.75 & 6.14 $\times 10^{-1}$ & 3.0 & 3.5 & 4.6 \\ \hline
4.75 : 6.00 & 3.19 $\times 10^{-1}$ & 3.1 & 3.3 & 4.5 \\ \hline
6.00 : 8.00 & 1.15 $\times 10^{-1}$ & 3.8 & 3.7 & 5.3 \\ \hline
8.00 : 11.00 & 3.32 $\times 10^{-2}$ & 5.4 & 3.7 & 6.5 \\ \hline
11.00 : 20.00 & 3.80 $\times 10^{-3}$ & 10.4 & 6.4 & 12.2 \\ \hline
\hline
$\eta(\Dstar)$ &$ \frac{d\sigma}{d\eta(\Dstar)}$ &$\delta_{unc}$  &$\delta_{cor}$&$\delta_{tot}$ \\
         & (nb)                         &$(\%)$            &$(\%)$ &$(\%)$ \\ 
\hline
-1.50 : -1.25 & 1.36 & 5.8 & 4.3 & 7.2 \\ \hline
-1.25 : -1.00 & 1.52 & 4.6 & 4.0 & 6.1 \\ \hline
-1.00 : -0.75 & 1.59 & 4.6 & 4.0 & 6.1 \\ \hline
-0.75 : -0.50 & 1.79 & 3.8 & 3.5 & 5.2 \\ \hline
-0.50 : -0.25 & 1.83 & 3.8 & 3.3 & 5.1 \\ \hline
-0.25 : 0.00 & 1.89 & 3.8 & 3.7 & 5.3 \\ \hline
0.00 : 0.25 & 1.86 & 4.0 & 3.4 & 5.2 \\ \hline
0.25 : 0.50 & 1.88 & 4.0 & 3.6 & 5.4 \\ \hline
0.50 : 0.75 & 1.91 & 4.1 & 3.5 & 5.4 \\ \hline
0.75 : 1.00 & 1.92 & 4.3 & 4.0 & 5.9 \\ \hline
1.00 : 1.25 & 2.08 & 4.7 & 4.0 & 6.1 \\ \hline
1.25 : 1.50 & 1.81 & 6.3 & 4.8 & 7.9 \\ \hline
\end{tabu}
\end{minipage}
\begin{minipage}[t]{0.5\textwidth}
\begin{tabu} to \columnwidth [t]{|X[c]|c|c|c|c|}
\hline
$z(\Dstar)$ &$ \frac{d\sigma}{dz(\Dstar)}$ &$\delta_{unc}$  &$\delta_{cor}$&$\delta_{tot}$ \\
         & (nb)                         &$(\%)$            &$(\%)$ &$(\%)$ \\ 
\hline
0.00 : 0.10 & 3.28 & 9.5 & 5.9 & 11.2 \\ \hline
0.10 : 0.20 & 7.35 & 4.8 & 6.3 & 7.9 \\ \hline
0.20 : 0.32 & 8.61 & 3.5 & 4.6 & 5.7 \\ \hline
0.32 : 0.45 & 8.92 & 2.7 & 3.9 & 4.7 \\ \hline
0.45 : 0.57 & 8.83 & 1.8 & 4.0 & 4.3 \\ \hline
0.57 : 0.80 & 4.78 & 2.4 & 5.1 & 5.6 \\ \hline
0.80 : 1.00 & 6.31 $\times 10^{-1}$ & 8.1 & 10.2 & 13.0 \\ \hline
\hline
$Q^2$ &$ \frac{d\sigma}{dQ^2}$ &$\delta_{unc}$  &$\delta_{cor}$&$\delta_{tot}$ \\
  ($\SI{}{GeV}^2$)       & ($\SI{}{nb/GeV}^2$)                         &$(\%)$            &$(\%)$ &$(\%)$ \\ 
\hline
5 : 8 & 4.74 $\times 10^{-1}$ & 4.0 & 5.0 & 6.4 \\ \hline
8 : 10 & 2.96 $\times 10^{-1}$ & 4.3 & 3.8 & 5.8 \\ \hline
10 : 13 & 2.12 $\times 10^{-1}$ & 3.8 & 4.0 & 5.6 \\ \hline
13 : 19 & 1.24 $\times 10^{-1}$ & 3.2 & 3.8 & 5.0 \\ \hline
19 : 28 & 7.26 $\times 10^{-2}$ & 3.5 & 3.6 & 5.0 \\ \hline
28 : 40 & 3.97 $\times 10^{-2}$ & 3.7 & 4.0 & 5.5 \\ \hline
40 : 60 & 1.64 $\times 10^{-2}$ & 4.4 & 4.7 & 6.4 \\ \hline
60 : 100 & 7.45 $\times 10^{-3}$ & 5.2 & 3.9 & 6.5 \\ \hline
100 : 158 & 2.08 $\times 10^{-3}$ & 7.2 & 5.3 & 9.0 \\ \hline
158 : 251 & 8.82 $\times 10^{-4}$ & 7.6 & 5.0 & 9.1 \\ \hline
251 : 1000 & 7.50 $\times 10^{-5}$ & 12.0 & 6.7 & 13.3 \\ \hline
\hline
$y$ &$ \frac{d\sigma}{dy}$ &$\delta_{unc}$  &$\delta_{cor}$&$\delta_{tot}$ \\
         & (nb)                         &$(\%)$            &$(\%)$ &$(\%)$ \\ 
\hline
0.02 : 0.05 & 12.13 & 5.8 & 9.1 & 10.8 \\ \hline
0.05 : 0.09 & 18.84 & 3.9 & 4.6 & 6.0 \\ \hline
0.09 : 0.13 & 16.99 & 3.4 & 4.3 & 5.5 \\ \hline
0.13 : 0.18 & 13.35 & 3.7 & 4.2 & 5.6 \\ \hline
0.18 : 0.26 & 11.19 & 3.4 & 3.7 & 5.0 \\ \hline
0.26 : 0.36 & 7.65 & 3.7 & 4.2 & 5.6 \\ \hline
0.36 : 0.50 & 4.78 & 4.0 & 5.3 & 6.6 \\ \hline
0.50 : 0.70 & 2.65 & 5.6 & 6.4 & 8.5 \\ \hline
\end{tabu}
\end{minipage}
\end{table*}

\begin{figure*}[tbp]
  \centering
  \begin{minipage}[t]{0.33\textwidth}
  \includegraphics[width=1.0\textwidth,trim=1mm 0 9mm 15mm,clip=true]{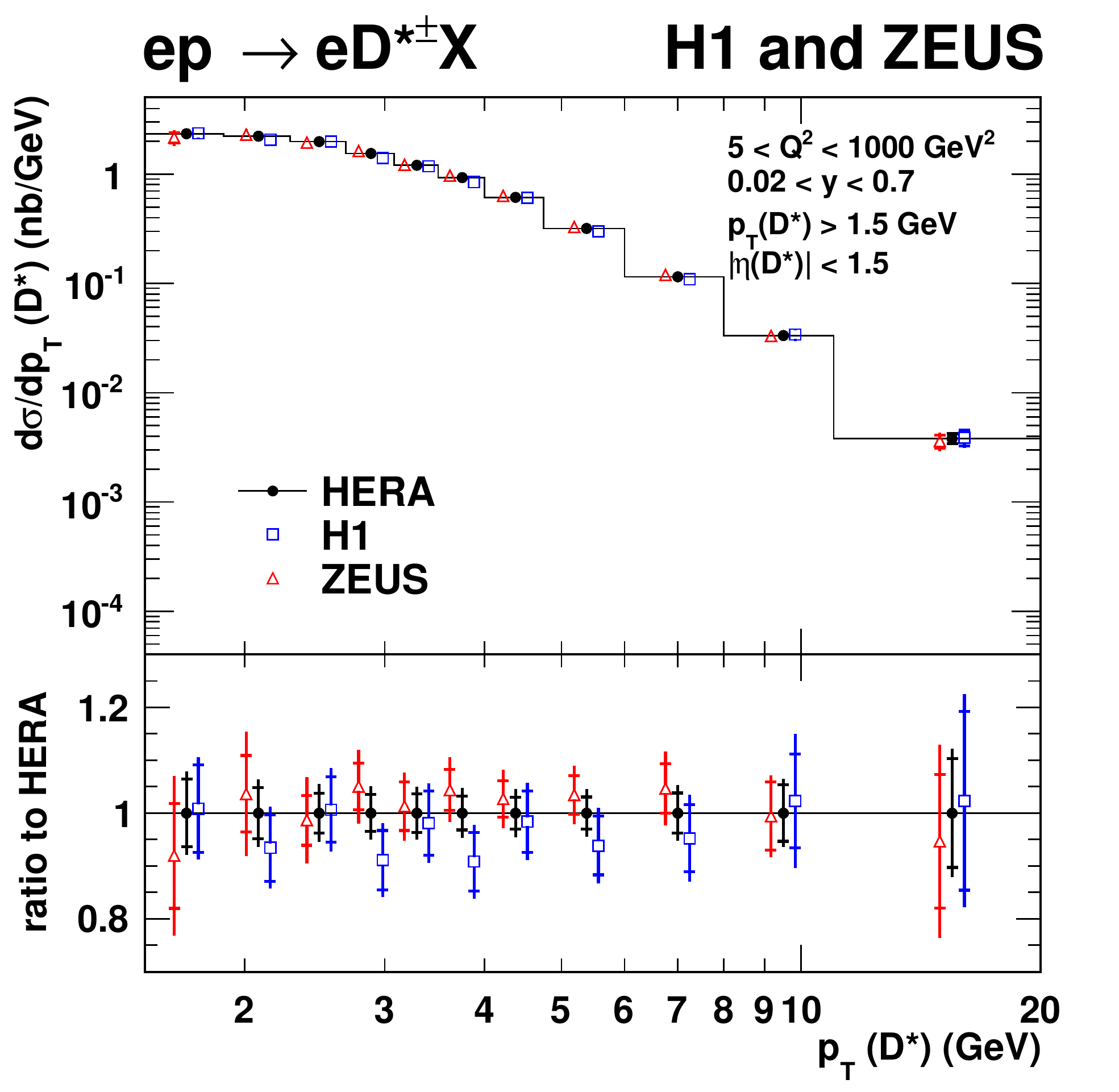}
  \put(-130,130){(a)}\\
  \includegraphics[width=1.0\textwidth,trim=1mm 0 9mm 15mm,clip=true]{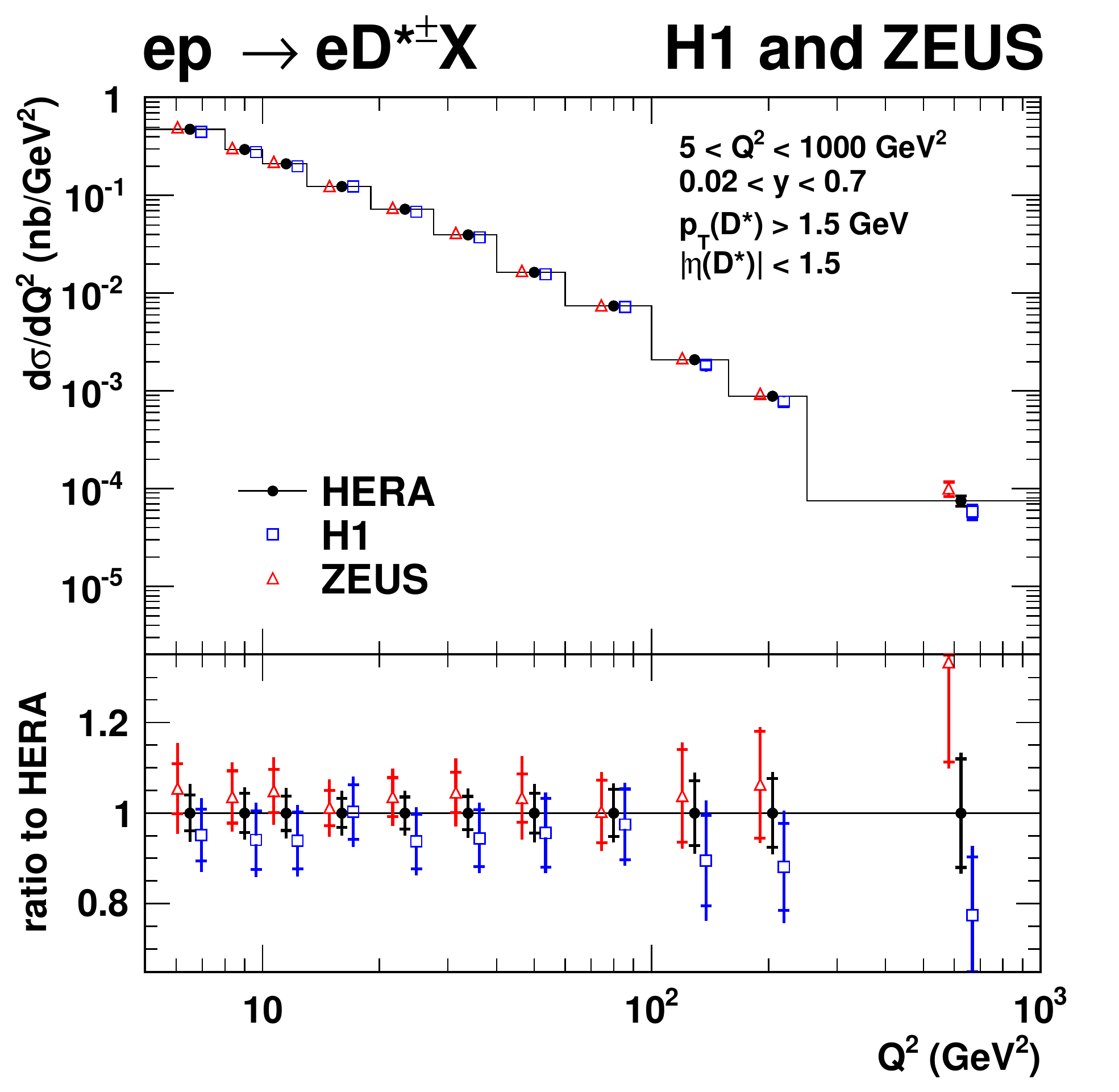}
  \put(-130,130){(d)}
  \end{minipage}
  \begin{minipage}[t]{0.33\textwidth}
  \includegraphics[width=1.0\textwidth,trim=1mm 0 9mm 15mm,clip=true]{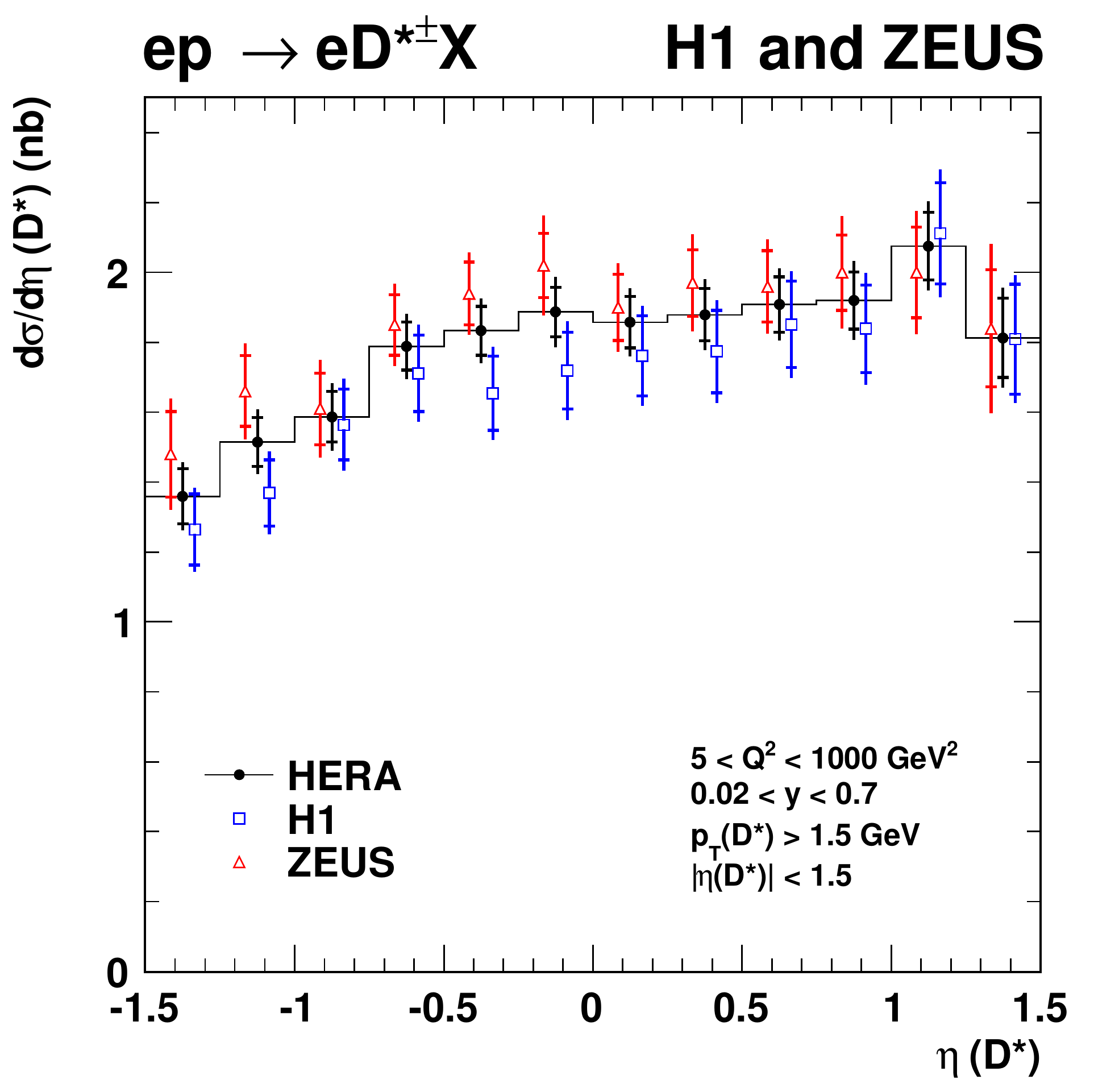}
  \put(-130,130){(b)}\\
  \includegraphics[width=1.0\textwidth,trim=1mm 0 9mm 15mm,clip=true]{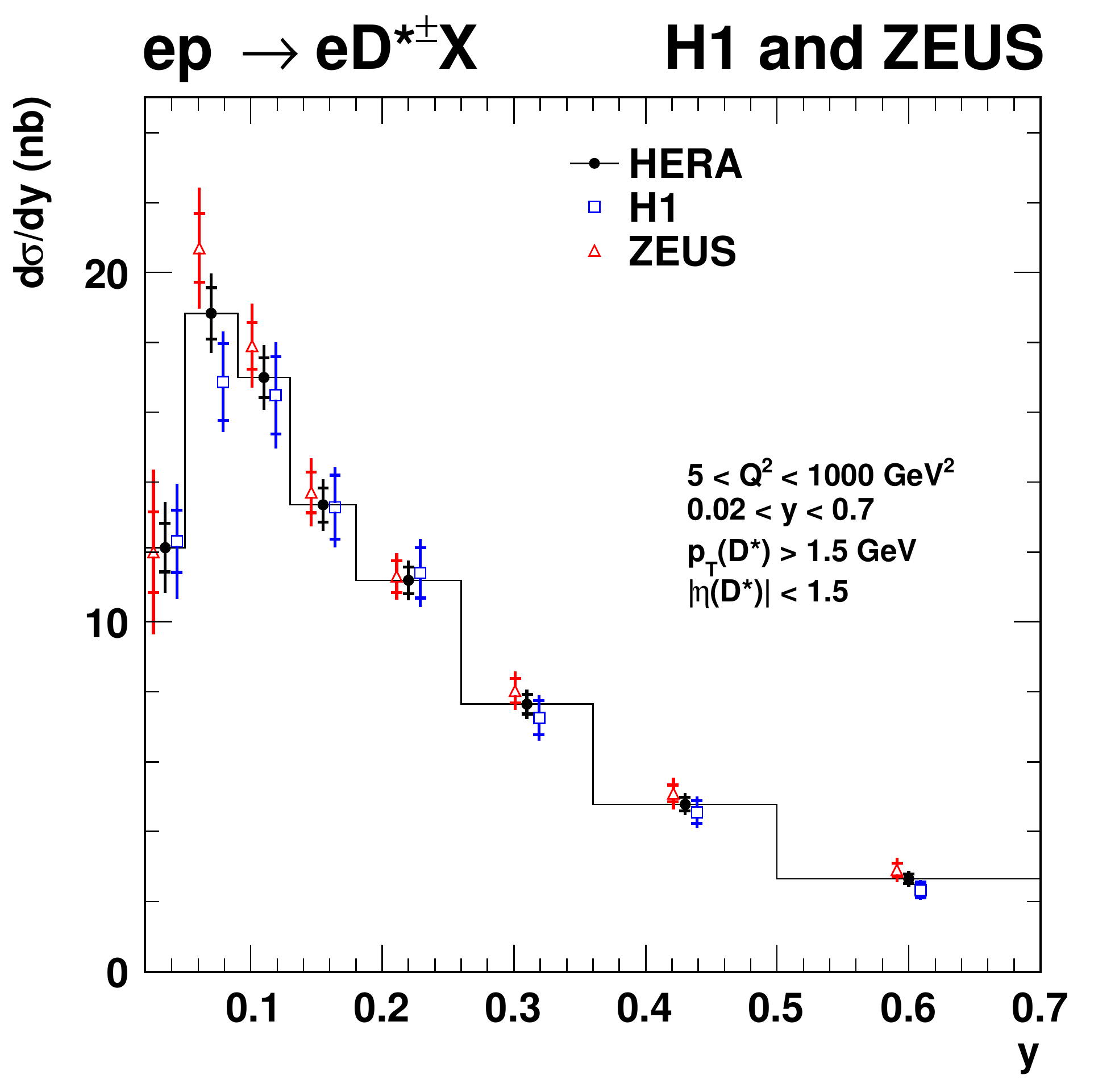}
  \put(-130,130){(e)}
  \end{minipage}
  \begin{minipage}[t]{0.33\textwidth}
  \includegraphics[width=1.0\textwidth,trim=1mm 0 9mm 15mm,clip=true]{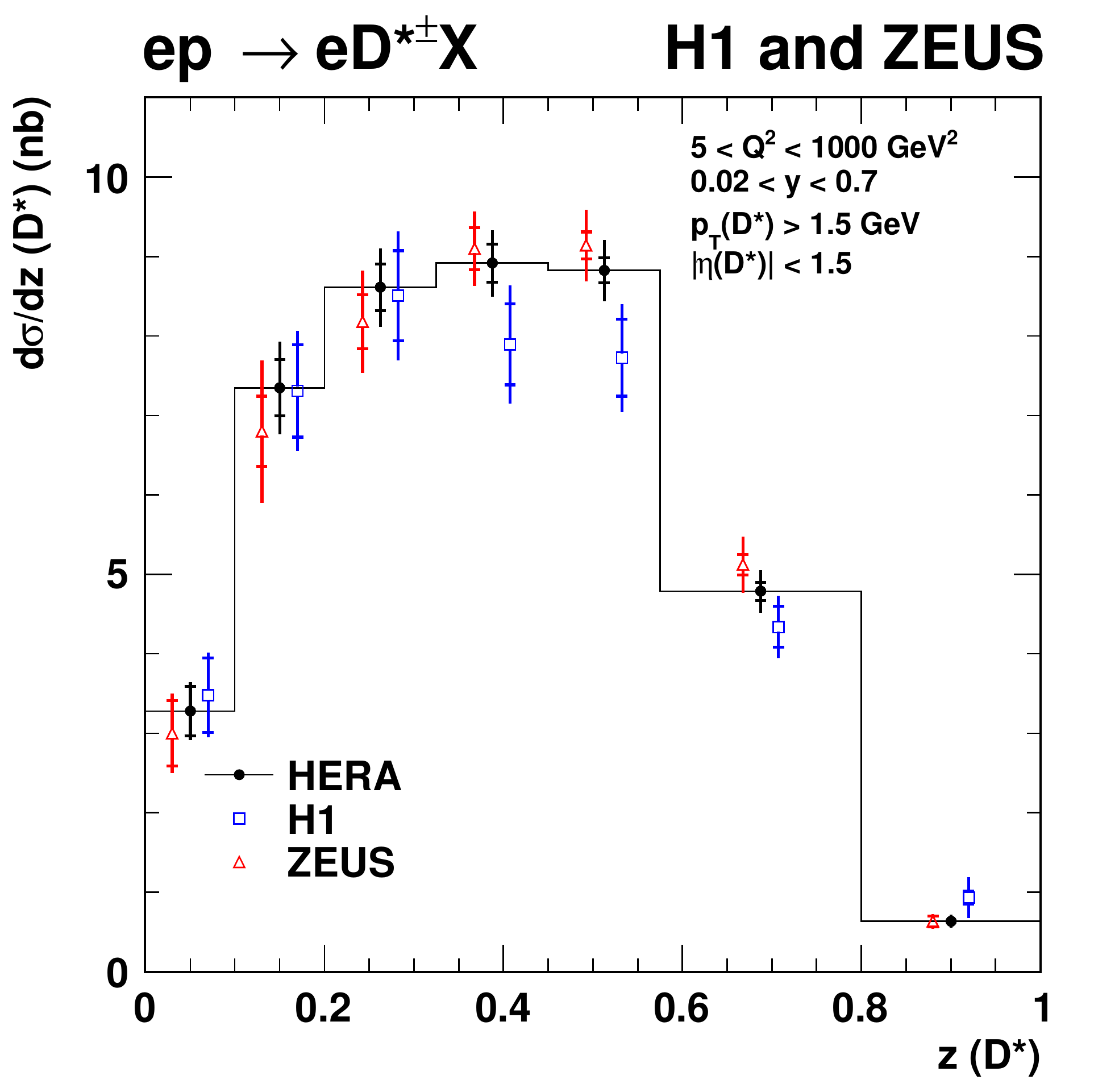}
  \put(-130,130){(c)}
  \caption[Single-differential \Dstar cross sections]
  {Single-differential \Dstar cross section as a function of $p_T(\Dstar)$ (a), $\eta(\Dstar)$ (b), $z(\Dstar)$ (c), $Q^2$ (d) and $y$ (e). 
	The triangles and open squares are the \Dstar cross sections before combination, shown with a small horizontal offset for better visibility. 
	The filled points are the combined \Dstar cross sections. The inner error bars indicate the uncorrelated part of the uncertainties. 
	The outer error bars represent the total uncertainties. The histogram indicates the binning used to calculate the \Dstar cross sections. 
	For (a) and (d), the bottom part shows the ratio of these \Dstar cross sections with respect to the central value of the combined \Dstar cross sections.}
	\label{fig:comb:dstar:single:combined}
  \end{minipage}
\end{figure*}

\begin{figure}[htbp]
  \centering
  \begin{minipage}[t]{0.495\columnwidth}
  \includegraphics[width=0.49\figwidth,trim=1mm 0 6mm 0mm,clip=true]{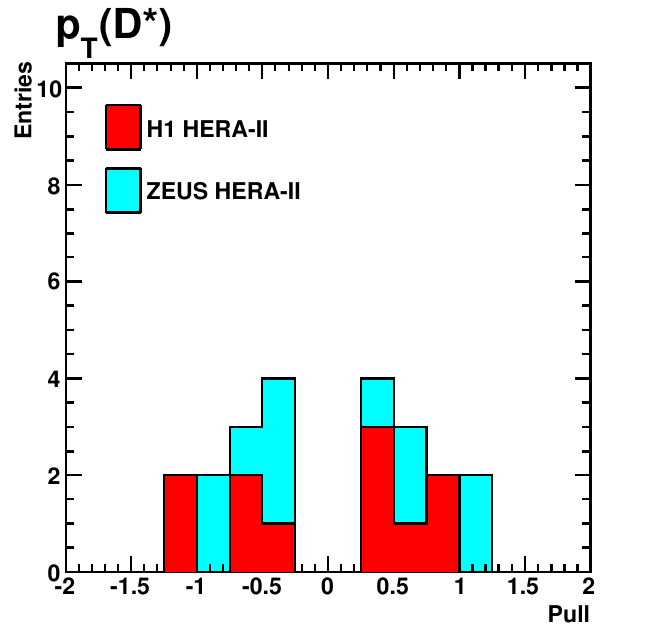}
  \put(-20,100){(a)}\\
  \includegraphics[width=0.49\figwidth,trim=1mm 0 6mm 0mm,clip=true]{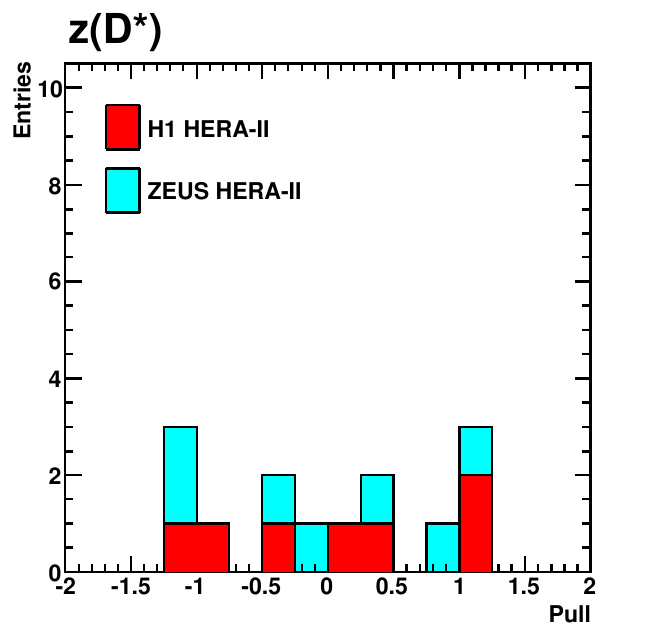}
  \put(-20,100){(c)}\\
  \includegraphics[width=0.49\figwidth,trim=1mm 0 6mm 0mm,clip=true]{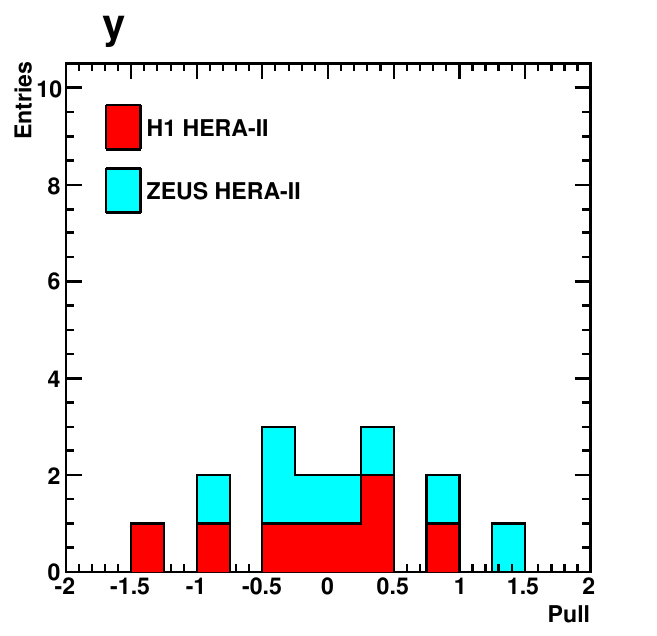}
  \put(-20,100){(e)}\\
  \end{minipage}
  \begin{minipage}[t]{0.495\columnwidth}
  \includegraphics[width=0.49\figwidth,trim=1mm 0 6mm 0mm,clip=true]{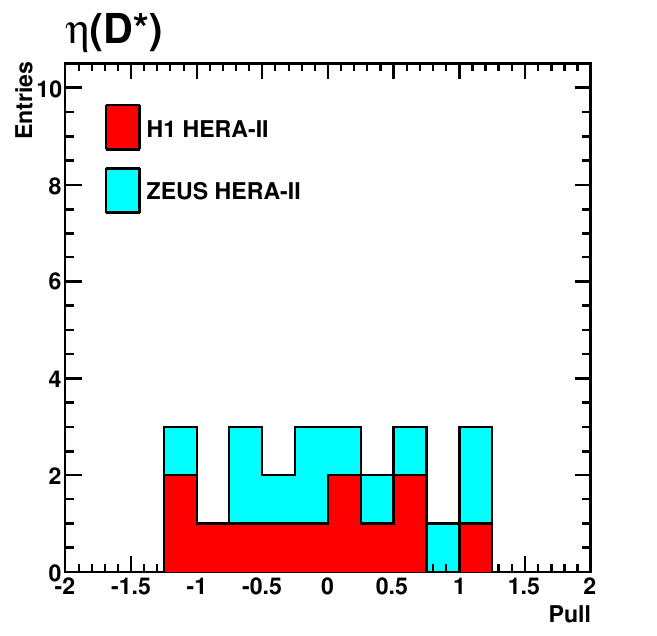}
  \put(-20,100){(b)}\\
  \includegraphics[width=0.49\figwidth,trim=1mm 0 6mm 0mm,clip=true]{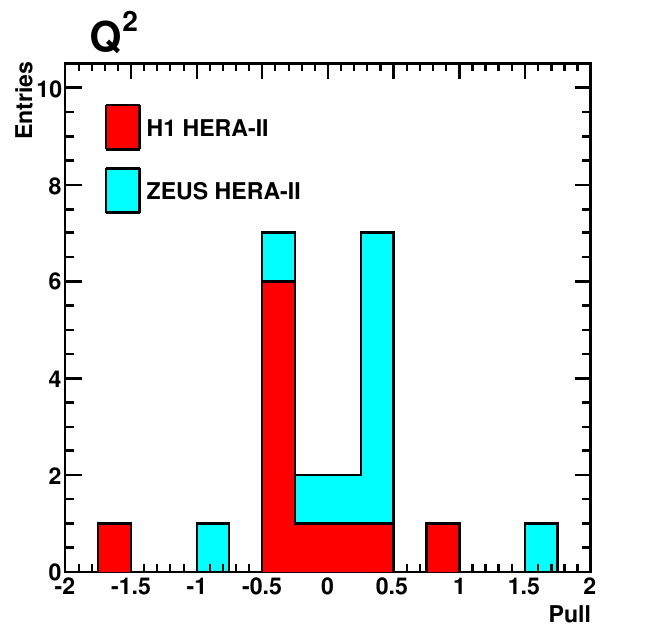}
  \put(-20,100){(d)}
  \caption[Pull distributions for single-differential \Dstar cross sections]
  {The pull distributions for the combination of the single-differential \Dstar cross sections as a function of $p_T(\Dstar)$ (a), $\eta(\Dstar)$ (b), $z(\Dstar)$ (c), $Q^2$ (d) and $y$ (e). 
  Contributions from the individual input datasets are shown separately.}
	\label{fig:comb:dstar:single:pulls}
  \end{minipage}
\end{figure}

The individual datasets as well as the results of the combination are shown
in Fig.~\ref{fig:comb:dstar:single:combined}.  
The consistency of the datasets as well as the reduction of the 
uncertainties are illustrated further for the steeply falling \Dstar cross sections 
as a function of $p_T(\Dstar)$ and $Q^2$ in the bottom parts. 
The input H1 and ZEUS datasets are similar in precision. 
The values of \chisq, \ndof and the corresponding $\chi^2$-probabilities for the combinations of the different distributions 
are reported in Table~\ref{tab:comb:dstar:single:chi2}. 
The combinations in the different variables have $\chi^2$-probability 
varying between 15\% and 87\%, i.e.\ the datasets are consistent.
The pull distributions are shown in Fig.~\ref{fig:comb:dstar:single:pulls}. 
Although Fig.~\ref{fig:comb:dstar:single:combined} indicates that the H1 data points lie on average below the ZEUS points, the pulls 
in Fig.~\ref{fig:comb:dstar:single:pulls} show an overall symmetric spread 
of the H1 and ZEUS input data around the combined results; this is explained by taking into account shifts of the correlated systematic uncertainties. 
The shifts and reductions of the correlated sources are consistent for the combinations of the \Dstar cross sections in different variables; 
they are provided in Appendix~\ref{sec:app:dstar} (Table~\ref{tab:comb:dstar:syst}). 

\begin{table}[htbp]
\caption[\chisq, \ndof and probabilities for combinations of single-differential \Dstar cross sections]
{The values of \chisq, \ndof and the corresponding $\chi^2$-probabilities for the combinations of the single-differential \Dstar cross sections as a function of different variables.}
\label{tab:comb:dstar:single:chi2}
\tabcolsep4.0mm
\begin{center}
\begin{tabularx}{\columnwidth}{|X|c|c|c|} \hline
	Cross section & \ndof & \chisq & p(\chisq,\ndof) \\ \hline
	${\rm d}\sigma/{\rm d}p_T(\Dstar)$ & 11    & 6.9    & 81\%            \\ \hline
	${\rm d}\sigma/{\rm d}\eta(\Dstar)$& 12    & 7.8    & 80\%            \\ \hline
	${\rm d}\sigma/{\rm d}z(\Dstar)$   & 7     &10.9    & 15\%            \\ \hline
	${\rm d}\sigma/{\rm d}Q^2$         & 11    & 6.1    & 87\%            \\ \hline
	${\rm d}\sigma/{\rm d}y$           & 8     & 5.8    & 67\%            \\ \hline
\end{tabularx}
\end{center}
\end{table}

The combined \Dstar cross sections exhibit significantly reduced uncertainties. 
While the effective doubling of the statistics of the combined result reduces 
the uncorrelated uncertainties (inner error bars in Fig.~\ref{fig:comb:dstar:single:combined}), 
the correlated uncertainties (quadratic difference of the outer and inner 
error bars) of the combined \Dstar cross sections are significantly reduced through 
cross-calibration effects between the two experiments. Typically, both effects 
contribute about equally to the reduction of the total uncertainty.

\paragraph*{Comparison with theoretical predictions\\}
\label{sec:comb:dstar:single:th}

\begin{figure*}[tbp]
  \centering
  \begin{minipage}[t]{0.33\textwidth}
  \includegraphics[width=1.0\textwidth,trim=1mm 0 9mm 15mm,clip=true]{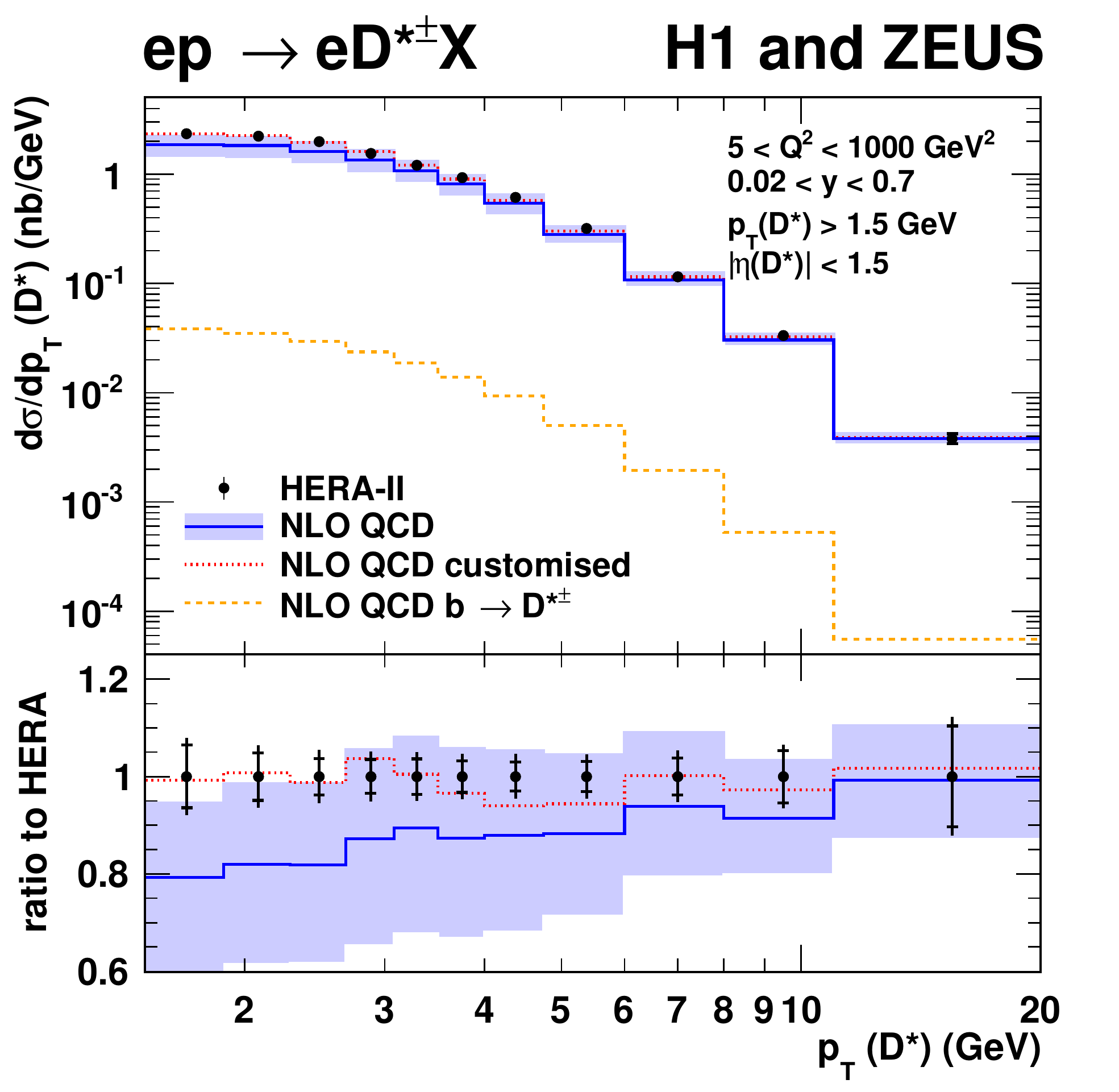}
  \put(-130,130){(a)}\\
  \includegraphics[width=1.0\textwidth,trim=1mm 0 9mm 15mm,clip=true]{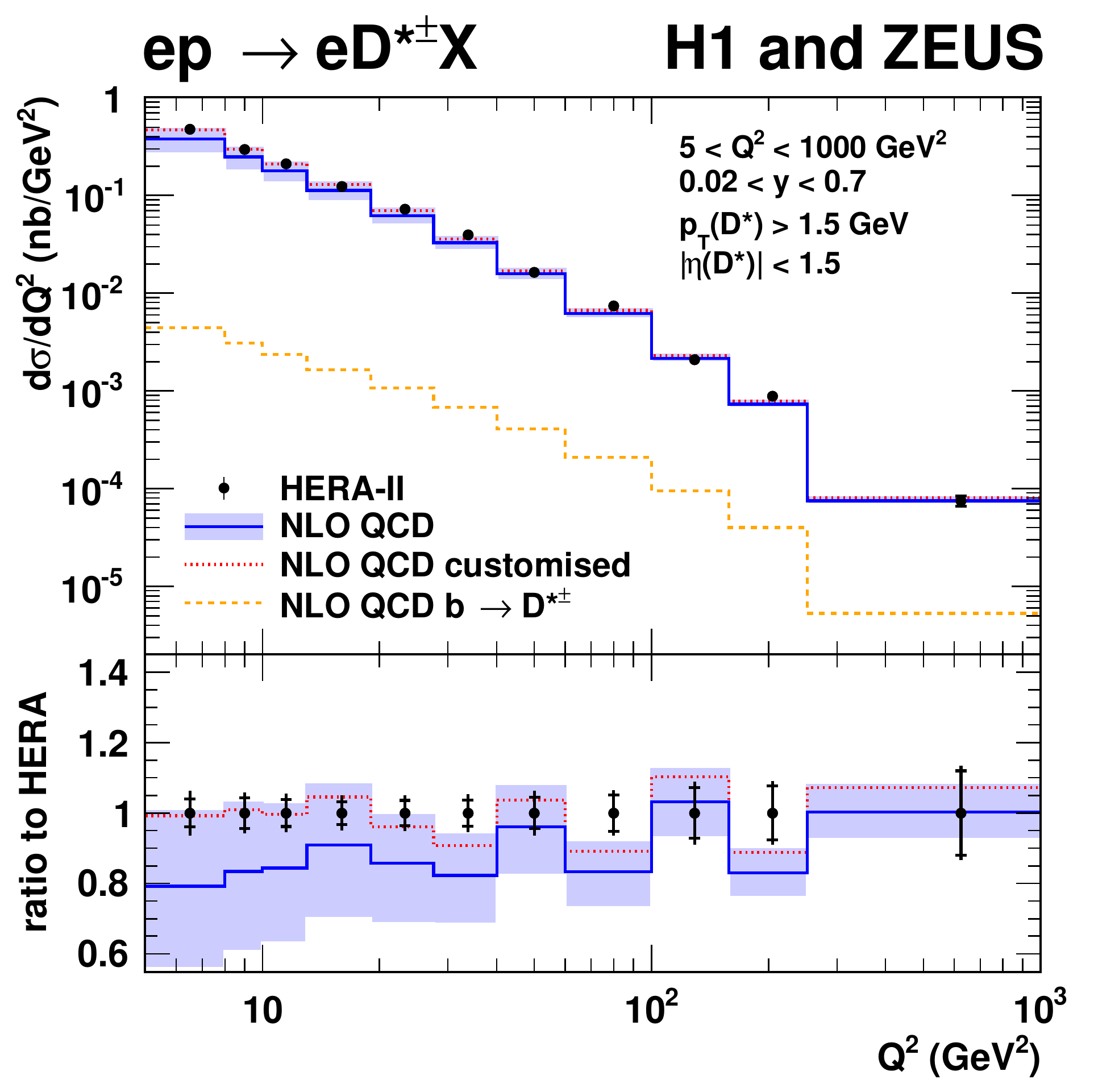}
  \put(-130,130){(d)}\\
  \end{minipage}
  \begin{minipage}[t]{0.33\textwidth}
  \includegraphics[width=1.0\textwidth,trim=1mm 0 9mm 15mm,clip=true]{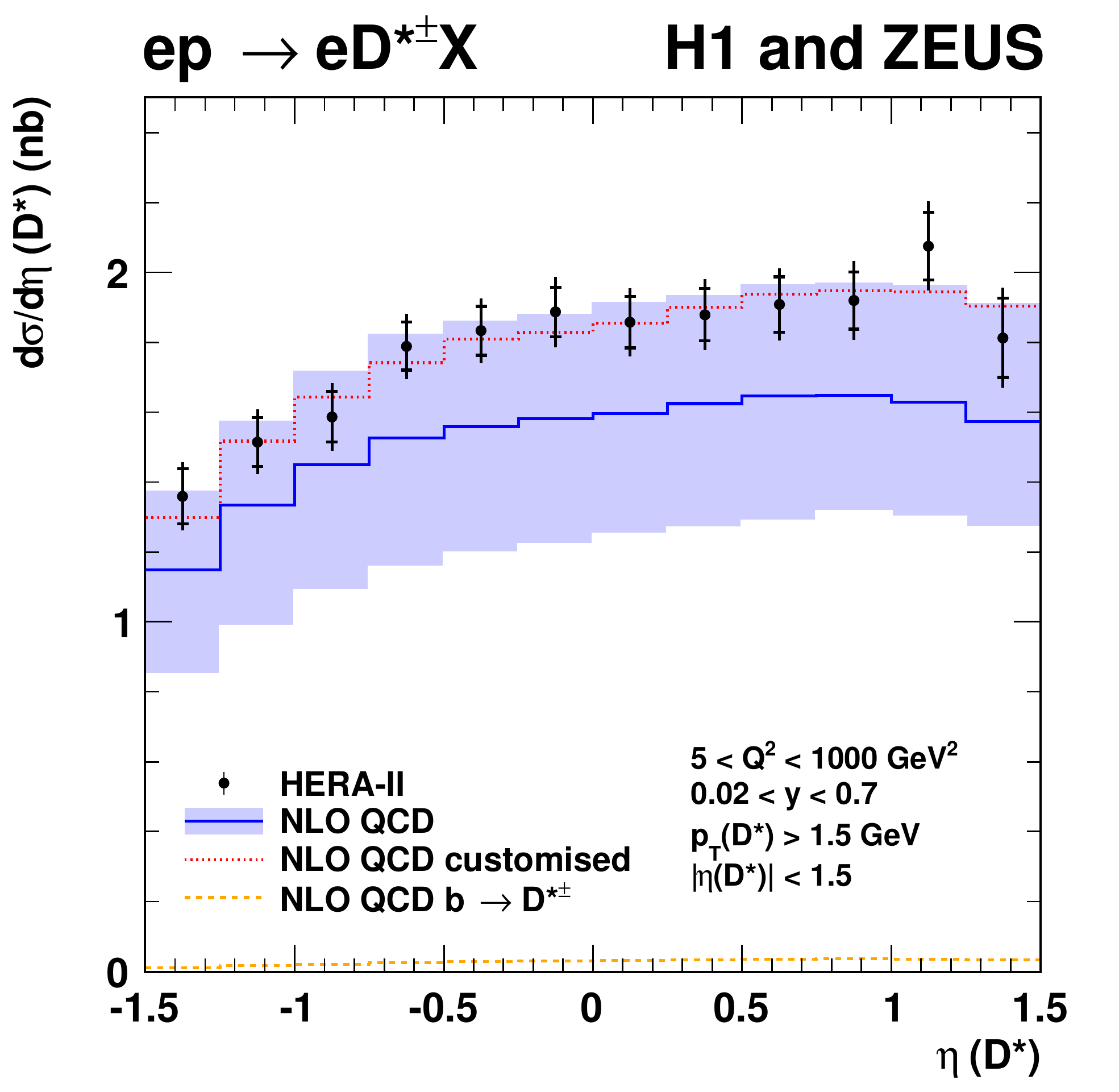}
  \put(-130,130){(b)}\\
  \includegraphics[width=1.0\textwidth,trim=1mm 0 9mm 15mm,clip=true]{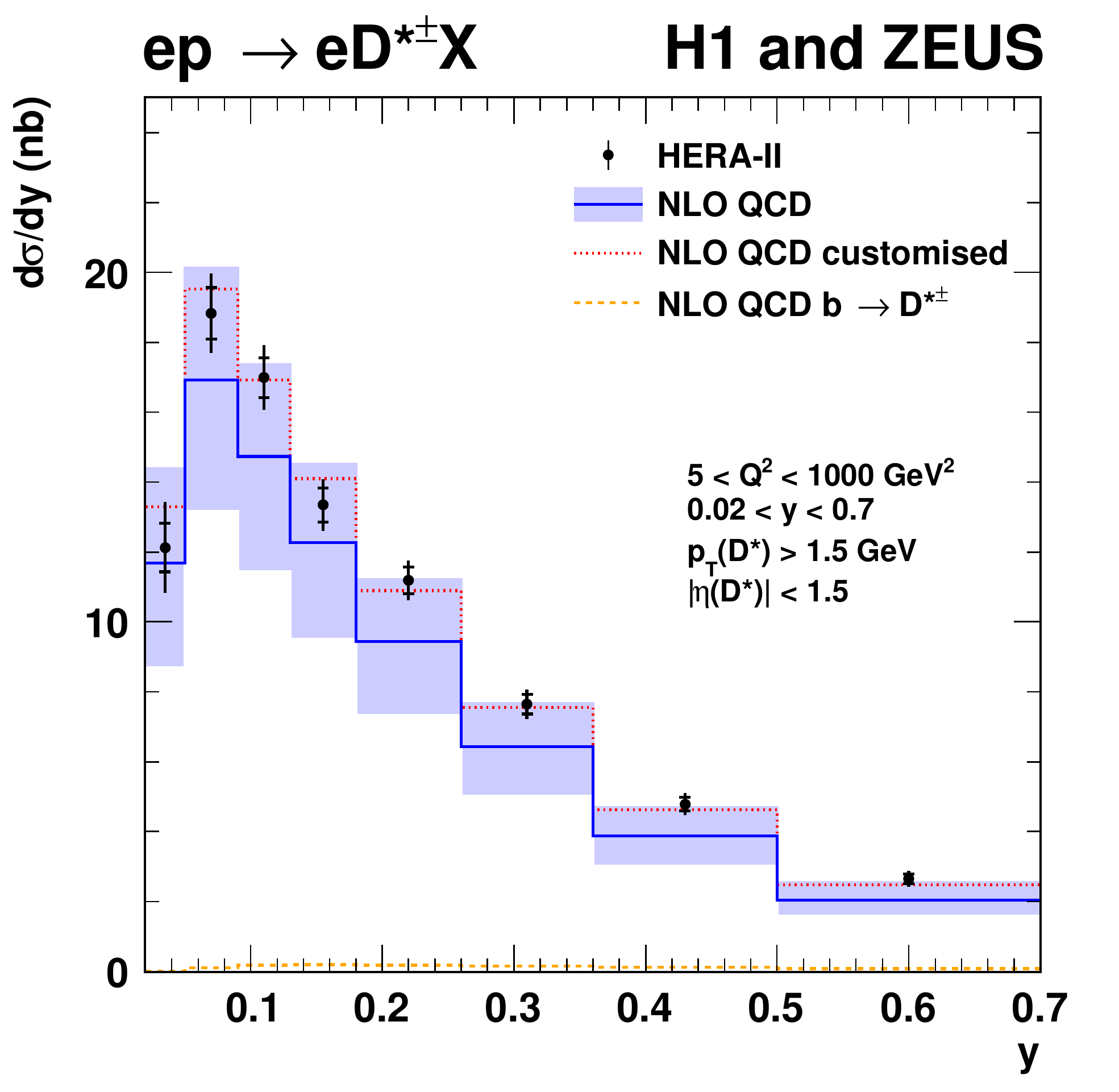}
  \put(-130,130){(e)}\\
  \end{minipage}
  \begin{minipage}[t]{0.33\textwidth}
  \includegraphics[width=1.0\textwidth,trim=1mm 0 9mm 15mm,clip=true]{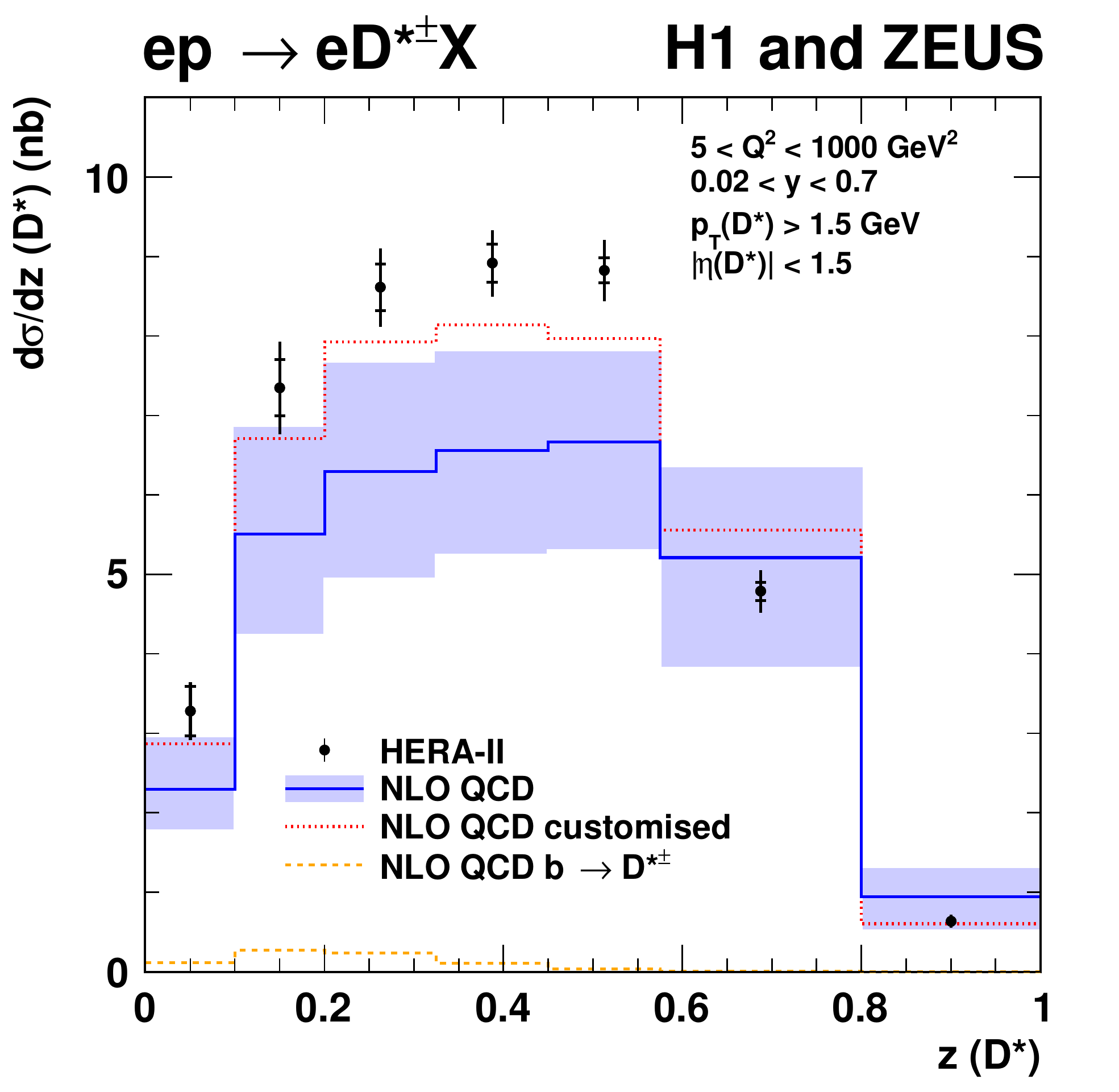}
  \put(-130,130){(c)}
  \caption[Single-differential \Dstar cross sections compared to NLO predictions]
  {Single-differential \Dstar cross section as a function of $p_T(\Dstar)$ (a), $\eta(\Dstar)$ (b), $z(\Dstar)$ (c), $Q^2$ (d) and $y$ (e). 
	The data points are the combined \Dstar cross sections. The inner 	error bars indicate the uncorrelated part of the uncertainties. 
	The outer error bars represent the total uncertainties. 
	Also shown are the NLO predictions from HVQDIS (including the beauty contribution), and their uncertainty band. 
	A customised NLO calculation (dotted line) is also shown.}
	\label{fig:comb:dstar:single:withth}
  \end{minipage}
\end{figure*}

The combined \Dstar cross sections as a function of 
$p_T(\Dstar)$, $\eta(\Dstar)$, $z(\Dstar)$, $Q^2$ and $y$ 
are compared to the NLO QCD predictions in the FFNS (described in Section~\ref{sec:comb:th}) in Fig.~\ref{fig:comb:dstar:single:withth}; there is also 
a dotted line referred to as `customised' NLO QCD predictions shown there, which will be discussed below. 
In general the predictions describe the data well. 
\ozmodN{The uncertainties of the data are as small as $5$\%} over a large fraction of the measured phase-space region, 
while the typical theory uncertainty ranges from 30\% at low $Q^2$ to 10\%
at high $Q^2$. The data points between the different distributions are 
statistically and systematically correlated, so they can be 
quantitatively compared to theory only on a one-by-one basis.

The theoretical predictions describe the combined data well within the corresponding uncertainty band, 
\ozmodN{however} the central theoretical curves underestimate the data normalisation. 
The central theoretical prediction shows a somewhat softer $y$ distribution than 
the data. The central prediction for $z(\Dstar)$ is \ozmodN{slightly} wider than the 
measured distribution.

\paragraph*{`Customised' theoretical predictions\\}
\label{sec:comb:dstar:single:customth}

\ozmodNN{As stated above, in overall the theoretical uncertainties are larger 
than the experimental uncertainties of the combined data.} 
Since the theoretical uncertainties \ozmodN{depend on} several correlated sources, 
it is rather difficult to make a strong statement about agreement between the theory and the data from Fig.~\ref{fig:comb:dstar:single:withth} itself.

\begin{figure*}[htbp]
\sidecaption
  \begin{minipage}[t]{0.4\textwidth}
  \includegraphics[width=1.0\textwidth,trim=1mm 0 15mm 15mm,clip=true]{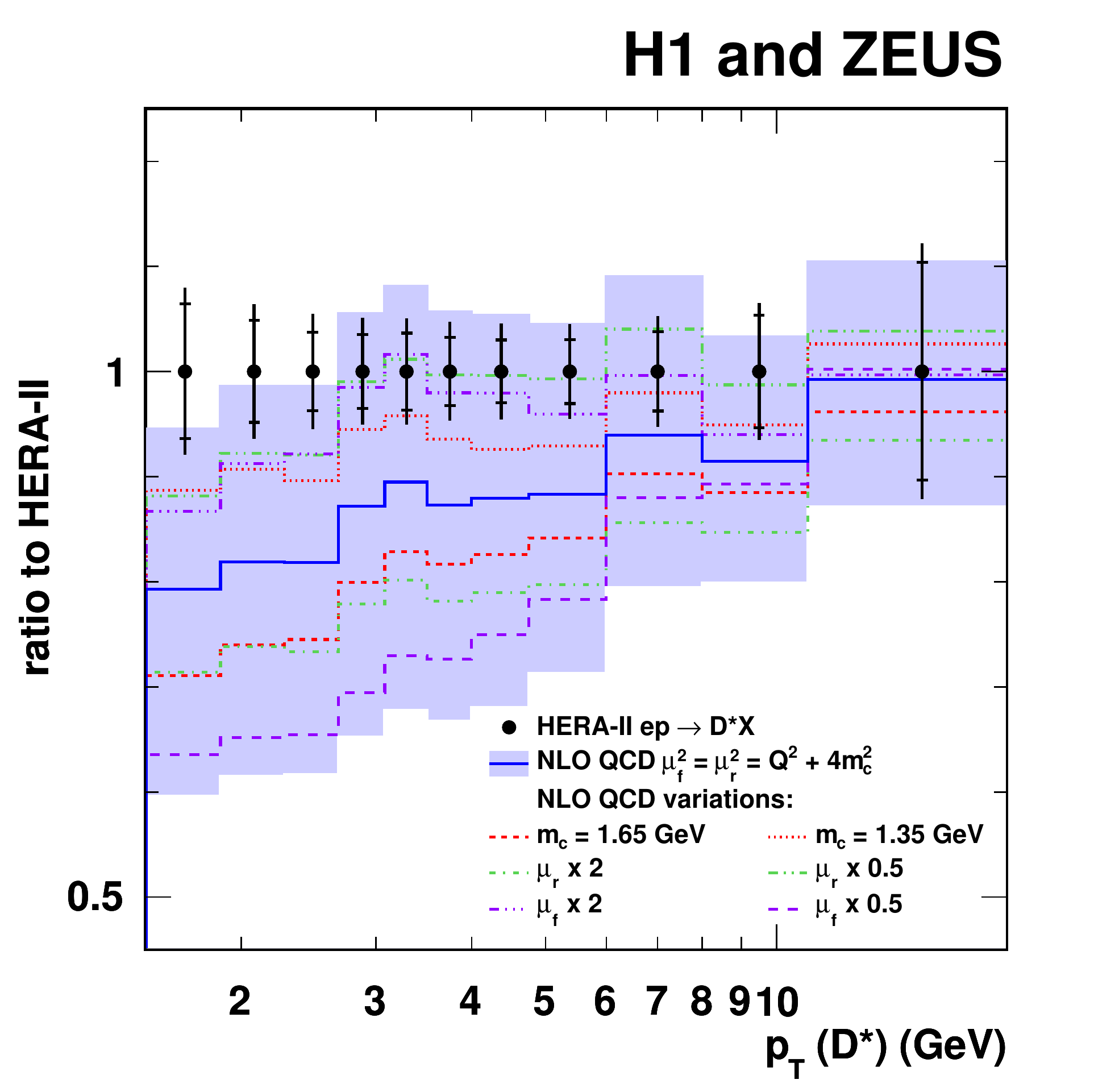}
  \put(-30,177){(a)}\\
  \includegraphics[width=1.0\textwidth,trim=1mm 0 15mm 15mm,clip=true]{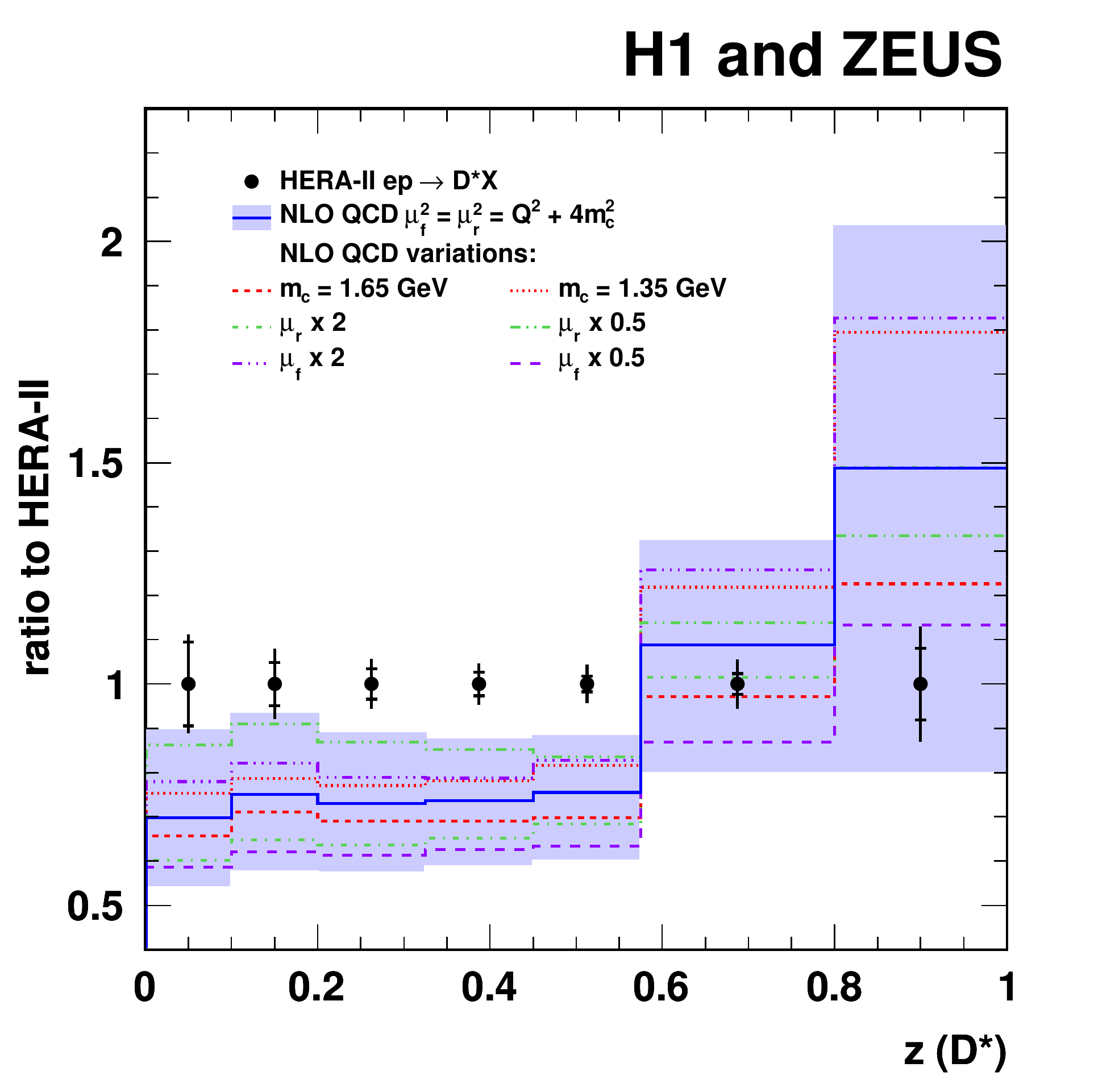}
  \put(-30,177){(c)}\\
  \end{minipage}
  \begin{minipage}[t]{0.4\textwidth}
  \includegraphics[width=1.0\textwidth,trim=1mm 0 15mm 15mm,clip=true]{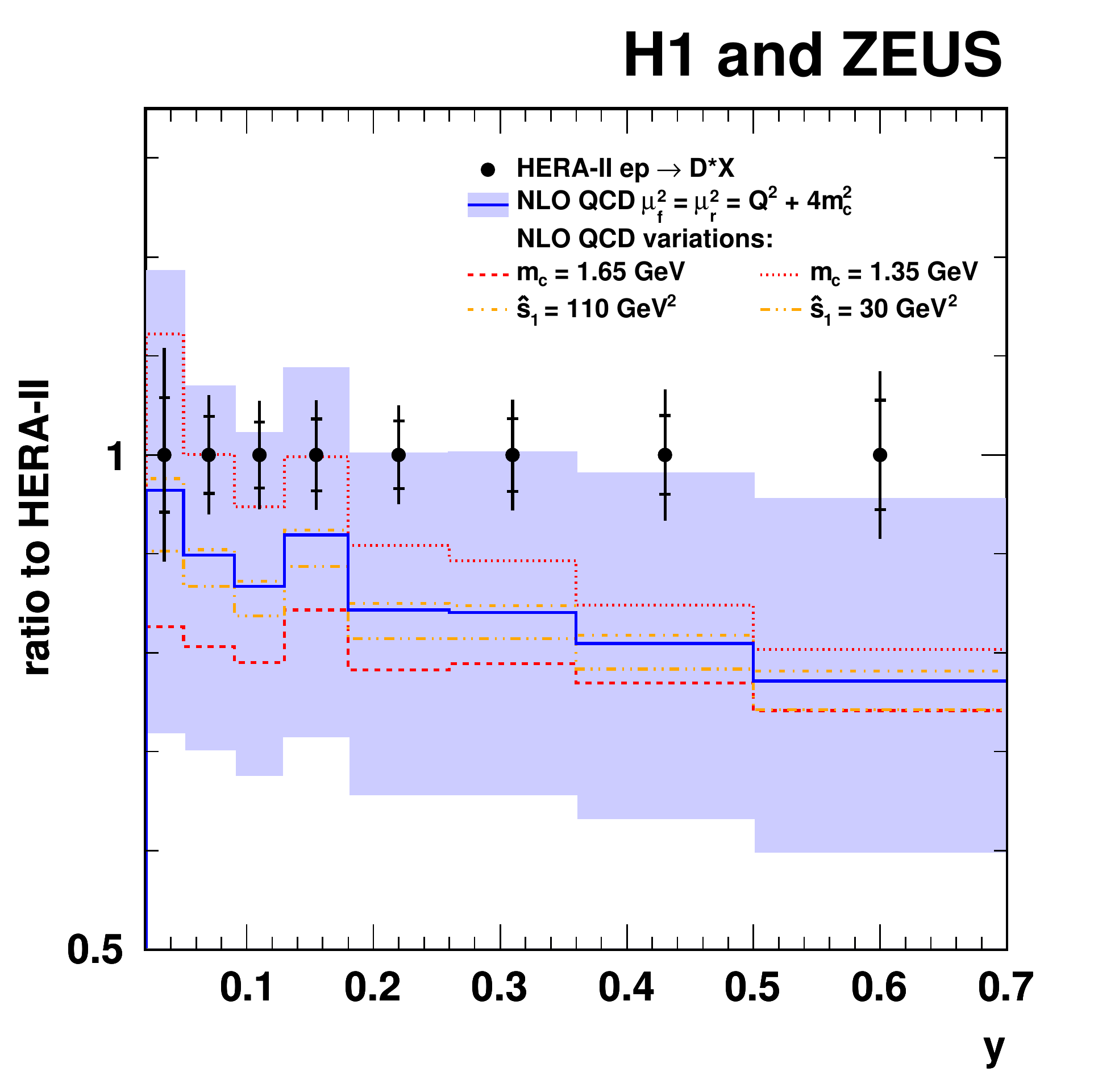}
  \put(-30,177){(b)}\\
  \includegraphics[width=1.0\textwidth,trim=1mm 0 15mm 15mm,clip=true]{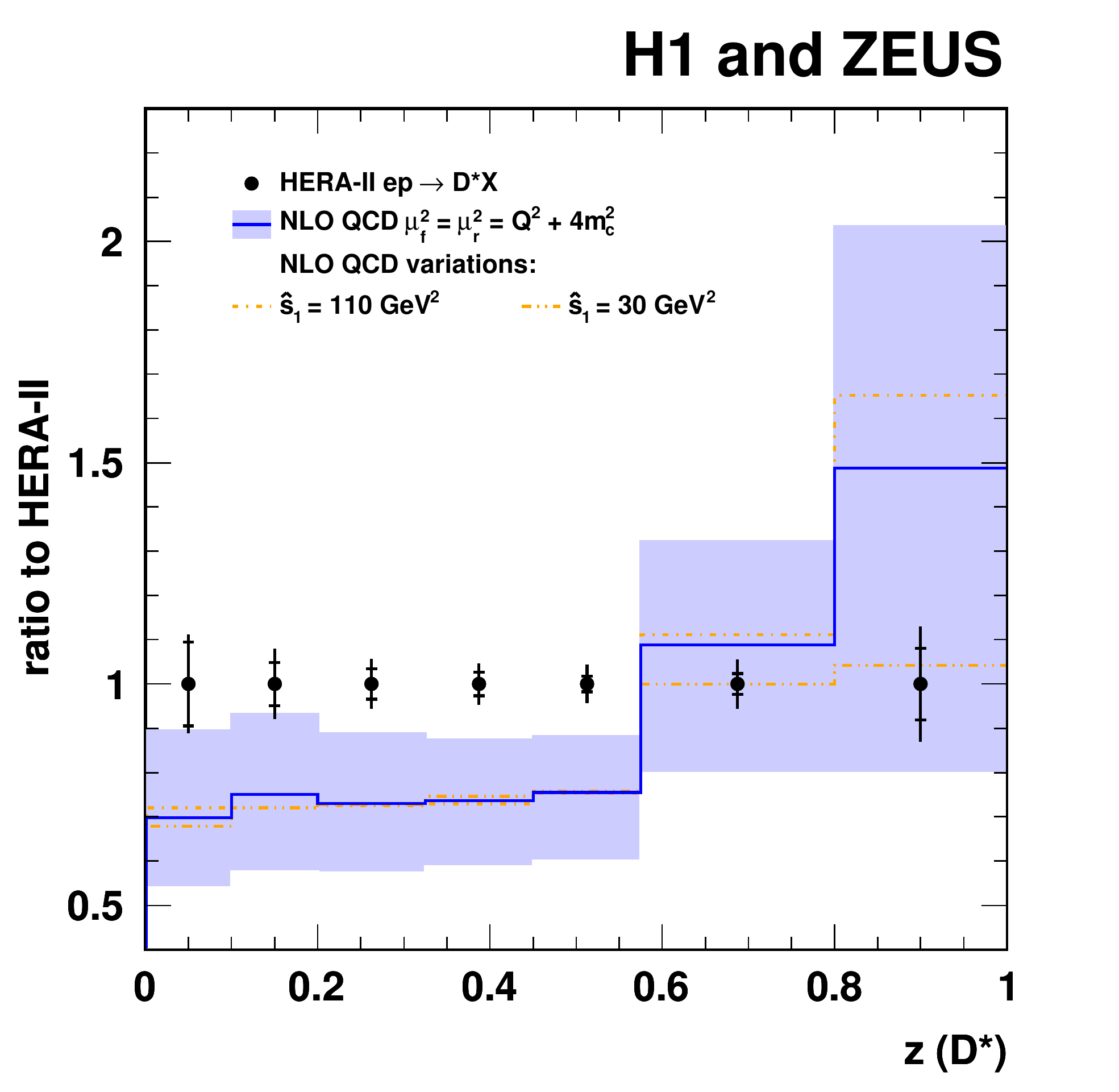}
  \put(-30,177){(d)}\\
  \end{minipage}
	\caption[Single-differential \Dstar cross sections compared to NLO variations]
	{Single-differential \Dstar cross section as a function of $p_T(\Dstar)$ (a), $y$ (b) and $z(\Dstar)$ (c, d) compared to NLO predictions with different variations:
	$c$-quark mass $m_{c}$, renormalisation scale $\mu_{r}$, 
	factorisation scale $\mu_{f}$ and fragmentation bin boundary $\hat s_1$.}
	\label{fig:comb:dstar:single:thvars}
\end{figure*}

In order to study the impact of the current theory uncertainties in more 
detail, the effect of each \ozmod{theoretical uncertainty} on the predictions was studied. 
The most conclusive variations on the predictions are shown separately 
in Fig.~\ref{fig:comb:dstar:single:thvars}, compared to the same data as in Fig.~\ref{fig:comb:dstar:single:withth}. 
Plots with all the variations are available in Appendix~\ref{sec:app:dstar} (Figs.~\ref{fig:comb:dstar:single:thvarpt} to~\ref{fig:comb:dstar:single:thvary}).

The NLO prediction as a function of $p_T(\Dstar)$ (Fig.~\ref{fig:comb:dstar:single:thvars}a) 
describes the data better if either
\begin{itemize}
\item the $c$-quark pole mass is reduced to 1.35 GeV; or
\item the renormalisation scale is reduced by a factor 2; or
\item the factorisation scale is increased by a factor 2.
\end{itemize}
Simultaneous variation of both scales will largely compensate and 
will therefore \ozmodN{result in} a much smaller effect.  

The prediction for the $z(\Dstar)$ distribution (Fig.~\ref{fig:comb:dstar:single:thvars}d) describes 
the shape of the data noticeably better if the fragmentation parameters 
are adjusted such that the bin boundary $\hat s_1$ between the two lowest 
fragmentation 
bins \cite{heracharmcomb} \ozmodN{is set to 30~GeV$^2$ (see Table~\ref{tab:comb:proc:pscorr:fragkart})}. 
This also slightly improves the shape of the $y$ distribution (Fig.~\ref{fig:comb:dstar:single:thvars}b).

The preference for a reduced renormalisation scale already observed for $p_T(\Dstar)$
is confirmed by the $z(\Dstar)$ distribution (Fig.~\ref{fig:comb:dstar:single:thvars}c).
However, the shape of the $z(\Dstar)$ distribution rather \ozmodNN{favours} variations of 
the charm mass and the factorisation scale in the opposite direction to those 
found for the $p_T(\Dstar)$ distribution.
The other kinematic variables do not contribute any additional information 
to these findings.

As stated before, within the large uncertainties indicated by the theory 
bands in Fig.~\ref{fig:comb:dstar:single:withth}, all distributions are \ozmodNN{reasonably well} 
described. However, the above study shows that the different contributions to 
these uncertainties do not only affect the normalisation but also change 
the shape of different distributions in \ozmodNN{various} ways. It is therefore 
\ozmod{not \textit{a priori} expected} that a variant of the prediction which gives a good description 
in one variable will also give a good description in another. 

Based on the above study, a `customised' calculation was hence performed with 
the goal to demonstrate that it is possible to obtain an acceptable description of 
the data in all variables at the same time, for both shape and normalisation, 
within the theoretical uncertainties quoted in Section \ref{sec:comb:th}.%
\footnote{Since several of the theory parameters (e.g.\ the renormalisation and factorisation scales) 
are not physical parameters, and hence their ``uncertainties'' have no physical relevance, 
a \ozmod{demonstration} rather than a detailed fit will suffice to clarify this point. 
Another reason not to perform a detailed fit is that the data are statistically correlated between the different distributions, 
therefore all the distributions must not be fitted simultaneously.}
For this calculation
\begin{itemize}
\item the renormalisation scale was reduced by a factor 2, with the 
factorisation scale unchanged;
\item the change of the fragmentation parameter $\hat s_1 = 30$~GeV$^2$ was applied;
\item at this stage, the resulting distributions were still found to underestimate the data normalisation. 
As the renormalisation and factorisation scales are recommended to differ by at
most a factor of two \cite{Baines:2006uw}, the only
significant remaining handle is the $c$-quark pole mass. This mass was set 
to $\SI{1.4}{GeV}$, a value which was also found to be compatible with the 
partially overlapping data used for a 
previous dedicated study \cite{heracharmcomb} of the $c$-quark mass;
\item all other parameters, which were found to \ozmodN{result in} a much smaller effect 
than those treated above, were left at their central settings as described in 
Section \ref{sec:comb:th}. 
\end{itemize} 

The result of this customised calculation is indicated as a dotted line in 
Fig.~\ref{fig:comb:dstar:single:withth}. Indeed a reasonable agreement with 
the data is achieved in all variables at the same time. 
This \textit{a posteriori} adjustment of theory parameters may be taken as a hint in which direction theoretical and 
phenomenological developments might need to \ozmodNN{move}: 
\begin{itemize}
\item
the strong improvement of the description of the data relative to the central 
prediction through the customisation of the renormalisation scale indicates 
that NNLO calculations, which might reduce the scale-related 
uncertainties to a level which matches the data precision, \ozmodN{are needed} to obtain
a more stringent statement concerning the agreement of the pQCD predictions with the data;
\item
the improvement from the customisation of one of the fragmentation parameters
and the still not fully satisfactory description of the $z(\Dstar)$ distribution 
indicate that further dedicated experimental and theoretical studies of the 
fragmentation 
treatment \ozmod{will} be helpful. 
\end{itemize}

In general, the precise single-differential distributions resulting from the 
combination, in particular those as a function of $p_T(\Dstar)$, $\eta(\Dstar)$ and $z(\Dstar)$, 
are sensitive to theoretical and phenomenological parameters in a way which 
complements the sensitivity of more inclusive variables like $Q^2$ and $y$. 

\subsubsection{Combination of double-differential cross section}
\label{sec:comb:dstar:double}

\ozmodN{This Section continues the \Dstar cross-section combination and presents a combination of double-differential cross section as a function of $Q^2$ and $y$.}

\paragraph*{Input measurements, phase-space region and combination details\\}
\label{sec:comb:dstar:double:det}

Since for the combination of the double-differential cross section as a function of $Q^2$ and $y$ 
the restriction to the same phase-space region in $Q^2$ does not apply, HERA-I \Dstar measurements can be included in the combination. 

Table~\ref{tab:comb:dstar:double:input} presents the datasets considered for the combination of the visible \Dstar double-differential cross section.%
\footnote{Same as in Table~\ref{tab:comb:dstar:single:input}, from the two sets of measurements in \cite{h1dstar_hera2}, the one 
compatible with the quoted cuts on $p_T(\Dstar)$ and $\eta(\Dstar)$, which are compatible with 
the phase space of the ZEUS measurement~\cite{zeusdstar_hera2}, was chosen and referred to as dataset I.}
Compared to Table~\ref{tab:comb:dstar:single:input}, Table~\ref{tab:comb:dstar:double:input} \ozmodNN{contains also} three most precise HERA-I measurements; 
it also has an additional column which reports the centre-of-mass energy, since the latter differs for one of the HERA-I measurements. 

\begin{table*}[tbp]
\caption[Datasets considered for combination of double-differential \Dstar cross section]
{Datasets considered for the combination of the visible \Dstar double-differential cross section. For each dataset the respective kinematic region, 
the integrated luminosity, ${\cal L}$, and the centre-of-mass energy, $\sqrt{s}$, are given.}
\label{tab:comb:dstar:double:input}
\begin{center}
\tabcolsep1.5mm
\renewcommand*{\arraystretch}{1.2}
\begin{tabularx}{\textwidth}{|X|l|l|l|l|l|l|}
\hline
\multirow{3}{*}{Dataset} & \multicolumn{4}{c|}{Kinematic range}      & ${\cal L}$ & ${\sqrt{s}}$  \\ \cline{2-5}
                          & $Q^2$        & $y$ & $p_T(\Dstar)$ & $\eta(\Dstar)$ & & \\ 
                          & [$\SI{}{GeV}^{2}$] &     & [$\SI{}{GeV}$]     &             & [$\text{pb}^{\text{-1}}$] & [$\SI{}{GeV}$]            \\ \hline
I:\ \ \  H1 $\Dstar$ HERA-II (med. $Q^2$) \cite{h1dstar_hera2} & $~~~~5:100$ & $0.02:0.70$ & $1.5:\infty$ & $-1.5:1.5$ & $348$ & $318$ \\ \hline
II:\ \  H1 $\Dstar$ HERA-II (high $Q^2$) \cite{h1dstarhighQ2} & $100:1000$ & $0.02:0.70$ & $1.5:\infty$ & $-1.5:1.5$ & $351$ & $318$ \\ \hline
III: ZEUS $\Dstar$ HERA-II \cite{zeusdstar_hera2} & $~~~~5:1000$ & $0.02:0.70$ & $1.5:20.0$ & $-1.5:1.5$ & $363$ & $318$ \\ \hline
IV:\ ZEUS $\Dstar$ HERA-I 98--00 \cite{zd00} & $~1.5:1000$ & $0.02:0.70$ & $1.5:15.0$ & $-1.5:1.5$ & $82$ & $318$ \\ \hline
\hline
V:\ ZEUS $\Dstar$ HERA-I 96--97 \cite{zd9697} & $~~~~1:600$ & $0.02:0.70$ & $1.5:15.0$ & $-1.5:1.5$ & $37$ & $300$ \\ \hline
VI:\ H1 $\Dstar$ HERA-I \cite{h1dstar_hera1} & $~~~~2:100$ & $0.05:0.70$ & $1.5:15.0$ & $-1.5:1.5$ & $47$ & $318$ \\ \hline
\end{tabularx}
\end{center}
\end{table*}

Inclusion of HERA-I measurements in the combination allows an extension of the kinematic range down to lower $Q^2$. 
Although all three HERA-I measurements have different lower $Q^2$ boundaries, 
a reasonable compromise between them was to choose the lower $Q^2$ equal to $Q^2=\SI{1.5}{GeV}^2$. 
Thus the overall phase-space region for the combined \Dstar cross sections is given by
\begin{equation}
\label{eq:comb:dstar:double:phasespace}
	\begin{split}
		&1.5< Q^2 < \SI{1000}{GeV}^2,\\
		&0.02<y<0.7,\\
		&p_T(\Dstar)>\SI{1.5}{GeV},\\
		&|\eta(\Dstar)|<1.5,\\
		&\sqrt{s}=\SI{318}{GeV}.
	\end{split}
\end{equation}

Some of the HERA-I measurements from Table~\ref{tab:comb:dstar:double:input} have a slightly different phase-space region and are performed at a different centre-of-mass energy. 
Moreover, their binning schemes for the double-differential cross section significantly differ from that which has been used for datasets I--III; 
dataset VI reports the double-differential cross section not as a function of $Q^2$ and $y$ but as a function of $Q^2$ and $x$. 
Thus inclusion of HERA-I measurements in the combination necessarily required applying swimming corrections. 

\begin{figure}[htbp]
  \centering
  \includegraphics[width=1.0\figwidth,trim=0mm 0mm 9mm 0mm,clip=true]{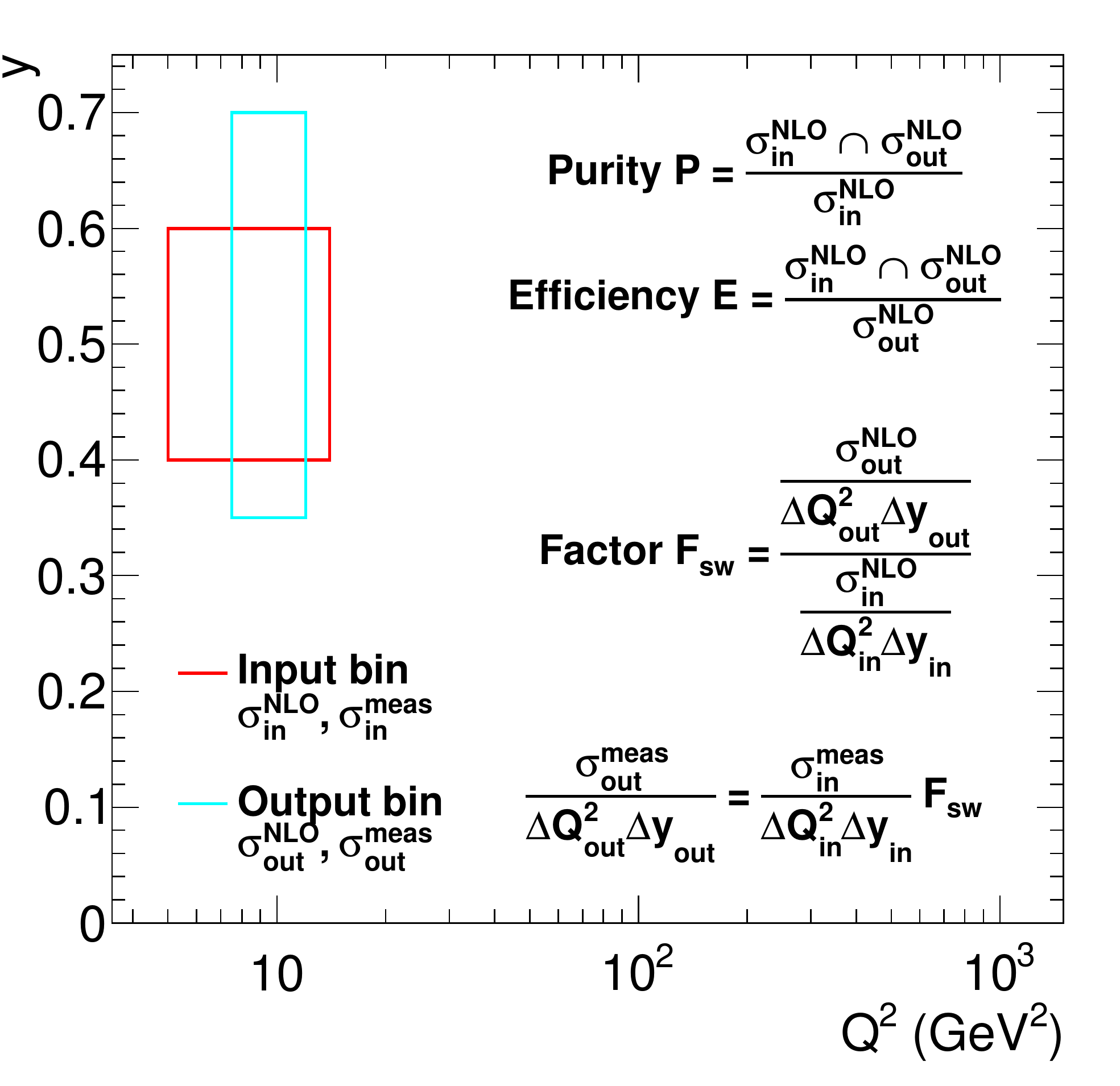}
  \caption[Definition of purity, efficiency and swimming factors]
  {Definition of purity, efficiency and swimming factors.}
	\label{fig:comb:dstar:double:defpe}
\end{figure}

A dedicated study was performed to select those HERA-I measurements which are reasonably compatible with the HERA-II ones. 
At first a common binning scheme had to be chosen. Since the HERA-II measurements still remain the most precise in the combination, 
the double-differential cross section as a function of $Q^2$ and $y$ was selected with the binning scheme which is based on datasets I--III 
(although slightly revised to improve consistency with the HERA-I measurements). 
It was extended at low $Q^2$ with the binning scheme based on the most precise HERA-I dataset IV. 
The new binning will be given together with the combined \Dstar cross sections in Table~\ref{tab:comb:dstar:double:combined}. 
\Dstar cross sections in the new bins (also referred to as \emph{destination}, or \emph{output}, bins) were obtained from the original bins (referred to also as \emph{input} bins) using the swimming procedure described in Section~\ref{sec:comb:proc:pscorr}. 
For each swum bin the following quantities were calculated:
\begin{itemize}
	\item the fraction of the cross section of the original bin contained in the new one, \emph{efficiency}, $E$;
	\item the fraction of the cross section of the new bin contained in the original one, \emph{purity}, $P$;
	\item the ratio of the swimming uncertainty to the experimental uncorrelated uncertainty in the corresponding bin, $R$.
\end{itemize}
Definition of purity, efficiency, and swimming factor, $F_{\rm sw}$, which were used to translate the differential cross section 
from the original bin to the destination one, is illustrated in Fig.~\ref{fig:comb:dstar:double:defpe}. 
Note that in some cases for a given original bin there can be several candidates for destination bins; 
in this case a destination bin with maximum $P$, $E$, and minimum $R$ was chosen. 
Sometimes it was \ozmodNN{advantageous} to combine two input bins before swimming.

\begin{figure*}[tbp]
  \centering
  \includegraphics[width=2.0\figwidth,trim=1mm 2mm 20mm 4mm,clip=true]{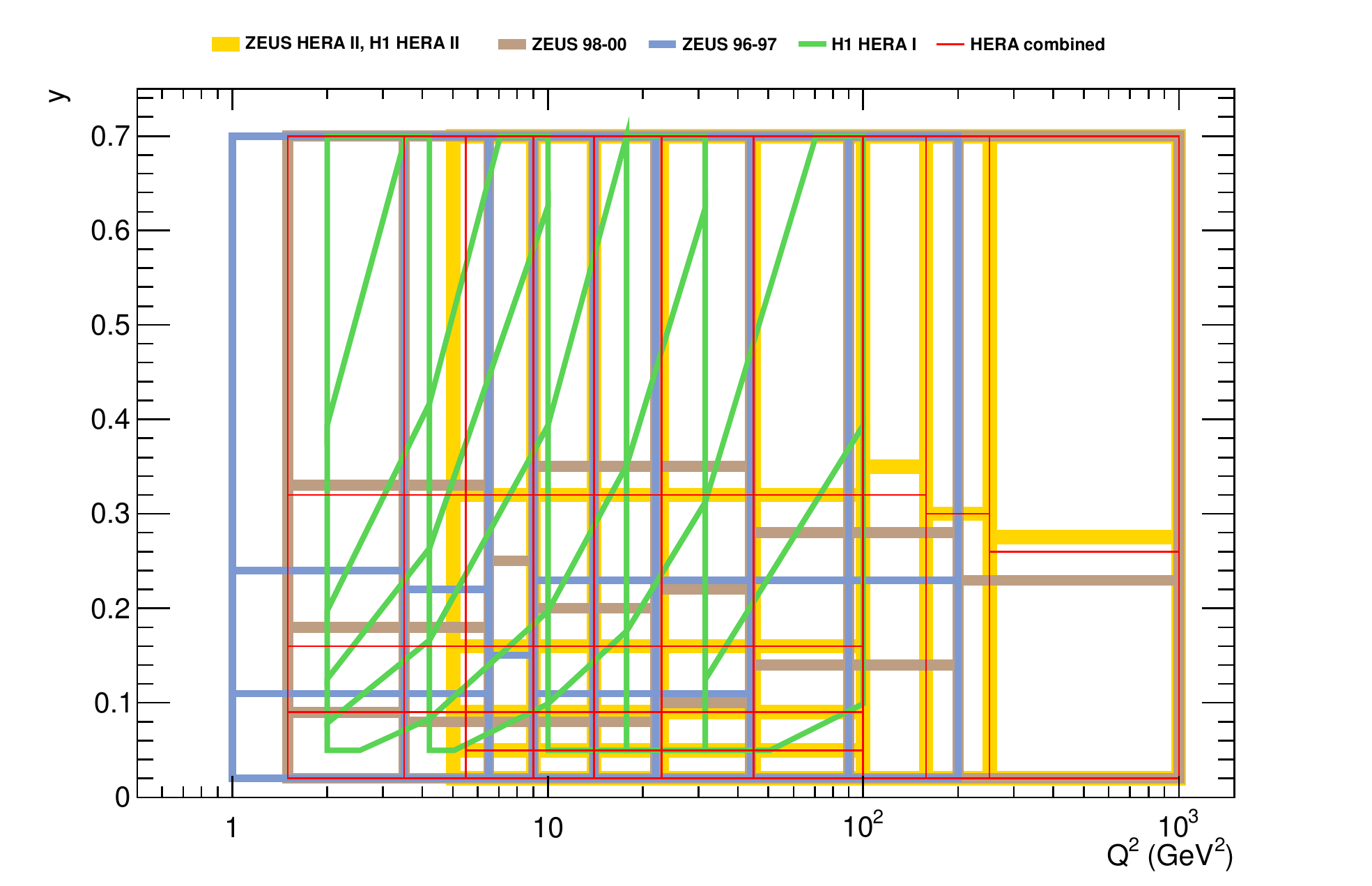}
  \caption[Overlap of binning schemes for \Dstar double-differential cross sections]
  {Overlap of the binning schemes for the \Dstar double-differential cross section from input measurements and the new binning scheme.}
	\label{fig:comb:dstar:double:grid}
\end{figure*}

\begin{figure*}[htbp]
  \sidecaption
  \centering
  \includegraphics[width=1.016\figwidth*\real{0.75},trim=0mm 2mm 7mm 0mm,clip=true]{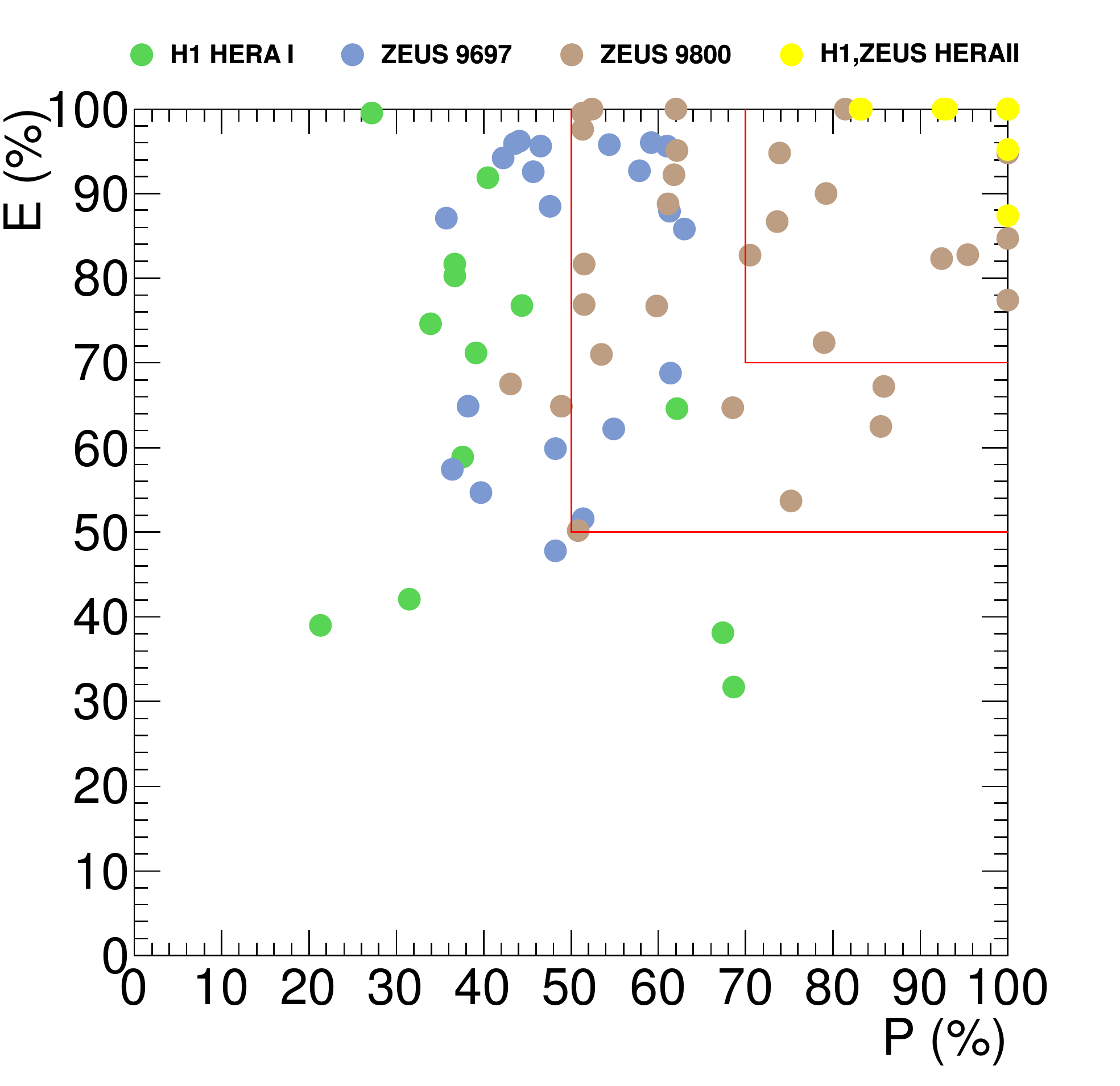}
  \includegraphics[width=0.984\figwidth*\real{0.75},trim=0mm 0mm 1mm 0mm,clip=true]{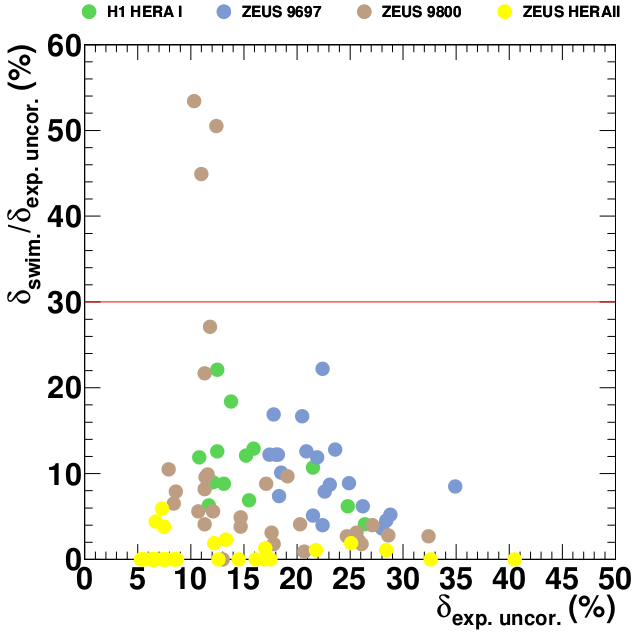}
  \caption[Purity vs.\ efficiency and $R$ for double-differential \Dstar cross sections]
  {Purity vs.\ efficiency (left) and the ratio of the swimming uncertainty to the experimental uncorrelated uncertainty vs.\ the latter (right) for combination of the double-differential \Dstar cross section. 
  Contributions from the individual input datasets are shown separately. The solid lines show cuts $P,E>50\%$, $P,E>70\%$ and $R<30\%$.}
	\label{fig:comb:dstar:double:inputpe}
\end{figure*}

The overlap of the binning schemes for the double-differential cross section from all input measurements and the new binning scheme is shown in Fig.~\ref{fig:comb:dstar:double:grid}. 
Fig.~\ref{fig:comb:dstar:double:inputpe} shows $P$ vs.\ $E$, and $R$ vs.\ its denominator, the experimental uncorrelated uncertainty, for all considered datasets. 
Since the binning scheme was chosen to be based on the HERA-II measurements, for all bins from datasets I--III purity, efficiency and the ratio satisfy $P,E>80\%$ and $R<10\%$ 
(note that in most bins $P=E=100\%$ and $R=0\%$, since the original and destination bins exactly coincide). 

Since the purpose of the combination is to provide the visible \Dstar cross sections, $P$ and $E$ should not be too low and $R$ not too large. 
Thus it is natural to introduce cuts on these quantities. 
Several possible values for the cuts on $P$ and $E$ are presented in Table~\ref{tab:comb:dstar:double:pecuts} 
together with the numbers of input bins which survive the cuts. A reasonable cut was chosen to be $P,E>50\%$. 
Then most of the input bins from dataset IV (29 of 31) survive this selection, although it eliminates most of the bins from datasets V and VI.%
\footnote{For dataset V the input bins are too large, while for dataset VI the main difficulty is the original differential cross section 
as a function of $Q^2$ and $x$.}
Therefore a decision was taken to include in the combination from the HERA-I measurements only dataset IV. 
In addition the cut $R<30\%$ was introduced. This eliminated 3 other input bins from dataset IV, so finally 26 of 31 original bins were kept. 
The data points removed from dataset IV mainly correspond to the low-$y$ region where larger bins were used for the HERA-I data; 
additionally they suffer more from the swimming uncertainties, since the NLO QCD predictions at low $y$ have a large mass dependence. 
All input bins from datasets I--III survived the above cuts on $P$, $E$ and $R$ and were kept.
The swimming procedure includes the contribution to dataset IV from the range $p_T(\Dstar)>\SI{15}{GeV}$. 
Similar to the case of the single-differential cross-section combination (Section~\ref{sec:comb:dstar:single:det}), 
the sensitivity of the shape to the beauty contribution was found to be negligible and thus was ignored.%

\begin{table}[tbp]
\caption[Possible values for cuts on $P$ and $E$]
{Possible values for the cuts on $P$ and $E$ together with numbers of input bins from different measurements which survive these cuts.}
\label{tab:comb:dstar:double:pecuts}
\footnotesize
\begin{center}
\tabcolsep0.15mm
\renewcommand*{\arraystretch}{1.2}
        \begin{tabular}{|c|c|c|c|c|} \hline
        min($P$,$E$)&    H1 HERA-I&   ZEUS 96--97&   ZEUS 98--00   & H1,ZEUS HERA-II\\
        (\%)&    &&&\\\hline
                   0&          17&          21&          31&  31\\\hline
                  30&          15&          21&          31&  31\\\hline
                  40&           5&          17&          31&  31\\\hline
                  50&           2&           9&          29&  31\\\hline
                  60&           2&           4&          20&  31\\\hline
                  70&           0&           0&          12&  31\\\hline
                  80&           0&           0&           6&  31\\\hline
        \end{tabular}
\end{center}
\end{table}
        
The branching ratios for datasets I, II and IV were updated to the PDG value~\cite{pdg2012}.
A full list of considered correlated sources is provided in Appendix~\ref{sec:app:dstar} (Table~\ref{tab:comb:dstar:syst}). 
Similar to the case of the single-differential \Dstar cross sections, 
all systematic uncertainties were treated as uncorrelated between H1 and ZEUS measurements, except for the branching-ratio uncertainty; although 
since the latter is fully correlated between all datasets, it is not changed in the combination and was not included in the combination 
but applied as an external uncertainty on the results.

\paragraph*{Combined \Dstar cross sections\\}
\label{sec:comb:dstar:double:res}

The combined double-differential cross section with the uncorrelated, correlated
and total uncertainties as a function of $Q^2$ and $y$ is given in Table~\ref{tab:comb:dstar:double:combined}. 
The total uncertainties were obtained by adding the uncorrelated and correlated uncertainties in quadrature. 
A detailed breakdown of the correlated uncertainties are provided in Appendix~\ref{sec:app:dstar} (Table~\ref{tab:comb:dstar:double:combinedfull3}).

\begin{table}[tbp]
\caption[Combined double-differential \Dstar cross section]
{The combined double-differential \Dstar cross section as a function of $Q^2$ and $y$, with its 
uncorrelated ($\delta_{unc}$), correlated ($\delta_{cor}$) and total ($\delta_{tot}$) uncertainties.
The cross sections are given in the kinematic region~\ref{eq:comb:dstar:double:phasespace}. 
}
\label{tab:comb:dstar:double:combined}
\begin{center}
\tabcolsep2.22mm
\renewcommand*{\arraystretch}{1.325}
\begin{tabularx}{\columnwidth}[t]{|X|X|c|c|c|c|}
\hline
$Q^2$ & $y$ & $ \frac{d^2\sigma}{dQ^2dy}$ &$\delta_{unc}$  &$\delta_{cor}$&$\delta_{tot}$ \\
  ($\SI{}{GeV}^{2}$)  &     & ($\SI{}{nb/GeV}^{2}$)                         &$(\%)$            &$(\%)$ &$(\%)$ \\ 
\hline
1.5 : 3.5 & 0.02 : 0.09 & 4.76 & 12.9 & 2.5 & 13.2 \\
        & 0.09 : 0.16 & 5.50 & 11.3 & 2.6 & 11.5 \\
        & 0.16 : 0.32 & 3.00 & 12.0 & 2.6 & 12.3 \\
        & 0.32 : 0.70 & 9.21 $\times 10^{-1}$ & 20.5 & 2.5 & 20.7 \\ \hline
3.5 : 5.5 & 0.02 : 0.09 & 2.22 & 11.3 & 2.8 & 11.6 \\
        & 0.09 : 0.16 & 1.98 & 7.9 & 2.7 & 8.3 \\
        & 0.16 : 0.32 & 1.09 & 20.2 & 2.7 & 20.4 \\
        & 0.32 : 0.70 & 3.47 $\times 10^{-1}$ & 14.6 & 2.6 & 14.8 \\ \hline
5.5 : 9 & 0.02 : 0.05 & 1.06 & 12.3 & 4.4 & 13.1 \\
        & 0.05 : 0.09 & 1.46 & 7.8 & 4.1 & 8.8 \\
        & 0.09 : 0.16 & 1.32 & 5.4 & 4.3 & 6.9 \\
        & 0.16 : 0.32 & 7.73 $\times 10^{-1}$ & 4.9 & 3.9 & 6.3 \\
        & 0.32 : 0.70 & 2.51 $\times 10^{-1}$ & 5.6 & 4.2 & 7.0 \\ \hline
9 : 14 & 0.02 : 0.05 & 5.20 $\times 10^{-1}$ & 13.0 & 6.6 & 14.6 \\
        & 0.05 : 0.09 & 7.68 $\times 10^{-1}$ & 6.6 & 3.9 & 7.7 \\
        & 0.09 : 0.16 & 5.69 $\times 10^{-1}$ & 4.6 & 2.8 & 5.4 \\
        & 0.16 : 0.32 & 4.12 $\times 10^{-1}$ & 4.6 & 3.1 & 5.6 \\
        & 0.32 : 0.70 & 1.51 $\times 10^{-1}$ & 5.6 & 4.0 & 6.9 \\ \hline
14 : 23 & 0.02 : 0.05 & 2.29 $\times 10^{-1}$ & 11.4 & 6.3 & 13.0 \\
        & 0.05 : 0.09 & 3.78 $\times 10^{-1}$ & 6.5 & 4.1 & 7.7 \\
        & 0.09 : 0.16 & 2.90 $\times 10^{-1}$ & 4.8 & 3.3 & 5.8 \\
        & 0.16 : 0.32 & 1.86 $\times 10^{-1}$ & 5.0 & 3.4 & 6.0 \\
        & 0.32 : 0.70 & 6.92 $\times 10^{-2}$ & 6.2 & 4.4 & 7.7 \\ \hline
23 : 45 & 0.02 : 0.05 & 6.91 $\times 10^{-2}$ & 14.8 & 8.2 & 16.7 \\
        & 0.05 : 0.09 & 1.23 $\times 10^{-1}$ & 5.9 & 3.6 & 6.9 \\
        & 0.09 : 0.16 & 1.14 $\times 10^{-1}$ & 4.4 & 3.0 & 5.3 \\
        & 0.16 : 0.32 & 7.42 $\times 10^{-2}$ & 4.3 & 3.0 & 5.2 \\
        & 0.32 : 0.70 & 3.21 $\times 10^{-2}$ & 5.2 & 3.7 & 6.4 \\ \hline
45 : 100 & 0.02 : 0.05 & 6.16 $\times 10^{-3}$ & 33.5 & 11.1 & 35.3 \\
        & 0.05 : 0.09 & 2.70 $\times 10^{-2}$ & 11.0 & 4.4 & 11.8 \\
        & 0.09 : 0.16 & 2.05 $\times 10^{-2}$ & 8.0 & 3.7 & 8.8 \\
        & 0.16 : 0.32 & 1.99 $\times 10^{-2}$ & 5.4 & 3.2 & 6.3 \\
        & 0.32 : 0.70 & 7.84 $\times 10^{-3}$ & 6.9 & 4.0 & 7.9 \\ \hline
100 : 158 & 0.02 : 0.32 & 4.12 $\times 10^{-3}$ & 8.2 & 4.1 & 9.2 \\
        & 0.32 : 0.70 & 2.18 $\times 10^{-3}$ & 11.1 & 4.1 & 11.9 \\ \hline
158 : 251 & 0.02 : 0.30 & 1.79 $\times 10^{-3}$ & 10.2 & 4.4 & 11.1 \\
        & 0.30 : 0.70 & 9.28 $\times 10^{-4}$ & 11.6 & 4.6 & 12.5 \\ \hline
251 : 1000 & 0.02 : 0.26 & 1.31 $\times 10^{-4}$ & 14.5 & 4.7 & 15.3 \\
        & 0.26 : 0.70 & 1.18 $\times 10^{-4}$ & 12.7 & 5.0 & 13.6 \\ \hline
\end{tabularx}
\end{center}
\end{table}

\begin{figure*}[tbp]
  \sidecaption
  \centering
  \includegraphics[width=1.65\figwidth,trim=7mm 8mm 13mm 16mm,clip=true]{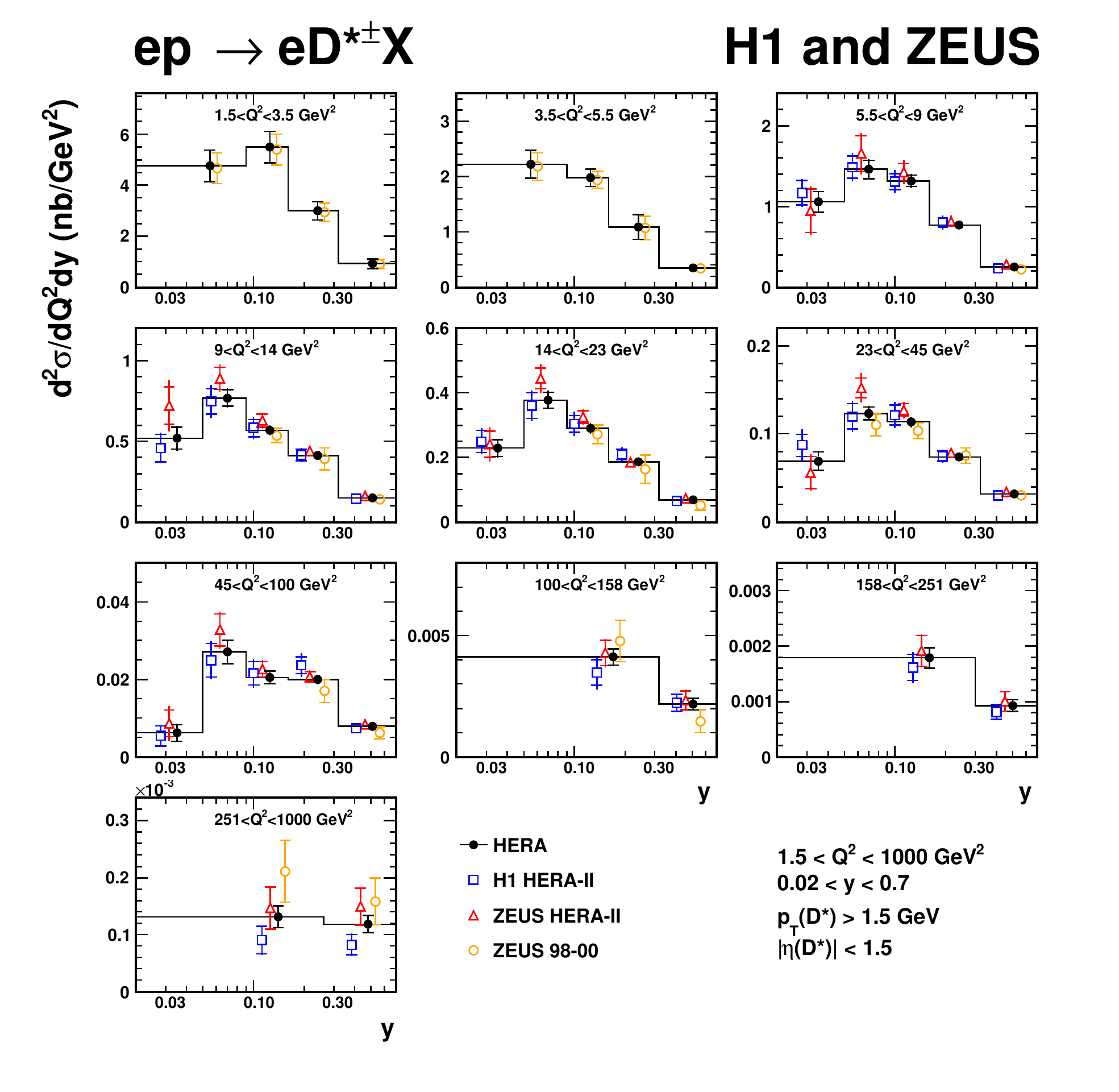}
  \caption[Double-differential \Dstar cross sections]
  {Double-differential $\Dstar$ cross sections as a function of $Q^2$ and $y$. 
	The triangles, open squares and open circles are the \Dstar cross sections before combination, shown with a small horizontal offset for better visibility. 
	The filled points are the combined \Dstar cross sections. The inner error bars indicate the uncorrelated part of the uncertainties. 
	The outer error bars represent the total uncertainties. The histogram indicates the binning used to calculate the \Dstar cross sections.}
	\label{fig:comb:dstar:double:combined}
\end{figure*}

The individual datasets as well as the results of the combination are shown
in Fig.~\ref{fig:comb:dstar:double:combined}. 
The combined \Dstar cross sections exhibit significantly reduced uncertainties. 
The input HERA-II H1 and ZEUS datasets are similar in precision. 
The precision of the ZEUS HERA-I data is smaller; however this sample also provides valuable input in some bins. 
In the first two $Q^2$ bins, the combination is based on the HERA-I data only; note that the uncertainty on the combined data in these bins is a bit reduced 
comparing to the original one because of reduction of the correlated systematic uncertainties. 

\begin{figure}[htbp]
  \centering
  \includegraphics[width=1.0\figwidth,trim=0mm 0 5mm 0mm,clip=true]{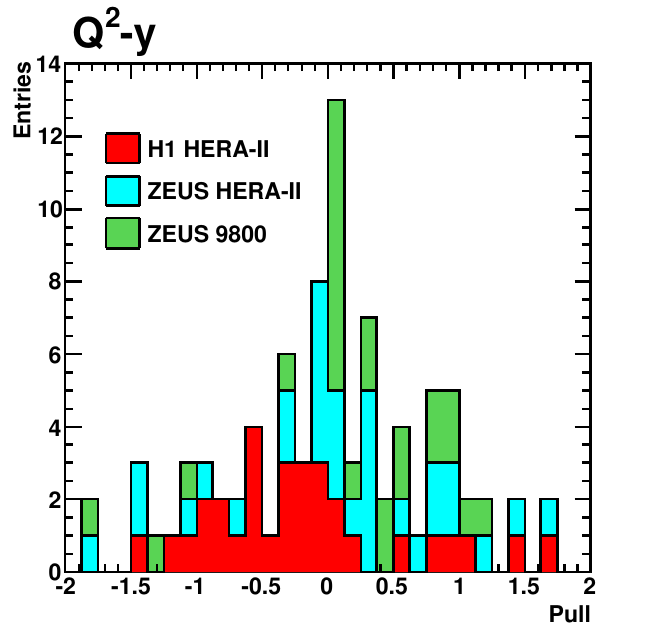}
  \caption[Pull distributions for double-differential \Dstar cross sections]
  {The pull distribution for the combination of the double-differential \Dstar cross sections as a function of $Q^2$ and $y$. 
  Contributions from the individual input datasets are shown separately.}
	\label{fig:comb:dstar:double:pull}
\end{figure}

The combination has $\chisqndof=38/48$; the corresponding probability is $85\%$, indicating consistency of the input measurements and, possibly, some overestimation 
of the experimental systematic uncertainties.
The pull distribution is shown in Fig.~\ref{fig:comb:dstar:double:pull}. It is \ozmodNN{close} to a unit Gaussian distribution. 
As was seen in the results of the single-differential cross section combination (Section~\ref{sec:comb:dstar:single:res}), 
although Fig.~\ref{fig:comb:dstar:double:combined} indicates that the H1 data points lie on average below the ZEUS points, the pull distribution 
in Fig.~\ref{fig:comb:dstar:double:pull} shows an overall symmetric spread 
of all input data around the combined results. 
The shifts and reductions of the correlated sources are provided in Appendix~\ref{sec:app:dstar} (Table~\ref{tab:comb:dstar:syst}).

\begin{figure*}[tbp]
  \sidecaption
  \centering
  \includegraphics[width=1.65\figwidth,trim=7mm 8mm 13mm 16mm,clip=true]{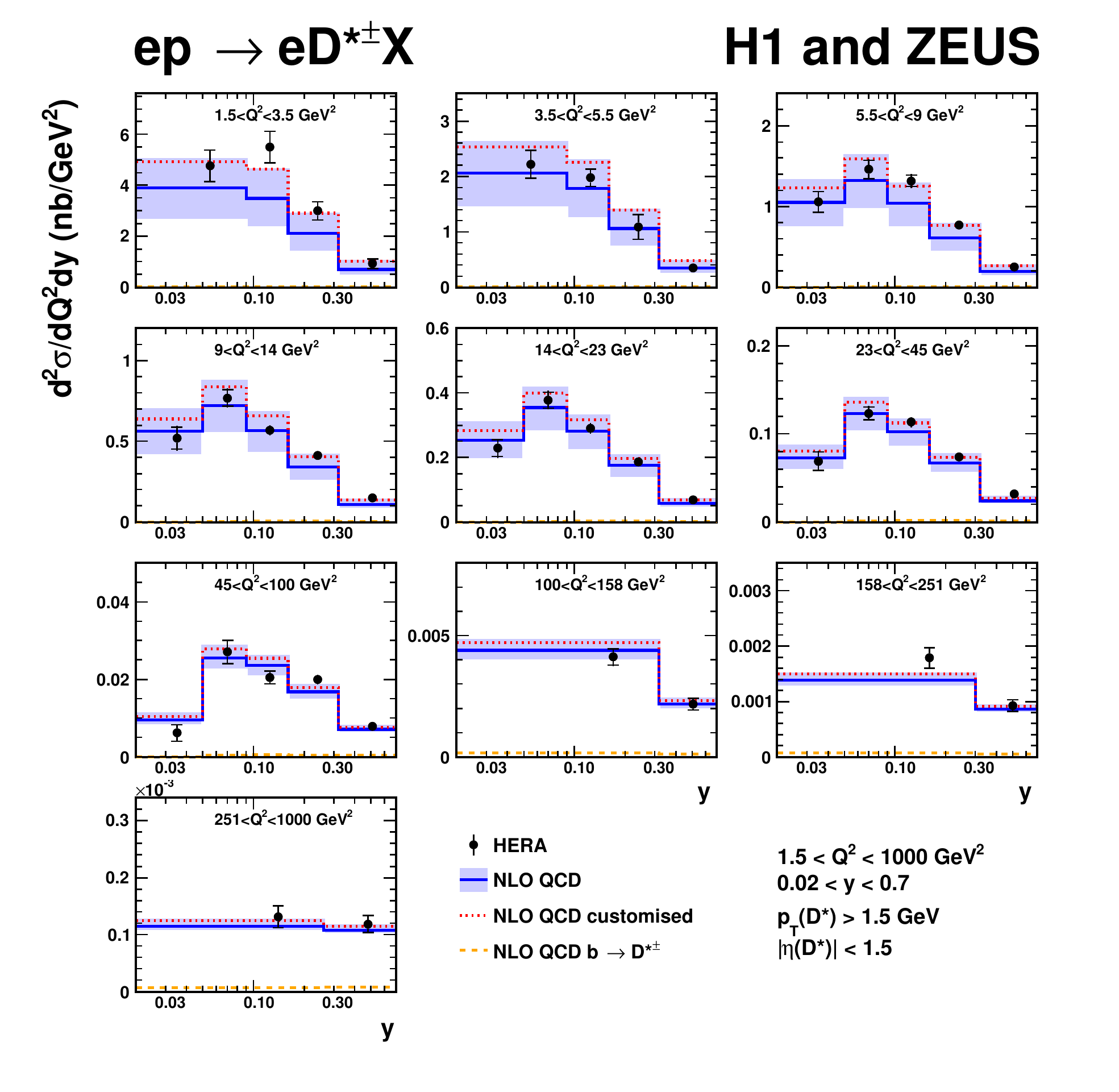}
  \caption[Double-differential \Dstar cross sections compared to NLO predictions]
	{Double-differential $\Dstar$ cross section as a function of $Q^2$ and $y$. 
	The data points are the combined \Dstar cross sections. The inner 	error bars indicate the uncorrelated part of the uncertainties. 
	The outer error bars represent the total uncertainties. 
	Also shown are the NLO predictions from HVQDIS (including the beauty contribution), and their uncertainty band. 
	A customised NLO calculation (dotted line) is also shown.}
	\label{fig:comb:dstar:double:withth}
\end{figure*}

The combined cross section is compared to the NLO QCD predictions in the FFNS (described in Section~\ref{sec:comb:th}) in Fig.~\ref{fig:comb:dstar:double:withth}. 
The customised calculation (see Section~\ref{sec:comb:dstar:single:customth}) is also shown. 
In general the predictions describe the data well. 
As seen before from the single-differential $y$ cross section, 
the central theory prediction shows a somewhat softer $y$ distribution than 
the data, in particular at low $Q^2$. 
The data reach a precision of about 5--10\% over a large fraction of the measured phase space,
while the typical theory uncertainty ranges from 30\% at low $Q^2$ to 10\% at high $Q^2$, 
so higher-order calculations would be very helpful to match the data precision. 
As well as the single-differential distributions, \ozmodNN{the} double-differential distribution gives \ozmodNN{additional} input to test
further theory improvements.

\subsection{Combination of reduced charm cross sections}
\label{sec:comb:red}

This Section describes the combination of HERA charm-production measurements in DIS at the level of the reduced cross 
sections (see Eq.~\ref{eq:th:hqcsred}). 
\ozmodNN{The information on charm production in DIS can be collected in a single dataset in the full phase space,}
integrated over $p_T$ and $\eta$ of the $c$ quark, 
because most theoretical predictions exist only for this inclusive quantity. 
Since all methods used to measure charm production at HERA, introduced in Section~\ref{sec:exp:hera:hfmeas}, have limited phase-space coverage 
and thus \ozmod{must} be corrected to the full phase space using theory, there are no reasons to restrict this combination to a specific phase-space region and binning scheme, 
and all datasets 
from H1~\cite{h1dstar_hera1,h1ltt_hera2,h1dstarhighQ2,h1dstar_hera2} and ZEUS~\cite{zd9697,zd00,zd0dp,zeus_muon,zeusdch_hera2,zeusdstar_hera2,zeussecvtx_hera2} 
were included for which the necessary information on systematic uncertainties needed for the combination is available 
and which have not been superseded \ozmod{by later measurements} \ozmodN{up to November 2016}. 

The results reported in this Section \ozmodNN{extend} the previous combination of H1 and ZEUS charm measurements~\cite{heracharmcomb} 
by \ozmodN{including} three recent ZEUS datasets~\cite{zeusdch_hera2,zeusdstar_hera2,zeussecvtx_hera2} which appeared after the combination~\cite{heracharmcomb} 
\ozmodNN{(referred to as `HERA 2012')} has been performed. 
The \ozmodN{combination procedure was kept} as close as possible to~\cite{heracharmcomb} to allow for a consistent comparison 
with the published results.

Section~\ref{sec:comb:red:det} introduces the combination details: the input datasets and treatment of their experimental uncertainties, 
definition of the reduced cross sections and the combination $Q^2\text{--}x$ grid 
and extraction of the reduced cross sections from the visible ones, 
needed to put the input measurements into the common grid. 
In Section~\ref{sec:comb:red:res} the results of the combination are presented and compared to the results from~\cite{heracharmcomb}. 
In Section~\ref{sec:comb:red:ffns} the combined data are compared to the theoretical predictions in the FFNS and the running and pole charm masses are extracted from the data.

\subsubsection{Combination details}
\label{sec:comb:red:det}

\paragraph*{Input data samples\\}
\label{sec:comb:red:det:data}

The datasets included in the combination are listed in Table~\ref{tab:comb:red:input} and correspond 
to $209$ individual cross-section measurements.%
\footnote{From the two sets of measurements in \cite{h1dstar_hera2}, the one 
in the wider $p_T(\Dstar)$ and $\eta(\Dstar)$ range was chosen and referred to as dataset I; this is another dataset from the one that was used in the combination of the \Dstar cross sections, described in Section~\ref{sec:comb:dstar}.} 
The combination includes measurements of charm production performed using
different tagging techniques: 
the reconstruction of particular 
decays of $D$-mesons (datasets 2--7, 9, 10), 
the inclusive analysis of 
tracks exploiting lifetime information (datasets 1, 11) and 
the reconstruction of muons from 
charm semi-leptonic decays (dataset 8). 

\begin{table*}[tbp]
\caption[Datasets used in combination of reduced charm cross sections]
{The most precise H1 and ZEUS measurements of charm production performed with various techniques. 
For each dataset the charm tagging method, 
the $Q^2$, $p_T$ ($E_T$) and $\eta$ range, the number of cross-section measurements, $N$, the integrated luminosity, 
${\cal L}$, and the centre-of-mass energy, $\sqrt{s}$, are given. The dataset with the $D^{0,\text{ no }D^{*+}}$ tagging method
is based on an analysis of $D^{0}$ mesons not originating from detectable $D^{*+}$
decays. 
}
\label{tab:comb:red:input}
\begin{center}
\tabcolsep1.2mm
\renewcommand*{\arraystretch}{1.2}
\begin{tabu} to \textwidth [t] {|l|l|l|l|l|l|r|r|r|}\hline 
       \multicolumn{2}{|l|}{Dataset} & Tagging method & $Q^2$ range & $p_T$ ($E_T$) range & $\eta$ range       & $N$ & ${\cal L}$& ${\sqrt{s}}$  \\ 
     & & & [GeV$^2$] & [GeV] & &  & [pb$^{-1}$] & [$\SI{}{GeV}$] \\ \hline
1&H1 VTX~\cite{h1ltt_hera2}    &Inclusive&  $5<Q^2<2000$ & not restricted & not restricted              & $29$ & $245$ & $318$ \\
2&H1 $\Dstar$ HERA-I~\cite{h1dstar_hera1}      &$D^{*+}$    &  $2<Q^2<100$ & $\SI{1.5}{}<p_T(\Dstar)<\SI{15}{}$ & $|\eta(\Dstar)|<1.5$                  & $17$ & $47$& $318$\\
3&H1 $\Dstar$ HERA-II (med. $Q^2$)~\cite{h1dstar_hera2}     &$D^{*+} $     &  $5<Q^2<100$ & $\SI{1.25}{}<p_T(\Dstar)<\SI{20}{}$ & $|\eta(\Dstar)|<1.8$                 &  $25$ & $348$& $318$\\
4&H1 $\Dstar$ HERA-II (high $Q^2$)~\cite{h1dstarhighQ2}     &$D^{*+}$     &  $100<Q^2<1000$ & $\SI{1.5}{}<p_T(\Dstar)<\SI{20}{}$ & $|\eta(\Dstar)|<1.8$           &  $6$ & $351$& $318$ \\
5&ZEUS $\Dstar$ 96-97~\cite{zd9697}           &$D^{*+}$      & $1<Q^2<200$ & $\SI{1.5}{}<p_T(\Dstar)<\SI{15}{}$ & $|\eta(\Dstar)|<1.5$                 & $2$1 & $37$& $300$ \\
6&ZEUS $\Dstar$ 98-00~\cite{zd00}           &$D^{*+}$      &  $1.5<Q^2<1000$ & $\SI{1.5}{}<p_T(\Dstar)<\SI{15}{}$ & $|\eta(\Dstar)|<1.5$           & $31$ & $82$& $318$ \\
7&ZEUS $D^0$ 2005~\cite{zd0dp}                  &$D^{0,{\rm no}D^{*+}}$ &$ 5<Q^2<1000$ & $\SI{1.5}{}<p_T(D^{0})<\SI{15}{}$ & $|\eta(\Dstar)|<1.6$   & $9$  & $134$& $318$\\
8&ZEUS $\mu$ 2005~\cite{zeus_muon}                    &Semi-leptonic         & $20<Q^2<10000$ & $p_T(\mu)>\SI{1.5}{}$ & $-1.6<\eta(\mu)<2.2$         & $8$  &  $126$& $318$  \\
9&ZEUS $D^+$ HERA-II~\cite{zeusdch_hera2}&$D^+$         & $5<Q^2<1000$ & $\SI{1.5}{}<p_T(\Dch)<\SI{15}{}$ & $|\eta(\Dstar)|<1.6$              & $14$   & $354$& $318$\\
10&ZEUS $\Dstar$ HERA-II~\cite{zeusdstar_hera2}                  &$\Dstar$         & $5<Q^2<1000$ & $p_T(\Dstar)>\SI{1.5}{}$ & $|\eta(\Dstar)|<1.5$              & $31$   & $363$& $318$\\
11&ZEUS VTX HERA-II~\cite{zeussecvtx_hera2}                  &Inclusive         & $5<Q^2<1000$ & $E_T^{\rm jet}>\SI{4.2}{}$ & $-1.6<\eta^{\rm jet}<2.2$              & $18$   & $354$& $318$\\ \hline
\end{tabu}
\end{center}
\end{table*}

Datasets 1--8 have been used in the previous `HERA 2012' combination, while datasets 9--11 were newly included. 
Note that dataset 9 replaced one of the datasets from `HERA 2012', 
which is its subset.

Correlations between systematic uncertainties of different measurements were 
accounted for as explained in Section~\ref{sec:comb:proc:unc}. 
All experimental systematic uncertainties were treated as independent between H1 and ZEUS. 
A full list of correlated sources is provided in Appendix~\ref{sec:app:red} (Table~\ref{tab:comb:red:syst}). 
The total uncorrelated systematic uncertainties were obtained by adding individual ones in quadrature.%
\footnote{For dataset 11 an additional uncorrelated systematic uncertainty was considered: an uncertainty
of 100\% on $\Delta_{\rm had} = C_{\rm had} - 1$ (Table~6 of~\cite{zeussecvtx_hera2}).}

\paragraph*{Reduced cross sections and common $Q^2\text{--}x$ grid\\}
\label{sec:comb:red:det:grid}

The quantities to be combined are the reduced charm cross sections, defined as follows:
\begin{eqnarray}
\sigma_{\rm red}^{c\bar{c}}&=&\frac{{\rm d}^2\sigma^{c\bar{c}}}{{\rm d}x {\rm d}Q^2} \cdot \frac{xQ^4}{2\pi\alpha^2\,(1+(1-y)^2)}\cr
&=&F_2^{c\bar{c}}-\frac{y^2}{1+(1-y)^2}F^{c\bar{c}}_L.
\label{eq:comb:red:red}
\end{eqnarray}
The cross
section ${{\rm d}^2 \sigma^{c\bar{c}}}/{{\rm d}x {\rm d}Q^2}$ is given at the
Born level without QED and electroweak radiative corrections, except for the running electromagnetic coupling $\alpha$. 

The reduced cross sections (and not the structure functions $F_2^{c\bar{c}}$) were chosen for the combination 
because they are proportional to the directly measured double-differential cross sections. 
$F_2^{c\bar{c}}$ and $F^{c\bar{c}}_L$ depend only on $Q^2$ and $x$. 
The presence of $y$ in definition~\ref{eq:comb:red:red} 
leads to \ozmodNN{a} dependence of \red on the centre-of-mass energy, $\sqrt{s}$. 
\ozmodNN{The combined reduced cross sections are provided at the centre-of-mass energy $\sqrt{s}=\SI{318}{GeV}$.}

The values of $\red$ for individual measurements were determined 
at the $52$ ($Q^2, x$) points of a common grid, chosen such that they are close to the centre-of-gravity in $Q^2$ and $x$ of the corresponding bins, taking advantage of the fact that 
the binning schemes used by the H1 and ZEUS experiments are similar (the grid points were kept the same as in the `HERA 2012' combination). 
For all but three grid points, at least 2 measurements entered into the combination; 
for points in the medium $Q^2$ bins, the number of input measurements \ozmodN{is as much as} 7. 
The phase-space region of the combined cross sections is \ozmodNN{determined as}
\begin{equation}
	\begin{split}
		2.5 \le Q^2 \le \SI{2000}{GeV^2},\\
		3 \times 10^{-5}\le x \le 5 \times 10^{-2}.
	\end{split}
\label{eq:comb:red:ps}
\end{equation}

\paragraph*{Extrapolation and corrections\\}
\label{sec:comb:red:det:extrcorr}

The results of the inclusive lifetime analysis (dataset 1) were directly taken from the 
original measurement in the form of \red. 
For all other measurements the inputs to the 
combination were visible cross sections, $\sigma_{\rm vis, bin}$, defined as the $D$-, $\mu$- or jet-production cross 
sections in a particular $p_T$ and $\eta$ range, in bins of $Q^2$ and $y$ or $x$. 

The reduced cross sections \red were obtained from the visible cross sections $\sigma_{\rm vis, bin}$ measured in a limited phase-space region \ozmodN{using theoretical predictions} 
according to the procedure described in Section~\ref{sec:comb:proc:pscorr}: 
the reduced charm cross section at a reference ($x, Q^2$) point is given by
\begin{equation}
\sigma_{\rm red}^{\rm c\bar{c}}(x,Q^2)=\sigma_{\rm vis, bin}
\frac{\sigma_{\rm red}^{\rm c\bar{c}, th}(x,Q^2)}{\sigma^{\rm th}_{\rm vis, bin}}.
\label{eq:comb:red:extr}
\end{equation}
To calculate $\sigma_{\rm red}^{\rm c\bar{c}, th}(x,Q^2)$ and the visible cross sections $\sigma^{\rm th}_{\rm vis, bin}$, 
the NLO QCD FFNS theory set-up was used, consistent with the previous `HERA 2012' combination.%
\footnote{The fully consistent theory set-up allowed for using existing input tables for \red for datasets 1--8, 
available from~\cite{heracharmcomb}, 
and straightforward comparison of new results of the combination with the `HERA 2012' results; 
also note that two of three newly included ZEUS measurements (datasets 10 and 11) already published 
\red extracted from the visible cross sections using exactly this theoretical set-up.}
This set-up is almost identical to the one described in Section~\ref{sec:comb:th}, except for the following minor changes:
\begin{enumerate}
	\item in the fragmentation process, calculation of the hadron energy and Lorentz boost were \ozmod{performed} by using \ozmodNN{the mass of the $c$ quark rather than the charmed hadron}%
		\footnote{Except for dataset 8.};
	\item if it needed to be subtracted, the beauty contribution was evaluated using the estimates of the corresponding \ozmod{publications} (based on MC re-normalised to data).
\end{enumerate}
The extrapolation factors, $R=\sigma^{\rm th}_{\rm bin} / \sigma^{\rm th}_{\rm vis, bin}$, where $\sigma^{\rm th}_{\rm bin}$ is the cross section in the full ($p_T$, $\eta$) phase-space region, 
vary in a wide range: from $R \gtrsim 1$ at high $Q^2$ to $R \sim 5$ at low $Q^2$. 
For dataset 5 the extrapolation procedure includes also the centre-of-mass energy correction.

The extrapolation uncertainties were estimated from the variations described in Section~\ref{sec:comb:th}, and were treated as correlated between datasets 2-11%
\footnote{The PDF uncertainties were neglected for the newly included datasets 9-11, since for the other ones they were found to be negligibly small (1\% on average)~\cite{heracharmcomb}.}. 
For dataset 1 the extrapolation uncertainties (except for the longitudinal fragmentation) do not appear explicitly and were covered by the experimental systematic uncertainties. 
The dominant contributions arise from the variation of the renormalisation and factorisation scales (average $5\text{--}6\%$, reaching $15\%$ at lowest $Q^2$) 
and from the variation of the fragmentation function (average $3\text{--}5\%$). 

Prior to the combination, datasets 1 and 11 were \ozmodN{corrected}, when needed, from the grids used in the original \ozmod{publications} 
to the common grid using the NLO FFNS calculation. 
The corrections were always smaller than $25\%$ and 
the associated uncertainties, obtained by varying the charm mass, the scales and the PDFs, were negligible. 
All $D$-meson cross sections were updated using the most recent branching ratios~\cite{pdg2012}. 

\subsubsection{Combined charm cross sections}
\label{sec:comb:red:res}

\begin{figure}[htbp]
  \centering
  \includegraphics[width=1.0\figwidth,trim=6mm 2mm 8mm 3mm,clip=true]{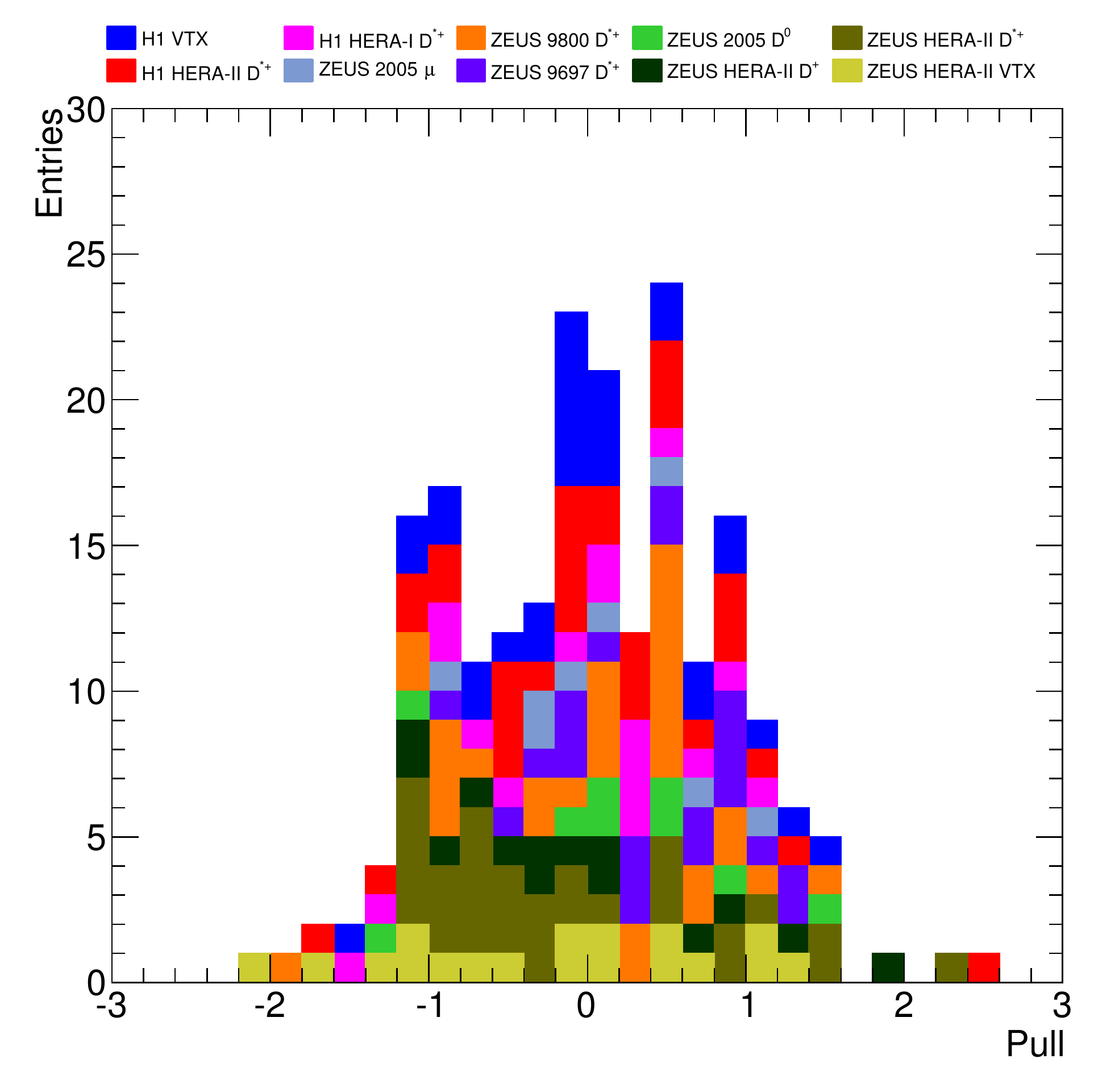}
  \caption[Pull distributions for reduced charm cross sections]
  {The pull distribution for the combination of the reduced charm cross sections. 
  Contributions from the individual input datasets are shown separately.}
	\label{fig:comb:red:pull}
\end{figure}

In total, $209$ measurements were combined to $52$ reduced cross-section measurements. 
\ozmod{The combination has} $\chisqndof=117/157$; 
the corresponding probability is $99.3\%$, indicating conservative estimation of the experimental systematic uncertainties 
of the input measurements. %
The pull distribution is shown in Fig.~\ref{fig:comb:red:pull}. 

The values of the combined cross section \red together with uncorrelated, correlated and total uncertainties are given in Table~\ref{tab:comb:red:combined}. 
A detailed breakdown of the correlated uncertainties is provided in Appendix~\ref{sec:app:red} (Table~\ref{tab:comb:red:combinedfull}).

\begin{table}[tbp]
\caption[Combined reduced charm cross sections]
{The combined reduced cross sections of charm production with their 
uncorrelated ($\delta_{unc}$), correlated ($\delta_{cor}$) and total ($\delta_{tot}$) uncertainties.} 
\label{tab:comb:red:combined}
\tabcolsep2.50mm
\renewcommand*{\arraystretch}{1.08}
\begin{tabularx}{\columnwidth}[t]{|X|X|c|c|c|c|}
\hline
$Q^2$ &$x$ &$\sigma_{\rm red}^{c\bar{c}}$&$\delta_{unc}$  &$\delta_{cor}$&$\delta_{tot}$ \\
  ($\SI{}{GeV^2}$)      & &                          &$(\%)$            &$(\%)$ &$(\%)$ \\ 
\hline
2.5 & 0.00003 & 0.1215 & 13.5 & 8.2 & 15.8 \\
2.5 & 0.00007 & 0.1144 & 8.7 & 8.2 & 11.9 \\
2.5 & 0.00013 & 0.0940 & 9.6 & 7.8 & 12.4 \\
2.5 & 0.00018 & 0.0966 & 9.2 & 7.2 & 11.7 \\
2.5 & 0.00035 & 0.0586 & 8.6 & 6.5 & 10.8 \\
5 & 0.00007 & 0.1568 & 14.9 & 7.7 & 16.8 \\
5 & 0.00018 & 0.1594 & 8.0 & 6.0 & 10.0 \\
5 & 0.00035 & 0.1208 & 6.9 & 5.6 & 8.9 \\
5 & 0.00100 & 0.0839 & 9.2 & 5.2 & 10.6 \\
7 & 0.00013 & 0.2333 & 5.5 & 6.7 & 8.7 \\
7 & 0.00018 & 0.2088 & 8.7 & 7.1 & 11.2 \\
7 & 0.00030 & 0.1833 & 3.7 & 5.2 & 6.3 \\
7 & 0.00050 & 0.1673 & 3.5 & 4.7 & 5.8 \\
7 & 0.00080 & 0.1249 & 6.1 & 4.4 & 7.5 \\
7 & 0.00160 & 0.0958 & 5.3 & 4.7 & 7.1 \\
12 & 0.00022 & 0.3279 & 5.8 & 5.9 & 8.3 \\
12 & 0.00032 & 0.3041 & 4.5 & 5.6 & 7.2 \\
12 & 0.00050 & 0.2470 & 3.3 & 4.1 & 5.3 \\
12 & 0.00080 & 0.1882 & 3.0 & 3.9 & 4.9 \\
12 & 0.00150 & 0.1586 & 4.2 & 4.3 & 6.0 \\
12 & 0.00300 & 0.1106 & 5.5 & 4.9 & 7.3 \\
18 & 0.00035 & 0.3306 & 6.3 & 5.4 & 8.3 \\
18 & 0.00050 & 0.3030 & 4.0 & 5.7 & 7.0 \\
18 & 0.00080 & 0.2685 & 3.1 & 3.6 & 4.7 \\
18 & 0.00135 & 0.2134 & 2.6 & 3.8 & 4.6 \\
18 & 0.00250 & 0.1723 & 2.7 & 3.7 & 4.5 \\
18 & 0.00450 & 0.1314 & 5.9 & 5.2 & 7.8 \\
32 & 0.00060 & 0.4348 & 14.6 & 4.9 & 15.4 \\
32 & 0.00080 & 0.3778 & 3.5 & 4.4 & 5.6 \\
32 & 0.00140 & 0.2874 & 2.7 & 3.2 & 4.2 \\
32 & 0.00240 & 0.2241 & 3.3 & 3.3 & 4.7 \\
32 & 0.00320 & 0.2136 & 5.3 & 3.8 & 6.5 \\
32 & 0.00550 & 0.1610 & 4.7 & 3.8 & 6.0 \\
32 & 0.00800 & 0.1022 & 9.8 & 5.4 & 11.2 \\
60 & 0.00140 & 0.3380 & 4.6 & 3.8 & 5.9 \\
60 & 0.00200 & 0.3440 & 3.9 & 2.6 & 4.7 \\
60 & 0.00320 & 0.2709 & 3.5 & 3.0 & 4.6 \\
60 & 0.00500 & 0.1993 & 3.6 & 3.0 & 4.7 \\
60 & 0.00800 & 0.1712 & 6.0 & 3.0 & 6.7 \\
60 & 0.01500 & 0.1014 & 9.9 & 4.2 & 10.8 \\
120 & 0.00200 & 0.3560 & 6.5 & 4.0 & 7.6 \\
120 & 0.00320 & 0.3619 & 9.5 & 2.7 & 9.9 \\
120 & 0.00550 & 0.2309 & 5.2 & 3.3 & 6.2 \\
120 & 0.01000 & 0.1605 & 4.7 & 2.9 & 5.5 \\
120 & 0.02500 & 0.0888 & 13.9 & 3.7 & 14.4 \\
200 & 0.00500 & 0.2510 & 6.6 & 3.9 & 7.7 \\
200 & 0.01300 & 0.1773 & 5.5 & 3.3 & 6.4 \\
350 & 0.01000 & 0.2264 & 8.1 & 4.0 & 9.0 \\
350 & 0.02500 & 0.1079 & 10.0 & 4.1 & 10.8 \\
650 & 0.01300 & 0.2124 & 9.5 & 5.6 & 11.1 \\
650 & 0.03200 & 0.0993 & 11.4 & 7.9 & 13.9 \\
2000 & 0.05000 & 0.0655 & 26.3 & 12.9 & 29.3 \\
\hline
\end{tabularx}
\end{table}

The individual datasets as well as the results of the combination are shown
in Fig.~\ref{fig:comb:red:combined}.%
\footnote{The same plots, but separately for each $Q^2$ bin, are available in Appendix~\ref{sec:app:red} (Figs.~\ref{fig:comb:red:comb_q2_1}to~\ref{fig:comb:red:comb_q2_12}).} 
The combined cross sections exhibit significantly reduced uncertainties. 
The input H1 and ZEUS data in total are similar in precision and contribute roughly equally to the averaged results. 
The combined data are significantly more precise than any of the individual 
input datasets. The uncertainty of the combined results is about $8\%$ on average and reaches 
$4\%$ in the region of small $x$ and medium $Q^2$. This is an improvement of about a factor of 2.5 with respect to each of the 
most precise datasets in the combination.

\begin{figure}[tbp]
  \centering
  \includegraphics[width=1.0\figwidth,trim=3mm 0 3mm 5mm,clip=true]{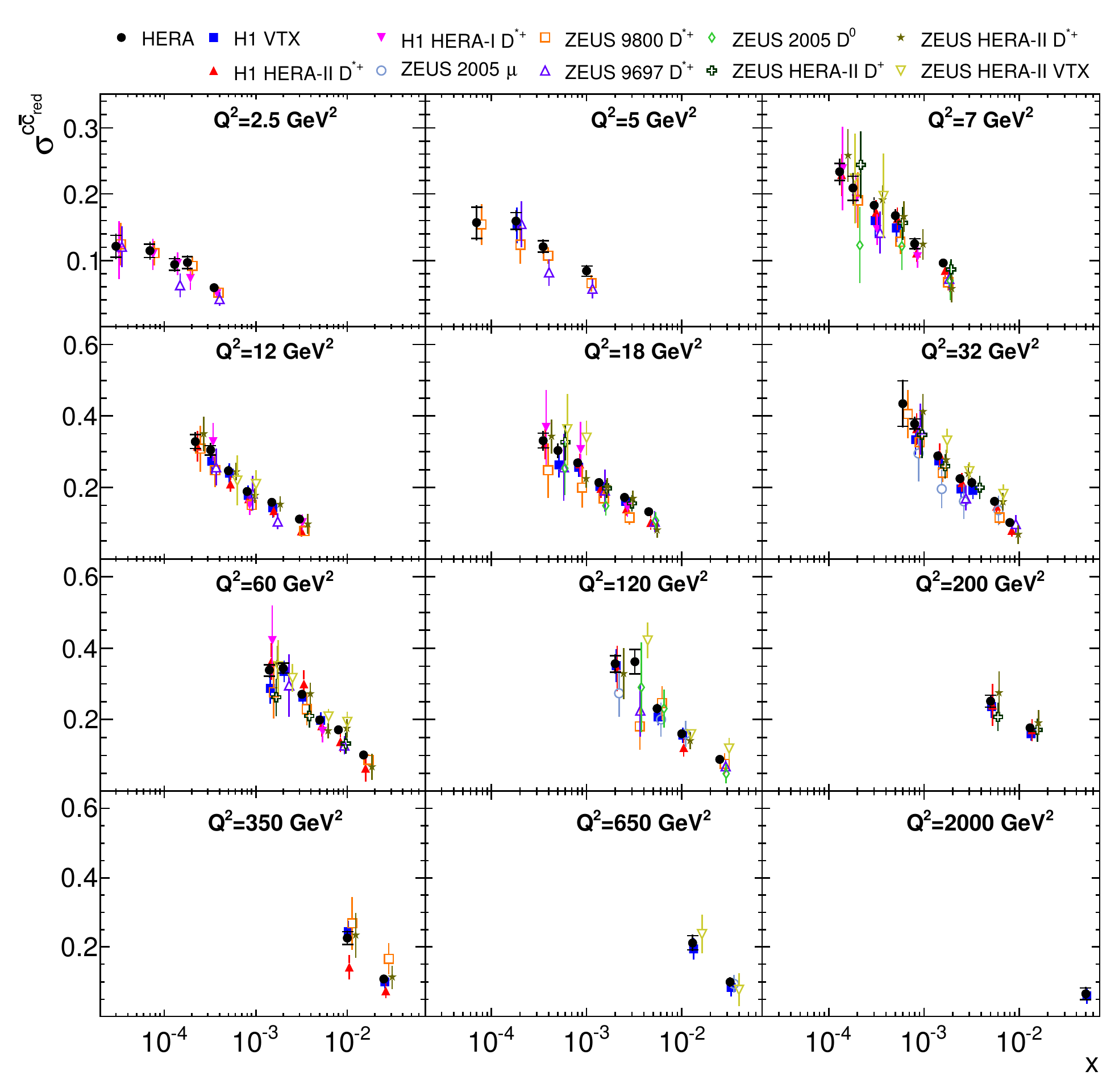}
  \caption[Combined reduced charm cross sections]
  {Combined measurements of $\sigma_{red}^{c\bar{c}}$ (closed circles) shown as a function of $x$ for given values of $Q^2$. 
  The input measurements are also shown with different markers. 
	For the combined data, the inner error bars indicate the uncorrelated part of the uncertainties and 
	the outer error bars represent the total uncertainties. 
  For presentation purposes each individual measurement is shifted in $x$.}  
	\label{fig:comb:red:combined}
\end{figure}

There are in total $78$ sources of correlated systematic uncertainty, including global normalisations, characterising the 
separate datasets. The shifts and the reduction of the correlated uncertainties are provided in Appendix~\ref{sec:app:red} (Table~\ref{tab:comb:red:syst}). 
None of these \ozmod{shifts exceeds} $1.3$ standard deviation. 
The influence of several correlated systematic uncertainties was reduced by more than a factor of two, while 
on average the reduction factors are about 20\% of the nominal standard deviation. 
The reductions \ozmodN{are due} to the different charm-tagging methods, and to the requirement that 
different measurements probe the same cross section at each ($x,Q^2$) point. 
The reduction of systematic uncertainties propagated to the other average 
points, including those which are based solely on the less precise measurements. 
Due to this propagation the uncertainty on the combined data in the points, to which only one input measurement contributes, was also reduced 
compared to the original one.
 
\paragraph*{Comparison to previous combination\\}
\label{sec:comb:red:rescomp2012}

Comparing $\chisqndof=117/157$ to the `HERA 2012' result, $\chisqndof=62/103$, the individual contributions from the newly included measurements are 
$11/14$, $30/31$ and $18/18$ for datasets 9, 10 and 11, respectively, and the total contribution from all three datasets is $59/63$, thus 
the new measurements are consistent. 

\begin{figure}[tbp]
  \centering
  \includegraphics[width=1.0\figwidth,trim=3.3mm 2mm 3.1mm 6mm,clip=true]{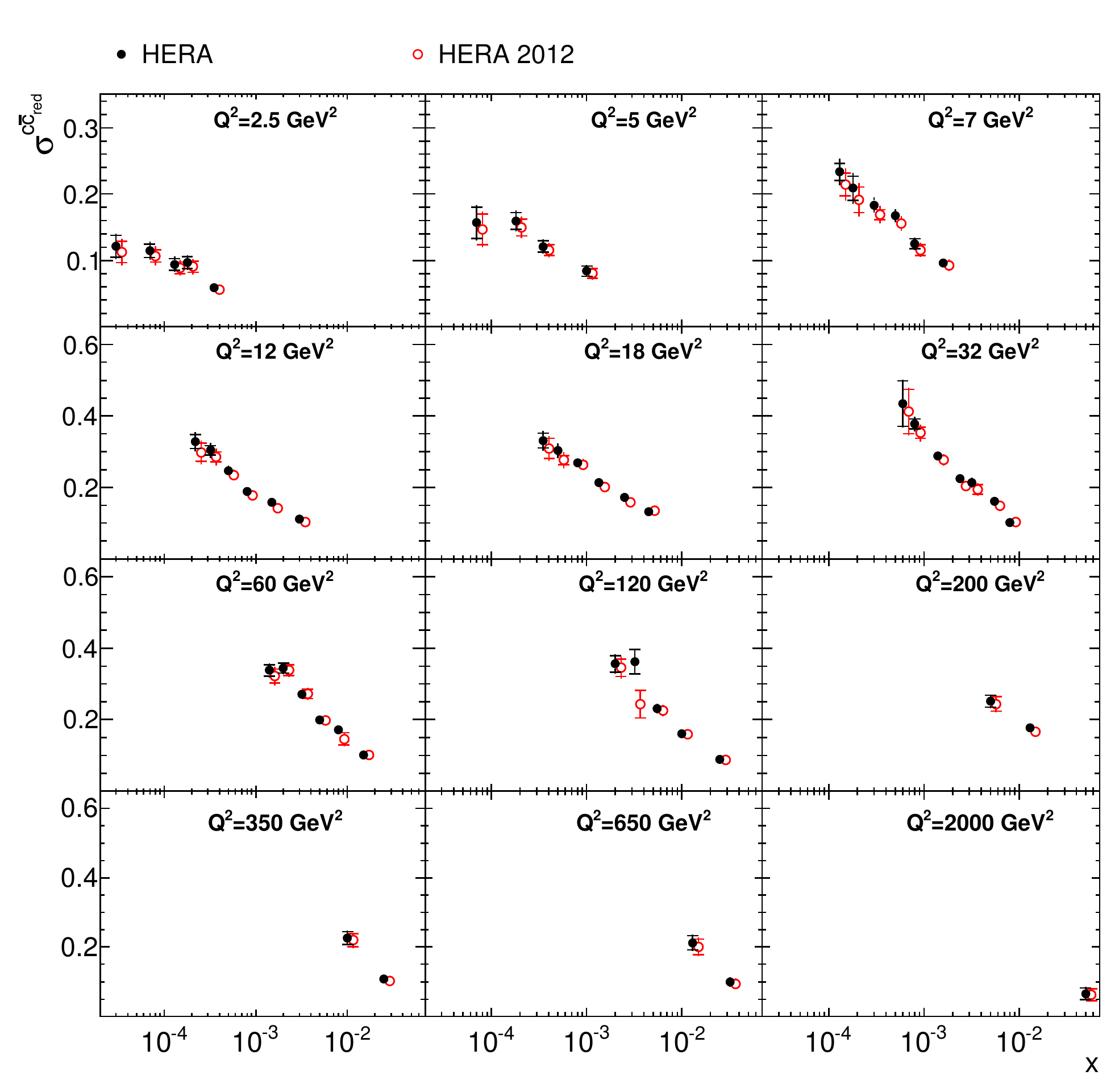}
  \caption[Reduced charm cross sections compared to `HERA 2012' results]
  {Combined reduced charm cross sections (closed circles) shown as a function of $x$ for given values of $Q^2$, compared to the `HERA 2012' results (open circles). 
  The error bars represent the total uncertainty. 
  The inner error bars indicate the uncorrelated part of the uncertainties. 
  For presentation purposes, the `HERA 2012' results are slightly shifted in $x$.}
	\label{fig:comb:red:combinedvspaper}
\end{figure}

\begin{figure}[tbp]
  \centering
  \includegraphics[width=1.0\figwidth,trim=3.3mm 2mm 3.1mm 6mm,clip=true]{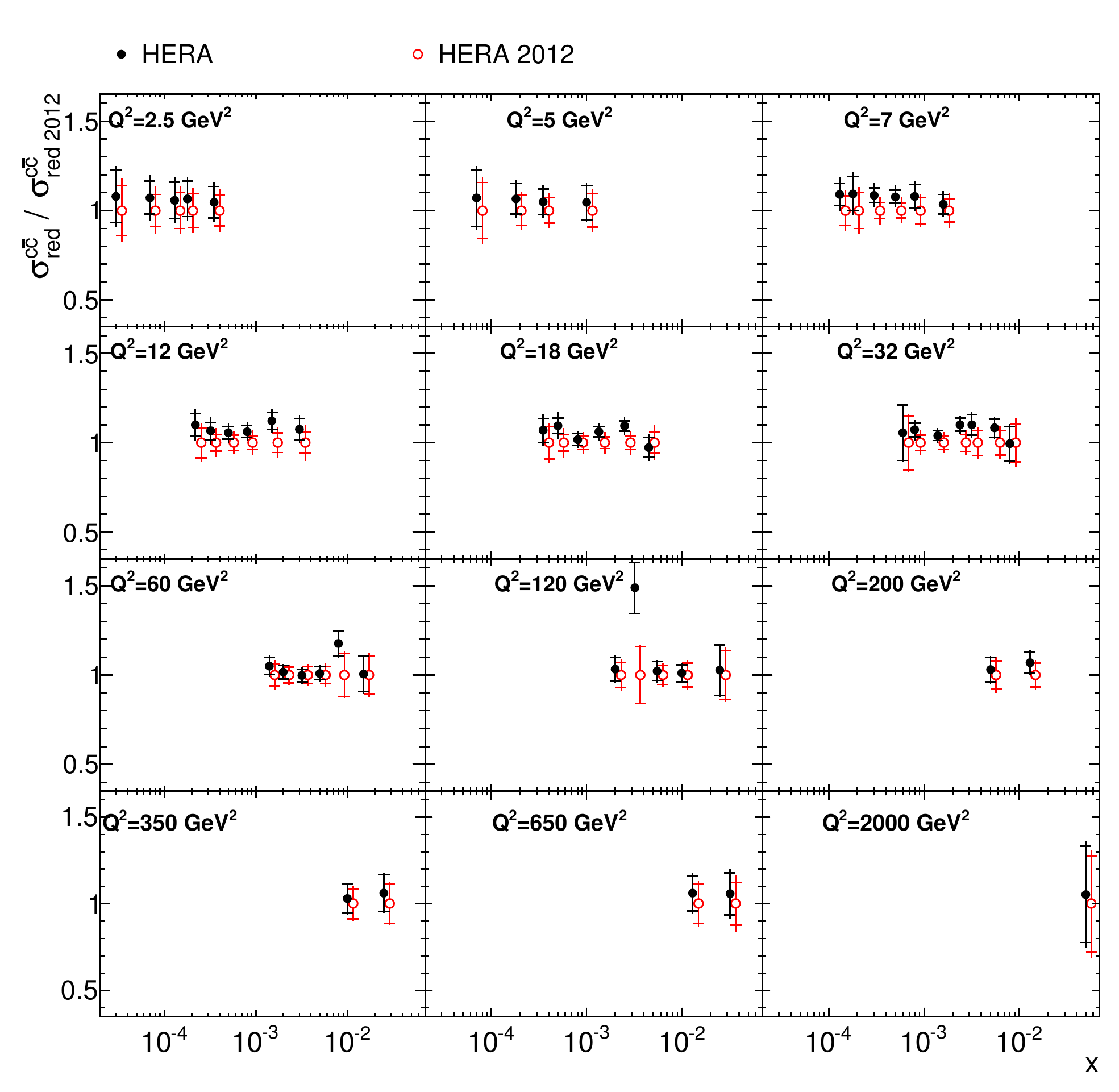}
  \caption[Reduced charm cross sections normalised to `HERA 2012' results]
  {Combined reduced charm cross sections (closed circles) shown as a function of $x$ for given values of $Q^2$, normalised to the `HERA 2012' results (open circles). 
  The error bars represent the total uncertainty. 
  The inner error bars indicate the uncorrelated part of the uncertainties. 
  For presentation purposes the `HERA 2012' results are slightly shifted in $x$.}
	\label{fig:comb:red:combinedvspaperrat}
\end{figure}

\begin{figure}[tbp]
  \centering
  \includegraphics[width=1.0\figwidth,trim=3.3mm 2mm 3.1mm 6mm,clip=true]{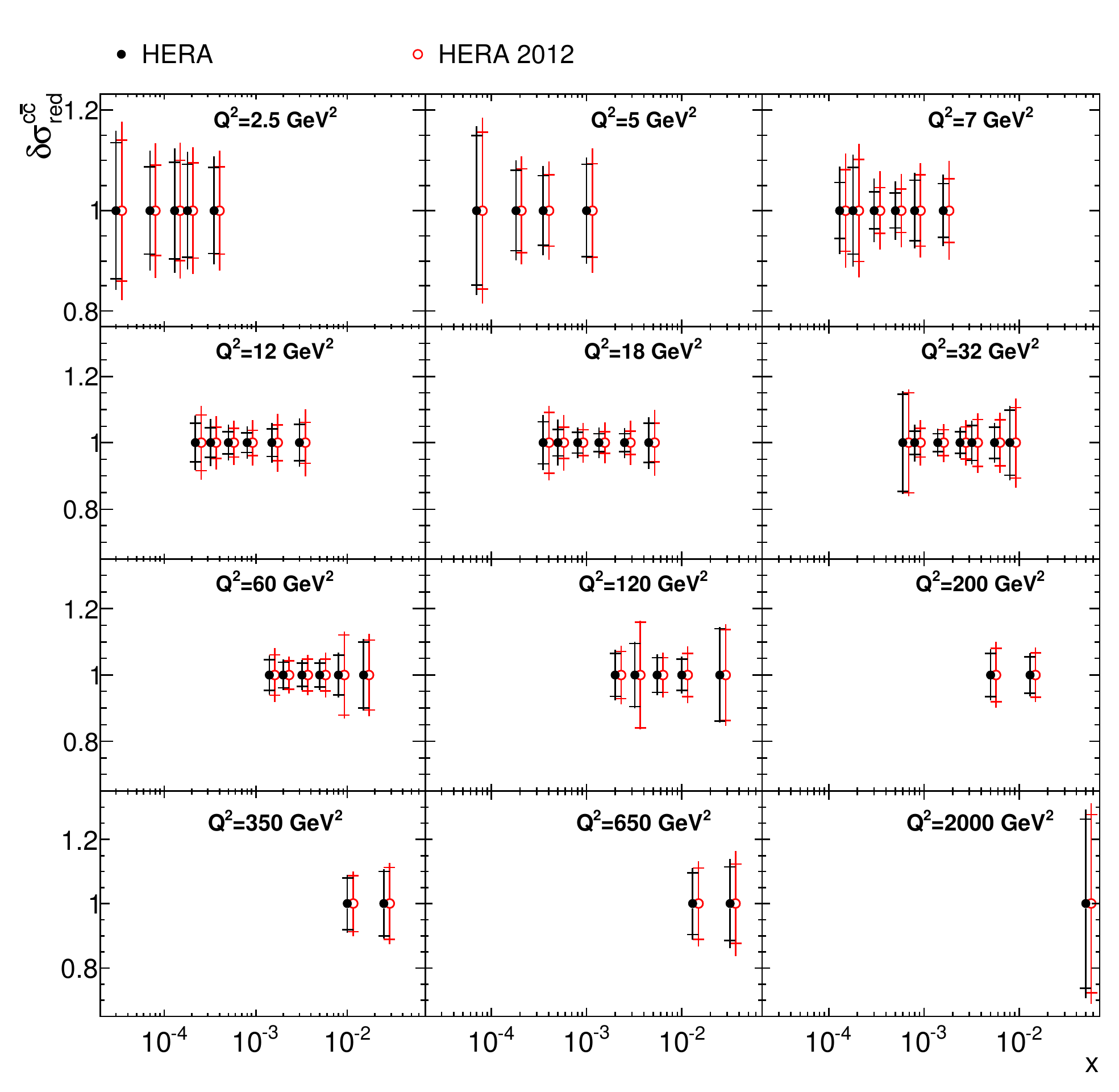}
  \caption[Uncertainties of reduced charm cross sections compared to `HERA 2012' results]
  {The relative uncertainties of the combined reduced charm cross sections (closed circles) shown as a function of $x$ for given values of $Q^2$, compared to the `HERA 2012' results (open circles). 
  The error bars represent the total uncertainty. 
  The inner error bars indicate the uncorrelated part of the uncertainties. 
  For presentation purposes the `HERA 2012' results are slightly shifted in $x$.}
	\label{fig:comb:red:combinedvspaperunc}
\end{figure}

The combined data are compared to the `HERA 2012' results in Fig.~\ref{fig:comb:red:combinedvspaper}; 
for a more detailed comparison Fig.~\ref{fig:comb:red:combinedvspaperrat} shows the same results normalised to the `HERA 2012' and 
Fig.~\ref{fig:comb:red:combinedvspaperunc} shows the comparison of the relative uncertainties. 
The new results are consistent with the previously published ones, although on average they lie slightly above. 
This is explained by taking into account the changes in the shifts of correlated systematic uncertainties, 
which affect all points simultaneously (mainly the theory-related sources and luminosity uncertainties, 
see Table~\ref{tab:comb:red:syst} in Appendix~\ref{sec:app:red}). 

The new combined cross sections exhibit reduced uncertainties. 
Typically, the reduction of the uncorrelated and correlated uncertainties contribute about equally to the total improvement. 
At medium $Q^2$, where new datasets 9, 10 and 11 contribute directly, the improvement is on average of the order of 20\% of the `HERA 2012' uncertainties, reaching 35\% in several points, 
and in the low and very high $Q^2$ bins the improvement is 5--15\% owing to the reduction of the correlated uncertainties only. 

\subsubsection{Comparison to theoretical predictions and QCD analysis}
\label{sec:comb:red:ffns}

\begin{figure}[tbp]
  \centering
  \includegraphics[width=1.0\figwidth,trim=3.3mm 2mm 3.1mm 6mm,clip=true]{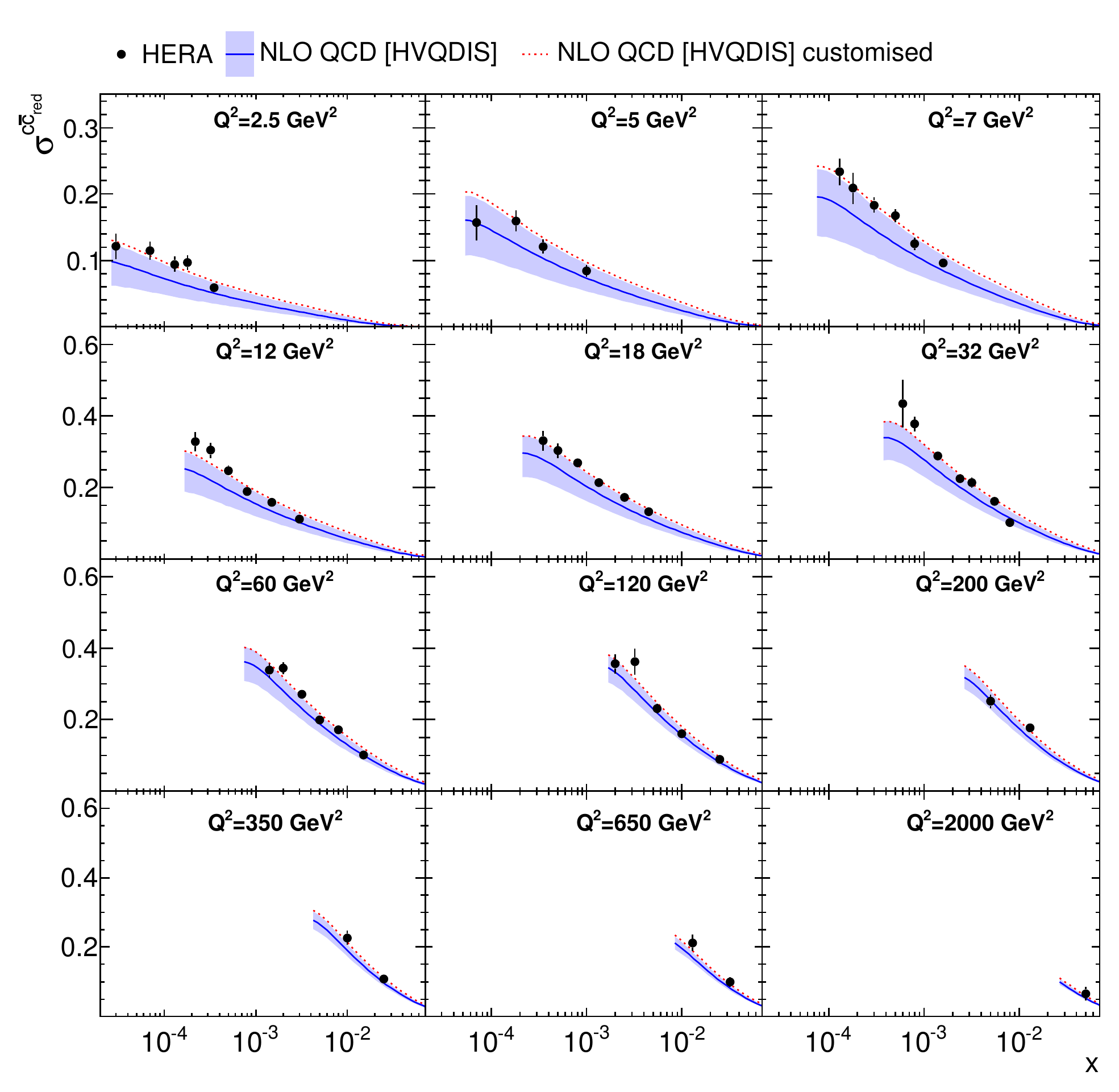}
  \caption[Reduced charm cross sections compared to NLO QCD FFNS predictions]
  {Combined measurements of $\sigma_{red}^{c\bar{c}}$ (closed circles) shown as a function of $x$ for given values of $Q^2$, compared to the NLO QCD FFNS theoretical predictions (solid line with band). 
	The customised NLO calculation (dotted line) is also shown.}
	\label{fig:comb:red:combinedvstheory}
\end{figure}

\begin{figure}[tbp]
  \centering
  \includegraphics[width=1.0\figwidth,trim=3.3mm 2mm 3.1mm 6mm,clip=true]{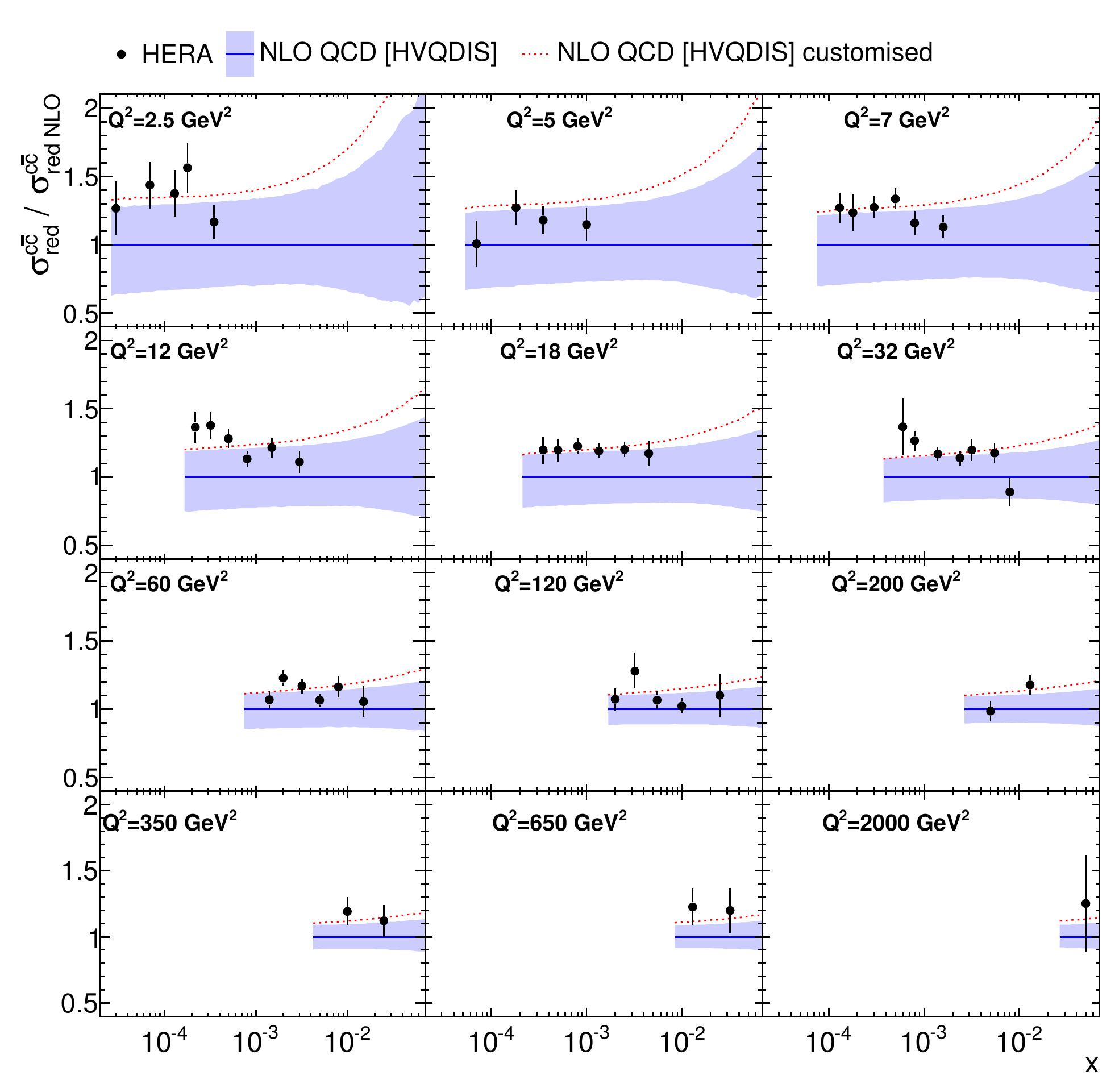}
  \caption[Reduced charm cross sections normalised to NLO QCD FFNS predictions]
  {Combined measurements of $\sigma_{red}^{c\bar{c}}$ (closed circles) shown as a function of $x$ for given values of $Q^2$, normalised to the NLO QCD FFNS theoretical predictions (solid line with band). 
	The customised NLO calculation is also shown (dotted line).}
	\label{fig:comb:red:combinedvstheoryrat}
\end{figure}

Fig.~\ref{fig:comb:red:combinedvstheory} presents a comparison of the NLO QCD predictions in the FFNS, calculated as described in Section~\ref{sec:comb:th}, to the combined data. 
This is more clearly seen in the ratio to the theoretical predictions, shown in Fig.~\ref{fig:comb:red:combinedvstheoryrat}. 
The predictions describe the data well within the uncertainties in the whole kinematic range of the combination, although the central theoretical curve underestimates the data normalisation, 
as observed also in the combination of the \Dstar cross sections (Section~\ref{sec:comb:dstar}). 
The `customised' NLO calculation (Section~\ref{sec:comb:dstar:single:customth}), while it was determined mainly from the exclusive \Dstar quantities in the restricted phase-space region, 
provides an improved description of the reduced cross-section normalisation, although it does not improve the description of the $x$ shape.

\begin{figure}[tbp]
  \centering
  \includegraphics[width=1.0\figwidth,trim=4.5mm 3mm 14.5mm 19mm,clip=true]{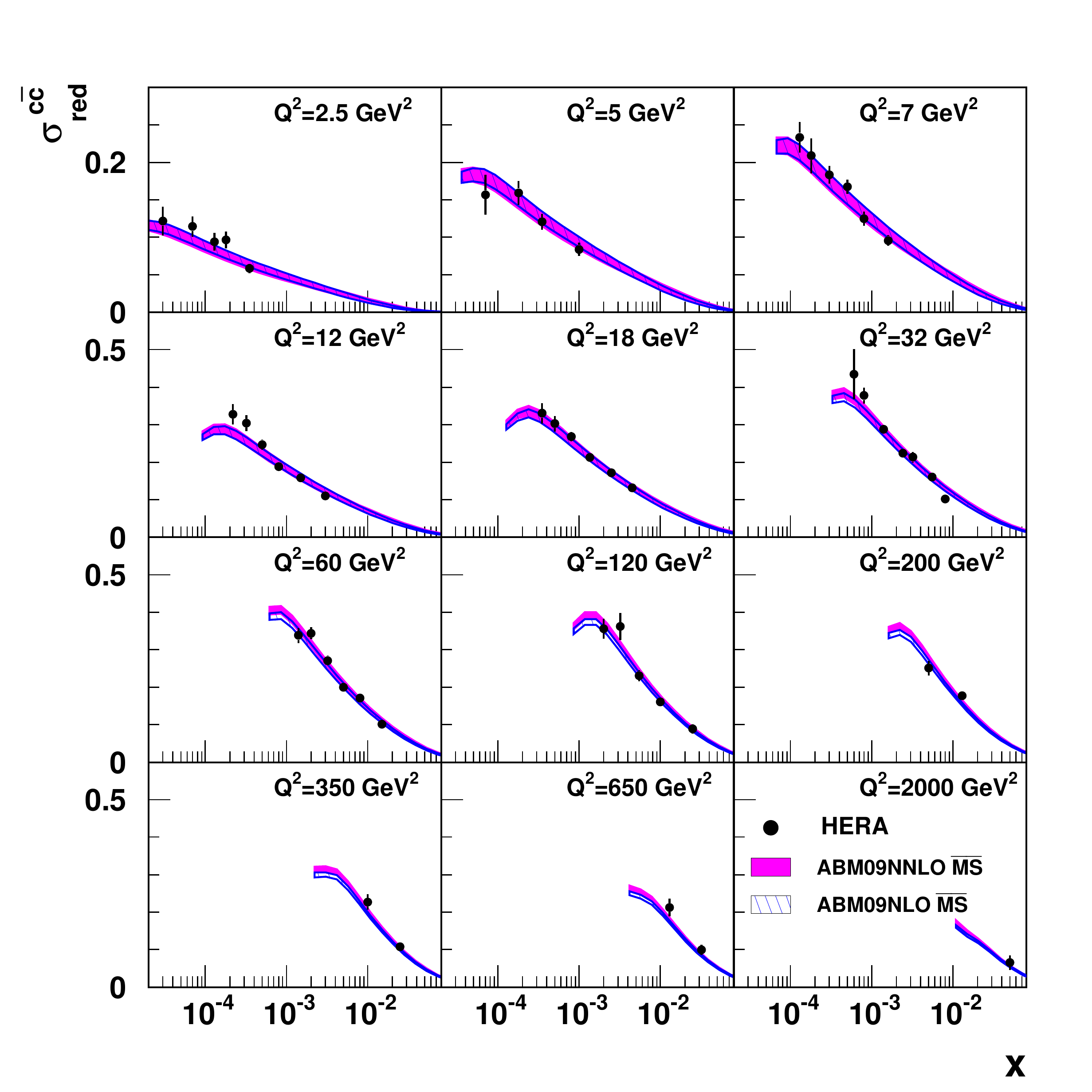}
  \caption[Reduced charm cross sections compared to ABM predictions]
  {Combined measurements of $\sigma_{red}^{c\bar{c}}$ (closed circles) shown as a function 
	of $x$ for particular $Q^2$, compared to the prediction by the ABM group at NLO (hashed band) and NNLO (shaded band) 
	in the FFNS using the $\overline{\text{MS}}$ definition for the $c$-quark mass.}
	\label{fig:comb:red:combinedvsabm}
\end{figure}

In Fig.~\ref{fig:comb:red:combinedvsabm} the data are compared to the predictions
by the ABM group in the FFNS at NLO and NNLO, based on the running-mass scheme~\cite{Alekhin:2010sv,Alekhin:2009ni}. 
The uncertainties on the predictions include the uncertainties on the charm mass, which dominate at small $Q^2$. 
The predictions at NLO and NNLO are very similar and describe the data well in the whole kinematic range of the measurement. 

The sensitivity of the theoretical predictions to the charm mass allows for  
the determination of its \ozmodN{best} value from the data in a QCD fit. 
The analysis was performed with the HERAFitter program~\cite{herafitter,HERAFitterPaper} , which is based on the NLO DGLAP 
evolution scheme~\cite{dglap_Gribov:1972ri,Dokshitzer:1977sg,dglap_Altarelli:1977zs,dglap_Curci:1980uw,dglap_Furmanski:1980cm,dglap_Moch:2004pa,dglap_Vogt:2004mw} as implemented in QCDNUM~\cite{qcdnum}. 
The strategy of the HERAPDF1.0 fit~\cite{DIScomb,heracharmcomb} was followed. 
The combined H1 and ZEUS inclusive $ep$ NC and CC DIS cross sections~\cite{DIScomb} were used to constrain the PDFs. 
The analysis was restricted to the inclusive data with $Q^2>Q^2_{min}=3.5$~GeV$^2$ to ensure the applicability of pQCD calculations; for the charm data this cut was not applied%
\footnote{For the charm data the applicability of pQCD calculations is ensured by the presence of a massive $c$ quark-antiquark pair in the final state; see also the scale choices.}. 
Theoretical predictions were obtained at NLO using the \ozmod{`FF ABM' and `FF ABM RUNM'} scheme for the heavy-quark pole- and running-mass treatment, respectively, 
as implemented in OPENQCDRAD~\cite{openqcdrad}. 
The factorisation and renormalisation scales were set to $\mu_f=\mu_r=Q$ for the light quarks and to $\mu_f=\mu_r=\sqrt{Q^2+4m_Q^2}$ for the heavy quarks. 
The number of active flavours in PDFs and $\alpha_s$ evolution was set to $n_f=3$. The strong coupling constant was set to $\alpha_s^{n_f=3}(M_Z)=0.105$, 
corresponding to the value $\alpha_s^{n_f=5}(M_Z) = 0.116$. 
The beauty mass was set to $m_b^{\rm pole}=\SI{4.75}{GeV}$ and $m_b(m_b)=\SI{4.18}{GeV}$~\cite{pdg2012} for the pole- and running-mass treatments, respectively.%
\footnote{For the calculation of the beauty contribution to the inclusive cross sections.}

The following combinations of PDFs were chosen in the fit procedure at the initial scale 
of the QCD evolution $Q^2_0= 1.4$~GeV$^2$: the valence-quark distributions $xu_{{v}}(x)$, 
$xd_{{v}}(x)$, the gluon distribution $xg(x)$ and the $u$-type and $d$-type 
anti-quark distributions (note that they are identical to the sea-quark distributions), 
$x\overline{{U}}(x)$, $x\overline{{D}}(x)$, where
$x\overline{{U}}(x) = x\overline{u}(x)$ and $x\overline{{D}}(x) = x\overline{d}(x) + x\overline{s}(x)$. 
At the scale $Q_0$, the PDFs are \ozmod{parametrised} by
\begin{equation}
\begin{array}{l}
		xg(x)=A_gx^{B_g}(1-x)^{C_g}-A'_gx^{B'_g}(1-x)^{C'_g},\\
		xu_v(x)=A_{u_v}x^{B_{u_v}}(1-x)^{C_{u_v}}(1+D_{u_v}x+E_{u_v}x^2),\\
		xd_v(x)=A_{d_v}x^{B_{d_v}}(1-x)^{C_{d_v}},\\
		x\overline{U}(x)=A_{\overline{U}}x^{B_{\overline{U}}}(1-x)^{C_{\overline{U}}}(1+D_{\overline{U}}x),\\
		x\overline{D}(x)=A_{\overline{D}}x^{B_{\overline{D}}}(1-x)^{C_{\overline{D}}}(1+D_{\overline{D}}x).\\
\end{array}
\label{eq:sec:comb:red:ffns:pdfpar}
\end{equation}
The normalisation parameters $A_{u_{\textrm{v}}}$, $A_{d_\textrm{v}}$, $A_g$ were determined by the QCD sum 
rules, \ozmodNN{the $B$ parameters determine the PDFs at small $x$, 
and the $C$ parameters describe the shape of the distributions as $x \to 1$.} 
A flexible form for the gluon distribution was 
adopted with the choice of $C'_g=25$ motivated by the approach of the MSTW group~\cite{Thorne:2006qt,Martin:2009iq}.
The $s$-quark distribution is expressed as
$x$-independent strangeness fraction, $f_s$, of the $d$-type sea, $x\overline{s} = f_sx\overline{D}$ at $Q^2_0$, 
where $f_s=0.31$ as in the analysis of~\cite{Martin:2009iq}. 
Additional constraints $B_{\overline{\textrm{U}}} = B_{\overline{\textrm{D}}}$ and 
$A_{\overline{\textrm{U}}} = A_{\overline{\textrm{D}}}(1 - f_s)$ were imposed, with $x\bar{u} \to x\bar{d}$ as $x \to 0$. 
The parameters $D_{u_v}$, $D_{\bar{U}}$ and $D_{\bar{D}}$ were set to $0$ for the nominal variant of the fit. 
In a compact way, these constraints can be summarised as
\begin{equation}
\begin{array}{l}
		A_{\overline{U}}=A_{\overline{D}}(1-f_s),~f_s=0.31,\\
		B_{\overline{U}}=B_{\overline{D}},\\
		C'_g=25,\\
		D_{u_v}=D_{\overline{U}}=D_{\overline{D}}=0,\\
		\int_{0}^{1}[\sum_{i}(q_i(x)+\overline{q}_i(x))+g(x)]xdx=1,\\
		\int_{0}^{1}[u(x)-\overline{u}(x)]dx=2,\\
		\int_{0}^{1}[d(x)-\overline{d}(x)]dx=1.
\end{array}
\label{eq:sec:comb:red:ffns:pdfparconstr}
\end{equation}
The analysis was performed by fitting the remaining 13 free parameters in~\ref{eq:sec:comb:red:ffns:pdfpar}%
\footnote{Note that a negative gluon distribution was allowed at the parametrisation scale.}. 
The charm mass was left free in the fit.

\ozmodN{The free parameters are determined in HERAFitter by minimisation of a $\chi^2$-function as implemented in the MINUIT package~\cite{minuit}. 
The $\chi^2$ function is similar to that described in Section~\ref{sec:comb:proc:def}: 
\begin{equation}
\begin{split}
	\label{eq:comb:proc:finalchi2fit}
	\chisq=&\sum_{e=1}^{N_e}\sum_{i=1}^{N_m}\frac{\left(m_i-\sum_{j=1}^{N_S}\gamma_i^{e,j} m_i b^{e,j}-\mu_i^e\right)^2}{{\delta_{stat,i}^e}^2 \mu_i^e m_i +{\delta_{uncor,i}^e m_i}^2}+\sum_{j=1}^{N_s}{b^{e,j}}^2\\ 
  +& \sum_{i=1}^{N_m} \ln \frac{ {\delta^e_{{\rm stat},i}}^2 \mu_i^e m_i + ({\delta^e_{{uncor},i}} m_i)^2}{ ({\delta^e_{{\rm stat},i}}^2 +{\delta^e_{{uncor},i}}^2)(\mu^e_i)^2},
\end{split}
\end{equation}
where the notation is equivalent to that in Eq.~\ref{eq:comb:proc:finalchi2}. 
The parameters $m_i$ are theoretical predictions \ozmodNN{which depend on the fitted parameters}. 
}
Systematic uncertainties are assumed to be proportional to the central prediction values, 
whereas statistical uncertainties scale with the square root of the predictions. 
Correlated uncertainties are treated using nuisance-parameter representation~\cite{HERAFitterPaper}. 
\ozmodNN{The $\chi^2$-function includes an additional logarithmic term which is relevant when the estimated statistical and uncorrelated systematic uncertainties of the data 
are rescaled during the fit~\cite{Aaron:2012qi}.}

The uncertainties were evaluated following the strategy of~\cite{DIScomb,heracharmcomb}. These include:
\begin{itemize}
	\item the fit uncertainty was evaluated~\cite{hessian,HERAFitterPaper,Pumplin:2000vx,Pumplin:2002vw} from a $\chi^2$ variation of 1;%
	\footnote{For the $c$-quark mass, the fit uncertainty was determined with the MINOS algorithm~\cite{minuitmanual}.}
	\item the model uncertainties from variation of theory model parameters:
		\begin{itemize}
			\item[$\circ$] $f_s$ was varied in the range $0.23<f_s<0.38$;
			\item[$\circ$] $m_b^{\rm pole}$ and $m_b(m_b)$ were varied in the ranges $4.5<m_b^{\rm pole}<\SI{5.0}{GeV}$ and $4.0<m_b(m_b)<\SI{4.4}{GeV}$ for the pole- and running-mass treatments, respectively;
			\item[$\circ$] $Q^2_{min}$ was varied in the range $2.5<Q^2_{min}<\SI{5.0}{GeV^2}$;
			\item[$\circ$] $\alpha_s^{n_f=3}(M_Z)$ was varied in the range $0.103<\alpha_s^{n_f=3}(M_Z)<0.107$, 
			corresponding to $0.114<\alpha_s^{n_f=5}(M_Z)<0.118$;
			\item[$\circ$] $\mu_f$ and $\mu_r$ for heavy-flavour production were varied simultaneously by a factor of two (the framework allows only their simultaneous variation); 
			the largest differences in this range were taken;
		\end{itemize}
	\item the parametrisation uncertainties:
		\begin{itemize}
			\item[$\circ$] $Q^2_0$ was varied in the range $1.0<Q^2_0<\SI{1.9}{GeV^2}$;
			\item[$\circ$] the parameter $D_{u_v}$ was released;%
				\footnote{I.e.~the fit was performed with 14 free parameters.}
			\item[$\circ$] the parameter $D_{\overline{U}}$ was released;
			\item[$\circ$] the parameter $D_{\overline{D}}$ was released.
		\end{itemize}
\end{itemize}

The fitted values of the pole and running charm masses are
\begin{equation}
	\begin{split}
		m_c^{\rm pole} = 1.334&~\substack{+0.039\\-0.043}{\rm (fit)}~\substack{+0.013\\-0.005}{\rm (mod)} ~\substack{+0.008\\-0.011}(\alpha_s)\\
		& ~\substack{+0.005\\-0.001}({\rm scale}) ~\substack{+0.020\\-0.001}{\rm (par)}~{\rm GeV},\\
		m_c(m_c) = 1.225&~\substack{+0.034\\-0.034}{\rm (fit)} ~\substack{+0.008\\-0.001}{\rm (mod)} ~\substack{+0.007\\-0.009}(\alpha_s)\\
		& ~\substack{+0.009\\-0.005}({\rm scale}) ~\substack{+0.015\\-0.000}{\rm (par)}~{\rm GeV}.
	\end{split}
\end{equation}
For the model uncertainties the $f_s$, $m_b$ and $Q^2_{min}$ variations were added in quadrature, while the $\alpha_s$ and scale uncertainties are quoted separately. 
For the parametrisation uncertainties, the largest differences of all variations was taken. 
The \chisqndof values are $\chisqndof=656/630$ and $\chisqndof=653/630$ for the pole- and running-mass treatments, respectively; 
the partial contribution from the combined charm data is $\chisqndof=66/52$ in both fits. 
These values indicate a good consistency of the fit, 
although they are slightly larger than obtained in the previous analysis~\cite{heracharmcomb} with the `HERA 2012' combined data 
(total $\chisqndof=628/626$ and charm $\chisqndof=44/47$ for the running-mass treatment). 
This might indicate that the more precise charm data require a somewhat more flexible PDF parametrisation. 

The determined $m_c(m_c)$ value is consistent with the `HERA 2012' result $m_c(m_c)=1.26~\pm 0.05{\rm (fit)} ~\pm 0.03{\rm (mod)} ~\pm 0.02{\rm (par)} ~\pm 0.02({\alpha_s})$ GeV 
and has a better accuracy owing to the more precise combined charm data used in the fit and to the usage of all $Q^2$ bins%
\footnote{Note, that for the previous `HERA 2012' result the lowest $Q^2$ bin of the charm data has not been included in the fit; 
repeating the fit with the `HERA 2012' data with the lowest $Q^2$ bin gives the closer value 
$m_c(m_c)=1.228~\substack{+0.048\\-0.038}{\rm (fit)}~\substack{+0.024\\-0.000}{\rm (mod)} ~\substack{+0.022\\-0.006}(\alpha_s) ~\substack{+0.025\\-0.010}({\rm scale}) ~\substack{+0.015\\-0.000}{\rm (par)}~{\rm GeV}$.}. 
The \ozmodN{improvement} of the model, $\alpha_s$ and scale uncertainties is attributed partially to the usage of all $Q^2$ bins 
\ozmodNN{and partially to better constraints on the gluon distribution coming from more accurate charm data, 
which stabilise the fit results against variations of external parameters.}
This value \ozmodN{is in agreement} with the other analyses of the `HERA 2012' charm data performed at NLO and partial NNLO~\cite{Gao:2013wwa,Alekhin:2012vu}%
\footnote{One of the four variants of the fitted $m_c(m_c)$ from~\cite{Gao:2013wwa} is quoted.}:
\begin{equation}
	\begin{split}
		m_c(m_c)=&1.15~\pm 0.04{\rm (exp)} ~\substack{+0.04\\-0.00}{\rm (scale)} {\rm ~GeV} \\
		&[{\rm NLO }~O(\alpha_s^2)~{\rm  (Alekhin)}],\\
		m_c(m_c)=&1.24~\pm 0.03{\rm (exp)} ~\substack{+0.03\\-0.02}{\rm (scale)} ~\substack{+0.00\\-0.07}{\rm (theory)} {\rm ~GeV} \\
		&[{\rm approx.~NNLO }~O(\alpha_s^3)~{\rm  (Alekhin)}],\\
		m_c(m_c)=&1.19 ~\substack{+0.08\\-0.15} {\rm ~GeV} ~~[{\rm NNLO }~O(\alpha_s^2)~{\rm  (CTEQ)}].\\
	\end{split}
\end{equation} 
Some differences between the results are attributed 
to different theoretical settings and procedures of uncertainty estimation 
(for more details see~\cite{Gao:2013wwa,Alekhin:2012vu} and references therein). 
The $m_c(m_c)$ value is also consistent with the world average of $m_c(m_c)=1.275\pm 0.025$~GeV~\cite{pdg2012} defined at two-loop QCD, based on lattice calculations and measurements of time-like processes. 

Finally, note that the fitted running-mass value $m_c(m_c)=\SI{1.225}{GeV}$ corresponds to $m_c^{\rm pole}=m_c(m_c)(1+4\alpha_s(m_c)/3\pi)=\SI{1.417}{GeV}$, 
calculated using the appropriate one-loop relation~\ref{eq:th:runpolmass}, 
which is consisted with the fitted value for $m_c^{\rm pole}=\SI{1.334}{GeV}$, 
\ozmodNN{but} the latter differs significantly from the world-average pole mass $m_c^{\rm pole}=1.67 \pm \SI{0.07}{GeV}$~\cite{pdg2012}, 
calculated from the world-average running mass using the two-loop relation. 
This illustrates one of the possible caveats in determination and usage of the pole mass in applications of pQCD, 
mentioned in Section~\ref{sec:th:qcd:pqcd:mass}. 
Since no attempt has been made to estimate the non-perturbative theoretical uncertainty on $m_c^{\rm pole}$, 
\ozmodN{the presented result should not be considered as a measurement, but rather as extraction of the value, which is optimal for these particular data.} 
\ozmodN{In order to show alternative ways to calculate predictions for charm production, Appendix~\ref{sec:comb:red:vfns} presents a comparison with the theoretical predictions in different VFNS and a determination of optimal $c$-quark mass parameters for these schemes.}

\subsection{Summary}
\label{sec:comb:summary}

Measurements of charm production by the H1 and ZEUS experiments were combined.
The combination was done separately for the single- and double-differential visible \Dstar cross sections, 
and for all available measurements of open charm production extrapolated to the full phase space.
The combination was performed in the kinematic region $1.5<Q^2<\SI{1000}{GeV^2}$ ($5<Q^2<\SI{1000}{GeV^2}$ for the single-differential cross sections), 
$0.02<y<0.7$, $p_T(\Dstar)>\SI{1.5}{GeV}$ and $|\eta(\Dstar)|<1.5$ 
for the visible \Dstar cross sections, and in the region $2.5 \le Q^2 \le \SI{2000}{GeV^2}$ and 
$3 \times 10^{-5}\le x \le 5 \times 10^{-2}$ for the reduced charm cross sections.
The procedure takes into account detailed information on correlations of the systematic uncertainties. 
For both combinations, the data were found to be consistent, and the combined sets exhibit significantly reduced uncertainties. 
The combination of visible \Dstar cross sections does not induce significant theory-related uncertainties, 
while the combination of the reduced charm cross sections presents the most precise charm dataset from HERA, 
\ozmodNN{however} is affected by the theory-related uncertainties \ozmodNN{in} the extrapolation procedure.

For the visible \Dstar cross sections, the combination was performed separately for the single-differential cross sections 
using the HERA-II data only, and for the double-differential cross section using the HERA-I and HERA-II data. 
Inclusion of the HERA-I data allowed an extension of the kinematic region in $Q^2$. 
NLO QCD predictions in the FFNS were compared to the combined \Dstar data. 
The predictions describe the data well within their uncertainties. 
Because the uncertainties of the combined data are smaller than the theoretical uncertainties, 
higher-order calculations and an improved treatment of the fragmentation process \ozmodNN{is needed} 
to reduce the theory uncertainty to a level comparable with the data precision. 
The \Dstar combined data can be used further as the most precise purely experimental charm measurement from HERA 
for tests of pQCD and phenomenological approaches, e.g.\ of the fragmentation process.

The combined reduced charm cross sections are consistent with the previous H1 and ZEUS charm combination and 
have an improved precision owing to the inclusion of new ZEUS measurements.
The combined data were compared to NLO QCD predictions in the FFNS and various VFNS. 
Most of the predictions describe the data well within their uncertainties. 
Similar to the \Dstar combination, the uncertainties of the combined data are smaller than the theoretical uncertainties, thus 
further improvement in the theoretical calculations \ozmodN{is required} to match the data precision. 
The best description of the data in the whole kinematic range is provided by the approximate NNLO FFNS predictions of the ABM group. 
The combined reduced charm cross sections were \ozmodNN{also} used as input for the QCD analysis to determine the optimal values of the \msbar running charm mass 
and $c$-quark mass parameters in different VFNS. The extracted value of the \msbar running charm mass is consistent with the world-average value 
and has competitive precision to other individual determinations in pQCD. 
These data can be used further as the most precise inclusive charm measurement from HERA 
for tests of pQCD and in QCD analyses to constrain the gluon distribution and to determine the $c$-quark mass.

\clearpage
\section{Heavy-flavour production at LHCb}
\label{sec:hflhcb}

This Section provides an overview of measurements of heavy-flavour production at the LHCb experiment. 
This overview is restricted to the selected measurements that are used for the QCD analysis presented in Section~\ref{sec:pdffit}, and 
describes their comparison with theoretical predictions, \ozmodN{including} a discussion of the theoretical uncertainties. 

\subsection{Introduction}
\label{sec:pdffit:intro}

As \ozmodN{outlined} in Section~\ref{sec:th:qcd:factoris}, 
PDFs are a necessary ingredient for QCD predictions in any process with incoming hadrons. 
Since they are not currently calculable from first principles, they must be extracted from data. 
At the present time several groups determine PDFs (for latest results see, 
e.g.\ \cite{Alekhin:2012ig,Accardi:2016qay,Dulat:2015mca,Jimenez-Delgado:2014twa,Abramowicz:2015mha,Harland-Lang:2014zoa,Ball:2014uwa}, 
\ozmodN{and for a recent review see}~\cite{Butterworth:2015oua,Accardi:2016ndt}). 

\begin{figure}[tbp]
  \centering
  \includegraphics[width=1.00\figwidth,trim=0 0 0 0mm,clip=true]{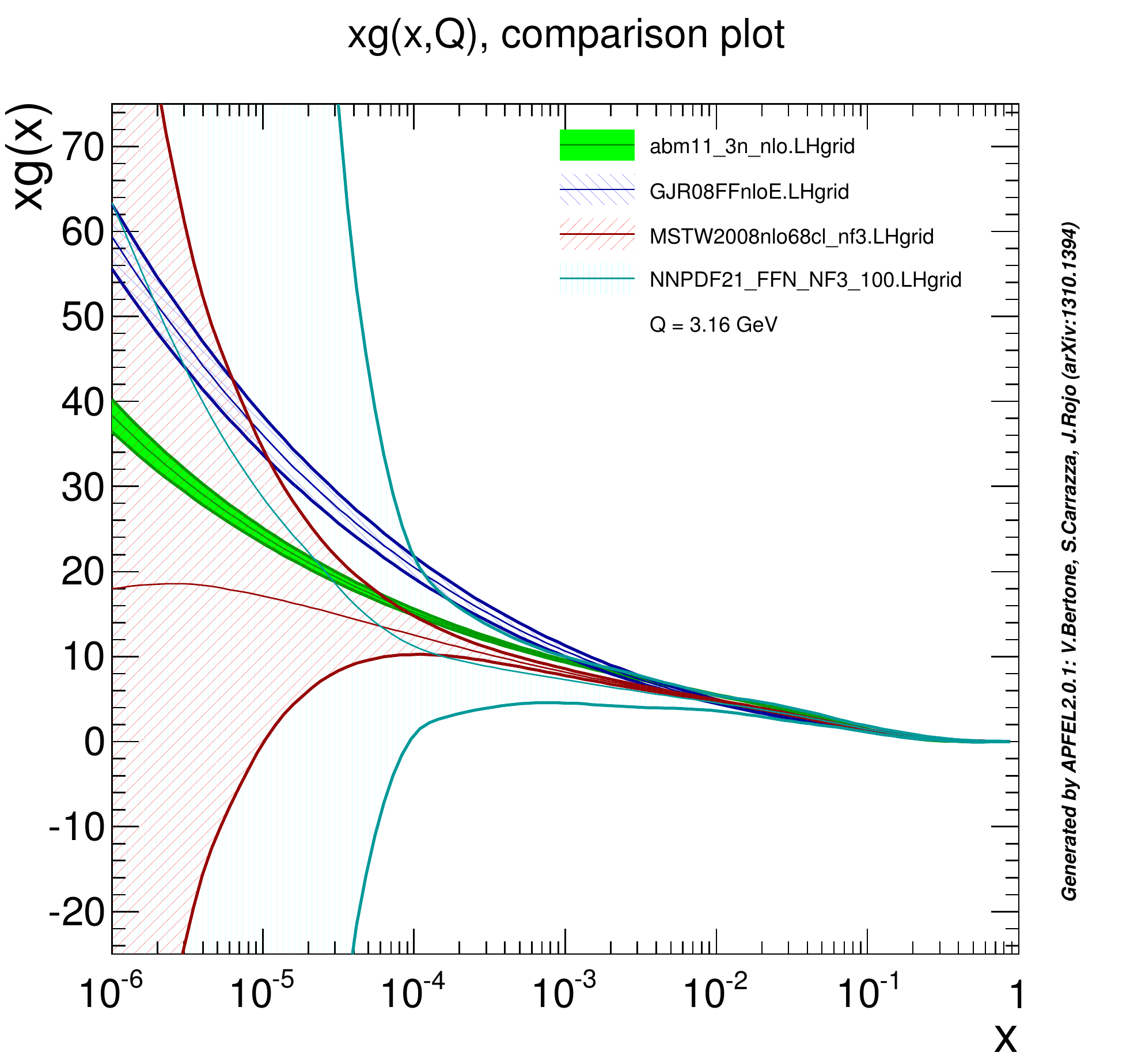}
  \caption[Gluon distributions at $Q^2=\SI{10}{GeV^2}$ from different PDF groups]
  {Gluon distributions at $Q^2=\SI{10}{GeV^2}$ from PDF groups~\cite{abm11,gjr,Martin:2010db,Ball:2011mu} with their uncertainties, represented by bands with different hatch styles. 
	The plot is obtained with the APFEL program~\cite{apfel}.}
	\label{fig:sec:pdffit:benchmark}
\end{figure}

In Fig.~\ref{fig:sec:pdffit:benchmark} the gluon distributions from several PDF sets~\cite{abm11,gjr,Martin:2010db,Ball:2011mu} are compared at the scale $Q^2=\SI{10}{GeV^2}$.%
\footnote{\ozmodNN{Note that} in this and the next Sections $Q^2$ denotes the PDF factorisation scale and not the virtuality of \ep scattering, 
$x$ denotes the longitudinal fraction of the proton momentum and not the Bjorken variable, 
$y$ denotes rapidity and not inelasticity, unless otherwise stated explicitly.} 
The FFNS variants of the fits with the number of active flavours $n_f=3$ were chosen. 
While being consistent with each other and well constrained in the region of medium $x$, 
\ozmodNN{the distributions} have a significant spread between the central values and \ozmodN{also} large uncertainties in the low-$x$ region%
\footnote{The region $x \lesssim 10^{-4}$ will be referred to as \emph{low x}.}, 
since \ozmodNN{presently no data constraining gluons in this region are included in the PDF fits.} 
Note also that within the uncertainty bands, some of the sets predict a negative gluon distribution in the region $x \lesssim 5 \times 10^{-5}$. 
\ozmod{This comparison illustrates that, in spite of using similar input data, 
depending on the assumptions made for a PDF extraction and the methods used to estimate the uncertainties 
a variety of predictions in the unmeasured region exists, and demonstrates the need of further experimental input.}

\ozmodNN{Such data exist from the LHCb Collaboration, which} has measured charm and beauty production in the forward rapidity region $2.0<y<4.5$ 
at the centre-of-mass energy $\sqrt{s}=\SI{7}{TeV}$~\cite{LHCbCharm,LHCbBeauty}%
\ozmod{\footnote{Later on LHCb measurements of charmed meson production at $\sqrt{s}=\SI{13}{TeV}$~\cite{Aaij:2015bpa} and $\sqrt{s}=\SI{5}{TeV}$ have been presented~\cite{Aaij:2016jht}. 
Also measurements of charm production in the central rapidity region were done by ATLAS~\cite{Aad:2015zix} 
and ALICE~\cite{ALICE:2011aa,Abelev:2012tca,Abelev:2012vra,Adam:2016ich}.}}. 
\ozmod{The measured quantities are one-particle inclusive charm- and beauty-hadron production cross sections 
in the $p_T$ ranges $0<p_T<\SI{8}{GeV}$ and $0<p_T<\SI{40}{GeV}$ for the charm and beauty measurements, respectively. 
Since the dominant process for heavy-flavour production in \pp collisions 
at these energies is gluon--gluon fusion (see Section~\ref{sec:th:hq:pp}), these data are sensitive to gluons at low $x$. 
Indeed, the $x$-range can be estimated (neglecting quark-to-hadron fragmentation effects) using the LO formulae of Eq.~\ref{eq:th:pp0x1x2}. 
For given values of $p_T$ and $y$, e.g.\ $y_1$ in Eq.~\ref{eq:th:pp0x1x2}, of one of the produced heavy quarks, 
the probed ranges of the two proton momentum fractions $x_1$ and $x_2$ are:
\begin{equation}
\begin{split}
  {x_1} = \frac{x_2{\rm e}^{y_1}}{\epsilon x_2-{\rm e}^{-{y_1}}},~
  \frac{{\rm e}^{-{y_1}}}{\epsilon - {\rm e}^{y_1}} \le x_2 \le 1, ~ \epsilon = \frac{\sqrt{s}}{m_T}, ~ m_T = \sqrt{M^2+p_T^2}.
\end{split}
\end{equation}
For the other momentum fraction $x_1$ the range is $\frac{{\rm e}^{y_1}}{\epsilon - {\rm e}^{-{y_1}}} \le x_1 \le 1$. 
Thus the lowest $x$ values probed by the LHCb charm data are 
$x \approx {{\rm e}^{-4.5}}/{(7000/1.4 - {\rm e}^{4.5})} \approx 2 \times 10^{-6}$. 
Equation~\ref{eq:th:pp0x1x2cs} \ozmodN{shows} that the cross section is suppressed when $|y_1-y_2|$ becomes large, 
\ozmodN{implying} that the quark and antiquark tend to be produced with the same rapidity. 
Assuming $y_1=y_2=y$ ($p_z=0$ of the produced heavy quark in the parton rest frame) in Eq.~\ref{eq:th:pp0x1x2}, 
the probed ranges of $x_1$ and $x_2$ become
\begin{equation}
\begin{split}
  x_{1,2} = 2{\rm e}^{\pm y} / \epsilon = 2e^{\pm y}\frac{m_T}{\sqrt{s}}.
\end{split}
\end{equation}
This estimation gives the lowest $x$ values probed by the LHCb charm data $x \approx 2e^{-4.5} \times {1.4}/{7000} \approx 4 \times 10^{-6}$. 
The low-$x$ data in the LHCb experiment provide new information to pin down
the gluons \ozmodNN{in the region unconstrained by HERA.} 
The corresponding LO theoretical predictions can be obtained by using Eq.~\ref{eq:th:pp0x1x2cs} (Section~\ref{sec:th:hq:pp}) 
and integrating over the rapidity of the unmeasured produced heavy quark over its kinematically allowed range 
$-{\rm ln}(\epsilon - {\rm e}^{-y_1}) \le y_2 \le {\rm ln}(\epsilon - {\rm e}^{y_1})$. 
The \ozmodNN{actually effective $(x_1,x_2)$ region} probed by these data is presented in Section~\ref{sec:pdffit:th:kinlowpt} using NLO calculations. 
}

\begin{figure*}[tbp]
  \centering
  \includegraphics[width=1.75\figwidth,trim=4mm 17mm 3mm 16mm,clip=true]{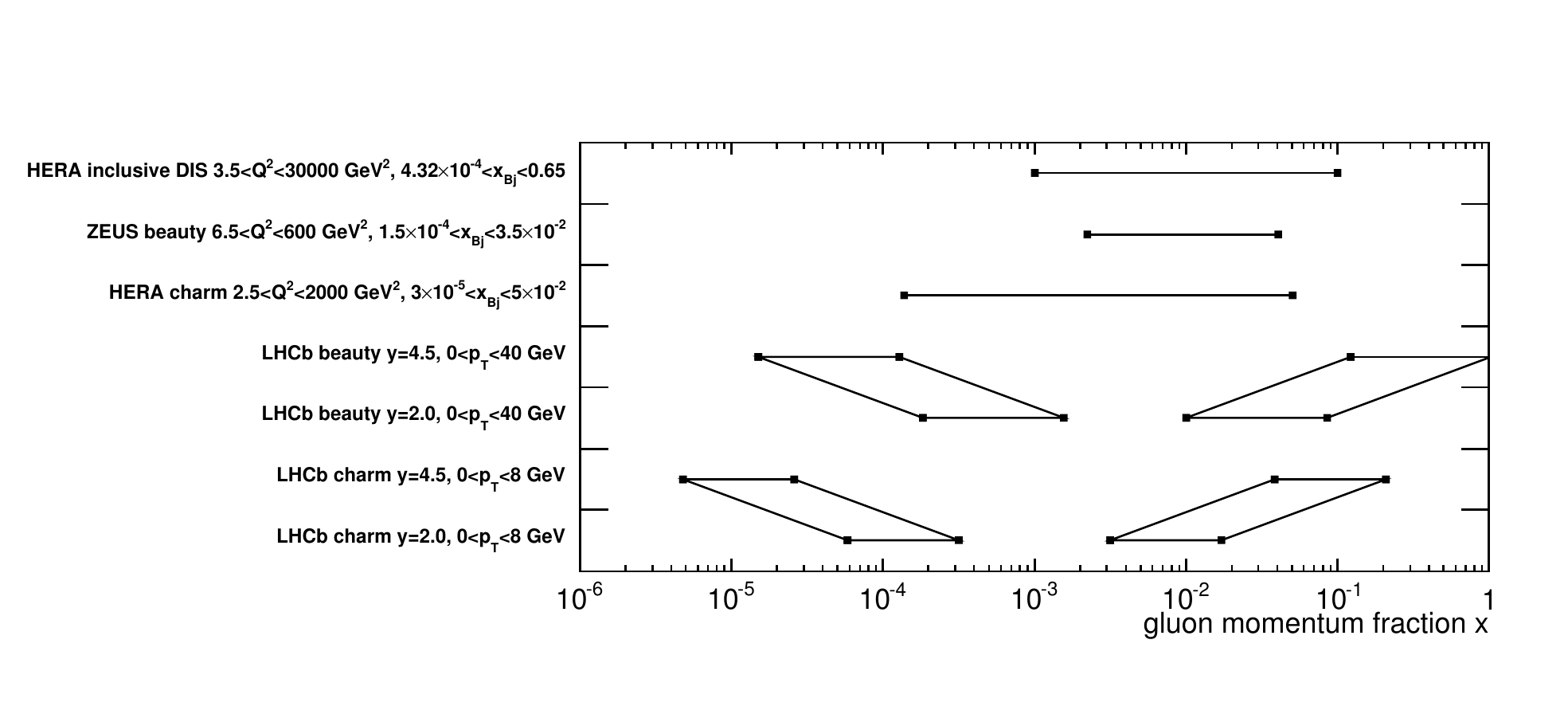}
  \caption[Kinematics in gluon $x$ as covered by HERA and LHCb data]
  {Kinematics in the gluon $x$ space as covered by the different HERA and LHCb data.}
	\label{fig:sec:pdffit:kinematics}
\end{figure*}

In Fig.~\ref{fig:sec:pdffit:kinematics} the kinematic regions which are covered by different HERA and LHCb data are plotted. 
The precise HERA DIS data~\cite{DIScomb} are only indirectly sensitive to gluons, so they constrain the gluon 
distribution well only in the region $10^{-3} \lesssim x \lesssim 10^{-1}$. 
The HERA heavy-flavour data~\cite{heracharmcomb,zeussecvtx_hera2} cover the region $10^{-4} \lesssim x \lesssim 10^{-2}$, 
while the LHCb data extend the coverage up to $x \lesssim 5 \times 10^{-6}$ (low-$p_T$ forward charm) and 
up to $x \lesssim 1$ (high-$p_T$ forward beauty).%
\footnote{The quoted regions are qualitatively determined in the following way: 
for the HERA DIS data, the $x$ range is indicated, where the gluon HERAPDF1.0~\cite{DIScomb} uncertainty 
at $Q^2=\SI{10}{GeV^2}$ is less than 10\%. 
For the HERA charm and beauty data, the LO formula $x=x_{\rm bj}(1+4m_Q^2/Q^2)$ is used, where $x_{\rm bj}$ is the Bjorken variable, 
$Q^2$ is boson virtuality and $m_Q$ is the heavy-quark pole mass. 
For the LHCb charm and beauty data, the LO estimation is used as described above.} 
Note that the LHCb data are sensitive to the product 
of gluon densities in two non-overlapping regions: forward and medium; since the latter is already well constrained by other 
data, the LHCb data \ozmod{will have} an impact mainly on the \ozmodN{low-$x$ region}. 
It is worth noting that using PDFs with \ozmodN{a strongly negative values of the gluon distribution} at low $x$ results in negative 
\ozmodN{and thus unphysical} predicted cross sections for the forward region of the LHCb charm data. 
This \ozmodNN{emphasises} the inclusion of the LHCb data in a PDF fit to constrain gluons at low $x$.

\subsection{\ozmod{Measurements of charm and beauty production at LHCb}}
\label{sec:exp:lhcb}

The Large Hadron Collider (LHC) is the world's largest and most powerful particle collider; 
its description can be found elsewhere~\cite{Evans:2008zzb}.
The LHCb detector at the LHC provides
unique \ozmodN{information} to the forward-rapidity region with a detector that is
tailored for flavour physics, 
therefore LHCb heavy-flavour data provide a unique access to the gluon distribution in the proton 
at very low values of the partonic momentum fraction  $x$. 
\ozmodN{In this review LHCb data published up to August 2014 are included.}

\begin{figure}[htbp]
  \centering
  \includegraphics[width=1.0\figwidth]{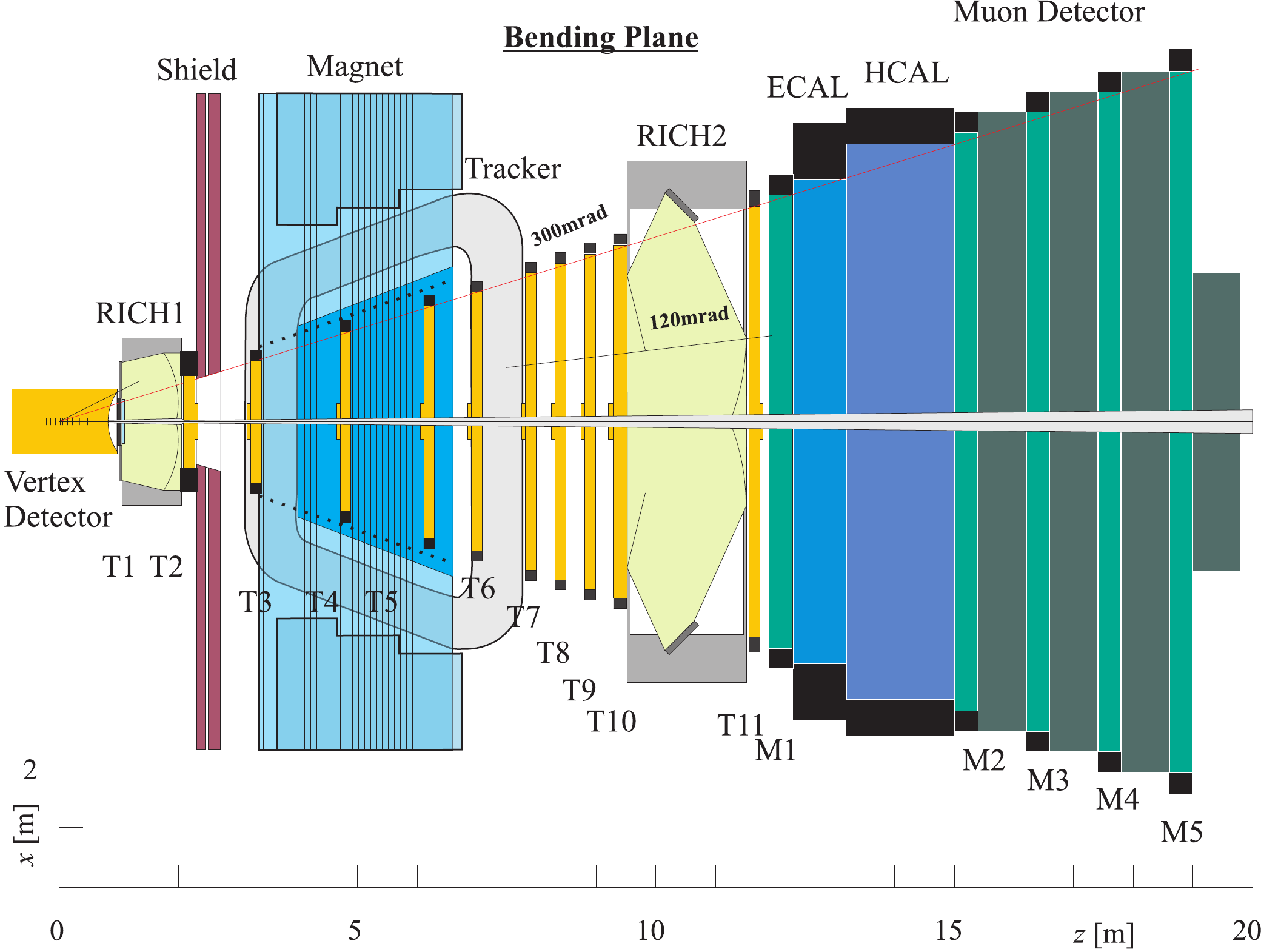}
  \caption[Schematic view of LHCb detector along beampipe]
	{A schematic view of the LHCb detector along the beampipe.}
  \label{fig:lhcb}
\end{figure}

The LHCb detector~\cite{Alves:2008zz} (Fig.~\ref{fig:lhcb}) is a single-arm forward spectrometer covering the pseudorapidity
range $2<\eta<5$, 
designed for the study of particles containing $b$ or $c$ quarks. 
The right-handed coordinate system adopted has the $\text{Z}$ axis along the beam. 
The detector \ozmodN{has} a high-precision tracking system consisting of a silicon-strip vertex detector
surrounding the \pp interaction region, a large-area silicon-strip detector located upstream
of a dipole magnet with a bending power of about $\SI{4}{Tm}$, and three stations of silicon-strip
detectors and straw drift-tubes placed downstream. The combined tracking system has a
momentum resolution ($\delta p/p$) that varies from 0.4\% at $\SI{5}{GeV}$ to 0.6\% at $\SI{100}{GeV}$ and
an impact-parameter resolution of $\SI{20}{\micro\metre}$ for tracks with high transverse momentum.
Charged hadrons are identified using two ring-imaging Cherenkov detectors (RICH). 
The RICH system~\cite{Adinolfi:2012qfa} of the LHCb experiment provides charged-particle identification over
a wide momentum range, from 2 to $\SI{100}{GeV}$%
\footnote{The typical momentum (in the laboratory frame) of the decay products in two-body $B$ decays is about $\SI{50}{GeV}$. 
The requirement of maintaining a high efficiency for the reconstruction of these decays leads
to the need for particle identification up to at least $\SI{100}{GeV}$. The lower momentum limit
of about $\SI{2}{GeV}$ follows from the need to identify decay products from high-multiplicity $B$
decays and also from the fact that particles below this momentum will not pass through the
dipole magnetic field ($\SI{4}{Tm}$) of the LHCb spectrometer~\cite{Adinolfi:2012qfa}.}. 
It consists of two RICH detectors that cover
between them the angular acceptance of the experiment, $15\text{--}\SI{300}{mrad}$ with respect to the beam axis.
Photon, electron, and hadron candidates are identified by a calorimeter system consisting of
scintillating-pad and pre-shower detectors, an electromagnetic calorimeter, and a hadronic
calorimeter. Muons are identified by a system composed of alternating layers of iron and
multi-wire proportional chambers.

\subsubsection{Measurement of prompt charm production}
\label{sec:exp:lhcb:charm}

\begin{figure}[htbp]
  \centering
  \includegraphics[width=0.495\figwidth,trim = 6mm 1mm 4mm 0,clip=true]{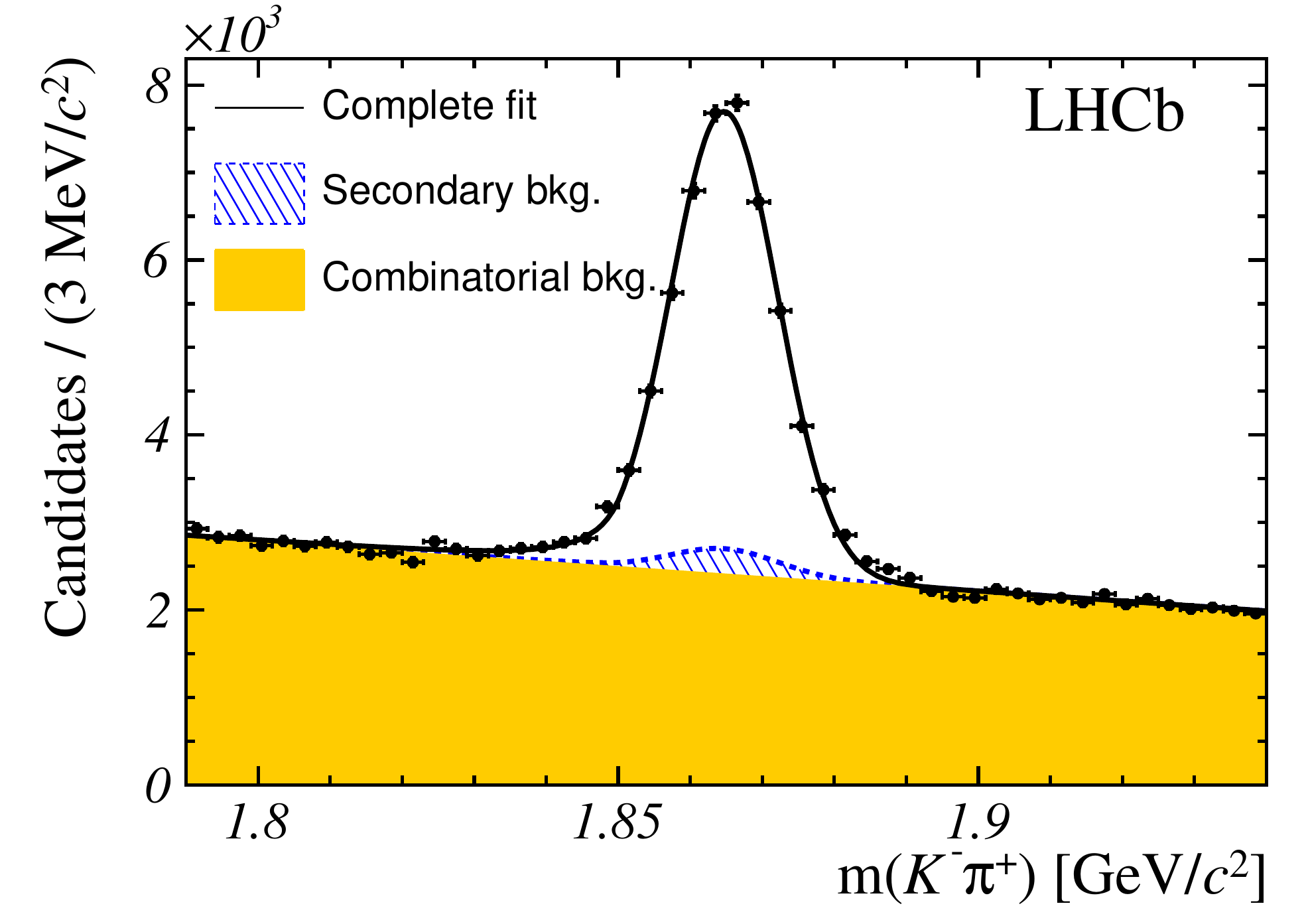}
  \includegraphics[width=0.495\figwidth,trim = 6mm 1mm 4mm 0,clip=true]{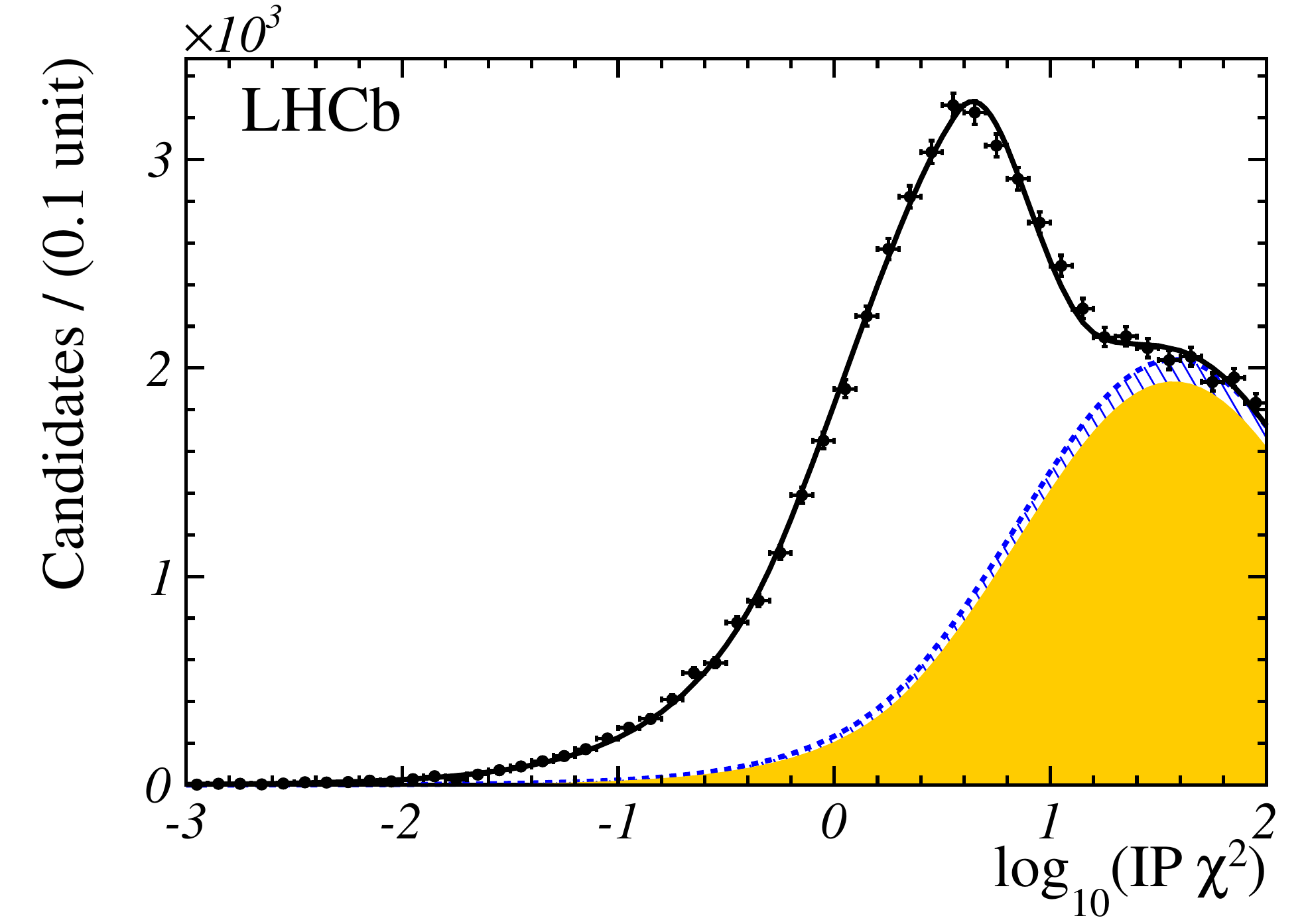}
  \includegraphics[width=0.495\figwidth,trim = 6mm 1mm 4mm 0,clip=true]{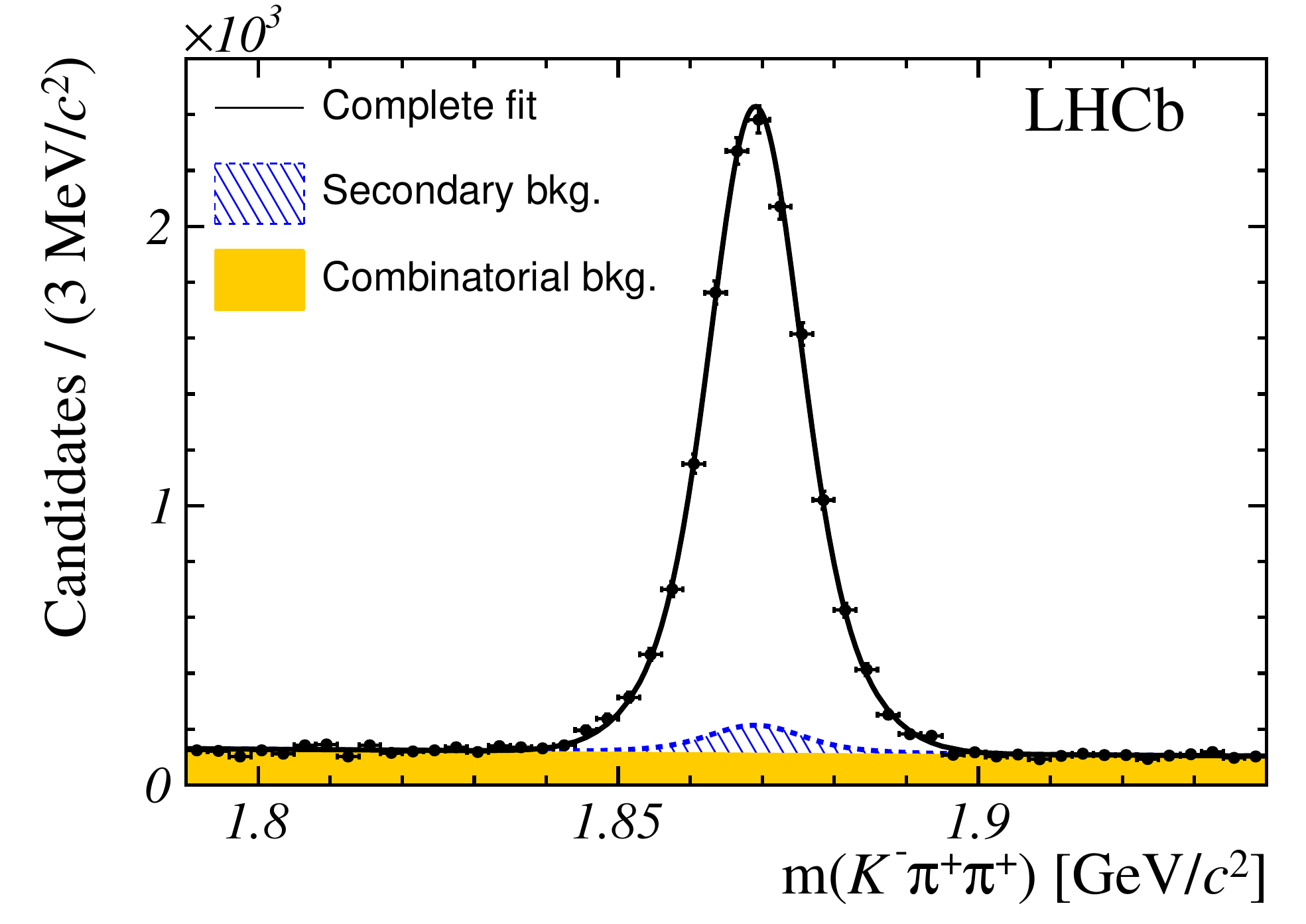}
  \includegraphics[width=0.495\figwidth,trim = 6mm 1mm 4mm 0,clip=true]{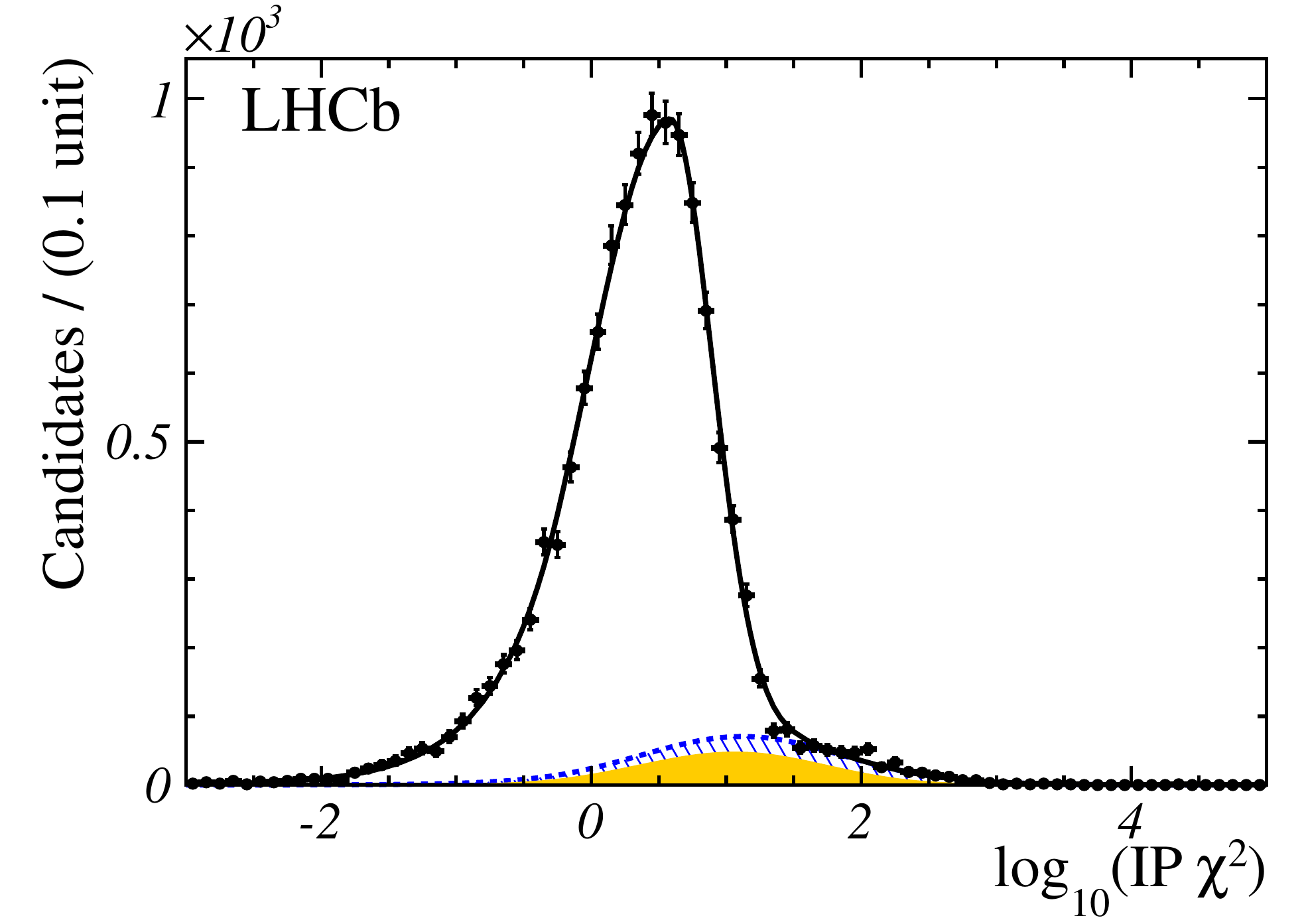}
  \caption[Mass and $\log_{10}({\rm IP}~\chisq)$ distributions from LHCb charm measurement] 
	{Mass and $\log_{10}({\rm IP} \chisq)$ distributions for selected $D^0 \to K^{-}\pi^{+}$ 
	and $D^{+} \to K^{-}\pi^{+}\pi^{+}$ candidates from the LHCb measurement of prompt charm production~\cite{LHCbCharm} 
	showing the masses of the $D^0$ candidates (top left), 
	$\log_{10}({\rm IP}~\chisq)$ distribution of $D^0$ candidates (top right), 
	masses of the $D^{+}$ candidates (bottom left) and $\log_{10}({\rm IP}~\chisq)$ distribution of $D^{+}$ candidates (bottom right).
	Projections of likelihood fits to the full data samples are shown with components as indicated in the legends.}
  \label{fig:lhcb:charm}
\end{figure}

LHCb measured $D^0$, $D^{+}$, $D_s^{+}$, $\Dstar$ and $\Lambda_c^{+}$ production using data 
corresponding to an integrated luminosity of $\SI{15}{nb^{-1}}$
in the region of rapidity $2.0 < y < 4.5$ and transverse momentum $0<p_T<\SI{8}{GeV}$ in \pp collisions 
at a centre-of-mass energy of $\SI{7}{TeV}$~\cite{LHCbCharm}. 
The analysis was based on fully reconstructed decays of charmed hadrons in the following
decay modes (Fig.~\ref{fig:lhcb:charm}): $D^0 \to K^{-}\pi^{+}$, $D^{+} \to K^{-}\pi^{+}\pi^{+}$, $\Dstar \to D^0(K^{-}\pi^{+})\pi^{+}$, 
$D_s^{+} \to \phi(K^{-}K^{+})\pi^{+}$ and $\Lambda_c^{+} \to pK^{-}\pi^{+}$.

Charmed hadrons may be produced at the \pp collision point either directly or as 
feed-down from the instantaneous decays of excited resonances. They may also \ozmodN{arise} 
in decays of beauty hadrons. The first two sources (direct production
and feed-down) are referred to as prompt. Charmed particles from beauty-hadron decays are
called secondary charmed hadrons. \ozmod{The result of the measurement is reported as} the production
cross sections of prompt charmed hadrons; secondary charmed hadrons were treated as
background. The measurement was performed in two-dimensional bins of $p_T$ and $y$. 
For the $\Lambda_c^{+}$ measurement, only single-differential cross sections 
as a function of $p_T$ and $y$ were measured. The prompt signal yields were selected using 
multi-dimensional extended maximum likelihood fits to the mass and $\log_{10}({\rm IP}~\chisq)$, 
where ${\rm IP}~\chisq$ is defined as the difference between the \chisq of the primary vertex, reconstructed 
with and without the considered particle~\cite{LHCbCharm} (Fig.~\ref{fig:lhcb:charm}). The dominant systematic uncertainty is 
the uncertainty on the tracking efficiency, which is $3\text{--}4$\% per one final-state track, 
thus resulting in $6\text{--}10$\% for the measured cross sections.

\begin{figure*}[htbp]
  \sidecaption
  \begin{minipage}[t]{0.30\textwidth}
  \includegraphics[width=1.0\textwidth,trim = 3mm 0 2mm 0,clip=true]{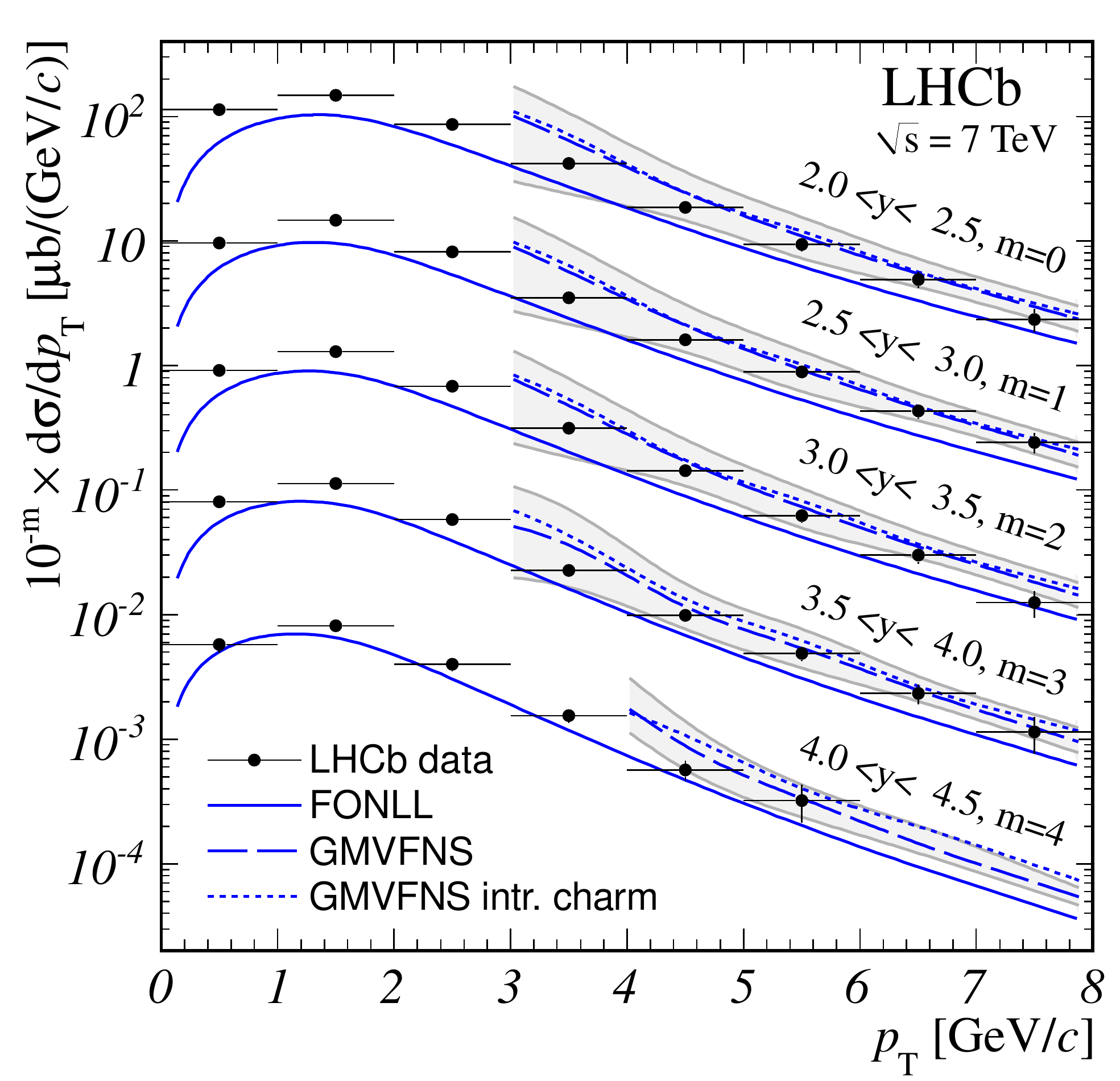}
  \includegraphics[width=1.0\textwidth,trim = 3mm 0 2mm 0,clip=true]{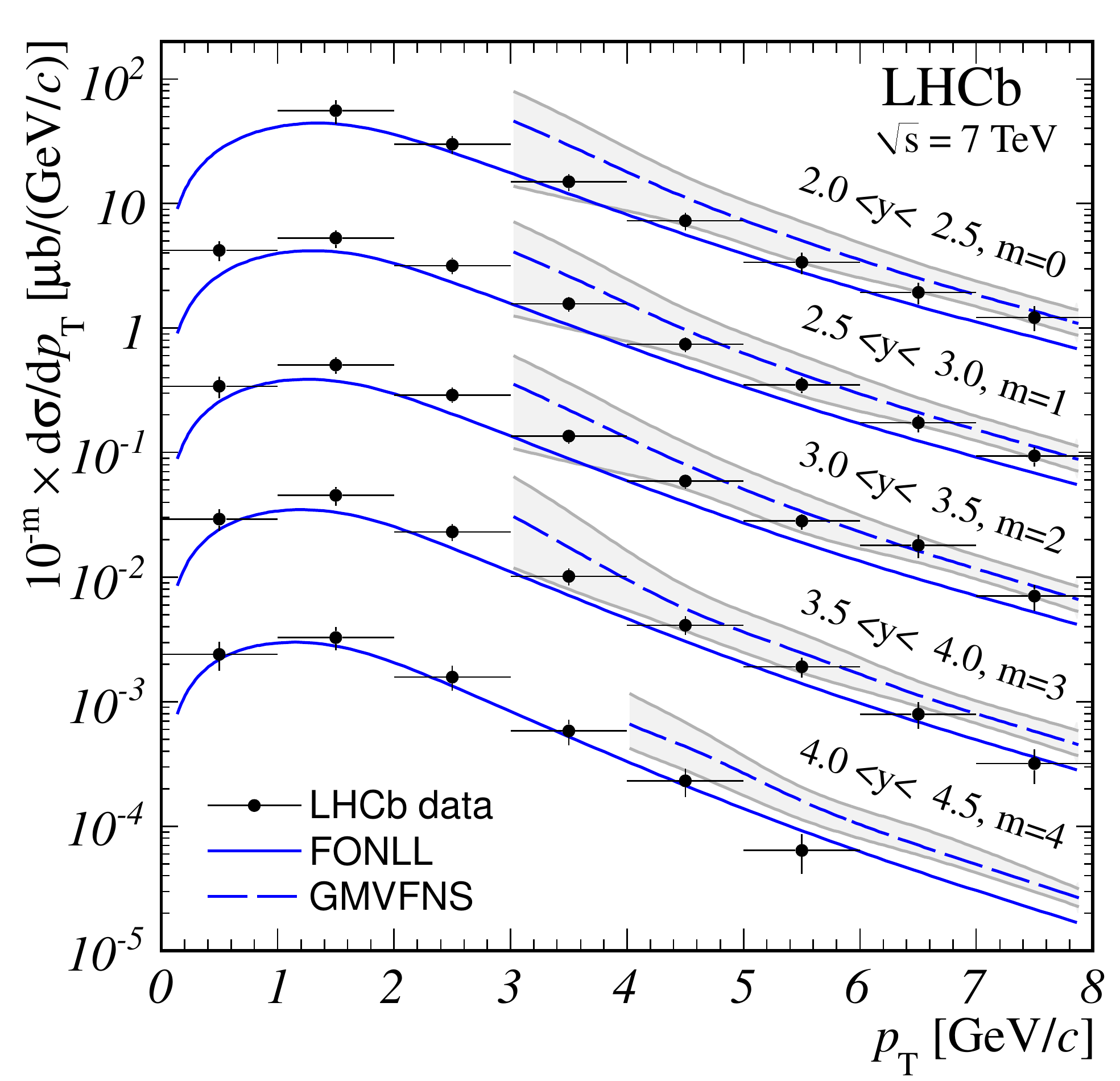}
  \end{minipage}
  \begin{minipage}[t]{0.30\textwidth}
  \includegraphics[width=1.0\textwidth,trim = 3mm 0 2mm 0,clip=true]{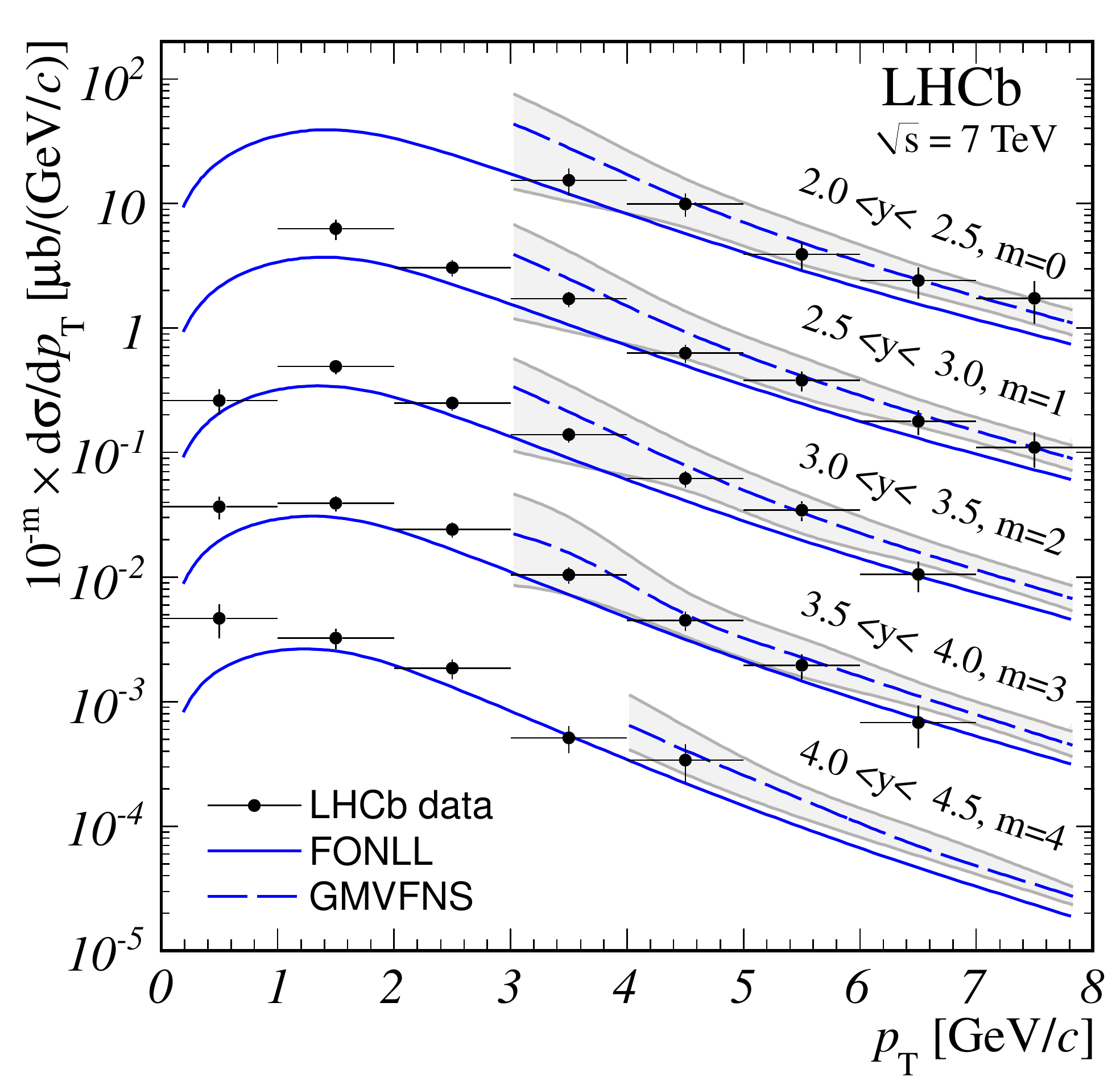}
  \includegraphics[width=1.0\textwidth,trim = 3mm 0 2mm 0,clip=true]{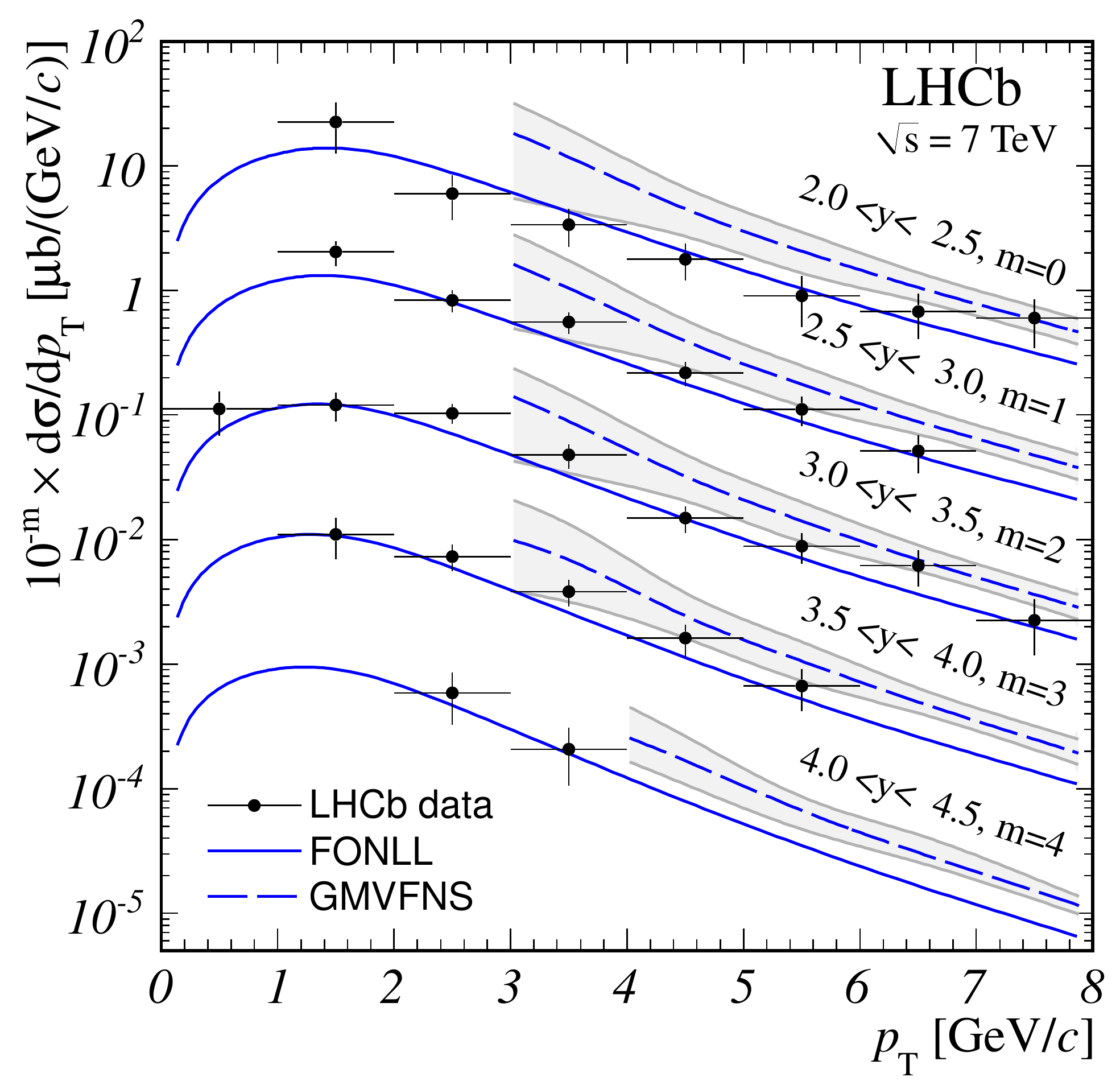}
  \end{minipage}
  \caption[Differential cross sections from LHCb charm measurement] 
	{Differential cross sections for $D^0$ (top left), $\Dstar$ (top right), $D^{+}$ (bottom left) 
	and $D_s^{+}$ (bottom right) from the LHCb measurement of prompt charm production~\cite{LHCbCharm} 
	compared to the theoretical predictions as provided by external groups at NLO in the FONLL~\cite{Cacciari:1998it,frag03,frag06,Cacciari:2012ny} 
	and other GM-VFNS approach~\cite{Kniehl:2004fy,Kniehl:2005de,Kniehl:2005ej,Kneesch:2007ey,Kniehl:2009ar,Kniehl:2012ti}. 
	The cross sections for different $y$ regions are shown as functions of $p_T$. 
	The $y$ ranges are shown as separate curves and associated sets of points scaled
	by factors $10^{-m}$, where the exponent $m$ is shown on the plot with the $y$ range.}
  \label{fig:lhcb:charmcs}
\end{figure*}

The measured double-differential cross sections of $D^0$, $D^{+}$, $D_s^{+}$ and $\Dstar$ production 
are shown in Fig.~\ref{fig:lhcb:charmcs} and compared to the theoretical predictions as provided by external groups at NLO 
in the FONLL~\cite{Cacciari:1998it,frag03,frag06,Cacciari:2012ny} 
and other GM-VFNS approach~\cite{Kniehl:2004fy,Kniehl:2005de,Kniehl:2005ej,Kneesch:2007ey,Kniehl:2009ar,Kniehl:2012ti} (see Sections~\ref{sec:th:hq:pp:fonll} and~\ref{sec:th:hq:pp:gmvfns}). 
The GM-VFNS predictions are shown for $p_T>\SI{3}{GeV}$. 
Predictions for $D^0$ mesons are also
compared with the GM-VFNS calculations using PDFs with intrinsic charm~\cite{Pumplin:2007wg}. 
As shown in Fig.~\ref{fig:lhcb:charmcs}, in the phase space of the present measurement the effect of intrinsic charm
is predicted to be small. All theoretical calculations describe the data well, 
although their uncertainties of the order of a factor $2$ 
significantly exceed the experimental uncertainties of the data.

\subsubsection{Measurement of beauty production}
\label{sec:exp:lhcb:beauty}

LHCb measured $B^{+}$, $B^{0}$ and $B_s^{0}$ production using data 
corresponding to an integrated luminosity of $\SI{0.36}{fb^{-1}}$ 
in the region of rapidity $2.0 < y < 4.5$ and transverse momentum $0<p_T<\SI{40}{GeV}$ in \pp collisions 
at a centre-of-mass energy of $\SI{7}{TeV}$~\cite{LHCbBeauty}. 
The analysis was based on fully reconstructed decays of beauty hadrons in the following
decay modes: $B^{+} \to J/\psi K^{+}$, $B^{0} \to J/\psi K^{*0}$ and $B_s^{0} \to J/\psi \phi$, with 
$J/\psi \to \mu^{+}\mu^{-}$, $K^{*0} \to K^{+}\pi^{-}$ and $\phi \to K^{+}K^{-}$. 
Similar \ozmodN{to the case of} the charm measurement~\cite{LHCbCharm}, 
\ozmodNN{also the beauty measurement} was performed in two-dimensional bins of $p_T$ and $y$.

\begin{figure}[htbp]
  \centering
  \begin{minipage}[t]{0.49\textwidth}
  \includegraphics[width=0.5\figwidth,trim = 1mm 0 6mm 0,clip=true]{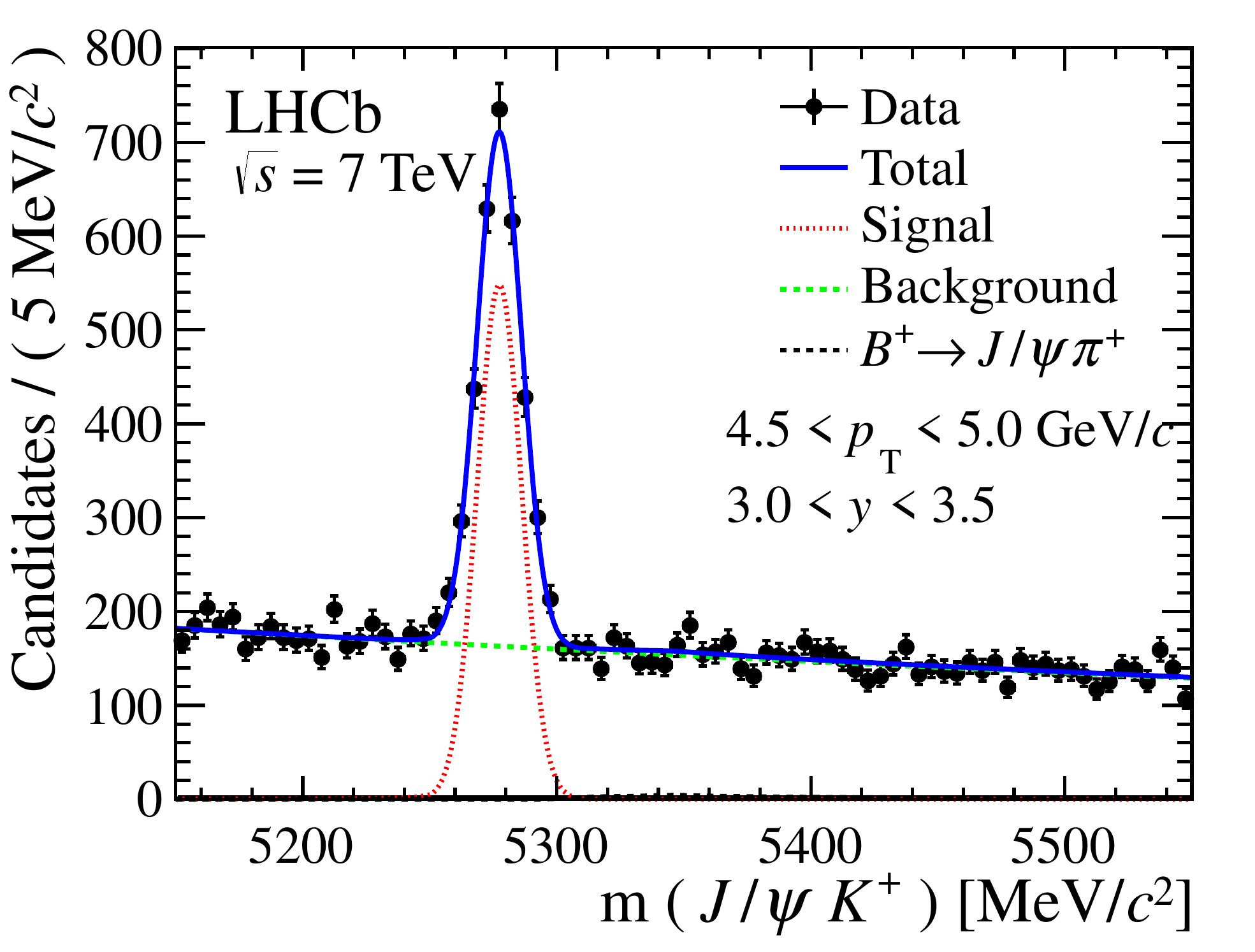}
  \includegraphics[width=0.5\figwidth,trim = 1mm 0 6mm 0,clip=true]{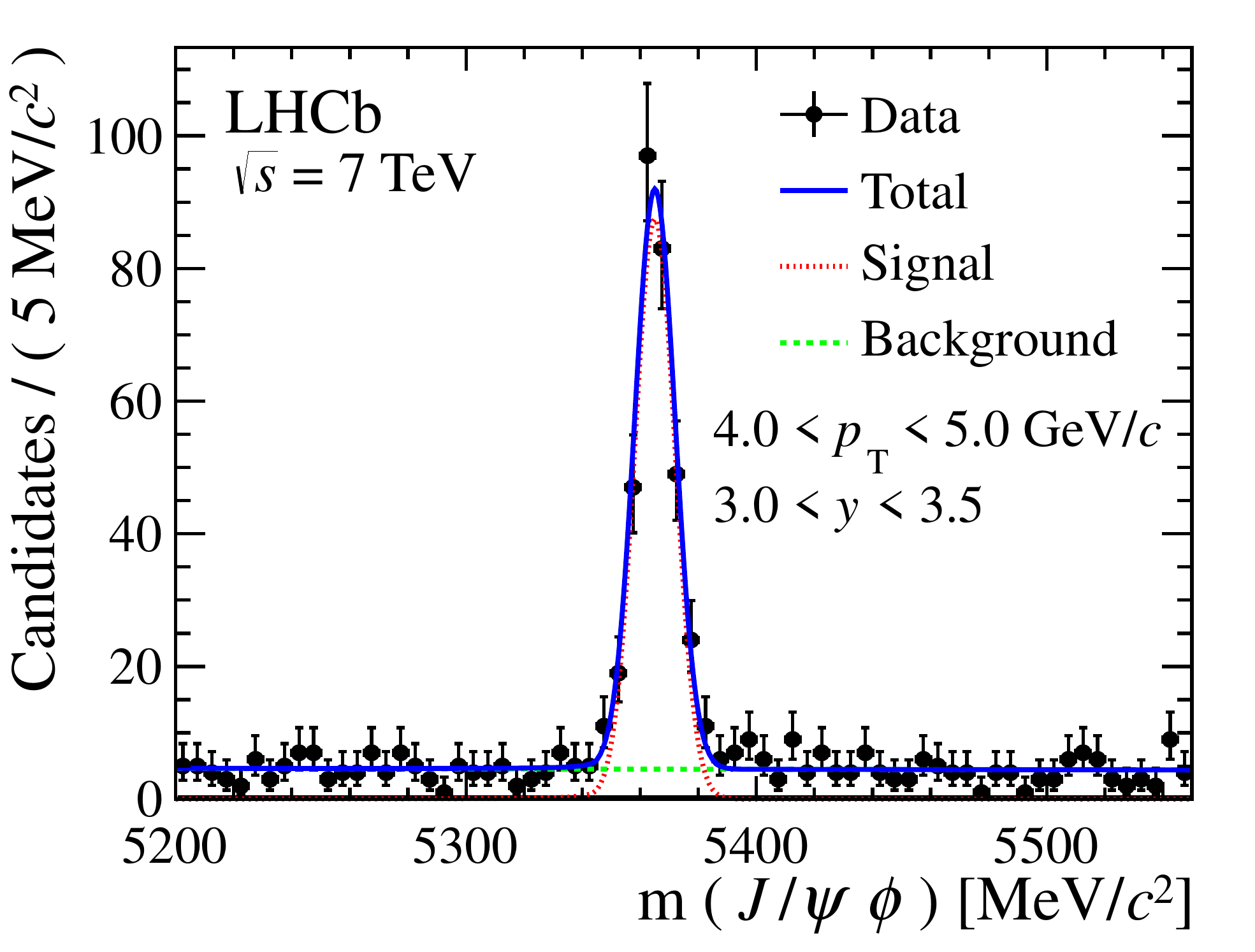}
  \end{minipage}
  \begin{minipage}[t]{0.49\textwidth}
  \includegraphics[width=0.5\figwidth,trim = 1mm 0 6mm 0,clip=true]{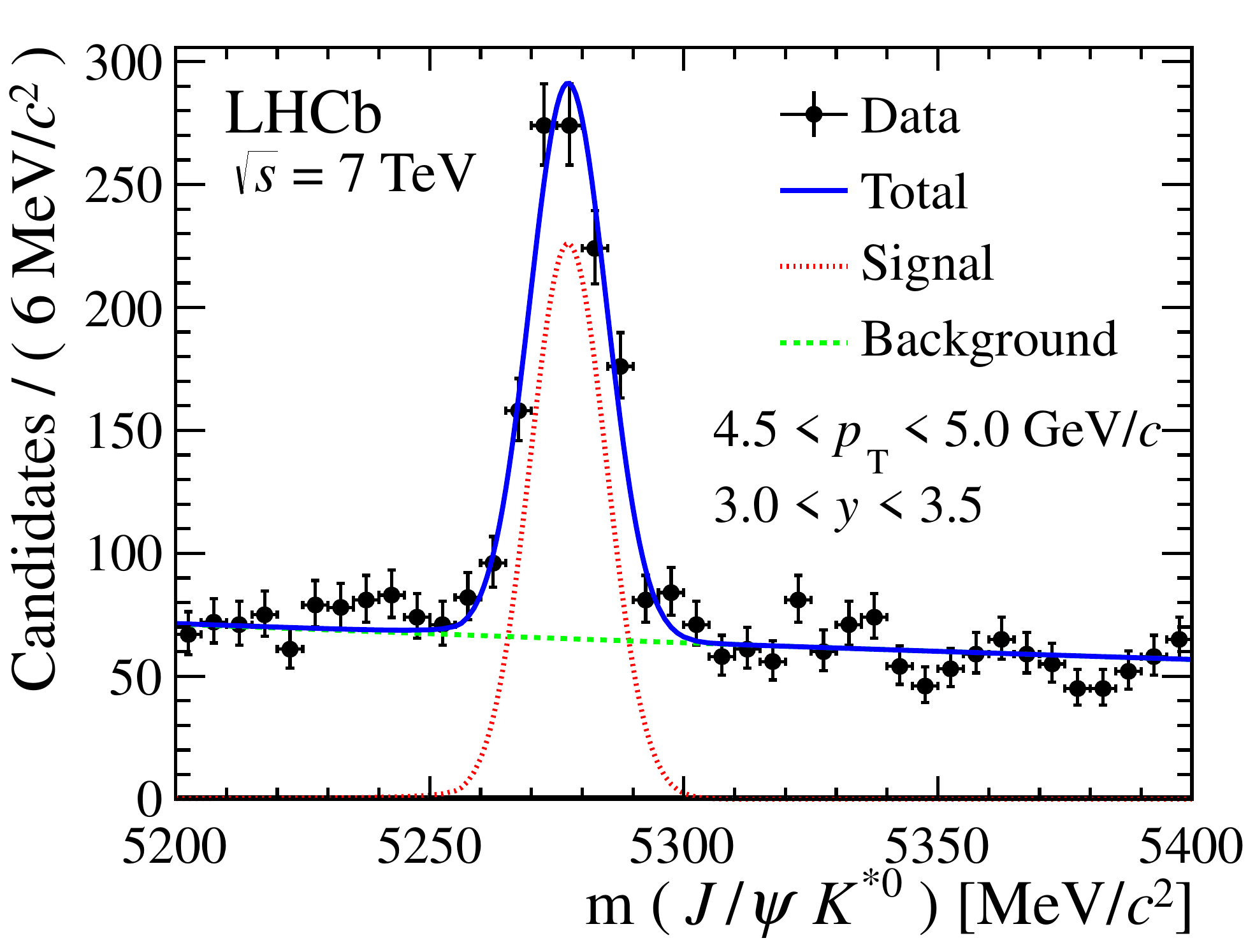}
  \caption[Mass distributions from LHCb beauty measurement] 
	{Invariant mass distributions of the selected candidates 
	from the LHCb measurement of beauty production~\cite{LHCbBeauty} for $B^{+}$ in the range $4.5 < p_T < \SI{5.0}{GeV}$, $3.0 < y < 3.5$ (top left), 
	 $B^0$ in the range $4.5 < p_T < \SI{5.0}{GeV}$, $3.0 < y < 3.5$ (top right), 
	 and $B_s^0$ in the range $4.0 < p_T < \SI{5.0}{GeV}$, $3.0 < y < 3.5$ (bottom).}
  \label{fig:lhcb:beauty}
  \end{minipage}
\end{figure}

The mass distributions of the selected candidates for one of the $p_T$ and $y$ bins 
are shown in Fig.~\ref{fig:lhcb:beauty}.
The dominant systematic uncertainty comes from the tracking ($2\text{--}9\%$) 
and trigger ($2\text{--}8\%$) efficiencies and finite size of the bins ($0\text{--}19\%$); 
for $B^0$ and $B_s^0$ the branching-ratio uncertainties are also sizeable ($\approx 10\%$).

\begin{figure}[htbp]
  \centering
  \begin{minipage}[t]{0.49\textwidth}
  \includegraphics[width=0.5\figwidth,trim = 0 0 6mm 7mm,clip=true]{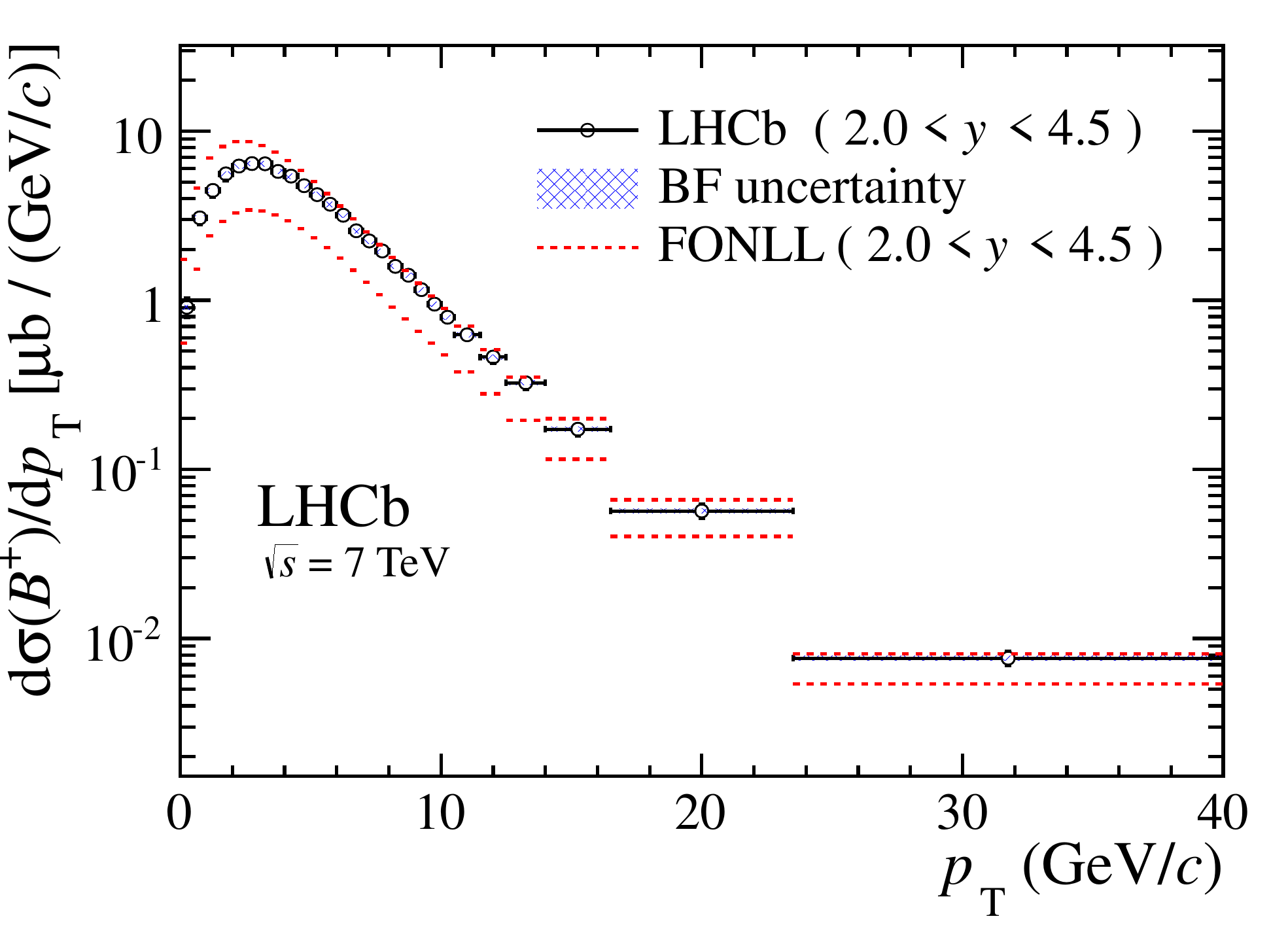}
  \includegraphics[width=0.5\figwidth,trim = 0 0 6mm 7mm,clip=true]{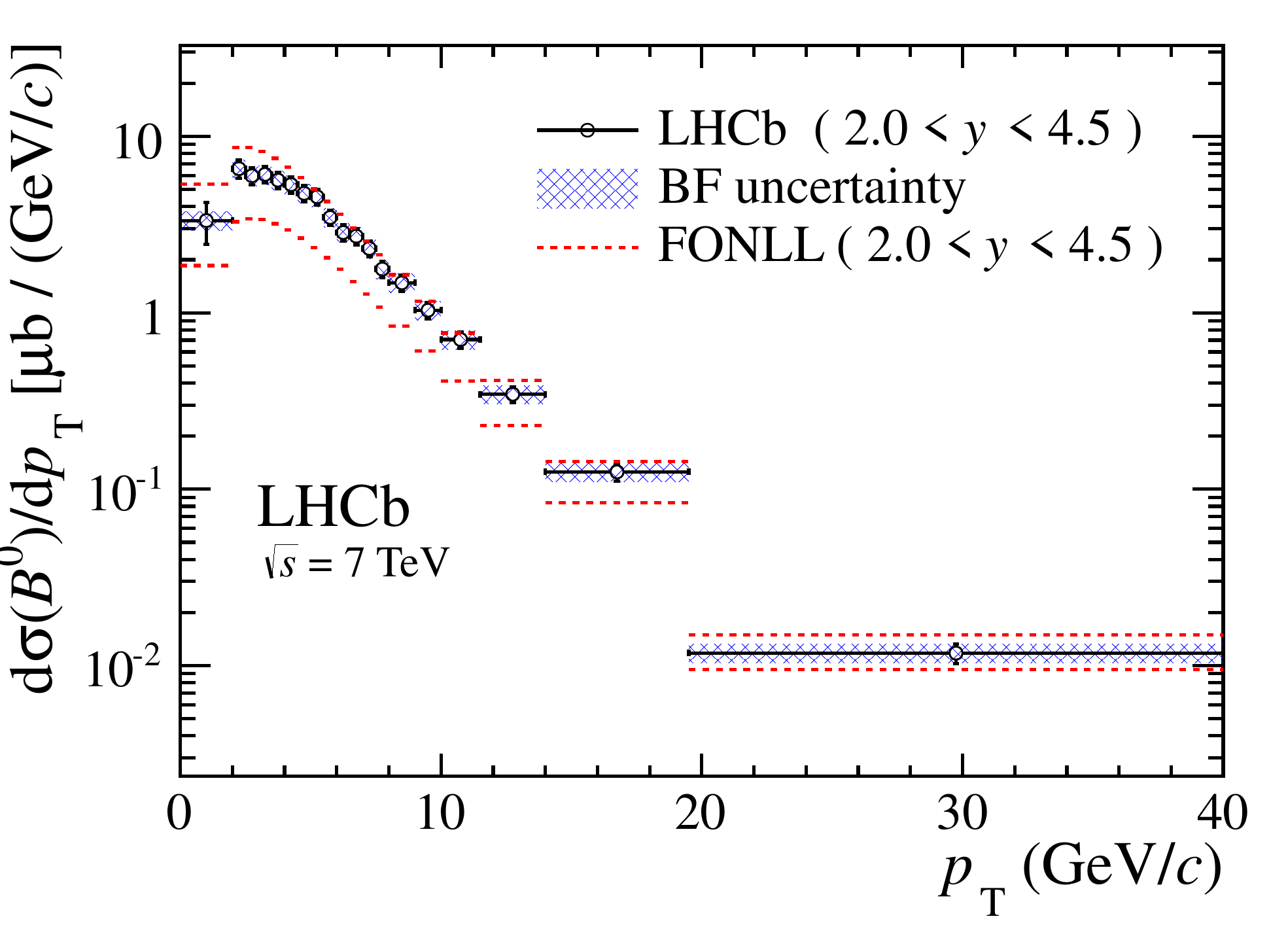}
  \end{minipage}
  \begin{minipage}[t]{0.49\textwidth}
  \includegraphics[width=0.5\figwidth,trim = 0 0 6mm 7mm,clip=true]{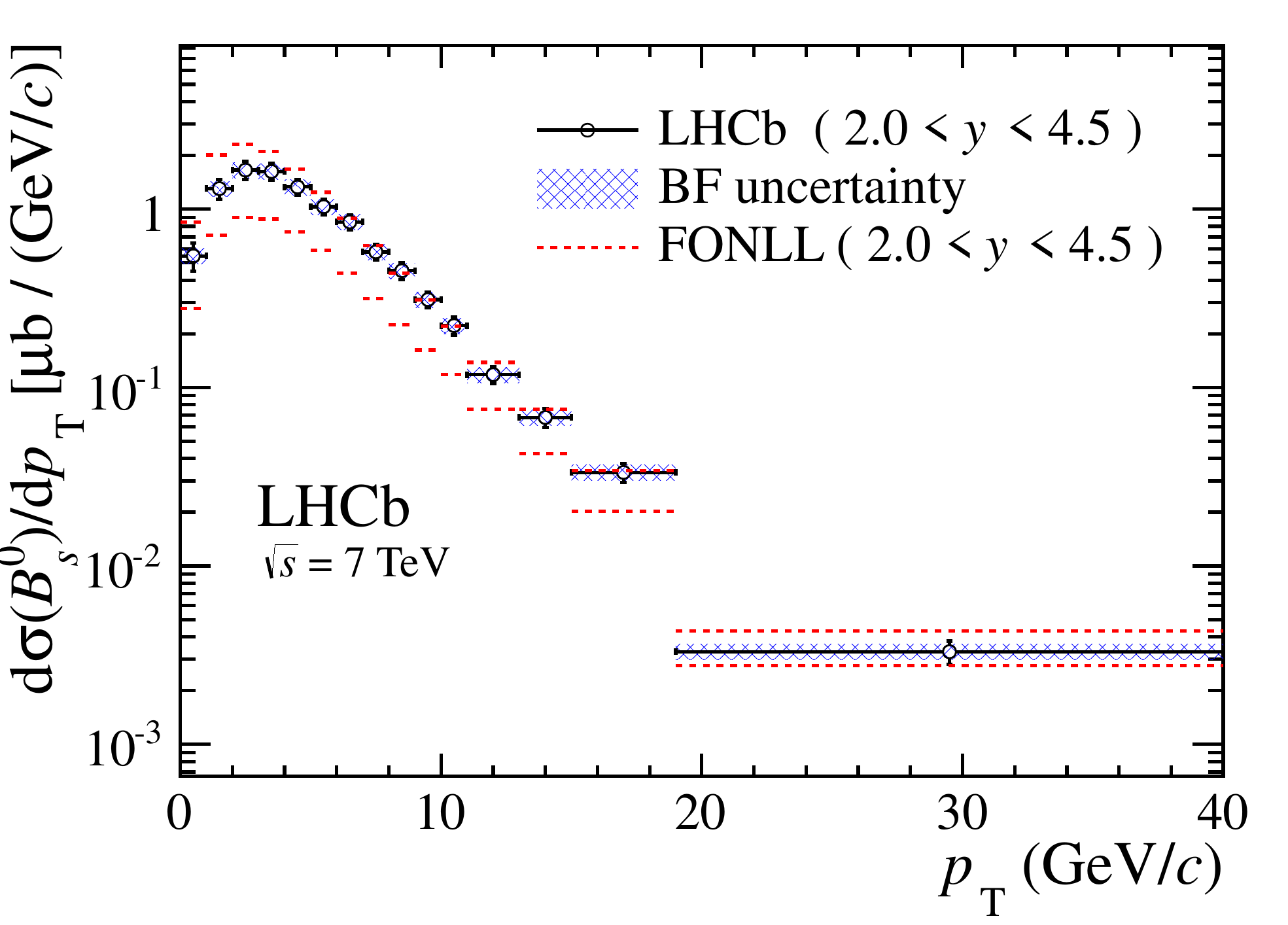}
  \caption[Differential cross sections as function of $p_T$ from LHCb beauty measurement] 
	{Differential cross sections as a function of $p_T$ for $B^{+}$ (top left), $B^{0}$  (top right) 
	and $B_s^0$ (bottom) mesons from the LHCb measurement of beauty production~\cite{LHCbBeauty} 
	compared to the FONLL theoretical predictions.}
  \label{fig:lhcb:beautycspt}
  \end{minipage}
\end{figure}

\begin{figure}[htbp]
  \centering
  \begin{minipage}[t]{0.49\textwidth}
  \includegraphics[width=0.5\figwidth,trim = 0 0 6mm 7mm,clip=true]{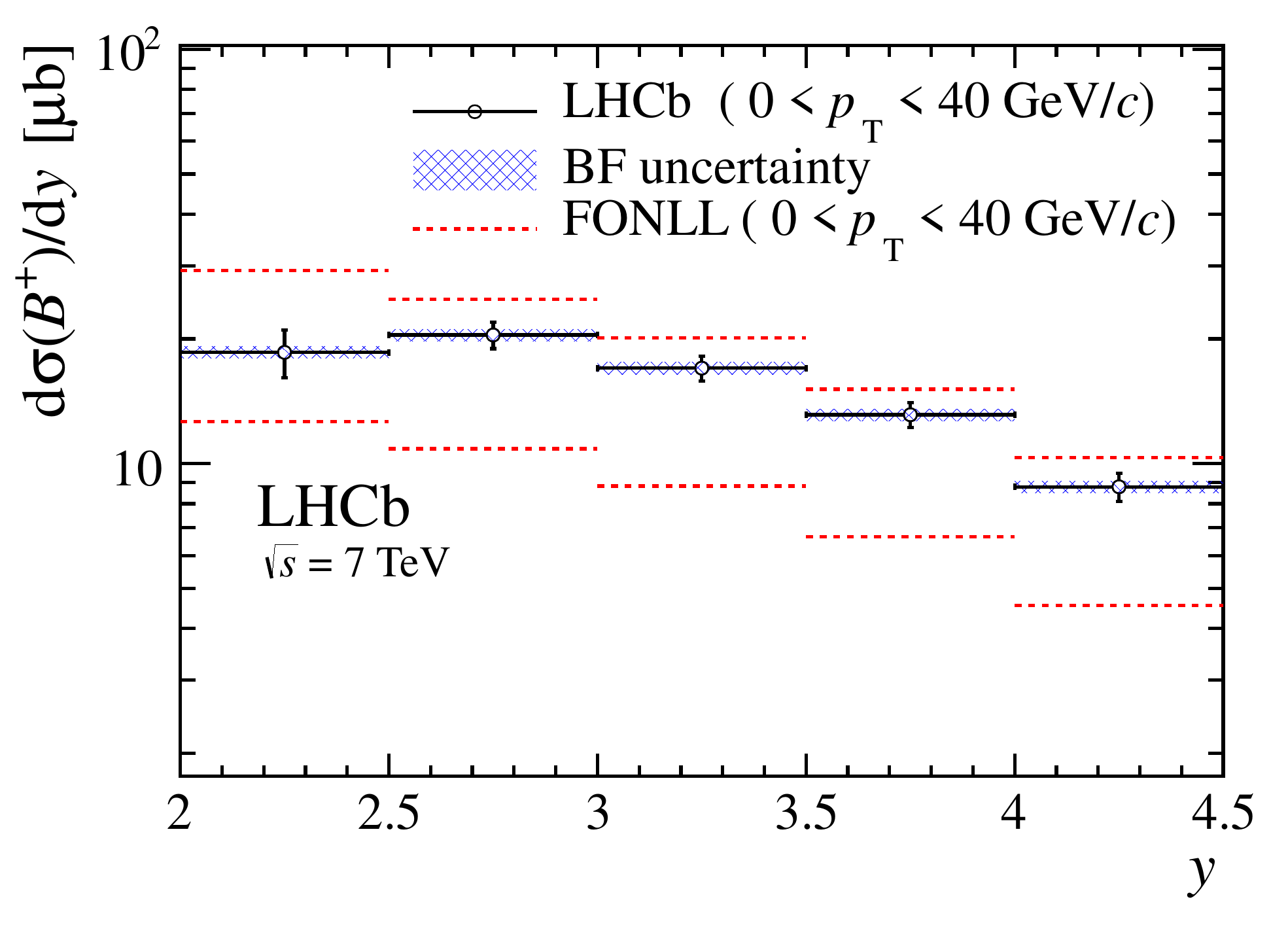}
  \includegraphics[width=0.5\figwidth,trim = 0 0 6mm 7mm,clip=true]{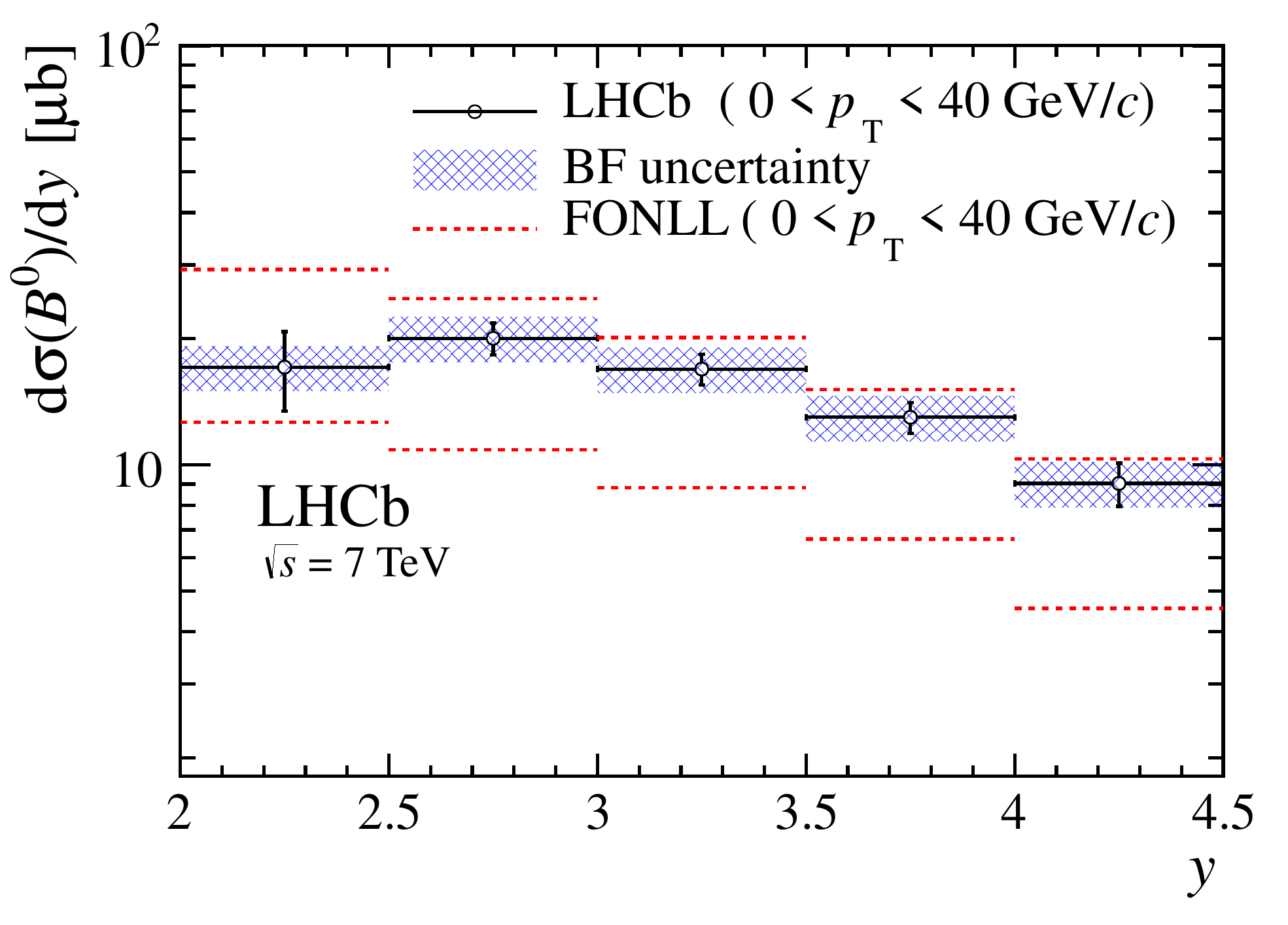}
  \end{minipage}
  \begin{minipage}[t]{0.49\textwidth}
  \includegraphics[width=0.5\figwidth,trim = 0 0 6mm 7mm,clip=true]{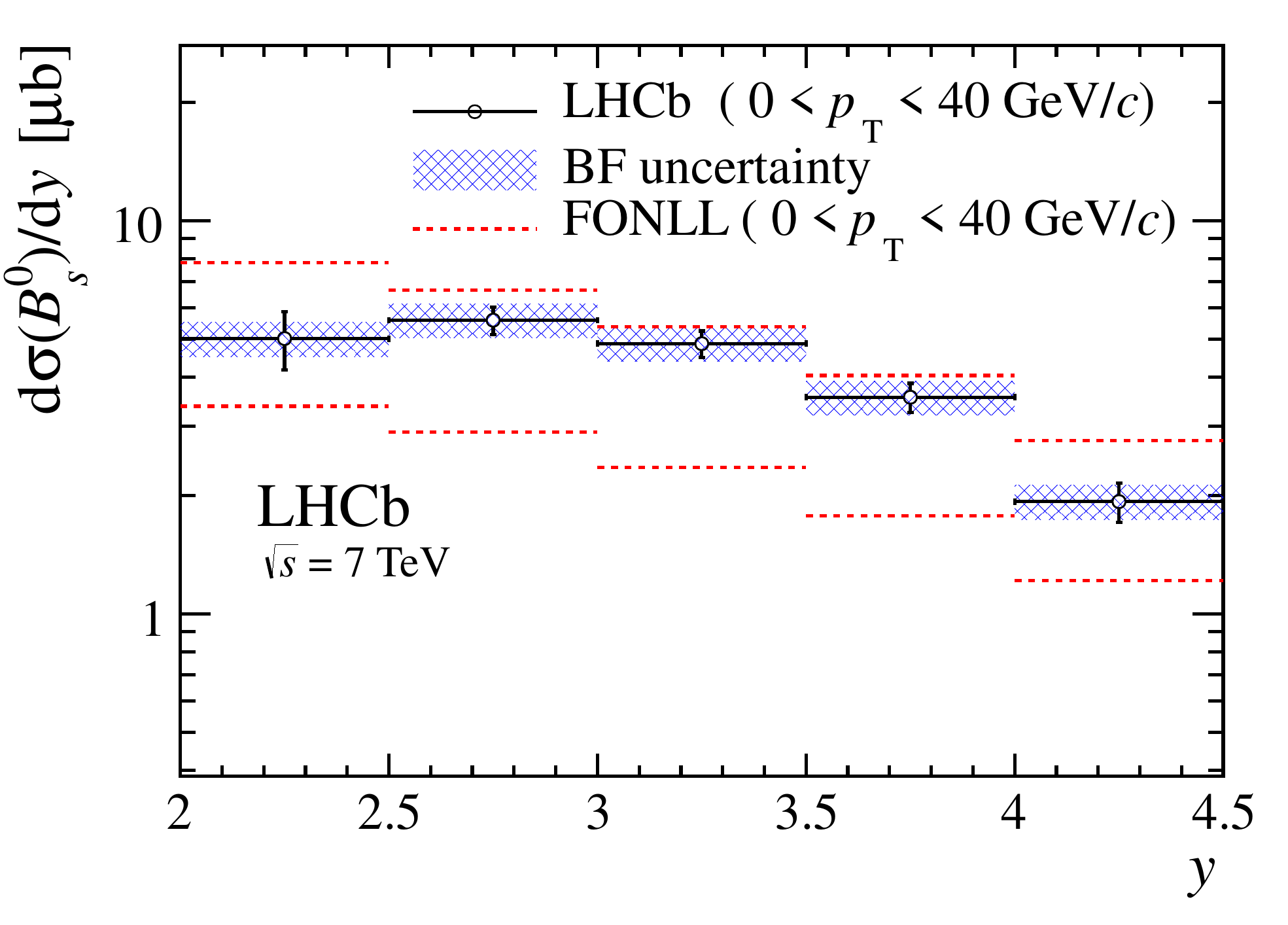}
  \caption[Differential cross sections as function of $y$ from LHCb beauty measurement] 
	{Differential cross sections as a function of $y$ for $B^{+}$ (top left), $B^{0}$  (top right) 
	and $B_s^0$ (bottom) mesons from the LHCb measurement of beauty production~\cite{LHCbBeauty} 
	compared to the FONLL theoretical predictions.}
  \label{fig:lhcb:beautycsy}
  \end{minipage}
\end{figure}

The measured cross sections, integrated over $p_T$ and $y$, are compared to the 
FONLL theoretical predictions in Figs.~\ref{fig:lhcb:beautycspt} and~\ref{fig:lhcb:beautycsy}, respectively.
Similar to the results of the charm measurement~\cite{LHCbCharm}, 
the FONLL calculations describe the data well, \ozmodN{albeit} within large uncertainties.

\subsection{\ozmodN{Comparison with theoretical predictions}}
\label{sec:pdffit:th}

Theoretical predictions for the charm and beauty data were obtained using the massive NLO $O(\alpha_s^3)$ calculations in the FFNS~\cite{mnrtotal,mnrsingle,mnrdouble} (see Section~\ref{sec:th:hq:pp:nlo}) using the MNR code~\cite{mnrplace}, 
implemented in the HERAFitter package~\cite{herafitter,HERAFitterPaper} for this purpose. 
Technical details of the implementation are described in Appendix~\ref{sec:pdffit:th:impl}. 
The parameters used in the calculations and the corresponding variations used to estimate the uncertainties are described below in Section~\ref{sec:pdffit:th:det}. 

\subsubsection{Details of MNR calculations}
\label{sec:pdffit:th:det}

\paragraph*{Parton-level cross sections\\}
\label{sec:pdffit:th:det:parton}

The parton-level cross sections were calculated using the one-particle inclusive option of the MNR calculations~\cite{mnrsingle} with the following settings:
\begin{itemize}
	\item {\bf the factorisation and renormalisation scales} were parametrised as $\mu_f=A^c_f\sqrt{p_T^2+m_c^2}$, $\mu_r=A^c_r\sqrt{p_T^2+m_c^2}$ for charm production 
		and similarly $\mu_f=A^b_f\sqrt{p_T^2+m_b^2}$, $\mu_r=A^b_r\sqrt{p_T^2+m_b^2}$ for beauty, 
		where $m_c$ and $m_b$ refer to the $c$- and $b$-quark pole masses, respectively. 
		The conventional choice for the coefficients $A^{c,b}_{f,r}$ is $A^c_f=A^b_f=A^c_r=A^b_r=1$ and the variations within the range [0.5;2] (independently or simultaneously). 
		Since the scale dependence of the predictions is of the order of a factor of $2$ \ozmod{(see Fig.~\ref{fig:th:hq:pp:scale} in Section~\ref{sec:th:hq:pp:nlo})}, the choice of the coefficients $A^{c,b}_{f,r}$ is crucial for a \ozmod{reliable} data description and has to be carefully studied; 
		the explicit details are given in Section~\ref{sec:pdffit:framework}; 
	\item {\bf the pole mass of the $c$ and $b$ quarks} were set to $m_c = 1.4$~GeV, $m_b = 4.5$~GeV, 
    or left free in the fit;
	\item {\bf strong coupling constant} $\alpha_s^{n_f=3}(M_Z) = 0.1059 \pm 0.0005$, corresponding to the PDG value $\alpha_s^{n_f=5}(M_Z) = 0.1185 \pm 0.0006$~\cite{pdg2012};
	\item {\bf the PDFs} were 3-flavour FFNS variants of HERAPDF1.0~\cite{DIScomb} or other global PDF groups, as specified later, 
    or left free in the fit. 
\end{itemize}

\paragraph*{Fragmentation\\}
\label{sec:pdffit:th:det:frag}

Non-perturbative fragmentation functions for charm and beauty were extracted from \epem and \ep data 
(see, e.g.~\cite{frag97,frag00,frag06,h1frag,zeusfrag}). 
So far no fragmentation measurements were done in \pp collisions. 
Universality of the fragmentation is often assumed; however it holds 
only if the perturbative part of the calculations is the same (see Section~\ref{sec:th:hq:frag}). 
Moreover, e.g.\ in~\cite{h1frag} the fragmentation-function parameters were shown to be different for two different kinematic regions. 

Since the kinematic region of the LHCb charm measurement is close to the HERA region where measurements were done 
by H1~\cite{h1frag} and ZEUS~\cite{zeusfrag}, the Kartvelishvili function~\cite{Kartvelishvili:1977pi} 
with $\alpha_k=4.4 \pm 1.7$ was used for the charm fragmentation, 
which covers the spread of the measurements~\cite{h1frag,zeusfrag}. 
The fragmentation was performed in the laboratory frame by rescaling the quark three-momentum, 
then the energy of the produced hadron was calculated using the hadron mass. 
The prescription \ozmod{for the fragmentation} described above was used for $\Dstar$, $D_s^{+}$ mesons and $\Lambda_c^{+}$ hadrons, 
while for $D^0$ and $D^{+}$ mesons the contribution from $D^{*+}$ and $D^{*0}$ mesons was treated as described in~\cite{frag03}. 
For beauty no fragmentation measurements at HERA exist; therefore the value $\alpha_k=11 \pm 4$ was used, extracted from measurements at LEP~\cite{frag00}. 
All beauty hadrons were treated equally. 
The fragmentation fractions for charmed hadrons were taken from~\cite{Lohrmann:2011np}. 
Their values are consistent with the recent determination from $ee$, $ep$ and $pp$ data~\cite{Lisovyi:2015uqa}. 
The fragmentation fractions for beauty hadrons are taken from~\cite{LHCbBeauty}.

\subsubsection{Kinematics of low-$p_T$ region}
\label{sec:pdffit:th:kinlowpt}

\begin{figure}[tbp]
  \centering
  \includegraphics[width=0.495\figwidth,trim=5mm 0 5mm 0mm,clip=true]{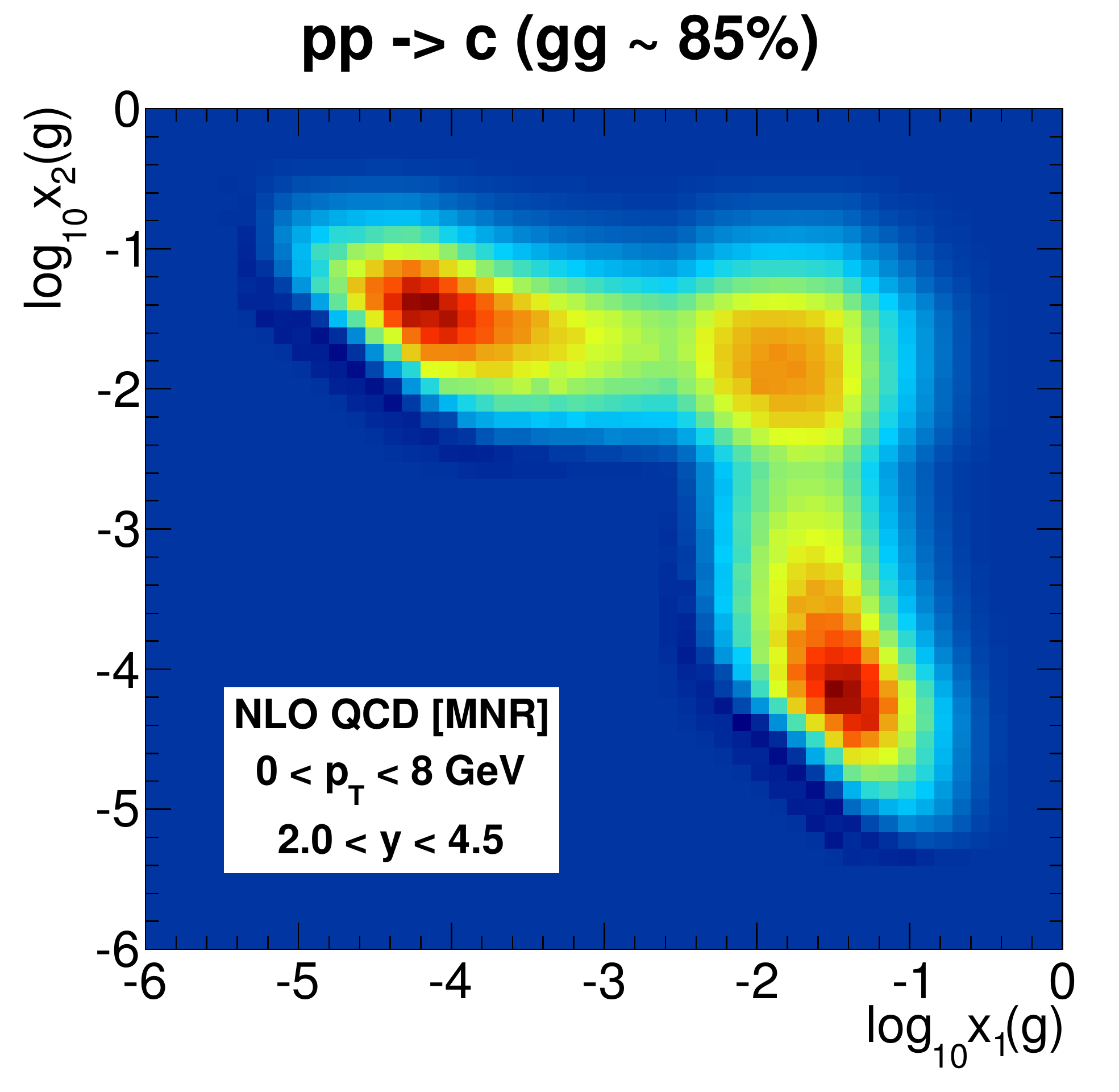}
  \includegraphics[width=0.495\figwidth,trim=5mm 0 5mm 0mm,clip=true]{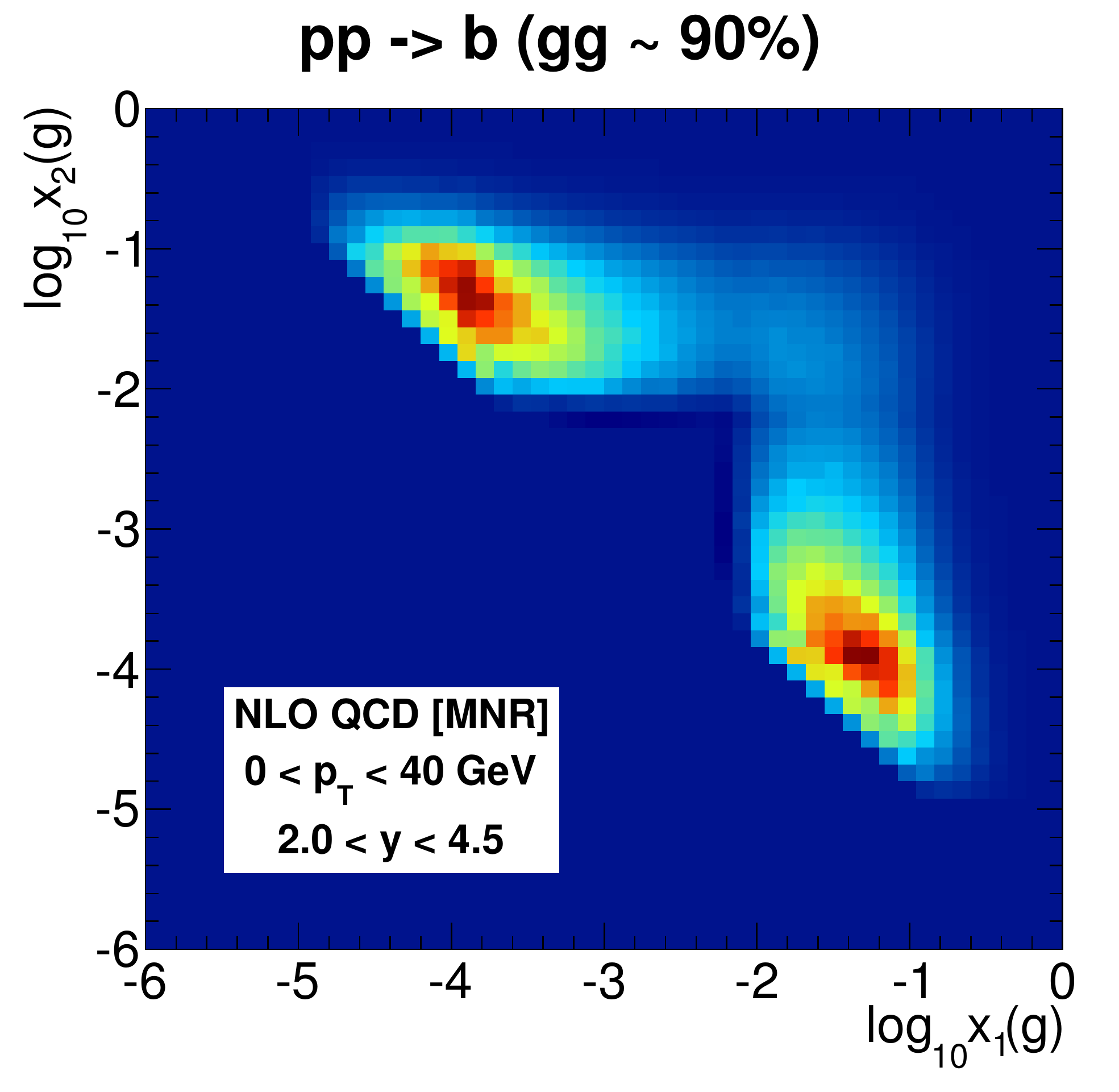}
  \caption[Two-dimensional $x$ distribution of two gluons for charm and beauty production]
  {The two-dimensional $x$ distribution of the two incoming gluons for charm (left) and beauty (right) production at LHCb via the $gg$ channel.}
	\label{fig:sec:pdffit:kinematicsnlo}
\end{figure}

\begin{figure}[htbp]
\centering
\vspace{0.4cm}
\begin{fmffile}{hqppnlofe1}
\begin{fmfgraph*}(90,40)
	\fmfpen{thin}
	\fmfleft{i1,i2}
	\fmfright{o1,o2}
	\fmf{gluon}{i1,v1}
	\fmflabel{$g$}{i1}
	\fmf{gluon}{i2,v2}
	\fmflabel{$g$}{i2}
	\fmf{quark}{v1,v2}
	\fmf{quark}{o1,v1}
	\fmflabel{$Q$}{o1}
	\fmf{quark}{v2,o2}
	\fmflabel{$\bar{Q}$}{o2}
	\fmfdot{v1}
	\fmfdot{v2}
	\fmffreeze
	\fmfshift{(0,-0.11w)}{v1}
	\fmfshift{(0,0.11w)}{v2}
\end{fmfgraph*}
\end{fmffile}
\hspace{0.75cm}
\begin{fmffile}{hqppnlofe2}
\begin{fmfgraph*}(90,60)
	\fmfpen{thin}
	\fmfleft{i1,i2}
	\fmfright{o1,o2,o3}
	\fmf{gluon}{i1,v1}
	\fmflabel{$g$}{i1}
	\fmf{gluon}{i2,v2}
	\fmflabel{$g$}{i2}
	\fmf{gluon}{v1,v3}
	\fmf{gluon}{v1,o1}
	\fmf{quark}{v3,v2}
	\fmf{quark}{v2,o3}
	\fmf{quark}{o2,v3}
	\fmflabel{$Q$}{o3}
	\fmflabel{$\bar{Q}$}{o2}
	\fmflabel{$g$}{o1}
	\fmfdot{v1}
	\fmfdot{v2}
	\fmfdot{v3}
	\fmffreeze
	\fmfshift{(0,-0.16h)}{v1}
	\fmfshift{(-0.14w,0.13h)}{v3}
	\fmfshift{(-0.20w,-0.21h)}{v2}
	\fmfshift{(0,-0.37h)}{i2}
	\fmfshift{(-0.03w,0.13h)}{o2}
	\fmfshift{(0.09w,0)}{o1}
	\fmfshift{(0.08w,0.05h)}{o3}
\end{fmfgraph*}
\end{fmffile}
\vspace{0.4cm}
  \caption[Example of flavour-excitation diagram]
  {\ozmod{An example of the LO heavy-flavour production diagram via the $gg$ channel (left) and NLO flavour excitation via initial-state gluon splitting (right).}}
  \label{fig:sec:pdffit:fep}
\end{figure}
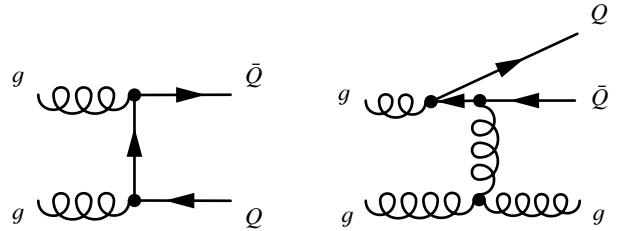

The dominant channel for heavy-flavour production at LHC is \ozmodN{$gg \to Q\bar{Q}$} (\ozmodNN{see} Fig.~\ref{fig:sec:pdffit:fep}, left). 
Fig.~\ref{fig:sec:pdffit:kinematicsnlo} shows the two-dimensional $x$ distribution of the two incoming gluons, 
as predicted for the LHCb data by the calculations described above. %
\ozmod{This more complete evaluation} confirms qualitatively the rough LO estimation (see Fig.~\ref{fig:sec:pdffit:kinematics}): 
the main contribution comes from the two separated regions $x_1 \approx 10^{-4.5}$, $x_2 \approx 10^{-1.5}$ for charm and $x_1 \approx 10^{-4.0}$, $x_2 \approx 10^{-1.2}$ for beauty; 
although for charm an additional concentrated region is observed at $x_1 \approx x_2 \approx 10^{-2.0}$-$10^{-1.5}$. %
The enhancement at medium $x$ comes from a class of $O(\alpha_s^3)$ corrections, given by the flavour-excitation diagrams (an example is given in Fig.~\ref{fig:sec:pdffit:fep}, right), 
which can be thought of as initial-state gluon-splitting processes~\cite{Mangano:1997ri}. 
The relevant region of the phase space in this case is the one with the heavy-quark propagator close to the mass shell (the low-$p_T$ region). 
In GM-VFNS approaches these effects are reabsorbed in the evolution of the PDFs 
by defining a heavy-quark density inside a proton. 
\ozmod{The inclusion} of the $gQ \to gQ$ process allows \ozmodNN{accounting for} the higher-order effects in the evolution equations, 
while \ozmod{the inclusion} of the NLO flavour-excitation diagrams reproduces instead more faithfully 
the exact kinematics and correlations of the flavour-creation process in the region close to the threshold~\cite{Mangano:1997ri}. 

\begin{figure*}[tbp]
  \sidecaption
  \centering
  \includegraphics[width=0.80\figwidth,trim=6mm 0 6mm 0mm,clip=true]{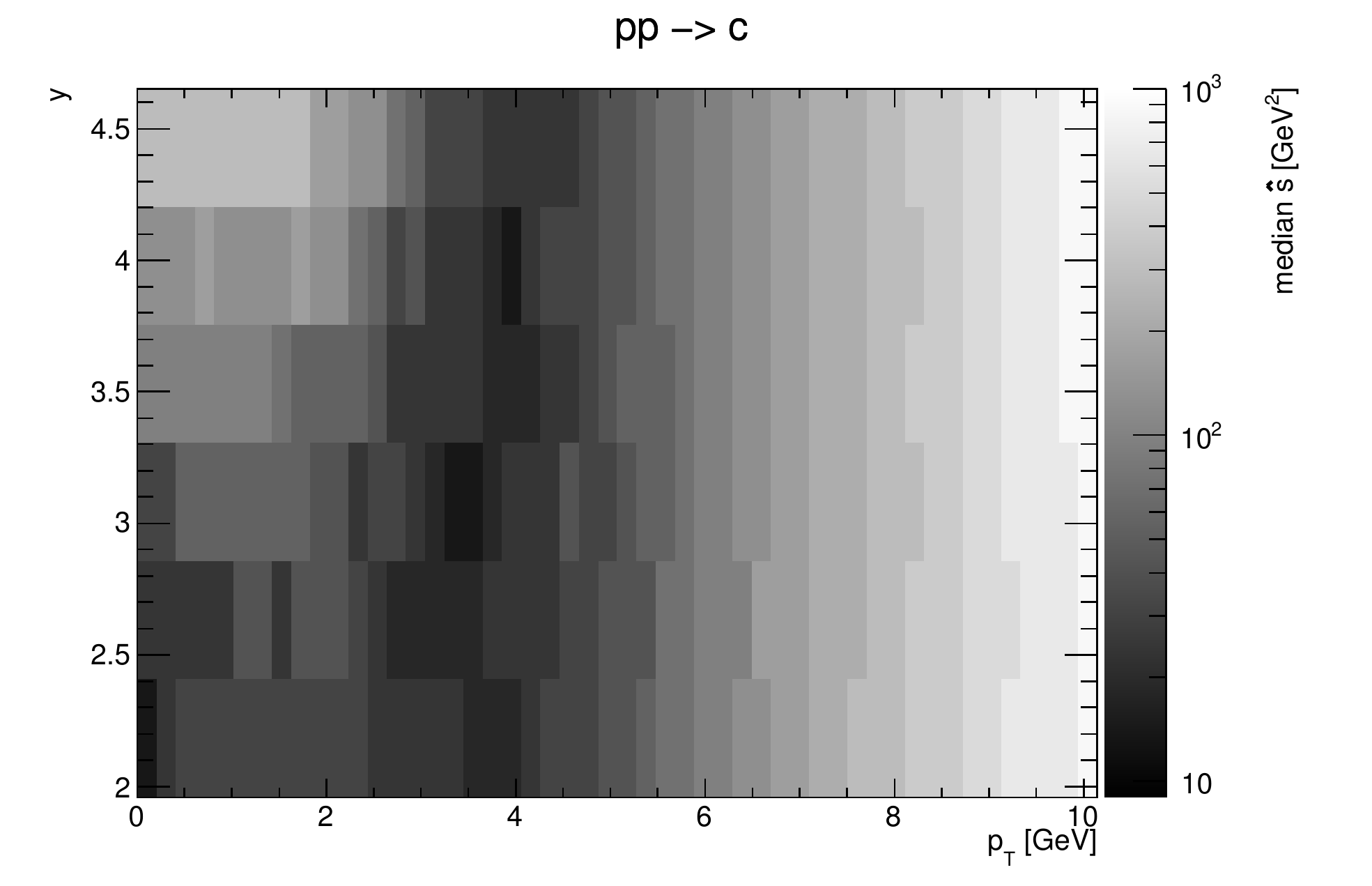}
  \includegraphics[width=0.80\figwidth,trim=5mm 0 7mm 0mm,clip=true]{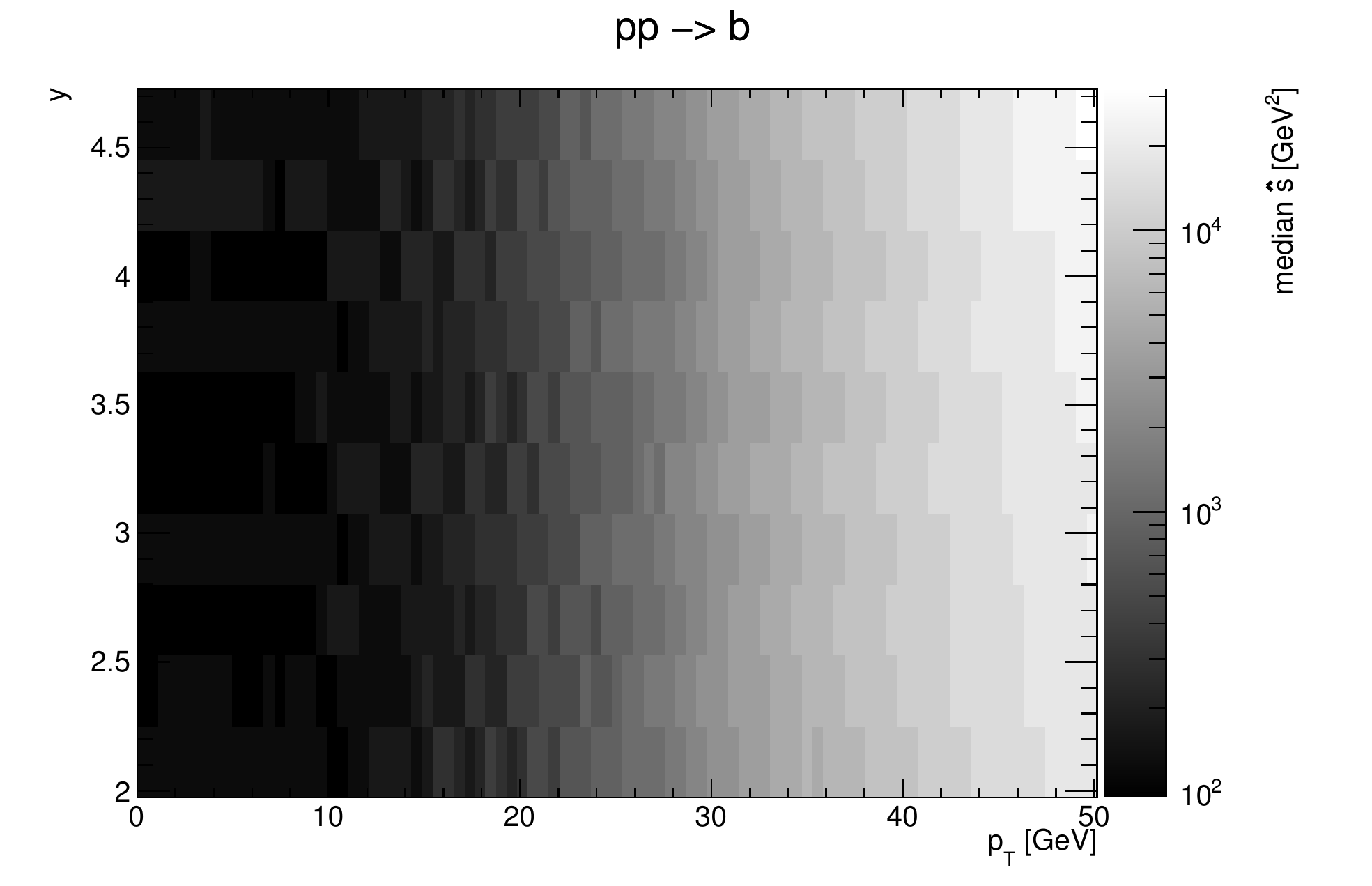}
  \caption[Median of $\sqrt{\hat{s}}$ vs.\ $p_T$ and $y$ for charm and beauty production at LHCb]
  {The median of the centre-of-mass energy $\sqrt{\hat{s}}$ in the parton-parton rest frame vs.\ transverse momentum $p_T$ and rapidity $y$ of the heavy quark for the charm (left) and beauty (right) LHCb data.}
	\label{fig:sec:pdffit:median}
\end{figure*}

The corrections from the flavour-excitation diagrams should in no way be thought of as a problem of the FFNS calculations, 
but rather as an important correction to the LO kinematics of the process. In order to demonstrate this, Fig.~\ref{fig:sec:pdffit:median} shows 
the median of the centre-of-mass energy in the parton-parton rest frame, $\sqrt{\hat{s}}$, vs.\ transverse momentum and rapidity of the heavy quark for the charm and beauty LHCb data. 
While in the case of beauty, the $\sqrt{\hat{s}}$ median smoothly increases with increasing $p_T$ \ozmod{almost independently} of $y$, 
for charm in the threshold region $0 < p_T \lesssim \SI{2}{GeV}$ a strong increase of the $\sqrt{\hat{s}}$ median with increasing $y$ is observed, 
\ozmod{This increase is correlated with the concentrated region at medium $x$ in Fig.~\ref{fig:sec:pdffit:kinematicsnlo}}. 
\ozmod{The results shown in these two Figs.~\ref{fig:sec:pdffit:kinematicsnlo}, \ref{fig:sec:pdffit:median}} 
indicate that the statement about the sensitivity of charm production at low $p_T$ and forward $y$ to the low-$x$ gluon region should be taken with some caution: 
in fact, about 50\% of contribution in the corner region $p_T \lesssim \SI{2}{GeV}$, $y \gtrsim 3.5 $ does not come from low-$x$ gluons; 
\ozmod{instead it comes from the \ozmodNN{medium}-$x$ region, and this particular contribution is described in the $O(\alpha_s^3)$ calculations at LO.}

\subsubsection{Comparison to FONLL calculations}
\label{sec:pdffit:th:fonll}

As mentioned in Section~\ref{sec:th:hq:pp}, one of the state of the art calculations for heavy-flavour production in hadron collisions is the FONLL approach~\cite{Cacciari:1998it}. 
Briefly reminding, the FONLL approach merges the massive NLO calculations (MNR) with massless ones using a phenomenologically chosen matching function. 
Owing to resummation of the NLL part the FONLL calculations are expected to have improved convergence of the perturbative expansion at high $p_T$.

\begin{figure}[tbp]
  \centering
  \includegraphics[width=0.495\figwidth,trim=10mm 6mm 25mm 10mm,clip=true]{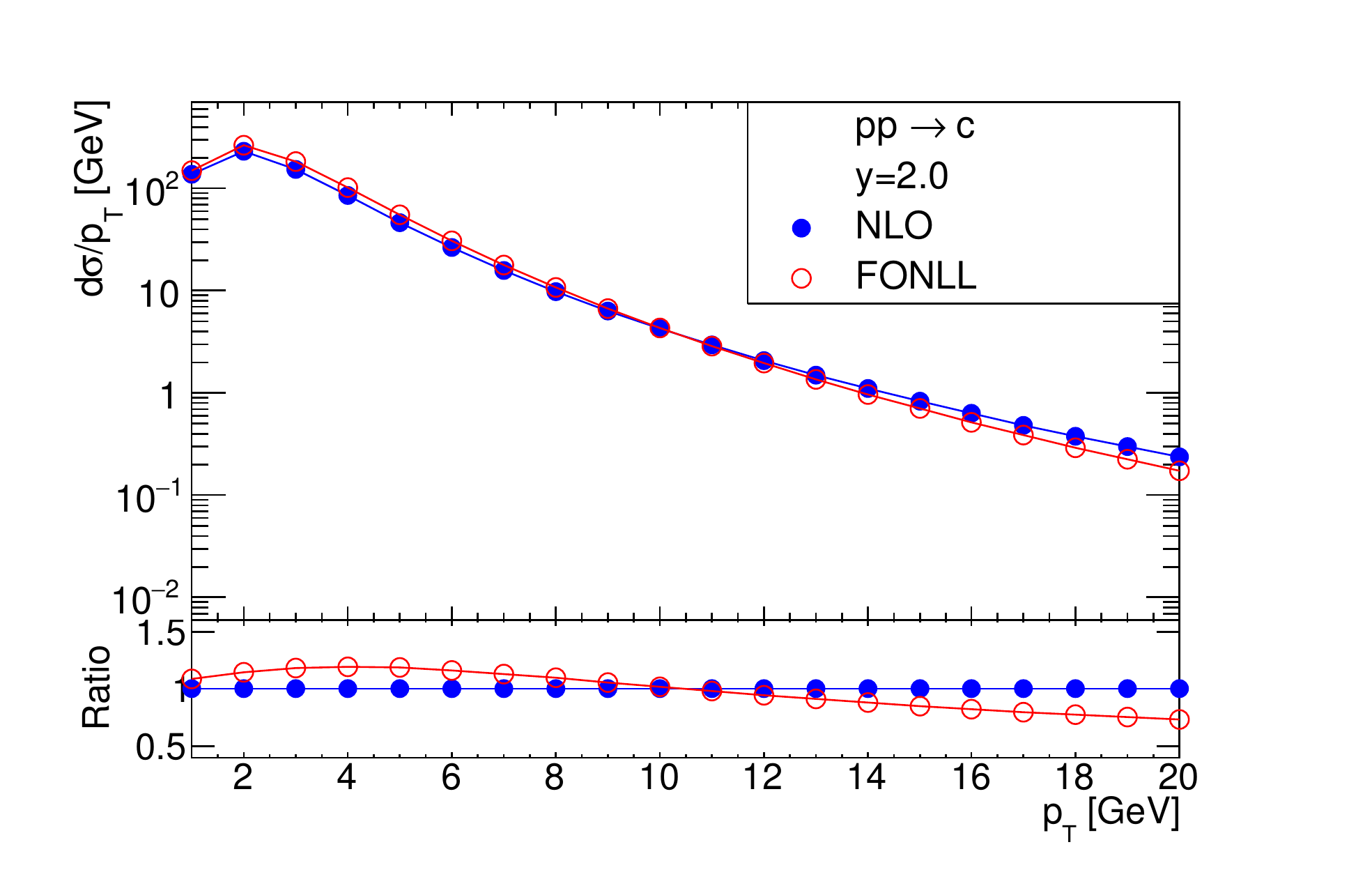}
  \includegraphics[width=0.495\figwidth,trim=10mm 6mm 25mm 10mm,clip=true]{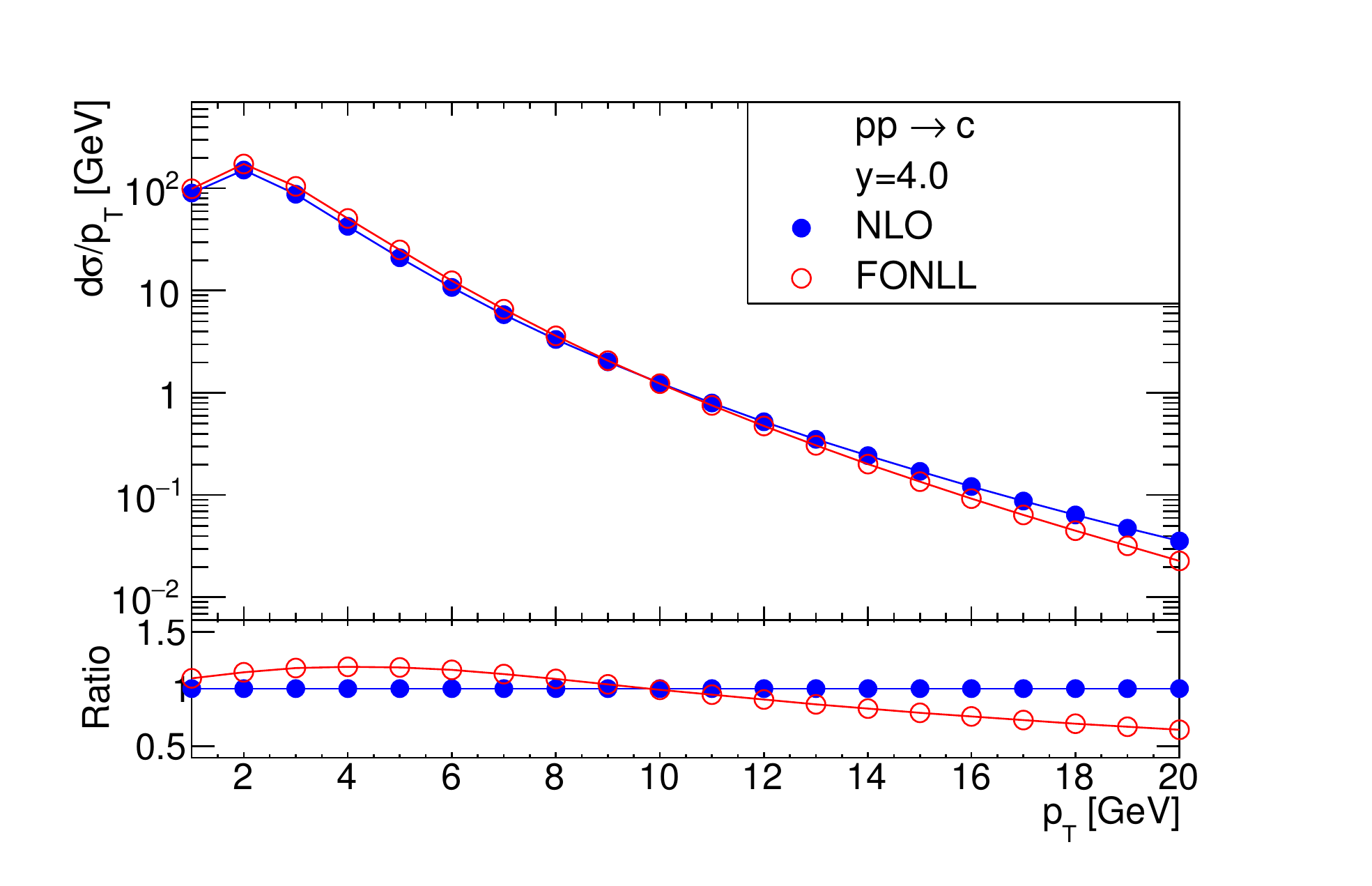}
  \includegraphics[width=0.495\figwidth,trim=10mm 6mm 25mm 10mm,clip=true]{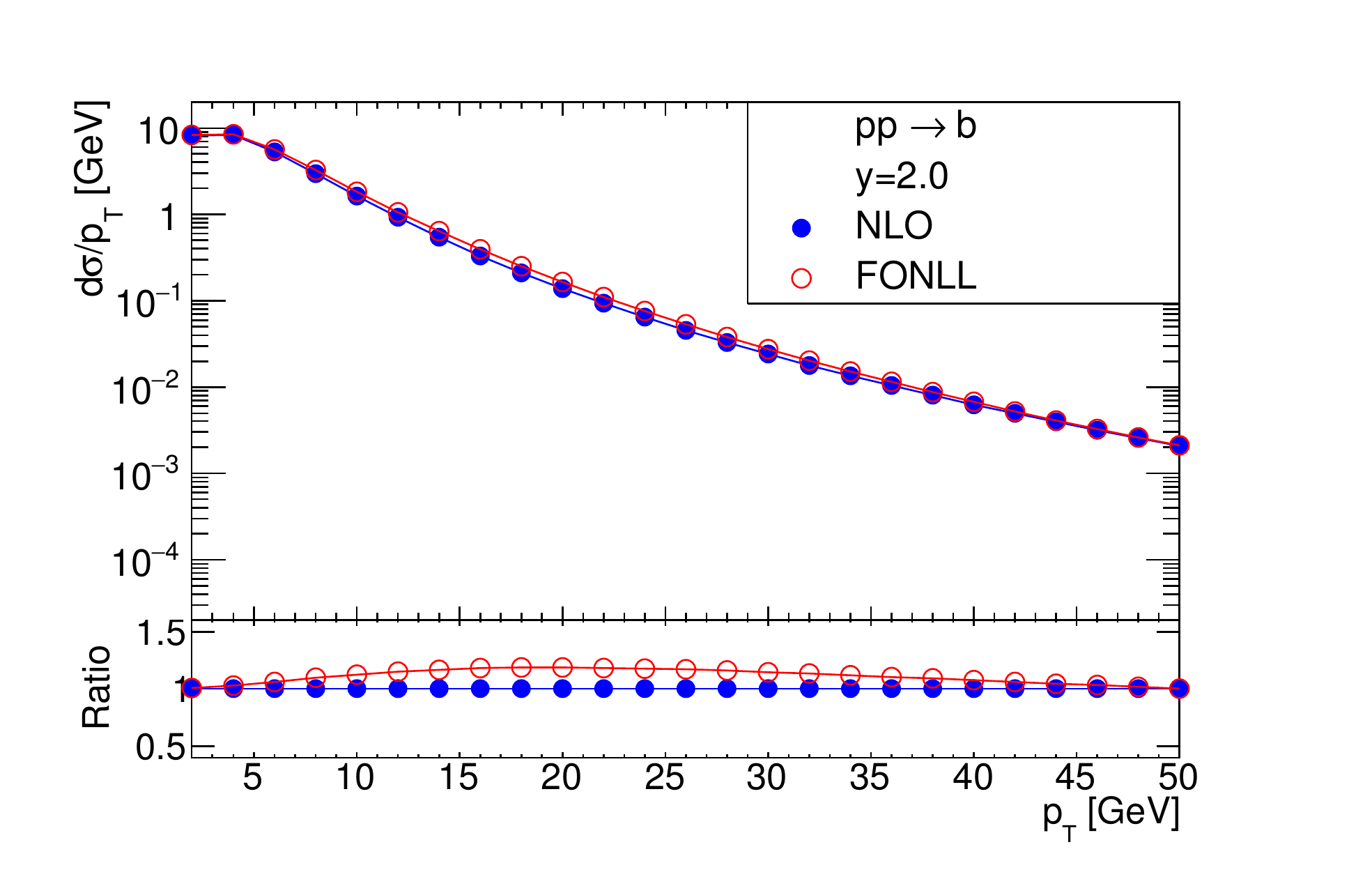}
  \includegraphics[width=0.495\figwidth,trim=10mm 6mm 25mm 10mm,clip=true]{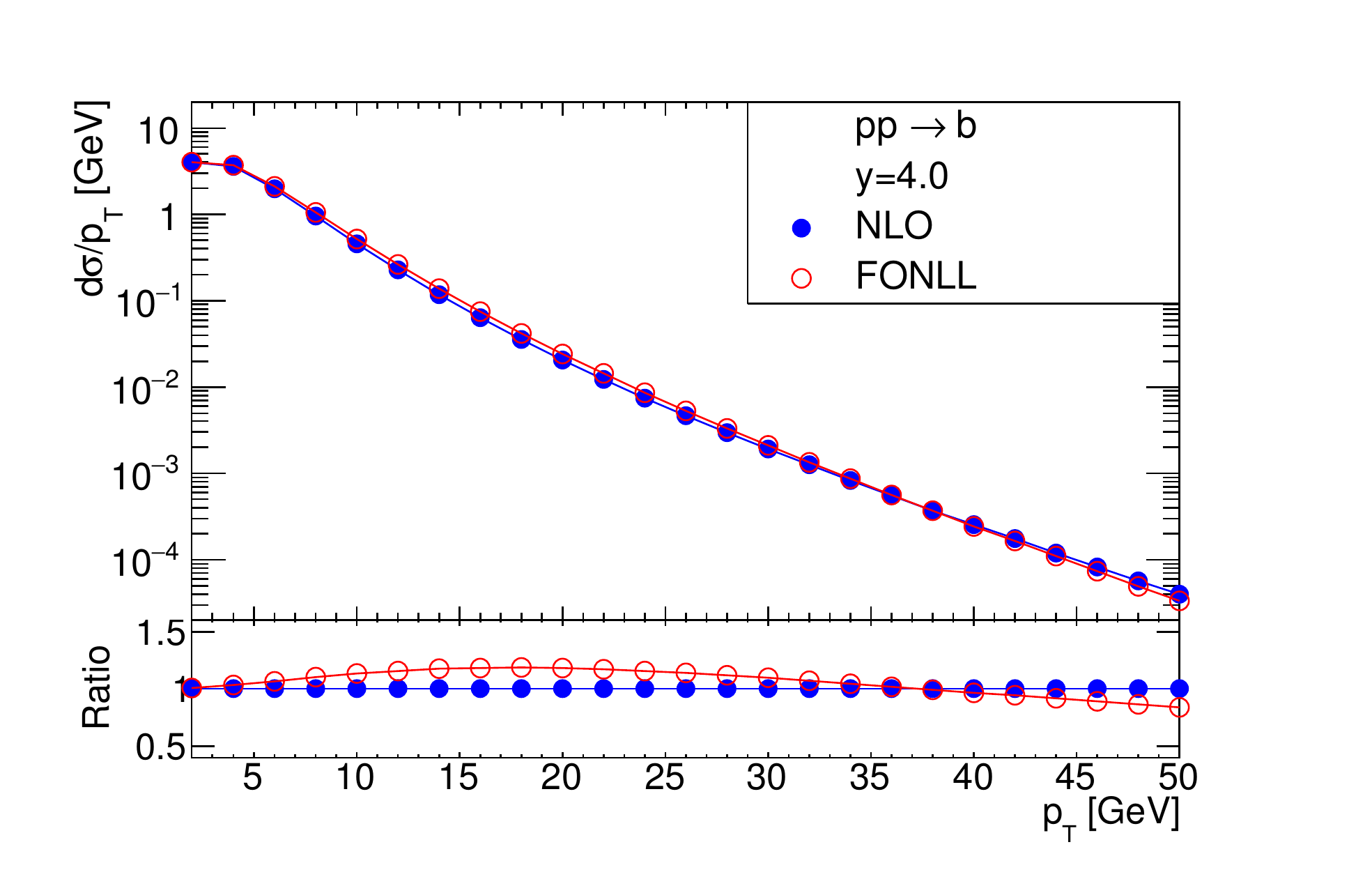}
  \caption[Comparison of NLO and FONLL predictions for charm and beauty at LHCb]
  {Comparison of NLO and FONLL predictions for the differential cross sections as a function of $p_T$, ${\rm d}\sigma / {\rm d}p_T$, 
  for charm at rapidity $y=2$ (top left) and $y=4$ (top right), and beauty at rapidity $y=2$ (bottom left) and $y=4$ (bottom right) at LHCb. The bottom pads show the ratio to the NLO predictions.}
	\label{fig:sec:pdffit:nlovsfonll}
\end{figure}

In Fig.~\ref{fig:sec:pdffit:nlovsfonll} the NLO predictions obtained with MNR as described in Section~\ref{sec:pdffit:th:det} are compared to the FONLL ones 
obtained using the public web interface~\cite{FONLLWeb}%
\footnote{The `FONLL' option of the FONLL program was used; the settings consistently used in MNR and FONLL calculations were: 
PDFs set is MSTW2008nlo68cl~\cite{Martin:2010db}, $\mu_f=\mu_r=\sqrt{m_Q^2+p_T^2}$, $m_c=\SI{1.5}{GeV}$, $m_b=\SI{4.75}{GeV}$.} 
for parton-level charm and beauty cross sections at LHCb. 
For the relevant regions of transverse momentum in the charm and beauty data, $0<p_T(c) \lesssim \SI{8}{GeV}$, $0<p_T(b) \lesssim \SI{40}{GeV}$%
\footnote{The fragmentation effects do not change significantly $p_T$ regions, since the heavy-flavour fragmentation functions are peaked near the scaling variable $z=1$ 
and the cross sections are steeply falling with increasing $p_T$.} 
the maximum deviations of the order of 20\% in the region $p_T \approx 3m_Q $ are observed. Note that these changes are well within the uncertainties from the scale variations. 

It is also instructive to look at Fig.~\ref{fig:th:hq:pp:fonll:nlovsfonllorig} (Section~\ref{sec:th:hq:pp:fonll}), taken from the original FONLL paper~\cite{Cacciari:1998it}, 
which shows the bands obtained from the scale variations for the NLO and FONLL calculations for beauty production at the Tevatron. 
The behaviour of the bands in Fig.~\ref{fig:th:hq:pp:fonll:nlovsfonllorig} 
is very similar to the change of the central values shown in Fig.~\ref{fig:sec:pdffit:nlovsfonll}. 
Note that in Fig.~\ref{fig:th:hq:pp:fonll:nlovsfonllorig} a significant reduction of the uncertainty band starts only at $p_T \gtrsim \SI{40}{GeV}$. 
\ozmod{Since the considered $p_T$ ranges of the LHCb data are not the high-$p_T$ region where the effects of the FONLL calculations become relevant, 
the usage of the NLO FFNS calculations as one of the currently best available theories for the LHCb kinematic region in the present study is justified.}%
\footnote{\ozmod{Theorists are continuously making progress, and 
the approximate NNLO $O(\alpha_s^4)$ predictions in the $gg$ and $q\bar{q}$ channels for differential cross sections 
for heavy-flavour production at hadron colliders became available recently~\cite{difftop}.}} 

\subsubsection{Predictions based on PDFs from HERA}
\label{sec:pdffit:th:predhera}

\ozmod{
In Fig.~\ref{fig:sec:pdffit:preddzerohera} (left) the measured $D^0$ cross sections~\cite{LHCbCharm} 
are compared to the NLO predictions.%
\footnote{The restriction to the $D^0$ final state for this comparison is motivated by the best experimental precision 
of the LHCb data~\cite{LHCbCharm} owing to the presence of two charged daughter tracks only 
and therefore the smallest systematic uncertainties on the tracking efficiency.} 
For these calculations the FFNS variant of the HERAPDF1.0 set~\cite{DIScomb} was used, 
\ozmod{when charm is treated as massive and does not present in the proton at all energy scales.} 
The factorisation and renormalisation scale were set to $\mu_f=\mu_r=\sqrt{m_Q^2+p_T^2}$ 
and varied independently up and down by a factor of $2$, 
and the heavy-quarks masses chosen to be $m_c=\SI{1.4 \pm 0.15}{GeV}$, $m_b=\SI{4.50 \pm 0.25}{GeV}$. 
The fragmentation functions were chosen and their uncertainties evaluated as described in Section~\ref{sec:pdffit:th:det:frag}. 
The theoretical uncertainties denoted as `MNR' in Fig.~\ref{fig:sec:pdffit:preddzerohera} include uncertainties from scale, 
heavy-quark mass and fragmentation function variations, while the total theoretical uncertainties include also those arising from the PDFs. 
The latter are dominant in the lowest $p_T$ bin $0<p_T<1$~GeV, as well as in the next one $1<p_T<2$~GeV for the highest $y$ values $3.5<y<4.5$. 
Although the `MNR' uncertainties are of the order of factor $\gtrsim 2$ and exceed experimental data uncertainties in all bins. 
\ozmod{In overall, the predictions based on PDFs from HERA describe the LHCb data well within very large theoretical uncertainties.} 
The right-hand side of Fig.~\ref{fig:sec:pdffit:preddzerohera} shows the same measurement, but normalised in $y$: 
each cross section is divided by the corresponding value in the central $y$ bin $3.0<y<3.5$. 
Such normalisation leads to a great reduction of the `MNR' uncertainties of the theoretical predictions 
(discussed in detail later in Section~\ref{sec:pdffit:det:strategy:norm}), making them of the same order or below the experimental uncertainties, 
and enables the PDF uncertainties being visible in all bins. 
Note that the PDF uncertainties for the cross sections in the lowest $p_T$ bins of Fig.~\ref{fig:sec:pdffit:preddzerohera} 
extend to unphysical values below zero (although not appearing on the left-hand side of Fig.~\ref{fig:sec:pdffit:preddzerohera} because of the logarithmic scale) 
due to the usage of PDFs with a negative gluon distribution at low $x$. 

\begin{figure}[htbp]
  \includegraphics[width=1.0\figwidth,trim = 11mm 9mm 11mm 9mm,clip=true]{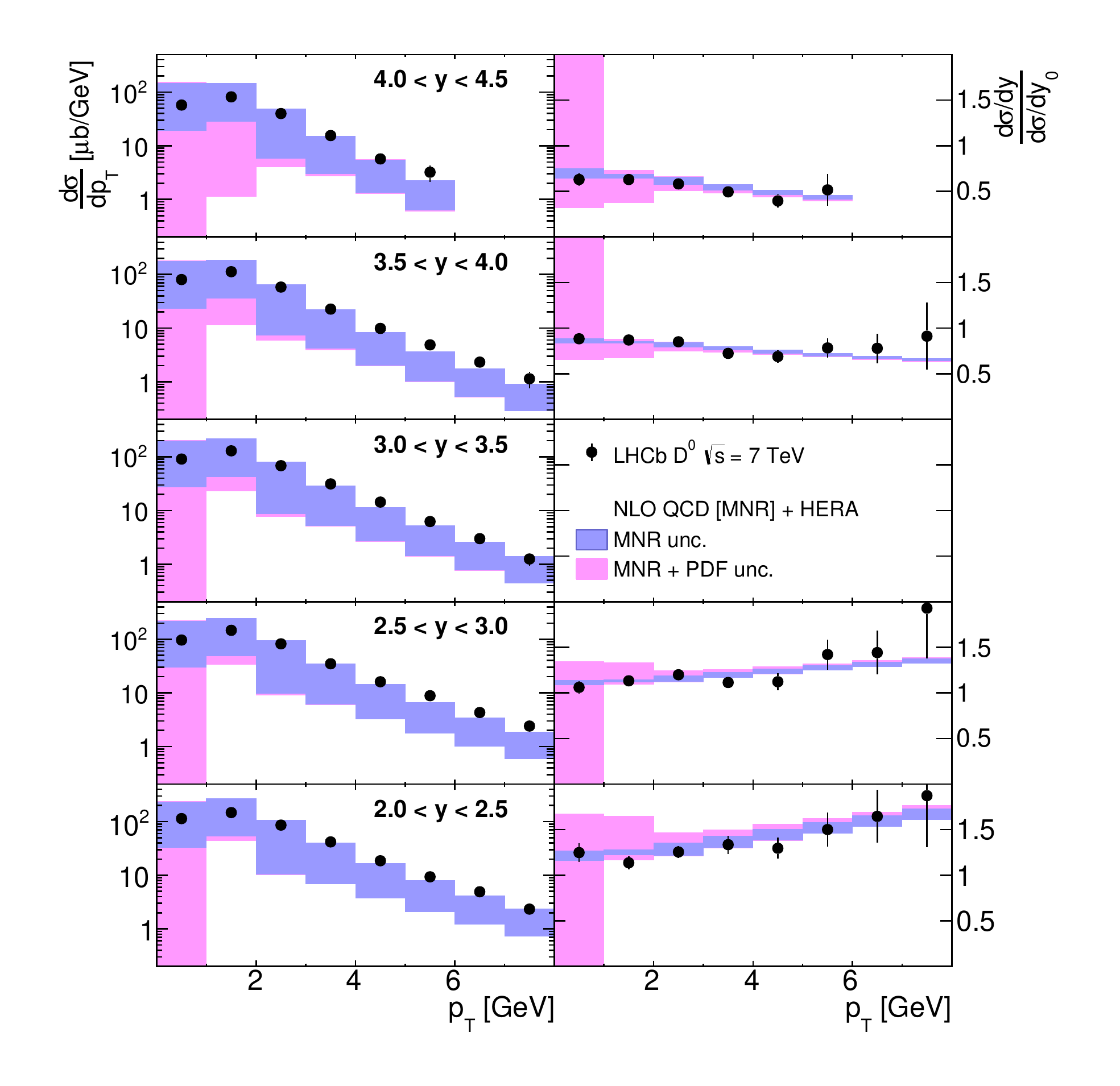}
  \caption[Differential D^{0} cross sections compared to HERA predictions] 
	{Differential cross sections for $D^0$ from the LHCb measurement of prompt charm production~\cite{LHCbCharm} 
	compared to theoretical predictions based on HERA PDFs as described in the text. 
  The cross sections for different $y$ regions are shown as	functions of $p_T$ (left) 
  and normalised to the central $y$ bin (right). 
  The theoretical uncertainties arising from different sources (see text) are shown separately.}
  \label{fig:sec:pdffit:preddzerohera}
\end{figure}

This illustrates the power of the LHCb charm data to eventially constrain existing PDFs. 
The \ozmodN{quantitative} analysis follows in Section~\ref{sec:pdffit}.
}

\clearpage
\section{\ozmodN{QCD analysis of HERA and LHCb heavy-flavour data}}
\label{sec:pdffit}

\ozmod{This Section describes a QCD analysis, extending the HERA fit by including the LHCb data presented in Section~\ref{sec:hflhcb}.} 
The general strategy of the study was to perform \ozmodN{first} a PDF fit with the HERA-only data, \ozmodN{which \ozmodNN{is}} close to HERAPDF1.0~\cite{DIScomb}, 
and then repeat the fit with the LHCb data included. 
\ozmodNN{In this way the net effect of the additional information is becoming obvious.} 
Results presented in this Section have been published by the PROSA Collaboration~\cite{Zenaiev:2015rfa}. 

\subsection{PDF fitting framework}
\label{sec:pdffit:framework}

The datasets used in the PDF fit are listed in Table~\ref{tab:pdffit:input}: 
\begin{itemize}
	\item the combined HERA-I inclusive $ep$ NC and CC DIS cross sections~\cite{DIScomb} (datasets 1--4) \ozmodN{serve} to constrain the core of the PDFs; 
		the analysis is restricted to the inclusive data with virtuality $Q^2>Q^2_{min}=\SI{3.5}{GeV^2}$ to ensure the applicability of pQCD;
	\item the combined HERA charm data~\cite{heracharmcomb} and ZEUS beauty~\cite{zeussecvtx_hera2} data (datasets 5--6) were used to constrain the gluon PDF and the charm and beauty masses;
	\item the LHCb charm~\cite{LHCbCharm} 
	and beauty~\cite{LHCbBeauty} data (datasets 7--14) ~--- \ozmodNN{the new ingredients} of the study ~--- \ozmodN{serve} to constrain the gluon PDF at low $x$.
\end{itemize}

\begin{table}[tbp]
\caption[Datasets used in PDF fit]
{Datasets used in the PDF fit. Similar entries are grouped together.}
\label{tab:pdffit:input}
\begin{center}
\tabcolsep6.1mm
\renewcommand*{\arraystretch}{1.2}
\begin{tabularx}{\columnwidth}{|X|l|l|}
\hline
\multicolumn{3}{|l|}{Dataset} \\\hline
 1 & NC \emp & \multirow{4}{*}{HERA-I DIS~\cite{DIScomb}} \\\cline{1-2}
 2 & NC \epp & \\\cline{1-2}
 3 & CC \emp & \\\cline{1-2}
 4 & CC \epp & \\\hline
 5 & \multicolumn{2}{l|}{HERA DIS Charm~\cite{heracharmcomb}} \\\hline
 6 & \multicolumn{2}{l|}{ZEUS Vertex DIS Beauty~\cite{zeussecvtx_hera2}} \\\hline
 7 & $D^{0}$ & \multirow{5}{*}{LHCb Charm~\cite{LHCbCharm}} \\\cline{1-2}
 8 & $D^{+}$ & \\\cline{1-2}
 9 & $D^{*+}$ & \\\cline{1-2}
10 & $D^{+}_{s}$ & \\\cline{1-2}
11 & $\Lambda^{+}_c$ & \\\hline
12 & $B^{+}$ & \multirow{3}{*}{LHCb Beauty~\cite{LHCbBeauty}} \\\cline{1-2}
13 & $B^{0}$ & \\\cline{1-2}
14 & $B_{s}$ & \\\hline
\end{tabularx}
\end{center}
\end{table}

The HERA data (datasets 1--6) were treated in the same way as in the original papers~\cite{DIScomb,heracharmcomb,zeussecvtx_hera2}.%
\footnote{Except that for dataset 5 in contrast to~\cite{heracharmcomb} all $Q^2$ bins were used including the lowest one $Q^2=\SI{2.5}{GeV^2}$; 
the applicability of pQCD for the charm data is ensured by the presence of a massive $c$ quark-antiquark pair in the final state.} 

For the LHCb data (datasets 7--14), the double-differential cross sections as a function of $p_T$ and $y$, 
$\ozmod{\frac{{\rm d}^2 \sigma(pp \to H_c X)}{dp_Tdy}}$, were used as published in~\cite{LHCbCharm,LHCbBeauty}%
\footnote{For the $\Lambda^{+}_c$ measurement from~\cite{LHCbCharm}, where no double-differential distribution is available, 
the single-differential cross sections $\ozmod{\frac{{\rm d} \sigma(pp \to \Lambda^{+}_c X)}{dp_T}}$ and $\ozmod{\frac{{\rm d} \sigma(pp \to \Lambda^{+}_c X)}{dy}}$ 
were used for the `LHCb Abs' and `LHCb Norm' approaches (see Section~\ref{sec:pdffit:det:strategy}), respectively.}. 
The correlations between the systematic uncertainties were taken into account as described 
in Section~3.3 of~\cite{LHCbCharm} and Section~4 of~\cite{LHCbBeauty}, treating as correlated those 
which are reported as single values for all ($p_T$, $y$) bins.%
\footnote{Except that for the sources `Bin size', `Trigger efficiency', `Tracking efficiency', `Muon identification' and 
`Angular distribution' from the beauty measurement the lowest boundaries from the respective ranges reported 
in Table~1 of~\cite{LHCbBeauty} were taken as a correlated part of the uncertainties.} 
The 3.5\% luminosity uncertainty was treated as correlated between the charm and beauty measurements. 
\ozmod{In addition} to the experimental uncertainties, the correlated fragmentation-fraction uncertainties were assigned to the data. 

\subsubsection{Details of PDF fit}
\label{sec:pdffit:det:framework}

The PDF fitting framework is \ozmod{the same} which has been used for the FFNS fit with the HERA charm combined data, 
described in Section~\ref{sec:comb:red:ffns}. The charm and beauty masses were left free in the fit and determined in HERAFitter by minimisation of a $\chi^2$-function. 
In variants of the fit with the LHCb heavy-flavour data additional uncertainties were evaluated, which are related to the uncertainties of the respective theoretical calculations; 
they are referred to as the `MNR uncertainties':
\begin{itemize}
	\item variations of the fragmentation and renormalisation scales, as described later in Section~\ref{sec:pdffit:det:strategy}, and
	\item variations of the fragmentation parameters $\alpha_k=4.4 \pm 1.7$ for charm and $\alpha_k=11 \pm 4$ for beauty.
\end{itemize}
The MNR uncertainties were obtained by adding these variations in quadrature.

\subsubsection{Strategy of QCD analysis}
\label{sec:pdffit:det:strategy}

The strategy of the QCD analysis was to perform several PDF fits, with and without the LHCb data, and then compare the results. 
Two approaches of fitting the LHCb data were studied: 
fitting the absolute cross section or the cross section normalised \ozmod{as described below.}

\paragraph*{Fitting absolute LHCb cross sections\\}
\label{sec:pdffit:det:strategy:abs}

	The absolute double-differential cross sections \dsigmadptdy measured by LHCb were fitted. 
	These quantities contain the maximum information; they are therefore sensitive to all physical and non-physical parameters of the theoretical calculations: the PDFs, heavy-quark mass, fragmentation function 
	and especially factorisation and renormalisation scales. The scale dependence of the predictions is of the order of a factor of 2, thus much exceeding the experimental data uncertainties, 
	although the PDF uncertainties from the \ozmod{available PDF sets} at very low $x$ and low $Q^2$ are even larger (see Fig.~\ref{fig:sec:pdffit:preddzerohera}).%
	This fact makes the study rather complicated: in order to account for the uncertainties of the perturbative predictions, the scales should be varied, 
	but these external variations change the description of the data and the fit results drastically, and thus are hard to control. 

	\ozmod{For this variant of the fit, the scale parameters $A_f^c$, $A_r^c$, $A_f^b$ and $A_r^b$ were therefore \ozmodNN{included in the fit}, 
	while for the estimation of the scale uncertainties the following procedure was used:} 
	the factorisation and renormalisation scales were varied independently one at a time in the ranges [$0.50$;$2.00$] and [$0.25$;$1.00$]%
	\footnote{The `common' range for the variations [$0.50$;$2.00$] for $\mu_r$ is \ozmod{ignored}, since the $\chisqndof$ for 
	$A^c_r=A^b_r=2.00$ variation was found to be unacceptably large ($\chisqndof=2497/1089$).}, 
	respectively, while the other scale was being refitted.
	\ozmod{To be specific}, the scale uncertainties include the following four variations:
	\begin{itemize}
		\item $A^c_f=A^b_f=2.00$ with $A^c_r$, $A^b_r$ free;
		\item $A^c_f=A^b_f=0.50$ with $A^c_r$, $A^b_r$ free;
		\item $A^c_r=A^b_r=1.00$ with $A^c_f$, $A^b_f$ free;
		\item $A^c_r=A^b_r=0.25$ with $A^c_f$, $A^b_f$ free.
	\end{itemize}
	In addition, for the variation $A^c_f=A^b_f=0.50$, the cut $p_T>\SI{2}{GeV}$ was applied for the charm LHCb data to ensure that the factorisation scale 
	is above $\SI{1}{GeV^2}$, since this is technically required in the framework. %
	This sophisticated procedure \ozmod{for the treatment of the scale variations} was necessary to give a reasonable description of the data for all variations of the fit.

\paragraph*{Fitting normalised LHCb cross sections\\}
\label{sec:pdffit:det:strategy:norm}

\begin{figure*}[tbp]
  \sidecaption
  \centering
  \includegraphics[width=1.43\figwidth,trim=3mm 4mm 3mm 12mm,clip=true]{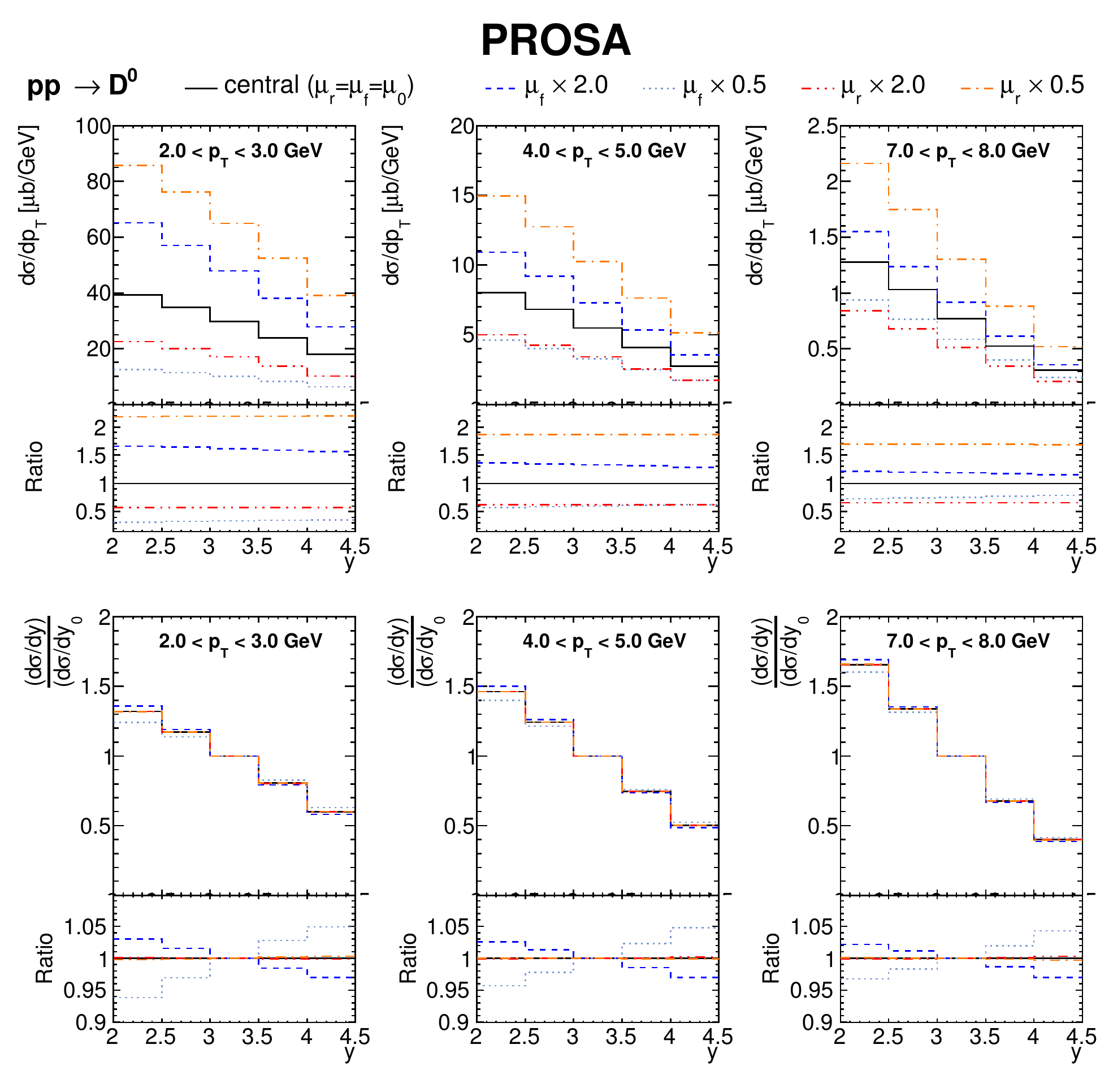}
  \caption[NLO predictions with scale variations for charm LHCb data]
  {NLO QCD predictions for charm LHCb data with different scale choices for absolute (top) and normalised (bottom) cross sections. 
	Bottom parts indicate the ratio of predictions to the central scale choice. 
	The predictions were obtained by using the FFNS variant of MSTW 2008 PDFs~\cite{Martin:2010db} with $n_f=3$; 
	the charm mass was set to $m_c=1.5$ GeV.}
	\label{fig:pdffit:ScaleDep_Charm}
\end{figure*}

\begin{figure*}[tbp]
  \sidecaption
  \centering
  \includegraphics[width=1.43\figwidth,trim=3mm 4mm 3mm 12mm,clip=true]{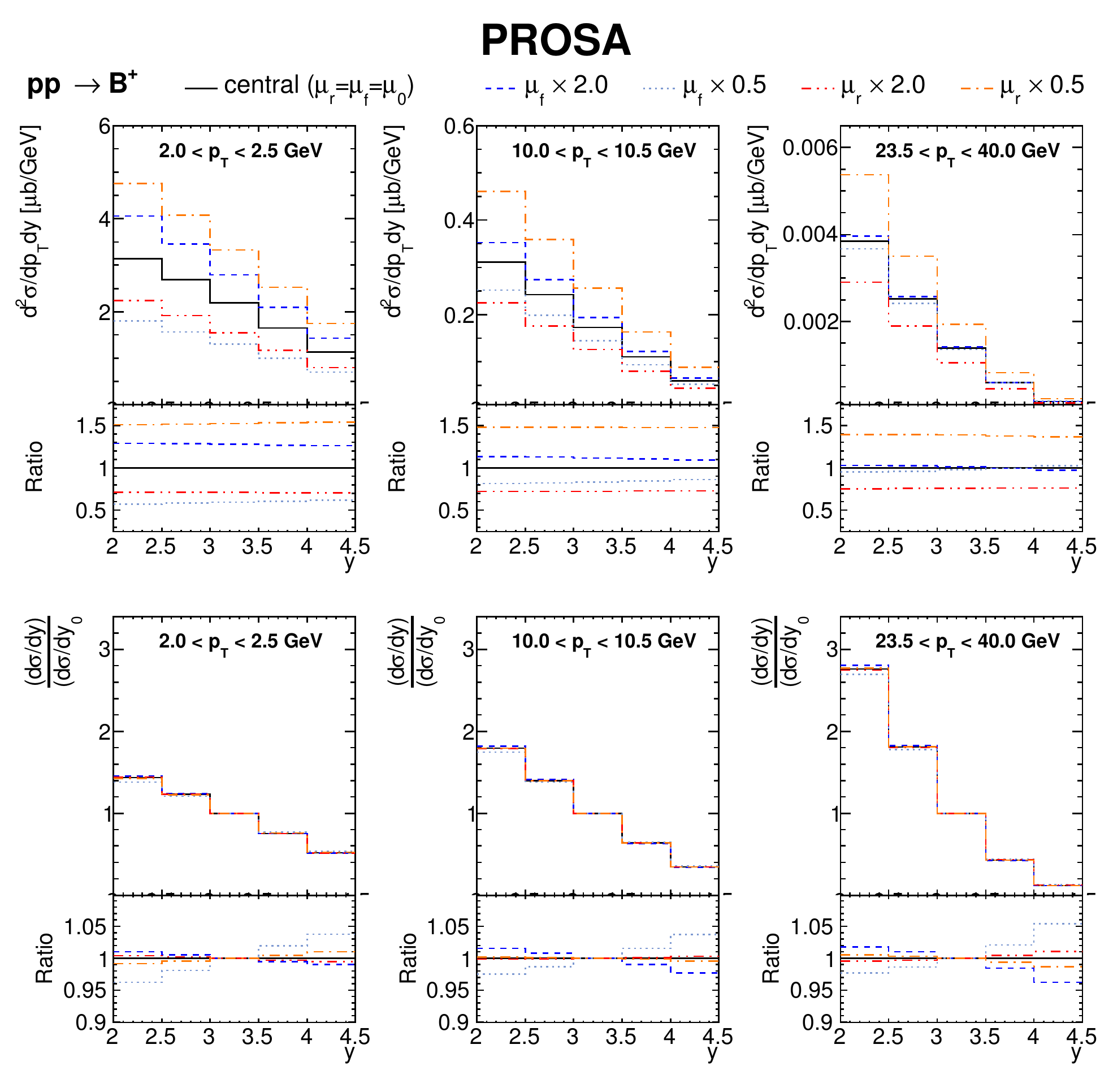}
  \caption[NLO predictions with scale variations for beauty LHCb data]
  {NLO QCD predictions for beauty LHCb data with different scale choices for absolute (top) and normalised (bottom) cross sections. 
	Bottom parts indicate the ratio of predictions to the central scale choice. 
	The predictions were obtained by using the FFNS variant of MSTW 2008 PDFs~\cite{Martin:2010db} with $n_f=3$; 
	the beauty mass was set to $m_b=4.5$ GeV.}
	\label{fig:pdffit:ScaleDep_Beauty}
\end{figure*}

	The $y$ shape of the cross-section ratio \dydyz in each $p_T$ bin was fitted, 
	where $\frac{{\rm d}\sigma}{{\rm d}y_0}$ is the cross section in the central rapidity bin $3.0<y_0<3.5$. 
	The virtue is that the observable defined in this way has a much reduced scale dependence, while it \ozmodNN{still remains} sensitive to the PDFs, namely to their $x$ shape. 
	This can be understood easily: the change in the production rate in neighboring bins of $y$ is driven mainly by the change in the input PDFs, while the 
	hard-scattering process remains essentially the same. Hence the $\mu_r$ dependence is reduced to $\sim 1\%$ and the $\mu_f$ to $\sim 5\text{--}10\%$ 
	(the renormalisation scale affects the matrix elements only, while the factorisation scale affects both the matrix elements and the PDFs). 
	Reduction of the scale dependence is illustrated in Figs.~\ref{fig:pdffit:ScaleDep_Charm},~\ref{fig:pdffit:ScaleDep_Beauty}.
	In addition, the dependence on the heavy-quark mass and the fragmentation function is also significantly reduced. For the mass, it is reduced almost to zero, 
	while the fragmentation effects are still sizeable at low transverse momentum, since the fragmentation is performed by rescaling the quark three-momentum and 
	thus it changes the rapidity of a massive particle.

	Owing to the greatly reduced dependence on the scales, in this variant of the fit the `common' scale choice and variations were used: 
	\ozmod{the scale parameters were fixed to $A^c_f=A^b_f=A^c_r=A^b_r=1$ and for the estimation of the scale uncertainties they were varied independently one at a time in the range [$0.5$;$2.0$].} 
	\ozmod{Explicitly}, the scale uncertainties include the following four variations:
	\begin{itemize}
		\item $A^c_f=A^b_f=2.0$, $A^c_r=A^b_r=1.0$;
		\item $A^c_f=A^b_f=0.5$, $A^c_r=A^b_r=1.0$;
		\item $A^c_f=A^b_f=1.0$, $A^c_r=A^b_r=2.0$;
		\item $A^c_f=A^b_f=1.0$, $A^c_r=A^b_r=0.5$.
	\end{itemize}
	Similar to the previous approach, 
	for the variation $A^c_f=A^b_f=0.50$, the cut $p_T>\SI{2}{GeV}$ was applied for the charm LHCb data to ensure that the factorisation scale 
	is above $\SI{1}{GeV^2}$.
	
	All correlated experimental systematic uncertainties and the fragmentation fractions cancel for the normalised cross sections, 
	while for a given $p_T$ bin the uncorrelated uncertainty of the central $\frac{{\rm d}\sigma}{{\rm d}y_0}$ bin was treated as correlated between the remaining $y$ bins 
	(because the cross sections in the remaining $y$ bins were divided by the same $\frac{d \sigma}{dy_0}$).

\subsection{Fit results}
\label{sec:pdffit:res}

The results of three fits are presented and discussed:
\begin{itemize}
	\item the fit with the HERA-only data, referred to as `HERA only' (Section~\ref{sec:pdffit:res:heraonly});
	\item the fit with the HERA and LHCb data using the absolute LHCb cross sections, referred to as `LHCb Abs' (Section~\ref{sec:pdffit:res:lhcbabs});
	\item the fit with the HERA and LHCb data using the normalised LHCb cross sections, referred to as `LHCb Norm' (Section~\ref{sec:pdffit:res:lhcbnorm}).
\end{itemize}
The direct comparison of all fitted PDFs \ozmodNN{follow in} Section~\ref{sec:pdffit:impact}.

\subsubsection{`HERA only'}
\label{sec:pdffit:res:heraonly}

\begin{figure}[htbp]
  \centering
  \includegraphics[width=0.75\figwidth,trim=0 2mm 0 9mm,clip=true]{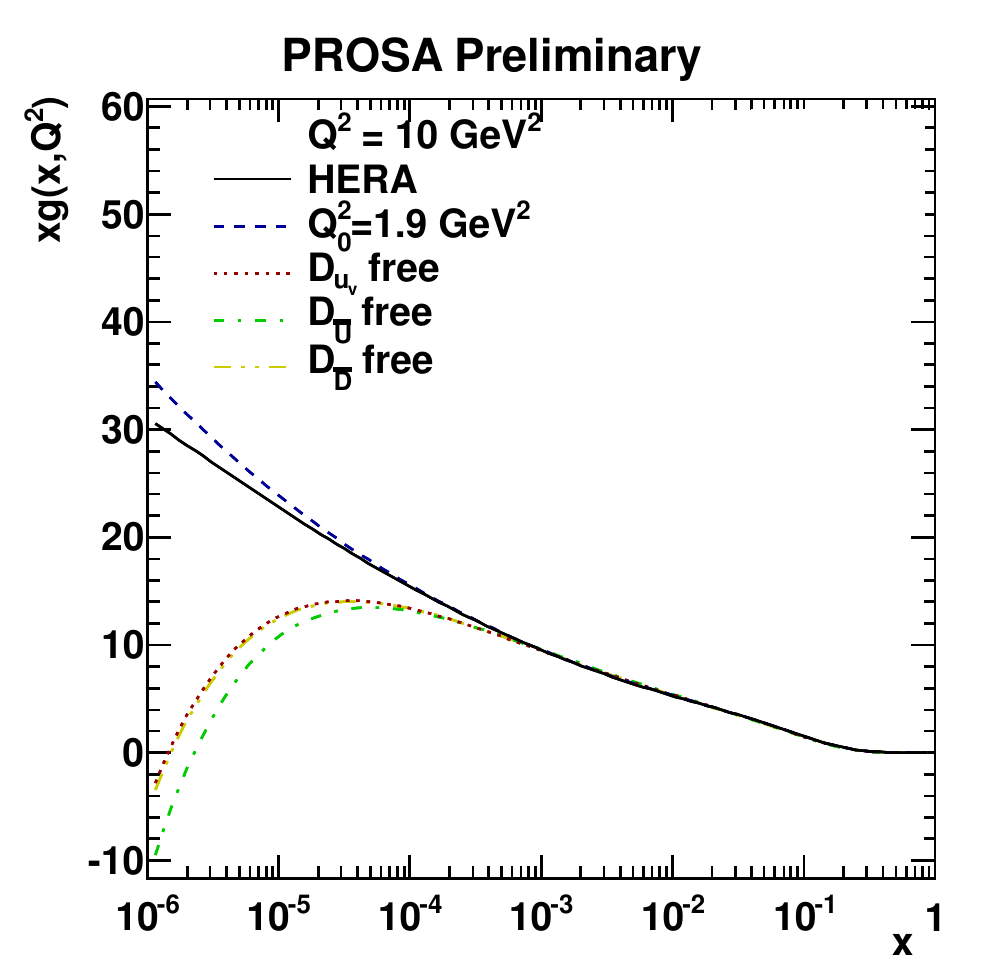}
  \caption[Parametrisation variations for gluon distribution in `HERA only' fit]
  {The parametrisation variations for the gluon distribution at $Q^2=\SI{10}{GeV^2}$ in the fit with the HERA-only data. 
  \ozmodNNN{`HERA' stands for the nominal fit, `$Q^2_0 = 1.9$ GeV$^2$' for the parametrisation parametrisation corresponding 
    to the change of the PDF evolution starting scale to $1.9$ GeV$^2$, and `$D_{u_v}$ free', `$D_{\bar{U}}$ free', `$D_{\bar{D}}$ free' 
    for the parametrisation variations corresponding to released parameters $D_{u_v}$, $D_{\bar{U}}$, $D_{\bar{D}}$, respectively, one at a time 
    (see Section~\ref{sec:comb:red:ffns} and Eq.~\ref{eq:sec:comb:red:ffns:pdfpar}).}}
	\label{fig:sec:pdffit:VarPar_HERAOnly_q2_10}
\end{figure}

Here the results of the fit with the HERA-only data (datasets 1--6) are presented. 
The total $\chi^2$ per degree of freedom for the fit is $\chisqndof=647/646$ \ozmodN{yields a} good consistency of the data. 
\ozmod{All individual $\chi^2$ contributions (see Eq.~\ref{eq:comb:proc:finalchi2fit} 
in Section~\ref{sec:comb:red:ffns}) are given in Table~\ref{tab:pdffit:chisqndof}.} 
\ozmod{For reference, Table~\ref{tab:pdffit:par} \ozmodN{lists} the fitted parameters (see Eq.~\ref{eq:sec:comb:red:ffns:pdfpar}) with their fit uncertainties only, 
while the total PDF uncertainties are shown below in Figs.~\ref{fig:pdffit:FittedPDFs_HERAOnly_q2_10} to~\ref{fig:pdffit:FittedPDFs_LHCbNorm_q2_10}.} 

\begin{table}[tbp]
\caption[\chisqndof for all datasets for three variants of fit]
{\chisqndof for all datasets for three variants of the fit. 
The contributions from correlated sources and logarithmic correction, total \chisqndof and the corresponding probability values are also given.}
\label{tab:pdffit:chisqndof}
\begin{center}
\footnotesize
\tabcolsep0.3mm
\renewcommand*{\arraystretch}{1.2}
\begin{tabularx}{\columnwidth}{|X|rcl|rcl|rcl|}
\hline
\multirow{2}{*}{Dataset} & \multicolumn{9}{c|}{$\chisqndof$}  \\ \cline{2-10}
 & \multicolumn{3}{c|}{HERA only} & \multicolumn{3}{c|}{LHCb Abs} & \multicolumn{3}{c|}{LHCb Norm} \\ \hline
   NC DIS HERA-I comb. $e^-p$ & $108$ &$/$& $145$ & $108$ &$/$& $145$ & $108$ &$/$& $145$  \\ \hline
   NC DIS HERA-I comb. $e^+p$ & $407$ &$/$& $379$ & $419$ &$/$& $379$ & $419$ &$/$& $379$  \\ \hline
   CC DIS HERA-I comb. $e^-p$ & $22$ &$/$& $34$ & $26$ &$/$& $34$ & $26$ &$/$& $34$  \\ \hline
   CC DIS HERA-I comb. $e^+p$ & $37$ &$/$& $34$ & $39$ &$/$& $34$ & $41$ &$/$& $34$  \\ \hline
   $c\bar{c}$ DIS HERA comb. & $50$ &$/$& $52$ & $78$ &$/$& $52$ & $47$ &$/$& $52$  \\ \hline
   $b\bar{b}$ DIS ZEUS Vertex & $12$ &$/$& $17$ & $16$ &$/$& $17$ & $12$ &$/$& $17$  \\ \hline
   LHCb $D^0$ & && & $68$ &$/$& $38$ & $17$ &$/$& $30$  \\ \hline
   LHCb $D^{+}$ & && & $53$ &$/$& $37$ & $18$ &$/$& $29$  \\ \hline
   LHCb $\Dstar$ & && & $50$ &$/$& $31$ & $19$ &$/$& $22$  \\ \hline
   LHCb $D_s^{+}$ & && & $24$ &$/$& $28$ & $11$ &$/$& $20$  \\ \hline
   LHCb $\Lambda_c^{+}$ & && & $5$ &$/$& $6$  & $5$ &$/$& $3$  \\ \hline
   LHCb $B^{+}$ & && & $99$ &$/$& $135$ & $81$ &$/$& $108$  \\ \hline
   LHCb $B^{0}$ & && & $66$ &$/$& $95$ & $35$ &$/$& $76$  \\ \hline
   LHCb $B_s^{0}$ & && & $78$ &$/$& $75$ & $23$ &$/$& $60$  \\ \hline
\hline
Correlated uncertainties & \multicolumn{3}{c|}{$9$} & \multicolumn{3}{c|}{$73$} & \multicolumn{3}{c|}{$49$} \\\hline
Logarithmic correction & \multicolumn{3}{c|}{$2$} & \multicolumn{3}{c|}{$-129$} & \multicolumn{3}{c|}{$48$} \\\hline
Total \chisqndof & $647$ &$/$& $646$ & $1073$ &$/$& $1087$ & $958$ &$/$& $994$ \\\hline
p(\chisq,\ndof) & \multicolumn{3}{c|}{$49\%$} & \multicolumn{3}{c|}{$61\%$} & \multicolumn{3}{c|}{$79\%$} \\
\hline
\end{tabularx}
\end{center}
\end{table}

\begin{table}[h]
\caption[Fitted parameters for QCD analysis]{
  \label{tab:pdffit:par}
	The fitted parameters for the QCD analysis (see Eq.~\ref{eq:sec:comb:red:ffns:pdfpar}). 
	The listed uncertainties are the fitting uncertainties only. Uncertainties are not quoted for parameters 
	that are fixed.}
\begin{center}
\tabcolsep0.5mm
\renewcommand*{\arraystretch}{1.25}
\begin{tabu} to \columnwidth {|X[l]|c|c|c|} \hline
 Parameter   & HERA only  &  LHCb Abs & LHCb Norm  \\ 
  \hline
  $B_g$ & $-0.08 \pm 0.14$& $-0.135 \pm 0.069$& $-0.075 \pm 0.095$  \\ 
  $C_g$ & $7.3 \pm 1.1$& $6.83 \pm 0.31$& $5.23 \pm 0.34$  \\ 
  $A'_{g}$ & $1.99 \pm 0.60$& $1.74 \pm 0.22$& $1.29 \pm 0.32$  \\ 
  $B'_{g}$ & $-0.15 \pm 0.11$& $-0.194 \pm 0.044$& $-0.155 \pm 0.050$  \\ 
  $B_{u_{{v}}}$ & $0.688 \pm 0.025$& $0.668 \pm 0.020$& $0.649 \pm 0.021$  \\ 
  $C_{u_{{v}}}$ & $4.75 \pm 0.24$& $4.99 \pm 0.23$& $4.98 \pm 0.23$  \\ 
  $E_{u_{{v}}}$ & $10.1 \pm 2.4$& $12.2 \pm 2.4$& $13.5 \pm 2.7$  \\ 
  $B_{d_{{v}}}$ & $0.86 \pm 0.10$& $0.928 \pm 0.093$& $0.959 \pm 0.088$  \\ 
  $C_{d_{{v}}}$ & $4.95 \pm 0.53$& $5.50 \pm 0.56$& $5.59 \pm 0.55$  \\ 
  ${C_{\overline{U}}}$ & $1.79 \pm 0.35$& $1.63 \pm 0.21$& $1.63 \pm 0.24$  \\ 
  ${A_{\overline{D}}}$ & $0.1466 \pm 0.0088$& $0.1727 \pm 0.0068$& $0.1579 \pm 0.0073$  \\ 
  ${B_{\overline{D}}}$ & $-0.1663 \pm 0.0081$& $-0.1462 \pm 0.0058$& $-0.1551 \pm 0.0067$  \\ 
  ${C_{\overline{D}}}$ & $4.6 \pm 1.8$& $10.4 \pm 2.5$& $15.1 \pm 4.2$  \\ 
  $m_c$ [GeV] & $1.344 \pm 0.055$& $1.709 \pm 0.024$& $1.257 \pm 0.014$  \\ 
  $m_b$ [GeV] & $4.31 \pm 0.16$& $4.673 \pm 0.079$& $4.19 \pm 0.13$  \\     
  $A_{f}^{c}$ && $0.659 \pm 0.020$& $ 1.0 $  \\ 
  $A_{f}^{b}$ && $0.262 \pm 0.007$&$ 1.0 $  \\ 
  $A_{r}^{c}$ && $0.444 \pm 0.021$& $ 1.0 $  \\ 
  $A_{r}^{b}$ && $0.335 \pm 0.024$& $ 1.0 $  \\ 
  \hline
    \end{tabu}
  \end{center}
\end{table}

The fit, model and parametrisation uncertainties of the gluon, sea%
\footnote{The sea-quark distribution is defined as $\Sigma = \overline{u} + \overline{d} + \overline{s}$.} 
and valence-quark distributions at the scale $Q^2=\SI{10}{GeV^2}$ are shown in 
Fig.~\ref{fig:pdffit:FittedPDFs_HERAOnly_q2_10}. 
Note the gluon uncertainties in the low-$x$ region: since this region is not 
covered directly by the HERA data, the dominant uncertainties are the parametrisation ones, namely those which 
\ozmodN{arise} from releasing the $D_{\bar{U}}$ parameter. 
This is illustrated in Fig.~\ref{fig:sec:pdffit:VarPar_HERAOnly_q2_10}, 
where all parametrisation variations are shown separately. 
\ozmodNN{At low values $x \lesssim 10^{-4}$ the gluon distribution is not directly constrained by the HERA data 
and should be considered as an extrapolation which relies on PDF parametrization assumptions.} 
Qualitatively it can be understood in the 
following way: since gluons in this region are not constrained by the data, they are \ozmodNN{indirectly} constrained via the sum 
rules for certain distributions of all other partons. 
When the parametrisation for other partons (in particular, for sea quarks) is changed, 
\ozmodN{new parameter values} via the sum rules \ozmodN{necessarily} result in a different distribution for gluons in the low-$x$ region. 
\ozmodNNN{However, the parametrisation variations do not result in a significant difference in the $x$ region constrained by the HERA data.} 
This also explains the large spread of the results obtained by the different PDF groups, which was observed in Fig.~\ref{fig:sec:pdffit:benchmark}: 
since the different groups use different parametrisations and none of them uses data which constrain gluons at low $x$, 
they \ozmodN{all} obtain different gluon distributions in this region. 

\begin{figure}[htbp]
  \centering
  \includegraphics[width=0.495\figwidth,trim=2mm 2mm 2mm 8.5mm,clip=true]{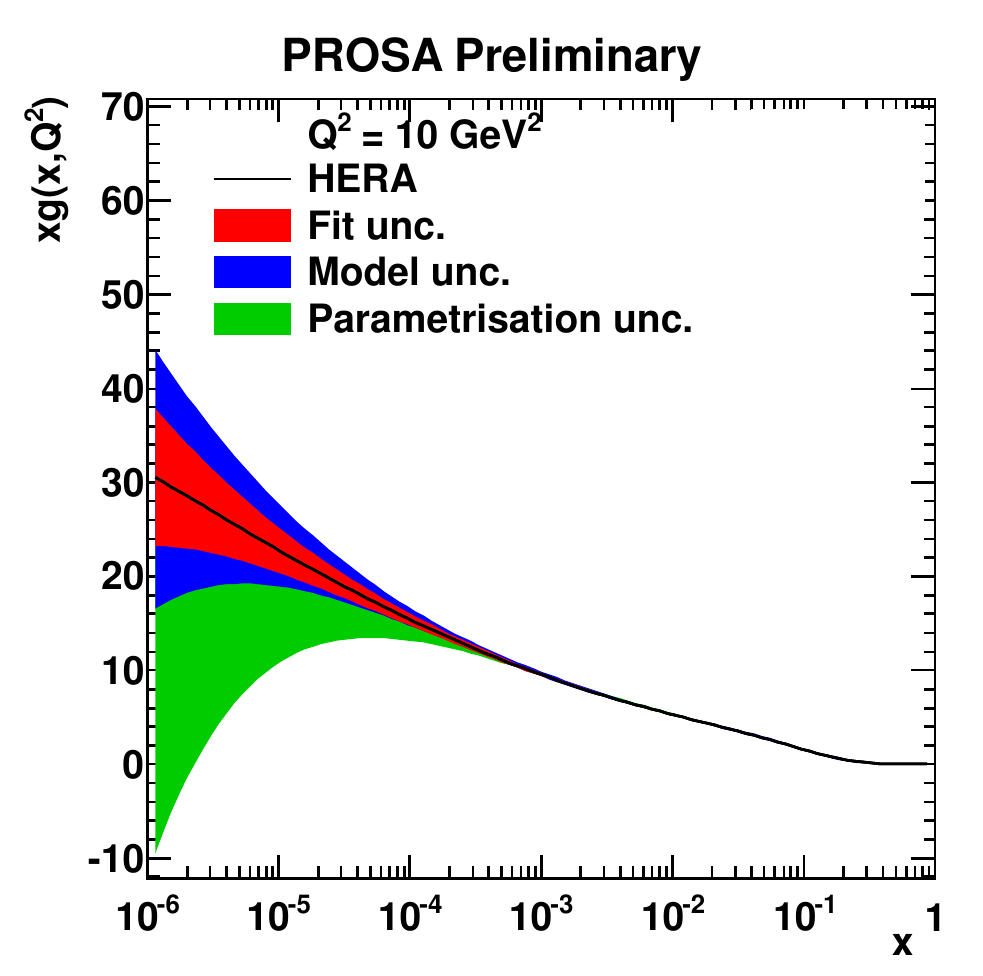}
  \includegraphics[width=0.495\figwidth,trim=2mm 2mm 2mm 8.5mm,clip=true]{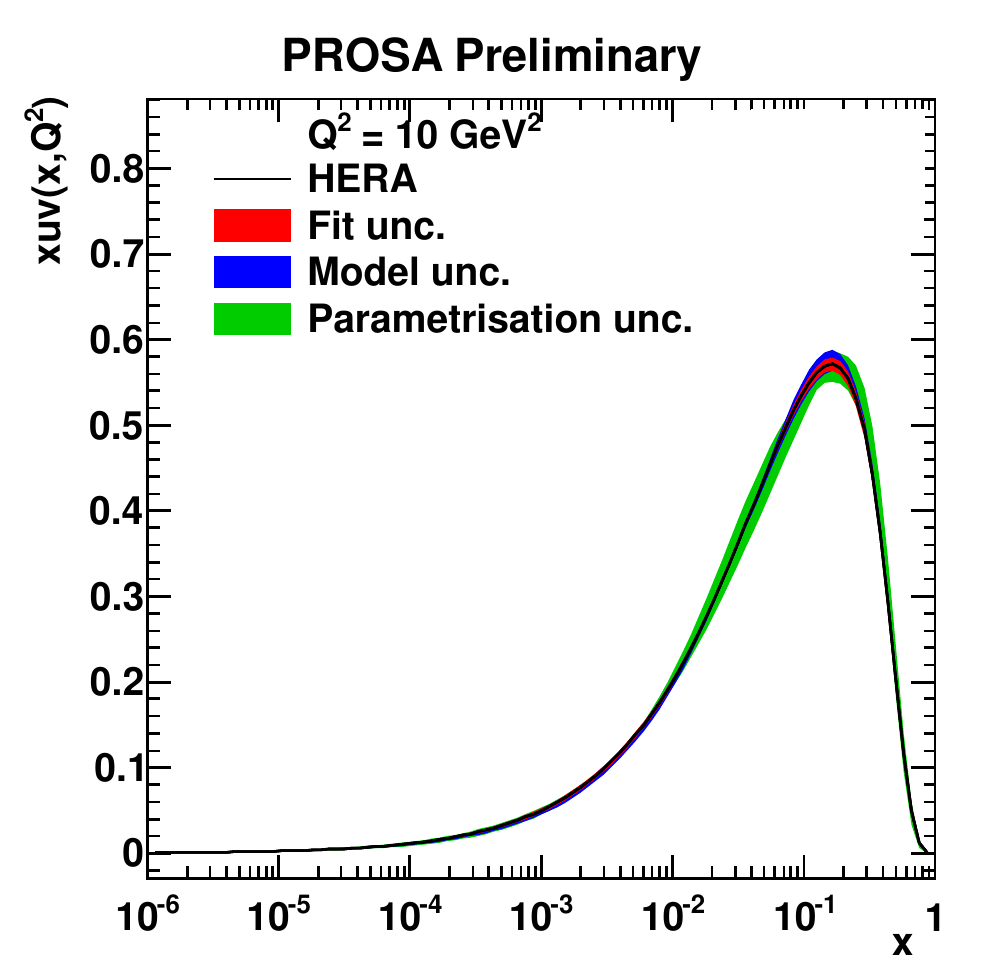}
  \includegraphics[width=0.495\figwidth,trim=2mm 2mm 2mm 8.5mm,clip=true]{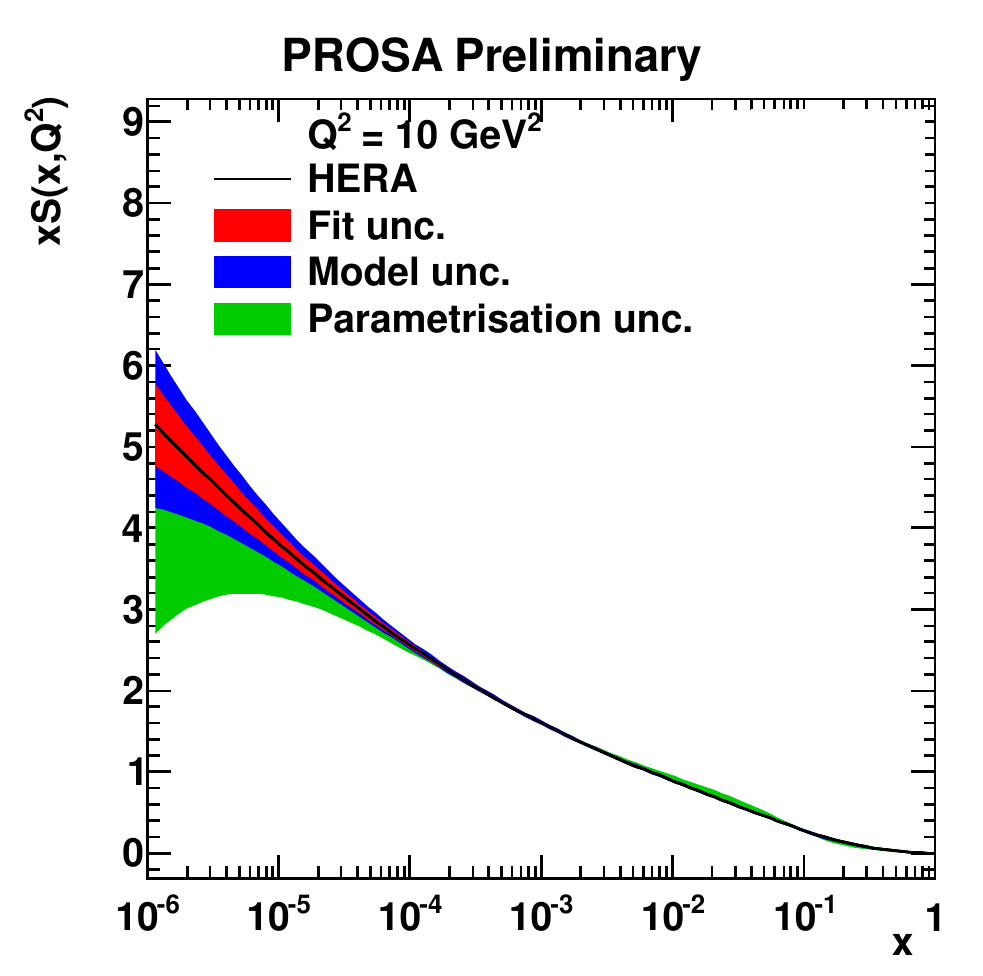}
  \includegraphics[width=0.495\figwidth,trim=2mm 2mm 2mm 8.5mm,clip=true]{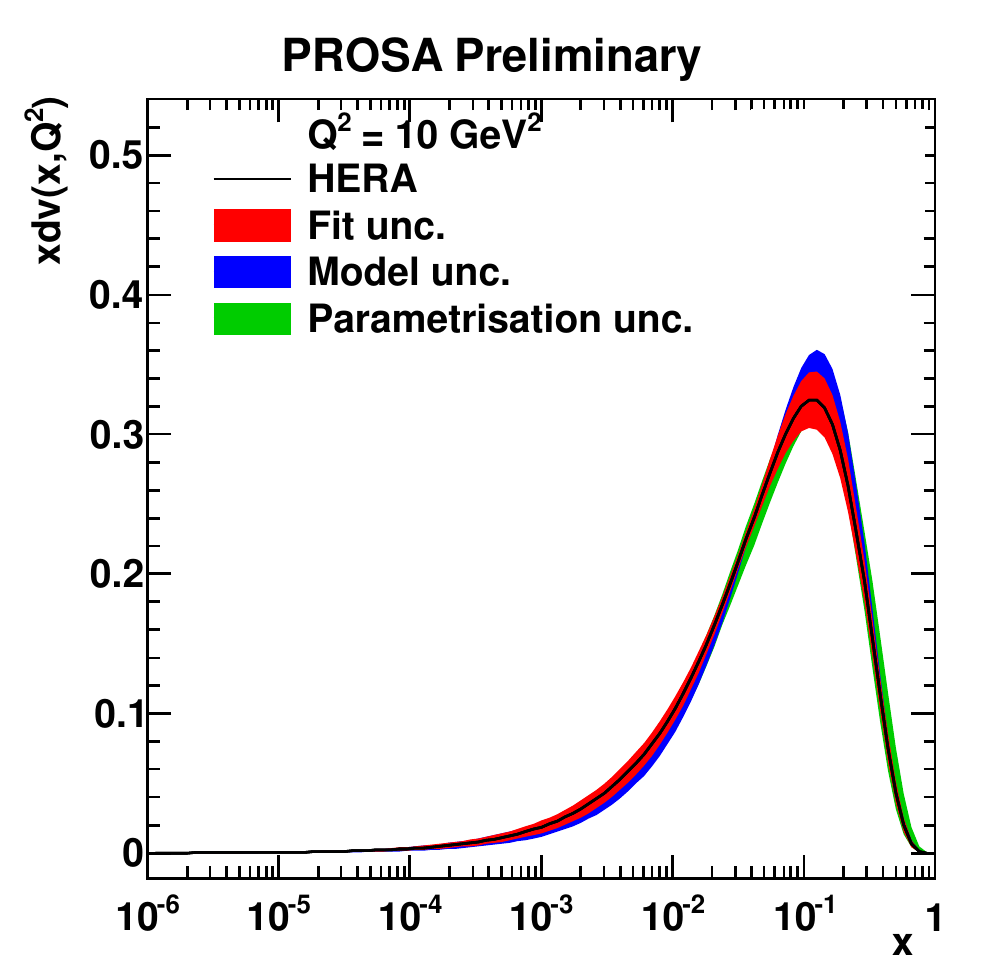}
  \caption[Individual contributions to PDF uncertainties for `HERA only' fit]
  {The individual contributions to the uncertainties of the gluon (top left), $u$-valence (top right), sea (bottom left) and $d$-valence (bottom right) distributions at $Q^2=\SI{10}{GeV^2}$ 
  obtained in the fit with the HERA-only data.}
	\label{fig:pdffit:FittedPDFs_HERAOnly_q2_10}
\end{figure}

\subsubsection{`LHCb Abs'}
\label{sec:pdffit:res:lhcbabs}

\begin{figure*}[htbp]
  \centering
  \includegraphics[width=0.75\figwidth,trim=1mm 0 4mm 8mm,clip=true]{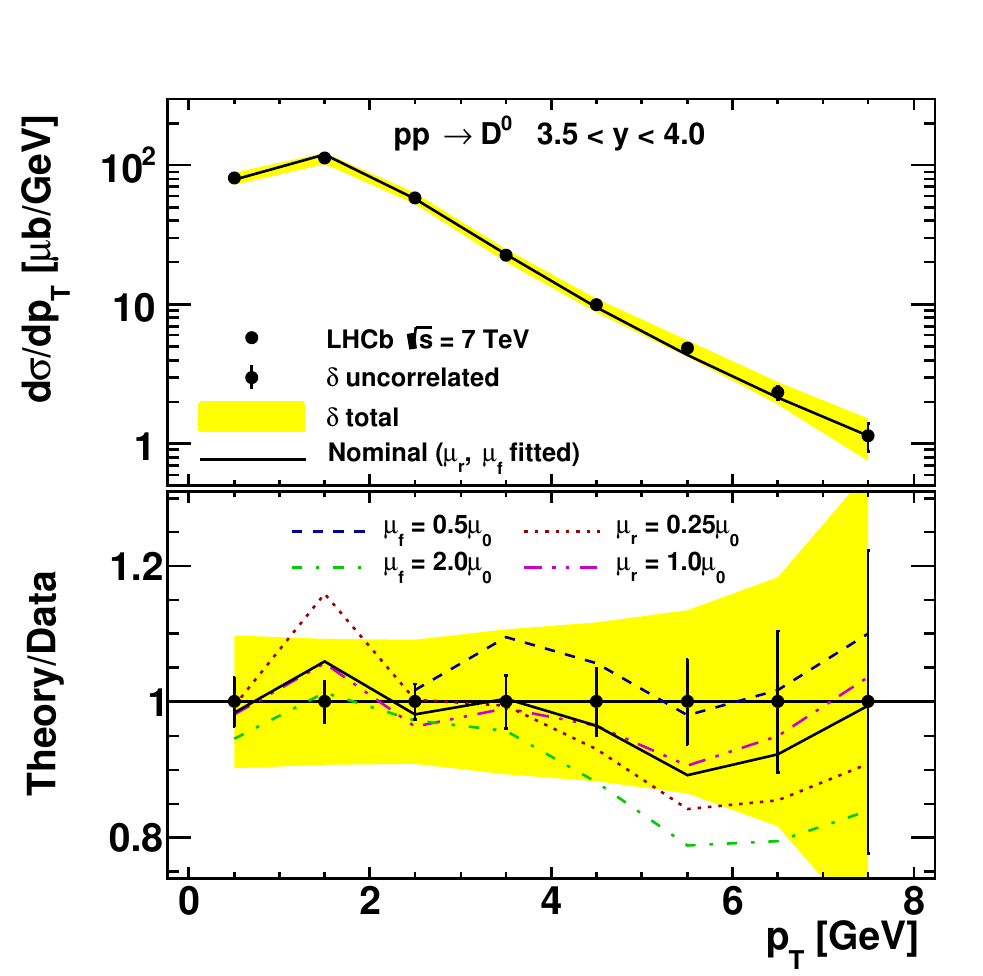}
  \includegraphics[width=0.75\figwidth,trim=1mm 0 4mm 8mm,clip=true]{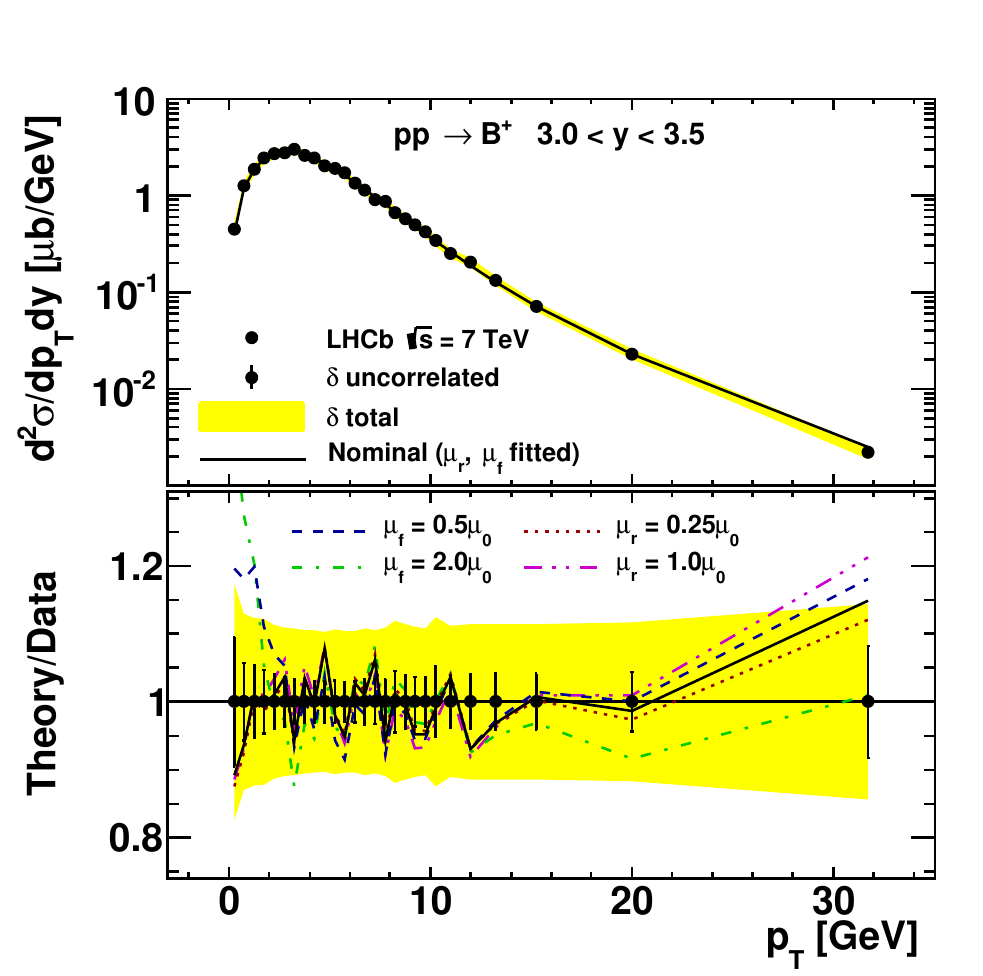}
  \caption[Data to theory comparison for LHCb absolute cross sections]
  {Data to theory comparison for a representative subset of the LHCb absolute cross sections: 
	$D^0$ mesons, bin $3.5<y<4.0$ (left); $B^{+}$ mesons, bin $3.0<y<3.5$ (right). 
	In the bottom panels the ratios theory/data for the nominal fit \ozmod{with fitted renormalisation and factorisation scales,} and the scale 
	variations are shown. For demonstration purpose, correlated shifts for data points obtained in the fit using nuisance 
	parameters are applied to theoretical predictions. Uncorrelated uncertainties for data points are shown as they are 
	rescaled in the fit, while total uncertainties are shown as not rescaled.}
	\label{fig:pdffit:FittedData_LHCbAbs}
\end{figure*}

Here the results of the fit with the HERA and LHCb data (datasets 1--14) using the `LHCb Abs' approach are presented. 
The total $\chisq$ per degree of freedom is $\chisqndof=1073/1087$. 
The partial \chisqndof for all datasets are given in Table~\ref{tab:pdffit:chisqndof}. 
For the LHCb charm and beauty datasets they vary from $0.9$ to $1.8$ and from $0.7$ to $1.0$, respectively, indicating an overall 
reasonable description of the charm and beauty data. 
\ozmod{The inclusion of the LHCb data do not change significantly the \ozmodNN{quality of the} description of the HERA datasets compared to the `HERA only` fit, 
\ozmodNN{assuring} consistency between HERA and LHCb data.}%
\footnote{\ozmod{With the only exception of the worsening of the HERA charm data description due to the 
value of $m_c$ which is in a tension with that preferred by the HERA charm data 
(see Tab.~\ref{tab:pdffit:par}), if only the criterion of a $\chi^2$ variation of $1$ is considered.}} 
As an example of the data description in the fit, in Fig.~\ref{fig:pdffit:FittedData_LHCbAbs} the cross sections for $D^{0}$ and $B^{+}$ mesons for one of the $y$ bins are shown. 

\ozmod{Adjusting the scales in the fit results in the following fitted values for the scale parameters (see Table~\ref{tab:pdffit:par}):}
\begin{equation}
\begin{split}
	A^c_f=0.66,\\
	A^c_r=0.44,\\
	A^b_f=0.26,\\
	A^b_r=0.33.
\end{split}
\label{eq:fittedscales}
\end{equation}
Note that all values are within the range [0.25;1.00]; also note the significant difference between the fitted scales for charm and beauty. 
Additionally, as expected, a positive correlation was found between $A^c_f$, $A^c_r$ and $A^b_f$, $A^b_r$, respectively.
\ozmod{The fitted pole masses of the $c$ and $b$ quarks are consistent with the ones obtained in the `HERA only' fit 
within the intrinsic theoretical systematic uncertainty of the pole mass definition (see Section~\ref{sec:th:qcd:pqcd:mass}).}

\begin{figure}[htbp]
  \centering
  \begin{minipage}[t]{0.49\columnwidth}
  \includegraphics[width=0.5\figwidth,trim=5 0 5 10mm,clip=true]{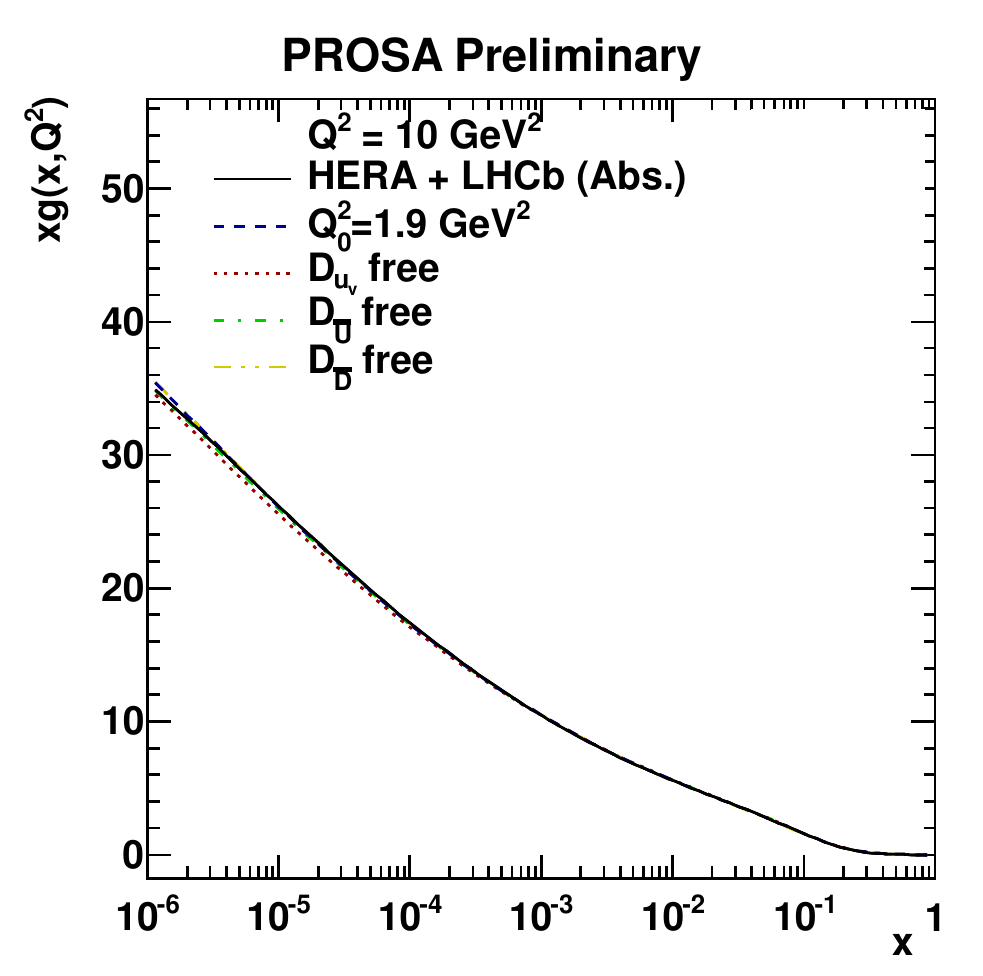}
  \includegraphics[width=0.5\figwidth,trim=5 0 5 0mm,clip=true]{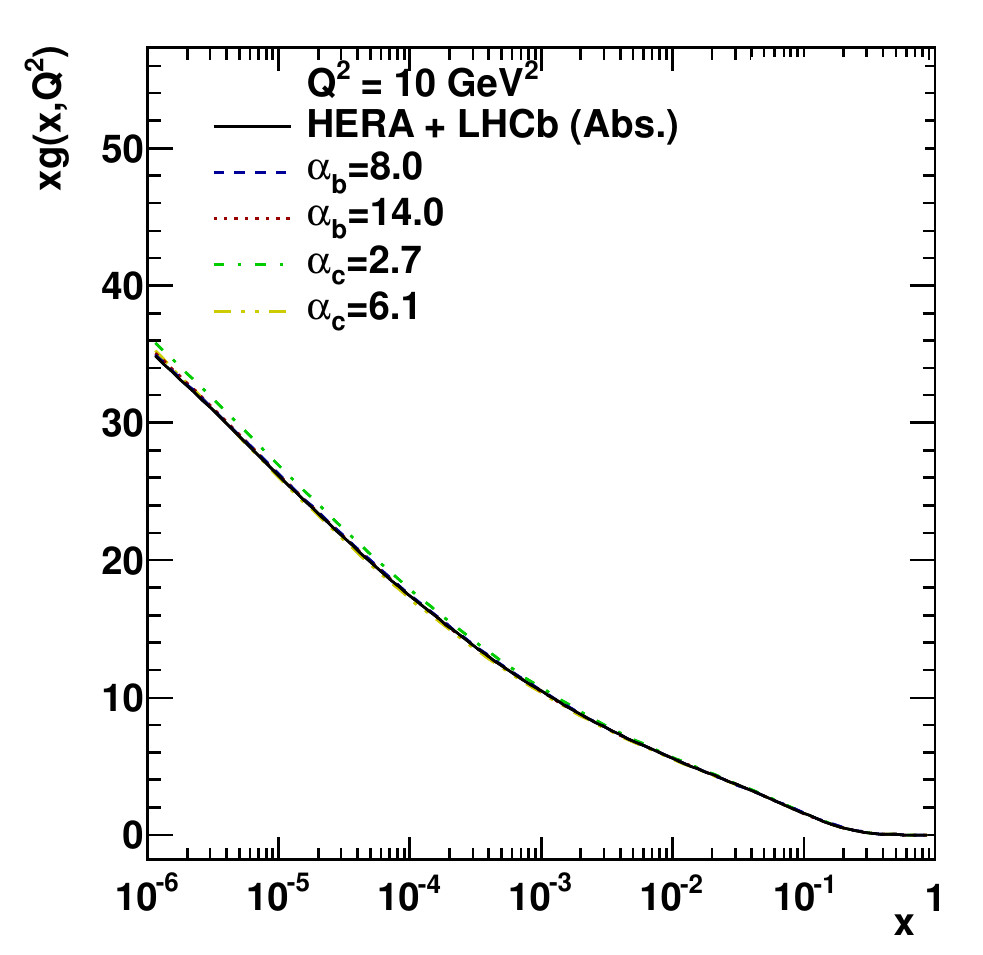}
  \end{minipage}
  \begin{minipage}[t]{0.49\columnwidth}
  \includegraphics[width=0.495\figwidth,trim=0 0 0 0mm,clip=true]{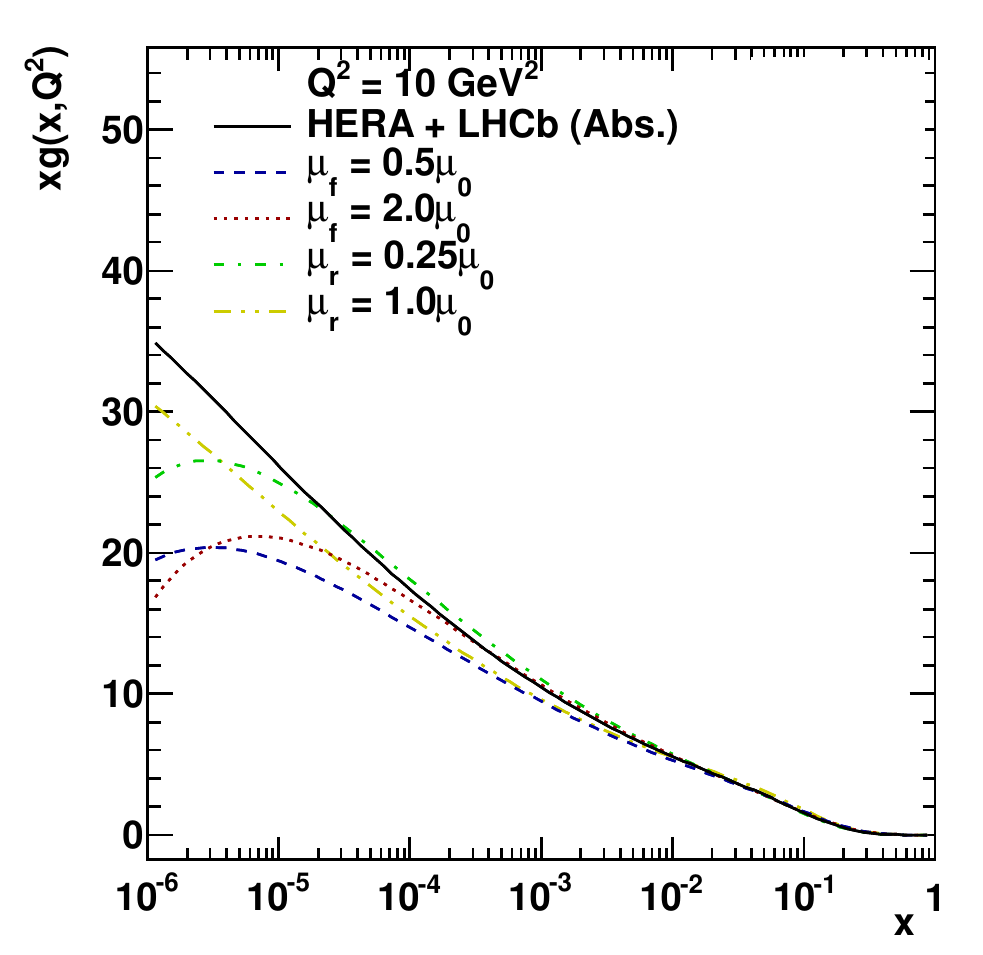}
  \begin{minipage}[t]{0.05\columnwidth}
  \end{minipage}
  \begin{minipage}[t]{0.95\columnwidth}
  \caption[Individual variations for gluon distribution in `LHCb Abs' fit]
  {The parametrisation (top left), scale (top right) and fragmentation (bottom) variations for the gluon distribution at $Q^2=\SI{10}{GeV^2}$ in the fit with the HERA and LHCb data using the `LHCb Abs' approach.}
	\label{fig:pdffit:Var_LHCbAbs_q2_10}
  \end{minipage}
  \end{minipage}
\end{figure}

\begin{figure}[htbp]
  \centering
  \includegraphics[width=0.495\figwidth,trim=2mm 2mm 2mm 8.5mm,clip=true]{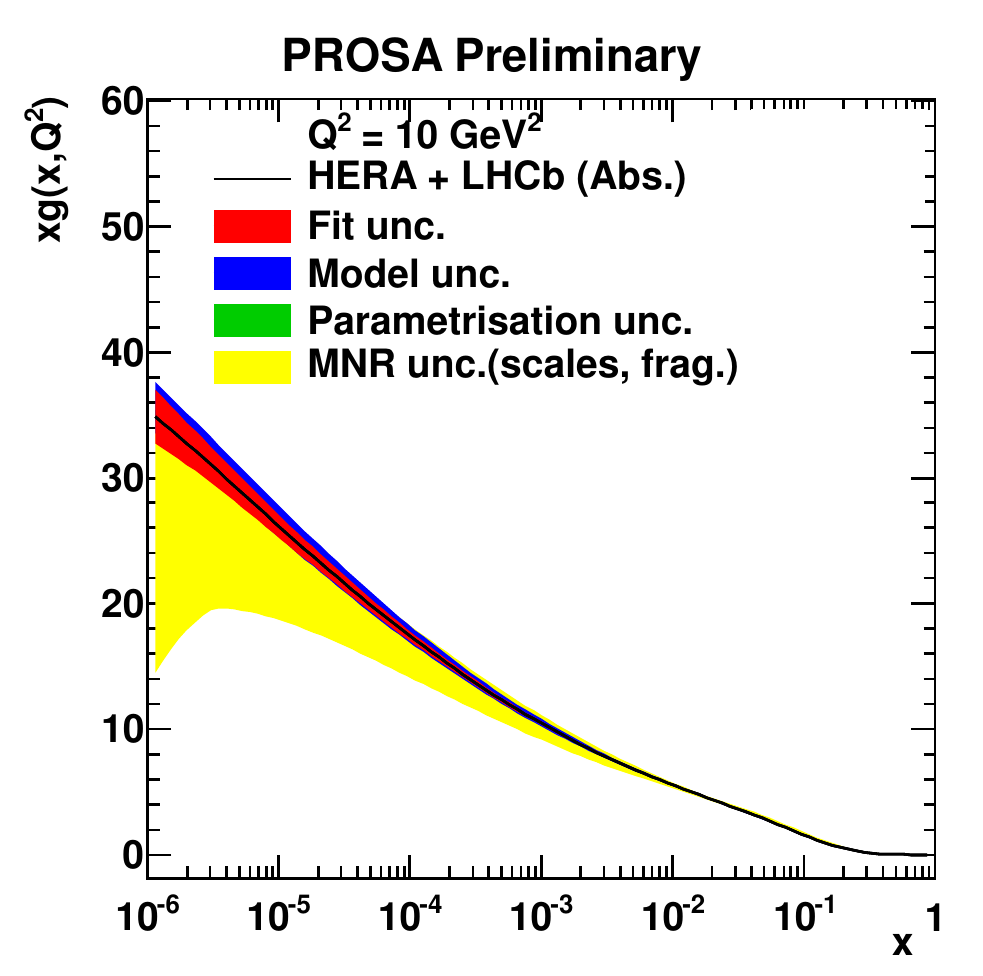}
  \includegraphics[width=0.495\figwidth,trim=2mm 2mm 2mm 8.5mm,clip=true]{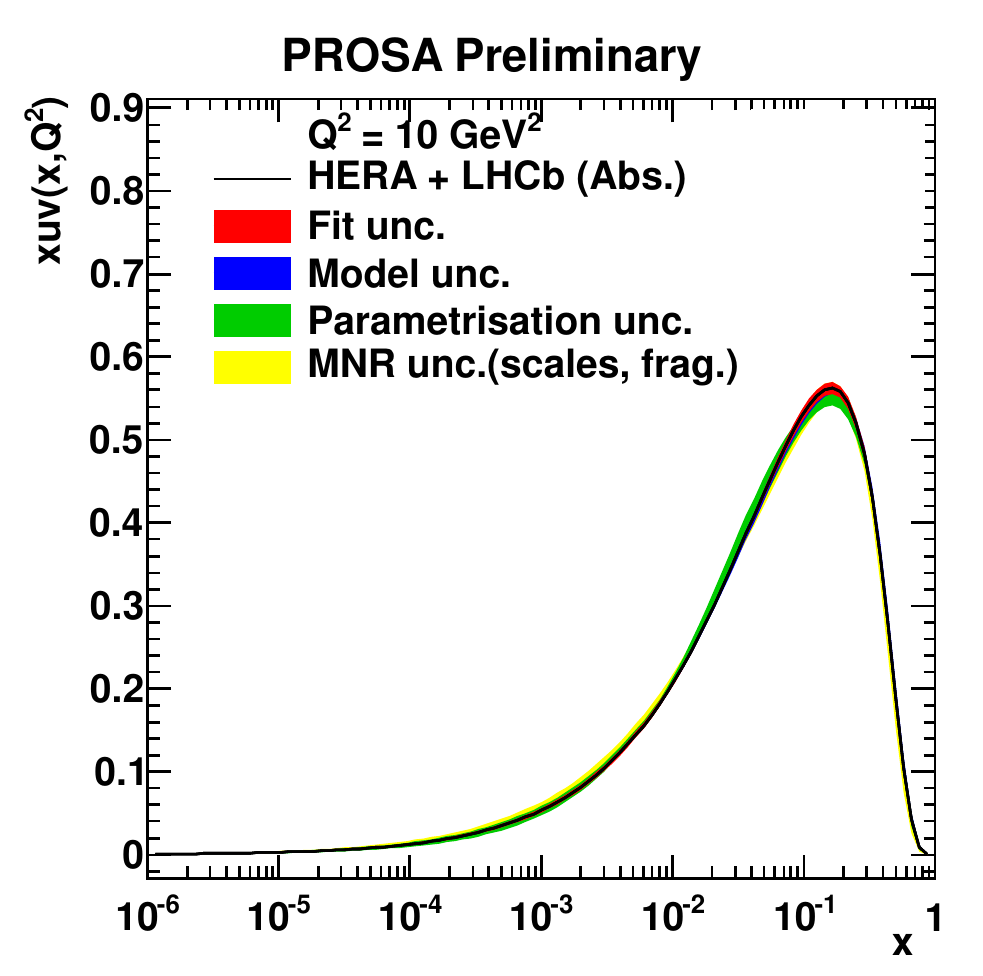}
  \includegraphics[width=0.495\figwidth,trim=2mm 2mm 2mm 8.5mm,clip=true]{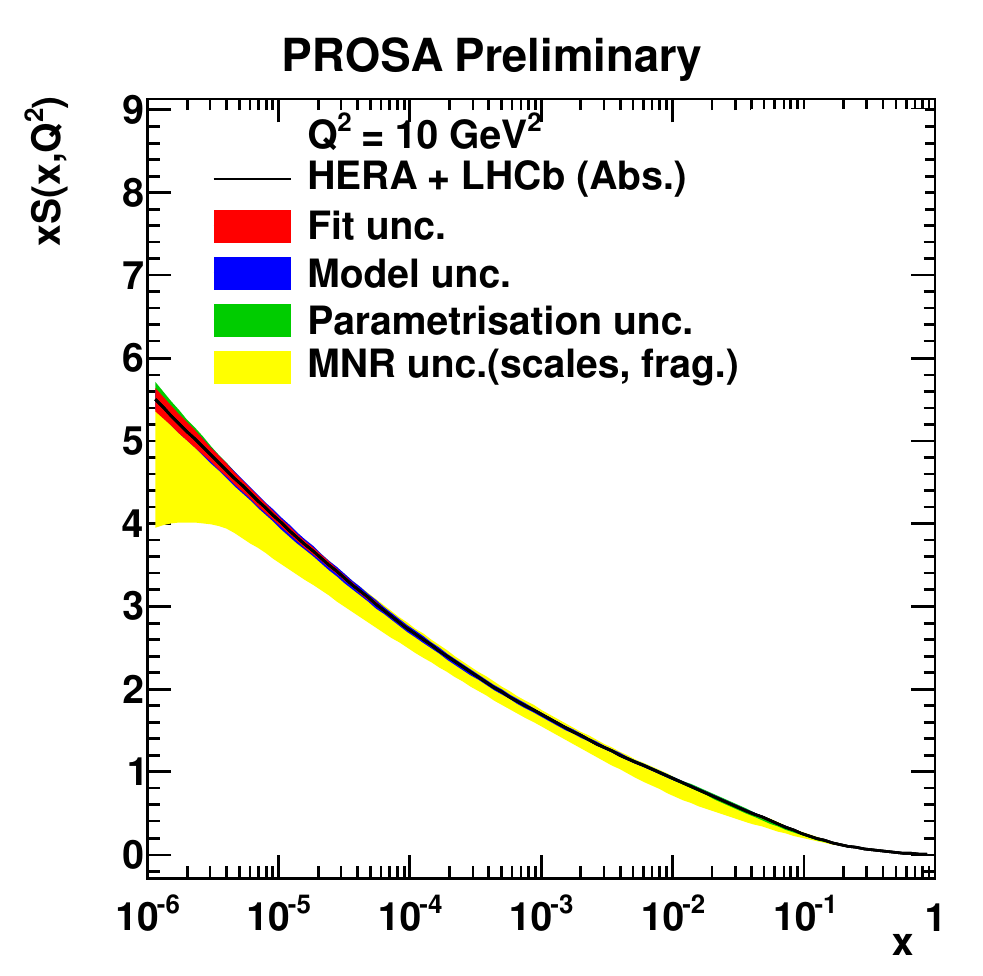}
  \includegraphics[width=0.495\figwidth,trim=2mm 2mm 2mm 8.5mm,clip=true]{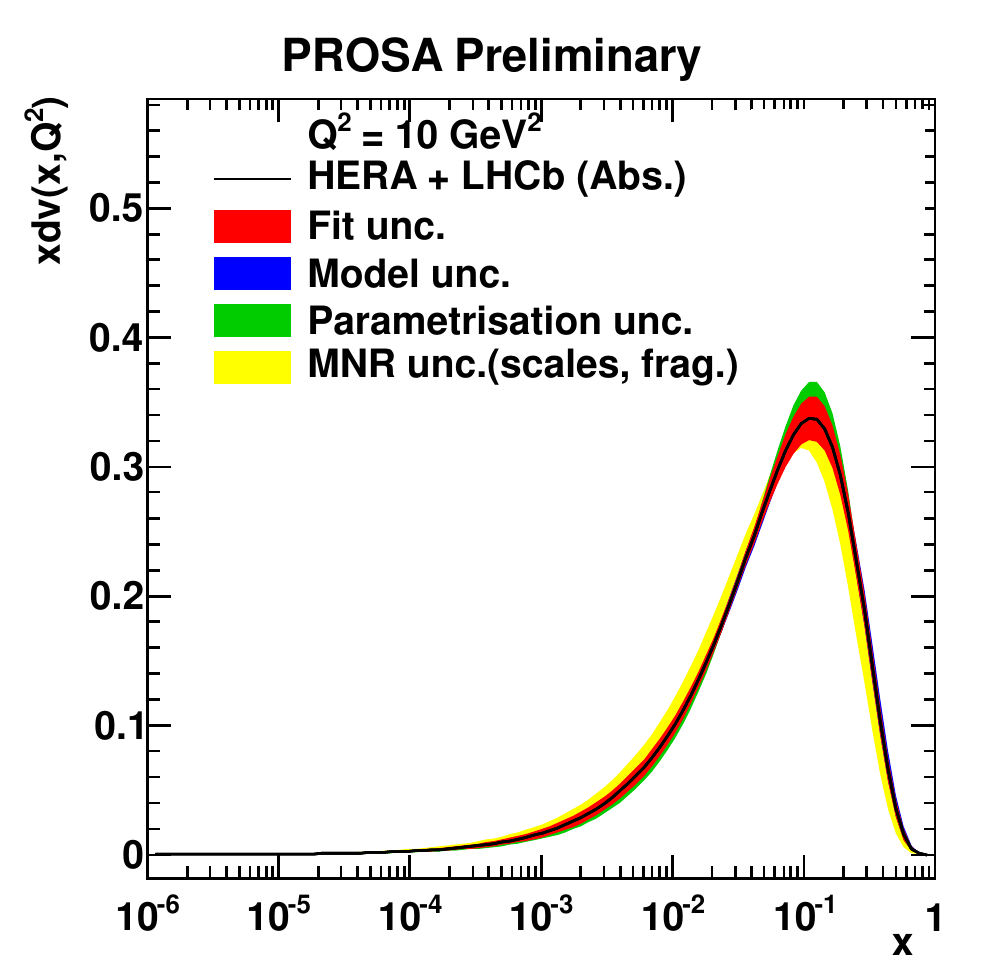}
  \caption[Individual contributions to PDF uncertainties for `LHCb Abs' fit]
  {The individual contributions to the uncertainties of the gluon (top left), $u$-valence (top right), sea (bottom left) and $d$-valence (bottom right) distributions at $Q^2=\SI{10}{GeV^2}$ 
  obtained in the fit with the HERA and LHCb data using the `LHCb Abs' approach.}
	\label{fig:pdffit:FittedPDFs_LHCbAbs_q2_10}
\end{figure}

The parametrisation variations are shown in Fig.~\ref{fig:pdffit:Var_LHCbAbs_q2_10} (top left). 
As expected, in contrast to the results obtained with the HERA-only data, gluons are now strongly constrained in the low-$x$ region.
\ozmod{The fit uncertainties for the parameters of the gluon distribution are significantly reduced, 
as can be seen from Table~\ref{tab:pdffit:par}.}
\ozmodNN{Note that Table~\ref{tab:pdffit:par} list only the fit uncertainties of the PDF parameters, 
while the other components of the PDF uncertainties are shown in Fig.~\ref{fig:pdffit:FittedPDFs_LHCbAbs_q2_10}.}

The effect of scale variations on the predictions in the fit is shown in Fig.~\ref{fig:pdffit:FittedData_LHCbAbs} and 
their effect on the fitted PDFs is shown in Fig.~\ref{fig:pdffit:Var_LHCbAbs_q2_10} (top right). 
The scale uncertainties are much larger than the parametrisation uncertainties, however compared 
to the fit with the HERA-only data they are a factor of 3 smaller than the total uncertainties. 
Another interesting observation from Figs.~\ref{fig:pdffit:FittedData_LHCbAbs} and~\ref{fig:pdffit:Var_LHCbAbs_q2_10} (top right) is 
that the changes for the predictions from the scale variations are predominantly changes in their normalisation 
(see also Figs.~\ref{fig:pdffit:ScaleDep_Charm},~\ref{fig:pdffit:ScaleDep_Beauty}); 
the fit handles them by adjusting the other scale and making large shifts for the correlated uncertainties of the data. 
\ozmodNN{It shows up in a formally bad $\chi^2$ value, 
however the PDFs are less affected and remain reasonable,} compared to the huge total `HERA only' uncertainty at low $x$.

In addition, variations of the fragmentation-function parameters were performed, as described in Section~\ref{sec:pdffit:th:det:frag}. 
The effect on the fitted PDFs is shown in Fig.~\ref{fig:pdffit:Var_LHCbAbs_q2_10} (bottom). 
All fragmentation variations \ozmodN{yield} in a good description of the data, since relatively small changes 
of the $p_T$ shape are easily compensated by adjusting the scales. 
The resulting uncertainties are much smaller compared to the scale variations. 
However, for charm a strong tendency was observed: the LHCb charm data \ozmodNN{favour} a harder fragmentation function; 
this is discussed in more detail in Appendix~\ref{sec:app:pdffit:frag}. 

Finally, all individual contributions to the uncertainties are shown in Fig.~\ref{fig:pdffit:FittedPDFs_LHCbAbs_q2_10}. 
The dominant uncertainties \ozmod{on the gluon distribution} in the low-$x$ region become the MNR ones, in particular coming from the scale variations. 

\subsubsection{`LHCb Norm'}
\label{sec:pdffit:res:lhcbnorm}

\begin{figure*}[htbp]
  \centering
  \includegraphics[width=0.75\figwidth,trim=0mm 0 4mm 8mm,clip=true]{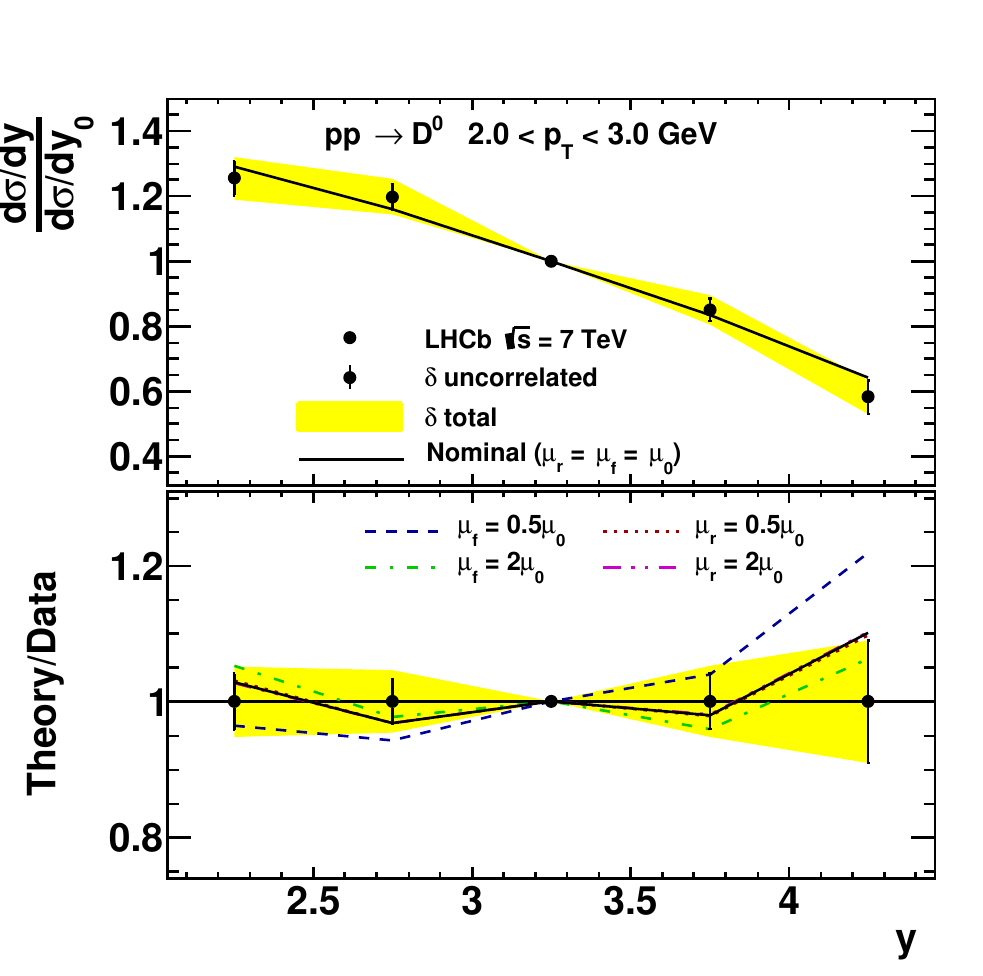}
  \includegraphics[width=0.75\figwidth,trim=0mm 0 4mm 8mm,clip=true]{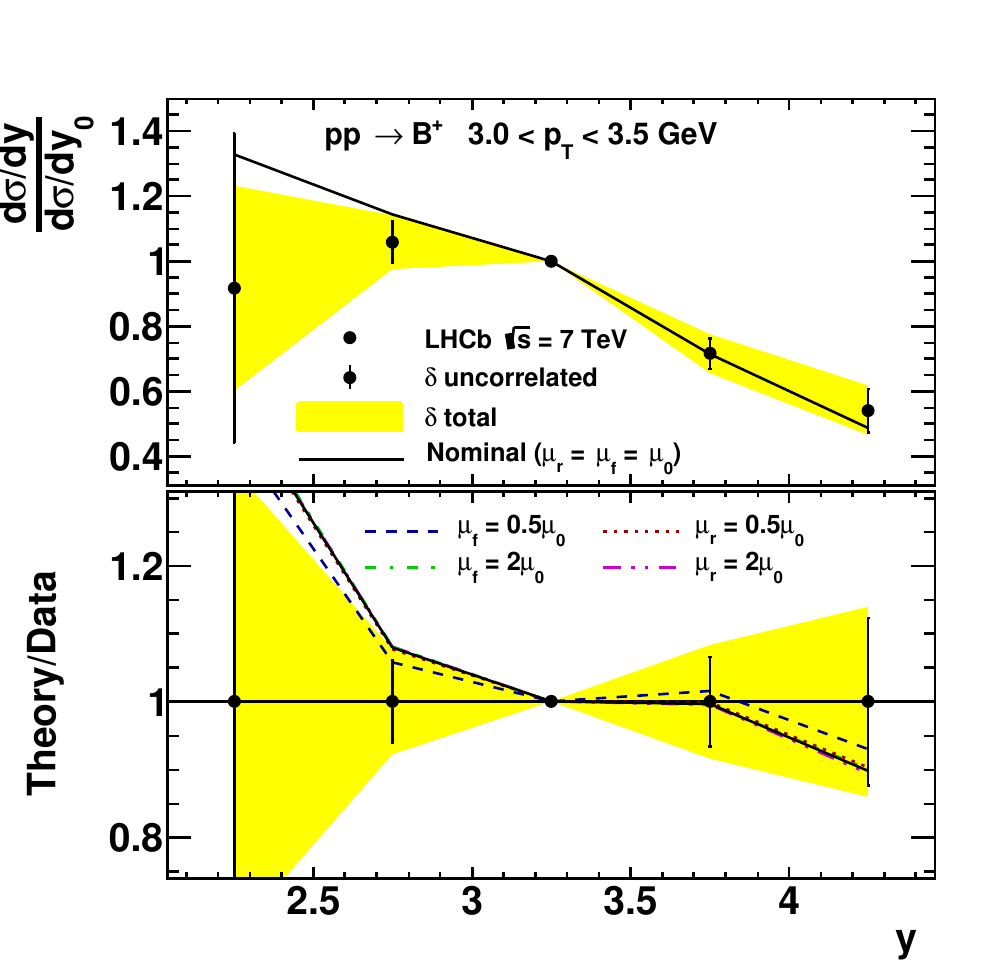}
  \caption[Data to theory comparison for LHCb normalised cross sections]
  {Data to theory comparison for a representative subset of the LHCb normalised cross sections: 
	$D^0$ mesons, bin $2.0 < p_T < 3.0$~GeV (left); $B^{+}$ mesons, bin $3.0 < p_T < 3.5$~GeV (right). 
	The central rapidity bins are fixed to $1$ by the definition of the normalised cross sections. 
	In the bottom panels the ratios theory/data for the nominal fit and the scale 
	variations are shown. For demonstration purpose, correlated shifts for data points obtained in the fit using nuisance 
	parameters are applied to theoretical predictions. Uncorrelated uncertainties for data points are shown as they are 
	rescaled in the fit, while total uncertainties are shown as not rescaled.}
	\label{fig:pdffit:FittedData_LHCbNorm}
\end{figure*}

Here the results of the fit with the HERA and LHCb data (datasets 1--14) using the `LHCb Norm' approach are presented. 
The total $\chisq$ per degree of freedom is $\chisqndof=958/994$. 
The partial \chisqndof for all datasets are given in Table~\ref{tab:pdffit:chisqndof}. 
For the LHCb charm and beauty datasets they vary from $0.4$ to $0.9$%
\footnote{Except for the low-statistics $\Lambda^{+}_c$ dataset, where $\chisqndof=4.9/3$.}, 
indicating description of the data and possible overestimation of the uncorrelated experimental uncertainties for the $y$ shape 
(this can be thought of rather as an underestimation of correlations of the systematics). 
\ozmod{Similar to the `LHCb abs' fit, the inclusion of the LHCb data do not change significantly \ozmodNN{the quality of} the description of the HERA datasets 
compared to the `HERA only` fit, \ozmodNN{assuring} consistency between HERA and LHCb data.}%
As an example of the data description in the fit, in Fig.~\ref{fig:pdffit:FittedData_LHCbNorm} the cross sections for $D^{0}$ and $B^{+}$ mesons for one of the $y$ bins are shown. 

\begin{figure}[htbp]
  \centering
  \begin{minipage}[t]{0.49\columnwidth}
  \includegraphics[width=0.5\figwidth,trim=5 0 5 10mm,clip=true]{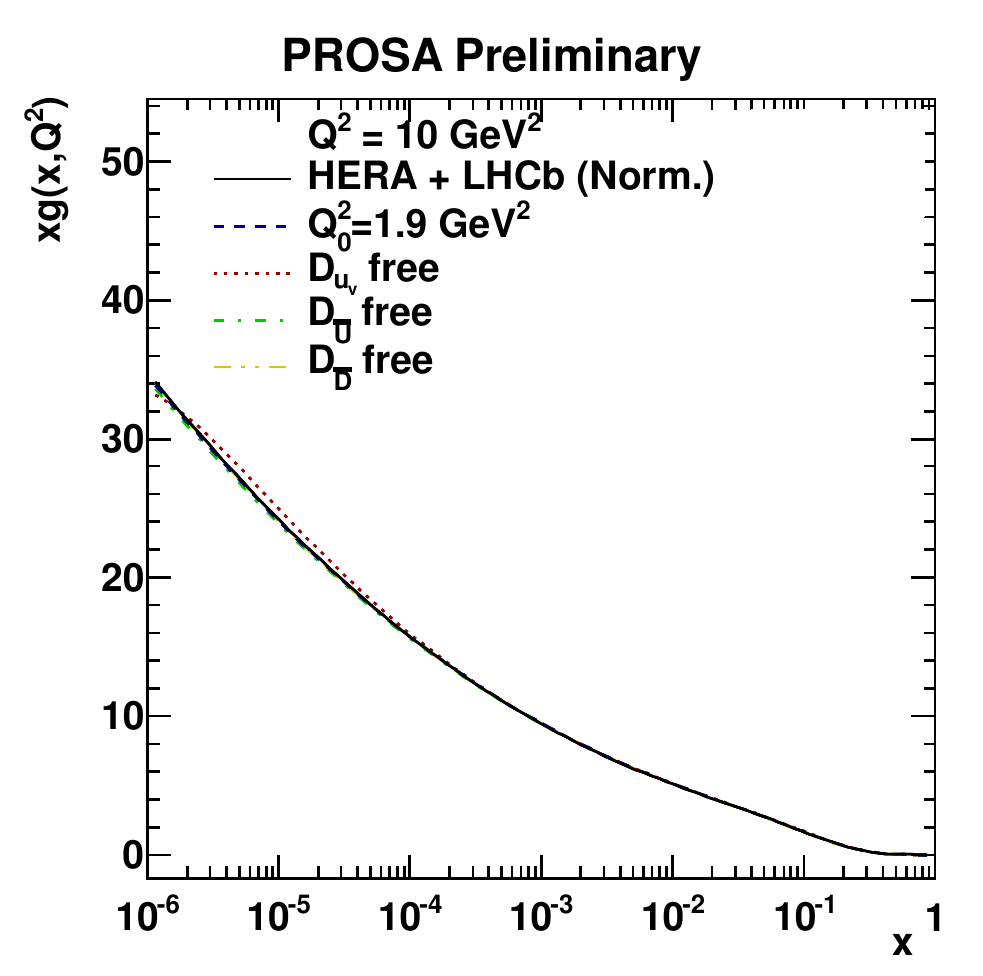}
  \includegraphics[width=0.5\figwidth,trim=5 0 5 0mm,clip=true]{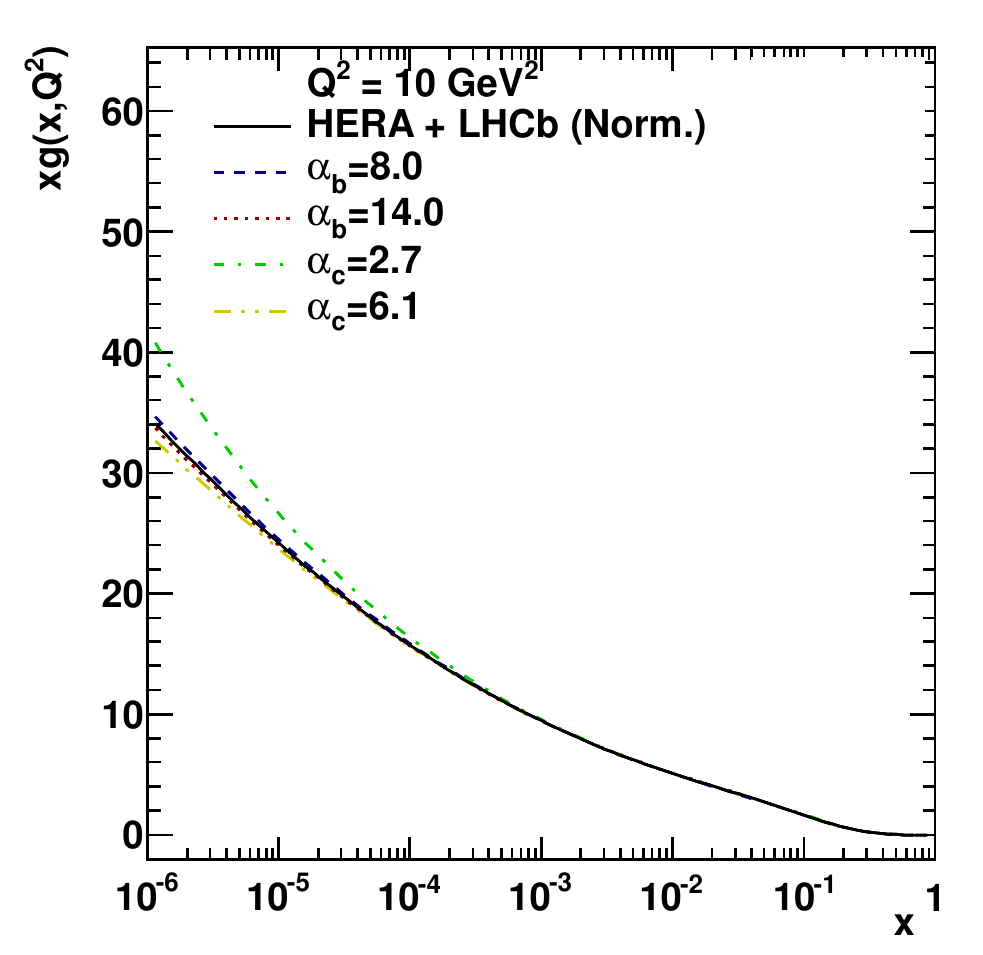}
  \end{minipage}
  \begin{minipage}[t]{0.49\columnwidth}
  \includegraphics[width=0.495\figwidth,trim=0 0 0 0mm,clip=true]{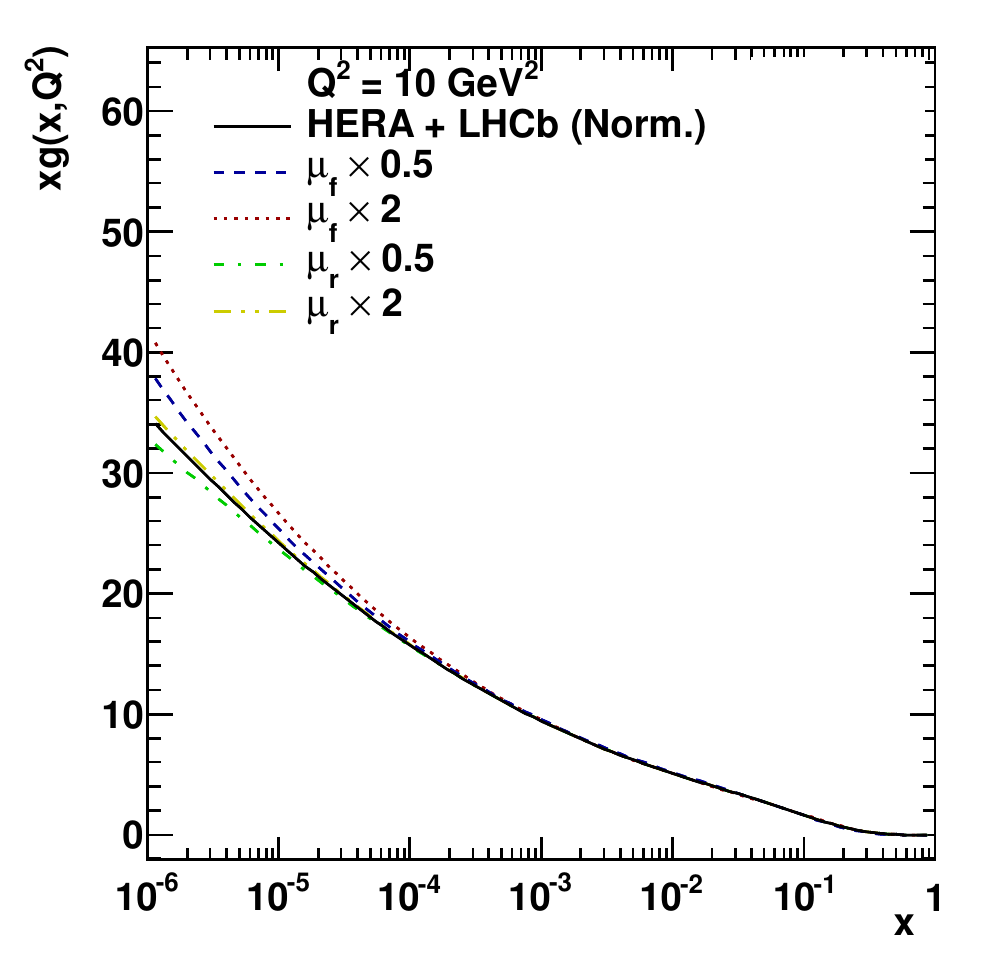}
  \begin{minipage}[t]{0.05\columnwidth}
  \end{minipage}
  \begin{minipage}[t]{0.95\columnwidth}
  \caption[Individual variations for gluon distribution in `LHCb Norm' fit]
  {The parametrisation (top left), scale (top right) and fragmentation (bottom) variations for the gluon distribution at $Q^2=\SI{10}{GeV^2}$ in the fit with the HERA and LHCb data using the `LHCb Norm' approach.}
	\label{fig:pdffit:Var_LHCbNorm_q2_10}
  \end{minipage}
  \end{minipage}
\end{figure}

The parametrisation variations are shown in Fig.~\ref{fig:pdffit:Var_LHCbNorm_q2_10} (top left). 
\ozmodNN{The gluon distribution in the low-$x$ region is constrained considerably compared to the `HERA only' results, 
although somewhat weaker than in the `LHCb Abs' approach.}

The effect of the scale variations on the predictions in the fit is shown in Fig.~\ref{fig:pdffit:FittedData_LHCbNorm} and 
their effect on the fitted PDFs is shown in Fig.~\ref{fig:pdffit:Var_LHCbNorm_q2_10} (top right). 
The effect of the fragmentation variations on the PDFs is shown in Fig.~\ref{fig:pdffit:Var_LHCbNorm_q2_10} (bottom). 
Note that fragmentation uncertainties obtained in this approach are larger than in the `LHCb Abs' one, 
since the fragmentation effects are not reabsorbed in the refitted scales. All these variations result in a reasonable data description.

\begin{figure}[htbp]
  \centering
  \includegraphics[width=0.495\figwidth,trim=2mm 2mm 2mm 8.5mm,clip=true]{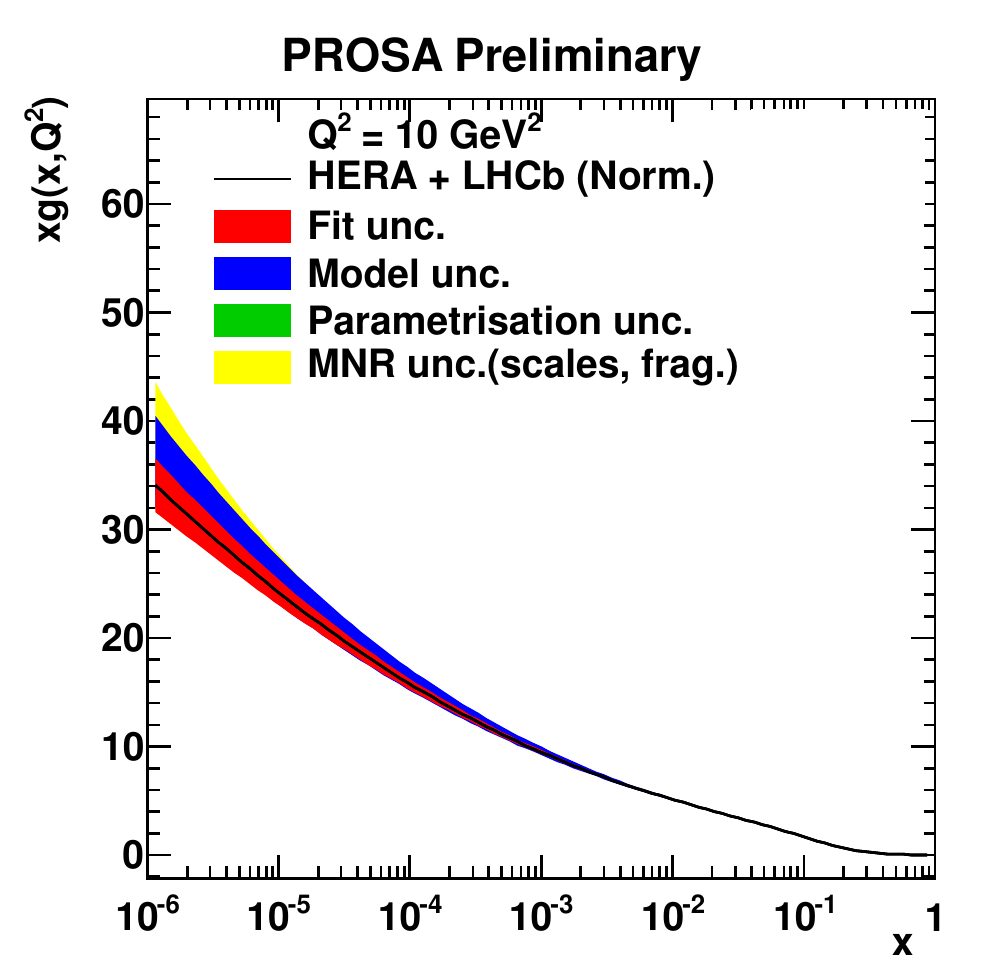}
  \includegraphics[width=0.495\figwidth,trim=2mm 2mm 2mm 8.5mm,clip=true]{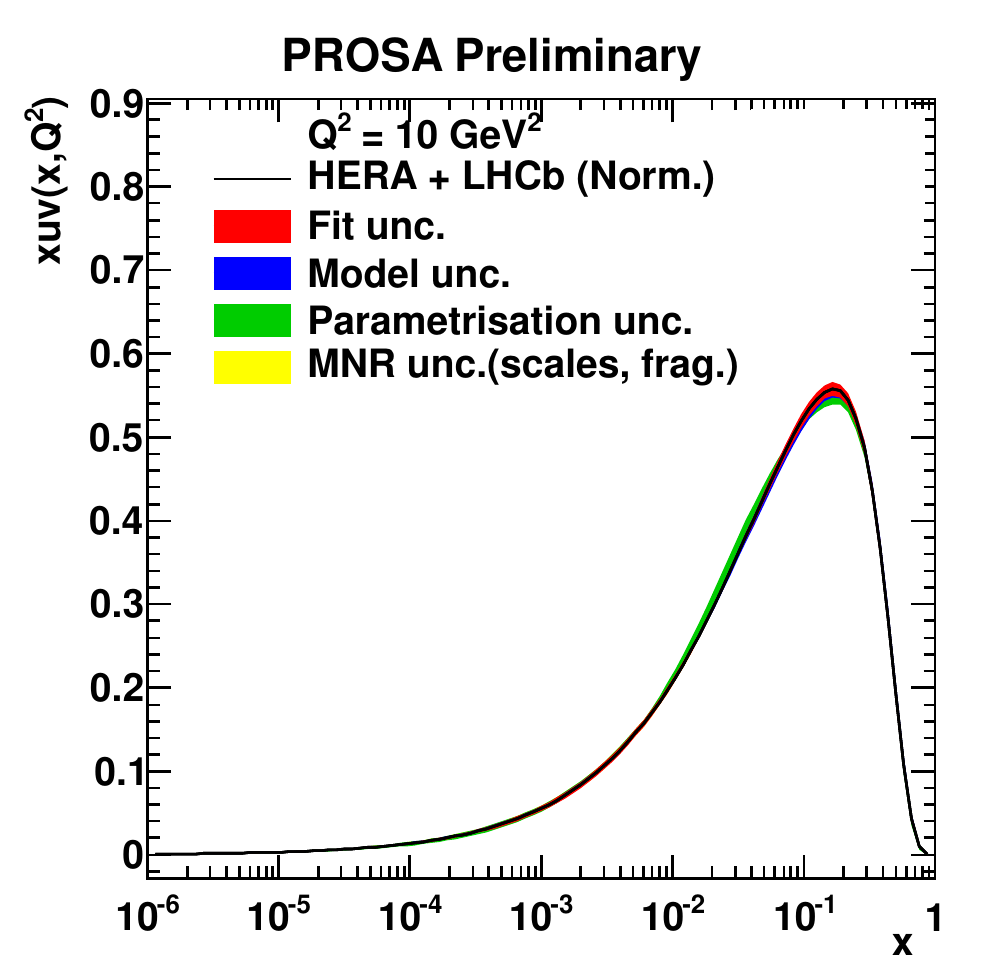}
  \includegraphics[width=0.495\figwidth,trim=2mm 2mm 2mm 8.5mm,clip=true]{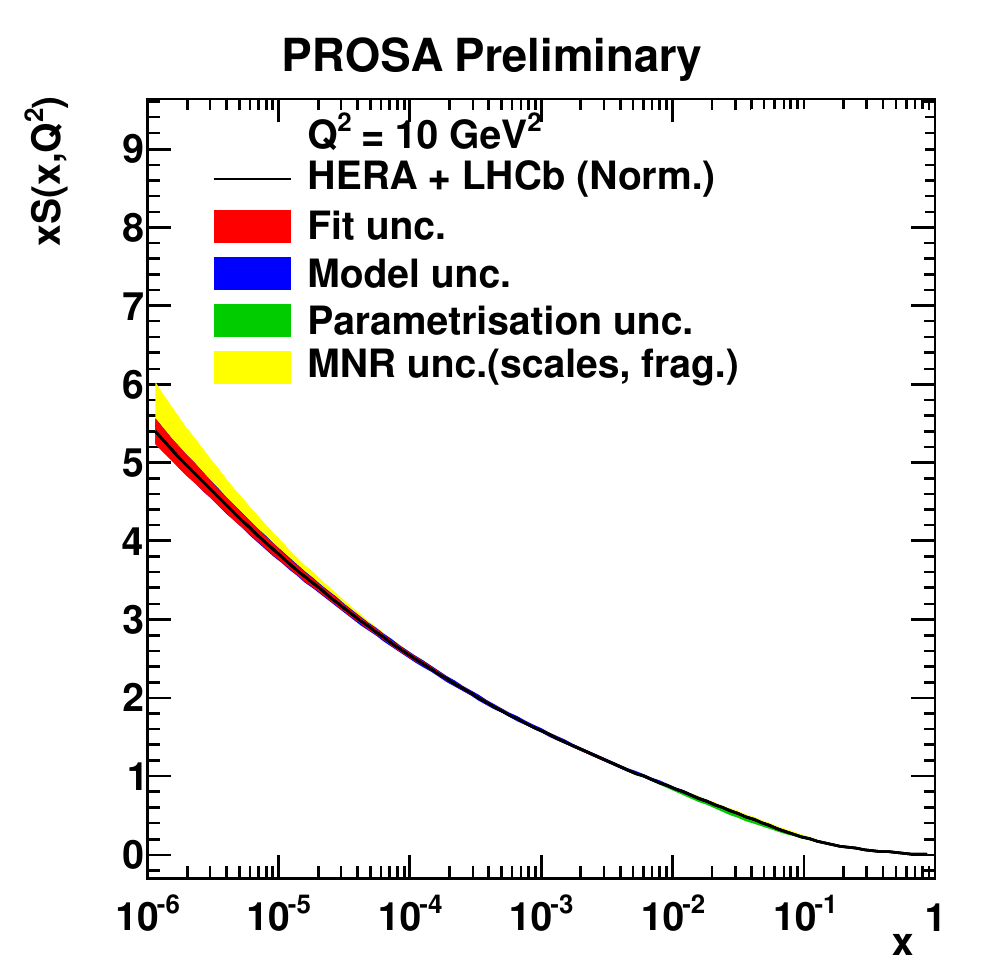}
  \includegraphics[width=0.495\figwidth,trim=2mm 2mm 2mm 8.5mm,clip=true]{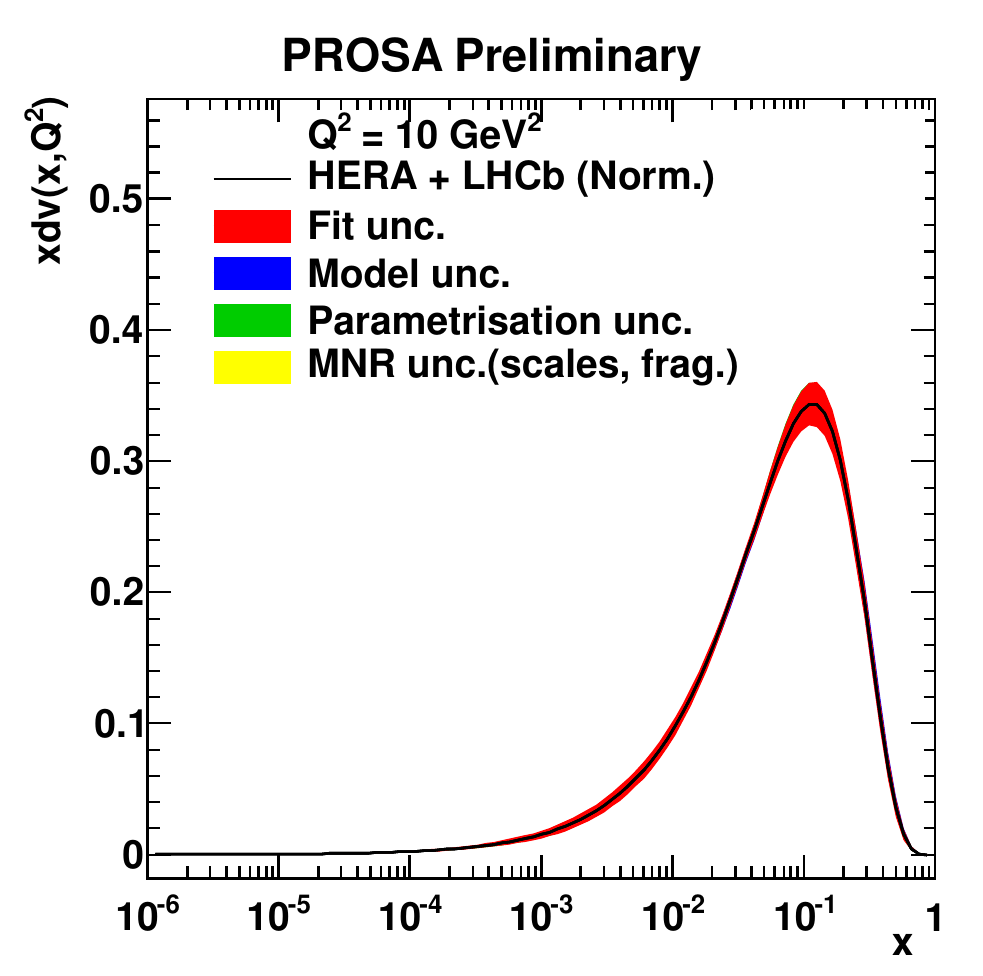}
  \caption[Individual contributions to PDF uncertainties for `LHCb Norm' fit]
  {The individual contributions to the uncertainties of the gluon (top left), $u$-valence (top right), sea (bottom left) and $d$-valence (bottom right) distributions at $Q^2=\SI{10}{GeV^2}$ 
  obtained in the fit with the HERA and LHCb data using the `LHCb Norm' approach.}
	\label{fig:pdffit:FittedPDFs_LHCbNorm_q2_10}
\end{figure}

Finally, all individual contributions to the uncertainties are shown in Fig.~\ref{fig:pdffit:FittedPDFs_LHCbNorm_q2_10}. 
The dominant uncertainties in the low-$x$ and low-$Q^2$ region still remain the MNR ones, although they are comparable in size with the uncertainties from other sources. 

\subsection{Impact of LHCb heavy-flavour data on PDFs}
\label{sec:pdffit:impact}

\begin{figure}[htbp]
  \centering
  \includegraphics[width=0.495\figwidth,trim=2mm 1mm 4mm 8.5mm,clip=true]{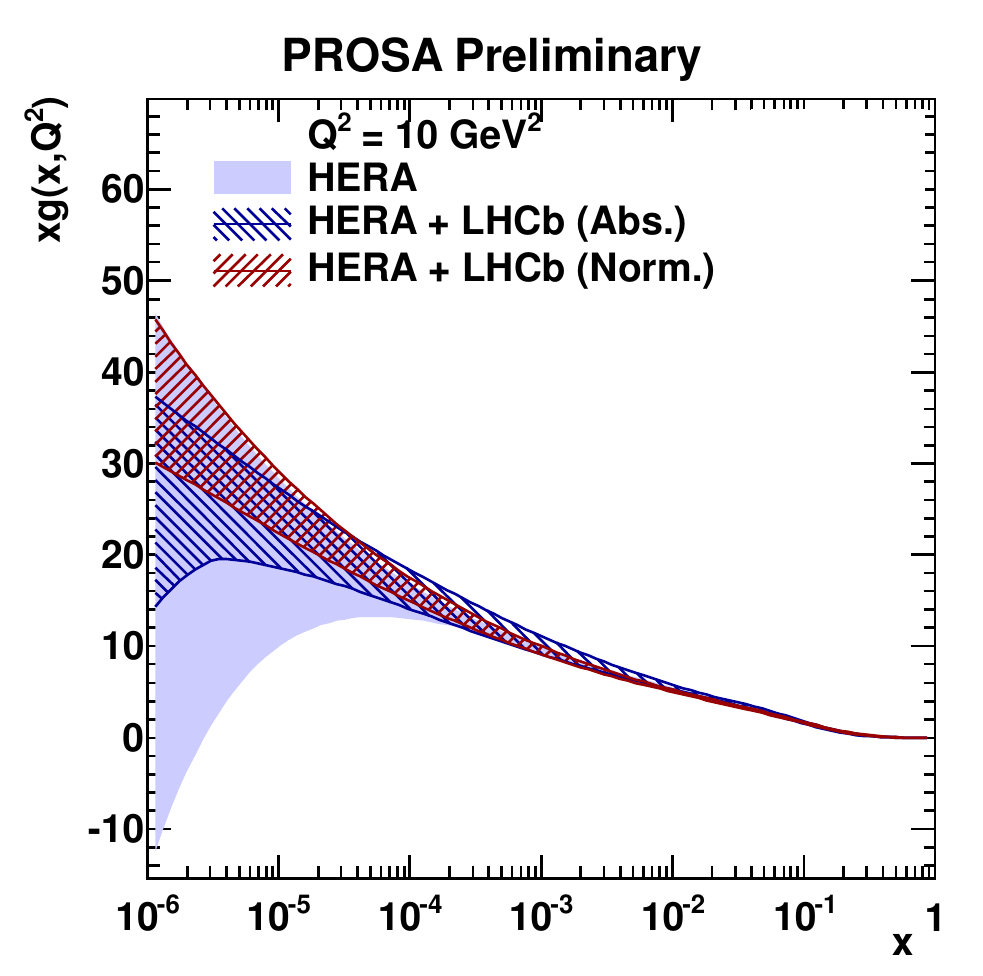}
  \includegraphics[width=0.495\figwidth,trim=2mm 1mm 4mm 8.5mm,clip=true]{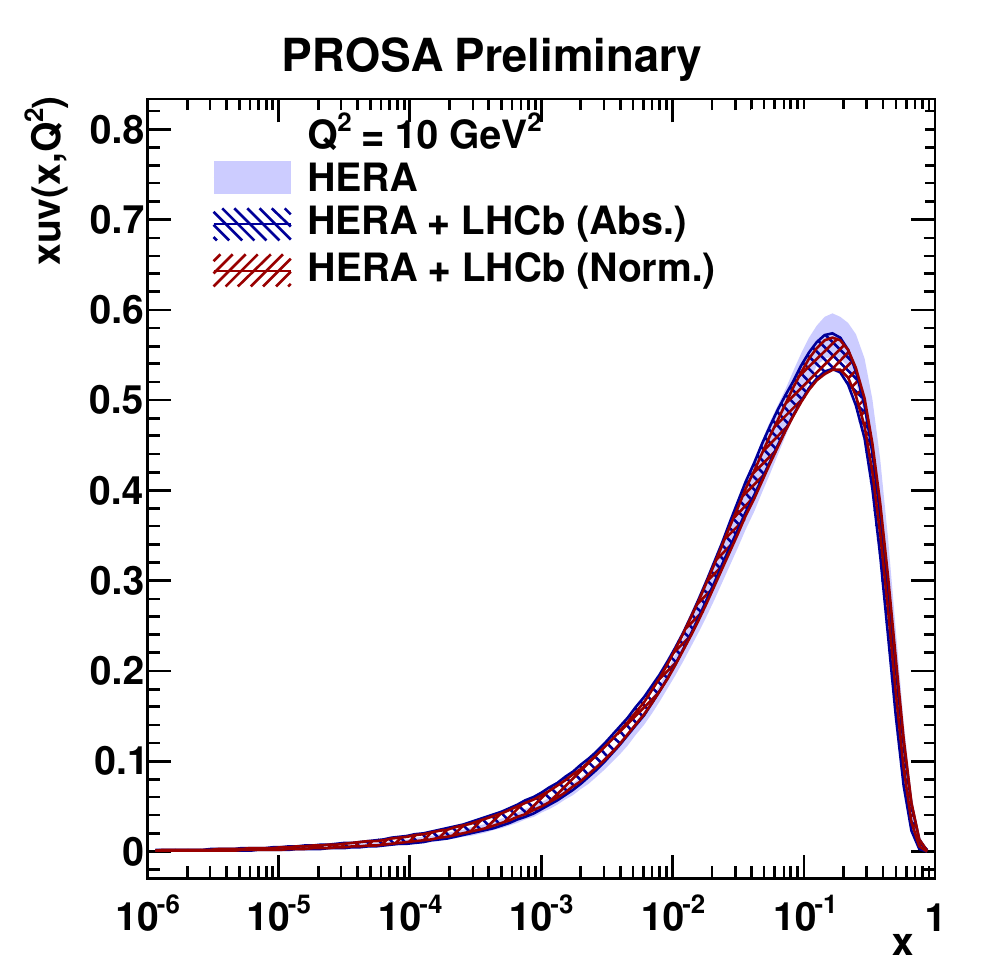}
  \includegraphics[width=0.495\figwidth,trim=2mm 1mm 4mm 8.5mm,clip=true]{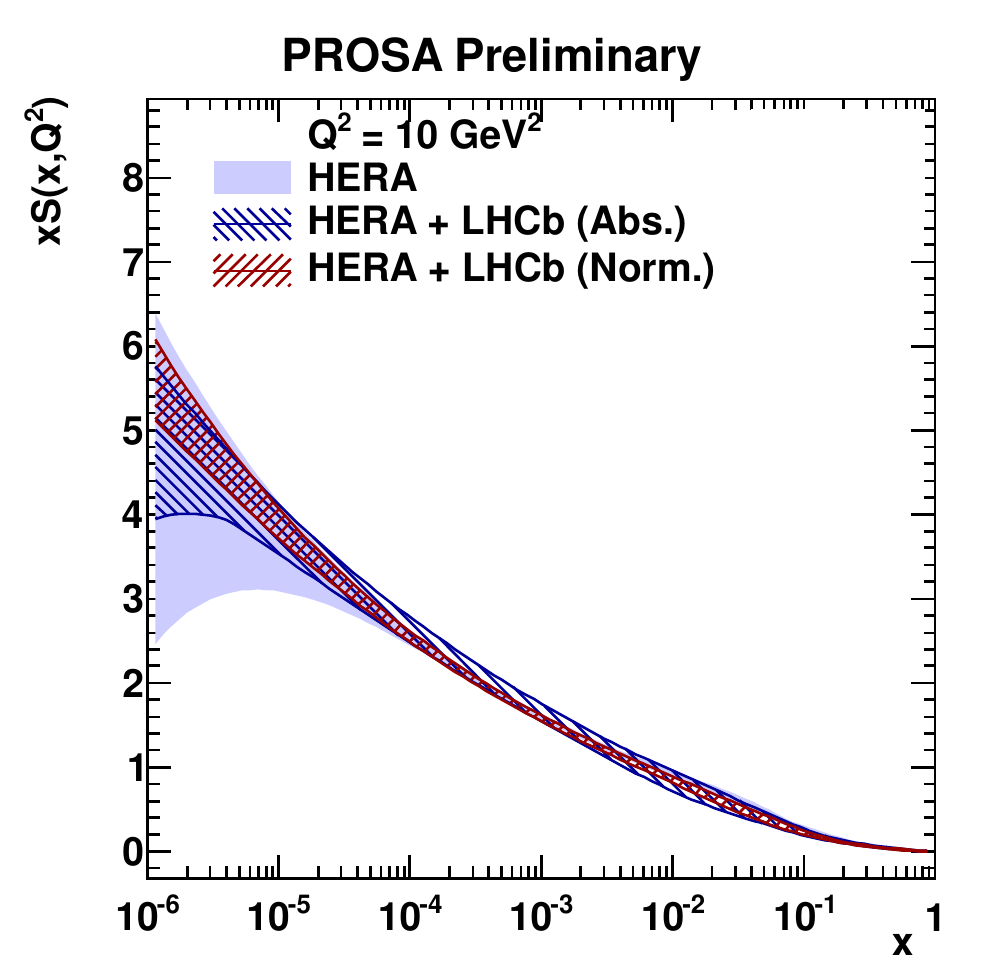}
  \includegraphics[width=0.495\figwidth,trim=2mm 1mm 4mm 8.5mm,clip=true]{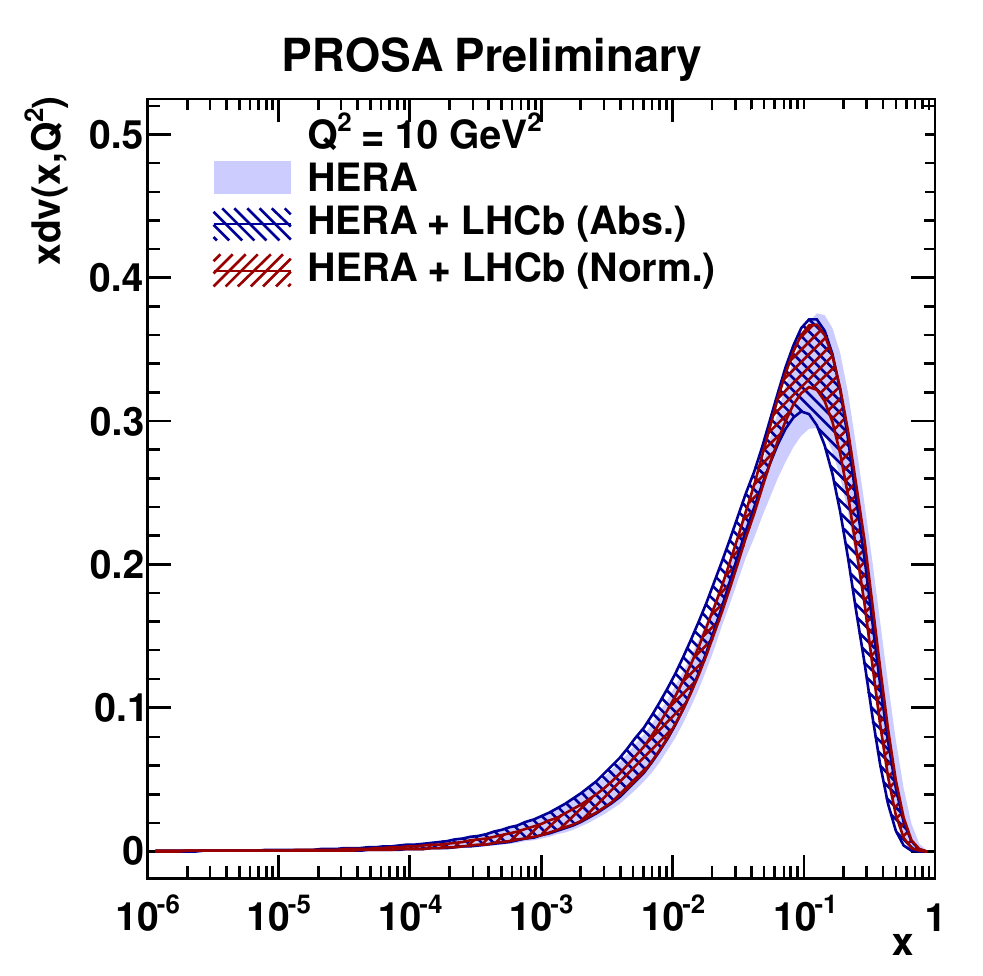}
  \caption[Comparison of PDFs at $Q^2=\SI{10}{GeV^2}$ from three fits]
  {The gluon (top left), $u$-valence (top right), sea (bottom left) and $d$-valence (bottom right) distributions at $Q^2=\SI{10}{GeV^2}$ obtained in the fit with the HERA-only, HERA and LHCb absolute, and HERA and LHCb normalised data.	The widths of the bands represent the total uncertainties.}
	\label{fig:pdffit:FittedPDFs_Comparison_q2_10}
\end{figure}

\begin{figure}[htbp]
  \centering
  \includegraphics[width=0.495\figwidth,trim=2mm 1mm 4mm 8.5mm,clip=true]{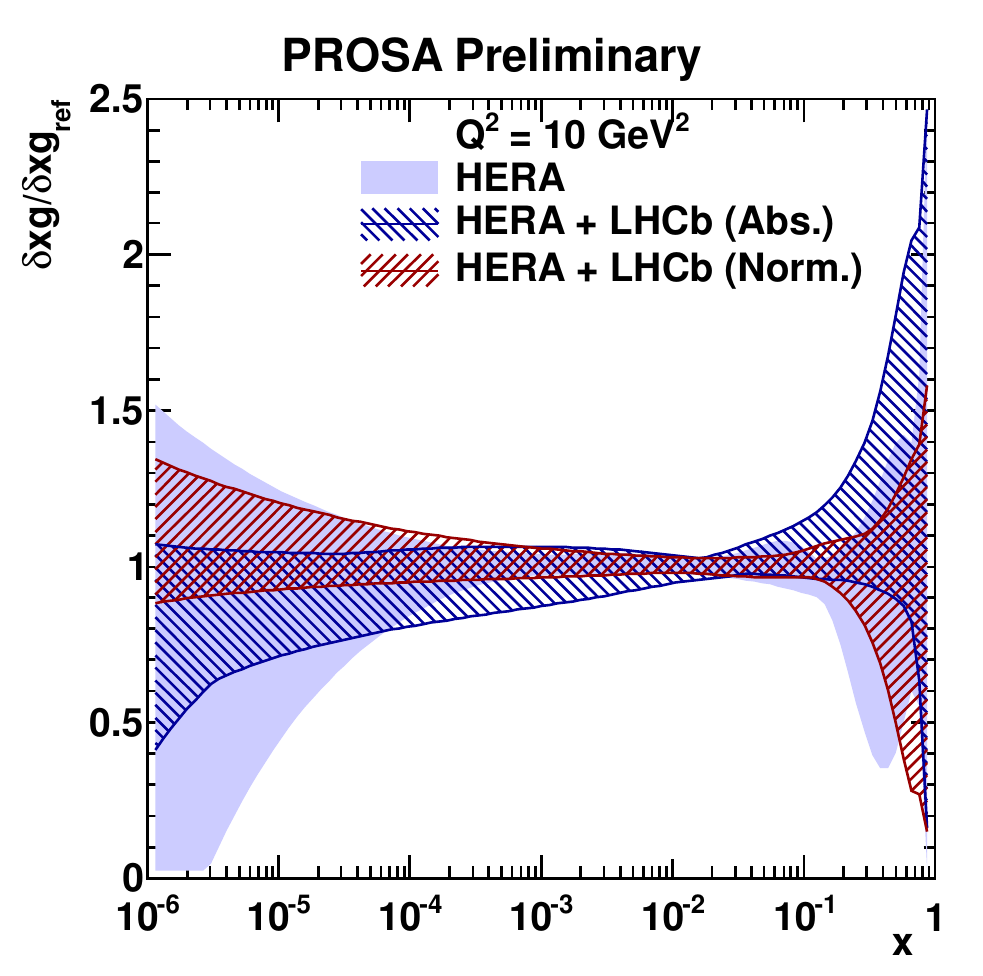}
  \includegraphics[width=0.495\figwidth,trim=2mm 1mm 4mm 8.5mm,clip=true]{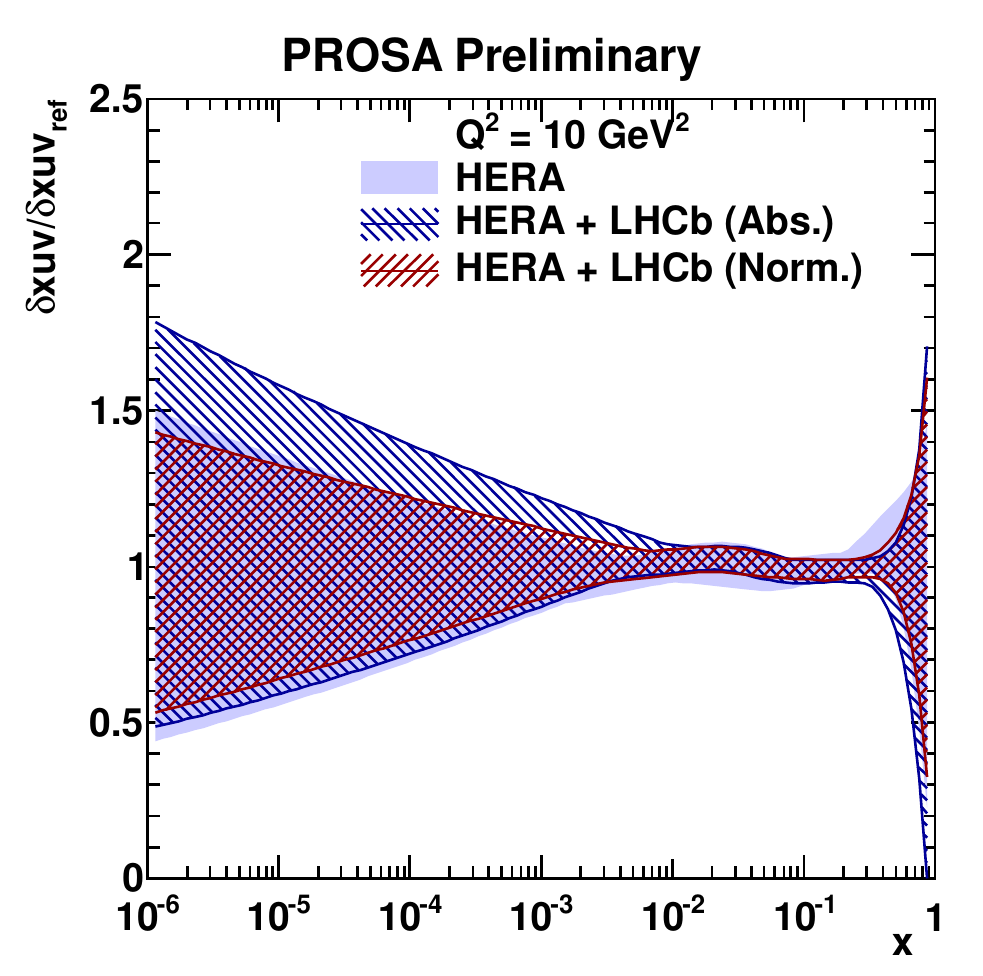}
  \includegraphics[width=0.495\figwidth,trim=2mm 1mm 4mm 8.5mm,clip=true]{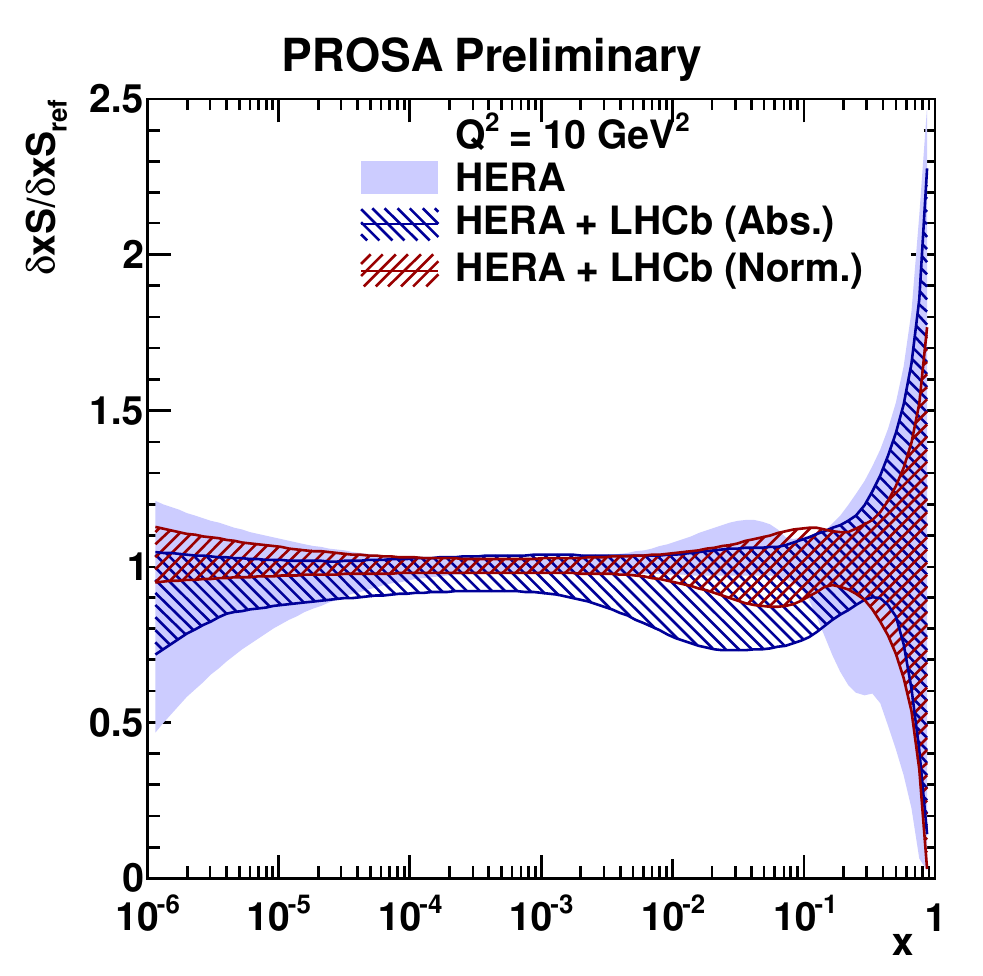}
  \includegraphics[width=0.495\figwidth,trim=2mm 1mm 4mm 8.5mm,clip=true]{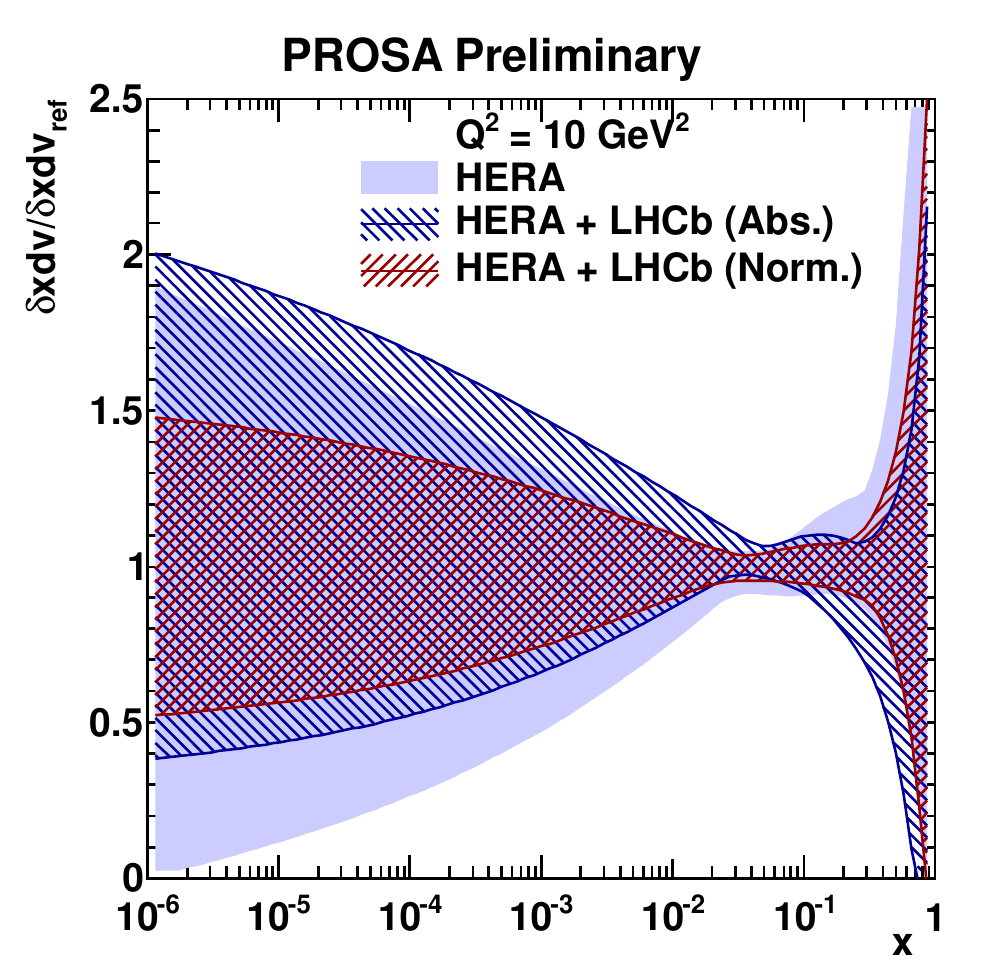}
  \caption[Comparison of relative PDF uncertainties at $Q^2=\SI{10}{GeV^2}$ from three fits]
  {The gluon (top left), $u$-valence (top right), sea (bottom left) and $d$-valence (bottom right) distributions at $Q^2=\SI{10}{GeV^2}$ obtained in the fit with the HERA-only, HERA and LHCb absolute, and HERA and LHCb normalised data, 
	normalised to one for a direct comparison of the uncertainties.
	The widths of the bands represent the total uncertainties.}
	\label{fig:pdffit:FittedPDFs_Comparison_q2_10_ratio}
\end{figure}

\ozmod{The main results of this study can be seen in Figs.~\ref{fig:pdffit:FittedPDFs_Comparison_q2_10} and~\ref{fig:pdffit:FittedPDFs_Comparison_q2_10_ratio}: 
the PDFs obtained in the `HERA only', `LHCb Abs' and `LHCb Norm' fits are compared at the scale $Q^2=\SI{10}{GeV^2}$ 
in Fig.~\ref{fig:pdffit:FittedPDFs_Comparison_q2_10}, while 
their relative uncertainties are compared in Fig.~\ref{fig:pdffit:FittedPDFs_Comparison_q2_10_ratio}.}%
\footnote{The \ozmodNN{corresponding} plots for $Q^2=\SI{100}{GeV^2}$ are available in Appendix~\ref{sec:app:pdffit:add} (Figs.~\ref{fig:pdffit:FittedPDFs_Comparison_q2_100},~\ref{fig:pdffit:FittedPDFs_Comparison_q2_100_ratio}).} 

The two approaches of fitting the LHCb data result in strong and consistent constraints on the gluon distribution at low $x$. 
Improvement is also observed for the sea density, while the valence-quark distributions\ozmod{, which are decoupled in the evolution,} remain essentially the same. 
Improvement of the sea distribution comes mainly from the \ozmodN{correlation} of gluons and sea quarks via the PDF evolution equations. 
Quantitatively, on average, the reduction of the total uncertainty for gluons and sea quarks in the region $10^{-6}<x<10^{-4}$, observed up to the scales $Q^2 \sim \SI{1000}{GeV^2}$, is of the order of a factor of $1.5\text{--}4$. 
The gluon distribution \ozmodN{is now} positive in the region directly covered by the data $x \gtrsim 4 \times 10^{-6}$, $Q^2 \gtrsim m_c^2 \approx \SI{2}{GeV^2}$. 
\ozmod{Compared to the gluon distributions provided by several modern PDF sets, shown in Fig.~\ref{fig:sec:pdffit:benchmark}, 
the shape of the gluon distribution favoured by the LHCb data is close to that in the ABM PDFs.} 
The distributions at medium $x$ mainly remain unchanged, 
although in the `LHCb Abs' approach some enlargement of the uncertainty is observed, 
explained by the inclusion of the scale uncertainties for the LHCb data. 
No such effect is observed in the `LHCb Norm' approach, where all variations describe the data well. 

In addition, a noticeable improvement is observed for all partons in the region of high $x$ \ozmod{also}. 
This is presumably \ozmod{due to} the constraints of the beauty LHCb data (see Fig.~\ref{fig:sec:pdffit:kinematics}) which cover this region, as well as a side effect of the improvement at low $x$, \ozmod{imposed} via the momentum sum rule. 

Obviously, in the fit with the absolute cross sections, more information \ozmodN{is available} to constrain the gluons: the absolute cross sections 
constrain the normalisation of the product of the gluon PDFs at low and medium $x$, leading to the \ozmodNN{correlation 
between the low-$x$ region and the medium one.} However the uncertainty of this calibration is of the order of 
a factor 2, propagated from the scale uncertainties of the absolute cross sections.

\begin{figure}[htbp]
  \includegraphics[width=1.0\figwidth,trim = 11mm 9mm 11mm 9mm,clip=true]{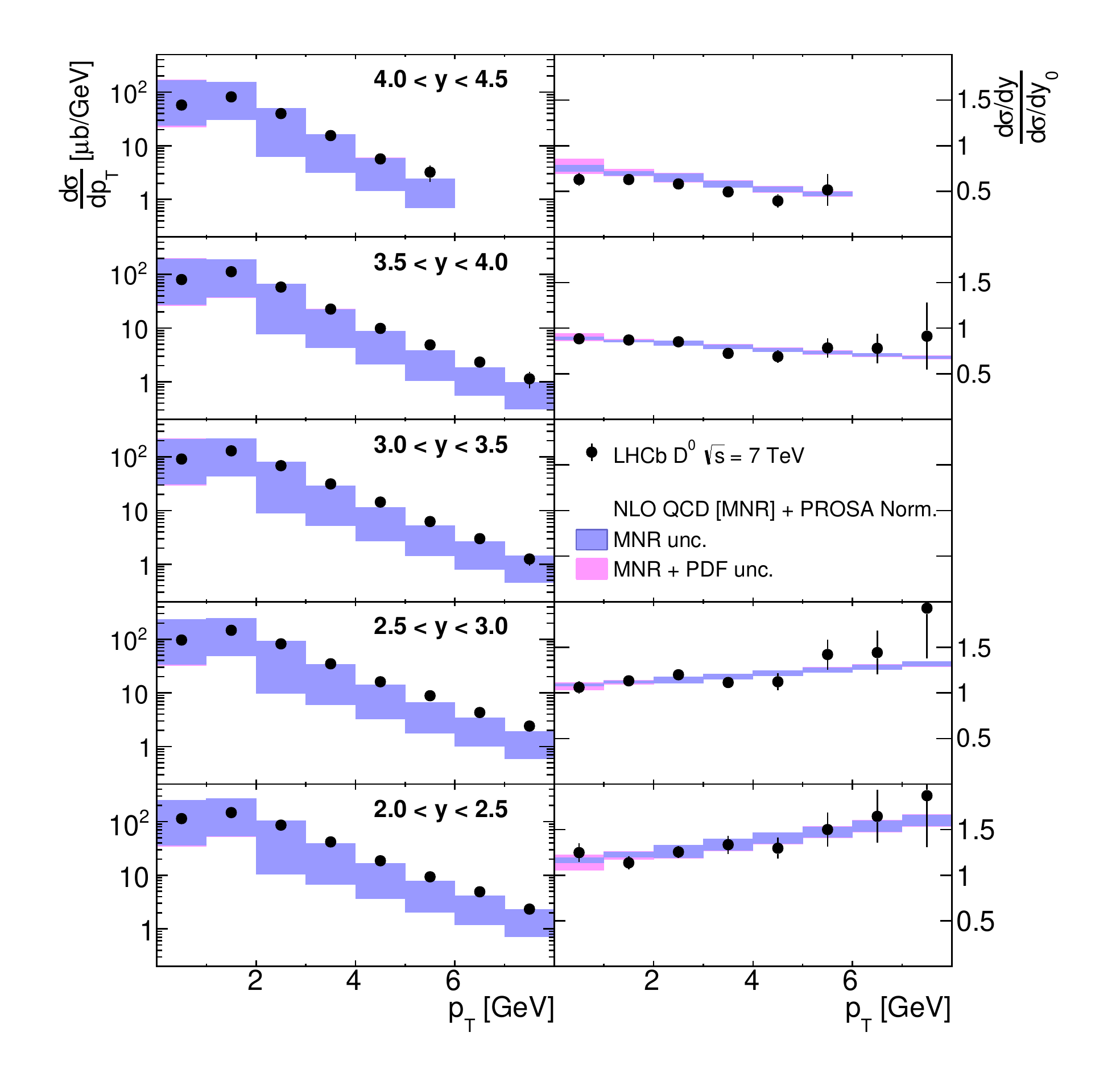}
  \caption[Differential D^{0} cross sections compared to HERA predictions] 
	{Differential cross sections for $D^0$ from the LHCb measurement of prompt charm production~\cite{LHCbCharm} 
	compared to theoretical predictions based the fitted `LHCb Norm` PDFs. 
  The other details are as in Fig.~\ref{fig:sec:pdffit:preddzerohera}.}
  \label{fig:sec:pdffit:preddzerolhcbnorm}
\end{figure}

In the fit with the normalised cross sections only the \ozmodN{information of the $y$ distribution} of the cross sections was used, 
\ozmod{which is sensitive to the slope of the gluon distribution.} 
In this approach therefore, the final impact of the LHCb data crucially depends 
on the presence of any $x$ region where \ozmod{the absolute values of the gluon distribution must be constrained} by other data (preferably at $x \sim 10^{-4}\text{--}10^{-3}$). 
Despite the reduced sensitivity, significantly smaller theoretical uncertainties \ozmodNN{are resulting} owing to the reduced scale dependence 
of the normalised cross sections, so the final results are more precise than in the `LHCb Abs' approach. 
\ozmod{\ozmodN{As opposed} to the absolute case, the theoretical calculations are 
data independent in this case. 
In Fig.~\ref{fig:sec:pdffit:preddzerolhcbnorm} the measured $D^0$ cross sections~\cite{LHCbCharm} 
are compared to the NLO predictions obtained using the fitted `LHCb Norm' PDFs. 
The theoretical uncertainties denoted as `MNR' in Fig.~\ref{fig:sec:pdffit:preddzerohera} include uncertainties from scale, 
heavy-quark mass and fragmentation function variations, while the total theoretical uncertainties include also those arising from the PDFs. 
Compared to the predicitons obtained using the PDFs from HERA only (see Fig.~\ref{fig:sec:pdffit:preddzerohera}), 
the PDF uncertainties are barely visible in this comparison, as expected, since PDFs are constrained by the data. 
It is important to note that the PDFs determined using the $y$ shape of the LHCb data provide a good description 
of the absolute cross sections, within large theoretical `MNR' uncertainties for the latter.} 

\subsection{\ozmod{Concluding remarks}}
\label{sec:pdffit:concl}

The observed strong impact of the charm and beauty LHCb measurements \ozmodNN{demonstrates} 
that these data \ozmod{are} a \ozmodNNNN{powerful} addition to the existing global PDF fits. 
Quantitatively the reduction of the gluon and sea quark distribution uncertainties 
in the $x$ range $x \lesssim 10^{-4}$, not probed by HERA, is of the order of a factor of $1.5\text{--}4$. 
\ozmodNNNN{The inclusion of the LHCb data results in a positive definite gluon distribution 
in the full phase space covered by the data.}

The study \ozmodNN{indicates} that the provided constraints are subject to sizeable theoretical uncertainties. 
Currently none of the PDF fitting groups estimate perturbative uncertainties of theoretical predictions (e.g.\ uncertainties from scale variations), 
\ozmod{therefore} the `LHCb Norm' approach should look more attractive since the uncertainties from the scale variations are not crucial, 
while in the `LHCb Abs' approach, results without the scale uncertainties \ozmodNN{will not be trustworthy.} 

\ozmodNN{Furthermore} the inclusion of perturbative theoretical uncertainties in the PDF fits is becoming a pressing issue. 
Once \ozmodN{an improved} strategy is developed, both considered approaches of fitting the LHCb heavy-flavour data can be used in the \ozmodN{future} global fits. 
Having the scale uncertainties under control will provide a possibility to exploit the LHCb heavy-flavour data also 
for a precise measurement of the charm and beauty masses\ozmodNN{, using the absolute cross sections.}

\ozmodNNNN{Improved} precision of the gluon distribution at low $x$ has implications in the physics of atmospheric showers, 
being important for calculations of prompt lepton fluxes in the atmosphere, see e.g.\ for recent studies~\cite{Garzelli:2015psa,Bhattacharya:2016jce,Gauld:2015kvh,Garzelli:2016xmx}. 
At some point astrophysical measurements of high-energy neutrinos \ozmodNNNN{may} provide an upper limit on charm hadroproduction in the atmosphere and 
complement collider-based measurements.

The final remark concerns the `LHCb Norm' approach. As previously mentioned,
the constraints at low $x$ obtained in this approach crucially depend on the presence of any medium-$x$ region already constrained by other data. 
In this context, the inclusion in the PDF fit of further datasets sensitive to gluons, e.g.\ jet measurements, should be very interesting. 
Another possible improvement may come from precise heavy-flavour measurements at the LHC near the kinematic threshold in the complementary rapidity region, 
which will extend the fitted $y$ shape to the central region $0<y<2$ and therefore to the medium range of gluon $x$. 

\clearpage
\section{Conclusions and outlook}
\label{sec:concl}
In this review the total set of charm production measurements in deep inelastic scattering at HERA, 
both inclusive and exclusive, has been presented and analysed. 
The data were obtained by the two Collaborations H1 and ZEUS 
with their multipurpose detectors during the two operation phases of HERA 
and constitute an essential achievement of the HERA program. 
The measurements rely on several techniques and exploit specially developed detector components. 
All individual data sets were shown to be consistent with each other 
and could therefore be combined into a precise common set. 
The experimental uncertainty of the combined results reaches 4\% in the region of medium $Q^2$, and is about 8\% on average. 
Also theoretical uncertainties entered, 
when extrapolations in unmeasured regions of phase space had to be performed.

Together with the HERA $ep$ data also the LHCb $pp$ data were included, 
since they provide complementary information about charm production. 
The constraining power of the precise double-differential LHCb data 
\ozmodNNNN{may be further improved by including in the QCD analysis 
also measurements of charm and beauty production at the LHC 
in the central rapidity phase-space region near the kinematic threshold, 
which can be obtained with the ATLAS and CMS detectors.}

The theoretical framework for predicting the charm cross sections has been outlined. 
An essential feature is the heavy mass of the $c$-quark and 
thus the presence of a hard scale such that the application 
of perturbative calculations is justified in the whole kinematic range. 
In the phase space of currently available experimental data 
the most rigorous theoretical calculations are performed in the fixed-flavor-number scheme (FFNS). 
In this scheme the $u$, $d$, $s$ quarks are massless, 
while the  mass effects of heavy quarks ($c$, $b$) are fully accounted for. 
Heavy-flavour production at HERA and LHC are dominated by processes involving gluons in the initial state. 
Therefore the investigation of the gluon distribution was a central issue 
in this review and showed the importance of the LHCb data.

Quantitative predictions of charm productions were possible 
once the parton distribution functions had been extracted from experimental data. 
The NLO QCD predictions were shown to agree in the whole phase space with the data on charm production. 
The sensitivity of the predictions to the mass of the $c$-quark made it possible 
to determine from the HERA data the optimal value of the $c$-quark \msbar mass 
with a precision of about 3\%. 
The extracted mass is consistent with the world-average value 
and has competitive precision to other individual determinations in perturbative QCD. 
The inclusion of the LHCb data to the HERA data improved considerably the accuracy of the gluon distribution at low $x$. 
The usage of the fixed-flavor-number scheme is of crucial importance in this analysis because it fully accounts for mass effects 
near the kinematic threshold without additional non-physical tunable parameters. 

It turned out that the experimental precision of the data exceeds the precision of the theoretical calculations. 
Uncertainties in the perturbative QCD predictions arise from missing higher orders, QCD parameters, 
and non-perturbative parton distribution functions and fragmentation functions \ozmodNNNN{which} must be extracted from data.
Dominant theoretical uncertainties of the NLO calculations for charm production 
in the bulk of the available phase space come from missing higher orders. 
Higher-order calculations are needed in order to reduce the theoretical uncertainty 
to the level of the current data precision and thus enabling stronger tests of QCD. 
New measurements\ozmodNN{, e.g.\ those from the LHC or future collider experiments,} 
with better precision or being performed \ozmodNNNN{in new regions of phase space,} 
are expected to further improve the understanding of QCD and development of high-energy physics.

\section*{Acknowledgements}
\label{sec:ack}
I would like to thank Dieter Haidt for the initiative to elaborate on my PhD thesis and his invaluable help in the preparation of the review. 
I thank my PhD supervisor Achim Geiser for several years of exciting collaboration, achieving the results described in this document. 
It is my pleasure to thank all members of ZEUS, H1 and PROSA Collaborations with whom I worked during these years.


%

\clearpage
\bibliography{refs/theory_refs,refs/hera_refs,refs/zevis_refs,refs/dplus_refs,refs/combination_refs,refs/pdffit_refs}

\providecommand{\href}[2]{#2}\begingroup\raggedright\begin{thebibliography}{100}

\bibitem{Abramowicz:2015mha}
H.~Abramowicz {\em et~al.}, ``{Combination of measurements of inclusive deep
  inelastic ${e^{\pm }p}$ scattering cross sections and QCD analysis of HERA
  data},'' \href{http://dx.doi.org/10.1140/epjc/s10052-015-3710-4}{{\em Eur.
  Phys. J.} {\bfseries C75} no.~12, (2015) 580},
\href{http://arxiv.org/abs/1506.06042}{{\ttfamily arXiv:1506.06042 [hep-ex]}}.

\bibitem{DIScomb}
F.~Aaron {\em et~al.}, ``{Combined Measurement and QCD Analysis of the
  Inclusive $e^{\pm} p$ Scattering Cross Sections at HERA},''
  \href{http://dx.doi.org/10.1007/JHEP01(2010)109}{{\em JHEP} {\bfseries 1001}
  (2010) 109},
\href{http://arxiv.org/abs/0911.0884}{{\ttfamily arXiv:0911.0884 [hep-ex]}}.

\bibitem{Zenaiev:2015qea}
O.~Zenaiev, {\em {Charm Production and QCD Analysis at HERA and LHC}}.
\newblock PhD thesis, Hamburg University, Hamburg, February, 2015.
\newblock
\url{http://bib-pubdb1.desy.de/search?of=hd&p=id:%22DESY-THESIS-2015-012%22}.
\newblock

\bibitem{zeusdch_hera2}
I.~Abt {\em et~al.}, ``{Measurement of $D^\pm$ production in deep inelastic
  $ep$ scattering with the ZEUS detector at HERA},''
  \href{http://dx.doi.org/10.1007/JHEP05(2013)023}{{\em JHEP} {\bfseries 1305}
  (2013) 023},
\href{http://arxiv.org/abs/1302.5058}{{\ttfamily arXiv:1302.5058 [hep-ex]}}.

\bibitem{dstarcombpaper}
H.~Abramowicz {\em et~al.}, ``{Combination of differential D$^{*\pm}$
  cross-section measurements in deep-inelastic ep scattering at HERA},''
  \href{http://dx.doi.org/10.1007/JHEP09(2015)149}{{\em JHEP} {\bfseries 09}
  (2015) 149},
\href{http://arxiv.org/abs/1503.06042}{{\ttfamily arXiv:1503.06042 [hep-ex]}}.

\bibitem{Zenaiev:2015rfa}
O.~Zenaiev {\em et~al.}, ``{Impact of heavy-flavour production cross sections
  measured by the LHCb experiment on parton distribution functions at low x},''
  \href{http://dx.doi.org/10.1140/epjc/s10052-015-3618-z}{{\em Eur. Phys. J.}
  {\bfseries C75} no.~8, (2015) 396},
\href{http://arxiv.org/abs/1503.04581}{{\ttfamily arXiv:1503.04581 [hep-ph]}}.

\bibitem{Bardeen:1978yd}
W.~A. Bardeen, A.~Buras, D.~Duke, and T.~Muta, ``{Deep Inelastic Scattering
  Beyond the Leading Order in Asymptotically Free Gauge Theories},''
\href{http://dx.doi.org/10.1103/PhysRevD.18.3998}{{\em Phys.Rev.} {\bfseries
  D18} (1978) 3998}.

\bibitem{pdg2014}
K.~Olive {\em et~al.}, ``{Review of Particle Physics},''
\href{http://dx.doi.org/10.1088/1674-1137/38/9/090001}{{\em Chin.Phys.}
  {\bfseries C38} (2014) 090001}.

\bibitem{Collins:1989gx}
J.~C. Collins, D.~E. Soper, and G.~F. Sterman, ``{Factorization of Hard
  Processes in QCD},'' {\em Adv.Ser.Direct.High Energy Phys.} {\bfseries 5}
  (1988) 1--91,
\href{http://arxiv.org/abs/hep-ph/0409313}{{\ttfamily arXiv:hep-ph/0409313
  [hep-ph]}}.

\bibitem{Behnke:2015qja}
O.~Behnke, A.~Geiser, and M.~Lisovyi, ``{Charm, Beauty and Top at HERA},''
  \href{http://dx.doi.org/10.1016/j.ppnp.2015.06.002}{{\em Prog. Part. Nucl.
  Phys.} {\bfseries 84} (2015) 1--72},
\href{http://arxiv.org/abs/1506.07519}{{\ttfamily arXiv:1506.07519 [hep-ex]}}.

\bibitem{Gluck:1993dpa}
M.~Gluck, E.~Reya, and M.~Stratmann, ``{Heavy quarks at high-energy
  colliders},''
\href{http://dx.doi.org/10.1016/0550-3213(94)00131-6}{{\em Nucl.Phys.}
  {\bfseries B422} (1994) 37--56}.

\bibitem{Kuprash:2014nfa}
O.~Kuprash, {\em {QCD analysis of Isolated Photon and Jet Data from $ep$, $pp$,
  and $p\bar{p}$ Collisions and Tests of $\alpha_s$ Running}}.
\newblock PhD thesis, Hamburg University,
December, 2014.
\newblock

\bibitem{qcdnum}
M.~Botje, ``{QCDNUM: Fast QCD Evolution and Convolution},''
  \href{http://dx.doi.org/10.1016/j.cpc.2010.10.020}{{\em Comput.Phys.Commun.}
  {\bfseries 182} (2011) 490--532},
\href{http://arxiv.org/abs/1005.1481}{{\ttfamily arXiv:1005.1481 [hep-ph]}}.

\bibitem{Collins:1986mp}
J.~C. Collins and W.-K. Tung, ``{Calculating Heavy Quark Distributions},''
\href{http://dx.doi.org/10.1016/0550-3213(86)90425-6}{{\em Nucl.Phys.}
  {\bfseries B278} (1986) 934}.

\bibitem{Aivazis:1993kh}
M.~Aivazis, F.~I. Olness, and W.-K. Tung, ``{Leptoproduction of heavy quarks.
  1. General formalism and kinematics of charged current and neutral current
  production processes},''
  \href{http://dx.doi.org/10.1103/PhysRevD.50.3085}{{\em Phys.Rev.} {\bfseries
  D50} (1994) 3085--3101},
\href{http://arxiv.org/abs/hep-ph/9312318}{{\ttfamily arXiv:hep-ph/9312318
  [hep-ph]}}.

\bibitem{Aivazis:1993pi}
M.~Aivazis, J.~C. Collins, F.~I. Olness, and W.-K. Tung, ``{Leptoproduction of
  heavy quarks. 2. A unified QCD formulation of charged and neutral current
  processes from fixed target to collider energies},''
  \href{http://dx.doi.org/10.1103/PhysRevD.50.3102}{{\em Phys.Rev.} {\bfseries
  D50} (1994) 3102--3118},
\href{http://arxiv.org/abs/hep-ph/9312319}{{\ttfamily arXiv:hep-ph/9312319
  [hep-ph]}}.

\bibitem{Buza:1995ie}
M.~Buza, Y.~Matiounine, J.~Smith, R.~Migneron, and W.~van Neerven, ``{Heavy
  quark coefficient functions at asymptotic values $Q^2 > m^2$},''
  \href{http://dx.doi.org/10.1016/0550-3213(96)00228-3}{{\em Nucl.Phys.}
  {\bfseries B472} (1996) 611--658},
\href{http://arxiv.org/abs/hep-ph/9601302}{{\ttfamily arXiv:hep-ph/9601302
  [hep-ph]}}.

\bibitem{Collins:1998rz}
J.~C. Collins, ``{Hard scattering factorization with heavy quarks: a general
  treatment},'' \href{http://dx.doi.org/10.1103/PhysRevD.58.094002}{{\em
  Phys.Rev.} {\bfseries D58} (1998) 094002},
\href{http://arxiv.org/abs/hep-ph/9806259}{{\ttfamily arXiv:hep-ph/9806259
  [hep-ph]}}.

\bibitem{Kramer:2000hn}
M.~Kramer, F.~I. Olness, and D.~E. Soper, ``{Treatment of heavy quarks in
  deeply inelastic scattering},''
  \href{http://dx.doi.org/10.1103/PhysRevD.62.096007}{{\em Phys.Rev.}
  {\bfseries D62} (2000) 096007},
\href{http://arxiv.org/abs/hep-ph/0003035}{{\ttfamily arXiv:hep-ph/0003035
  [hep-ph]}}.

\bibitem{Tung:2001mv}
W.-K. Tung, S.~Kretzer, and C.~Schmidt, ``{Open heavy flavor production in QCD:
  conceptual framework and implementation issues},''
  \href{http://dx.doi.org/10.1088/0954-3899/28/5/321}{{\em J.Phys.} {\bfseries
  G28} (2002) 983--996},
\href{http://arxiv.org/abs/hep-ph/0110247}{{\ttfamily arXiv:hep-ph/0110247
  [hep-ph]}}.

\bibitem{Thorne:2006qt}
R.~Thorne, ``{A variable-flavor number scheme for NNLO},''
  \href{http://dx.doi.org/10.1103/PhysRevD.73.054019}{{\em Phys.Rev.}
  {\bfseries D73} (2006) 054019},
\href{http://arxiv.org/abs/hep-ph/0601245}{{\ttfamily arXiv:hep-ph/0601245
  [hep-ph]}}.

\bibitem{Forte:2010ta}
S.~Forte, E.~Laenen, P.~Nason, and J.~Rojo, ``{Heavy quarks in deep-inelastic
  scattering},'' \href{http://dx.doi.org/10.1016/j.nuclphysb.2010.03.014}{{\em
  Nucl.Phys.} {\bfseries B834} (2010) 116--162},
\href{http://arxiv.org/abs/1001.2312}{{\ttfamily arXiv:1001.2312 [hep-ph]}}.

\bibitem{Bertone:2016pbr}
V.~Bertone {\em et~al.}, ``{A determination of $m_c(m_c)$ from HERA data using
  a matched heavy-flavor scheme},''
  \href{http://dx.doi.org/10.1007/JHEP08(2016)050}{{\em JHEP} {\bfseries 08}
  (2016) 050},
\href{http://arxiv.org/abs/1605.01946}{{\ttfamily arXiv:1605.01946 [hep-ph]}}.

\bibitem{Pumplin:2007wg}
J.~Pumplin, H.~Lai, and W.~Tung, ``{The Charm Parton Content of the Nucleon},''
  \href{http://dx.doi.org/10.1103/PhysRevD.75.054029}{{\em Phys.Rev.}
  {\bfseries D75} (2007) 054029},
\href{http://arxiv.org/abs/hep-ph/0701220}{{\ttfamily arXiv:hep-ph/0701220
  [hep-ph]}}.

\bibitem{Lykasov:2012hf}
G.~Lykasov, V.~Bednyakov, A.~Pikelner, and N.~Zimine, ``{Forward Heavy Flavour
  Production in p-p Collisions at LHC and Intrinsic Quark Components in
  Proton},'' \href{http://dx.doi.org/10.1209/0295-5075/99/21002}{{\em
  Europhys.Lett.} {\bfseries 99} (2012) 21002},
\href{http://arxiv.org/abs/1205.1131}{{\ttfamily arXiv:1205.1131 [hep-ph]}}.

\bibitem{Ball:2016neh}
R.~D. Ball, V.~Bertone, M.~Bonvini, S.~Carrazza, S.~Forte, A.~Guffanti, N.~P.
  Hartland, J.~Rojo, and L.~Rottoli, ``{A Determination of the Charm Content of
  the Proton},''
\href{http://arxiv.org/abs/1605.06515}{{\ttfamily arXiv:1605.06515 [hep-ph]}}.

\bibitem{Bigi:1994em}
I.~I. Bigi, M.~A. Shifman, N.~Uraltsev, and A.~Vainshtein, ``{The pole mass of
  the heavy quark. Perturbation theory and beyond},''
  \href{http://dx.doi.org/10.1103/PhysRevD.50.2234}{{\em Phys.Rev.} {\bfseries
  D50} (1994) 2234--2246},
\href{http://arxiv.org/abs/hep-ph/9402360}{{\ttfamily arXiv:hep-ph/9402360
  [hep-ph]}}.

\bibitem{Alekhin:2010sv}
S.~Alekhin and S.~Moch, ``{Heavy-quark deep-inelastic scattering with a running
  mass},'' \href{http://dx.doi.org/10.1016/j.physletb.2011.04.026}{{\em
  Phys.Lett.} {\bfseries B699} (2011) 345--353},
\href{http://arxiv.org/abs/1011.5790}{{\ttfamily arXiv:1011.5790 [hep-ph]}}.

\bibitem{Abdallah:2008ac}
J.~Abdallah {\em et~al.}, ``{Study of $b$-quark mass effects in multijet
  topologies with the DELPHI detector at LEP},''
  \href{http://dx.doi.org/10.1140/epjc/s10052-008-0631-5}{{\em Eur.Phys.J.}
  {\bfseries C55} (2008) 525--538},
\href{http://arxiv.org/abs/0804.3883}{{\ttfamily arXiv:0804.3883 [hep-ex]}}.

\bibitem{zeussecvtx_hera2}
H.~Abramowicz {\em et~al.}, ``{Measurement of beauty and charm production in
  deep inelastic scattering at HERA and measurement of the beauty-quark
  mass},'' \href{http://dx.doi.org/10.1007/JHEP09(2014)127}{{\em JHEP}
  {\bfseries 1409} (2014) 127},
\href{http://arxiv.org/abs/1405.6915}{{\ttfamily arXiv:1405.6915 [hep-ex]}}.

\bibitem{mcrunprel}
``Determination of charm mass running from an analysis of combined hera charm
  data.'' H1-prelim-14-071, ZEUS-prel-14-006.

\bibitem{Gray:1990yh}
N.~Gray, D.~J. Broadhurst, W.~Grafe, and K.~Schilcher, ``{Three Loop Relation
  of Quark $\overline{\rm MS}$ and Pole Masses},''
\href{http://dx.doi.org/10.1007/BF01614703}{{\em Z.Phys.} {\bfseries C48}
  (1990) 673--680}.

\bibitem{Broadhurst:1991fy}
D.~J. Broadhurst, N.~Gray, and K.~Schilcher, ``{Gauge invariant on-shell $Z_2$
  in QED, QCD and the effective field theory of a static quark},''
\href{http://dx.doi.org/10.1007/BF01412333}{{\em Z.Phys.} {\bfseries C52}
  (1991) 111--122}.

\bibitem{Chetyrkin:1999ys}
K.~Chetyrkin and M.~Steinhauser, ``{Short distance mass of a heavy quark at
  order $\alpha_s^3$},''
  \href{http://dx.doi.org/10.1103/PhysRevLett.83.4001}{{\em Phys.Rev.Lett.}
  {\bfseries 83} (1999) 4001--4004},
\href{http://arxiv.org/abs/hep-ph/9907509}{{\ttfamily arXiv:hep-ph/9907509
  [hep-ph]}}.

\bibitem{Melnikov:2000qh}
K.~Melnikov and T.~v. Ritbergen, ``{The three loop relation between the
  $\overline{\rm MS}$ and the pole quark masses},''
  \href{http://dx.doi.org/10.1016/S0370-2693(00)00507-4}{{\em Phys.Lett.}
  {\bfseries B482} (2000) 99--108},
\href{http://arxiv.org/abs/hep-ph/9912391}{{\ttfamily arXiv:hep-ph/9912391
  [hep-ph]}}.

\bibitem{zd9697}
J.~Breitweg {\em et~al.}, ``{Measurement of $D^{*\pm}$ production and the charm
  contribution to $F_2$ in deep inelastic scattering at HERA},''
  \href{http://dx.doi.org/10.1007/s100529900244}{{\em Eur.Phys.J.} {\bfseries
  C12} (2000) 35--52},
\href{http://arxiv.org/abs/hep-ex/9908012}{{\ttfamily arXiv:hep-ex/9908012
  [hep-ex]}}.

\bibitem{vanNeerven:2001tb}
W.~L. van Neerven, ``{Production of heavy quarks in deep inelastic lepton
  hadron scattering},'' \href{http://arxiv.org/abs/hep-ph/0107193}{{\ttfamily
  arXiv:hep-ph/0107193 [hep-ph]}}.
[AIP Conf. Proc.601,40(2001)].

\bibitem{sf}
A.~M. Cooper-Sarkar, R.~Devenish, and A.~De~Roeck, ``{Structure functions of
  the nucleon and their interpretation},''
  \href{http://dx.doi.org/10.1142/S0217751X98001670}{{\em Int.J.Mod.Phys.}
  {\bfseries A13} (1998) 3385--3586},
\href{http://arxiv.org/abs/hep-ph/9712301}{{\ttfamily arXiv:hep-ph/9712301
  [hep-ph]}}.

\bibitem{Diemoz:1987xu}
M.~Diemoz, F.~Ferroni, E.~Longo, and G.~Martinelli, ``{Parton Densities from
  Deep Inelastic Scattering to Hadronic Processes at Super Collider
  Energies},''
\href{http://dx.doi.org/10.1007/BF01560387}{{\em Z. Phys.} {\bfseries C39}
  (1988) 21}.

\bibitem{Allaby:1987vs}
J.~V. Allaby {\em et~al.}, ``{Parametrization of Proton Structure Functions},''
\href{http://dx.doi.org/10.1016/0370-2693(87)90384-4}{{\em Phys. Lett.}
  {\bfseries B197} (1987) 281--284}.

\bibitem{dglap_Gribov:1972ri}
V.~Gribov and L.~Lipatov, ``{Deep inelastic $ep$ scattering in perturbation
  theory},''
{\em Sov.J.Nucl.Phys.} {\bfseries 15} (1972) 438--450.

\bibitem{Dokshitzer:1977sg}
Y.~L. Dokshitzer, ``{Calculation of the Structure Functions for Deep Inelastic
  Scattering and $e^{+} e^{-}$ Annihilation by Perturbation Theory in Quantum
  Chromodynamics.},'' {\em Sov. Phys. JETP} {\bfseries 46} (1977) 641--653.
[Zh. Eksp. Teor. Fiz.73,1216(1977)].

\bibitem{dglap_Altarelli:1977zs}
G.~Altarelli and G.~Parisi, ``{Asymptotic Freedom in Parton Language},''
\href{http://dx.doi.org/10.1016/0550-3213(77)90384-4}{{\em Nucl.Phys.}
  {\bfseries B126} (1977) 298}.

\bibitem{dglap_Curci:1980uw}
G.~Curci, W.~Furmanski, and R.~Petronzio, ``{Evolution of Parton Densities
  Beyond Leading Order: The Nonsinglet Case},''
\href{http://dx.doi.org/10.1016/0550-3213(80)90003-6}{{\em Nucl.Phys.}
  {\bfseries B175} (1980) 27}.

\bibitem{dglap_Furmanski:1980cm}
W.~Furmanski and R.~Petronzio, ``{Singlet Parton Densities Beyond Leading
  Order},''
\href{http://dx.doi.org/10.1016/0370-2693(80)90636-X}{{\em Phys.Lett.}
  {\bfseries B97} (1980) 437}.

\bibitem{dglap_Moch:2004pa}
S.~Moch, J.~Vermaseren, and A.~Vogt, ``{The Three-Loop Splitting Functions in
  QCD: The Nonsinglet Case},''
  \href{http://dx.doi.org/10.1016/j.nuclphysb.2004.03.030}{{\em Nucl.Phys.}
  {\bfseries B688} (2004) 101--134},
\href{http://arxiv.org/abs/hep-ph/0403192}{{\ttfamily arXiv:hep-ph/0403192
  [hep-ph]}}.

\bibitem{dglap_Vogt:2004mw}
A.~Vogt, S.~Moch, and J.~Vermaseren, ``{The Three-Loop Splitting Functions in
  QCD: The Singlet Case},''
  \href{http://dx.doi.org/10.1016/j.nuclphysb.2004.04.024}{{\em Nucl.Phys.}
  {\bfseries B691} (2004) 129--181},
\href{http://arxiv.org/abs/hep-ph/0404111}{{\ttfamily arXiv:hep-ph/0404111
  [hep-ph]}}.

\bibitem{Kuraev:1976ge}
E.~A. Kuraev, L.~N. Lipatov, and V.~S. Fadin, ``{Multi-Reggeon Processes in the
  Yang-Mills Theory},''
{\em Sov.Phys.JETP} {\bfseries 44} (1976) 443--450.

\bibitem{Kuraev:1977fs}
E.~Kuraev, L.~Lipatov, and V.~S. Fadin, ``{The Pomeranchuk Singularity in
  Nonabelian Gauge Theories},''
{\em Sov.Phys.JETP} {\bfseries 45} (1977) 199--204.

\bibitem{Balitsky:1978ic}
I.~Balitsky and L.~Lipatov, ``{The Pomeranchuk Singularity in Quantum
  Chromodynamics},''
{\em Sov.J.Nucl.Phys.} {\bfseries 28} (1978) 822--829.

\bibitem{Ciafaloni:1987ur}
M.~Ciafaloni, ``{Coherence Effects in Initial Jets at Small $q^2$ / s},''
\href{http://dx.doi.org/10.1016/0550-3213(88)90380-X}{{\em Nucl. Phys.}
  {\bfseries B296} (1988) 49--74}.

\bibitem{Catani:1989yc}
S.~Catani, F.~Fiorani, and G.~Marchesini, ``{QCD Coherence in Initial State
  Radiation},''
\href{http://dx.doi.org/10.1016/0370-2693(90)91938-8}{{\em Phys.Lett.}
  {\bfseries B234} (1990) 339}.

\bibitem{Catani:1989sg}
S.~Catani, F.~Fiorani, and G.~Marchesini, ``{Small $x$ Behavior of Initial
  State Radiation in Perturbative QCD},''
\href{http://dx.doi.org/10.1016/0550-3213(90)90342-B}{{\em Nucl.Phys.}
  {\bfseries B336} (1990) 18}.

\bibitem{Marchesini:1994wr}
G.~Marchesini, ``{QCD coherence in the structure function and associated
  distributions at small $x$},''
  \href{http://dx.doi.org/10.1016/0550-3213(95)00149-M}{{\em Nucl.Phys.}
  {\bfseries B445} (1995) 49--80},
\href{http://arxiv.org/abs/hep-ph/9412327}{{\ttfamily arXiv:hep-ph/9412327
  [hep-ph]}}.

\bibitem{Witten:1975bh}
E.~Witten, ``{Heavy Quark Contributions to Deep Inelastic Scattering},''
\href{http://dx.doi.org/10.1016/0550-3213(76)90111-5}{{\em Nucl. Phys.}
  {\bfseries B104} (1976) 445--476}.

\bibitem{Babcock:1977fi}
J.~Babcock, D.~W. Sivers, and S.~Wolfram, ``{QCD Estimates for Heavy Particle
  Production},''
\href{http://dx.doi.org/10.1103/PhysRevD.18.162}{{\em Phys. Rev.} {\bfseries
  D18} (1978) 162}.

\bibitem{Novikov:1977yc}
V.~A. Novikov, M.~A. Shifman, A.~I. Vainshtein, and V.~I. Zakharov, ``{Charm
  Photoproduction and Quantum Chromodynamics},''
  \href{http://dx.doi.org/10.1016/0550-3213(78)90019-6}{{\em Nucl. Phys.}
  {\bfseries B136} (1978) 125}.
[Yad. Fiz.27,771(1978)].

\bibitem{Leveille:1978px}
J.~P. Leveille and T.~J. Weiler, ``{Characteristics of Heavy Quark
  Leptoproduction in QCD},''
\href{http://dx.doi.org/10.1016/0550-3213(79)90420-6}{{\em Nucl. Phys.}
  {\bfseries B147} (1979) 147--173}.

\bibitem{Gluck:1979aw}
M.~Gluck and E.~Reya, ``{Deep Inelastic Quantum Chromodynamic Charm
  Leptoproduction},''
\href{http://dx.doi.org/10.1016/0370-2693(79)90898-0}{{\em Phys. Lett.}
  {\bfseries B83} (1979) 98--102}.

\bibitem{heracharmcomb}
H.~Abramowicz {\em et~al.}, ``{Combination and QCD Analysis of Charm Production
  Cross Section Measurements in Deep-Inelastic $ep$ Scattering at HERA},''
  \href{http://dx.doi.org/10.1140/epjc/s10052-013-2311-3}{{\em Eur.Phys.J.}
  {\bfseries C73} (2013) 2311},
\href{http://arxiv.org/abs/1211.1182}{{\ttfamily arXiv:1211.1182 [hep-ex]}}.

\bibitem{Laenen:1992zk}
E.~Laenen, S.~Riemersma, J.~Smith, and W.~van Neerven, ``{Complete
  $O(\alpha_s)$ corrections to heavy flavor structure functions in
  electroproduction},''
\href{http://dx.doi.org/10.1016/0550-3213(93)90201-Y}{{\em Nucl.Phys.}
  {\bfseries B392} (1993) 162--228}.

\bibitem{Laenen:1992xs}
E.~Laenen, S.~Riemersma, J.~Smith, and W.~van Neerven, ``{$O(\alpha_s)$
  corrections to heavy flavor inclusive distributions in electroproduction},''
\href{http://dx.doi.org/10.1016/0550-3213(93)90202-Z}{{\em Nucl.Phys.}
  {\bfseries B392} (1993) 229--250}.

\bibitem{hvqdis}
B.~Harris and J.~Smith, ``{Charm quark and $D^{*\pm}$ cross-sections in deeply
  inelastic scattering at HERA},''
  \href{http://dx.doi.org/10.1103/PhysRevD.57.2806}{{\em Phys.Rev.} {\bfseries
  D57} (1998) 2806--2812},
\href{http://arxiv.org/abs/hep-ph/9706334}{{\ttfamily arXiv:hep-ph/9706334
  [hep-ph]}}.

\bibitem{Ablinger:2014nga}
J.~Ablinger, A.~Behring, J.~Bl{\"u}mlein, A.~De~Freitas, A.~von Manteuffel, and
  C.~Schneider, ``{The 3-loop pure singlet heavy flavor contributions to the
  structure function $F_2(x,Q^2)$ and the anomalous dimension},''
  \href{http://dx.doi.org/10.1016/j.nuclphysb.2014.10.008}{{\em Nucl. Phys.}
  {\bfseries B890} (2014) 48--151},
\href{http://arxiv.org/abs/1409.1135}{{\ttfamily arXiv:1409.1135 [hep-ph]}}.

\bibitem{Behring:2014eya}
A.~Behring, I.~Bierenbaum, J.~Bl{\"u}mlein, A.~De~Freitas, S.~Klein, and
  F.~Wißbrock, ``{The logarithmic contributions to the $O(\alpha^3_s)$
  asymptotic massive Wilson coefficients and operator matrix elements in deeply
  inelastic scattering},''
  \href{http://dx.doi.org/10.1140/epjc/s10052-014-3033-x}{{\em Eur. Phys. J.}
  {\bfseries C74} no.~9, (2014) 3033},
\href{http://arxiv.org/abs/1403.6356}{{\ttfamily arXiv:1403.6356 [hep-ph]}}.

\bibitem{Ablinger:2010ty}
J.~Ablinger, J.~Bl{\"u}mlein, S.~Klein, C.~Schneider, and F.~Wissbrock, ``{The
  $O(\alpha_s^3)$ Massive Operator Matrix Elements of $O(n_f)$ for the
  Structure Function $F_2(x,Q^2)$ and Transversity},''
  \href{http://dx.doi.org/10.1016/j.nuclphysb.2010.10.021}{{\em Nucl. Phys.}
  {\bfseries B844} (2011) 26--54},
\href{http://arxiv.org/abs/1008.3347}{{\ttfamily arXiv:1008.3347 [hep-ph]}}.

\bibitem{Blumlein:2012vq}
J.~Bl{\"u}mlein, A.~Hasselhuhn, S.~Klein, and C.~Schneider, ``{The
  $O(\alpha_s^3 n_f T_F^2 C_{A,F})$ Contributions to the Gluonic Massive
  Operator Matrix Elements},''
  \href{http://dx.doi.org/10.1016/j.nuclphysb.2012.09.001}{{\em Nucl. Phys.}
  {\bfseries B866} (2013) 196--211},
\href{http://arxiv.org/abs/1205.4184}{{\ttfamily arXiv:1205.4184 [hep-ph]}}.

\bibitem{Ablinger:2014vwa}
J.~Ablinger, A.~Behring, J.~Bl{\"u}mlein, A.~De~Freitas, A.~Hasselhuhn, A.~von
  Manteuffel, M.~Round, C.~Schneider, and F.~Wißbrock, ``{The 3-Loop
  Non-Singlet Heavy Flavor Contributions and Anomalous Dimensions for the
  Structure Function $F_2(x,Q^2)$ and Transversity},''
  \href{http://dx.doi.org/10.1016/j.nuclphysb.2014.07.010}{{\em Nucl. Phys.}
  {\bfseries B886} (2014) 733--823},
\href{http://arxiv.org/abs/1406.4654}{{\ttfamily arXiv:1406.4654 [hep-ph]}}.

\bibitem{Kawamura:2012cr}
H.~Kawamura, N.~Lo~Presti, S.~Moch, and A.~Vogt, ``{On the
  next-to-next-to-leading order QCD corrections to heavy-quark production in
  deep-inelastic scattering},''
  \href{http://dx.doi.org/10.1016/j.nuclphysb.2012.07.001}{{\em Nucl.Phys.}
  {\bfseries B864} (2012) 399--468},
\href{http://arxiv.org/abs/1205.5727}{{\ttfamily arXiv:1205.5727 [hep-ph]}}.

\bibitem{Bierenbaum:2009mv}
I.~Bierenbaum, J.~Bl{\"u}mlein, and S.~Klein, ``{Mellin Moments of the
  $O(\alpha^3_s)$ Heavy Flavor Contributions to unpolarized Deep-Inelastic
  Scattering at $Q^2 \gg m^2$ and Anomalous Dimensions},''
  \href{http://dx.doi.org/10.1016/j.nuclphysb.2009.06.005}{{\em Nucl. Phys.}
  {\bfseries B820} (2009) 417--482},
\href{http://arxiv.org/abs/0904.3563}{{\ttfamily arXiv:0904.3563 [hep-ph]}}.

\bibitem{Bierenbaum:2007qe}
I.~Bierenbaum, J.~Bl{\"u}mlein, and S.~Klein, ``{Two-Loop Massive Operator
  Matrix Elements and Unpolarized Heavy Flavor Production at Asymptotic Values
  $Q^2 \gg m^2$},''
  \href{http://dx.doi.org/10.1016/j.nuclphysb.2007.04.030}{{\em Nucl. Phys.}
  {\bfseries B780} (2007) 40--75},
\href{http://arxiv.org/abs/hep-ph/0703285}{{\ttfamily arXiv:hep-ph/0703285
  [HEP-PH]}}.

\bibitem{Bierenbaum:2008yu}
I.~Bierenbaum, J.~Bl{\"u}mlein, S.~Klein, and C.~Schneider, ``{Two-Loop Massive
  Operator Matrix Elements for Unpolarized Heavy Flavor Production to
  O(epsilon)},'' \href{http://dx.doi.org/10.1016/j.nuclphysb.2008.05.016}{{\em
  Nucl. Phys.} {\bfseries B803} (2008) 1--41},
\href{http://arxiv.org/abs/0803.0273}{{\ttfamily arXiv:0803.0273 [hep-ph]}}.

\bibitem{Heinrich:2004kj}
G.~Heinrich and B.~A. Kniehl, ``{Next-to-leading-order predictions for
  $D^{*\pm}$ plus jet photoproduction at DESY HERA},''
  \href{http://dx.doi.org/10.1103/PhysRevD.70.094035}{{\em Phys. Rev.}
  {\bfseries D70} (2004) 094035},
\href{http://arxiv.org/abs/hep-ph/0409303}{{\ttfamily arXiv:hep-ph/0409303
  [hep-ph]}}.

\bibitem{Sandoval:2009dw}
C.~Sandoval, ``{Inclusive Single Hadron Production in Neutral Current
  Deep-Inelastic Scattering at Next-to-Leading Order},'' in {\em {Proceedings,
  17th International Workshop on Deep-Inelastic Scattering and Related Subjects
  (DIS 2009): Madrid, Spain, April 26-30, 2009}}.
\newblock \href{http://arxiv.org/abs/0908.0824}{{\ttfamily arXiv:0908.0824
  [hep-ph]}}.
\newblock
\url{https://inspirehep.net/record/827997/files/arXiv:0908.0824.pdf}.
\newblock

\bibitem{Sandoval:2009zz}
C.~Sandoval, {\em {Inclusive production of hadrons in neutral and charged
  current deep inelastic scattering}}.
\newblock PhD thesis, Hamburg University,
November, 2009.
\newblock

\bibitem{h1dstarhighQ2}
F.~Aaron {\em et~al.}, ``{Measurement of the $D^{*\pm}$ Meson Production Cross
  Section and $F_2^{c\bar{c}}$, at High $Q^2$, in $ep$ Scattering at HERA},''
  \href{http://dx.doi.org/10.1016/j.physletb.2010.02.024}{{\em Phys.Lett.}
  {\bfseries B686} (2010) 91--100},
\href{http://arxiv.org/abs/0911.3989}{{\ttfamily arXiv:0911.3989 [hep-ex]}}.

\bibitem{h1dstar_hera2}
F.~Aaron {\em et~al.}, ``{Measurement of $D^{*\pm}$ Meson Production and
  Determination of $F_{2}^{c\bar{c}}$ at low $Q^2$ in Deep-Inelastic Scattering
  at HERA},'' \href{http://dx.doi.org/10.1140/epjc/s10052-011-1769-0,
  10.1140/epjc/s10052-012-2252-2}{{\em Eur.Phys.J.} {\bfseries C71} (2011)
  1769},
\href{http://arxiv.org/abs/1106.1028}{{\ttfamily arXiv:1106.1028 [hep-ex]}}.

\bibitem{Thorne:1997ga}
R.~S. Thorne and R.~G. Roberts, ``{An Ordered analysis of heavy flavor
  production in deep inelastic scattering},''
  \href{http://dx.doi.org/10.1103/PhysRevD.57.6871}{{\em Phys. Rev.} {\bfseries
  D57} (1998) 6871--6898},
\href{http://arxiv.org/abs/hep-ph/9709442}{{\ttfamily arXiv:hep-ph/9709442
  [hep-ph]}}.

\bibitem{Thorne:2012az}
R.~Thorne, ``{Effect of changes of variable flavor number scheme on parton
  distribution functions and predicted cross sections},''
  \href{http://dx.doi.org/10.1103/PhysRevD.86.074017}{{\em Phys.Rev.}
  {\bfseries D86} (2012) 074017},
\href{http://arxiv.org/abs/1201.6180}{{\ttfamily arXiv:1201.6180 [hep-ph]}}.

\bibitem{Mangano:1997ri}
M.~L. Mangano, ``{Two lectures on heavy quark production in hadronic
  collisions},''
\href{http://arxiv.org/abs/hep-ph/9711337}{{\ttfamily arXiv:hep-ph/9711337
  [hep-ph]}}.

\bibitem{Guzzi:2014wia}
M.~Guzzi, K.~Lipka, and S.-O. Moch, ``{Top-quark pair production at hadron
  colliders: differential cross section and phenomenological applications with
  DiffTop},'' \href{http://dx.doi.org/10.1007/JHEP01(2015)082}{{\em JHEP}
  {\bfseries 1501} (2015) 082},
\href{http://arxiv.org/abs/1406.0386}{{\ttfamily arXiv:1406.0386 [hep-ph]}}.

\bibitem{Gauld:2015yia}
R.~Gauld, J.~Rojo, L.~Rottoli, and J.~Talbert, ``{Charm production in the
  forward region: constraints on the small-x gluon and backgrounds for neutrino
  astronomy},'' \href{http://dx.doi.org/10.1007/JHEP11(2015)009}{{\em JHEP}
  {\bfseries 11} (2015) 009},
\href{http://arxiv.org/abs/1506.08025}{{\ttfamily arXiv:1506.08025 [hep-ph]}}.

\bibitem{Alekhin:2014sya}
S.~Alekhin, J.~Bl{\"u}mlein, L.~Caminadac, K.~Lipka, K.~Lohwasser, {\em
  et~al.}, ``{Determination of Strange Sea Quark Distributions from
  Fixed-target and Collider Data},''
  \href{http://dx.doi.org/10.1103/PhysRevD.91.094002}{{\em Phys.Rev.}
  {\bfseries D91} no.~9, (2015) 094002},
\href{http://arxiv.org/abs/1404.6469}{{\ttfamily arXiv:1404.6469 [hep-ph]}}.

\bibitem{Klasen:2014dba}
M.~Klasen, C.~Klein-Bösing, K.~Kovarik, G.~Kramer, M.~Topp, {\em et~al.},
  ``{NLO Monte Carlo predictions for heavy-quark production at the LHC: $pp$
  collisions in ALICE},'' \href{http://dx.doi.org/10.1007/JHEP08(2014)109}{{\em
  JHEP} {\bfseries 1408} (2014) 109},
\href{http://arxiv.org/abs/1405.3083}{{\ttfamily arXiv:1405.3083 [hep-ph]}}.

\bibitem{Gluck:1977zm}
M.~Gluck, J.~F. Owens, and E.~Reya, ``{Gluon Contribution to Hadronic J/psi
  Production},''
\href{http://dx.doi.org/10.1103/PhysRevD.17.2324}{{\em Phys. Rev.} {\bfseries
  D17} (1978) 2324}.

\bibitem{Combridge:1978kx}
B.~L. Combridge, ``{Associated Production of Heavy Flavor States in p p and
  anti-p p Interactions: Some QCD Estimates},''
\href{http://dx.doi.org/10.1016/0550-3213(79)90449-8}{{\em Nucl. Phys.}
  {\bfseries B151} (1979) 429--456}.

\bibitem{Hagiwara:1978hw}
K.~Hagiwara and T.~Yoshino, ``{Hadroproduction of Heavy Quark Flavors in
  {QCD}},''
\href{http://dx.doi.org/10.1016/0370-2693(79)90217-X}{{\em Phys. Lett.}
  {\bfseries B80} (1979) 282--285}.

\bibitem{Jones:1977di}
L.~M. Jones and H.~W. Wyld, ``{On Hadronic Charm Production by Gluon Fusion},''
\href{http://dx.doi.org/10.1103/PhysRevD.17.1782}{{\em Phys. Rev.} {\bfseries
  D17} (1978) 1782}.

\bibitem{Georgi:1978kx}
H.~M. Georgi, S.~L. Glashow, M.~E. Machacek, and D.~V. Nanopoulos, ``{Charmed
  Particles From Two - Gluon Annihilation in Proton Proton Collisions},''
\href{http://dx.doi.org/10.1016/0003-4916(78)90270-1}{{\em Annals Phys.}
  {\bfseries 114} (1978) 273}.

\bibitem{mnrtotal}
P.~Nason, S.~Dawson, and R.~K. Ellis, ``{The Total Cross-Section for the
  Production of Heavy Quarks in Hadronic Collisions},''
\href{http://dx.doi.org/10.1016/0550-3213(88)90422-1}{{\em Nucl.Phys.}
  {\bfseries B303} (1988) 607}.

\bibitem{Beenakker:1988bq}
W.~Beenakker, H.~Kuijf, W.~van Neerven, and J.~Smith, ``{QCD Corrections to
  Heavy Quark Production in $p\bar{p}$ Collisions},''
\href{http://dx.doi.org/10.1103/PhysRevD.40.54}{{\em Phys.Rev.} {\bfseries D40}
  (1989) 54--82}.

\bibitem{mnrsingle}
P.~Nason, S.~Dawson, and R.~K. Ellis, ``{The One Particle Inclusive
  Differential Cross-Section for Heavy Quark Production in Hadronic
  Collisions},''
\href{http://dx.doi.org/10.1016/0550-3213(89)90286-1}{{\em Nucl.Phys.}
  {\bfseries B327} (1989) 49--92}.

\bibitem{Beenakker:1990maa}
W.~Beenakker, W.~van Neerven, R.~Meng, G.~Schuler, and J.~Smith, ``{QCD
  corrections to heavy quark production in hadron-hadron collisions},''
\href{http://dx.doi.org/10.1016/S0550-3213(05)80032-X}{{\em Nucl.Phys.}
  {\bfseries B351} (1991) 507--560}.

\bibitem{Ellis:1988sb}
R.~K. Ellis and P.~Nason, ``{QCD Radiative Corrections to the Photoproduction
  of Heavy Quarks},''
\href{http://dx.doi.org/10.1016/0550-3213(89)90571-3}{{\em Nucl.Phys.}
  {\bfseries B312} (1989) 551}.

\bibitem{Smith:1991pw}
J.~Smith and W.~van Neerven, ``{QCD corrections to heavy flavor photoproduction
  and electroproduction},''
\href{http://dx.doi.org/10.1016/0550-3213(92)90476-R}{{\em Nucl.Phys.}
  {\bfseries B374} (1992) 36--82}.

\bibitem{mnrdouble}
M.~L. Mangano, P.~Nason, and G.~Ridolfi, ``{Heavy quark correlations in hadron
  collisions at next-to-leading order},''
\href{http://dx.doi.org/10.1016/0550-3213(92)90435-E}{{\em Nucl.Phys.}
  {\bfseries B373} (1992) 295--345}.

\bibitem{Frixione:1993dg}
S.~Frixione, M.~L. Mangano, P.~Nason, and G.~Ridolfi, ``{Heavy quark
  correlations in photon-hadron collisions},''
  \href{http://dx.doi.org/10.1016/0550-3213(94)90501-0}{{\em Nucl.Phys.}
  {\bfseries B412} (1994) 225--259},
\href{http://arxiv.org/abs/hep-ph/9306337}{{\ttfamily arXiv:hep-ph/9306337
  [hep-ph]}}.

\bibitem{mnrplace}
``A package of fortran routines for the computation of heavy quark cross
  sections and distributions in hadron-hadron collisions in qcd at
  next-to-leading order.''
\newblock \url{http://www.ge.infn.it/~ridolfi/hvqlibx.tgz}.

\bibitem{Czakon:2015owf}
M.~Czakon, D.~Heymes, and A.~Mitov, ``{High-precision differential predictions
  for top-quark pairs at the LHC},''
  \href{http://dx.doi.org/10.1103/PhysRevLett.116.082003}{{\em Phys. Rev.
  Lett.} {\bfseries 116} no.~8, (2016) 082003},
\href{http://arxiv.org/abs/1511.00549}{{\ttfamily arXiv:1511.00549 [hep-ph]}}.

\bibitem{Czakon:2016dgf}
M.~Czakon, D.~Heymes, and A.~Mitov, ``{Dynamical scales for multi-TeV top-pair
  production at the LHC},''
\href{http://arxiv.org/abs/1606.03350}{{\ttfamily arXiv:1606.03350 [hep-ph]}}.

\bibitem{Cacciari:1998it}
M.~Cacciari, M.~Greco, and P.~Nason, ``{The $p_T$ spectrum in heavy flavor
  hadroproduction},''
  \href{http://dx.doi.org/10.1088/1126-6708/1998/05/007}{{\em JHEP} {\bfseries
  9805} (1998) 007},
\href{http://arxiv.org/abs/hep-ph/9803400}{{\ttfamily arXiv:hep-ph/9803400
  [hep-ph]}}.

\bibitem{Cacciari:2001td}
M.~Cacciari, S.~Frixione, and P.~Nason, ``{The $p_T$ spectrum in heavy flavor
  photoproduction},''
  \href{http://dx.doi.org/10.1088/1126-6708/2001/03/006}{{\em JHEP} {\bfseries
  0103} (2001) 006},
\href{http://arxiv.org/abs/hep-ph/0102134}{{\ttfamily arXiv:hep-ph/0102134
  [hep-ph]}}.

\bibitem{Cacciari:2012ny}
M.~Cacciari, S.~Frixione, N.~Houdeau, M.~L. Mangano, P.~Nason, {\em et~al.},
  ``{Theoretical predictions for charm and bottom production at the LHC},''
  \href{http://dx.doi.org/10.1007/JHEP10(2012)137}{{\em JHEP} {\bfseries 1210}
  (2012) 137},
\href{http://arxiv.org/abs/1205.6344}{{\ttfamily arXiv:1205.6344 [hep-ph]}}.

\bibitem{FONLLWeb}
``Program to calculate heavy quark transverse momentum and rapidity
  distributions in hadron-hadron and photon-hadron collisions.''
\newblock \url{http://cacciari.web.cern.ch/cacciari/fonll}.

\bibitem{Kniehl:2005mk}
B.~Kniehl, G.~Kramer, I.~Schienbein, and H.~Spiesberger, ``{Collinear
  subtractions in hadroproduction of heavy quarks},''
  \href{http://dx.doi.org/10.1140/epjc/s2005-02200-7}{{\em Eur.Phys.J.}
  {\bfseries C41} (2005) 199--212},
\href{http://arxiv.org/abs/hep-ph/0502194}{{\ttfamily arXiv:hep-ph/0502194
  [hep-ph]}}.

\bibitem{Baines:2006uw}
J.~Baines, S.~Baranov, O.~Behnke, J.~Bracinik, M.~Cacciari, {\em et~al.},
  ``{Heavy quarks (Working Group 3): Summary Report for the HERA-LHC Workshop
  Proceedings},''
\href{http://arxiv.org/abs/hep-ph/0601164}{{\ttfamily arXiv:hep-ph/0601164
  [hep-ph]}}.

\bibitem{Mueller:1978xu}
A.~H. Mueller, ``{Cut Vertices and their Renormalization: A Generalization of
  the Wilson Expansion},''
\href{http://dx.doi.org/10.1103/PhysRevD.18.3705}{{\em Phys.Rev.} {\bfseries
  D18} (1978) 3705}.

\bibitem{Collins:1981ta}
J.~C. Collins and G.~F. Sterman, ``{Soft Partons in {QCD}},''
\href{http://dx.doi.org/10.1016/0550-3213(81)90370-9}{{\em Nucl.Phys.}
  {\bfseries B185} (1981) 172}.

\bibitem{Collins:1985ue}
J.~C. Collins, D.~E. Soper, and G.~F. Sterman, ``{Factorization for Short
  Distance Hadron-Hadron Scattering},''
\href{http://dx.doi.org/10.1016/0550-3213(85)90565-6}{{\em Nucl.Phys.}
  {\bfseries B261} (1985) 104}.

\bibitem{Collins:1988ig}
J.~C. Collins, D.~E. Soper, and G.~F. Sterman, ``{Soft Gluons and
  Factorization},''
\href{http://dx.doi.org/10.1016/0550-3213(88)90130-7}{{\em Nucl.Phys.}
  {\bfseries B308} (1988) 833}.

\bibitem{Bodwin:1984hc}
G.~T. Bodwin, ``{Factorization of the Drell-Yan Cross-Section in Perturbation
  Theory},''
\href{http://dx.doi.org/10.1103/PhysRevD.34.3932,
  10.1103/PhysRevD.31.2616}{{\em Phys.Rev.} {\bfseries D31} (1985) 2616}.

\bibitem{Mele:1990cw}
B.~Mele and P.~Nason, ``{The Fragmentation function for heavy quarks in QCD},''
\href{http://dx.doi.org/10.1016/0550-3213(91)90597-Q}{{\em Nucl.Phys.}
  {\bfseries B361} (1991) 626--644}.

\bibitem{Bjorken:1977md}
J.~Bjorken, ``{Properties of Hadron Distributions in Reactions Containing Very
  Heavy Quarks},''
\href{http://dx.doi.org/10.1103/PhysRevD.17.171}{{\em Phys.Rev.} {\bfseries
  D17} (1978) 171--173}.

\bibitem{Suzuki:1977km}
M.~Suzuki, ``{Fragmentation of Hadrons from Heavy Quark Partons},''
\href{http://dx.doi.org/10.1016/0370-2693(77)90761-4}{{\em Phys.Lett.}
  {\bfseries B71} (1977) 139}.

\bibitem{Jaffe:1993ie}
R.~Jaffe and L.~Randall, ``{Heavy quark fragmentation into heavy mesons},''
  \href{http://dx.doi.org/10.1016/0550-3213(94)90495-2}{{\em Nucl.Phys.}
  {\bfseries B412} (1994) 79--105},
\href{http://arxiv.org/abs/hep-ph/9306201}{{\ttfamily arXiv:hep-ph/9306201
  [hep-ph]}}.

\bibitem{Nason:1996pk}
P.~Nason and B.~Webber, ``{Nonperturbative corrections to heavy quark
  fragmentation in $e^{+}e^{-}$ annihilation},''
  \href{http://dx.doi.org/10.1016/S0370-2693(97)00129-9}{{\em Phys.Lett.}
  {\bfseries B395} (1997) 355--363},
\href{http://arxiv.org/abs/hep-ph/9612353}{{\ttfamily arXiv:hep-ph/9612353
  [hep-ph]}}.

\bibitem{Cacciari:2002xb}
M.~Cacciari and E.~Gardi, ``{Heavy quark fragmentation},''
  \href{http://dx.doi.org/10.1016/S0550-3213(03)00435-8}{{\em Nucl.Phys.}
  {\bfseries B664} (2003) 299--340},
\href{http://arxiv.org/abs/hep-ph/0301047}{{\ttfamily arXiv:hep-ph/0301047
  [hep-ph]}}.

\bibitem{Kartvelishvili:1977pi}
V.~Kartvelishvili, A.~Likhoded, and V.~Petrov, ``{On the Fragmentation
  Functions of Heavy Quarks Into Hadrons},''
\href{http://dx.doi.org/10.1016/0370-2693(78)90653-6}{{\em Phys.Lett.}
  {\bfseries B78} (1978) 615}.

\bibitem{Bowler:1981sb}
M.~Bowler, ``{$e^{+}e^{-}$ Production of Heavy Quarks in the String Model},''
\href{http://dx.doi.org/10.1007/BF01574001}{{\em Z.Phys.} {\bfseries C11}
  (1981) 169}.

\bibitem{Peterson:1982ak}
C.~Peterson, D.~Schlatter, I.~Schmitt, and P.~M. Zerwas, ``{Scaling Violations
  in Inclusive $e^{+}e^{-}$ Annihilation Spectra},''
\href{http://dx.doi.org/10.1103/PhysRevD.27.105}{{\em Phys.Rev.} {\bfseries
  D27} (1983) 105}.

\bibitem{Collins:1984ms}
P.~Collins and T.~Spiller, ``{The Fragmentation of Heavy Quarks},''
\href{http://dx.doi.org/10.1088/0305-4616/11/12/006}{{\em J.Phys.} {\bfseries
  G11} (1985) 1289}.

\bibitem{h1frag}
F.~Aaron {\em et~al.}, ``{Study of Charm Fragmentation into $D^{*\pm}$ Mesons
  in Deep-Inelastic Scattering at HERA},''
  \href{http://dx.doi.org/10.1140/epjc/s10052-008-0792-2}{{\em Eur.Phys.J.}
  {\bfseries C59} (2009) 589--606},
\href{http://arxiv.org/abs/0808.1003}{{\ttfamily arXiv:0808.1003 [hep-ex]}}.

\bibitem{tevatronbexcess}
M.~Cacciari and P.~Nason, ``{Is there a significant excess in bottom
  hadroproduction at the Tevatron?},''
  \href{http://dx.doi.org/10.1103/PhysRevLett.89.122003}{{\em Phys.Rev.Lett.}
  {\bfseries 89} (2002) 122003},
\href{http://arxiv.org/abs/hep-ph/0204025}{{\ttfamily arXiv:hep-ph/0204025
  [hep-ph]}}.

\bibitem{Klein:2008di}
M.~Klein and R.~Yoshida, ``{Collider Physics at HERA},''
  \href{http://dx.doi.org/10.1016/j.ppnp.2008.05.002}{{\em
  Prog.Part.Nucl.Phys.} {\bfseries 61} (2008) 343--393},
\href{http://arxiv.org/abs/0805.3334}{{\ttfamily arXiv:0805.3334 [hep-ex]}}.

\bibitem{hera}
``{HERA - A Proposal for a Large Electron Proton Colliding Beam Facility at
  DESY},'' Tech. Rep. DESY-HERA-81-10, 1981.

\bibitem{verenapictures}
``Pictures database on the verena schoenberg's zeus homepage.''
\newblock \url{http://www-zeus.desy.de/~vschoenb/}.

\bibitem{Abt:1996hi}
I.~Abt {\em et~al.}, ``{The H1 detector at HERA},''
\href{http://dx.doi.org/10.1016/S0168-9002(96)00893-5}{{\em Nucl.Instrum.Meth.}
  {\bfseries A386} (1997) 310--347}.

\bibitem{zeusbb}
{ZEUS Coll.}, ``The zeus detector. status report,'' tech. rep., 1993.
\newblock \url{http://www-zeus.desy.de/bluebook/bluebook.html}.

\bibitem{HERMES:1993aa}
{HERMES Coll.}, ``{HERMES technical design report},'' Tech. Rep.
  DESY-PRC-93-06, 1993.

\bibitem{Hartouni:1995cf}
E.~P. Hartouni, M.~Kreisler, G.~Van~Apeldoorn, H.~van~der Graaf, W.~Ruckstuhl,
  {\em et~al.}, ``{HERA-B: An experiment to study CP violation in the $B$
  system using an internal target at the HERA proton ring. Design report},''
  Tech. Rep. DESY-PRC-95-01, 1995.

\bibitem{HERAUpgrade}
U.~Schneekloth, ``{The HERA luminosity upgrade},'' Tech. Rep. DESY-HERA-98-05,
  July, 1998.

\bibitem{Abt:1996xv}
I.~Abt {\em et~al.}, ``{The tracking, calorimeter and muon detectors of the H1
  experiment at HERA},''
\href{http://dx.doi.org/10.1016/S0168-9002(96)00894-7}{{\em Nucl.Instrum.Meth.}
  {\bfseries A386} (1997) 348--396}.

\bibitem{Appuhn:1996na}
R.~Appuhn {\em et~al.}, ``{The H1 lead / scintillating fiber calorimeter},''
\href{http://dx.doi.org/10.1016/S0168-9002(96)01171-0}{{\em Nucl.Instrum.Meth.}
  {\bfseries A386} (1997) 397--408}.

\bibitem{Pitzl:2000wz}
D.~Pitzl, O.~Behnke, M.~Biddulph, K.~Bosiger, R.~Eichler, {\em et~al.}, ``{The
  H1 silicon vertex detector},''
  \href{http://dx.doi.org/10.1016/S0168-9002(00)00488-5}{{\em
  Nucl.Instrum.Meth.} {\bfseries A454} (2000) 334--349},
\href{http://arxiv.org/abs/hep-ex/0002044}{{\ttfamily arXiv:hep-ex/0002044
  [hep-ex]}}.

\bibitem{Nozicka:2003ru}
M.~Nozicka, ``{The forward and backward silicon trackers of H1},''
\href{http://dx.doi.org/10.1016/S0168-9002(02)02010-7}{{\em Nucl.Instrum.Meth.}
  {\bfseries A501} (2003) 54--59}.

\bibitem{Becker:2007ms}
J.~Becker, K.~Bosiger, L.~Lindfeld, K.~Muller, P.~Robmann, {\em et~al.}, ``{A
  Vertex trigger based on cylindrical multiwire proportional chambers},''
  \href{http://dx.doi.org/10.1016/j.nima.2007.11.024}{{\em Nucl.Instrum.Meth.}
  {\bfseries A586} (2008) 190--203},
\href{http://arxiv.org/abs/physics/0701002}{{\ttfamily arXiv:physics/0701002
  [physics]}}.

\bibitem{h1ctd_res}
{C. Kleinwort}, ``{H1 Alignment Experience, in Proceedings of the First LHC
  Detector Alignment Workshop, CERN-2007-004, p. 41},''.

\bibitem{Wolff:1992rj}
T.~Wolff, J.~Riedberger, M.~Arpagaus, R.~Bernet, R.~Eichler, {\em et~al.}, ``{A
  Drift chamber track finder for the first level trigger of the H1
  experiment},''
\href{http://dx.doi.org/10.1016/0168-9002(92)90345-5}{{\em Nucl.Instrum.Meth.}
  {\bfseries A323} (1992) 537--541}.

\bibitem{Baird:2001xc}
A.~Baird, E.~Elsen, Y.~Fleming, M.~Kolander, S.~Kolya, {\em et~al.}, ``{A Fast
  high resolution track trigger for the H1 experiment},''
  \href{http://dx.doi.org/10.1109/23.958765}{{\em IEEE Trans.Nucl.Sci.}
  {\bfseries 48} (2001) 1276--1285},
\href{http://arxiv.org/abs/hep-ex/0104010}{{\ttfamily arXiv:hep-ex/0104010
  [hep-ex]}}.

\bibitem{Meer:2001im}
D.~Meer, D.~Muller, J.~Muller, A.~Schoning, and C.~Wissing, ``{A
  Multifunctional processing board for the fast track trigger of the H1
  experiment},'' \href{http://dx.doi.org/10.1109/TNS.2002.1003736}{{\em IEEE
  Trans.Nucl.Sci.} {\bfseries 49} (2002) 357--361},
\href{http://arxiv.org/abs/hep-ex/0107010}{{\ttfamily arXiv:hep-ex/0107010
  [hep-ex]}}.

\bibitem{Andrieu:1993kh}
B.~Andrieu {\em et~al.}, ``{The H1 liquid argon calorimeter system},''
\href{http://dx.doi.org/10.1016/0168-9002(93)91257-N}{{\em Nucl.Instrum.Meth.}
  {\bfseries A336} (1993) 460--498}.

\bibitem{Andrieu:1993xn}
B.~Andrieu {\em et~al.}, ``{Electron / pion separation with the H1 LAr
  calorimeters},''
\href{http://dx.doi.org/10.1016/0168-9002(94)90870-2}{{\em Nucl.Instrum.Meth.}
  {\bfseries A344} (1994) 492--506}.

\bibitem{Andrieu:1994yn}
B.~Andrieu {\em et~al.}, ``{Beam tests and calibration of the H1 liquid argon
  calorimeter with electrons},''
\href{http://dx.doi.org/10.1016/0168-9002(94)91155-X}{{\em Nucl.Instrum.Meth.}
  {\bfseries A350} (1994) 57--72}.

\bibitem{Nicholls:1995di}
T.~Nicholls {\em et~al.}, ``{Performance of an electromagnetic lead /
  scintillating fiber calorimeter for the H1 detector},''
\href{http://dx.doi.org/10.1016/0168-9002(95)01443-8}{{\em Nucl.Instrum.Meth.}
  {\bfseries A374} (1996) 149--156}.

\bibitem{Bethe:1934za}
H.~Bethe and W.~Heitler, ``{On the stopping of fast particles and on the
  creation of positive electrons},''
\href{http://dx.doi.org/10.1098/rspa.1934.0140}{{\em Proc.Roy.Soc.Lond.}
  {\bfseries A146} (1934) 83--112}.

\bibitem{Aaron:2012kn}
F.~Aaron {\em et~al.}, ``{Determination of the Integrated Luminosity at HERA
  using Elastic QED Compton Events},''
  \href{http://dx.doi.org/10.1140/epjc/s10052-014-2733-6,
  10.1140/epjc/s10052-012-2163-2}{{\em Eur.Phys.J.} {\bfseries C72} (2012)
  2163},
\href{http://arxiv.org/abs/1205.2448}{{\ttfamily arXiv:1205.2448 [hep-ex]}}.

\bibitem{h1trigger_Eichler}
{R. A. Eichler}, ``{Triggering With Short Bunch Distances: The H1 Trigger at
  HERA as an Example", invited talk given at 5th Int. Conf. on Instrumentation
  for Colliding Beam Physics, Novosibirsk, USSR, Mar 15--21, 1990},''.

\bibitem{h1trigger_Elsen}
{E.~Elsen}, ``{Aspects of the H1 trigger and data acquisition system, prepared
  for 2nd Annual Conference on Electronics for Future Colliders, Chestnut
  Ridge, N.Y., 19--21 May 1992},''.

\bibitem{Harnew:1988ye}
N.~Harnew, G.~Heath, M.~Jeffs, J.~Nash, G.~Salmon, {\em et~al.}, ``{Vertex
  Triggering Using Time Difference Measurements in the ZEUS Central Tracking
  Detector},''
\href{http://dx.doi.org/10.1016/0168-9002(89)91096-6}{{\em Nucl.Instrum.Meth.}
  {\bfseries A279} (1989) 290--296}.

\bibitem{Foster:1992mj}
B.~Foster, G.~Heath, T.~Llewellyn, D.~Gingrich, N.~Harnew, {\em et~al.}, ``{The
  performance of the ZEUS central tracking detector $z$-by-timing electronics
  in a transputer based data acquisition system},''
\href{http://dx.doi.org/10.1016/0920-5632(93)90023-Y}{{\em
  Nucl.Phys.Proc.Suppl.} {\bfseries 32} (1993) 181--188}.

\bibitem{Foster:1993ja}
B.~Foster {\em et~al.}, ``{The design and construction of the ZEUS central
  tracking detector},''
\href{http://dx.doi.org/10.1016/0168-9002(94)91313-7}{{\em Nucl.Instrum.Meth.}
  {\bfseries A338} (1994) 254--283}.

\bibitem{Acerbi:1987cw}
E.~Acerbi, F.~Alessandria, G.~Baccaglioni, E.~Fabrici, L.~Rossi, {\em et~al.},
  ``{Thin and compensating solenoids for ZEUS detector},''
\newblock 1987.
\newblock
In ``Boston 1987, Proceedings, Magnet technology'' 1354--1357.

\bibitem{Polini:2007sw}
A.~Polini {\em et~al.}, ``{The design and performance of the ZEUS Micro Vertex
  detector},'' \href{http://dx.doi.org/10.1016/j.nima.2007.08.167}{{\em
  Nucl.Instrum.Meth.} {\bfseries A581} (2007) 656--686},
\href{http://arxiv.org/abs/0708.3011}{{\ttfamily arXiv:0708.3011
  [physics.ins-det]}}.

\bibitem{Fourletov:2004iu}
S.~Fourletov, ``{Straw tube tracking detector (STT) for ZEUS},''
\href{http://dx.doi.org/10.1016/j.nima.2004.07.212}{{\em Nucl.Instrum.Meth.}
  {\bfseries A535} (2004) 191--196}.

\bibitem{Bamberger:1997fg}
A.~Bamberger {\em et~al.}, ``{The small angle rear tracking detector of
  ZEUS},''
\href{http://dx.doi.org/10.1016/S0168-9002(97)01029-2}{{\em Nucl.Instrum.Meth.}
  {\bfseries A401} (1997) 63--80}.

\bibitem{Derrick:1991tq}
M.~Derrick, D.~Gacek, N.~Hill, B.~Musgrave, R.~Noland, {\em et~al.}, ``{Design
  and construction of the ZEUS barrel calorimeter},''
\href{http://dx.doi.org/10.1016/0168-9002(91)90094-7}{{\em Nucl.Instrum.Meth.}
  {\bfseries A309} (1991) 77--100}.

\bibitem{Andresen:1991ph}
A.~Andresen {\em et~al.}, ``{Construction and beam test of the ZEUS forward and
  rear calorimeter},''
\href{http://dx.doi.org/10.1016/0168-9002(91)90095-8}{{\em Nucl.Instrum.Meth.}
  {\bfseries A309} (1991) 101--142}.

\bibitem{Caldwell:1992wc}
A.~Caldwell, I.~Gialas, S.~Mishra, J.~Parsons, S.~Ritz, {\em et~al.}, ``{Design
  and implementation of a high precision readout system for the ZEUS
  calorimeter},''
\href{http://dx.doi.org/10.1016/0168-9002(92)90413-X}{{\em Nucl.Instrum.Meth.}
  {\bfseries A321} (1992) 356--364}.

\bibitem{Bernstein:1993kj}
A.~Bernstein {\em et~al.}, ``{Beam tests of the ZEUS barrel calorimeter},''
\href{http://dx.doi.org/10.1016/0168-9002(93)91078-2}{{\em Nucl.Instrum.Meth.}
  {\bfseries A336} (1993) 23--52}.

\bibitem{Abbiendi:1993mi}
G.~Abbiendi, M.~Bonori, R.~Brugnera, R.~Carlin, V.~Chiaratti, {\em et~al.},
  ``{The ZEUS barrel and rear muon detector},''
\href{http://dx.doi.org/10.1016/0168-9002(93)91176-N}{{\em Nucl.Instrum.Meth.}
  {\bfseries A333} (1993) 342--354}.

\bibitem{Andruszkow:2001jy}
J.~Andruszkow {\em et~al.}, ``{Luminosity measurement in the ZEUS
  experiment},''
{\em Acta Phys.Polon.} {\bfseries B32} (2001) 2025--2058.

\bibitem{Helbich:2005qf}
M.~Helbich, Y.~Ning, S.~Paganis, Z.~Ren, W.~Schmidke, {\em et~al.}, ``{The
  spectrometer system for measuring ZEUS luminosity at HERA},''
  \href{http://dx.doi.org/10.1016/j.nima.2006.06.049}{{\em Nucl.Instrum.Meth.}
  {\bfseries A565} (2006) 572--588},
\href{http://arxiv.org/abs/physics/0512153}{{\ttfamily arXiv:physics/0512153
  [physics]}}.

\bibitem{Smith:1994nx}
W.~Smith, I.~Ali, B.~Behrens, C.~Fordham, C.~Foudas, {\em et~al.}, ``{The ZEUS
  calorimeter first level trigger},''
\href{http://dx.doi.org/10.1016/0168-9002(94)01163-X}{{\em Nucl.Instrum.Meth.}
  {\bfseries A355} (1995) 278--294}.

\bibitem{Heath:1991rk}
G.~Heath, B.~Foster, T.~Short, S.~Wilson, M.~Lancaster, {\em et~al.}, ``{The
  ZEUS first level tracking trigger},''
\href{http://dx.doi.org/10.1016/0168-9002(92)90741-L}{{\em Nucl.Instrum.Meth.}
  {\bfseries A315} (1992) 431--435}.

\bibitem{Quadt:1999hr}
A.~Quadt, R.~Devenish, S.~Topp-Jorgensen, M.~Sutton, H.~Uijterwaal, {\em
  et~al.}, ``{The design and performance of the ZEUS central tracking detector
  second level trigger},''
\href{http://dx.doi.org/10.1016/S0168-9002(99)00826-8}{{\em Nucl.Instrum.Meth.}
  {\bfseries A438} (1999) 472--501}.

\bibitem{Bhadra:1989kz}
S.~Bhadra, M.~Crombie, D.~P. Kirkby, and R.~Orr, ``{The ZEUS third level
  trigger system},''
\href{http://dx.doi.org/10.1016/0010-4655(89)90236-1}{{\em Comput.Phys.Commun.}
  {\bfseries 57} (1989) 321--324}.

\bibitem{h1ltt_hera2}
F.~Aaron {\em et~al.}, ``{Measurement of the Charm and Beauty Structure
  Functions using the H1 Vertex Detector at HERA},''
  \href{http://dx.doi.org/10.1140/epjc/s10052-009-1190-0}{{\em Eur.Phys.J.}
  {\bfseries C65} (2010) 89--109},
\href{http://arxiv.org/abs/0907.2643}{{\ttfamily arXiv:0907.2643 [hep-ex]}}.

\bibitem{h1dstar_hera1}
A.~Aktas {\em et~al.}, ``{Production of $D^{*\pm}$ Mesons with Dijets in
  Deep-Inelastic Scattering at HERA},''
  \href{http://dx.doi.org/10.1140/epjc/s10052-007-0296-5}{{\em Eur.Phys.J.}
  {\bfseries C51} (2007) 271--287},
\href{http://arxiv.org/abs/hep-ex/0701023}{{\ttfamily arXiv:hep-ex/0701023
  [hep-ex]}}.

\bibitem{zd00}
S.~Chekanov {\em et~al.}, ``{Measurement of $D^{*\pm}$ production in deep
  inelastic $e^{\pm}p$ scattering at HERA},''
  \href{http://dx.doi.org/10.1103/PhysRevD.69.012004}{{\em Phys.Rev.}
  {\bfseries D69} (2004) 012004},
\href{http://arxiv.org/abs/hep-ex/0308068}{{\ttfamily arXiv:hep-ex/0308068
  [hep-ex]}}.

\bibitem{zd0dp}
S.~Chekanov {\em et~al.}, ``{Measurement of $D^{\pm}$ and $D^0$ production in
  deep inelastic scattering using a lifetime tag at HERA},''
  \href{http://dx.doi.org/10.1140/epjc/s10052-009-1088-x}{{\em Eur.Phys.J.}
  {\bfseries C63} (2009) 171--188},
\href{http://arxiv.org/abs/0812.3775}{{\ttfamily arXiv:0812.3775 [hep-ex]}}.

\bibitem{zeus_muon}
S.~Chekanov {\em et~al.}, ``{Measurement of charm and beauty production in deep
  inelastic $ep$ scattering from decays into muons at HERA},''
  \href{http://dx.doi.org/10.1140/epjc/s10052-009-1193-x}{{\em Eur.Phys.J.}
  {\bfseries C65} (2010) 65--79},
\href{http://arxiv.org/abs/0904.3487}{{\ttfamily arXiv:0904.3487 [hep-ex]}}.

\bibitem{zeusdstar_hera2}
H.~Abramowicz {\em et~al.}, ``{Measurement of $ D^{*\pm}$ production in deep
  inelastic scattering at HERA},''
  \href{http://dx.doi.org/10.1007/JHEP05(2013)097}{{\em JHEP} {\bfseries 1305}
  (2013) 097},
\href{http://arxiv.org/abs/1303.6578}{{\ttfamily arXiv:1303.6578 [hep-ex]}}.

\bibitem{Andrews:1975xt}
D.~E. Andrews, J.~R. Harvey, F.~Lobkowicz, E.~N. May, C.~A. Nelson, Jr., E.~H.
  Thorndike, and M.~E. Nordberg, Jr., ``{An Improved Upper Limit for
  Photoproduction of psi (3105) Near Threshold},''
\href{http://dx.doi.org/10.1103/PhysRevLett.34.1134}{{\em Phys. Rev. Lett.}
  {\bfseries 34} (1975) 1134--1136}.

\bibitem{Nussinov:1975ay}
S.~Nussinov, ``{On Possible Effects of Decays of Charmed Particle
  Resonances},''
\href{http://dx.doi.org/10.1103/PhysRevLett.35.1672}{{\em Phys.Rev.Lett.}
  {\bfseries 35} (1975) 1672}.

\bibitem{Feldman:1977ir}
G.~Feldman, I.~Peruzzi, M.~Piccolo, G.~Abrams, M.~Alam, {\em et~al.},
  ``{Observation of the Decay $D^{*+} \to D^{0} \pi^{+}$},''
\href{http://dx.doi.org/10.1103/PhysRevLett.38.1313}{{\em Phys.Rev.Lett.}
  {\bfseries 38} (1977) 1313}.

\bibitem{Gibbons:1997ag}
L.~Gibbons {\em et~al.}, ``{The Inclusive decays $B \to DX$ and $B \to D^{*}
  X$},'' \href{http://dx.doi.org/10.1103/PhysRevD.56.3783}{{\em Phys.Rev.}
  {\bfseries D56} (1997) 3783--3802},
\href{http://arxiv.org/abs/hep-ex/9703006}{{\ttfamily arXiv:hep-ex/9703006
  [hep-ex]}}.

\bibitem{LHCbCharm}
R.~Aaij {\em et~al.}, ``{Prompt charm production in $pp$ collisions at
  $\sqrt{s}=7$ TeV},''
  \href{http://dx.doi.org/10.1016/j.nuclphysb.2013.02.010}{{\em Nucl.Phys.}
  {\bfseries B871} (2013) 1--20},
\href{http://arxiv.org/abs/1302.2864}{{\ttfamily arXiv:1302.2864 [hep-ex]}}.

\bibitem{Aktas:2004az}
A.~Aktas {\em et~al.}, ``{Measurement of $F_2^{c \bar{c}}$ and $F_2^{b\bar{b}}$
  at high $Q^{2}$ using the H1 vertex detector at HERA},''
  \href{http://dx.doi.org/10.1140/epjc/s2005-02154-8}{{\em Eur.Phys.J.}
  {\bfseries C40} (2005) 349--359},
\href{http://arxiv.org/abs/hep-ex/0411046}{{\ttfamily arXiv:hep-ex/0411046
  [hep-ex]}}.

\bibitem{Aktas:2005iw}
A.~Aktas {\em et~al.}, ``{Measurement of $F_2^{c\bar{c}}$ and $F_2^{b\bar{b}}$
  at low $Q^2$ and $x$ using the H1 vertex detector at HERA},''
  \href{http://dx.doi.org/10.1140/epjc/s2005-02415-6}{{\em Eur.Phys.J.}
  {\bfseries C45} (2006) 23--33},
\href{http://arxiv.org/abs/hep-ex/0507081}{{\ttfamily arXiv:hep-ex/0507081
  [hep-ex]}}.

\bibitem{Aktas:2004ka}
A.~Aktas {\em et~al.}, ``{Inclusive production of $D^{+}$, $D^{0}$, $D^{+}_{s}$
  and $D^{*+}$ mesons in deep inelastic scattering at HERA},''
  \href{http://dx.doi.org/10.1140/epjc/s2004-02069-x}{{\em Eur. Phys. J.}
  {\bfseries C38} (2005) 447--459},
\href{http://arxiv.org/abs/hep-ex/0408149}{{\ttfamily arXiv:hep-ex/0408149
  [hep-ex]}}.

\bibitem{mishath}
M.~Lisovyi, {\em {Measurement of charm production in deep inelastic scattering
  using lifetime tagging for $D^{\pm}$ meson decays with the ZEUS detector at
  HERA}}.
\newblock PhD thesis, Hamburg University,
October, 2011.
\newblock

\bibitem{kalman}
R.~Kalman, ``{A New Approach to Linear Filtering and Prediction Problems},''
  {\em Trans. ASME J. Basic Engineering} {\bfseries D82} (1960) 35.

\bibitem{trk_avery}
P.~Avery, ``{Applied Fitting Theory V: Track Fitting Using the Kalman
  Filter}.'' unpublished, 1992.

\bibitem{verbth}
A.~Verbytskyi, {\em {Production of the excited charm mesons $D_{1}$ and
  $D^{*}_{2}$ at HERA}}.
\newblock PhD thesis, Hamburg University,
February, 2013.
\newblock

\bibitem{Fruhwirth:1999tya}
R.~Frühwirth and A.~Strandlie, ``{Track fitting with ambiguities and noise: A
  study of elastic tracking and nonlinear filters},''
\href{http://dx.doi.org/10.1016/S0010-4655(99)00231-3}{{\em
  Comput.Phys.Commun.} {\bfseries 120} no.~2-3, (1999) 197--214}.

\bibitem{Briskin:1998sv}
G.~M. Briskin, {\em {Diffractive Dissociation in $ep$ Deep Inelastic
  Scattering}}.
\newblock PhD thesis, Tel Aviv U., 1998.
\newblock
\url{http://www-library.desy.de/cgi-bin/showprep.pl?desy-thesis98-036}.
\newblock

\bibitem{Abramowicz:1995zi}
H.~Abramowicz, A.~Caldwell, and R.~Sinkus, ``{Neural network based electron
  identification in the ZEUS calorimeter},''
  \href{http://dx.doi.org/10.1016/0168-9002(95)00612-5}{{\em
  Nucl.Instrum.Meth.} {\bfseries A365} (1995) 508--517},
\href{http://arxiv.org/abs/hep-ex/9505004}{{\ttfamily arXiv:hep-ex/9505004
  [hep-ex]}}.

\bibitem{Kappes:2001jy}
A.~Kappes, {\em {Measurement of $e^{-}p \to e^{-} X$ differential
  cross-sections at high $Q^2$ and of the structure function $xF_3$ with ZEUS
  at HERA}}.
\newblock PhD thesis, Bonn University,
2001.
\newblock

\bibitem{Amaldi:1979qp}
{F. Jacquet and A. Blondel}, ``{Study of an $ep$ Facility for Europe DESY,
  Hamburg, April 2-3, 1979},''
No.~DESY-79-48.
\newblock

\bibitem{Buchmuller:1992rq1}
{S. Bentvelsen et al.}, ``{Physics at HERA. Proceedings, Workshop, Hamburg,
  Germany, October 29-30, 1991. Vol. 1, p.23},''
\newblock
1992.
\newblock

\bibitem{Buchmuller:1992rq2}
{K. Hoeger}, ``{Physics at HERA. Proceedings, Workshop, Hamburg, Germany,
  October 29-30, 1991. Vol. 1, p.43},''
\newblock
1992.
\newblock

\bibitem{pythia}
T.~Sjostrand, L.~Lonnblad, and S.~Mrenna, ``{PYTHIA 6.2: Physics and manual},''
\href{http://arxiv.org/abs/hep-ph/0108264}{{\ttfamily arXiv:hep-ph/0108264
  [hep-ph]}}.

\bibitem{ariadne}
L.~Lonnblad, ``{ARIADNE version 4: A program for simulation of QCD cascades
  implementing the color dipole model},''
\href{http://dx.doi.org/10.1016/0010-4655(92)90068-A}{{\em Comput.Phys.Commun.}
  {\bfseries 71} (1992) 15--31}.

\bibitem{rapgap}
H.~Jung, ``{Hard diffractive scattering in high-energy $ep$ collisions and the
  Monte Carlo generator RAPGAP},''
\href{http://dx.doi.org/10.1016/0010-4655(94)00150-Z}{{\em Comput.Phys.Commun.}
  {\bfseries 86} (1995) 147--161}.

\bibitem{Shehzadi:2011sxa}
R.~Shehzadi, {\em {Measurement of beauty production in deep inelastic
  scattering at HERA using decays into electrons}}.
\newblock PhD thesis, Bonn University,
January, 2011.
\newblock

\bibitem{geant}
R.~Brun, F.~Bruyant, M.~Maire, A.~McPherson, and P.~Zanarini, ``{GEANT3}.''
  CERN-DD-EE-84-1,
1987.

\bibitem{heracles}
A.~Kwiatkowski, H.~Spiesberger, and H.~Mohring, ``{Heracles: An Event Generator
  for $e p$ Interactions at {HERA} Energies Including Radiative Processes:
  Version 1.0},''
\href{http://dx.doi.org/10.1016/0010-4655(92)90136-M}{{\em Comput.Phys.Commun.}
  {\bfseries 69} (1992) 155--172}.

\bibitem{cteq5l}
H.~Lai {\em et~al.}, ``{Global QCD analysis of parton structure of the nucleon:
  CTEQ5 parton distributions},''
  \href{http://dx.doi.org/10.1007/s100529900196}{{\em Eur.Phys.J.} {\bfseries
  C12} (2000) 375--392},
\href{http://arxiv.org/abs/hep-ph/9903282}{{\ttfamily arXiv:hep-ph/9903282
  [hep-ph]}}.

\bibitem{pdg2012}
J.~Beringer {\em et~al.}, ``{Review of Particle Physics (RPP)},''
\href{http://dx.doi.org/10.1103/PhysRevD.86.010001}{{\em Phys.Rev.} {\bfseries
  D86} (2012) 010001}.

\bibitem{minuit}
F.~James and M.~Roos, ``{Minuit: A System for Function Minimization and
  Analysis of the Parameter Errors and Correlations},''
\href{http://dx.doi.org/10.1016/0010-4655(75)90039-9}{{\em Comput.Phys.Commun.}
  {\bfseries 10} (1975) 343--367}.

\bibitem{minuitmanual}
F.~James, ``Cern program library long writeup d506. minuit reference manual.
  version 94.1.''.

\bibitem{Chekanov:2009kj}
S.~Chekanov {\em et~al.}, ``{Measurement of charm and beauty production in deep
  inelastic $ep$ scattering from decays into muons at HERA},''
  \href{http://dx.doi.org/10.1140/epjc/s10052-009-1193-x}{{\em Eur.Phys.J.}
  {\bfseries C65} (2010) 65--79},
\href{http://arxiv.org/abs/0904.3487}{{\ttfamily arXiv:0904.3487 [hep-ex]}}.

\bibitem{Abramowicz:2010zq}
H.~Abramowicz {\em et~al.}, ``{Measurement of beauty production in DIS and
  $F_2^{b\bar{b}}$ extraction at ZEUS},''
  \href{http://dx.doi.org/10.1140/epjc/s10052-010-1423-2}{{\em Eur.Phys.J.}
  {\bfseries C69} (2010) 347--360},
\href{http://arxiv.org/abs/1005.3396}{{\ttfamily arXiv:1005.3396 [hep-ex]}}.

\bibitem{Abramowicz:2011kj}
H.~Abramowicz {\em et~al.}, ``{Measurement of beauty production in deep
  inelastic scattering at HERA using decays into electrons},''
  \href{http://dx.doi.org/10.1140/epjc/s10052-011-1573-x}{{\em Eur.Phys.J.}
  {\bfseries C71} (2011) 1573},
\href{http://arxiv.org/abs/1101.3692}{{\ttfamily arXiv:1101.3692 [hep-ex]}}.

\bibitem{ozmaster}
O.~Zenaiev, ``{Measurement of total and differential cross sections of \Dch
  meson production in deep inelastic scattering with the ZEUS detector at
  HERA},'' Master's thesis, National Taras Shevchenko University of Kyiv, June,
  2011.
\newblock In Ukrainian.

\bibitem{libovth}
V.~Libov, {\em {Measurement of Charm and Beauty Production in Deep Inelastic
  Scattering at HERA and Test Beam Studies of ATLAS Pixel Sensors}}.
\newblock PhD thesis, Hamburg University,
December, 2012.
\newblock

\bibitem{zeuss}
S.~Chekanov {\em et~al.}, ``{A ZEUS next-to-leading-order QCD analysis of data
  on deep inelastic scattering},''
  \href{http://dx.doi.org/10.1103/PhysRevD.67.012007}{{\em Phys.Rev.}
  {\bfseries D67} (2003) 012007},
\href{http://arxiv.org/abs/hep-ex/0208023}{{\ttfamily arXiv:hep-ex/0208023
  [hep-ex]}}.

\bibitem{zeusfrag}
S.~Chekanov {\em et~al.}, ``{Measurement of the charm fragmentation function in
  $D^{*\pm}$ photoproduction at HERA},''
  \href{http://dx.doi.org/10.1088/1126-6708/2009/04/082}{{\em JHEP} {\bfseries
  0904} (2009) 082},
\href{http://arxiv.org/abs/0901.1210}{{\ttfamily arXiv:0901.1210 [hep-ex]}}.

\bibitem{frag06}
M.~Cacciari, P.~Nason, and C.~Oleari, ``{A study of heavy flavored meson
  fragmentation functions in $e^{+}e^{-}$ annihilation},''
  \href{http://dx.doi.org/10.1088/1126-6708/2006/04/006}{{\em JHEP} {\bfseries
  0604} (2006) 006},
\href{http://arxiv.org/abs/hep-ph/0510032}{{\ttfamily arXiv:hep-ph/0510032
  [hep-ph]}}.

\bibitem{Seuster:2005tr}
R.~Seuster {\em et~al.}, ``{Charm hadrons from fragmentation and $B$ decays in
  $e^{+}e^{-}$ annihilation at $\sqrt{s} = 10.6$ GeV},''
  \href{http://dx.doi.org/10.1103/PhysRevD.73.032002}{{\em Phys.Rev.}
  {\bfseries D73} (2006) 032002},
\href{http://arxiv.org/abs/hep-ex/0506068}{{\ttfamily arXiv:hep-ex/0506068
  [hep-ex]}}.

\bibitem{Lohrmann:2011np}
E.~Lohrmann, ``{A Summary of Charm Hadron Production Fractions},''
\href{http://arxiv.org/abs/1112.3757}{{\ttfamily arXiv:1112.3757 [hep-ex]}}.

\bibitem{Glazov:2005rn}
A.~Glazov, ``{Averaging of DIS cross section data},''
\href{http://dx.doi.org/10.1063/1.2122026}{{\em AIP Conf.Proc.} {\bfseries 792}
  (2005) 237--240}.

\bibitem{heraverager}
``{HERAverager: an averaging tool developed for the H1 and ZEUS data
  combination (version 0.0.2)}.''
\newblock \url{https://wiki-zeuthen.desy.de/HERAverager}.

\bibitem{hessian}
J.~Pumplin, D.~Stump, R.~Brock, D.~Casey, J.~Huston, {\em et~al.},
  ``{Uncertainties of predictions from parton distribution functions. 2. The
  Hessian method},'' \href{http://dx.doi.org/10.1103/PhysRevD.65.014013}{{\em
  Phys.Rev.} {\bfseries D65} (2001) 014013},
\href{http://arxiv.org/abs/hep-ph/0101032}{{\ttfamily arXiv:hep-ph/0101032
  [hep-ph]}}.

\bibitem{barlow}
R.~Barlow, {\em {Statistics: a guide to use of statistical methods in the
  physical sciences}}.
\newblock Manchester Physics Series, 1989.
\newblock ISBN: 978-0-471-92295-7.

\bibitem{blobel_lohrmann}
E.~L. V.~Blobel, {\em {Statistical and numerical methods in data analysis}}.
\newblock 2012.
\newblock \url{http://www-library.desy.de/preparch/books/BloLoBuch.pdf}.
\newblock (in German) ISBN 978-3-935702-66-9 (e-Buch).

\bibitem{Stump:2001gu}
D.~Stump, J.~Pumplin, R.~Brock, D.~Casey, J.~Huston, {\em et~al.},
  ``{Uncertainties of predictions from parton distribution functions. 1. The
  Lagrange multiplier method},''
  \href{http://dx.doi.org/10.1103/PhysRevD.65.014012}{{\em Phys.Rev.}
  {\bfseries D65} (2001) 014012},
\href{http://arxiv.org/abs/hep-ph/0101051}{{\ttfamily arXiv:hep-ph/0101051
  [hep-ph]}}.

\bibitem{offset}
M.~Botje, ``{Error estimates on parton density distributions},''
  \href{http://dx.doi.org/10.1088/0954-3899/28/5/305}{{\em J.Phys.} {\bfseries
  G28} (2002) 779--790},
\href{http://arxiv.org/abs/hep-ph/0110123}{{\ttfamily arXiv:hep-ph/0110123
  [hep-ph]}}.

\bibitem{abm11}
S.~Alekhin, J.~Bl{\"u}mlein, and S.~Moch, ``{Parton Distribution Functions and
  Benchmark Cross Sections at NNLO},''
  \href{http://dx.doi.org/10.1103/PhysRevD.86.054009}{{\em Phys.Rev.}
  {\bfseries D86} (2012) 054009},
\href{http://arxiv.org/abs/1202.2281}{{\ttfamily arXiv:1202.2281 [hep-ph]}}.

\bibitem{Martin:2010db}
A.~Martin, W.~Stirling, R.~Thorne, and G.~Watt, ``{Heavy-quark mass dependence
  in global PDF analyses and 3- and 4-flavour parton distributions},''
  \href{http://dx.doi.org/10.1140/epjc/s10052-010-1462-8}{{\em Eur.Phys.J.}
  {\bfseries C70} (2010) 51--72},
\href{http://arxiv.org/abs/1007.2624}{{\ttfamily arXiv:1007.2624 [hep-ph]}}.

\bibitem{Adam:2006nu}
N.~Adam {\em et~al.}, ``{Absolute Branching Fraction Measurements for $D^{+}$
  and $D^{0}$ Inclusive Semileptonic Decays},''
  \href{http://dx.doi.org/10.1103/PhysRevLett.97.251801}{{\em Phys.Rev.Lett.}
  {\bfseries 97} (2006) 251801},
\href{http://arxiv.org/abs/hep-ex/0604044}{{\ttfamily arXiv:hep-ex/0604044
  [hep-ex]}}.

\bibitem{Chekanov:2004tk}
S.~Chekanov {\em et~al.}, ``{Measurement of beauty production in deep inelastic
  scattering at HERA},''
  \href{http://dx.doi.org/10.1016/j.physletb.2004.08.048}{{\em Phys.Lett.}
  {\bfseries B599} (2004) 173--189},
\href{http://arxiv.org/abs/hep-ex/0405069}{{\ttfamily arXiv:hep-ex/0405069
  [hep-ex]}}.

\bibitem{evtgen}
D.~Lange, ``{The EvtGen particle decay simulation package},''
\href{http://dx.doi.org/10.1016/S0168-9002(01)00089-4}{{\em Nucl.Instrum.Meth.}
  {\bfseries A462} (2001) 152--155}.

\bibitem{Albrecht:1995yb}
H.~Albrecht {\em et~al.}, ``{Two measurements of $B^0\bar{B}^0$ mixing using
  kaon tagging},''
\href{http://dx.doi.org/10.1016/0370-2693(96)00206-7}{{\em Phys.Lett.}
  {\bfseries B374} (1996) 256--264}.

\bibitem{frag00}
P.~Nason and C.~Oleari, ``{A phenomenological study of heavy quark
  fragmentation functions in $e^{+}e^{-}$ annihilation},''
  \href{http://dx.doi.org/10.1016/S0550-3213(99)00673-2}{{\em Nucl.Phys.}
  {\bfseries B565} (2000) 245--266},
\href{http://arxiv.org/abs/hep-ph/9903541}{{\ttfamily arXiv:hep-ph/9903541
  [hep-ph]}}.

\bibitem{Alekhin:2009ni}
S.~Alekhin, J.~Bl{\"u}mlein, S.~Klein, and S.~Moch, ``{The 3, 4, and 5-flavor
  NNLO Parton from Deep-Inelastic-Scattering Data and at Hadron Colliders},''
  \href{http://dx.doi.org/10.1103/PhysRevD.81.014032}{{\em Phys.Rev.}
  {\bfseries D81} (2010) 014032},
\href{http://arxiv.org/abs/0908.2766}{{\ttfamily arXiv:0908.2766 [hep-ph]}}.

\bibitem{herafitter}
``Herafitter: an open source qcd fit framework (version 1.0.0).''
\newblock \url{https://wiki-zeuthen.desy.de/HERAFitter}.

\bibitem{HERAFitterPaper}
S.~Alekhin {\em et~al.}, ``{HERAFitter},''
  \href{http://dx.doi.org/10.1140/epjc/s10052-015-3480-z}{{\em Eur. Phys. J.}
  {\bfseries C75} no.~7, (2015) 304},
\href{http://arxiv.org/abs/1410.4412}{{\ttfamily arXiv:1410.4412 [hep-ph]}}.

\bibitem{openqcdrad}
``{S. Alekhin, ``OPENQCDRAD-1.5''}.''
\newblock \url{http://www-zeuthen.desy.de/~alekhin/OPENQCDRAD/}.

\bibitem{Martin:2009iq}
A.~Martin, W.~Stirling, R.~Thorne, and G.~Watt, ``{Parton distributions for the
  LHC},'' \href{http://dx.doi.org/10.1140/epjc/s10052-009-1072-5}{{\em
  Eur.Phys.J.} {\bfseries C63} (2009) 189--285},
\href{http://arxiv.org/abs/0901.0002}{{\ttfamily arXiv:0901.0002 [hep-ph]}}.

\bibitem{Aaron:2012qi}
F.~Aaron {\em et~al.}, ``{Inclusive Deep Inelastic Scattering at High $Q^2$
  with Longitudinally Polarised Lepton Beams at HERA},''
  \href{http://dx.doi.org/10.1007/JHEP09(2012)061}{{\em JHEP} {\bfseries 1209}
  (2012) 061},
\href{http://arxiv.org/abs/1206.7007}{{\ttfamily arXiv:1206.7007 [hep-ex]}}.

\bibitem{Pumplin:2000vx}
J.~Pumplin, D.~Stump, and W.~Tung, ``{Multivariate fitting and the error matrix
  in global analysis of data},''
  \href{http://dx.doi.org/10.1103/PhysRevD.65.014011}{{\em Phys.Rev.}
  {\bfseries D65} (2001) 014011},
\href{http://arxiv.org/abs/hep-ph/0008191}{{\ttfamily arXiv:hep-ph/0008191
  [hep-ph]}}.

\bibitem{Pumplin:2002vw}
J.~Pumplin, D.~R. Stump, J.~Huston, H.~L. Lai, P.~M. Nadolsky, and W.~K. Tung,
  ``{New generation of parton distributions with uncertainties from global QCD
  analysis},'' \href{http://dx.doi.org/10.1088/1126-6708/2002/07/012}{{\em
  JHEP} {\bfseries 07} (2002) 012},
\href{http://arxiv.org/abs/hep-ph/0201195}{{\ttfamily arXiv:hep-ph/0201195
  [hep-ph]}}.

\bibitem{Gao:2013wwa}
J.~Gao, M.~Guzzi, and P.~M. Nadolsky, ``{Charm quark mass dependence in a
  global QCD analysis},''
  \href{http://dx.doi.org/10.1140/epjc/s10052-013-2541-4}{{\em Eur.Phys.J.}
  {\bfseries C73} (2013) 2541},
\href{http://arxiv.org/abs/1304.3494}{{\ttfamily arXiv:1304.3494 [hep-ph]}}.

\bibitem{Alekhin:2012vu}
S.~Alekhin, J.~Bl{\"u}mlein, K.~Daum, K.~Lipka, and S.~Moch, ``{Precise
  charm-quark mass from deep-inelastic scattering},''
  \href{http://dx.doi.org/10.1016/j.physletb.2013.02.010}{{\em Phys.Lett.}
  {\bfseries B720} (2013) 172--176},
\href{http://arxiv.org/abs/1212.2355}{{\ttfamily arXiv:1212.2355 [hep-ph]}}.

\bibitem{Alekhin:2012ig}
S.~Alekhin, J.~Bl{\"u}mlein, and S.~Moch, ``{Parton Distribution Functions and
  Benchmark Cross Sections at NNLO},''
  \href{http://dx.doi.org/10.1103/PhysRevD.86.054009}{{\em Phys. Rev.}
  {\bfseries D86} (2012) 054009},
\href{http://arxiv.org/abs/1202.2281}{{\ttfamily arXiv:1202.2281 [hep-ph]}}.

\bibitem{Accardi:2016qay}
A.~Accardi, L.~T. Brady, W.~Melnitchouk, J.~F. Owens, and N.~Sato,
  ``{Constraints on large-$x$ parton distributions from new weak boson
  production and deep-inelastic scattering data},''
  \href{http://dx.doi.org/10.1103/PhysRevD.93.114017}{{\em Phys. Rev.}
  {\bfseries D93} no.~11, (2016) 114017},
\href{http://arxiv.org/abs/1602.03154}{{\ttfamily arXiv:1602.03154 [hep-ph]}}.

\bibitem{Dulat:2015mca}
S.~Dulat, T.-J. Hou, J.~Gao, M.~Guzzi, J.~Huston, P.~Nadolsky, J.~Pumplin,
  C.~Schmidt, D.~Stump, and C.~P. Yuan, ``{New parton distribution functions
  from a global analysis of quantum chromodynamics},''
  \href{http://dx.doi.org/10.1103/PhysRevD.93.033006}{{\em Phys. Rev.}
  {\bfseries D93} no.~3, (2016) 033006},
\href{http://arxiv.org/abs/1506.07443}{{\ttfamily arXiv:1506.07443 [hep-ph]}}.

\bibitem{Jimenez-Delgado:2014twa}
P.~Jimenez-Delgado and E.~Reya, ``{Delineating parton distributions and the
  strong coupling},'' \href{http://dx.doi.org/10.1103/PhysRevD.89.074049}{{\em
  Phys. Rev.} {\bfseries D89} no.~7, (2014) 074049},
\href{http://arxiv.org/abs/1403.1852}{{\ttfamily arXiv:1403.1852 [hep-ph]}}.

\bibitem{Harland-Lang:2014zoa}
L.~A. Harland-Lang, A.~D. Martin, P.~Motylinski, and R.~S. Thorne, ``{Parton
  distributions in the LHC era: MMHT 2014 PDFs},''
  \href{http://dx.doi.org/10.1140/epjc/s10052-015-3397-6}{{\em Eur. Phys. J.}
  {\bfseries C75} no.~5, (2015) 204},
\href{http://arxiv.org/abs/1412.3989}{{\ttfamily arXiv:1412.3989 [hep-ph]}}.

\bibitem{Ball:2014uwa}
R.~D. Ball {\em et~al.}, ``{Parton distributions for the LHC Run II},''
  \href{http://dx.doi.org/10.1007/JHEP04(2015)040}{{\em JHEP} {\bfseries 04}
  (2015) 040},
\href{http://arxiv.org/abs/1410.8849}{{\ttfamily arXiv:1410.8849 [hep-ph]}}.

\bibitem{Butterworth:2015oua}
J.~Butterworth {\em et~al.}, ``{PDF4LHC recommendations for LHC Run II},''
  \href{http://dx.doi.org/10.1088/0954-3899/43/2/023001}{{\em J. Phys.}
  {\bfseries G43} (2016) 023001},
\href{http://arxiv.org/abs/1510.03865}{{\ttfamily arXiv:1510.03865 [hep-ph]}}.

\bibitem{Accardi:2016ndt}
A.~Accardi {\em et~al.}, ``{A Critical Appraisal and Evaluation of Modern
  PDFs},'' \href{http://dx.doi.org/10.1140/epjc/s10052-016-4285-4}{{\em Eur.
  Phys. J.} {\bfseries C76} no.~8, (2016) 471},
\href{http://arxiv.org/abs/1603.08906}{{\ttfamily arXiv:1603.08906 [hep-ph]}}.

\bibitem{gjr}
M.~Gluck, P.~Jimenez-Delgado, and E.~Reya, ``{Dynamical parton distributions of
  the nucleon and very small-x physics},''
  \href{http://dx.doi.org/10.1140/epjc/s10052-007-0462-9}{{\em Eur.Phys.J.}
  {\bfseries C53} (2008) 355--366},
\href{http://arxiv.org/abs/0709.0614}{{\ttfamily arXiv:0709.0614 [hep-ph]}}.

\bibitem{Ball:2011mu}
R.~D. Ball, V.~Bertone, F.~Cerutti, L.~Del~Debbio, S.~Forte, {\em et~al.},
  ``{Impact of Heavy Quark Masses on Parton Distributions and LHC
  Phenomenology},''
  \href{http://dx.doi.org/10.1016/j.nuclphysb.2011.03.021}{{\em Nucl.Phys.}
  {\bfseries B849} (2011) 296--363},
\href{http://arxiv.org/abs/1101.1300}{{\ttfamily arXiv:1101.1300 [hep-ph]}}.

\bibitem{apfel}
V.~Bertone, S.~Carrazza, and J.~Rojo, ``{APFEL: A PDF evolution library with
  QED corrections},'' \href{http://dx.doi.org/10.1016/j.cpc.2014.03.007}{{\em
  Comput.Phys.Commun.} {\bfseries 185} (2014) 1647--1668},
\href{http://arxiv.org/abs/1310.1394}{{\ttfamily arXiv:1310.1394 [hep-ph]}}.

\bibitem{LHCbBeauty}
R.~Aaij {\em et~al.}, ``{Measurement of $B$ meson production cross-sections in
  proton-proton collisions at $\sqrt{s}$ = 7 TeV},''
  \href{http://dx.doi.org/10.1007/JHEP08(2013)117}{{\em JHEP} {\bfseries 1308}
  (2013) 117},
\href{http://arxiv.org/abs/1306.3663}{{\ttfamily arXiv:1306.3663}}.

\bibitem{Aaij:2015bpa}
R.~Aaij {\em et~al.}, ``{Measurements of prompt charm production cross-sections
  in $pp$ collisions at $ \sqrt{s}=13 $ TeV},''
  \href{http://dx.doi.org/10.1007/JHEP03(2016)159}{{\em JHEP} {\bfseries 03}
  (2016) 159},
\href{http://arxiv.org/abs/1510.01707}{{\ttfamily arXiv:1510.01707 [hep-ex]}}.

\bibitem{Aaij:2016jht}
R.~Aaij {\em et~al.}, ``{Measurements of prompt charm production cross-sections
  in $pp$ collisions at $\sqrt{s} = 5\,$TeV},''
\href{http://arxiv.org/abs/1610.02230}{{\ttfamily arXiv:1610.02230 [hep-ex]}}.

\bibitem{Aad:2015zix}
G.~Aad {\em et~al.}, ``{Measurement of $D^{*\pm}$, $D^\pm$ and $D_s^\pm$ meson
  production cross sections in $pp$ collisions at $\sqrt{s}=7$ TeV with the
  ATLAS detector},''
  \href{http://dx.doi.org/10.1016/j.nuclphysb.2016.04.032}{{\em Nucl. Phys.}
  {\bfseries B907} (2016) 717--763},
\href{http://arxiv.org/abs/1512.02913}{{\ttfamily arXiv:1512.02913 [hep-ex]}}.

\bibitem{ALICE:2011aa}
B.~Abelev {\em et~al.}, ``{Measurement of charm production at central rapidity
  in proton-proton collisions at $\sqrt{s} = 7$ TeV},''
  \href{http://dx.doi.org/10.1007/JHEP01(2012)128}{{\em JHEP} {\bfseries 1201}
  (2012) 128},
\href{http://arxiv.org/abs/1111.1553}{{\ttfamily arXiv:1111.1553 [hep-ex]}}.

\bibitem{Abelev:2012tca}
B.~Abelev {\em et~al.}, ``{$D_{s}^+$ meson production at central rapidity in
  proton--proton collisions at $\sqrt{s}=7$ TeV},''
  \href{http://dx.doi.org/10.1016/j.physletb.2012.10.049}{{\em Phys. Lett.}
  {\bfseries B718} (2012) 279--294},
\href{http://arxiv.org/abs/1208.1948}{{\ttfamily arXiv:1208.1948 [hep-ex]}}.

\bibitem{Abelev:2012vra}
B.~Abelev {\em et~al.}, ``{Measurement of charm production at central rapidity
  in proton-proton collisions at $\sqrt{s}=2.76$ TeV},''
  \href{http://dx.doi.org/10.1007/JHEP07(2012)191}{{\em JHEP} {\bfseries 07}
  (2012) 191},
\href{http://arxiv.org/abs/1205.4007}{{\ttfamily arXiv:1205.4007 [hep-ex]}}.

\bibitem{Adam:2016ich}
J.~Adam {\em et~al.}, ``{$D$-meson production in $p$-Pb collisions at
  $\sqrt{s_{\rm NN}}=$5.02 TeV and in $pp$ collisions at $\sqrt{s}=$7 TeV},''
  \href{http://dx.doi.org/10.1103/PhysRevC.94.054908}{{\em Phys. Rev.}
  {\bfseries C94} no.~5, (2016) 054908},
\href{http://arxiv.org/abs/1605.07569}{{\ttfamily arXiv:1605.07569 [nucl-ex]}}.

\bibitem{Evans:2008zzb}
L.~R. Evans and P.~Bryant, ``{LHC Machine},''
  \href{http://dx.doi.org/10.1088/1748-0221/3/08/S08001}{{\em J. Instrum.}
  {\bfseries 3} (2008) S08001. 164 p}.
This report is an abridged version of the LHC Design Report (CERN-2004-003).

\bibitem{Alves:2008zz}
J.~Alves, A.~Augusto {\em et~al.}, ``{The LHCb Detector at the LHC},''
  \href{http://dx.doi.org/10.1088/1748-0221/3/08/S08005}{{\em J. Instrum.}
  {\bfseries 3} no.~LHCb-DP-2008-001. CERN-LHCb-DP-2008-001, (2008) S08005}.
Also published by CERN Geneva in 2010.

\bibitem{Adinolfi:2012qfa}
M.~Adinolfi {\em et~al.}, ``{Performance of the LHCb RICH detector at the
  LHC},'' \href{http://dx.doi.org/10.1140/epjc/s10052-013-2431-9}{{\em
  Eur.Phys.J.} {\bfseries C73} (2013) 2431},
\href{http://arxiv.org/abs/1211.6759}{{\ttfamily arXiv:1211.6759
  [physics.ins-det]}}.

\bibitem{frag03}
M.~Cacciari and P.~Nason, ``{Charm cross-sections for the Tevatron Run II},''
  \href{http://dx.doi.org/10.1088/1126-6708/2003/09/006}{{\em JHEP} {\bfseries
  0309} (2003) 006},
\href{http://arxiv.org/abs/hep-ph/0306212}{{\ttfamily arXiv:hep-ph/0306212
  [hep-ph]}}.

\bibitem{Kniehl:2004fy}
B.~A. Kniehl, G.~Kramer, I.~Schienbein, and H.~Spiesberger, ``{Inclusive
  $D^{*\pm}$ production in p anti-p collisions with massive charm quarks},''
  \href{http://dx.doi.org/10.1103/PhysRevD.71.014018}{{\em Phys. Rev.}
  {\bfseries D71} (2005) 014018},
\href{http://arxiv.org/abs/hep-ph/0410289}{{\ttfamily arXiv:hep-ph/0410289
  [hep-ph]}}.

\bibitem{Kniehl:2005de}
B.~A. Kniehl and G.~Kramer, ``{$D^0$, $D^{+}$, $D^{+}_{s}$, and
  $\Lambda^{+}_{c}$ fragmentation functions from CERN LEP1},''
  \href{http://dx.doi.org/10.1103/PhysRevD.71.094013}{{\em Phys. Rev.}
  {\bfseries D71} (2005) 094013},
\href{http://arxiv.org/abs/hep-ph/0504058}{{\ttfamily arXiv:hep-ph/0504058
  [hep-ph]}}.

\bibitem{Kniehl:2005ej}
B.~A. Kniehl, G.~Kramer, I.~Schienbein, and H.~Spiesberger, ``{Reconciling open
  charm production at the Fermilab Tevatron with QCD},''
  \href{http://dx.doi.org/10.1103/PhysRevLett.96.012001}{{\em Phys. Rev. Lett.}
  {\bfseries 96} (2006) 012001},
\href{http://arxiv.org/abs/hep-ph/0508129}{{\ttfamily arXiv:hep-ph/0508129
  [hep-ph]}}.

\bibitem{Kneesch:2007ey}
T.~Kneesch, B.~A. Kniehl, G.~Kramer, and I.~Schienbein, ``{Charmed-meson
  fragmentation functions with finite-mass corrections},''
  \href{http://dx.doi.org/10.1016/j.nuclphysb.2008.02.015}{{\em Nucl. Phys.}
  {\bfseries B799} (2008) 34--59},
\href{http://arxiv.org/abs/0712.0481}{{\ttfamily arXiv:0712.0481 [hep-ph]}}.

\bibitem{Kniehl:2009ar}
B.~A. Kniehl, G.~Kramer, I.~Schienbein, and H.~Spiesberger, ``{Open charm
  hadroproduction and the charm content of the proton},''
  \href{http://dx.doi.org/10.1103/PhysRevD.79.094009}{{\em Phys. Rev.}
  {\bfseries D79} (2009) 094009},
\href{http://arxiv.org/abs/0901.4130}{{\ttfamily arXiv:0901.4130 [hep-ph]}}.

\bibitem{Kniehl:2012ti}
B.~A. Kniehl, G.~Kramer, I.~Schienbein, and H.~Spiesberger, ``{Inclusive
  Charmed-Meson Production at the CERN LHC},''
  \href{http://dx.doi.org/10.1140/epjc/s10052-012-2082-2}{{\em Eur. Phys. J.}
  {\bfseries C72} (2012) 2082},
\href{http://arxiv.org/abs/1202.0439}{{\ttfamily arXiv:1202.0439 [hep-ph]}}.

\bibitem{frag97}
M.~Cacciari and M.~Greco, ``{$D^{*}$ production from $e^{+} e^{-}$ to $e p$
  collisions in NLO QCD},''
  \href{http://dx.doi.org/10.1103/PhysRevD.55.7134}{{\em Phys.Rev.} {\bfseries
  D55} (1997) 7134--7143},
\href{http://arxiv.org/abs/hep-ph/9702389}{{\ttfamily arXiv:hep-ph/9702389
  [hep-ph]}}.

\bibitem{Lisovyi:2015uqa}
M.~Lisovyi, A.~Verbytskyi, and O.~Zenaiev, ``{Combined analysis of charm-quark
  fragmentation-fraction measurements},''
  \href{http://dx.doi.org/10.1140/epjc/s10052-016-4246-y}{{\em Eur. Phys. J.}
  {\bfseries C76} no.~7, (2016) 397},
\href{http://arxiv.org/abs/1509.01061}{{\ttfamily arXiv:1509.01061 [hep-ex]}}.

\bibitem{difftop}
``The fortran-based package, which allows the user to calculate the
  differential and total cross section for heavy-quark pair production at
  hadron colliders in one-particle inclusive kinematics.''
\newblock \url{http://difftop.hepforge.org/}.

\bibitem{Garzelli:2015psa}
M.~V. Garzelli, S.~Moch, and G.~Sigl, ``{Lepton fluxes from atmospheric charm
  revisited},'' \href{http://dx.doi.org/10.1007/JHEP10(2015)115}{{\em JHEP}
  {\bfseries 10} (2015) 115},
\href{http://arxiv.org/abs/1507.01570}{{\ttfamily arXiv:1507.01570 [hep-ph]}}.

\bibitem{Bhattacharya:2016jce}
A.~Bhattacharya, R.~Enberg, Y.~S. Jeong, C.~S. Kim, M.~H. Reno, I.~Sarcevic,
  and A.~Stasto, ``{Prompt atmospheric neutrino fluxes: perturbative QCD models
  and nuclear effects},'' \href{http://dx.doi.org/10.1007/JHEP11(2016)167}{{\em
  JHEP} {\bfseries 11} (2016) 167},
\href{http://arxiv.org/abs/1607.00193}{{\ttfamily arXiv:1607.00193 [hep-ph]}}.

\bibitem{Gauld:2015kvh}
R.~Gauld, J.~Rojo, L.~Rottoli, S.~Sarkar, and J.~Talbert, ``{The prompt
  atmospheric neutrino flux in the light of LHCb},''
  \href{http://dx.doi.org/10.1007/JHEP02(2016)130}{{\em JHEP} {\bfseries 02}
  (2016) 130},
\href{http://arxiv.org/abs/1511.06346}{{\ttfamily arXiv:1511.06346 [hep-ph]}}.

\bibitem{Garzelli:2016xmx}
M.~V. Garzelli, S.~Moch, O.~Zenaiev, A.~Cooper-Sarkar, A.~Geiser, K.~Lipka,
  R.~Placakyte, and G.~Sigl, ``{Prompt neutrino fluxes in the atmosphere with
  PROSA parton distribution functions},''
\href{http://arxiv.org/abs/1611.03815}{{\ttfamily arXiv:1611.03815 [hep-ph]}}.

\bibitem{Aaron:2009bp}
F.~Aaron {\em et~al.}, ``{Measurement of the Inclusive $ep$ Scattering Cross
  Section at Low $Q^2$ and x at HERA},''
  \href{http://dx.doi.org/10.1140/epjc/s10052-009-1128-6}{{\em Eur.Phys.J.}
  {\bfseries C63} (2009) 625--678},
\href{http://arxiv.org/abs/0904.0929}{{\ttfamily arXiv:0904.0929 [hep-ex]}}.

\bibitem{Belov:2013oda}
P.~Belov, {\em {Combination of the H1 and ZEUS inclusive cross-section
  measurements at proton beam energies of 460 GeV and 575 GeV and tests of low
  Bjorken-$x$ phenomenological models}}.
\newblock PhD thesis, Hamburg University,
May, 2013.
\newblock

\bibitem{Nicholass:2008zz}
D.~Nicholass, \href{http://dx.doi.org/10.3204/DESY-THESIS-2008-046}{{\em {The
  study of $D^{\pm}$ and $D^0$ meson production in deep inelastic scattering at
  HERA II with the ZEUS detector}}}.
\newblock PhD thesis, University Coll. London,
August, 2008.
\newblock

\bibitem{Radescu:2010zz}
V.~Radescu, ``{Combination and QCD analysis of the HERA inclusive cross
  sections},'' {\em PoS} {\bfseries ICHEP2010} (2010) 168.
\url{http://www.desy.de/h1zeus/combined_results/index.php?do=proton_structure}.

\bibitem{Ball:2011uy}
R.~D. Ball {\em et~al.}, ``{Unbiased global determination of parton
  distributions and their uncertainties at NNLO and at LO},''
  \href{http://dx.doi.org/10.1016/j.nuclphysb.2011.09.024}{{\em Nucl.Phys.}
  {\bfseries B855} (2012) 153--221},
\href{http://arxiv.org/abs/1107.2652}{{\ttfamily arXiv:1107.2652 [hep-ph]}}.

\bibitem{ct10f3}
H.-L. Lai, M.~Guzzi, J.~Huston, Z.~Li, P.~M. Nadolsky, {\em et~al.}, ``{New
  parton distributions for collider physics},''
  \href{http://dx.doi.org/10.1103/PhysRevD.82.074024}{{\em Phys.Rev.}
  {\bfseries D82} (2010) 074024},
\href{http://arxiv.org/abs/1007.2241}{{\ttfamily arXiv:1007.2241 [hep-ph]}}.

\bibitem{ct12nnlo}
P.~Nadolsky, J.~Gao, M.~Guzzi, J.~Huston, H.-L. Lai, {\em et~al.}, ``{Progress
  in CTEQ-TEA PDF Analysis},''
\href{http://arxiv.org/abs/1206.3321}{{\ttfamily arXiv:1206.3321 [hep-ph]}}.

\bibitem{fastnlo_Kluge:2006xs}
T.~Kluge, K.~Rabbertz, and M.~Wobisch, ``{FastNLO: Fast pQCD calculations for
  PDF fits},''
\href{http://arxiv.org/abs/hep-ph/0609285}{{\ttfamily arXiv:hep-ph/0609285
  [hep-ph]}}.

\bibitem{fastnlo_Wobisch:2011ij}
M.~Wobisch, D.~Britzger, T.~Kluge, K.~Rabbertz, and F.~Stober, ``{Theory-Data
  Comparisons for Jet Measurements in Hadron-Induced Processes},''
\href{http://arxiv.org/abs/1109.1310}{{\ttfamily arXiv:1109.1310 [hep-ph]}}.

\bibitem{fastnlo_Britzger:2012bs}
D.~Britzger, K.~Rabbertz, F.~Stober, and M.~Wobisch, ``{New features in version
  2 of the fastNLO project},''
\href{http://arxiv.org/abs/1208.3641}{{\ttfamily arXiv:1208.3641 [hep-ph]}}.

\bibitem{applgrid_Carli:2010rw}
T.~Carli, D.~Clements, A.~Cooper-Sarkar, C.~Gwenlan, G.~P. Salam, {\em et~al.},
  ``{A posteriori inclusion of parton density functions in NLO QCD final-state
  calculations at hadron colliders: The APPLGRID Project},''
  \href{http://dx.doi.org/10.1140/epjc/s10052-010-1255-0}{{\em Eur.Phys.J.}
  {\bfseries C66} (2010) 503--524},
\href{http://arxiv.org/abs/0911.2985}{{\ttfamily arXiv:0911.2985 [hep-ph]}}.

\bibitem{amcfast_Bertone:2014zva}
V.~Bertone, R.~Frederix, S.~Frixione, J.~Rojo, and M.~Sutton, ``{aMCfast:
  automation of fast NLO computations for PDF fits},''
  \href{http://dx.doi.org/10.1007/JHEP08(2014)166}{{\em JHEP} {\bfseries 1408}
  (2014) 166},
\href{http://arxiv.org/abs/1406.7693}{{\ttfamily arXiv:1406.7693 [hep-ph]}}.

\bibitem{vegas_Lepage:1977sw}
G.~P. Lepage, ``{A New Algorithm for Adaptive Multidimensional Integration},''
\href{http://dx.doi.org/10.1016/0021-9991(78)90004-9}{{\em J.Comput.Phys.}
  {\bfseries 27} (1978) 192}.

\bibitem{Braaten:1994bz}
E.~Braaten, K.-m. Cheung, S.~Fleming, and T.~C. Yuan, ``{Perturbative QCD
  fragmentation functions as a model for heavy quark fragmentation},''
  \href{http://dx.doi.org/10.1103/PhysRevD.51.4819}{{\em Phys.Rev.} {\bfseries
  D51} (1995) 4819--4829},
\href{http://arxiv.org/abs/hep-ph/9409316}{{\ttfamily arXiv:hep-ph/9409316
  [hep-ph]}}.

\end{thebibliography}\endgroup
\bibliographystyle{refs/utphys}  
\addcontentsline{toc}{section}{References}


\begin{appendix}
\appendix
%


\clearpage
\section{Measurement of \Dch production: additional information}
\label{sec:app:dch}

In this Appendix additional information on the measurement of \Dch production (see Section~\ref{sec:dch}) is provided. 

\begin{figure}[htbp]
  \centering
  \includegraphics[width=0.495\figwidth,trim=0 0mm 1mm 0,clip=true]{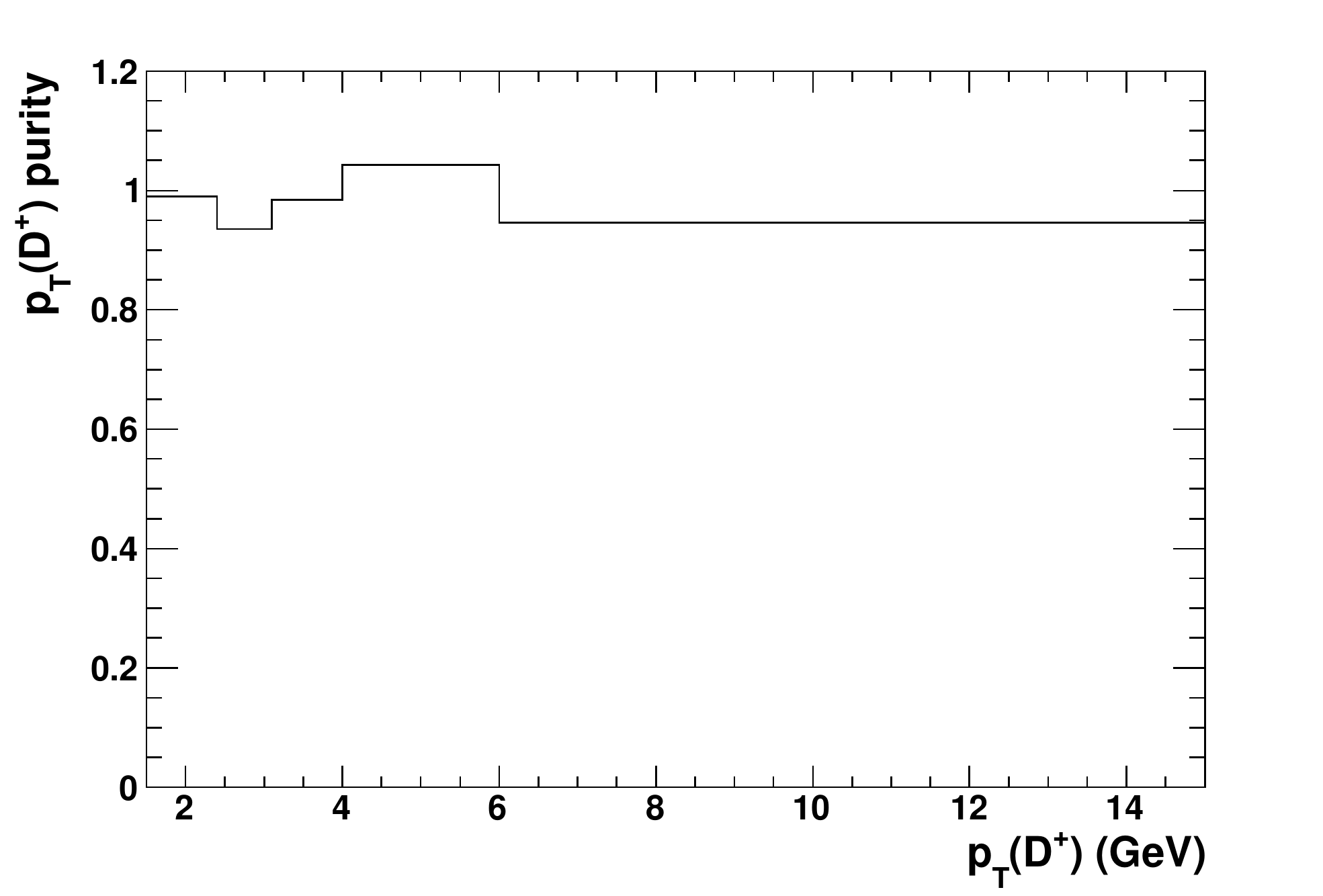}
  \includegraphics[width=0.495\figwidth,trim=0 0mm 1mm 0,clip=true]{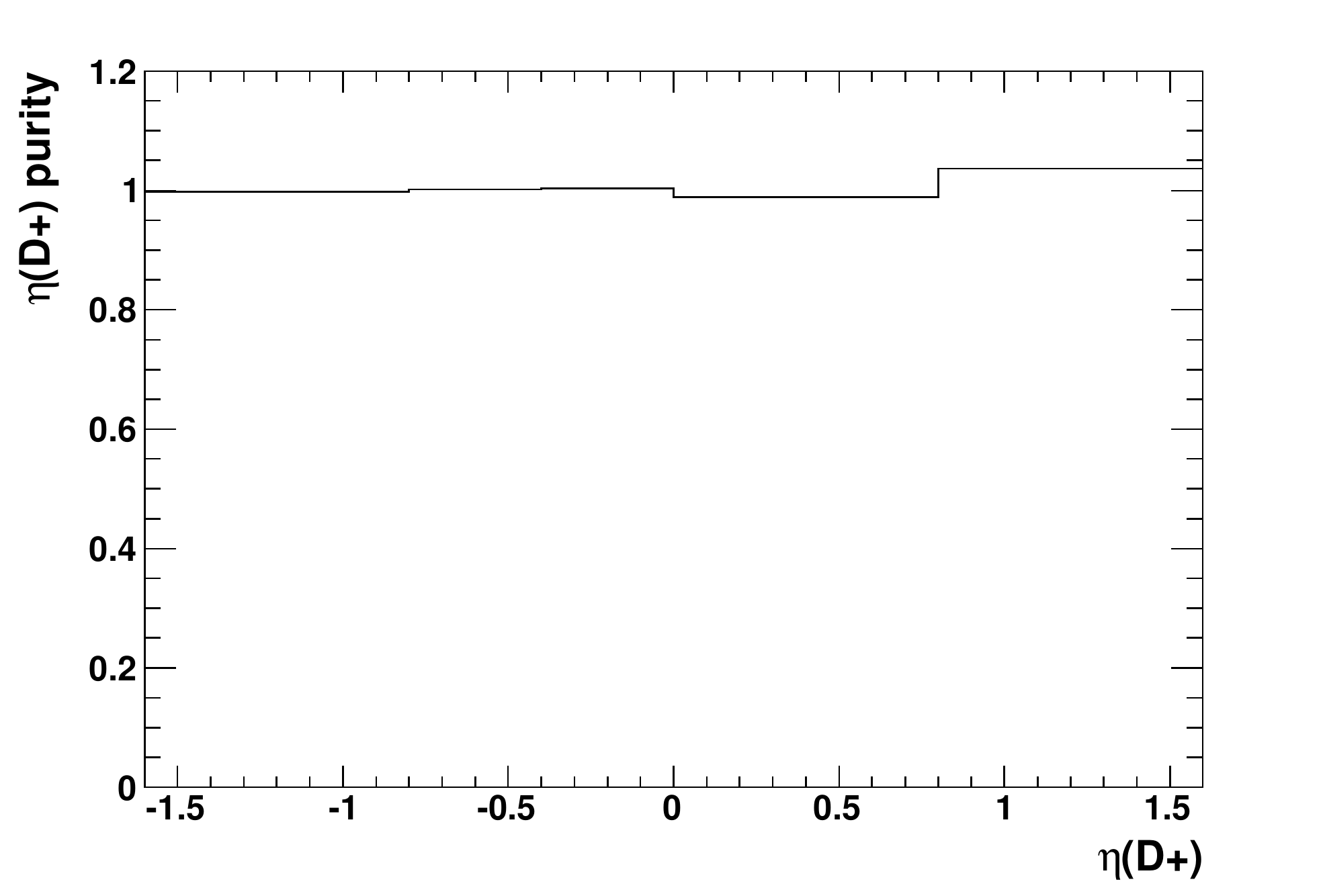}
  \includegraphics[width=0.495\figwidth,trim=0 0mm 1mm 0,clip=true]{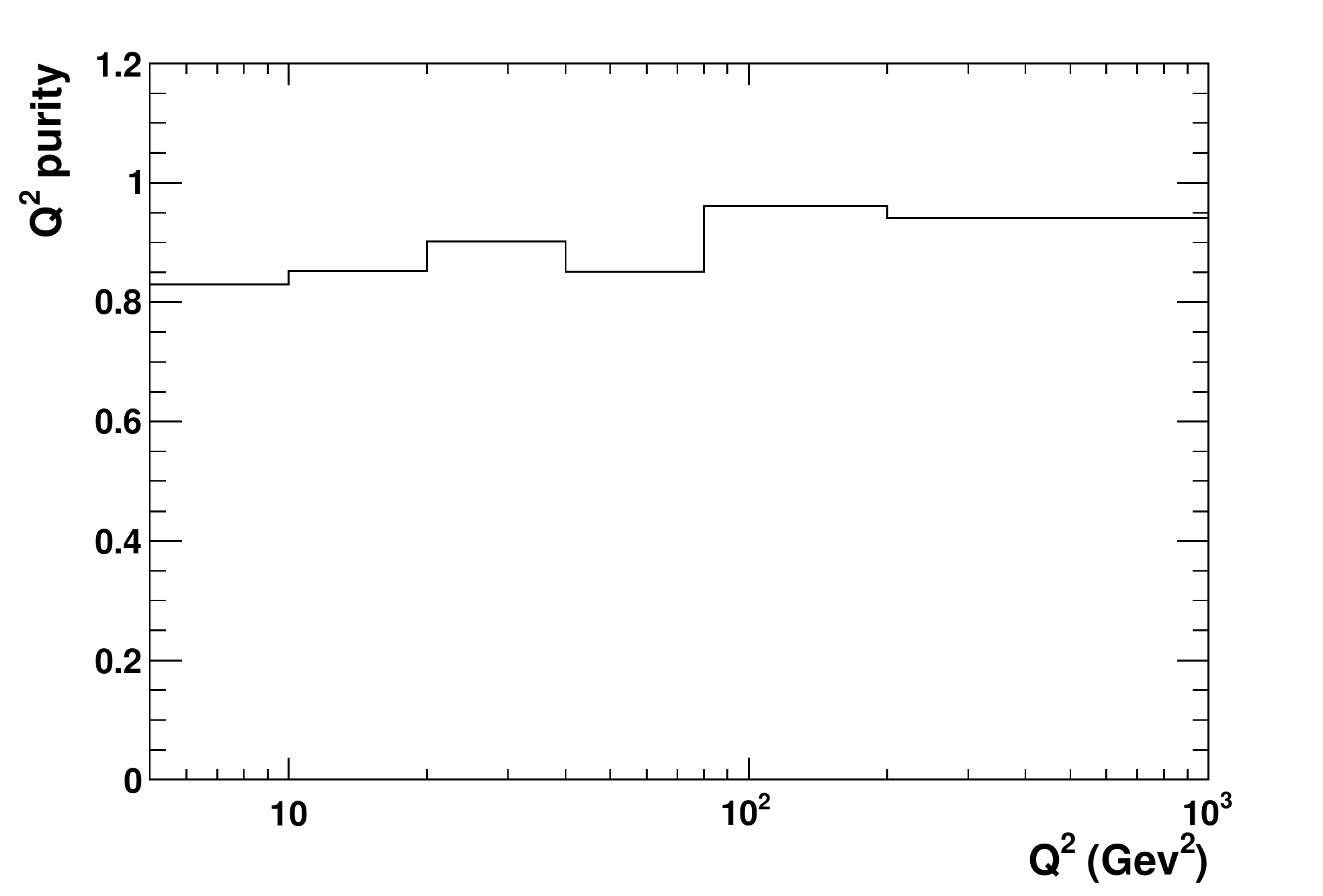}
  \includegraphics[width=0.495\figwidth,trim=0 0mm 1mm 0,clip=true]{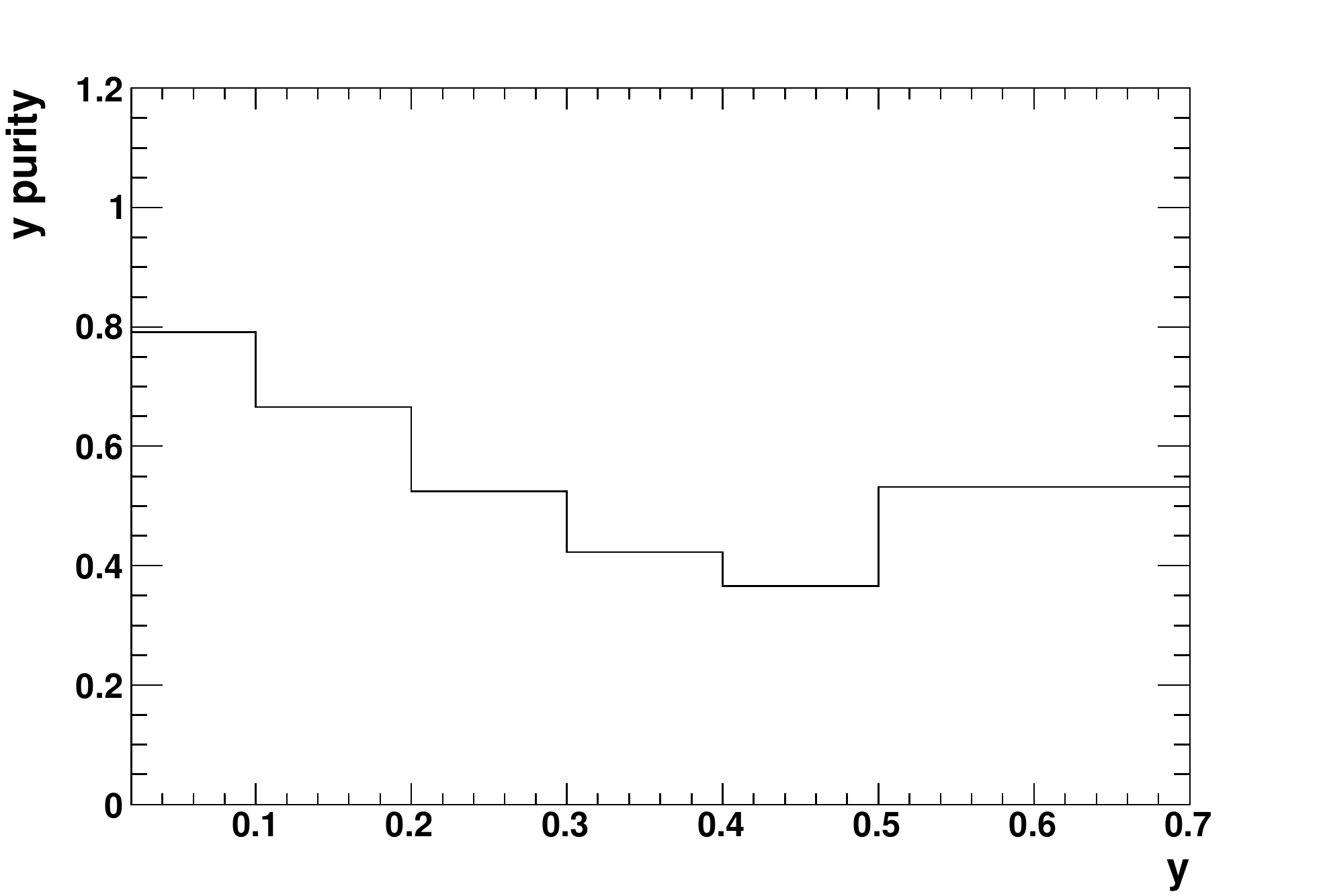}
  \caption[Purity as function of $p_T(\Dch)$, $\eta(\Dch)$, $Q^2$ and $y$]
	{Purity as a function of $p_T(\Dch)$ (top left), $\eta(\Dch)$ (top right), $Q^2$ (bottom left) and $y$ (bottom right).}
  \label{fig:dch:pur}
\end{figure}

Fig.~\ref{fig:dch:pur} shows purity as a function of $p_T(\Dch)$, $\eta(\Dch)$, $Q^2$ and $y$. 

\begin{figure}[htbp]
  \centering
  \includegraphics[width=0.495\figwidth,trim=0 0mm 1mm 0,clip=true]{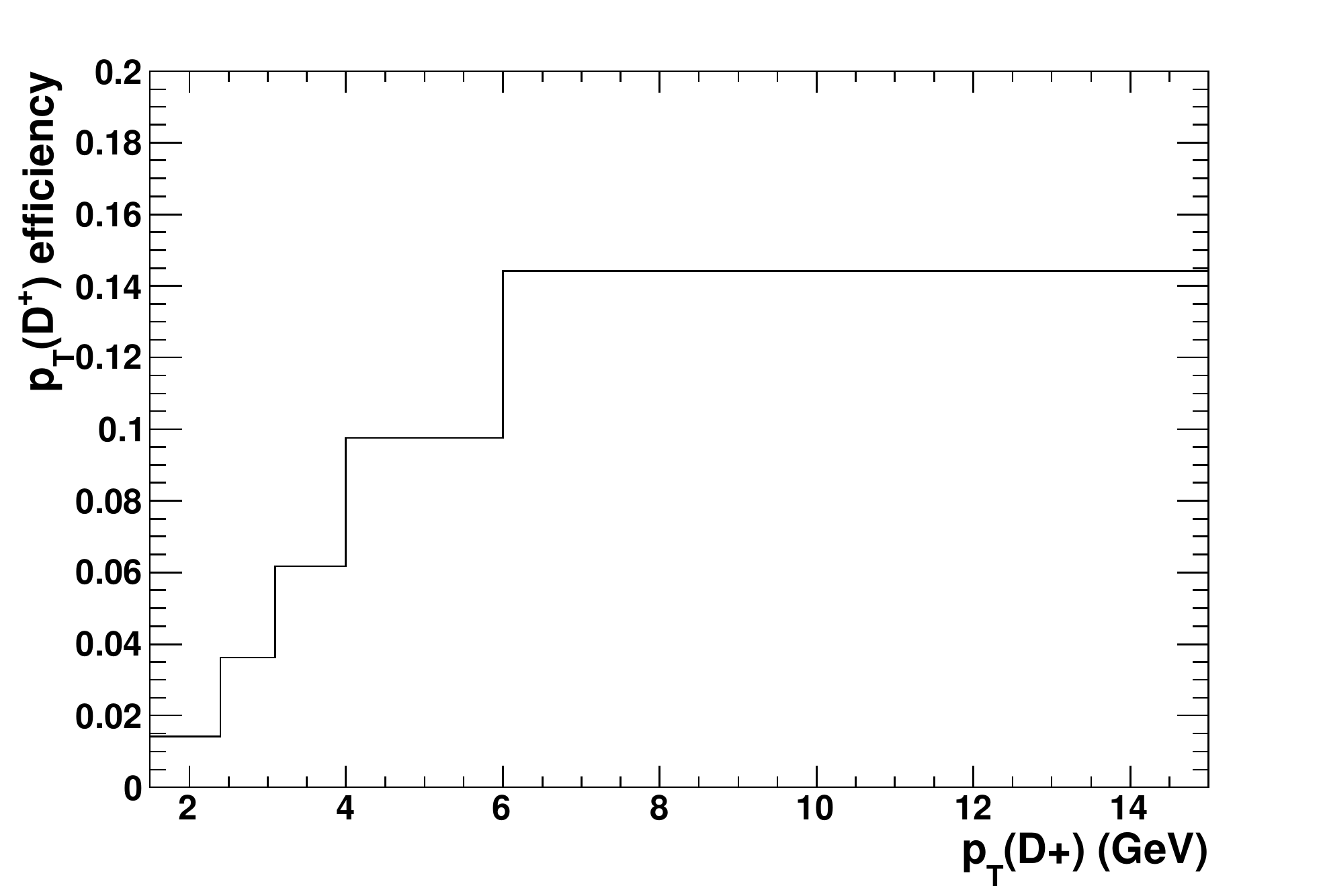}
  \includegraphics[width=0.495\figwidth,trim=0 0mm 1mm 0,clip=true]{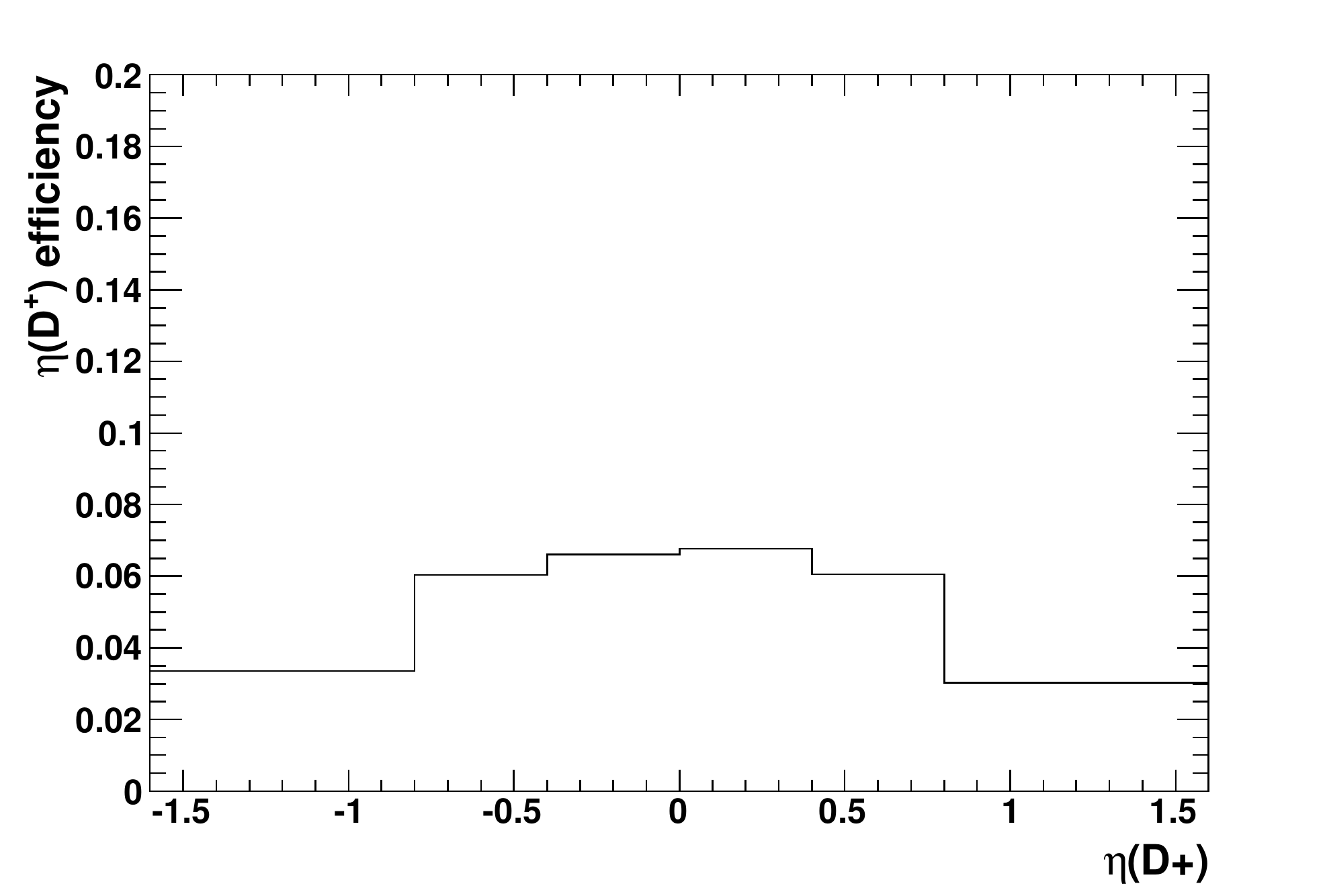}
  \includegraphics[width=0.495\figwidth,trim=0 0mm 1mm 0,clip=true]{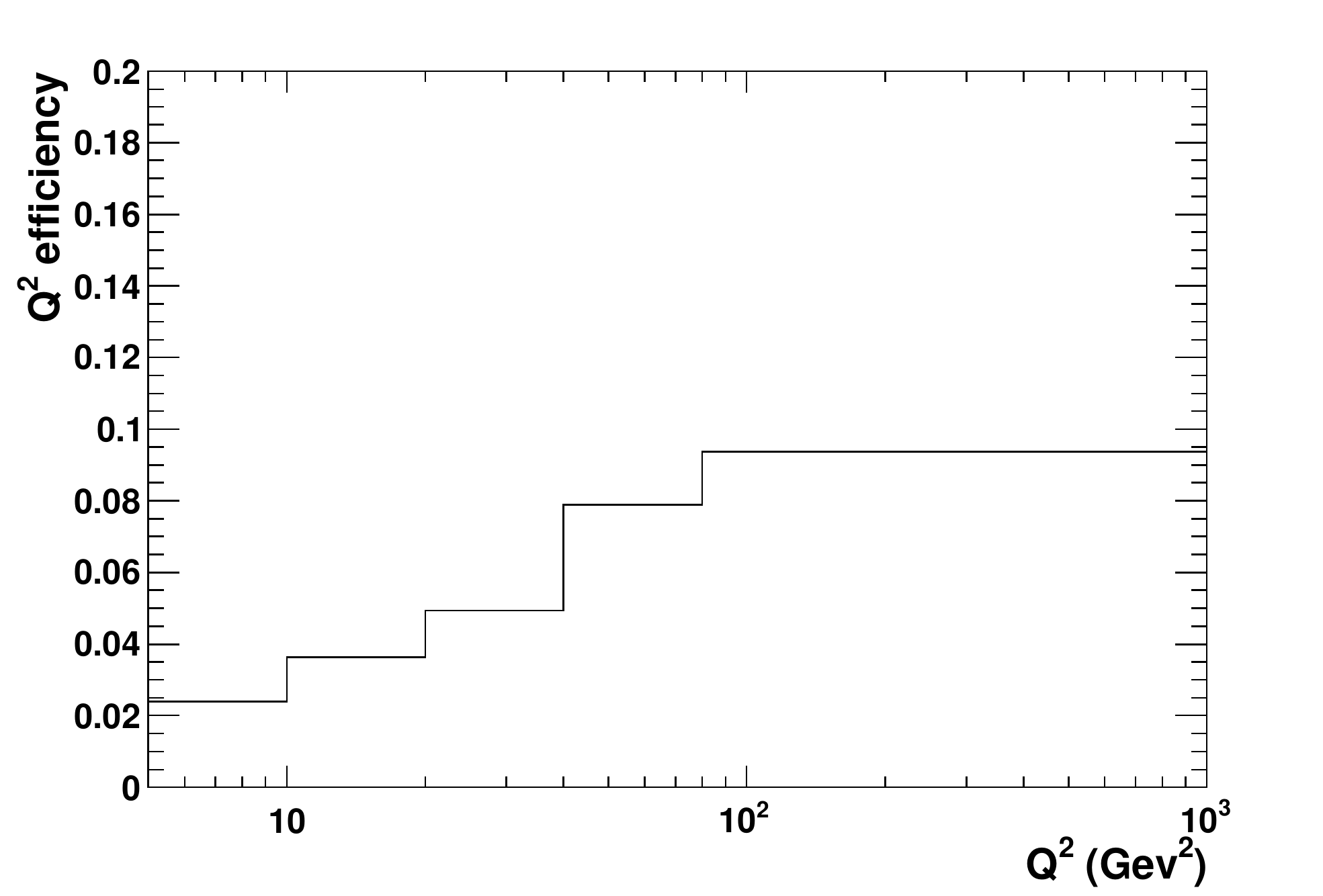}
  \includegraphics[width=0.495\figwidth,trim=0 0mm 1mm 0,clip=true]{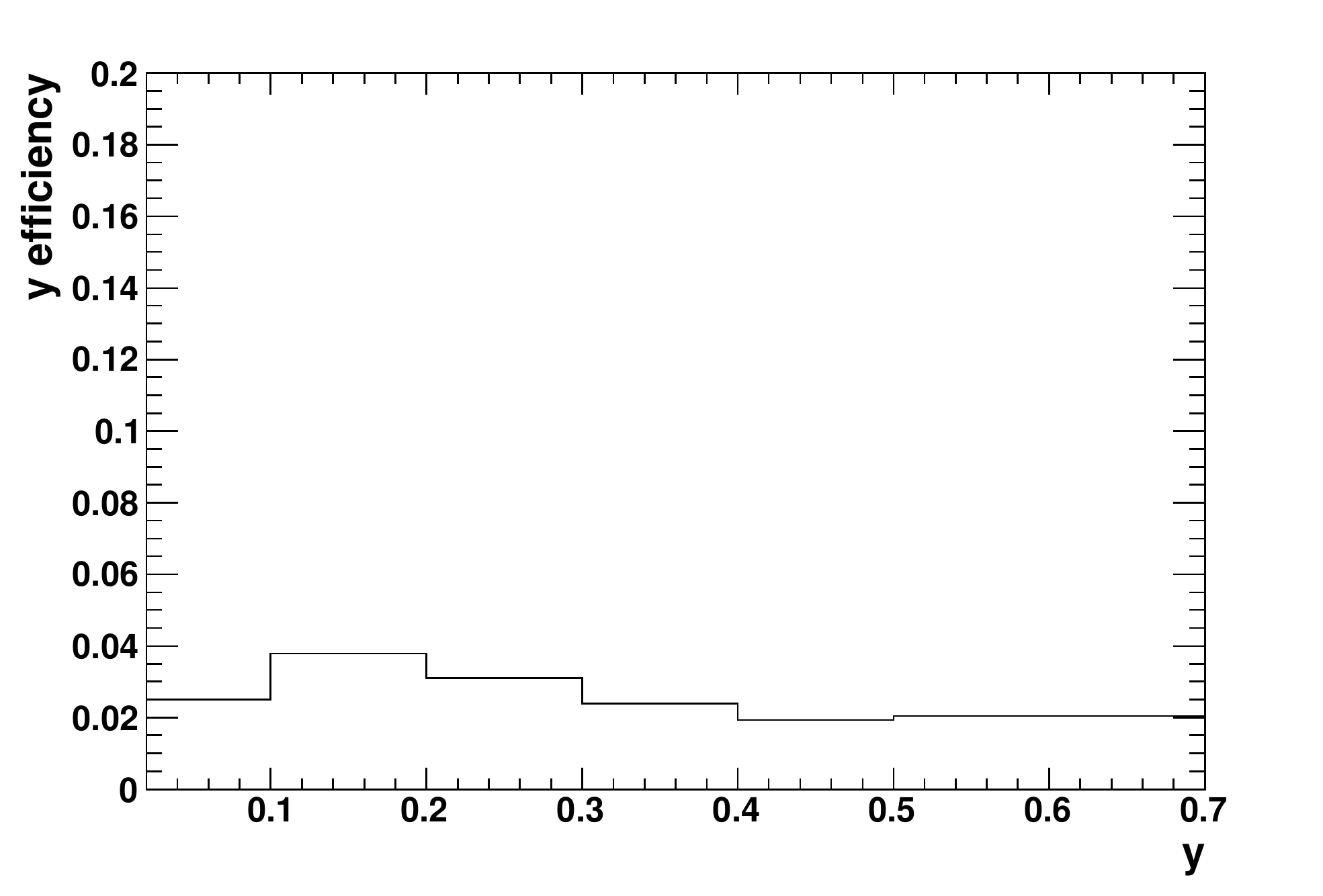}
  \caption[Efficiency as function of $p_T(\Dch)$, $\eta(\Dch)$, $Q^2$ and $y$]
	{Efficiency as a function of $p_T(\Dch)$ (top left), $\eta(\Dch)$ (top right), $Q^2$ (bottom left) and $y$ (bottom right).}
  \label{fig:dch:eff}
\end{figure}

Fig.~\ref{fig:dch:eff} shows efficiency as a function of $p_T(\Dch)$, $\eta(\Dch)$, $Q^2$ and $y$. 

\begin{figure*}[htbp]
  \centering
  \begin{minipage}[t]{0.33\textwidth}
  \includegraphics[width=1.0\textwidth,trim=2mm 0mm 12mm 0,clip=true]{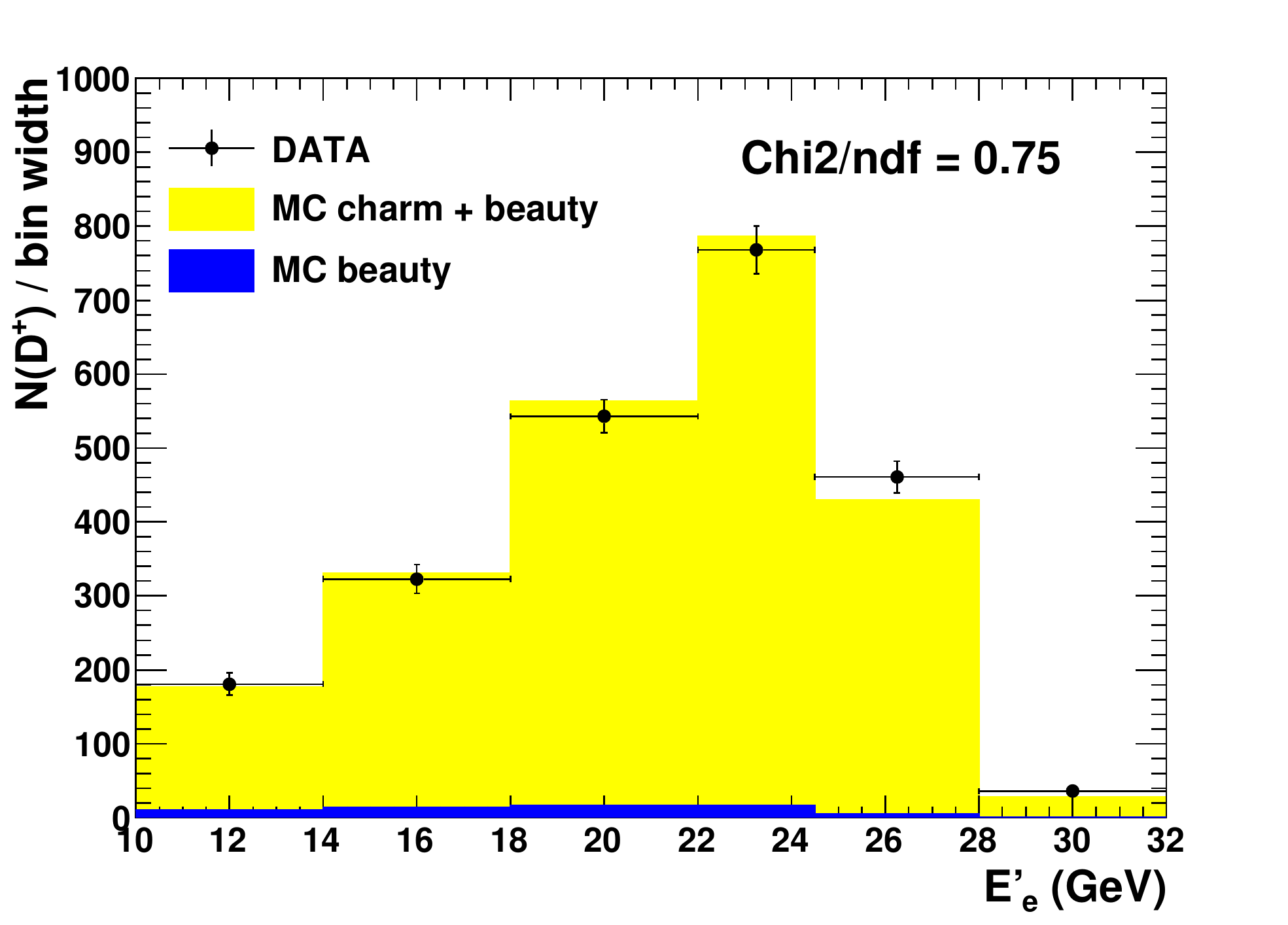}
  \put(-25,65){(a)}\\
  \includegraphics[width=1.0\textwidth,trim=2mm 0mm 12mm 0,clip=true]{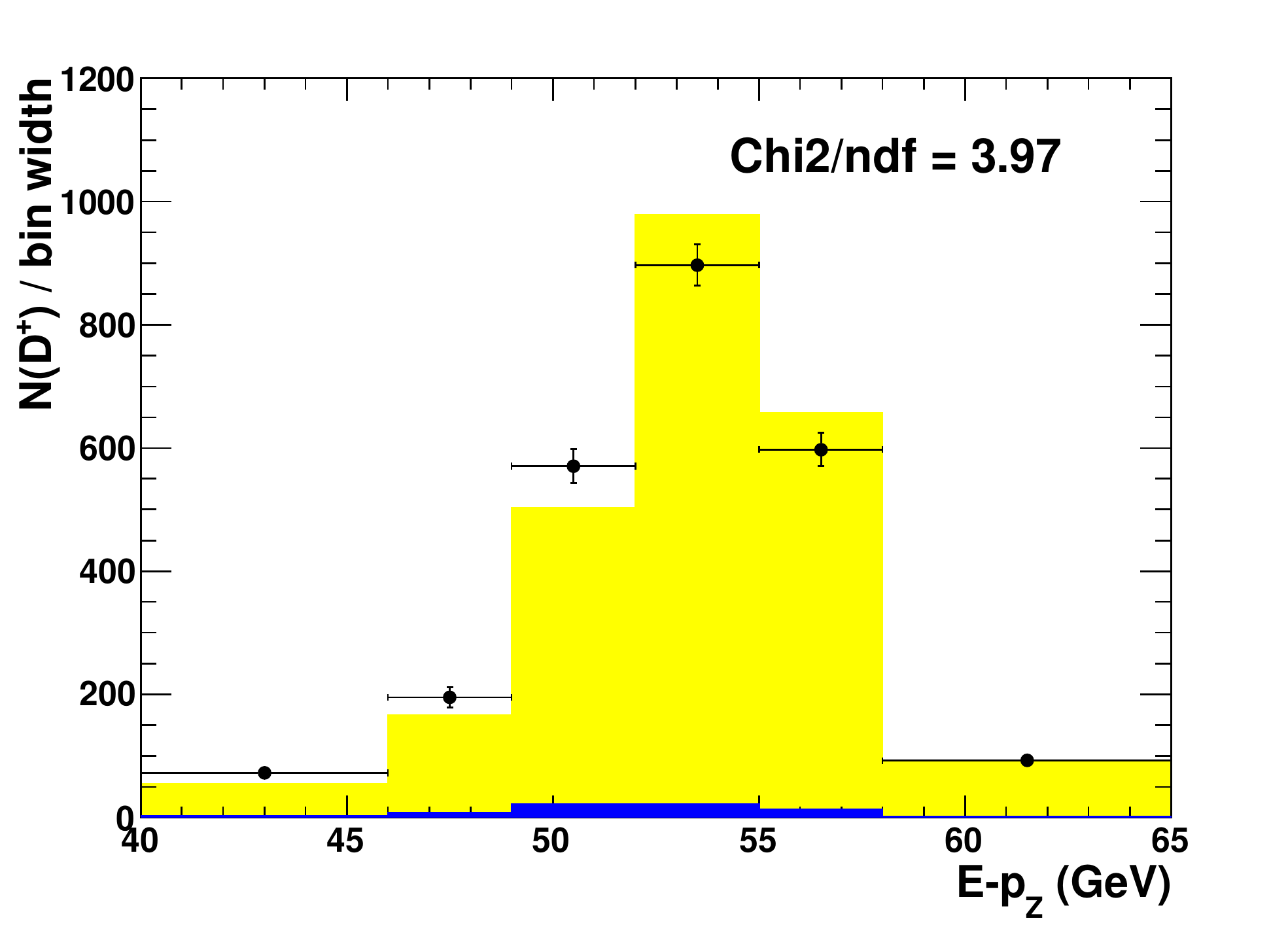}
  \put(-25,65){(d)}\\
  \end{minipage}
  \begin{minipage}[t]{0.33\textwidth}
  \includegraphics[width=1.0\textwidth,trim=2mm 0mm 12mm 0,clip=true]{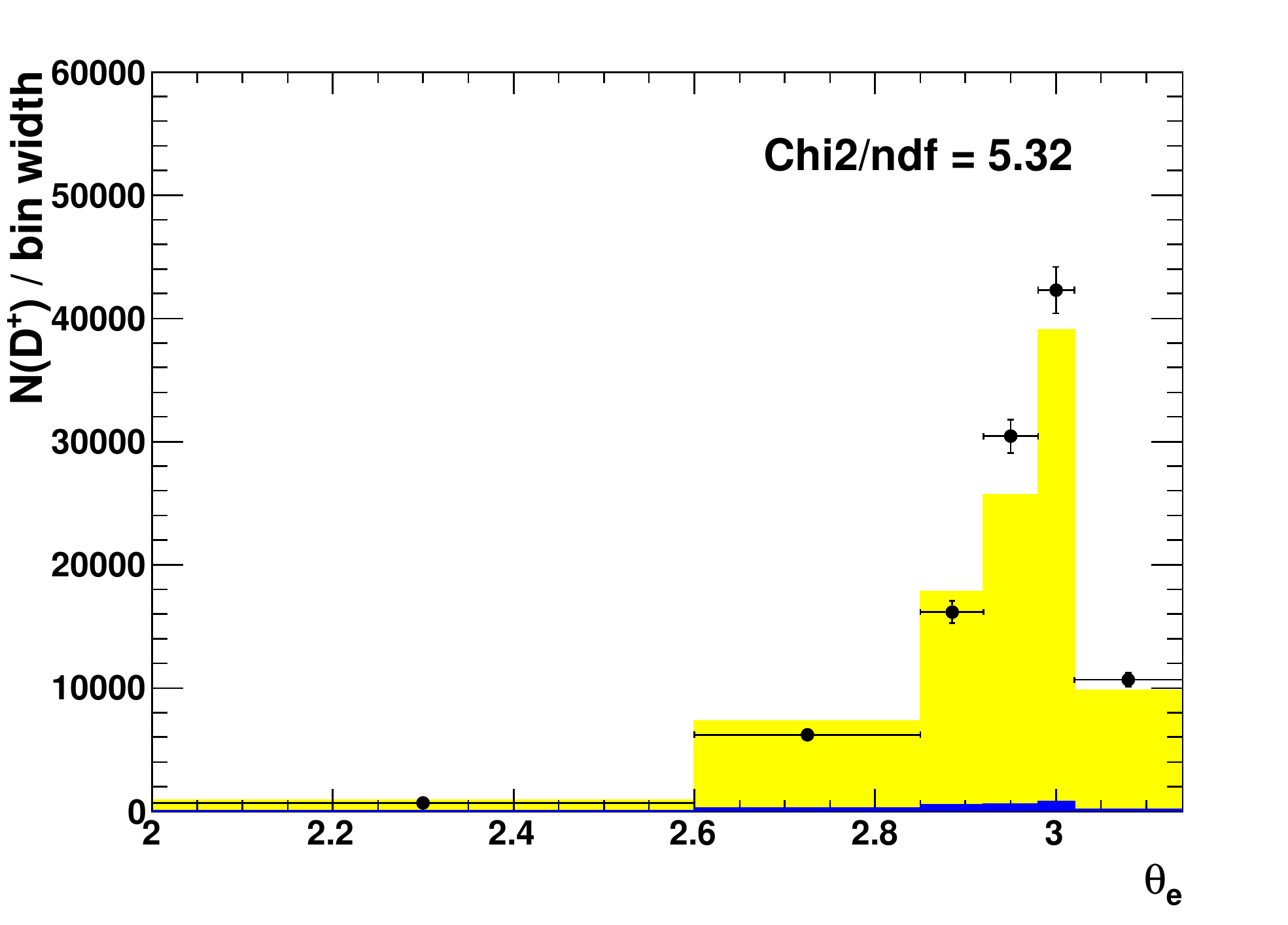}
  \put(-135,65){(b)}\\
  \includegraphics[width=1.0\textwidth,trim=2mm 0mm 12mm 0,clip=true]{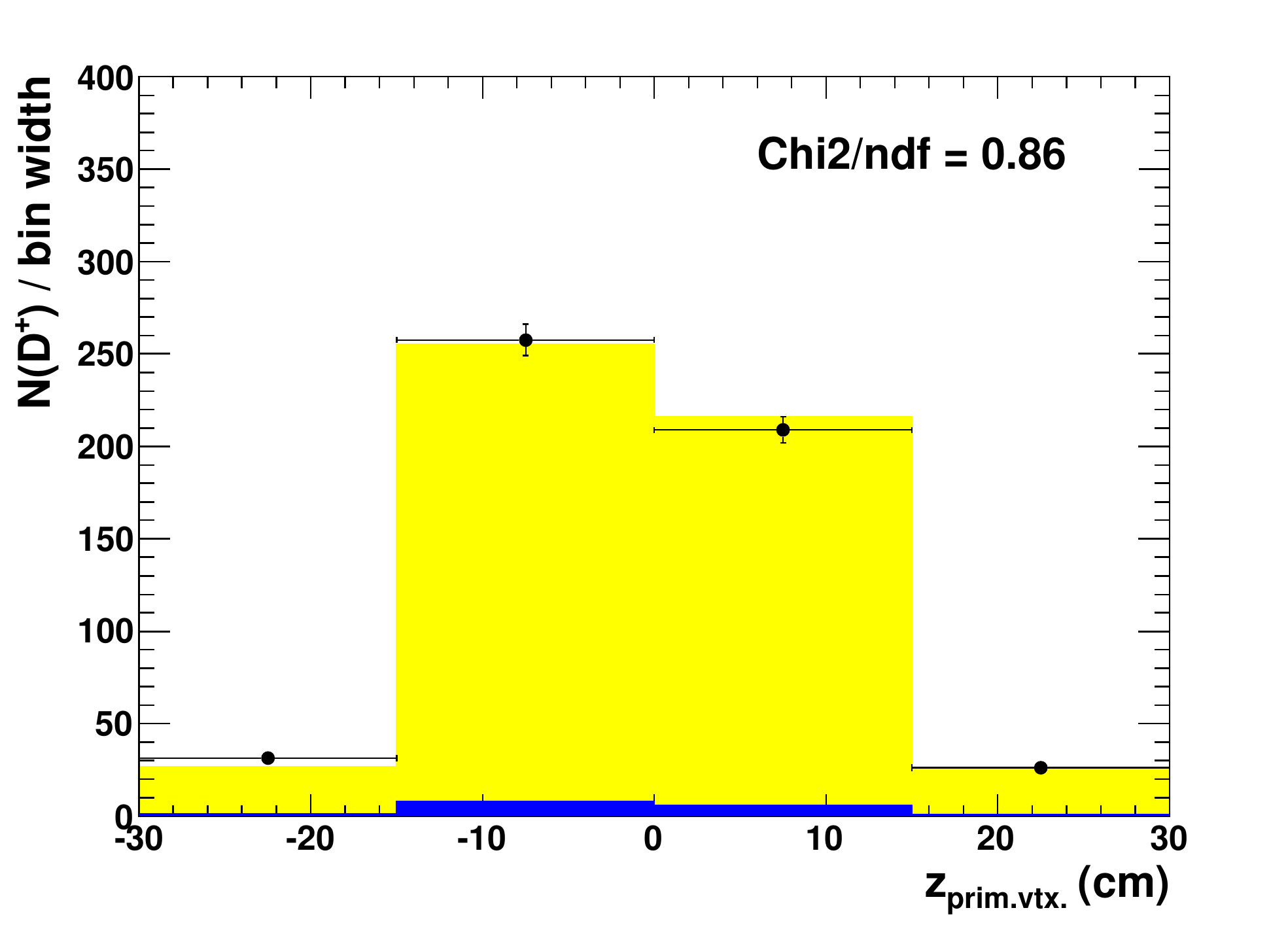}
  \put(-25,65){(e)}\\
  \end{minipage}
  \begin{minipage}[t]{0.33\textwidth}
  \includegraphics[width=1.0\textwidth,trim=2mm 0mm 12mm 0,clip=true]{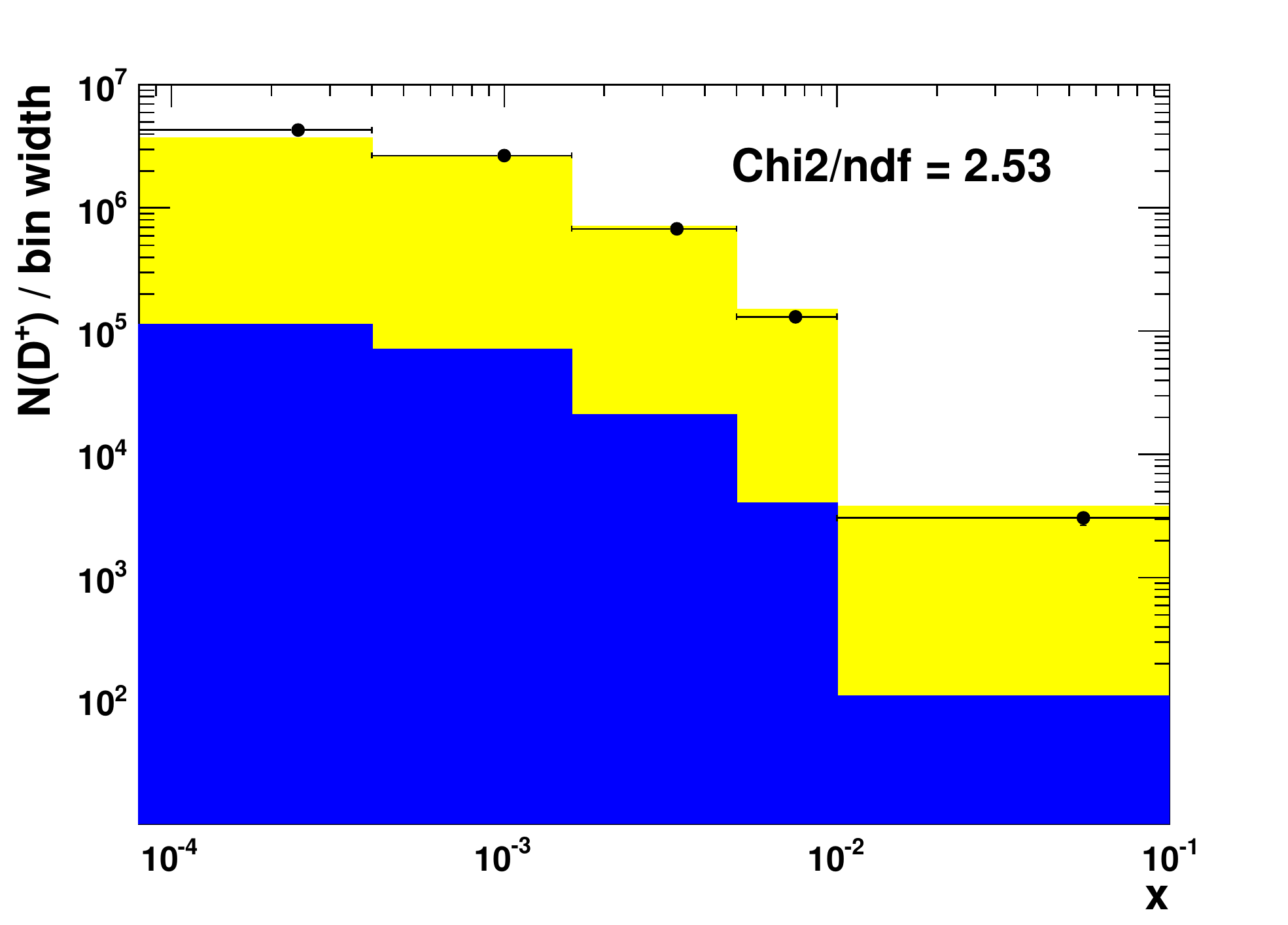}
  \put(-25,65){(c)}
  \caption[Additional control plots]
	{Control plots for $E_e^{\prime}$ (a), polar angle of the scattered electron, $\theta_e$, (b), $x$ (c), 
	$\delta_{\rm had}$ (d) and $Z_{\rm vtx}$ (e). 
	The data are shown as points, with bars representing the statistical uncertainty. 
	The sum of charm and beauty MC is shown as the light shaded area; the beauty contribution is shown separately as the dark shaded area.}
  \label{fig:dch:cpadd}
  \end{minipage}
\end{figure*}

Fig.~\ref{fig:dch:cpadd} shows control plots for $E_e^{\prime}$, polar angle of the scattered electron, $x$, $\delta_{\rm had}$ and $Z_{\rm vtx}$.

\begin{figure*}[htbp]
  \sidecaption
  \centering
  \includegraphics[width=1.65\figwidth,trim=0 1mm 2mm 0,clip=true]{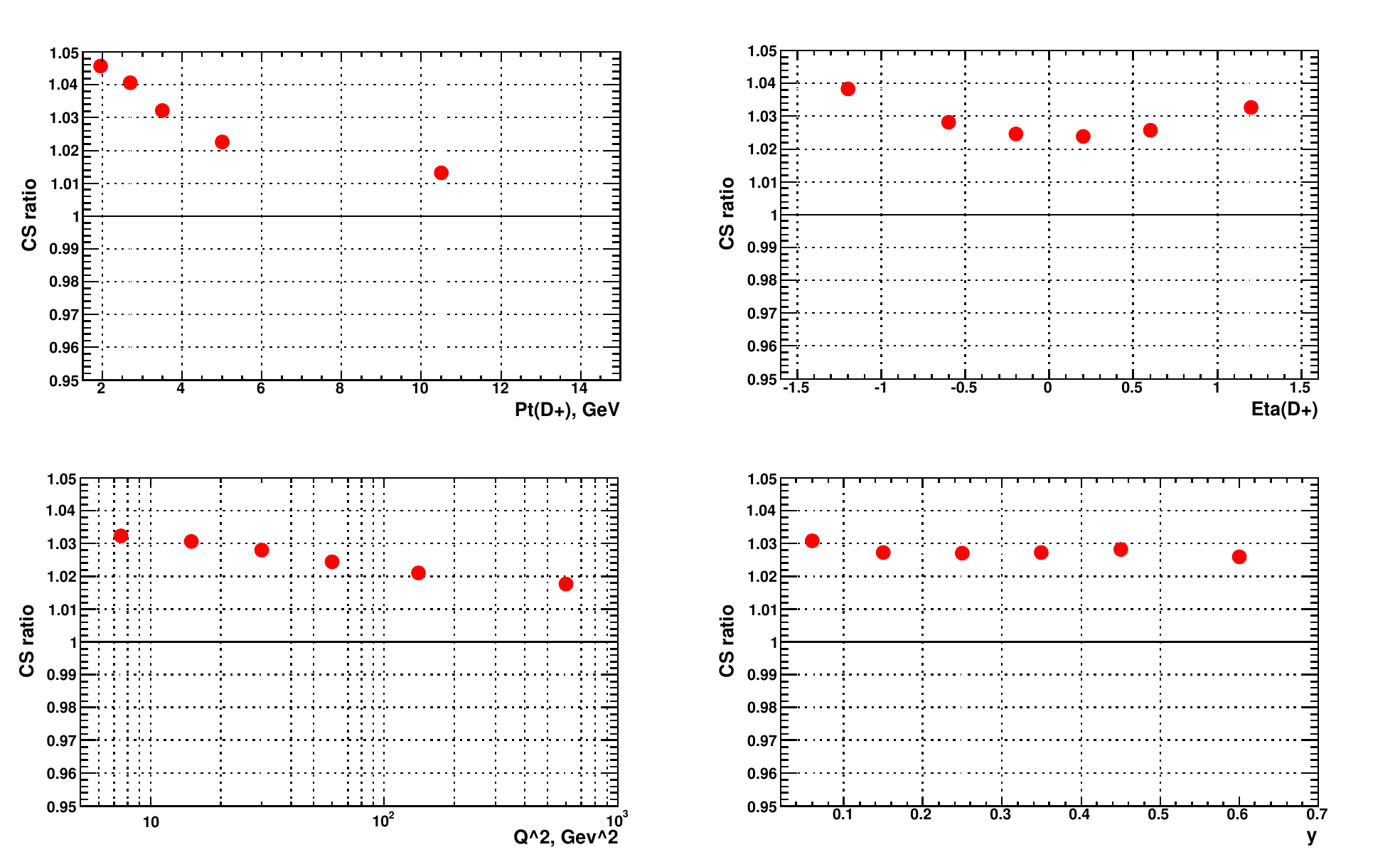}
  \caption[Tracking inefficiency correction as function of $p_T(\Dch)$, $\eta(\Dch)$, $Q^2$ and $y$]
	{Effect of the tracking inefficiency correction as a function of $p_T(\Dch)$ (top left), $\eta(\Dch)$ (top right), $Q^2$ (bottom left) and $y$ (bottom right).}
  \label{fig:dch:trackeff_cs}
\end{figure*}

Fig.~\ref{fig:dch:trackeff_cs} shows effect of the tracking inefficiency correction as a function of $p_T(\Dch)$, $\eta(\Dch)$, $Q^2$ and $y$

\clearpage
\section{Combination procedure: additional details}
\label{sec:app:comb}

In this Appendix additional information on the combination procedure (Section~\ref{sec:comb:proc}) is provided. 
\subsection{Minimisation method}
\label{sec:comb:proc:min}

The minimisation method described below~\cite{Aaron:2009bp,Belov:2013oda} is applicable if the uncertainties of the measurements do not depend on the central values (the additive treatment); 
in the case of the multiplicative treatment this method is extended with an iteration procedure described in the next Section~\ref{sec:comb:proc:it}. 
In this case the \chisq-function~\ref{eq:comb:proc:extchi2} can be considered. 
Since \chisq in~\ref{eq:comb:proc:extchi2} is a quadratic form of $\mathbf{m}$ and $\mathbf{b}$, it may be rearranged such 
that it takes a simpler form. 
To show this explicitly, \chisq can be written as a Taylor series up to its second derivatives 
near the minimum, $(\mathbf{m_0},\mathbf{b_0})$:
\begin{equation}
\begin{split}
	\label{eq:comb:proc:extchi2:taylor}
	\chisq(\mathbf{m},&\mathbf{b})=\sum_{e=1}^{N_e}\sum_{i=1}^{N_m}\frac{\left(m_i-\sum_{j=1}^{N_s}\Gamma_i^{e,j}b^{e,j}-\mu_i^e\right)^2}{{\sigma_i^e}^2}+\sum_{j=1}^{N_s}{b^{e,j}}^2\\
	&=\left.\chisq\right|_{0}+\left.\frac{\partial \chisq}{\partial \mathbf{m}}\right|_{0}(\mathbf{m}-\mathbf{m_0})
	+\left.\frac{\partial \chisq}{\partial \mathbf{b}}\right|_{0}(\mathbf{b}-\mathbf{b_0})\\
	&+\frac{1}{2}\left.\frac{\partial^2 \chisq}{\partial \mathbf{m}^2}\right|_{0}(\mathbf{m}-\mathbf{m_0})^2
	+\frac{1}{2}\left.\frac{\partial^2 \chisq}{\partial \mathbf{b}^2}\right|_{0}(\mathbf{b}-\mathbf{b_0})^2\\
	&+\left.\frac{\partial^2 \chisq}{\partial\mathbf{m}\partial\mathbf{b}}\right|_{0}(\mathbf{m}-\mathbf{m_0})(\mathbf{b}-\mathbf{b_0}).
\end{split}
\end{equation}
Notation $\left.\right|_{0}$ indicates that the expression is evaluated at $\mathbf{m}=\mathbf{m_0}$, $\mathbf{b}=\mathbf{b_0}$.
Note, that this is an \emph{exact} expression, because \chisq is a quadratic form; 
moreover the second derivatives are constant, i.e.\ 
$\left.\frac{\partial^2 \chisq}{\partial \mathbf{m}^2}\right|_{0}=\frac{\partial^2 \chisq}{\partial \mathbf{m}^2}$, 
$\left.\frac{\partial^2 \chisq}{\partial \mathbf{b}^2}\right|_{0}=\frac{\partial^2 \chisq}{\partial \mathbf{b}^2}$, 
$\left.\frac{\partial^2 \chisq}{\partial \mathbf{m} \partial \mathbf{b}}\right|_{0}=\frac{\partial^2 \chisq}{\partial \mathbf{m} \partial \mathbf{b}}$.

It is useful to give explicit expressions for the second derivatives and to introduce the following matrix notations:
\begin{equation}
\begin{split}
	\frac{\partial^2 \chisq}{\partial m_i \partial m_j}&=2\delta_{ij}\sum_{e=1}^{N_e}\frac{1}{{\sigma_i^e}^2}=2[\mathbf{A_M}]_{ij},~~1 \le i,j \le N_m,\\
	\frac{\partial^2 \chisq}{\partial b_i \partial b_j}&=2\left(\delta_{ij}+\sum_{e=1}^{N_e}\sum_{k=1}^{N_m}\frac{\Gamma_k^{e,i}\Gamma_k^{e,j}}{{\sigma_k^e}^2}\right)=2[\mathbf{A_S}]_{ij},~~1 \le i,j \le N_s,\\
	\frac{\partial^2 \chisq}{\partial m_i \partial b_j}&=-2\sum_{e=1}^{N_e}\frac{\Gamma_i^{e,j}}{{\sigma_i^e}^2}=2[\mathbf{A_{SM}}]_{ij},~~1 \le i \le N_m,~~1 \le j \le N_s,
\end{split}
\end{equation}
where $\delta_{ij}$ is the Kronecker delta.

The minimum $\chisq_{\rm min}=\left.\chisq\right|_{0}$ is found by solving a system of linear equations:
\begin{equation}
\begin{split}
	\left.\frac{\partial \chisq}{\partial \mathbf{m}}\right|_{0}&=0 \Longrightarrow \mathbf{A_M}\mathbf{m_0}+\mathbf{A_{SM}}\mathbf{b_0}-\mathbf{C_M}=0,\\
	\left.\frac{\partial \chisq}{\partial \mathbf{b}}\right|_{0}&=0 \Longrightarrow \mathbf{{A_{SM}}}^T\mathbf{m_0}+\mathbf{A_{S}}\mathbf{b_0}-\mathbf{C_S}=0,
\label{eq:comb:proc:extremum}
\end{split}
\end{equation}
where
\begin{equation}
\begin{split}
	[\mathbf{C_M}]_{i}&=\sum_{e=1}^{N_e}\frac{\mu_i^e}{{\sigma_i^e}^2},\\
	[\mathbf{C_S}]_{i}&=-\sum_{e=1}^{N_e}\sum_{j=1}^{N_m}\frac{\mu_j^e\Gamma_j^{e,i}}{{\sigma_j^e}^2}.
\end{split}
\end{equation}

System of linear equations~\ref{eq:comb:proc:extremum} can be written in a form of one matrix equation:
\begin{equation}
\label{eq:comb:proc:matrixeq}
	\begin{pmatrix}
	\mathbf{A_M} & \mathbf{A_{SM}} \\ \mathbf{A_{SM}}^T & \mathbf{A_S}
	\end{pmatrix}
	\begin{pmatrix}
	{\mathbf{m_0}} \\ {\mathbf{b_0}}
	\end{pmatrix}
	=
	\begin{pmatrix}
	\mathbf{C_m} \\ \mathbf{C_s}
	\end{pmatrix}	.
\end{equation}
Although solving of this system can be performed directly by inversion of the whole matrix, 
it is more convenient to take an advantage of the diagonal structure of the block $\mathbf{A_M}$ 
and solve the system using the method of the Schur complement:
\begin{equation}
\label{eq:comb:proc:matrixsol}
\begin{split}
	\mathbf{b_0}&=\mathbf{A_S^{\prime}}^{-1}(\mathbf{C_S}-\mathbf{A_{SM}}^T\mathbf{A_{M}}^{-1} \mathbf{C_{M}}),\\
	\mathbf{m_0}&=\mathbf{A_M}^{-1}(\mathbf{C_M}-\mathbf{A_{SM}}{\mathbf{b_0}}),
\end{split}	
\end{equation}
where
\begin{equation}
	\mathbf{A_S^{\prime}}=\mathbf{A_{S}}-\mathbf{A_{SM}}^T\mathbf{A_{M}}^{-1} \mathbf{A_{SM}}.
\end{equation}
This method benefits from the fact, that the only non-trivial inversion to be performed is the inversion 
of the block $\mathbf{A_S^{\prime}}$. The size of this block is $N_s \times N_s$ and usually much 
smaller than the total size of the system~\ref{eq:comb:proc:matrixeq}, $(N_m+N_s) \times (N_m+N_s)$, 
therefore this method is preferable for computation.

Obtained solution ($\mathbf{m_0}$,$\mathbf{b_0}$) solves the minimisation problem for the central values.
To find the uncertainties on ($\mathbf{m_0}$,$\mathbf{b_0}$), the $\chisq$ expansion in Eq.~\ref{eq:comb:proc:extchi2:taylor} can be written, 
taking into account that ($\mathbf{m_0}$,$\mathbf{b_0}$) is its minimum:
\begin{equation}
\begin{split}
	\chisq(\mathbf{m},\mathbf{b})&=\chisq_{\rm min}+\braket{\mathbf{m}-\mathbf{m_0}|\mathbf{A_M}|\mathbf{m}-\mathbf{m_0}}\\
	&+2\braket{\mathbf{m}-\mathbf{m_0}|\mathbf{A_{SM}}|\mathbf{b}-\mathbf{b_0}}+\braket{\mathbf{b}-\mathbf{b_0}|\mathbf{A_S}|\mathbf{b}-\mathbf{b_0}}.
\end{split}
\end{equation}
Denoting $\mathbf{m}-\mathbf{m_0}=\mathbf{\tilde{m}}$, $\mathbf{b}-\mathbf{b_0}=\mathbf{\tilde{b}}$:
\begin{equation}
\begin{split}
	\chisq(\mathbf{m},\mathbf{b})&=\chisq_{\rm min}+\braket{\mathbf{\tilde{m}}|\mathbf{A_M}|\mathbf{\tilde{m}}}+2\braket{\mathbf{\tilde{m}}|\mathbf{A_{SM}}|\mathbf{\tilde{b}}}+\braket{\mathbf{\tilde{b}}|\mathbf{A_S}|\mathbf{\tilde{b}}}.
\end{split}
\end{equation}
To separate contributions from $\mathbf{\tilde{m}}$ and $\mathbf{\tilde{b}}$ in the term $2\braket{\mathbf{\tilde{m}}|\mathbf{A_{SM}}|\mathbf{\tilde{b}}}$, 
introduce a variable substitution $\ket{\mathbf{\tilde{m}^{\prime}}}=\ket{\mathbf{\tilde{m}}}-\mathbf{X}\ket{\mathbf{\tilde{b}}}$:
\begin{equation}
\begin{split}
	\chisq(\mathbf{m},\mathbf{b})&=
	\chisq_{\rm min}+\braket{\mathbf{\tilde{m}^{\prime}}+\mathbf{\tilde{b}}\mathbf{X}^T|\mathbf{A_M}|\mathbf{\tilde{m}^{\prime}}+\mathbf{X}\mathbf{\tilde{b}}}\\
	&+2\braket{\mathbf{\tilde{m}^{\prime}}+\mathbf{\tilde{b}}\mathbf{X}^T|\mathbf{A_{SM}}|\mathbf{\tilde{b}}}+\braket{\mathbf{\tilde{b}}|\mathbf{A_S}|\mathbf{\tilde{b}}}\\
	&=\chisq_{\rm min}+\braket{\mathbf{\tilde{m}^{\prime}}|\mathbf{A_M}|\mathbf{\tilde{m}^{\prime}}}+
	\braket{\mathbf{\tilde{b}}|\mathbf{A_S}+2\mathbf{X}^T\mathbf{A_{SM}}+\mathbf{X}^T\mathbf{A_M}\mathbf{X}|\mathbf{\tilde{b}}}\\
	&+\braket{\mathbf{\tilde{m}^{\prime}}|2\mathbf{A_M}\mathbf{X}+2\mathbf{A_{SM}}|\mathbf{\tilde{b}}},
\end{split}
\end{equation}
thus choosing $\mathbf{X}=-\mathbf{A_M}^{-1}\mathbf{A_{SM}}$:
\begin{equation}
\begin{split}
	\chisq(\mathbf{m},\mathbf{b})&=
	\chisq_{\rm min}+\braket{\mathbf{\tilde{m}^{\prime}}|\mathbf{A_M}|\mathbf{\tilde{m}^{\prime}}}+
	\braket{\mathbf{\tilde{b}}|\mathbf{A_S}-\mathbf{A_{SM}}^T\mathbf{A_M}^{-1}\mathbf{A_{SM}}|\mathbf{\tilde{b}}}\\
	&=\chisq_{\rm min}+\braket{\mathbf{\tilde{m}^{\prime}}|\mathbf{A_M}|\mathbf{\tilde{m}^{\prime}}}+\braket{\mathbf{\tilde{b}}|\mathbf{A_S^{\prime}}|\mathbf{\tilde{b}}}.
\label{eq:comb:proc:finalorig}
\end{split}
\end{equation}
Here $\mathbf{A_M}=\mathbf{({A_M})}^T$, $\mathbf{A_M}^{-1}=\mathbf{({A_M}}^{-1})^T$ were used.

Eq.~\ref{eq:comb:proc:finalorig} allows interpretation of matrices $\mathbf{A_M}$ and $\mathbf{A_S^{\prime}}$ 
in terms of uncertainties on $\mathbf{\tilde{m}}$ and $\mathbf{\tilde{b}}$. 
Since variation of \chisq-function of 1 corresponds to one standard deviation, 
diagonal elements of matrix $\mathbf{A_M}$ gives the uncertainty on $\mathbf{\tilde{m}^{\prime}}$ and 
therefore the uncorrelated uncertainty on $\mathbf{m_0}$:
\begin{equation}
	\delta^{\rm uncor} {{m_0}_i}=\left([\mathbf{A_M}]_{ii}\right)^{-1/2}=\left(\sum_{e=1}^{N_e}\frac{1}{{\sigma_i^e}^2}\right)^{-1/2}
\label{eq:comb:proc:uncor}
\end{equation}
and diagonal elements of matrix $\mathbf{A_S^{\prime}}$ gives the uncertainties on the fitted values of the 
nuisance parameter $\mathbf{b_0}$:
\begin{equation}
	\delta {b_0}_i=\left([\mathbf{A_S^{\prime}}]_{ii}\right)^{-1/2},
\end{equation}
which are also referred to as the \emph{reduction factors} of correlated uncertainties.
Propagating them to $\ket{\mathbf{\tilde{m}}}=\ket{\mathbf{\tilde{m}^{\prime}}}+\mathbf{X}\ket{\mathbf{\tilde{b}}}$ gives the correlated uncertainties on $\mathbf{m_0}$:
\begin{equation}
	\delta^{{\rm cor}~j} {{m_0}_i}=\left([-\mathbf{A_M}^{-1}\mathbf{A_{SM}A_S^{\prime}}]_{ij}\right)^{-1/2},~~1 \le i \le N_m,~~1 \le j \le N_s.
\end{equation}

In Eq.~\ref{eq:comb:proc:finalorig} variables $\mathbf{\tilde{b}}$ are still mixed because of the 
non-diagonal structure of matrix $\mathbf{A_S^{\prime}}$, so 
the correlated uncertainties $\delta^{{\rm cor}~j} {{m_0}_i}$ are not independent. 
It is possible to decompose them, 
diagonilising this matrix
\begin{equation}
	\mathbf{UA_S^{\prime}U}^{-1}=\mathbf{DD}
\label{eq:comb:proc:sdiag}
\end{equation}
and introducing new independent (diagonalised) correlated error sources, $\tilde{b^{\prime}}$:
\begin{equation}
	\mathbf{\tilde{b^{\prime}}}=\mathbf{DU\tilde{b}}.
\label{eq:comb:proc:bdiag}
\end{equation}
Here $\mathbf{U}$ is an orthogonal matrix, composed of the eigenvectors of $\mathbf{A_S^{\prime}}$, and 
$\mathbf{D}$ is a diagonal matrix, composed of the corresponding square roots of eigenvalues. 
Using~\ref{eq:comb:proc:sdiag} and~\ref{eq:comb:proc:bdiag}, \chisq-function from Eq.~\ref{eq:comb:proc:finalorig} 
can be written as:
\begin{equation}
\begin{split}
	&\chisq(\mathbf{m},\mathbf{b})=\chisq_{\rm min}+\braket{\mathbf{\tilde{m^{\prime}}|\mathbf{A_M}|\mathbf{\tilde{m}^{\prime}}}}
	+\braket{\mathbf{\tilde{b^{\prime}}}|\mathbf{{I}}|\mathbf{\tilde{b^{\prime}}}},
\label{eq:comb:proc:finalindep}
\end{split}
\end{equation}
where $\mathbf{{I}}$ is the unit matrix. 
Thus diagonalised correlated uncertainty sources are independent variables distributed according to the unit Gaussian 
distribution around zero. Propagating them to $\mathbf{m}=\mathbf{m^{\prime}}+\mathbf{X\tilde{b}}$ gives the correlated uncertainties on $\mathbf{m_0}$:
\begin{equation}
	\delta^{{\rm cor}~j} {{m_0}_i}=\left([\mathbf{-A_M}^{-1}\mathbf{A_{SM}D}^{-1}\mathbf{U}^{-1}]_{ij}\right)^{-1/2}, 1 \le i \le N_m, 1 \le j \le N_s.
\label{eq:comb:proc:cor}
\end{equation}

Summarising results of~\ref{eq:comb:proc:uncor} and~\ref{eq:comb:proc:cor}, averaged quantities can be written as:
\begin{equation}
	m_i={m_0}_i+\delta^{\rm uncor} {{m_0}_i} a_i - \sum_{j=1}^{N_s}\delta^{{\rm cor}~j} {{m_0}_i} b_j,~~1 \le i \le N_m
\label{eq:comb:proc:final}
\end{equation}
with $a_i$ and $b_j$ being independently distributed according to the unit Gaussian distribution around zero.

\subsection{Iterative procedure}
\label{sec:comb:proc:it}

If some of the uncertainties are treated multiplicatively, the extremum conditions~\ref{eq:comb:proc:extremum} do not 
produce a system of linear equations, since $\Gamma_i^{e,j}$ are functions of unknown $\mathbf{m_0}$. 
In this case the averaging technique described in Section~\ref{sec:comb:proc:min} still can be used, but 
the average has to be found in an iterative procedure~\cite{Aaron:2009bp,Belov:2013oda}: 
first equation~\ref{eq:comb:proc:extchi2} is used to get an initial approximation for ${\mathbf{m_0}}$ and ${\mathbf{b_0}}$ 
which are used to recalculate the uncertainties as $\Gamma_i^{e,j}=\gamma_i^{e,j}{{m_0}_i}$ and 
${\sigma_i^e}^2={\delta_{i,stat}^e}^2{{m_0}_i}^2+{\delta_{i,uncor}^e}^2{{m_0}_i}^2$. Then the 
determination of ${{m_0}_i}$ is repeated. Typically convergence is observed after two iterations and the iteration 
procedure is terminated.

Note that this iterative procedure does not give the exact minimum of the \chisq-function~\ref{eq:comb:proc:finalchi2}. 
Although there are arguments~\cite{Aaron:2012qi,herafitter} that the exact minimum of~\ref{eq:comb:proc:finalchi2} 
is biased, while the iterative procedure described above gives an unbiased result.

\clearpage
\section{Combination of visible \Dstar cross sections: additional information}
\label{sec:app:dstar}

In this Appendix additional information on the combination of the visible \Dstar cross sections (see Section~\ref{sec:comb:dstar}) is provided. 

Table~\ref{tab:comb:dstar:syst} provides information on the fitted nuisance parameters.

The combined data with all correlations are provided in Tables~\ref{tab:comb:dstar:single:combinedfull12} to~\ref{tab:comb:dstar:double:combinedfull3}.

\begin{figure*}[htbp]
  \centering
  \begin{minipage}[t]{0.33\textwidth}
	  \includegraphics[width=1.0\textwidth,trim=0 0 0 15mm,clip=true]{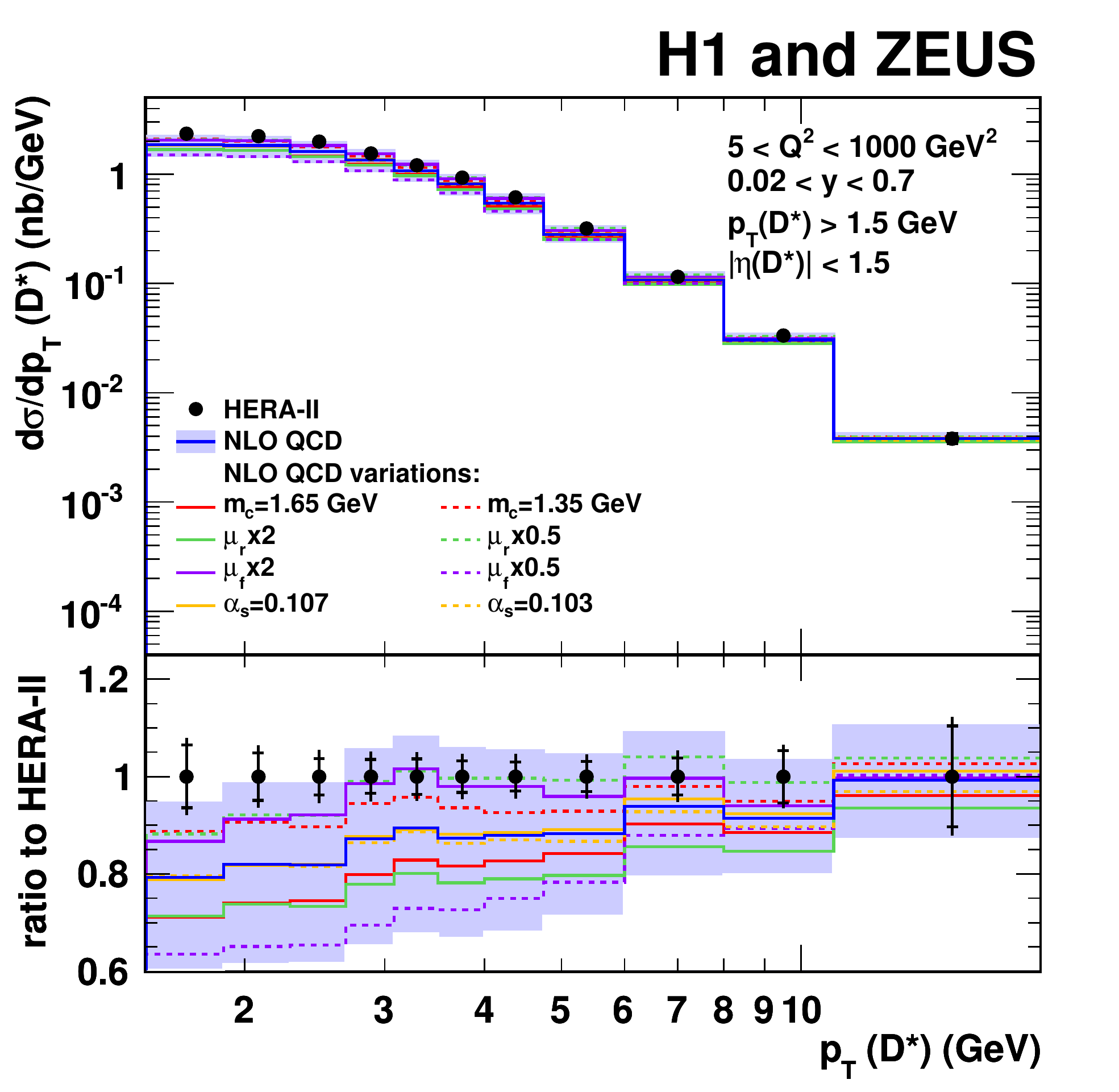}
  \end{minipage}
  \begin{minipage}[t]{0.33\textwidth}
	  \includegraphics[width=1.0\textwidth,trim=0 0 0 15mm,clip=true]{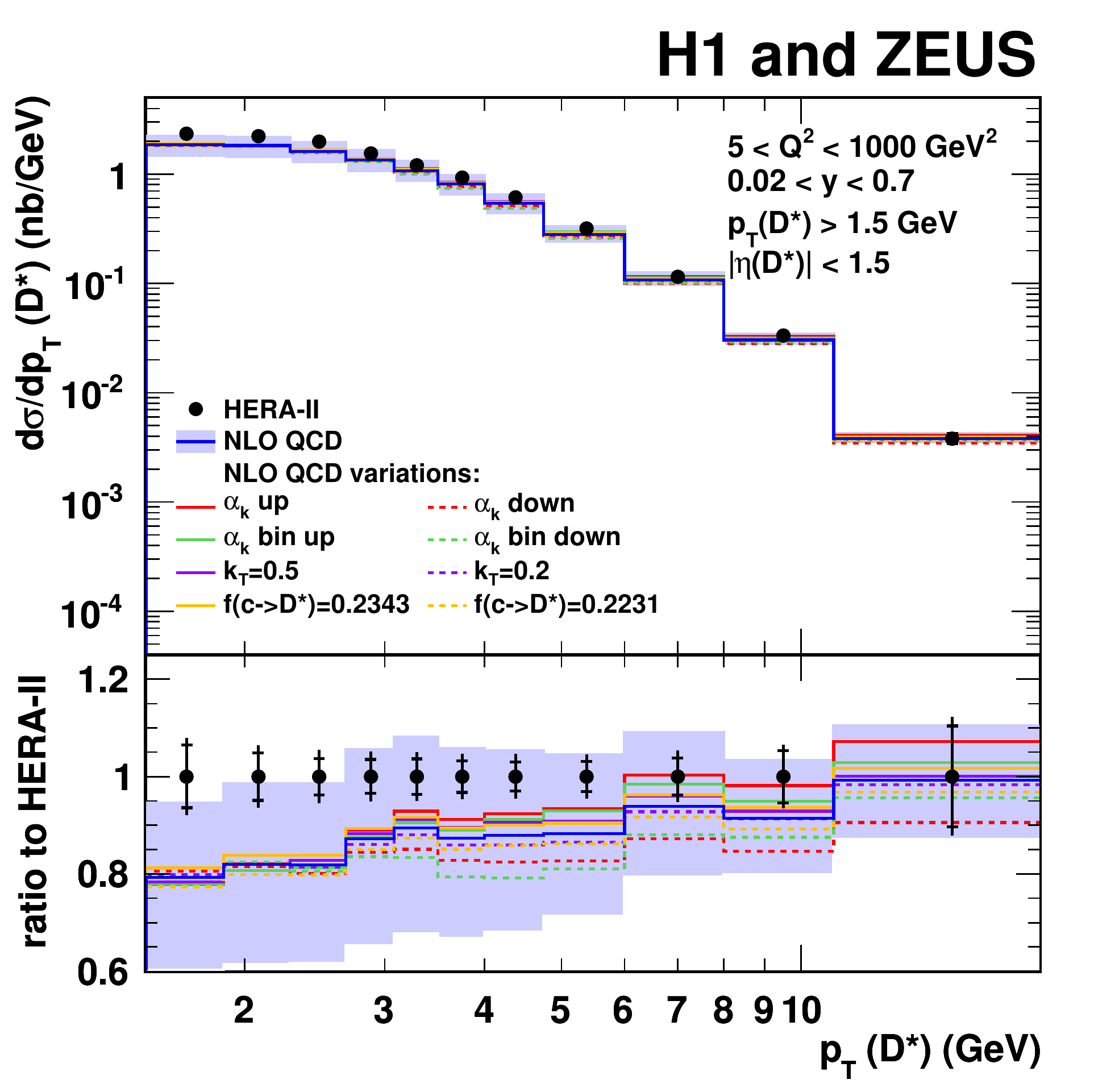}
  \end{minipage}
  \begin{minipage}[t]{0.33\textwidth}
	  \includegraphics[width=1.0\textwidth,trim=0 0 0 15mm,clip=true]{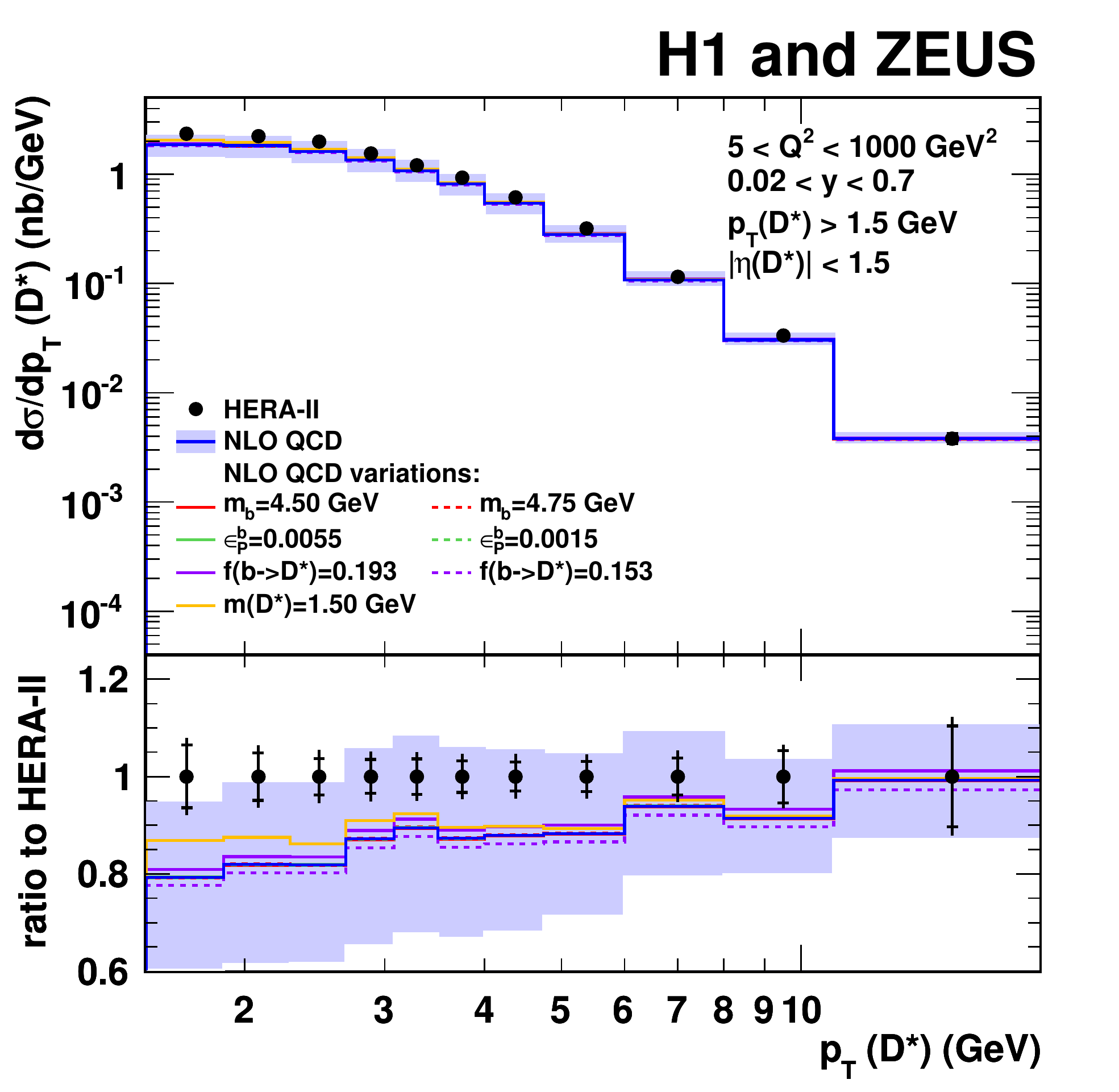}
  \end{minipage}
	  \caption[\Dstar cross section as function of $p_T(\Dstar)$ compared to NLO variations]
	  {Single-differential \Dstar cross section as a function of $p_T(\Dstar)$ compared to NLO predictions with individual variations.}
		\label{fig:comb:dstar:single:thvarpt}
\end{figure*}

\begin{figure*}[htbp]
  \centering
  \begin{minipage}[t]{0.33\textwidth}
	  \includegraphics[width=1.0\textwidth,trim=0 0 0 15mm,clip=true]{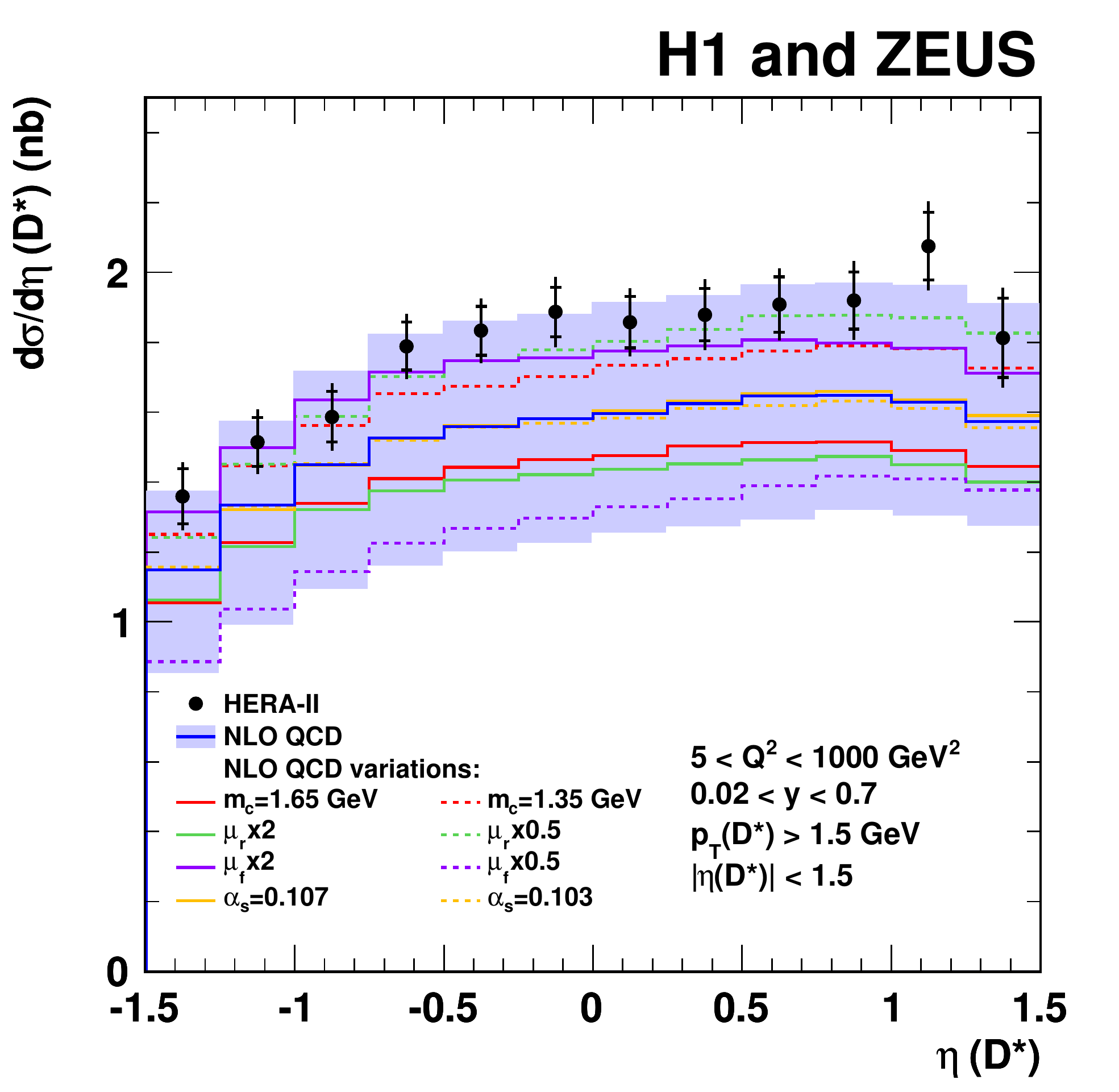}
  \end{minipage}
  \begin{minipage}[t]{0.33\textwidth}
	  \includegraphics[width=1.0\textwidth,trim=0 0 0 15mm,clip=true]{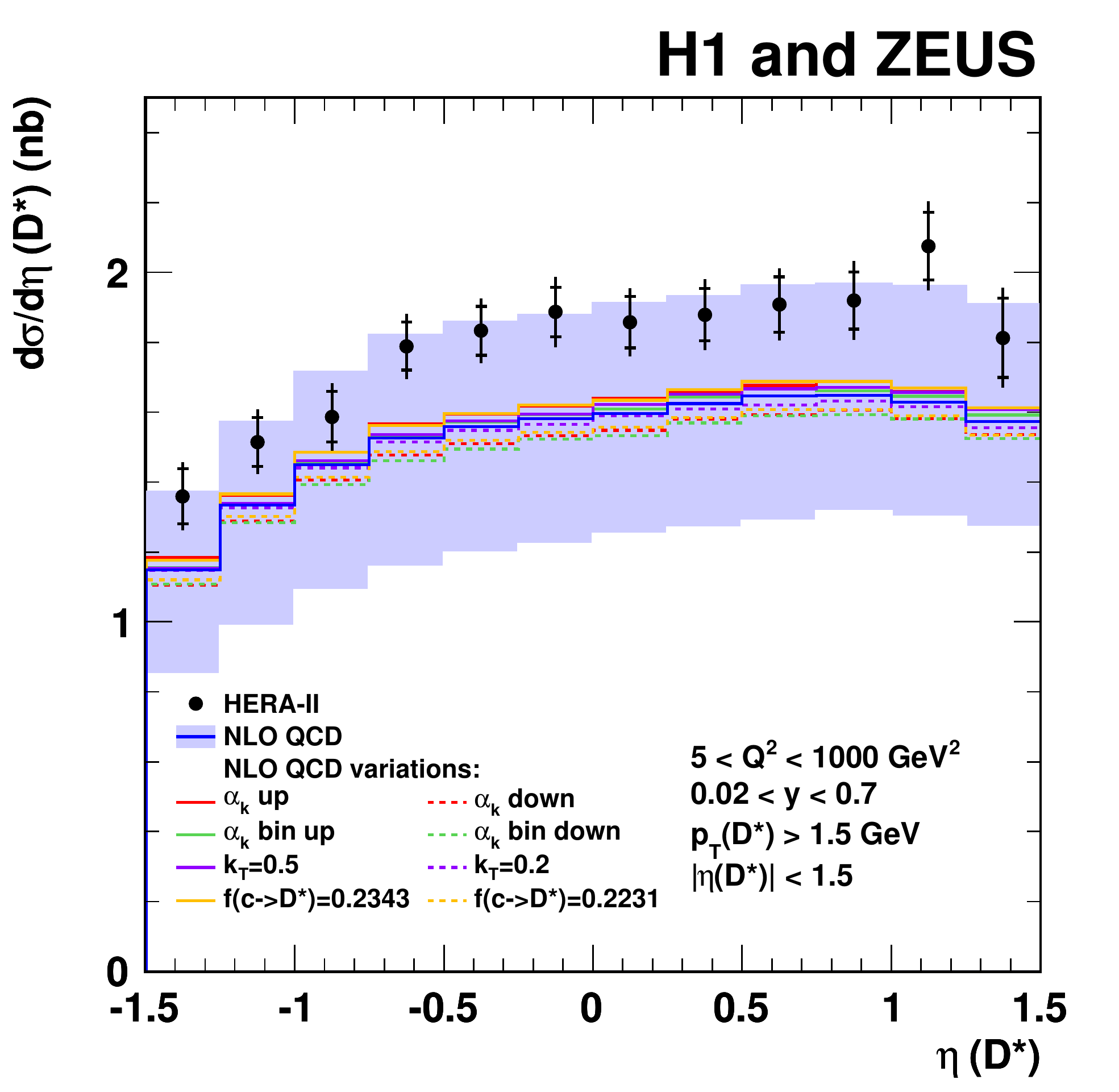}
  \end{minipage}
  \begin{minipage}[t]{0.33\textwidth}
	  \includegraphics[width=1.0\textwidth,trim=0 0 0 15mm,clip=true]{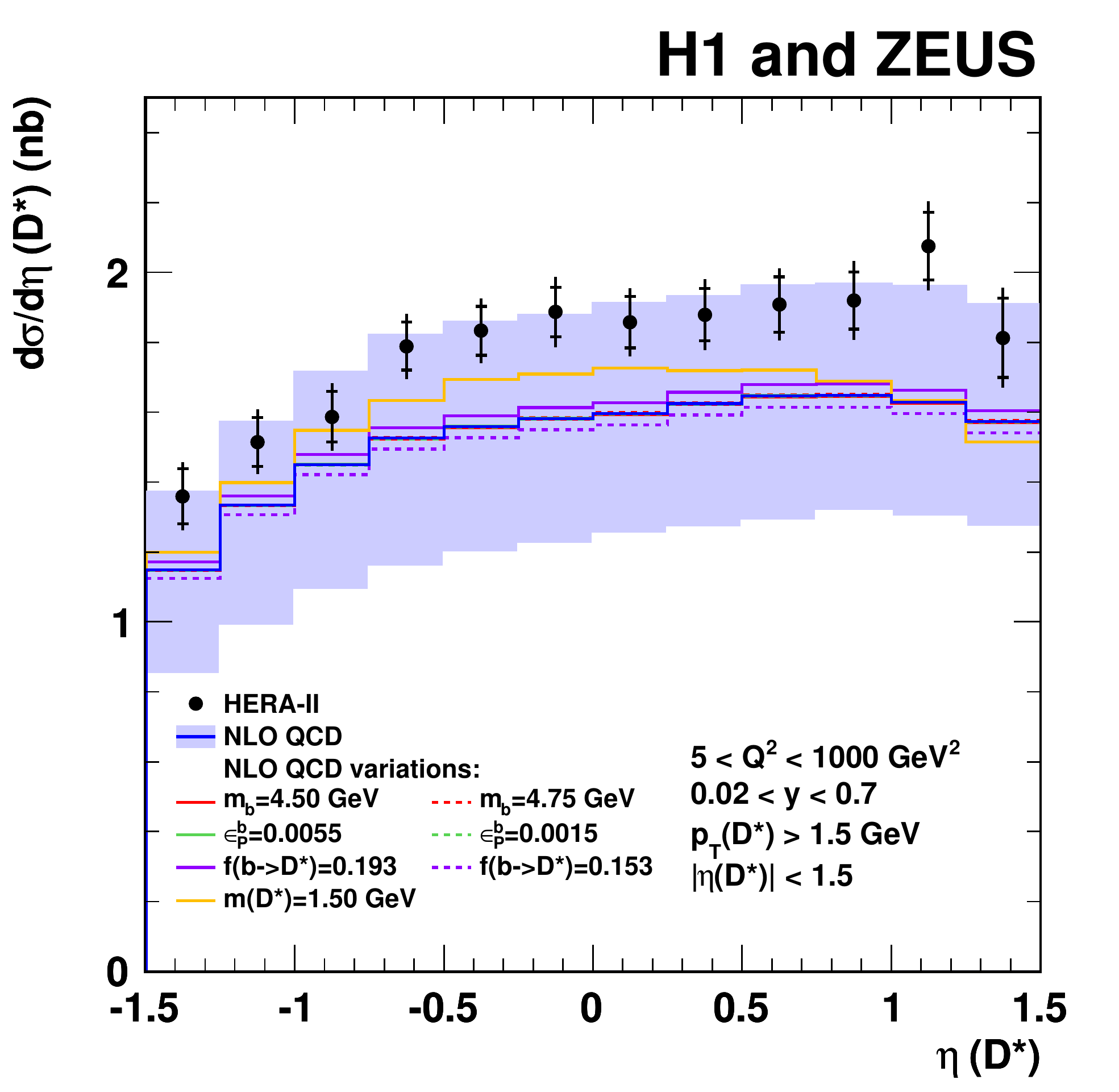}
  \end{minipage}
	  \caption[\Dstar cross section as function of $\eta(\Dstar)$ compared to NLO variations]
	  {Single-differential \Dstar cross section as a function of $\eta(\Dstar)$ compared to NLO predictions with individual variations.}
		\label{fig:comb:dstar:single:thvareta}
\end{figure*}

\begin{figure*}[htbp]
  \centering
  \begin{minipage}[t]{0.33\textwidth}
	  \includegraphics[width=1.0\textwidth,trim=0 0 0 15mm,clip=true]{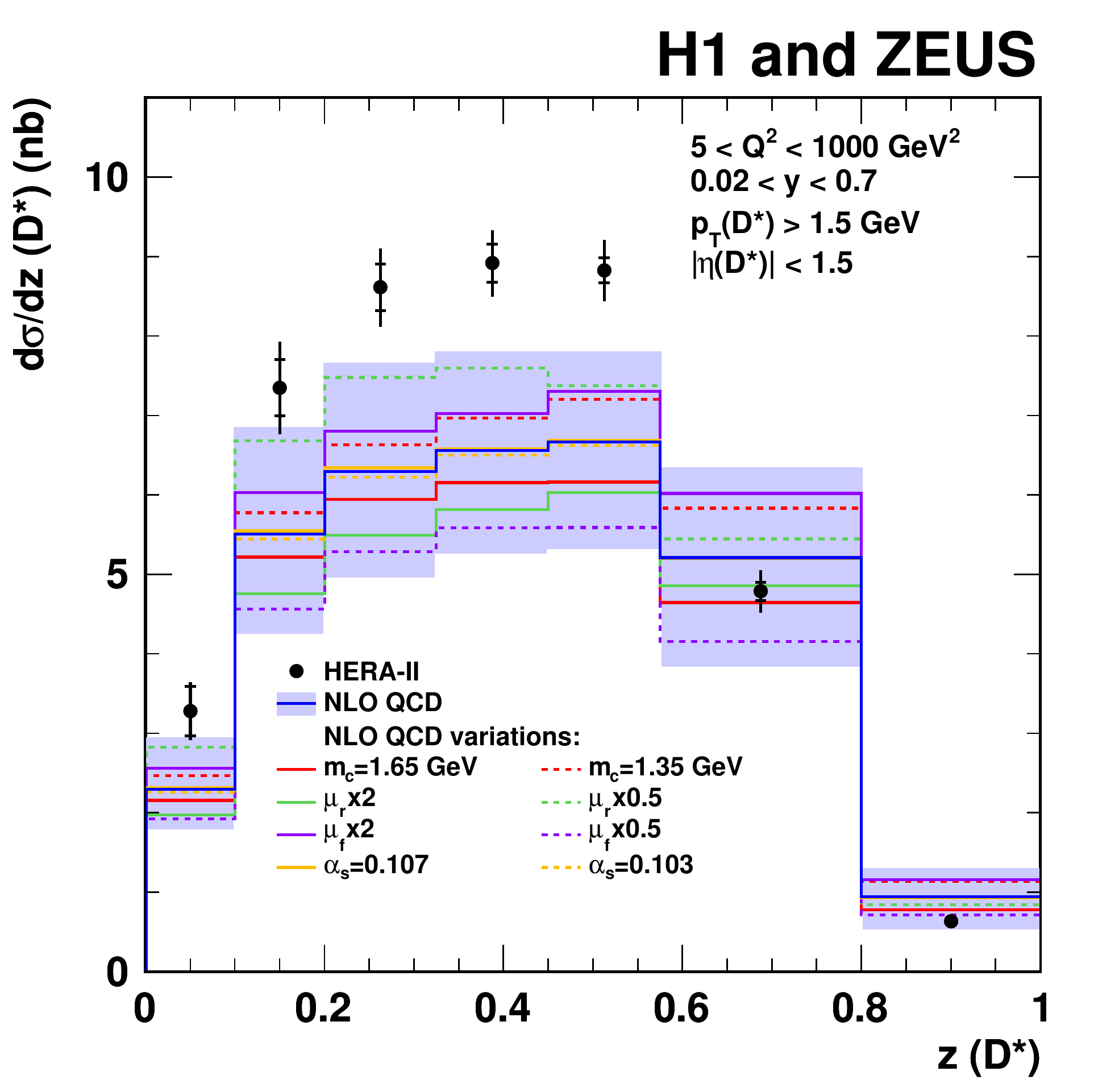}
  \end{minipage}
  \begin{minipage}[t]{0.33\textwidth}
	  \includegraphics[width=1.0\textwidth,trim=0 0 0 15mm,clip=true]{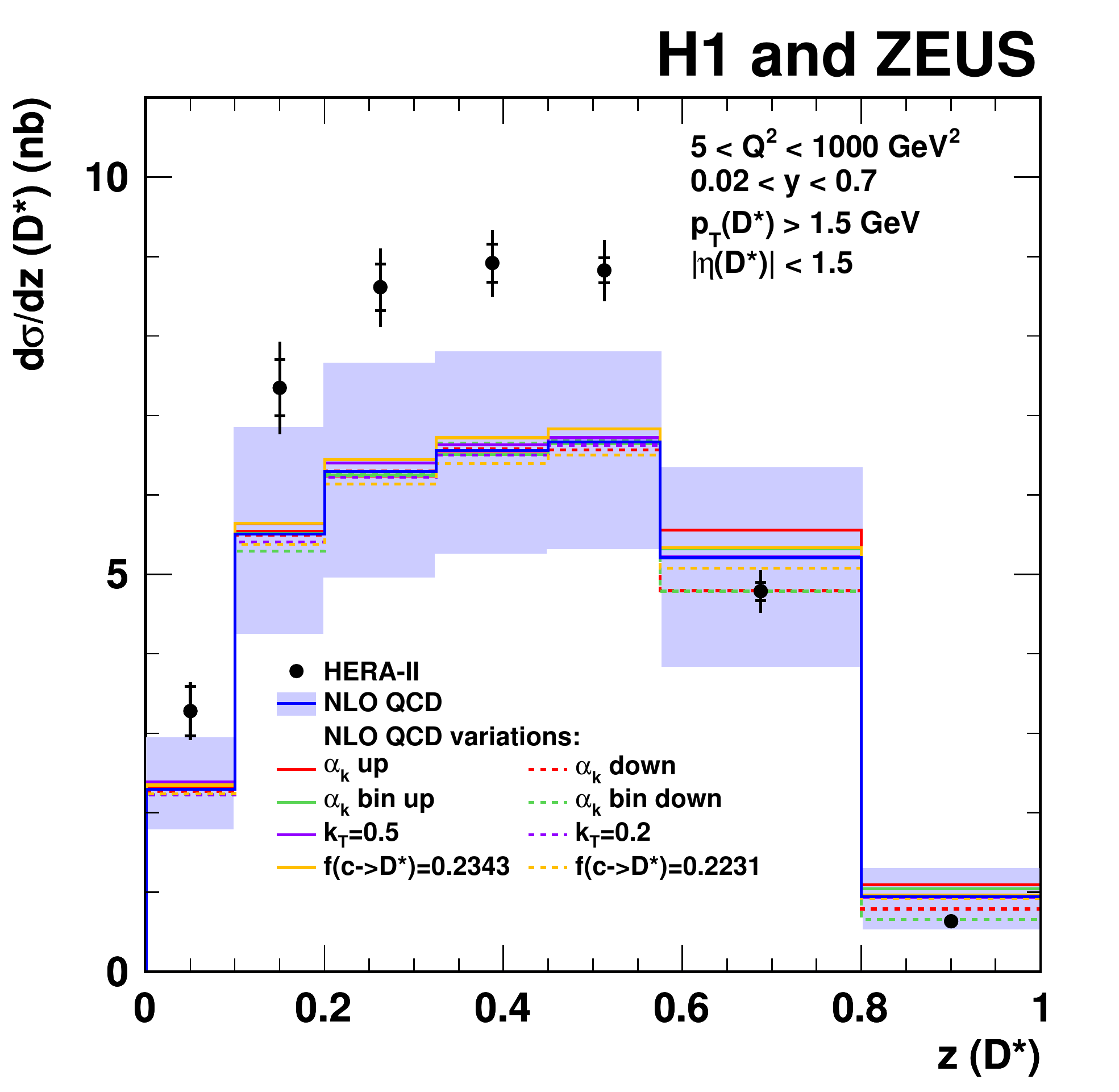}
  \end{minipage}
  \begin{minipage}[t]{0.33\textwidth}
	  \includegraphics[width=1.0\textwidth,trim=0 0 0 15mm,clip=true]{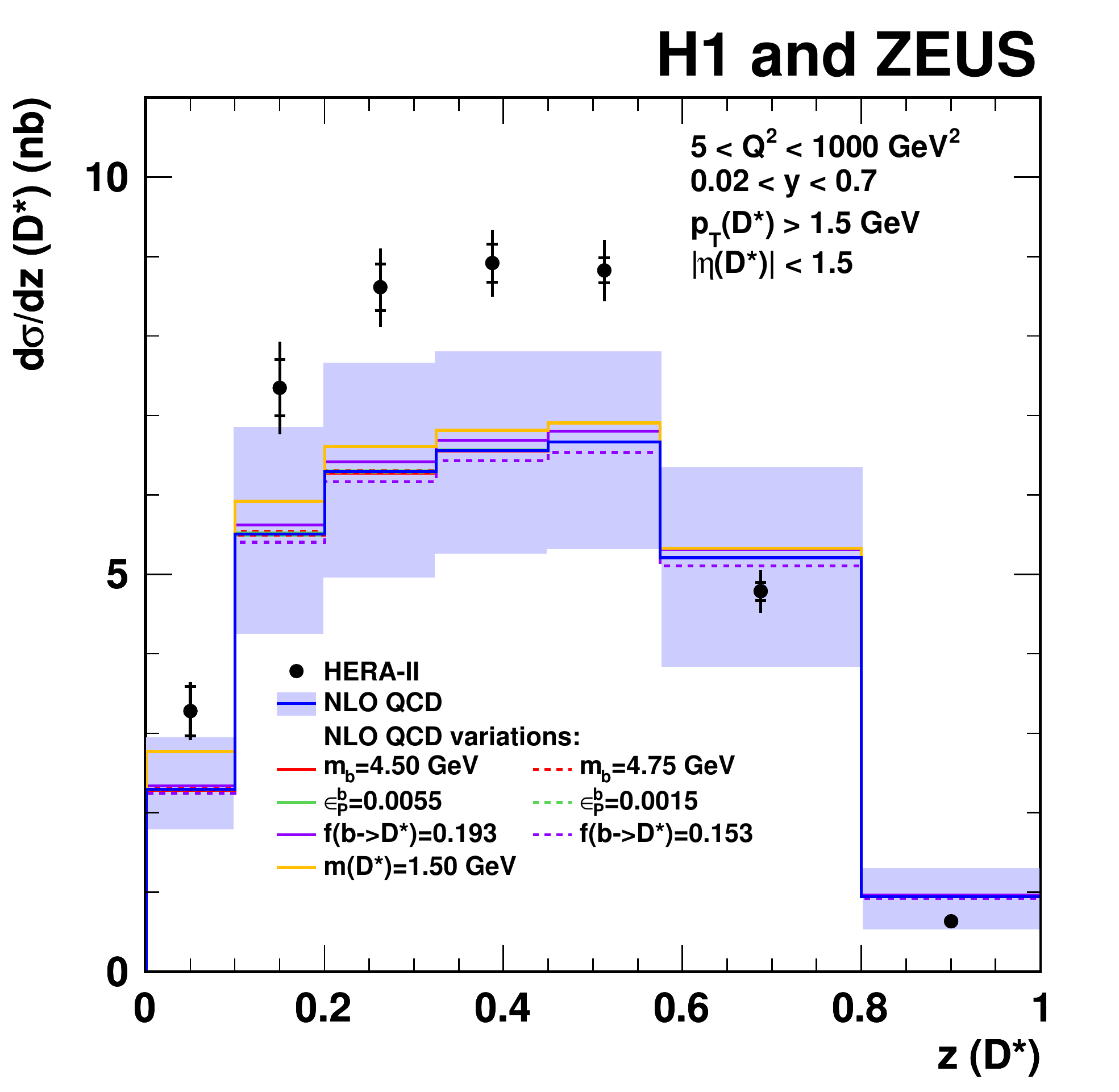}
  \end{minipage}
	  \caption[\Dstar cross section as function of $z(\Dstar)$ compared to NLO variations]
	  {Single-differential \Dstar cross section as a function of $z(\Dstar)$ compared to NLO predictions with individual variations.}
		\label{fig:comb:dstar:single:thvarz}
\end{figure*}

\begin{figure*}[htbp]
  \centering
  \begin{minipage}[t]{0.33\textwidth}
	  \includegraphics[width=1.0\textwidth,trim=0 0 0 15mm,clip=true]{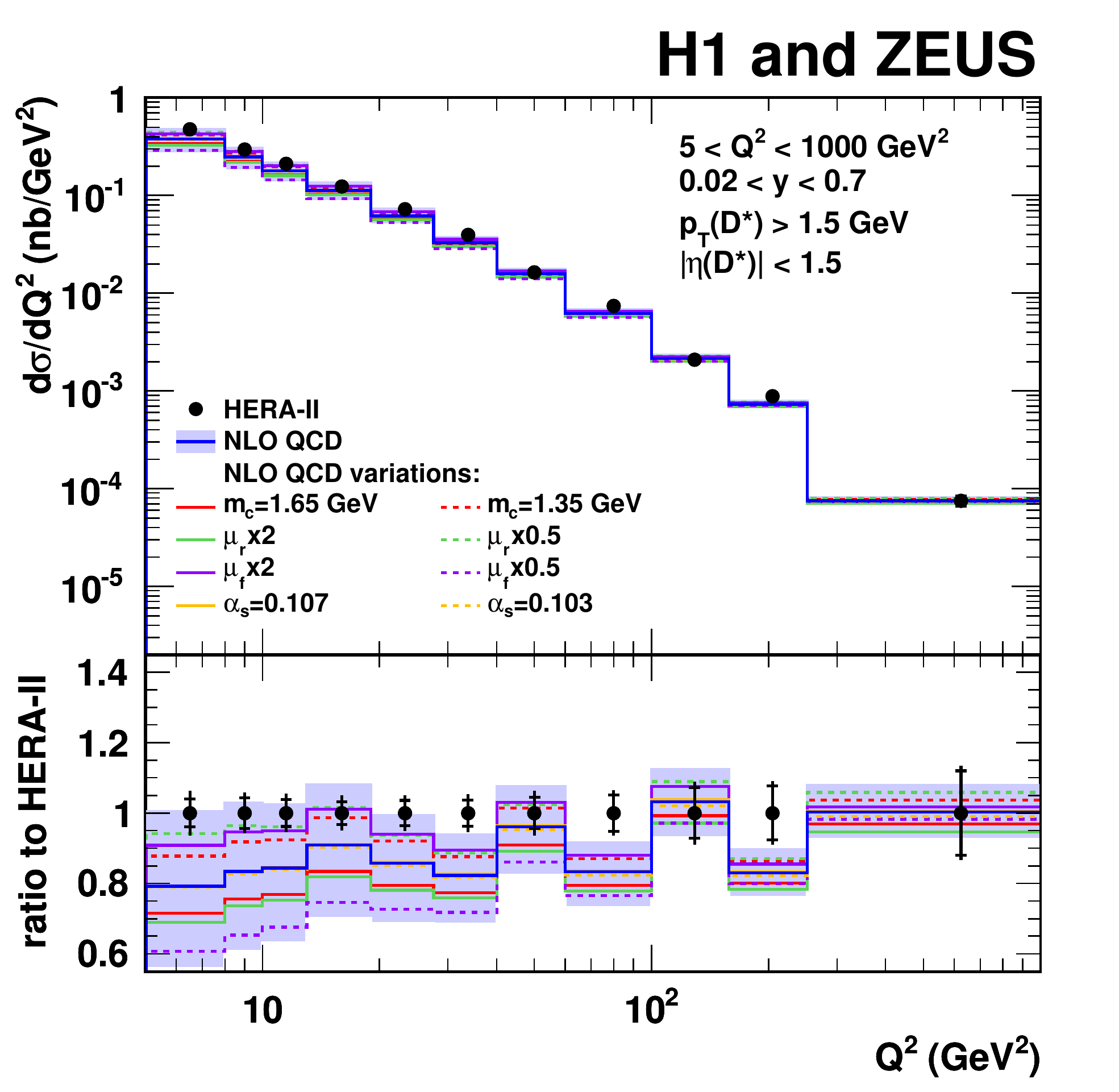}
  \end{minipage}
  \begin{minipage}[t]{0.33\textwidth}
	  \includegraphics[width=1.0\textwidth,trim=0 0 0 15mm,clip=true]{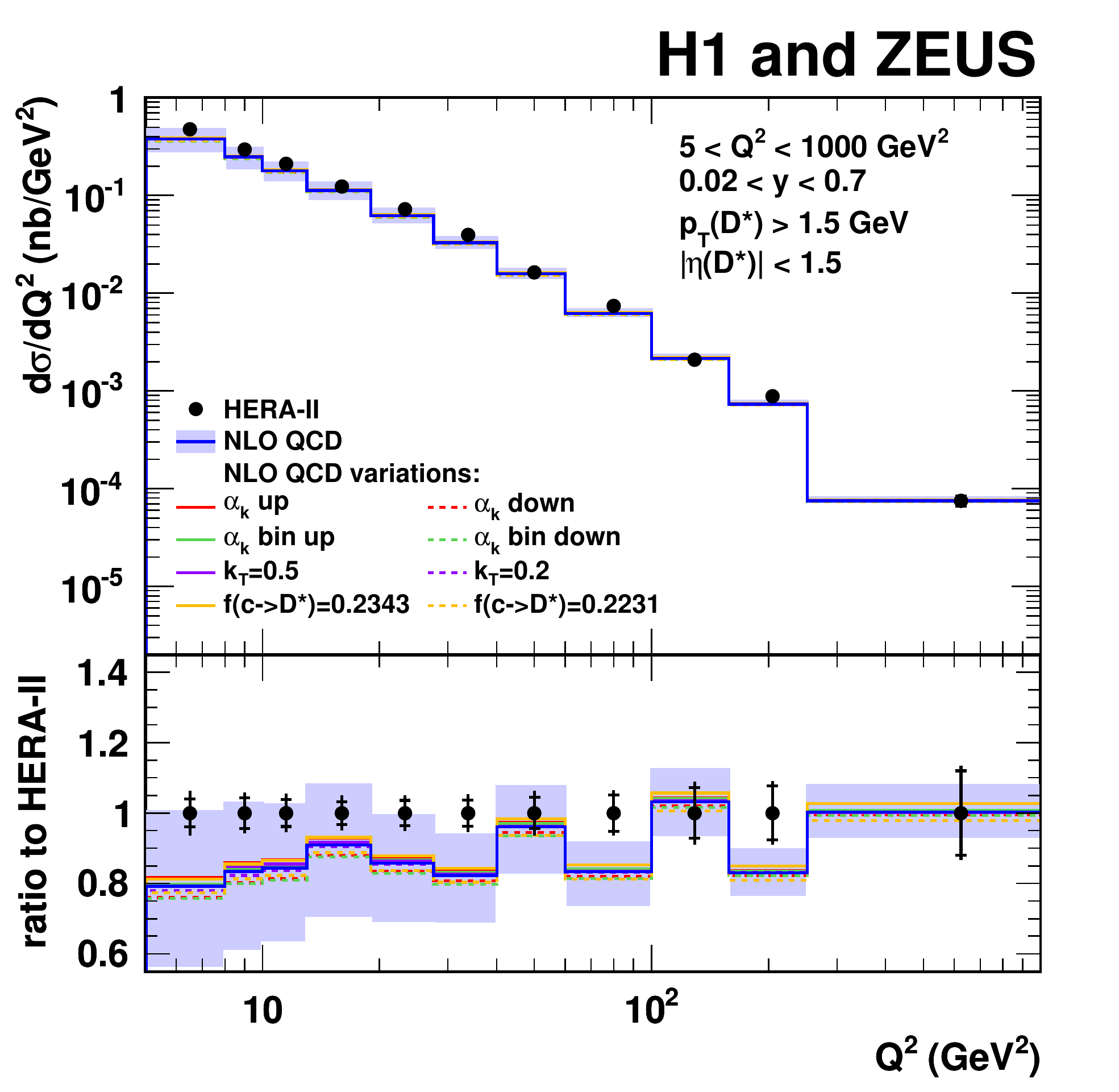}
  \end{minipage}
  \begin{minipage}[t]{0.33\textwidth}
	  \includegraphics[width=1.0\textwidth,trim=0 0 0 15mm,clip=true]{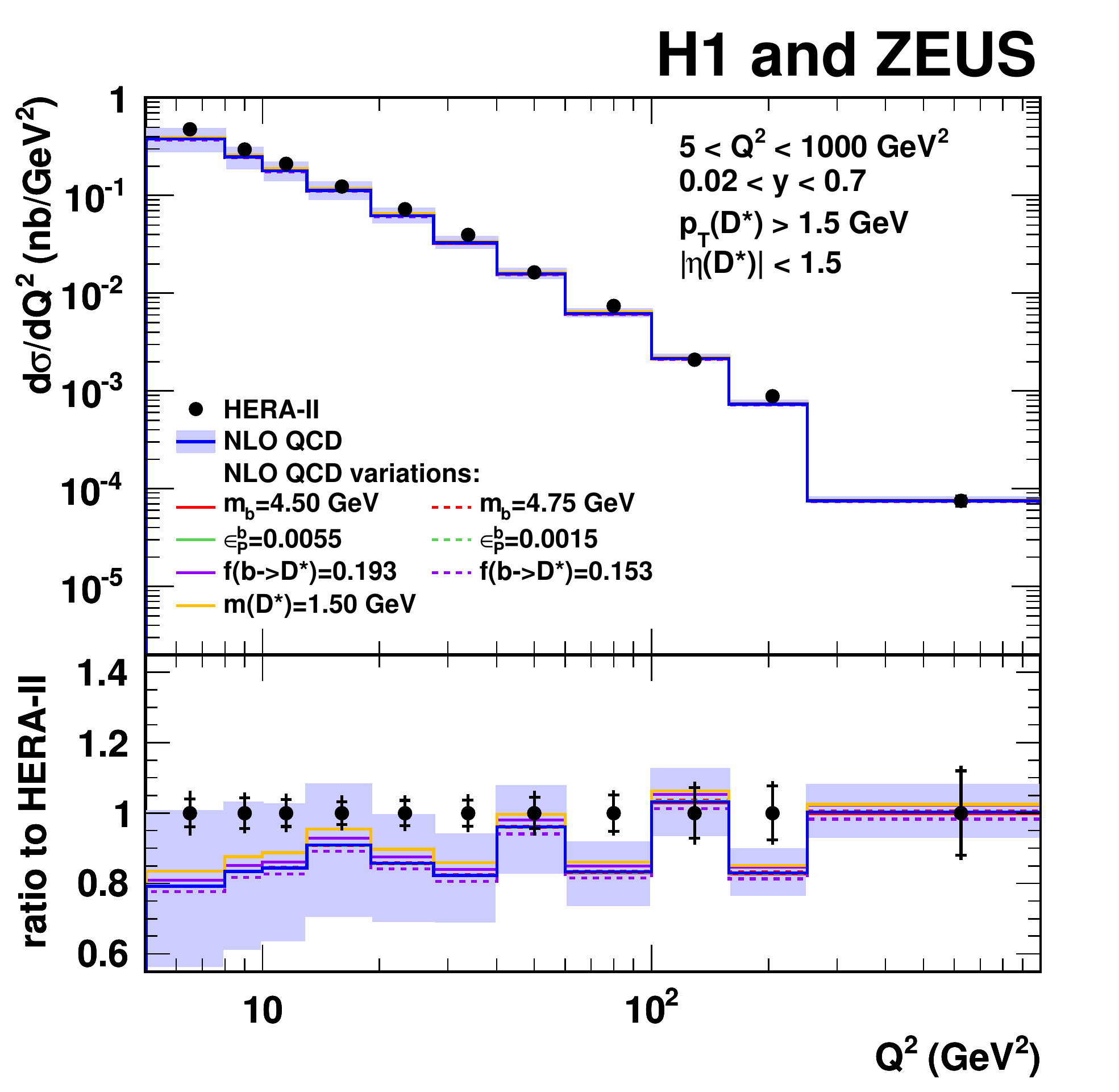}
  \end{minipage}
	  \caption[\Dstar cross section as function of $Q^2$ compared to NLO variations]
	  {Single-differential \Dstar cross section as a function of $Q^2$ compared to NLO predictions with individual variations.}
		\label{fig:comb:dstar:single:thvarq2}
\end{figure*}

\begin{figure*}[htbp]
  \centering
  \begin{minipage}[t]{0.33\textwidth}
	  \includegraphics[width=1.0\textwidth,trim=0 0 0 15mm,clip=true]{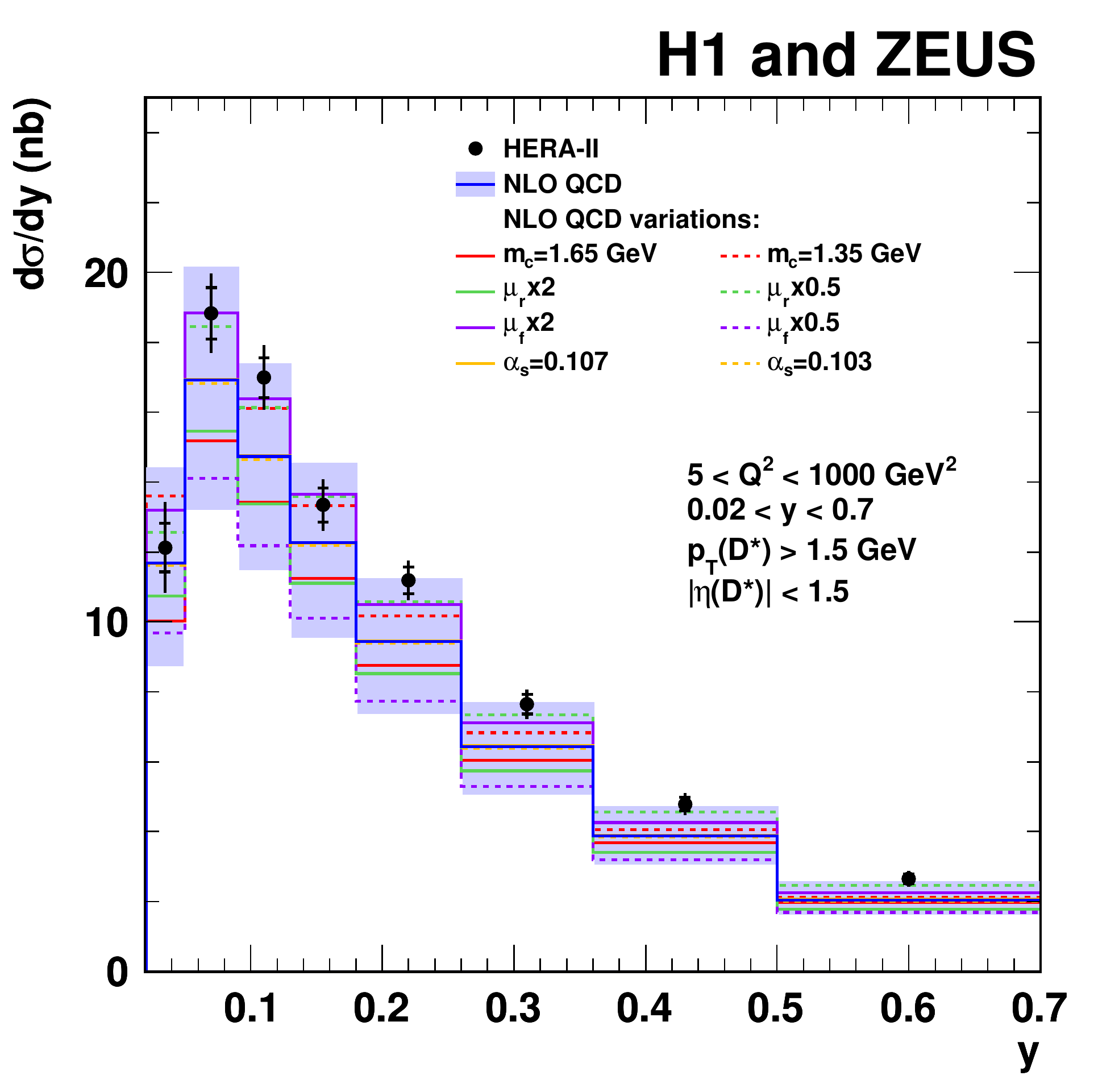}
  \end{minipage}
  \begin{minipage}[t]{0.33\textwidth}
	  \includegraphics[width=1.0\textwidth,trim=0 0 0 15mm,clip=true]{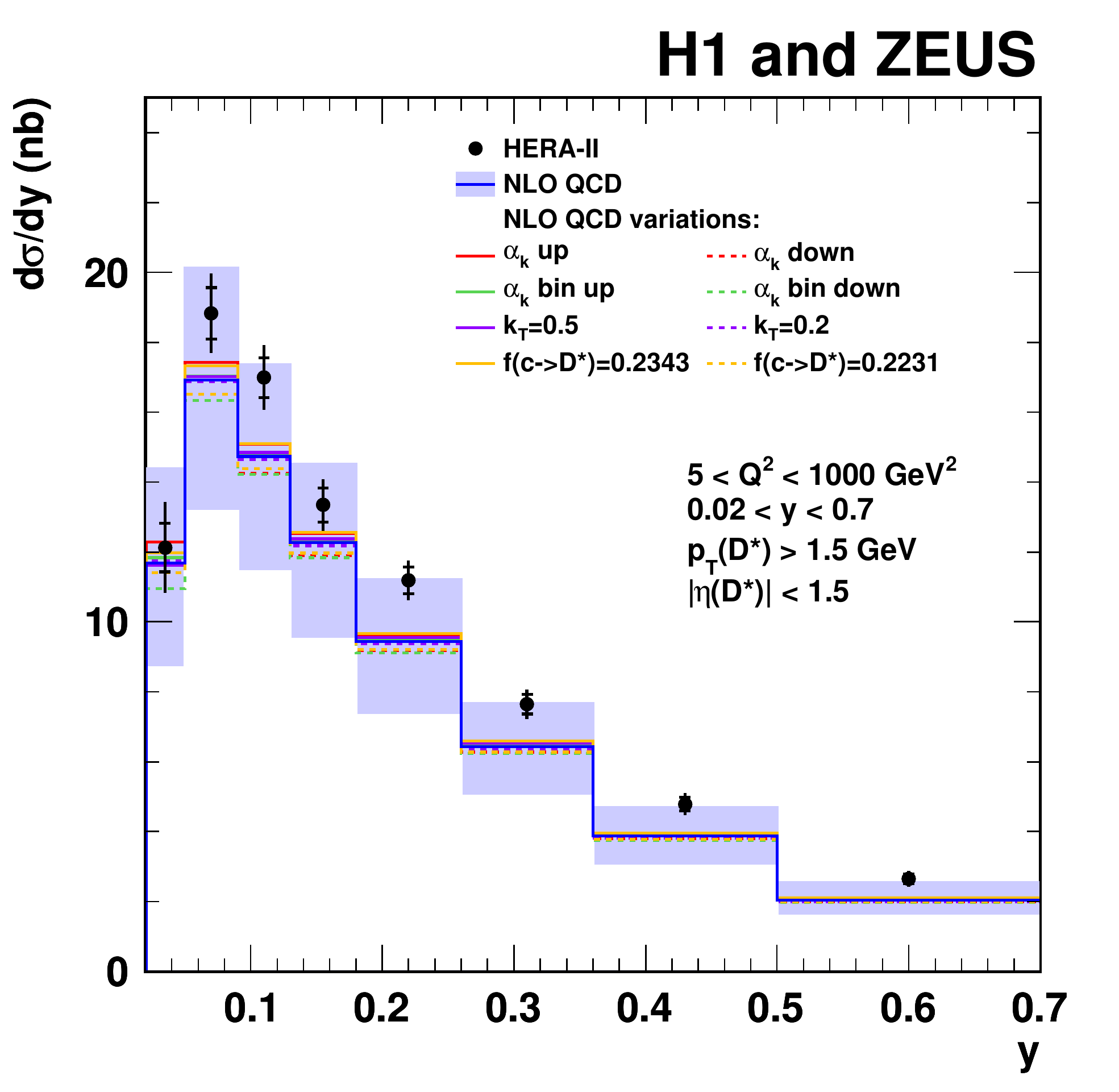}
  \end{minipage}
  \begin{minipage}[t]{0.33\textwidth}
	  \includegraphics[width=1.0\textwidth,trim=0 0 0 15mm,clip=true]{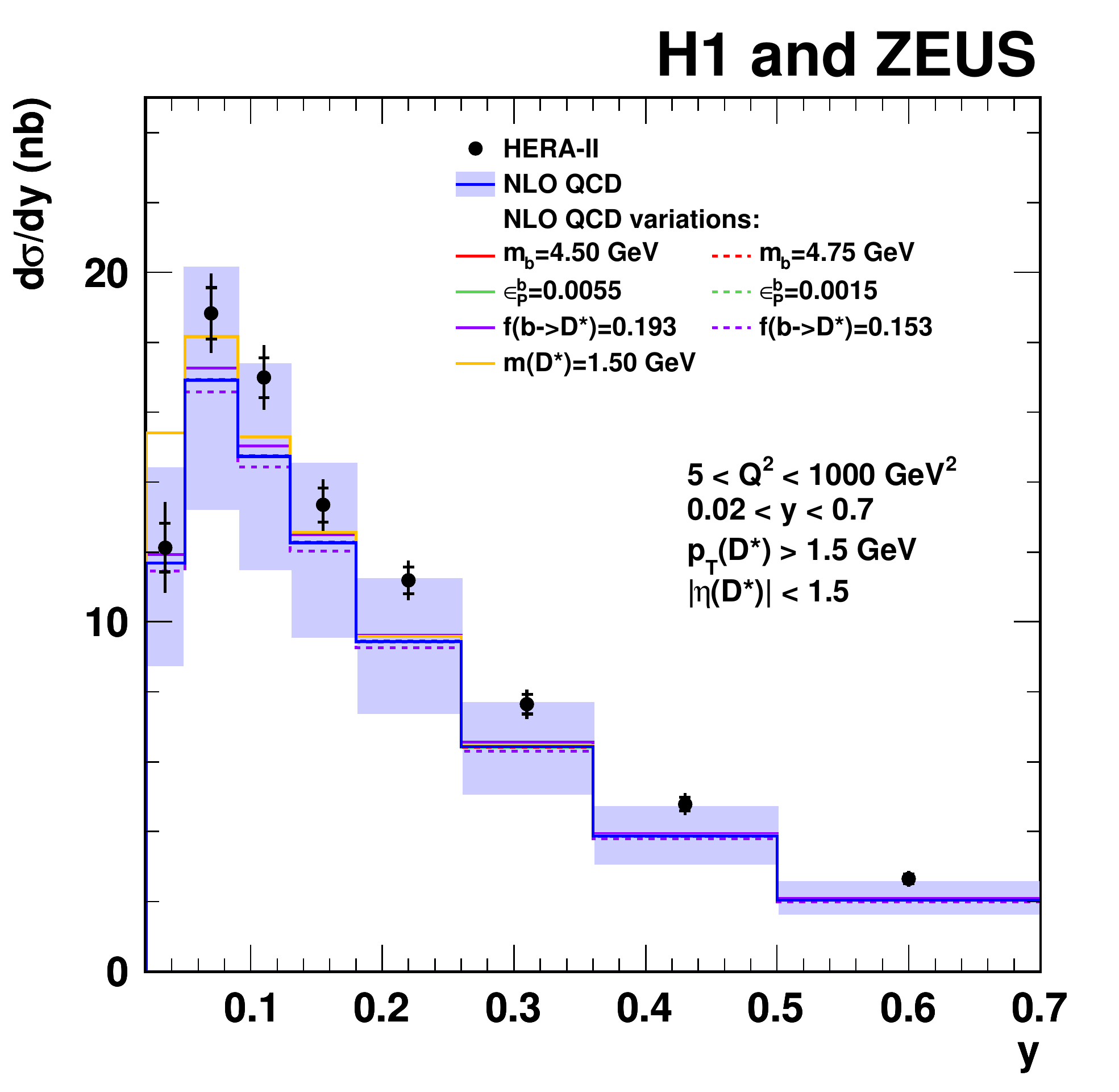}
  \end{minipage}
	  \caption[\Dstar cross section as function of $y$ compared to NLO variations]
	  {Single-differential \Dstar cross section as a function of $y$ compared to NLO predictions with individual variations.}
		\label{fig:comb:dstar:single:thvary}
\end{figure*}

Figs.~\ref{fig:comb:dstar:single:thvarpt} to~\ref{fig:comb:dstar:single:thvary} show comparison of the combined data to the individual theoretical variations 
(see Section~\ref{sec:comb:dstar:single:customth}).

\begin{figure*}[htbp]
  \centering
  \begin{minipage}[t]{0.33\textwidth}
	  \includegraphics[width=1.0\textwidth,trim=0 0 0 15mm,clip=true]{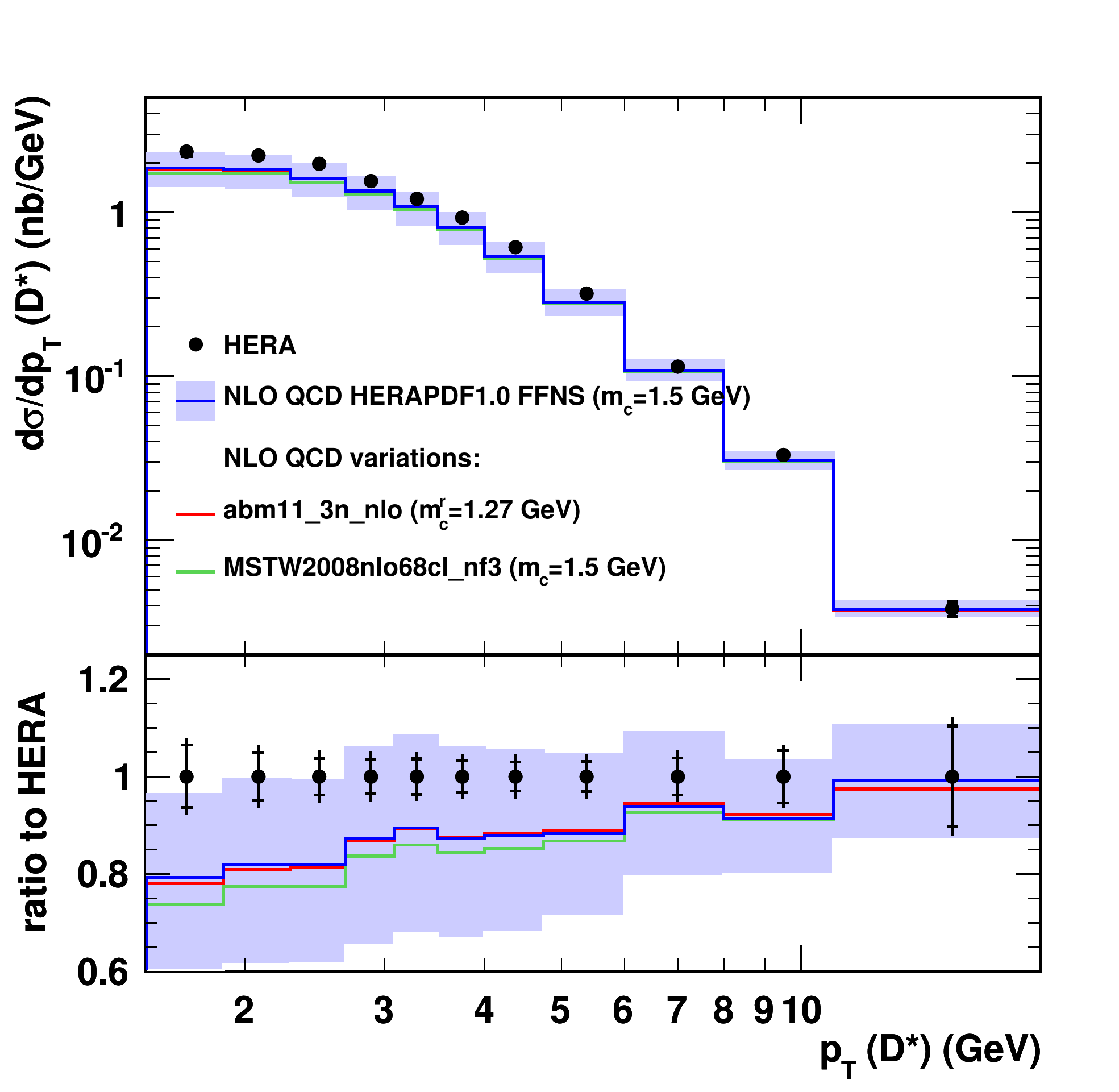}
	  \put(-35,80){(a)}\\
	  \includegraphics[width=1.0\textwidth,trim=0 0 0 15mm,clip=true]{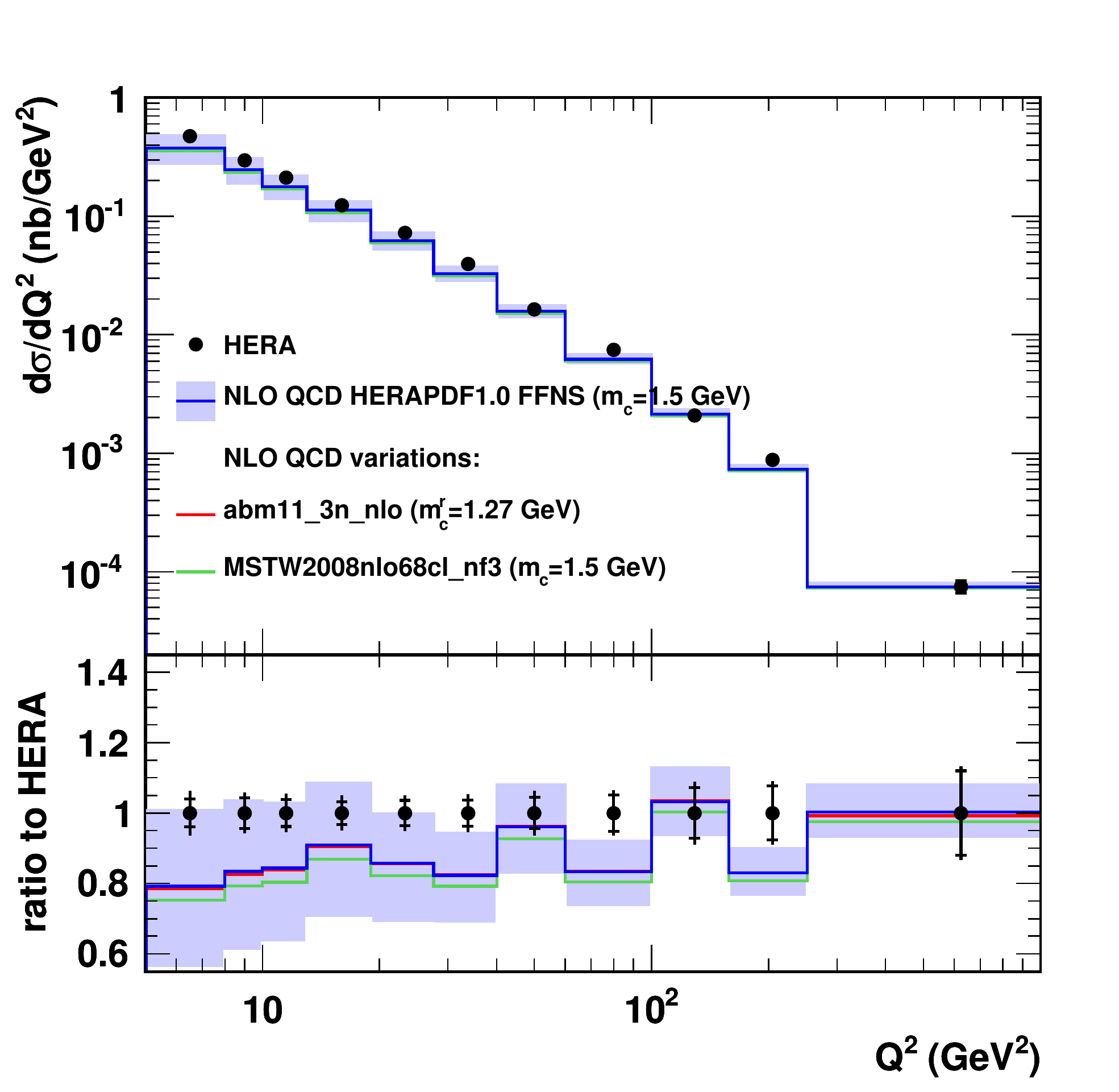}
	  \put(-35,80){(d)}\\
  \end{minipage}
  \begin{minipage}[t]{0.33\textwidth}
	  \includegraphics[width=1.0\textwidth,trim=0 0 0 15mm,clip=true]{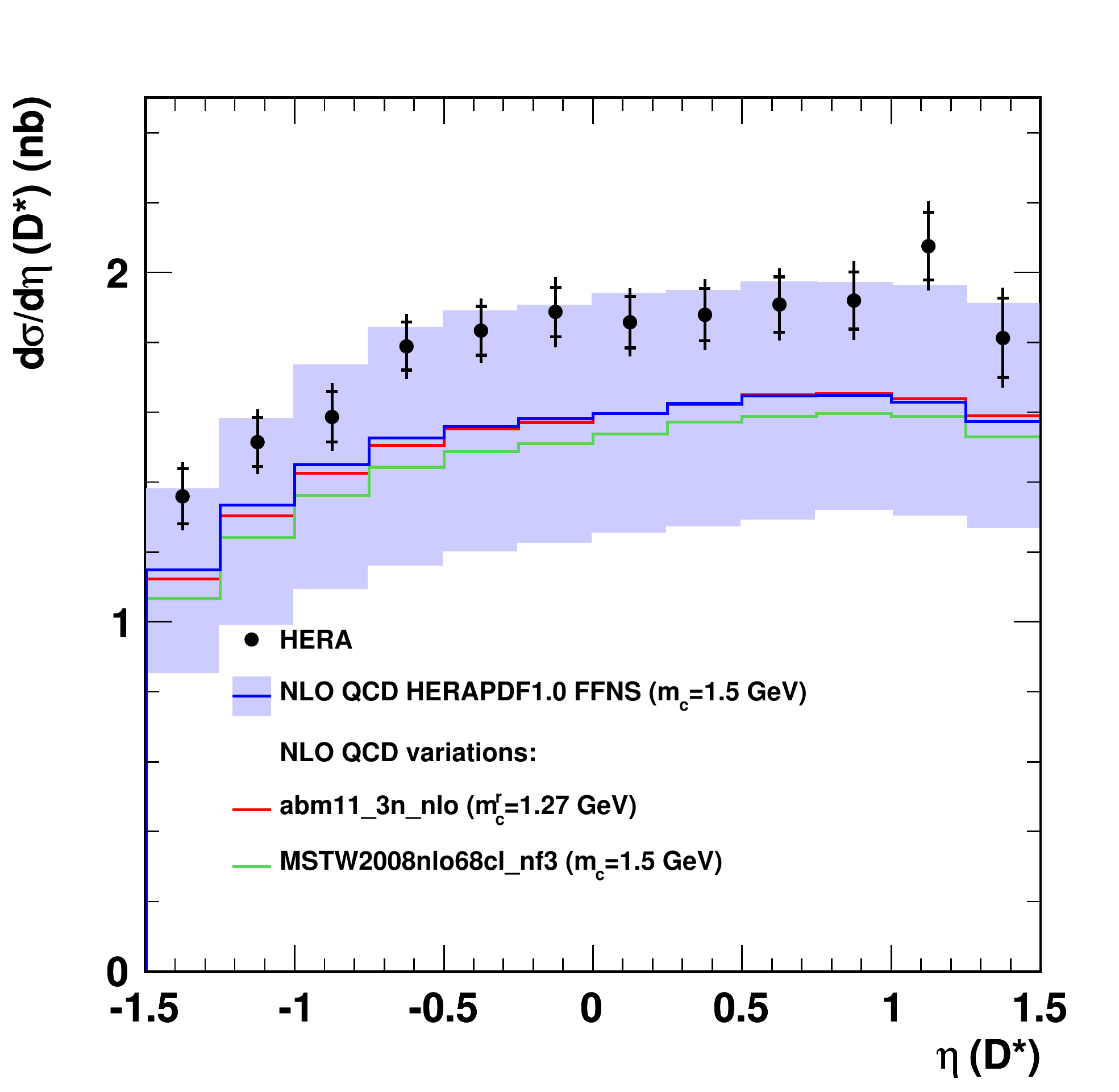}
	  \put(-35,80){(b)}\\
	  \includegraphics[width=1.0\textwidth,trim=0 0 0 15mm,clip=true]{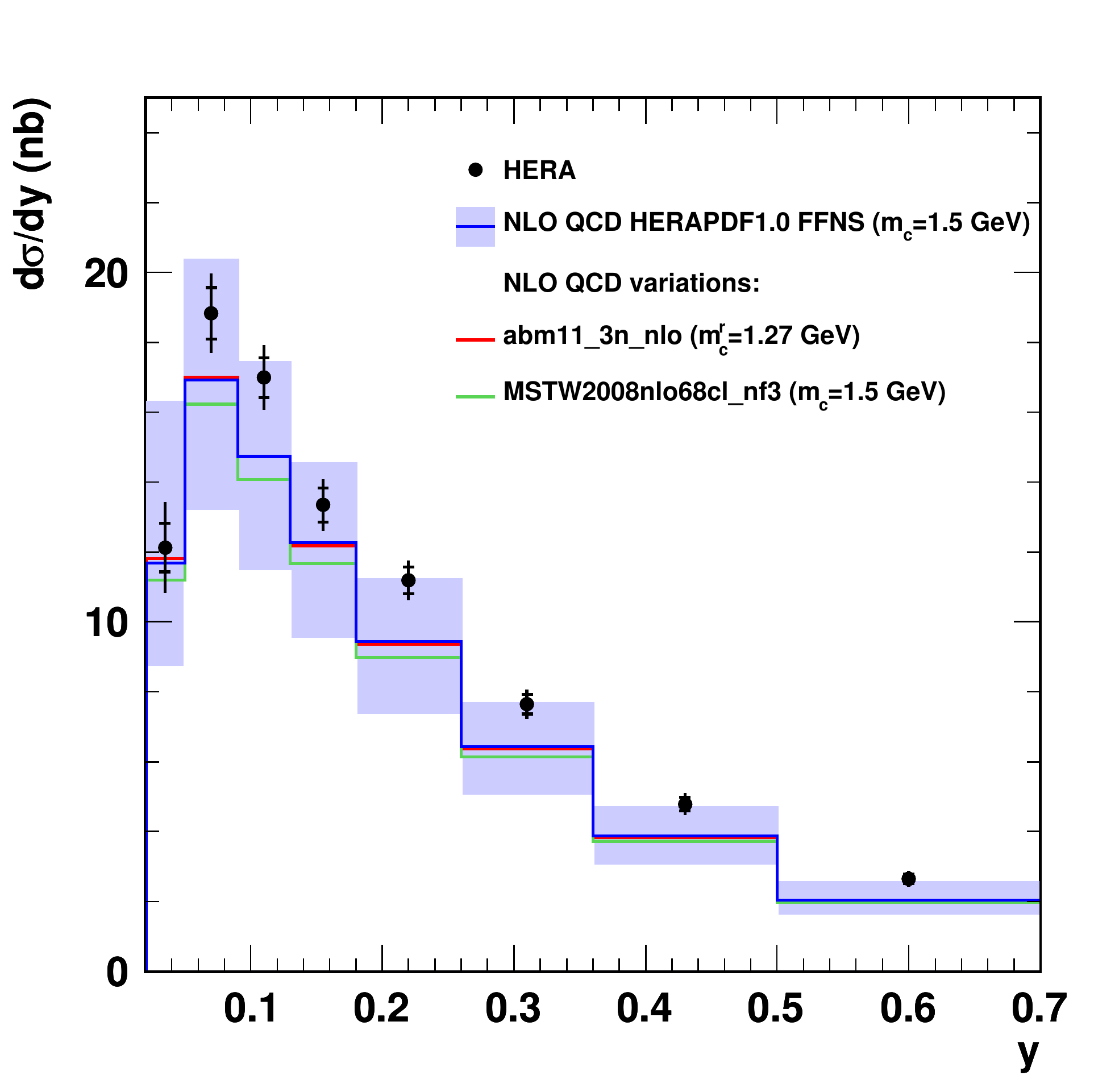}
	  \put(-35,80){(e)}\\
  \end{minipage}
  \begin{minipage}[t]{0.33\textwidth}
	  \includegraphics[width=1.0\textwidth,trim=0 0 0 15mm,clip=true]{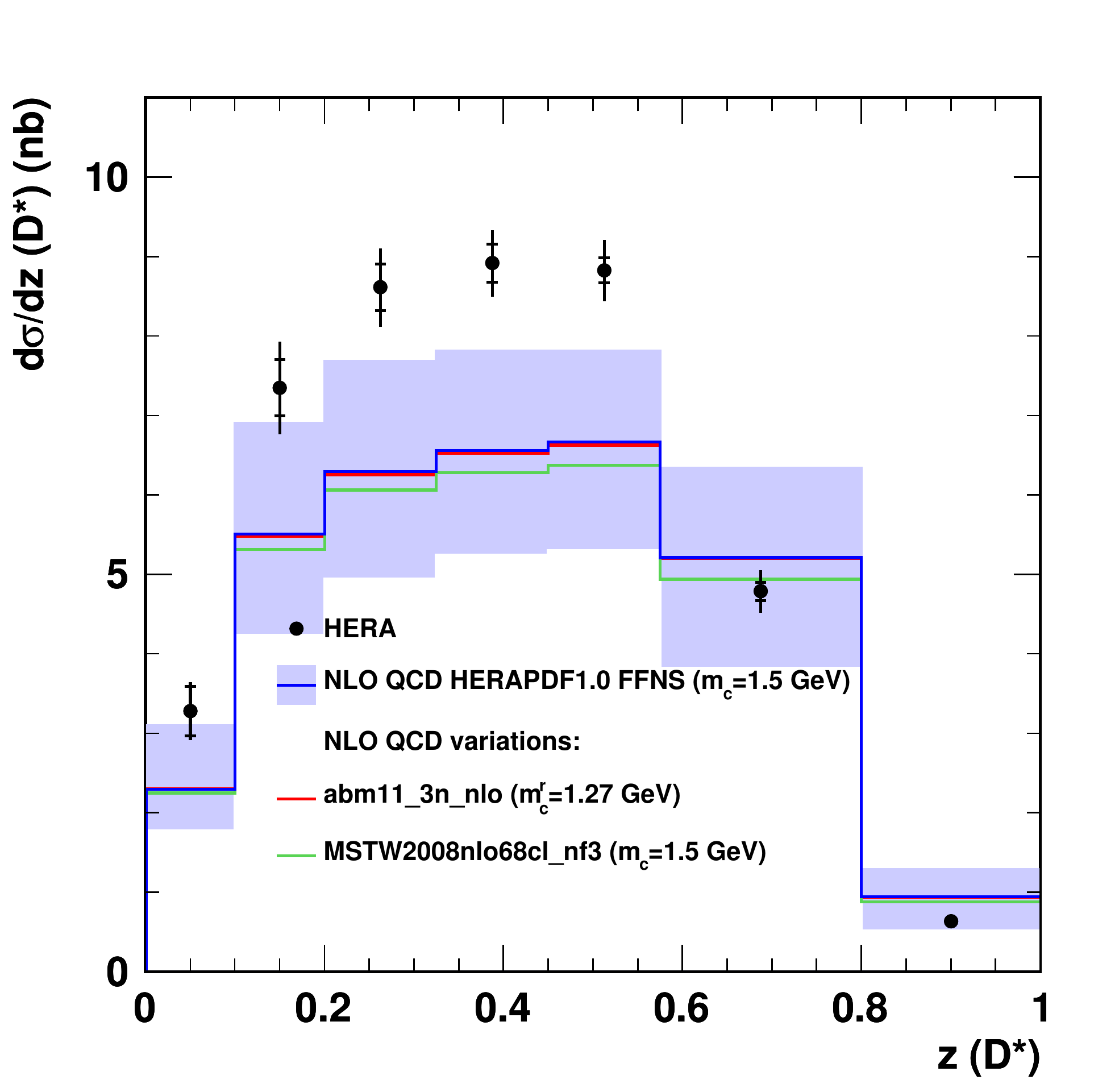}
	  \put(-35,80){(c)}
	  \caption[\Dstar cross sections compared to NLO predictions with different PDFs]
	  {Single-differential \Dstar cross section as a function of $p_T(\Dstar)$ (a), $\eta(\Dstar)$ (b), $z(\Dstar)$ (c), $Q^2$ (d) and $y$ (e) compared to NLO predictions with various PDFs~\cite{DIScomb,abm11,Martin:2010db}.}
	  \label{fig:comb:dstar:single:pdfvar}
  \end{minipage}
\end{figure*}

Fig.~\ref{fig:comb:dstar:single:pdfvar} shows comparison of the combined data to the theoretical variations obtained with different PDFs 
(see Section~\ref{sec:comb:th}).

\begin{table*}[tbp]
\caption[Sources of  correlated uncertainties in combination of visible \Dstar cross sections]
{Sources of bin-to-bin correlated uncertainties
  considered in the combination of the visible \Dstar cross sections. For each source the affected datasets, name, type (see Section~\ref{sec:comb:proc:unc}) and reference to the place, 
  where information can be found, are given, together with the shift (sh) and reduction factor (red) in the combination obtained after the first iteration. For sources which do not affect the combination of a given differential cross section, no shifts and reductions are quoted.}
\label{tab:comb:dstar:syst}
\footnotesize
\begin{center}
\tabcolsep1.7mm
\renewcommand*{\arraystretch}{1.2}
\begin{tabu} to \textwidth{|l|l|X[l]|l|l|>{$}r<{$}|>{$}r<{$}|>{$}r<{$}|>{$}r<{$}|>{$}r<{$}|>{$}r<{$}|>{$}r<{$}|>{$}r<{$}|>{$}r<{$}|>{$}r<{$}|>{$}r<{$}|>{$}r<{$}|} \hline
Desc- & Da- & \multirow{2}{*}{Name}&Ty-& Refe- & \multicolumn{2}{c}{$\frac{{\rm d}\sigma}{{\rm d}Q^2}$} & \multicolumn{2}{|c}{$\frac{{\rm d}\sigma}{{\rm d}y}$} & \multicolumn{2}{|c}{$\frac{{\rm d}\sigma}{{\rm d}p_T(\Dstar)}$} & \multicolumn{2}{|c}{$\frac{{\rm d}\sigma}{{\rm d}\eta(\Dstar)}$} & \multicolumn{2}{|c}{$\frac{{\rm d}\sigma}{{\rm d}z(\Dstar)}$} & \multicolumn{2}{|c|}{$\frac{{\rm d}^2\sigma}{{\rm d}Q^2{\rm d}y}$} \\ \cline{6-17}
riptor&taset&&pe&rence&\text{sh}&\text{red}&\text{sh}&\text{red}&\text{sh}&\text{red}&\text{sh}&\text{red}&\text{sh}&\text{red}&\text{sh}&\text{red} \\ \hline
          $\delta_{1}$& I,II     &H1 CJC efficiency                  &S& H1 Col. &   0.8&  0.9&   0.3&  0.9&   0.5&  0.9&   0.5&  0.9&   0.4&  0.9&   0.6&  0.8 \\ \hline
    			$\delta_{2}$&  I,II     &H1 luminosity               &N& H1 Col.  & 0.5&  0.9&   0.4&  0.9&   0.6&  0.9&   0.6&  0.9&   0.4&  0.9&   0.1&  0.9 \\ \hline
          $\delta_{3}$& I,II     & H1 MC PDF                         &S& H1 Col.  &  0.1&  1.0&   0.1&  1.0&   0.2&  1.0&   0.2&  1.0&   0.1&  1.0&   0.0&  1.0 \\ \hline
          $\delta_{4}$& I,II     & H1 electron energy                &S& H1 Col.  &  0.2&  1.0&   0.0&  1.0&   0.0&  1.0&   0.0&  1.0&   0.7&  0.9&   0.0&  0.8 \\ \hline
          $\delta_{5}$& I,II     & H1 electron polar angle           &S& H1 Col.  &  0.2&  1.0&   0.1&  1.0&   0.1&  1.0&   0.2&  1.0&   0.2&  1.0&   0.3&  0.9 \\ \hline
          $\delta_{6}$& I,II     & H1 hadronic en. scale          &S& H1 Col.  &  0.1&  1.0&   0.2&  0.9&   0.0&  1.0&   0.0&  1.0&  $-1.0$&  0.7&   0.0&  1.0 \\ \hline
          $\delta_{7}$& II       & H1 frag. thres., high $Q^2$  &S& H1 Col. &  0.0&  1.0&      &     &      &     &      &     &      &     &   0.0&  1.0 \\ \hline
          $\delta_{8}$& I,II     & H1 alternat. MC model           &S& H1 Col.  &  0.4&  0.9&   0.4&  0.9&   0.1&  1.0&   0.0&  1.0&  -1.0&  0.8&   1.2&  0.7 \\ \hline
          $\delta_{9}$& I,II     & H1 alternat. MC frag.           &S& H1 Col.  &  0.0&  1.0&   0.0&  1.0&   0.0&  1.0&  -0.1&  1.0&   0.2&  1.0&   0.3&  0.9 \\ \hline
         $\delta_{10}$& I,II     & H1 frag. thresh.        &S& H1 Col.  &  0.0&  1.0&  -0.4&  0.9&   0.2&  1.0&   0.0&  1.0&   0.6&  0.9&   0.2&  0.8 \\ \hline
         $\delta_{11}$& I        & H1 high $Q^2$ uncertainty            &N& \cite{h1dstarhighQ2}  &     &     &   0.1&  1.0&   0.0&  0.9&   0.1&  1.0&   0.1&  1.0&    &        \\ \hline
         $\delta_{12}$&  III       &  ZEUS hadron. en. scale     &S& \cite{zeusdstar_hera2} & 0.0&  1.0&  -0.1&  0.8&   0.0&  1.0&   0.0&  1.0&  -0.9&  0.9&  -0.5&  0.7 \\ \hline
         $\delta_{13}$&  III       &   ZEUS electron en. scale    &S& \cite{zeusdstar_hera2}      & 0.1&  0.9&   0.2&  0.9&   0.0&  1.0&   0.2&  1.0&   0.0&  1.0&   0.4&  0.7 \\ \hline
         $\delta_{14}$&  III       &   ZEUS $p_T(\pi_s)$ correction   &S& \cite{zeusdstar_hera2}      & -0.1&  1.0&  -0.1&  1.0&  -0.1&  1.0&  -0.3&  1.0&   0.0&  1.0&  -0.7&  0.9 \\ \hline
         $\delta_{15}$&  III       &   ZEUS $M(K\pi)$ cut var.&S& \cite{zeusdstar_hera2}     & -0.3&  0.8&  -0.7&  0.8&   0.4&  0.6&  -0.3&  0.7&   0.5&  0.8&  -0.7&  0.9 \\ \hline
         $\delta_{16}$&  III       &   ZEUS track. efficiency      &S& \cite{zeusdstar_hera2}      & -0.2&  0.9&  -0.4&  0.9&  -0.4&  0.9&  -0.2&  0.9&  -0.2&  0.9&  -0.7&  1.0 \\ \hline
         $\delta_{17}$&  III       &   ZEUS $b$ MC norm.           &S& \cite{zeusdstar_hera2}& 0.0&  1.0&   0.0&  1.0&   0.0&  1.0&   0.0&  1.0&   0.1&  1.0&   0.0&  1.0 \\ \hline
         $\delta_{18}$&  III       &   ZEUS PHP MC norm.           &S& \cite{zeusdstar_hera2}& 0.0&  1.0&  -0.1&  1.0&   0.0&  1.0&  -0.1&  1.0&   0.1&  1.0&  -0.3&  1.0 \\ \hline
         $\delta_{19}$&  III       &   ZEUS diffr. MC norm.   &S& \cite{zeusdstar_hera2}& 0.0&  1.0&   0.1&  0.9&   0.2&  1.0&   0.0&  1.0&   0.0&  1.0&   0.7&  0.9 \\ \hline
         $\delta_{20}$&  III       &    ZEUS MC rew. $p_T$, $Q^2$&S& \cite{zeusdstar_hera2}& 0.3&  0.9&   0.0&  1.0&  -0.1&  1.0&   0.0&  1.0&   0.0&  1.0&   0.6&  0.9 \\ \hline
         $\delta_{21}$&  III       &   ZEUS MC rew. $\eta$   &S& \cite{zeusdstar_hera2}     & 0.0&  1.0&   0.0&  0.8&  -0.2&  1.0&  -0.3&  1.0&  -0.2&  1.0&   0.4&  0.8 \\ \hline
         $\delta_{22}$&  III       &  ZEUS lum. (HERA-II)      &N& \cite{zeusdstar_hera2}     &-0.2&  1.0&  -0.1&  1.0&  -0.2&  1.0&  -0.2&  1.0&  -0.1&  1.0&  -0.7&  0.9 \\ \hline
         $\delta_{23}$&  IV       &	ZEUS lum. (98-00)          &N& \cite{zd00}   &   &     &      &     &      &     &      &     &      &     &  0.8 &  0.9 \\ \hline
         $\delta_{24}$& 			I-IV       &        Theory $m_c$ variation            &T& Theory  & &   &     &    &     &    &     &    &     &    &    0.0&  1.0 \\ \hline
         $\delta_{25}$& 			I-IV       &        Theory $\mu_r$, $\mu_f$ variation &T& Theory  & &   &     &    &     &    &     &    &     &    &    0.0&  1.0 \\ \hline
         $\delta_{26}$& 			I-IV       &        Theory $\alpha_s$ variation       &T& Theory  & &   &     &    &     &    &     &    &     &    &    0.0&  1.0 \\ \hline
         $\delta_{27}$& 			I-IV       &        Theory longitud. frag.         &T& Theory & &    &     &    &     &    &     &    &     &    &    0.1&  1.0 \\ \hline
         $\delta_{28}$& 			I-IV       &        Theory transverse frag.           &T& Theory & &    &     &    &     &    &     &    &     &    &    0.0&  1.0 \\ \hline
\end{tabu}
\end{center}
\end{table*}

\begin{table*}[tbp]
\caption[Combined \Dstar cross section with correlations as function of $p_T(\Dstar)$ and $\eta(\Dstar)$]
{The combined single-differential \Dstar cross sections as a function of $p_T(\Dstar)$, $\eta(\Dstar)$, $z(\Dstar)$, $Q^2$ and $y$, with their 
uncorrelated ($\delta_{unc}$), correlated ($\delta_1$ to $\delta_{22}$) and external branching-ratio ($\delta_{br}$) uncertainties.
The cross sections are given in the kinematic region~\ref{eq:comb:dstar:single:phasespace}. 
} 
\label{tab:comb:dstar:single:combinedfull12}
\footnotesize
\tabcolsep0.15mm
\renewcommand*{\arraystretch}{1.0}
\begin{tabu} to \textwidth{|>{$}c<{$}|>{$}l<{$}|>{$}r<{$}|>{$}r<{$}|>{$}r<{$}|>{$}r<{$}|>{$}r<{$}|>{$}r<{$}|>{$}r<{$}|>{$}r<{$}|>{$}r<{$}|>{$}r<{$}|>{$}r<{$}|>{$}r<{$}|>{$}r<{$}|>{$}r<{$}|>{$}r<{$}|>{$}r<{$}|>{$}r<{$}|>{$}r<{$}|>{$}r<{$}|>{$}r<{$}|>{$}r<{$}|>{$}r<{$}|>{$}r<{$}|>{$}r<{$}|} \hline
p_T(\Dstar)&\frac{{\rm d}\sigma}{{\rm d}p_T(\Dstar)}&\delta_{unc}&\delta_1&\delta_2&\delta_3&\delta_4&\delta_5&\delta_6&\delta_7&\delta_8&\delta_9&\delta_{10}&\delta_{11}&\delta_{12}&\delta_{13}&\delta_{14}&\delta_{15}&\delta_{16}&\delta_{17}&\delta_{18}&\delta_{19}&\delta_{20}&\delta_{21}&\delta_{22}&\delta_{br} \\
$[\SI{}{GeV}]$&$[\SI{}{nb/GeV}]$&[\%]&[\%]&[\%]&[\%]&[\%]&[\%]&[\%]&[\%]&[\%]&[\%]&[\%]&[\%]&[\%]&[\%]&[\%]&[\%]&[\%]&[\%]&[\%]&[\%]&[\%]&[\%]&[\%]&[\%] \\ \hline
 1.50: 1.88&   2.358&   6.4&  -0.4&   0.8&   0.8&  -1.0&   0.4&   0.0&  -1.8&   0.3&   0.1&  -1.0&   0.6&   1.0&  -0.1&   0.7&   0.6&  -0.4&  -1.4&   0.2&  -0.5&  -0.8&  -2.7&   0.4&   1.5 \\ 
 1.88: 2.28&   2.227&   4.8&  -0.5&   0.5&   0.7&  -0.9&   0.3&   0.0&  -1.9&   0.3&   0.2&  -1.0&   0.6&   0.8&  -0.2&   0.6&   0.5&  -0.4&  -1.0&   0.5&  -0.3&  -0.4&  -2.1&   0.3&   1.5 \\ 
 2.28: 2.68&   1.984&   3.7&  -0.4&   0.6&   0.5&  -0.9&   0.3&   0.0&  -1.9&   0.3&   0.2&  -1.0&   0.7&   1.0&  -0.1&   0.7&   0.4&  -0.3&  -0.9&   0.5&  -0.5&  -0.0&  -1.7&   0.6&   1.5 \\ 
 2.68: 3.08&   1.559&   3.5&  -0.4&   0.6&   0.5&  -0.8&   0.3&   0.0&  -1.9&   0.3&   0.2&  -1.0&   0.6&   1.0&  -0.2&   0.7&   0.4&  -0.8&  -0.5&   0.2&  -0.2&   0.2&  -0.9&   0.4&   1.5 \\ 
 3.08: 3.50&   1.209&   3.6&  -0.4&   0.6&   0.5&  -0.8&   0.3&   0.0&  -1.8&   0.2&   0.1&  -0.9&   0.7&   1.1&  -0.3&   0.5&   0.3&  -0.4&  -0.4&   0.3&  -0.1&   0.1&  -0.7&   0.3&   1.5 \\ 
 3.50: 4.00&   0.9328&   3.2&  -0.4&   0.7&   0.3&  -0.8&   0.3&   0.0&  -1.7&   0.2&   0.0&  -0.9&   0.7&   1.0&  -0.2&   0.6&   0.3&  -0.4&  -0.3&   0.0&  -0.2&   0.1&  -0.7&   0.4&   1.5 \\ 
 4.00: 4.75&   0.6161&   3.0&  -0.4&   0.6&   0.3&  -0.7&   0.3&   0.0&  -1.7&   0.1&  -0.0&  -0.9&   0.8&   1.1&  -0.1&   0.5&   0.5&  -0.4&  -0.2&   0.4&   0.0&   0.6&   0.6&   0.0&   1.5 \\ 
 4.75: 6.00&   0.3204&   3.0&  -0.4&   0.6&   0.3&  -0.7&   0.3&   0.0&  -1.6&   0.1&  -0.1&  -0.9&   0.8&   1.0&  -0.2&   0.6&   0.3&  -0.3&  -0.0&   0.1&  -0.2&   0.5&  -0.4&   0.3&   1.5 \\ 
 6.00: 8.00&   0.1152&   3.8&  -0.3&   0.7&   0.2&  -0.6&   0.2&   0.0&  -1.6&   0.1&  -0.0&  -0.9&   0.6&   1.1&  -0.2&   0.5&   0.2&  -0.0&  -0.2&   0.4&   0.5&   1.3&   1.0&  -0.3&   1.5 \\ 
 8.00:11.00&   0.03334&   5.3&  -0.3&   0.7&   0.3&  -0.7&   0.3&   0.0&  -1.5&   0.1&  -0.1&  -0.9&   0.7&   1.0&  -0.1&   0.1&   0.4&   0.5&  -0.3&  -0.1&  -0.6&   1.4&  -1.1&   0.2&   1.5 \\ 
11.00:20.00&   0.003819&  10.3&  -0.3&   0.6&   0.5&  -0.8&   0.4&   0.0&  -1.4&   0.1&  -0.0&  -0.8&   0.5&   1.1&   0.1&   0.1&   0.1&   0.7&   0.2&  -1.6&  -0.6&   0.8&  -5.1&   1.4&   1.5 \\ \hline
\hline
\eta(\Dstar)&\frac{{\rm d}\sigma}{{\rm d}\eta(\Dstar)}&\delta_{unc}&\delta_1&\delta_2&\delta_3&\delta_4&\delta_5&\delta_6&\delta_7&\delta_8&\delta_9&\delta_{10}&\delta_{11}&\delta_{12}&\delta_{13}&\delta_{14}&\delta_{15}&\delta_{16}&\delta_{17}&\delta_{18}&\delta_{19}&\delta_{20}&\delta_{21}&\delta_{22}&\delta_{br} \\ 
&[$\SI{}{nb}$]&[\%]&[\%]&[\%]&[\%]&[\%]&[\%]&[\%]&[\%]&[\%]&[\%]&[\%]&[\%]&[\%]&[\%]&[\%]&[\%]&[\%]&[\%]&[\%]&[\%]&[\%]&[\%]&[\%]&[\%] \\ \hline
-1.50:-1.25&   1.360&   5.8&   1.1&   1.2&  -0.5&  -1.1&  -0.9&  -0.0&  -0.6&  -0.5&   0.3&  -0.2&  -0.3&  -0.0&  -0.1&   0.1&   0.6&   0.6&  -0.9&  -0.9&   1.2&  -2.5&  -0.6&  -0.1&   1.5 \\ 
-1.25:-1.00&   1.515&   4.6&   1.0&   1.3&  -0.4&  -0.9&  -0.7&  -0.0&  -0.4&  -0.5&   0.4&  -0.2&  -0.3&  -0.2&   0.4&   0.5&   0.7&   0.8&  -0.7&  -1.1&   0.9&  -1.9&  -0.8&   0.0&   1.5 \\ 
-1.00:-0.75&   1.587&   4.6&   1.0&   1.3&  -0.4&  -0.9&  -0.7&  -0.0&  -0.4&  -0.5&   0.5&  -0.3&  -0.3&  -0.2&   0.4&   0.7&   0.7&   0.6&  -1.0&  -1.1&   0.8&  -1.6&  -1.3&   0.0&   1.5 \\ 
-0.75:-0.50&   1.789&   3.8&   1.1&   1.3&  -0.3&  -1.0&  -0.6&  -0.0&  -0.4&  -0.5&   0.4&  -0.2&  -0.3&  -0.1&   0.2&   0.4&   0.5&   0.4&  -0.7&  -0.7&   0.6&  -1.7&  -0.6&   0.2&   1.5 \\ 
-0.50:-0.25&   1.833&   3.8&   1.1&   1.2&  -0.3&  -1.0&  -0.6&  -0.0&  -0.3&  -0.6&   0.3&  -0.1&  -0.2&  -0.1&   0.1&   0.3&   0.5&   0.2&  -0.6&  -0.7&   0.5&  -1.6&   0.3&  -0.0&   1.5 \\
-0.25: 0.00&   1.887&   3.8&   1.2&   1.3&  -0.2&  -1.0&  -0.5&  -0.0&  -0.3&  -0.5&   0.4&  -0.1&  -0.2&  -0.1&   0.0&   0.2&   0.6&   0.4&  -0.7&  -0.6&   0.9&  -1.4&  -1.5&   0.3&   1.5 \\ 
 0.00: 0.25&   1.857&   4.0&   1.2&   1.3&  -0.2&  -1.0&  -0.5&  -0.0&  -0.3&  -0.6&   0.4&  -0.1&  -0.1&  -0.1&   0.0&  -0.0&   0.1&   0.4&  -0.3&  -0.5&   0.5&  -1.6&   1.0&  -0.1&   1.5 \\ 
 0.25: 0.50&   1.879&   4.0&   1.1&   1.3&  -0.2&  -1.0&  -0.5&  -0.0&  -0.2&  -0.5&   0.5&  -0.1&  -0.1&  -0.1&   0.2&   0.1&   0.5&   0.7&  -0.6&  -0.5&   0.8&  -1.4&  -1.4&   0.3&   1.5 \\
 0.50: 0.75&   1.909&   4.1&   1.1&   1.3&  -0.2&  -1.0&  -0.5&  -0.0&  -0.3&  -0.5&   0.5&  -0.1&  -0.1&  -0.1&   0.2&   0.3&   0.5&   0.4&  -0.8&  -0.6&   0.5&  -1.6&  -0.9&   0.2&   1.5 \\ 
 0.75: 1.00&   1.920&   4.3&   1.1&   1.3&  -0.2&  -1.0&  -0.5&  -0.0&  -0.3&  -0.5&   0.4&  -0.1&  -0.1&  -0.1&   0.2&   0.3&   0.4&   0.8&  -1.3&  -0.6&   0.6&  -1.8&  -1.6&   0.2&   1.5 \\ 
 1.00: 1.25&   2.075&   4.7&   1.1&   1.3&  -0.2&  -0.9&  -0.5&  -0.0&  -0.3&  -0.6&   0.4&  -0.2&  -0.1&  -0.2&   0.2&   0.2&   0.7&   0.7&  -1.0&  -0.4&   0.3&  -2.3&   1.3&  -0.3&   1.5 \\ 
 1.25: 1.50&   1.813&   6.3&   1.0&   1.4&  -0.3&  -0.9&  -0.5&  -0.0&  -0.3&  -0.5&   0.4&  -0.2&  -0.2&  -0.2&   0.4&   0.6&   0.7&   0.7&  -1.4&  -1.3&   0.4&  -2.0&  -2.5&   0.2&   1.5 \\ \hline
\hline
z(\Dstar)&\frac{{\rm d}\sigma}{{\rm d}z(\Dstar)}&\delta_{unc}&\delta_1&\delta_2&\delta_3&\delta_4&\delta_5&\delta_6&\delta_7&\delta_8&\delta_9&\delta_{10}&\delta_{11}&\delta_{12}&\delta_{13}&\delta_{14}&\delta_{15}&\delta_{16}&\delta_{17}&\delta_{18}&\delta_{19}&\delta_{20}&\delta_{21}&\delta_{22}&\delta_{br} \\ 
&$[\SI{}{nb}]$&[\%]&[\%]&[\%]&[\%]&[\%]&[\%]&[\%]&[\%]&[\%]&[\%]&[\%]&[\%]&[\%]&[\%]&[\%]&[\%]&[\%]&[\%]&[\%]&[\%]&[\%]&[\%]&[\%]&[\%] \\ \hline
0.000:0.100&   3.277&   9.5&   1.7&   1.9&  -0.0&   1.1&   0.5&  -1.0&   1.0&   0.5&  -0.9&   0.4&   0.2&   0.7&   0.1&   0.1&  -0.1&  -1.1&   3.4&  -2.0&   1.1&  -1.3&   0.7&   0.6&   1.5 \\ 
0.100:0.200&   7.346&   4.8&   1.8&   1.2&  -0.2&   1.1&   0.2&  -1.0&   0.8&   0.4&  -0.7&   0.1&   0.3&   0.7&   0.4&   0.1&   0.3&  -1.4&   2.5&  -1.5&   2.9&  -2.9&   0.9&   0.8&   1.5 \\ 
0.200:0.325&   8.612&   3.5&   2.0&   1.0&  -0.1&   0.9&   0.2&  -0.6&   0.6&   0.5&  -0.2&   0.1&   0.3&   0.5&   0.6&   0.1&   0.3&  -1.2&   2.0&   0.3&   0.9&  -2.0&   0.6&   0.5&   1.5 \\ 
0.325:0.450&   8.918&   2.7&   2.1&   1.0&  -0.1&   0.9&   0.2&  -0.4&   0.4&   0.5&  -0.0&   0.0&   0.4&   0.4&   0.6&  -0.0&   0.3&  -0.7&   1.6&   1.2&  -0.7&  -0.3&   0.0&   0.3&   1.5 \\ 
0.450:0.575&   8.827&   1.8&   2.3&   1.2&   0.0&   0.8&   0.4&  -0.0&   0.3&   0.5&   0.3&   0.0&   0.4&   0.3&   0.4&  -0.1&   0.3&  -0.4&   0.7&   1.6&   0.9&  -0.1&  -0.1&   0.8&   1.5 \\ 
0.575:0.800&   4.785&   2.4&   2.2&   1.4&   0.1&   0.9&   0.6&  -0.0&   0.6&   0.5&   0.2&   0.1&   0.6&   0.4&   0.1&  -0.1&   0.2&   0.8&  -0.5&   2.5&   2.5&   0.4&  -0.5&   0.7&   1.5 \\ 
0.800:1.000&   0.6308&   8.1&   1.6&   0.8&  -0.3&   2.4&   1.0&  -1.5&   1.5&   0.2&  -1.3&  -0.2&   1.3&   1.0&  -0.3&  -0.1&   0.9&   3.8&  -2.8&   3.8&   4.4&   4.0&   3.1&  -0.4&   1.5 \\ \hline
\hline
Q^2&\frac{{\rm d}\sigma}{{\rm d}Q^2}&\delta_{unc}&\delta_1&\delta_2&\delta_3&\delta_4&\delta_5&\delta_6&\delta_7&\delta_8&\delta_9&\delta_{10}&\delta_{11}&\delta_{12}&\delta_{13}&\delta_{14}&\delta_{15}&\delta_{16}&\delta_{17}&\delta_{18}&\delta_{19}&\delta_{20}&\delta_{21}&\delta_{22}&\delta_{br} \\
$[\SI{}{GeV^2}]$&$[\SI{}{nb/GeV^2}]$&[\%]&[\%]&[\%]&[\%]&[\%]&[\%]&[\%]&[\%]&[\%]&[\%]&[\%]&[\%]&[\%]&[\%]&[\%]&[\%]&[\%]&[\%]&[\%]&[\%]&[\%]&[\%]&[\%]&[\%] \\ \hline
   5:   8&   0.4735&   4.0&   1.4&  -1.4&  -0.9&   0.9&  -0.9&   0.6&  -0.9&  -0.3&   1.1&  -0.2&   0.9&  -1.3&   0.7&   0.7&   0.8&   1.5&  -1.0&  -0.6&   0.7&  -2.1&  -1.2&   0.3&   1.5 \\ 
   8:  10&   0.2964&   4.3&   1.2&  -1.3&  -0.7&   0.7&  -1.0&   0.8&  -0.6&  -0.4&   1.0&   0.0&   0.8&  -1.0&   0.7&   0.4&   1.1&   1.1&  -0.2&  -0.1&  -0.1&   0.5&  -0.5&  -0.1&   1.5 \\ 
  10:  13&   0.2117&   3.8&   0.8&  -1.9&  -0.8&   0.8&  -0.6&   0.8&  -0.5&  -0.6&   1.0&  -0.1&   0.6&  -0.6&   0.4&  -0.1&   0.9&   1.0&  -0.9&   0.7&  -0.3&  -1.2&  -0.5&   0.4&   1.5 \\ 
  13:  19&   0.1236&   3.2&   0.8&  -1.8&  -0.8&   0.8&  -0.7&   0.8&  -0.4&  -0.7&   1.0&  -0.1&   0.6&  -0.6&   0.6&   0.0&   0.8&   0.8&  -0.0&   0.0&  -1.0&  -0.9&  -0.3&   0.5&   1.5 \\
  19:  28&   0.07263&   3.5&   0.8&  -1.8&  -0.7&   0.7&  -0.7&   0.8&  -0.3&  -0.7&   0.9&  -0.1&   0.6&  -0.6&   0.5&   0.2&   0.7&   0.7&  -0.0&   0.2&  -0.9&  -0.6&  -0.3&   0.4&   1.5 \\ 
  28:  40&   0.03970&   3.7&   0.7&  -1.9&  -0.8&   0.8&  -0.6&   0.8&  -0.3&  -0.8&   1.0&  -0.2&   0.5&  -0.5&   0.3&   0.0&   0.8&   0.6&  -0.6&   0.5&  -0.7&  -1.7&  -0.3&   0.6&   1.5 \\ 
  40:  60&   0.01635&   4.4&   0.7&  -2.0&  -0.8&   0.8&  -0.5&   0.7&  -0.3&  -0.7&   0.8&  -0.3&   0.6&  -0.4&   0.6&  -0.4&   0.5&   0.9&   0.5&  -0.0&  -1.5&  -2.6&  -0.6&   0.7&   1.5 \\ 
  60: 100&   0.007445&   5.2&   0.8&  -1.9&  -0.8&   0.9&  -0.5&   0.7&  -0.4&  -0.6&   0.9&  -0.3&   0.6&  -0.4&   0.7&  -0.3&   0.1&   0.2&   0.2&   0.9&  -1.6&  -0.4&   0.3&  -0.0&   1.5 \\ 
 100: 158&   0.002081&   7.2&   0.1&  -2.3&  -0.4&  -0.6&  -0.8&   0.3&  -0.2&  -0.8&   1.0&   0.0&   0.6&  -0.0&  -0.4&  -1.3&   0.7&   0.8&  -1.6&   1.8&  -2.5&  -0.8&  -1.3&  -0.5&   1.5 \\ 
 158: 251&   0.0008817&   7.6&   0.3&  -1.8&  -0.5&  -0.4&  -0.9&   0.2&  -0.4&  -0.8&   1.1&  -0.0&   0.7&  -0.3&  -0.0&  -1.0&   1.1&   0.0&  -0.2&   2.1&  -2.1&  -0.8&  -2.0&  -0.6&   1.5 \\ 
 251:1000&   0.0000749&  11.4&  -0.1&  -2.3&  -0.6&  -0.4&  -0.5&   0.2&  -0.2&  -0.9&   0.9&  -0.0&   0.6&   0.6&   0.2&  -1.6&   0.3&  -0.1&   0.5&   3.7&  -2.9&  -0.6&  -3.1&  -0.9&   1.5 \\ \hline
\hline
y&\frac{{\rm d}\sigma}{{\rm d}y}&\delta_{unc}&\delta_1&\delta_2&\delta_3&\delta_4&\delta_5&\delta_6&\delta_7&\delta_8&\delta_9&\delta_{10}&\delta_{11}&\delta_{12}&\delta_{13}&\delta_{14}&\delta_{15}&\delta_{16}&\delta_{17}&\delta_{18}&\delta_{19}&\delta_{20}&\delta_{21}&\delta_{22}&\delta_{br} \\
&$[\SI{}{nb}]$&[\%]&[\%]&[\%]&[\%]&[\%]&[\%]&[\%]&[\%]&[\%]&[\%]&[\%]&[\%]&[\%]&[\%]&[\%]&[\%]&[\%]&[\%]&[\%]&[\%]&[\%]&[\%]&[\%]&[\%] \\ \hline
0.02:0.05&  12.13&   5.8&  -0.6&   0.6&  -0.2&   0.8&   1.0&  -0.6&  -0.1&   1.4&   1.0&  -0.1&   0.0&  -2.3&  -0.2&   0.8&  -2.6&  -0.7&   5.4&   4.8&   2.7&   1.4&   0.4&  -0.6&   1.5 \\ 
0.05:0.09&  18.84&   3.9&  -1.3&   1.5&   0.2&   0.4&   1.0&   0.4&  -0.2&  -0.1&  -0.2&  -0.2&  -0.4&  -1.9&  -0.8&   0.1&   0.2&  -0.7&   1.6&   0.4&   2.2&  -0.7&   0.7&   0.2&   1.5 \\ 
0.09:0.13&  16.99&   3.4&  -1.3&   1.7&   0.2&   0.4&   0.8&   0.8&  -0.3&  -0.5&  -0.6&  -0.1&  -0.7&  -1.7&  -0.9&  -0.2&   0.4&  -0.9&   0.7&  -0.3&   1.9&  -0.7&  -0.1&   0.7&   1.5 \\ 
0.13:0.18&  13.35&   3.7&  -1.2&   1.3&   0.2&   0.4&   0.9&   0.7&  -0.4&   0.1&  -0.2&   0.0&  -0.6&  -1.9&  -0.7&   0.0&   0.4&  -0.4&  -0.5&   0.1&   1.4&  -1.8&  -0.0&   0.5&   1.5 \\ 
0.18:0.26&  11.19&   3.4&  -1.4&   1.5&   0.0&   0.5&   0.6&   0.9&  -0.4&  -0.2&  -0.5&   0.1&  -0.8&  -1.7&  -0.8&  -0.3&   0.2&  -0.5&  -0.2&  -0.2&   0.2&   0.6&  -0.2&   0.3&   1.5 \\
0.26:0.36&   7.649&   3.7&  -1.4&   1.5&   0.4&   0.2&   0.7&   1.0&  -0.5&  -0.1&  -0.5&   0.0&  -0.7&  -1.9&  -0.7&  -0.4&   1.0&  -1.0&  -0.5&  -1.0&   0.5&  -0.1&  -0.3&   0.4&   1.5 \\ 
0.36:0.50&   4.783&   4.0&  -1.7&   1.7&   0.6&   0.2&   0.7&   1.1&  -0.5&  -0.6&  -0.7&  -0.1&  -0.7&  -1.8&  -0.7&  -0.6&   2.1&  -0.2&  -0.4&  -2.4&   0.8&  -0.2&  -0.7&   0.7&   1.5 \\ 
0.50:0.70&   2.648&   5.6&  -2.2&   1.7&   0.9&  -0.4&   0.2&   0.8&  -0.4&  -0.4&  -1.0&  -0.6&  -0.5&  -2.1&  -0.5&  -0.8&   2.8&  -0.4&   0.3&  -2.8&  -1.4&   1.7&  -0.9&   0.3&   1.5 \\ \hline
\end{tabu}
\end{table*}

\begin{table*}[tbp]
\caption[Combined double-differential \Dstar cross section with correlations]
{The combined double-differential \Dstar cross section as a function of $Q^2$ and $y$, with its 
uncorrelated ($\delta_{unc}$), correlated ($\delta_1$ to $\delta_{28}$) and external branching-ratio ($\delta_{br}$) uncertainties.
The cross sections are given in the kinematic region~\ref{eq:comb:dstar:double:phasespace}. 
} 
\label{tab:comb:dstar:double:combinedfull3}
\scriptsize
\tabcolsep0.10mm
\renewcommand*{\arraystretch}{1.0}
\begin{tabu} to \textwidth{|>{$}c<{$}|>{$}c<{$}|>{$}l<{$}|>{$}r<{$}|>{$}r<{$}|>{$}r<{$}|>{$}r<{$}|>{$}r<{$}|>{$}r<{$}|>{$}r<{$}|>{$}r<{$}|>{$}r<{$}|>{$}r<{$}|>{$}r<{$}|>{$}r<{$}|>{$}r<{$}|>{$}r<{$}|>{$}r<{$}|>{$}r<{$}|>{$}r<{$}|>{$}r<{$}|>{$}r<{$}|>{$}r<{$}|>{$}r<{$}|>{$}r<{$}|>{$}r<{$}|>{$}r<{$}|>{$}r<{$}|>{$}r<{$}|>{$}r<{$}|>{$}r<{$}|>{$}r<{$}|>{$}r<{$}|} \hline
Q^2&y& \frac{d^2\sigma}{dQ^2dy}&\delta_{unc}&\delta_1&\delta_2&\delta_3&\delta_4&\delta_5&\delta_6&\delta_7&\delta_8&\delta_9&\delta_{10}&\delta_{11}&\delta_{12}&\delta_{13}&\delta_{14}&\delta_{15}&\delta_{16}&\delta_{17}&\delta_{18}&\delta_{19}&\delta_{20}&\delta_{21}&\delta_{22}&\delta_{23}&\delta_{24}&\delta_{25}&\delta_{26}&\delta_{27}&\delta_{28}&\delta_{br} \\ 
$[\SI{}{GeV^2}]$&&[\SI{}{nb/GeV^2}]&[\%]&[\%]&[\%]&[\%]&[\%]&[\%]&[\%]&[\%]&[\%]&[\%]&[\%]&[\%]&[\%]&[\%]&[\%]&[\%]&[\%]&[\%]&[\%]&[\%]&[\%]&[\%]&[\%]&[\%]&[\%]&[\%]&[\%]&[\%]&[\%]&[\%] \\ \hline
   1.5:   3.5&  0.02:0.09&   4.761&  12.9&   0.0&  -1.6&   0.0&   0.0&  -0.1&  -0.1&  -0.1&  -0.1&  -0.2&   0.1&   0.8&   0.0&   0.1&   0.3&  -0.2&  -0.0&  -0.0&   0.&5   0.0&  -0.2&   0.1&  -0.0&   0.1&  -0.5&   0.5&  -0.3&   0.0&   0.0&   1.5 \\
   1.5:   3.5&  0.09:0.16&   5.498&  11.3&   0.0&  -1.6&   0.0&   0.0&  -0.1&  -0.2&   0.1&   0.1&  -0.2&   0.1&   0.7&  -0.1&   0.1&   0.3&  -0.3&   0.3&   0.2&   0.5&   0.0&  -0.2&   0.1&  -0.0&   0.1&  -0.5&   0.6&  -0.3&   0.0&   0.0&   1.5 \\
   1.5:   3.5&  0.16:0.32&   2.994&  12.0&   0.0&  -1.6&   0.0&   0.0&  -0.1&  -0.1&   0.0&  -0.3&  -0.2&  -0.1&   0.7&  -0.2&   0.1&   0.3&  -0.3&   0.4&   0.3&   0.5&   0.0&  -0.3&   0.1&  -0.0&   0.1&  -0.5&   0.6&  -0.3&   0.0&   0.0&   1.5 \\
   1.5:   3.5&  0.32:0.70&   0.9211&  20.5&   0.0&  -1.6&   0.0&   0.0&  -0.1&  -0.2&   0.0&  -0.1&  -0.2&   0.0&   0.7&  -0.1&   0.1&   0.3&  -0.2&   0.1&   0.1&   0.5&   0.0&  -0.2&   0.1&  -0.0&   0.1&  -0.5&   0.5&  -0.3&   0.0&   0.0&   1.5 \\
   3.5:   5.5&  0.02:0.09&   2.2219&  11.3&   0.0&  -1.6&   0.0&   0.0&  -0.1&  -0.1&   0.1&  -0.2&  -0.2&   0.2&   0.8&   0.3&  -0.0&   0.3&   0.1&  -0.5&  -0.8&   0.7&   0.0&  -0.0&   0.0&  -0.1&   0.0&  -0.5&   0.5&  -0.3&   0.0&   0.0&   1.5 \\
   3.5:   5.5&  0.09:0.16&   1.9759&   7.9&   0.0&  -1.6&   0.0&   0.0&  -0.1&  -0.2&   0.2&   0.1&  -0.1&  -0.2&   0.8&   0.3&   0.3&   0.4&  -0.2&  -0.7&   0.1&   0.4&  -0.0&  -0.2&   0.1&  -0.0&   0.1&  -0.5&   0.5&  -0.3&   0.0&   0.0&   1.5 \\
   3.5:   5.5&  0.16:0.32&   1.0890&  20.2&   0.0&  -1.6&   0.0&   0.0&  -0.1&  -0.2&   0.4&  -0.4&  -0.2&  -0.0&   0.7&  -0.2&   0.2&   0.3&  -0.4&   0.2&   0.5&   0.4&  -0.0&  -0.3&   0.1&   0.0&   0.1&  -0.5&   0.6&  -0.3&   0.0&   0.0&   1.5 \\
   3.5:   5.5&  0.32:0.70&   0.3468&  14.6&   0.0&  -1.6&   0.0&   0.0&  -0.1&  -0.1&   0.2&  -0.2&  -0.1&  -0.4&   0.8&   0.1&   0.2&   0.4&  -0.2&  -0.2&   0.1&   0.5&  -0.0&  -0.2&   0.1&  -0.0&   0.1&  -0.5&   0.5&  -0.3&   0.0&   0.0&   1.5 \\
   5.5:   9.0&  0.02:0.05&   1.057&  12.3&   0.0&  -1.6&   0.0&   0.0&  -0.1&  -0.4&   0.3&   0.3&  -0.5&   0.1&   0.5&  -0.0&  -0.4&  -0.2&   0.6&  -0.2&  -0.0&   0.0&  -0.0&  -0.3&   0.4&   1.5&   0.5&   2.6&   0.8&   1.5&   0.4&  -1.0&   1.5 \\
   5.5:   9.0&  0.05:0.09&   1.461&   7.8&   0.0&  -1.6&   0.0&   0.0&  -0.1&  -0.3&   0.3&   0.1&  -0.2&   0.3&   0.8&  -0.1&  -0.3&   0.0&   0.4&   0.2&  -0.2&   0.0&  -0.6&  -0.3&  -0.7&   0.8&  -0.6&   2.5&   0.1&   0.9&   0.6&  -1.0&   1.5 \\
   5.5:   9.0&  0.09:0.16&   1.317&   5.4&   0.0&  -1.6&   0.0&   0.0&  -0.1&  -0.2&   0.3&  -0.0&  -0.3&   0.2&   0.8&  -0.0&   0.2&   0.2&  -0.3&   0.5&   0.1&   0.5&   0.5&  -1.3&  -1.3&  -0.2&  -0.7&   2.6&  -0.5&   0.6&  -0.3&  -0.7&   1.5 \\
   5.5:   9.0&  0.16:0.32&   0.7733&   4.9&   0.0&  -1.6&   0.0&   0.0&  -0.1&  -0.2&   0.2&   0.1&  -0.3&   0.1&   0.7&   0.1&   0.1&   0.1&  -0.3&   0.2&   0.2&   0.9&   0.3&  -0.4&  -0.8&  -0.8&  -0.0&   2.2&  -0.7&   0.9&   0.3&  -0.8&   1.5 \\
   5.5:   9.0&  0.32:0.70&   0.2509&   5.6&   0.0&  -1.6&   0.0&   0.0&  -0.2&  -0.2&   0.2&   0.1&  -0.2&   0.0&   0.8&   0.7&  -0.1&  -0.0&  -0.5&   0.4&   0.1&   0.6&   0.3&   0.1&  -0.5&  -1.9&   1.4&   1.6&  -1.1&   0.5&  -0.8&  -0.6&   1.5 \\
   9.0:  14.0&  0.02:0.05&   0.5201&  13.0&   0.0&  -1.6&   0.0&   0.0&  -0.1&  -0.3&   0.2&   0.1&  -0.9&  -0.1&   0.3&   0.0&  -1.0&  -0.4&   1.2&  -0.2&  -0.6&  -0.6&   0.6&  -2.4&   0.9&   4.0&   1.6&   1.6&   0.8&   2.5&  -0.0&  -0.2&   1.5 \\
   9.0:  14.0&  0.05:0.09&   0.7677&   6.6&   0.0&  -1.6&   0.0&   0.0&  -0.1&  -0.2&   0.2&   0.0&  -0.3&   0.1&   0.8&  -0.0&   0.1&   0.1&  -0.0&   0.1&  -0.3&   0.3&  -0.2&  -1.0&  -0.3&   1.0&  -0.4&   1.6&  -0.2&   2.1&  -0.1&  -0.4&   1.5 \\
   9.0:  14.0&  0.09:0.16&   0.5686&   4.6&   0.0&  -1.6&   0.0&   0.0&  -0.1&  -0.2&   0.2&  -0.0&  -0.2&   0.1&   0.8&  -0.1&   0.2&   0.3&  -0.4&   0.2&   0.0&   0.4&  -0.2&  -0.4&  -0.1&   0.1&  -0.7&   0.6&  -0.7&   0.7&   0.1&  -0.2&   1.5 \\
   9.0:  14.0&  0.16:0.32&   0.4121&   4.6&   0.0&  -1.6&   0.0&   0.0&  -0.1&  -0.2&   0.1&  -0.0&  -0.3&  -0.1&   0.7&  -0.4&   0.0&   0.1&  -0.4&   0.1&   0.2&   0.5&   0.3&   0.3&   0.1&  -0.5&   0.2&   0.9&  -1.3&   0.8&  -0.1&  -0.1&   1.5 \\
   9.0:  14.0&  0.32:0.70&   0.1506&   5.6&   0.0&  -1.6&   0.0&   0.0&  -0.1&  -0.2&   0.1&  -0.0&  -0.2&  -0.0&   0.7&   0.1&   0.1&   0.3&  -0.4&   0.3&   0.3&   0.4&  -0.7&   0.4&  -0.1&  -1.9&   1.1&   0.6&  -2.0&   0.2&   0.0&  -0.4&   1.5 \\
  14.0:  23.0&  0.02:0.05&   0.2289&  11.4&   0.0&  -1.6&   0.0&   0.0&  -0.1&  -0.4&   0.2&   0.0&  -0.4&   0.1&   0.9&  -0.2&  -1.0&  -0.4&   1.3&  -0.2&  -0.2&  -1.3&   0.4&   0.1&  -1.0&   4.0&   1.3&   1.5&   1.2&   2.3&  -0.3&   0.1&   1.5 \\
  14.0:  23.0&  0.05:0.09&   0.3779&   6.5&   0.0&  -1.6&   0.0&   0.0&  -0.1&  -0.3&   0.2&   0.0&  -0.4&   0.0&   0.9&   0.1&  -0.2&   0.2&   0.2&   0.1&  -0.1&   0.3&   0.3&  -0.5&   0.3&   1.4&  -0.3&   1.3&  -0.2&   2.5&  -0.5&  -0.2&   1.5 \\
  14.0:  23.0&  0.09:0.16&   0.2902&   4.8&   0.0&  -1.6&   0.0&   0.0&  -0.1&  -0.2&   0.2&   0.0&  -0.2&  -0.1&   0.8&   0.1&  -0.0&   0.4&  -0.2&   0.1&   0.1&   0.6&   0.3&  -0.1&   0.1&   0.4&  -0.9&   1.2&  -0.3&   1.3&   0.2&  -0.2&   1.5 \\
  14.0:  23.0&  0.16:0.32&   0.1862&   5.0&   0.0&  -1.6&   0.0&   0.0&  -0.1&  -0.3&   0.1&   0.0&  -0.3&  -0.0&   0.7&  -0.2&   0.1&   0.5&  -0.4&   0.1&   0.2&   0.6&   0.8&   0.2&   0.1&  -0.6&   0.2&   0.9&  -1.6&   0.9&  -0.0&  -0.2&   1.5 \\
  14.0:  23.0&  0.32:0.70&   0.06920&   6.2&   0.0&  -1.6&   0.0&   0.0&  -0.1&  -0.2&   0.2&   0.0&  -0.2&  -0.0&   0.7&   0.3&  -0.0&   0.6&  -0.6&   0.4&   0.4&   0.4&  -0.2&   0.3&  -0.9&  -2.0&   1.4&   0.8&  -2.3&  -0.1&  -0.4&  -0.4&   1.5 \\
  23.0:  45.0&  0.02:0.05&   0.06911&  14.8&   0.0&  -1.6&   0.0&   0.0&  -0.1&  -0.3&   0.3&  -0.0&  -0.6&  -0.1&   0.6&   0.4&  -1.2&  -1.2&   1.3&  -0.2&   0.2&   0.5&   1.5&  -0.8&   2.2&   4.2&  -0.2&   2.0&   2.0&   4.0&   2.5&  -0.9&   1.5 \\
  23.0:  45.0&  0.05:0.09&   0.1233&   5.9&   0.0&  -1.6&   0.0&   0.0&  -0.1&  -0.3&   0.2&   0.0&  -0.2&   0.1&   0.8&   0.0&  -0.5&   0.2&   0.4&   0.3&  -0.7&   0.4&  -0.1&  -0.5&   0.3&   1.1&  -0.0&   0.6&   0.2&   2.0&  -0.4&  -0.1&   1.5 \\
  23.0:  45.0&  0.09:0.16&   0.1135&   4.4&   0.0&  -1.6&   0.0&   0.0&  -0.1&  -0.2&   0.2&   0.0&  -0.2&   0.1&   0.8&   0.1&   0.3&   0.3&  -0.1&  -0.0&   0.0&   0.3&   0.0&  -0.4&   0.0&   0.6&  -0.7&   0.6&  -0.4&   1.2&  -0.0&  -0.1&   1.5 \\
  23.0:  45.0&  0.16:0.32&   0.07418&   4.3&   0.0&  -1.6&   0.0&   0.0&  -0.2&  -0.3&   0.1&   0.0&  -0.2&  -0.0&   0.6&  -0.4&   0.2&   0.3&  -0.2&   0.1&   0.2&   0.3&  -0.0&   0.1&   0.3&  -0.6&   0.1&   0.9&  -1.3&   0.5&   0.1&  -0.1&   1.5 \\
  23.0:  45.0&  0.32:0.70&   0.03209&   5.2&   0.0&  -1.6&   0.0&   0.0&  -0.1&  -0.2&   0.1&  -0.0&  -0.2&  -0.0&   0.7&  -0.1&   0.0&   0.1&  -0.7&   0.3&   0.2&   0.3&  -0.3&   0.5&  -0.3&  -1.3&   1.2&  -0.0&  -2.1&   0.5&  -0.1&  -0.2&   1.5 \\
  45.0: 100.0&  0.02:0.05&   0.006162&  33.5&   0.0&  -1.6&   0.0&   0.0&   0.0&  -0.3&   0.3&   0.1&  -0.4&  -0.0&  -0.1&   1.0&  -1.8&  -0.0&   2.3&  -0.3&   0.6&  -0.3&   0.1&  -1.8&  -4.1&   3.8&   1.0&   7.3&   4.3&  -0.4&   0.3&   0.0&   1.5 \\
  45.0: 100.0&  0.05:0.09&   0.02703&  11.0&   0.0&  -1.6&   0.0&   0.0&  -0.1&  -0.3&   0.1&  -0.0&  -0.2&   0.1&   0.7&   0.1&  -0.0&   0.1&   0.2&   0.1&  -0.3&   0.1&   0.7&  -0.7&  -0.4&   2.2&  -0.4&   1.2&   0.1&   2.4&   0.1&  -0.4&   1.5 \\
  45.0: 100.0&  0.09:0.16&   0.02051&   8.0&   0.0&  -1.6&   0.0&   0.0&  -0.1&  -0.2&   0.1&  -0.0&  -0.1&   0.0&   0.6&   0.2&   0.2&   0.3&   0.0&   0.2&   0.1&   0.7&   0.3&  -0.2&   0.5&   0.6&  -1.1&   1.0&  -0.7&   2.1&   0.0&  -0.2&   1.5 \\
  45.0: 100.0&  0.16:0.32&   0.01995&   5.4&   0.0&  -1.6&   0.0&   0.0&  -0.1&  -0.2&   0.1&   0.1&  -0.2&   0.0&   0.7&  -0.2&   0.3&   0.6&   0.1&   0.1&   0.1&   0.7&   0.1&   0.0&  -0.2&  -0.1&  -0.0&   0.6&  -1.1&   1.6&   0.3&  -0.2&   1.5 \\
  45.0: 100.0&  0.32:0.70&   0.007841&   6.9&   0.0&  -1.6&   0.0&   0.0&  -0.1&  -0.2&   0.1&   0.0&  -0.3&  -0.0&   0.6&  -0.1&  -0.0&   0.5&  -0.0&   0.1&   0.3&   0.3&   0.2&   0.4&  -0.3&  -1.2&   1.5&   0.0&  -2.4&   0.2&  -0.1&  -0.0&   1.5 \\
 100.0: 158.0&  0.02:0.32&   0.004124&   8.2&   0.0&  -1.6&   0.0&   0.0&  -0.0&  -0.2&   0.1&  -0.0&   0.0&  -0.0&   0.6&  -0.2&   0.0&  -0.1&   0.0&  -0.1&  -0.0&   0.5&   0.8&  -1.0&   1.2&   1.0&   1.3&   0.1&  -1.9&   0.4&  -1.4&  -0.3&   1.5 \\
 100.0: 158.0&  0.32:0.70&   0.002181&  11.1&   0.0&  -1.5&   0.0&   0.0&  -0.1&  -0.2&  -0.0&   0.0&  -0.2&  -0.0&   0.6&  -0.2&   1.3&   0.0&   0.7&   0.3&   0.2&   0.7&  -0.2&   1.1&   0.7&   0.8&   1.4&  -0.1&  -2.2&  -0.1&   0.1&  -0.4&   1.5 \\
 158.0: 251.0&  0.02:0.30&   0.001788&  10.2&   0.0&  -1.6&   0.0&   0.0&  -0.0&  -0.2&   0.0&  -0.0&   0.0&  -0.0&   0.5&  -0.1&   0.6&  -0.4&   0.5&   0.2&   0.2&   0.6&   0.2&   1.1&   0.2&   1.9&   1.3&   0.1&  -2.2&   0.1&  -1.3&  -0.3&   1.5 \\
 158.0: 251.0&  0.30:0.70&   0.0009276&  11.6&   0.0&  -1.5&   0.0&   0.0&  -0.2&  -0.1&   0.0&  -0.0&  -0.0&  -0.1&   0.3&  -0.2&   0.8&  -0.2&   0.2&   0.2&   0.1&   0.8&  -0.2&   0.6&  -0.2&   1.3&   2.0&   0.7&  -2.6&  -0.9&  -0.6&  -0.5&   1.5 \\
 251.0:1000.0&  0.02:0.26&   0.0001315&  14.5&   0.0&  -1.5&   0.0&   0.0&   0.1&  -0.2&   0.2&  -0.1&  -0.0&  -0.0&   0.5&   0.0&   0.6&  -0.3&   0.4&  -0.1&   0.1&   0.3&  -0.1&   0.5&   0.2&   1.7&   1.7&  -0.3&  -2.3&  -0.6&  -2.1&  -0.3&   1.5 \\
 251.0:1000.0&  0.26:0.70&   0.0001184&  12.7&   0.0&  -1.5&   0.0&   0.0&  -0.1&  -0.1&   0.2&   0.1&  -0.1&   0.0&   0.5&   0.1&   0.6&  -0.2&   0.1&   0.2&  -0.0&   0.5&  -0.7&   0.6&  -1.2&   2.0&   1.6&  -0.4&  -2.8&  -1.0&  -1.3&  -0.4&   1.5 \\ \hline
\end{tabu}
\end{table*}

\clearpage
\section{Combination of reduced charm cross sections}
\label{sec:app:red}

In this Appendix additional information on the combination of the reduced charm cross sections (see Section~\ref{sec:comb:red}) is provided. 
Section~\ref{sec:comb:red:vfns} presents a comparison with the theoretical predictions in different VFNS and a determination of optimal $c$-quark mass parameters for these schemes.

\subsection{Additional tables and plots}

Table~\ref{tab:comb:red:syst} provides information on the fitted nuisance parameters.

The combined data with all correlations are provided in Table~\ref{tab:comb:red:combinedfull}.

\begin{figure}[tbp]
  \centering
  \includegraphics[width=1.0\figwidth,trim=0 0 0 0mm,clip=true]{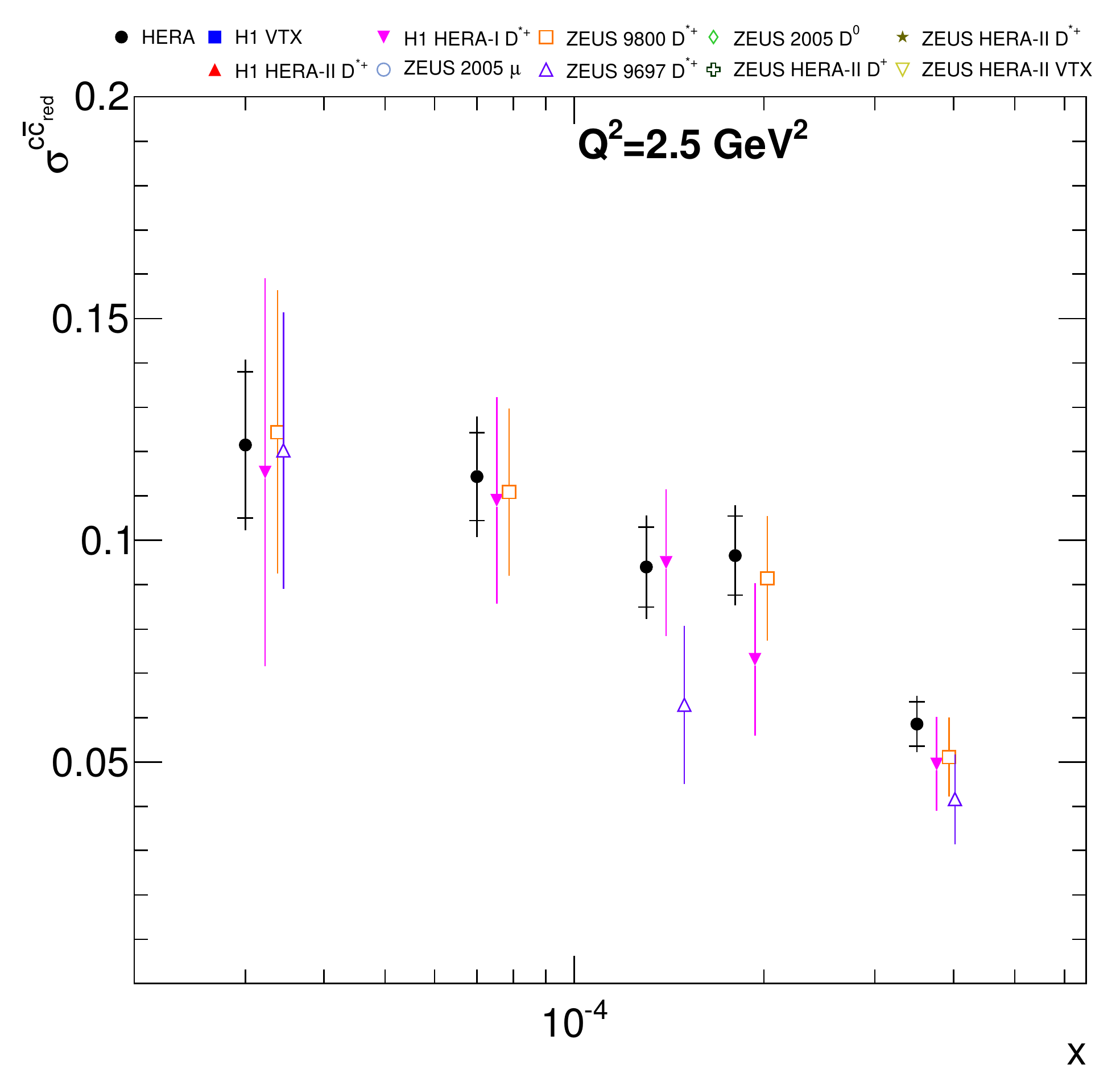}
  \caption[Combined reduced charm cross sections for $Q^2=\SI{2.5}{GeV^2}$]
  {Combined measurements of $\sigma_{red}^{c\bar{c}}$ (closed circles) shown as a function of $x$ for $Q^2=\SI{2.5}{GeV^2}$. 
  The input measurements are also shown with different markers. 
	For the combined data, the inner error bars indicate the uncorrelated part of the uncertainties and 
	the outer error bars represent the total uncertainties. 
  For presentation purposes each individual measurement is shifted in $x$.}  
	\label{fig:comb:red:comb_q2_1}
\end{figure}

\begin{figure}[tbp]
  \centering
  \includegraphics[width=1.0\figwidth,trim=0 0 0 0mm,clip=true]{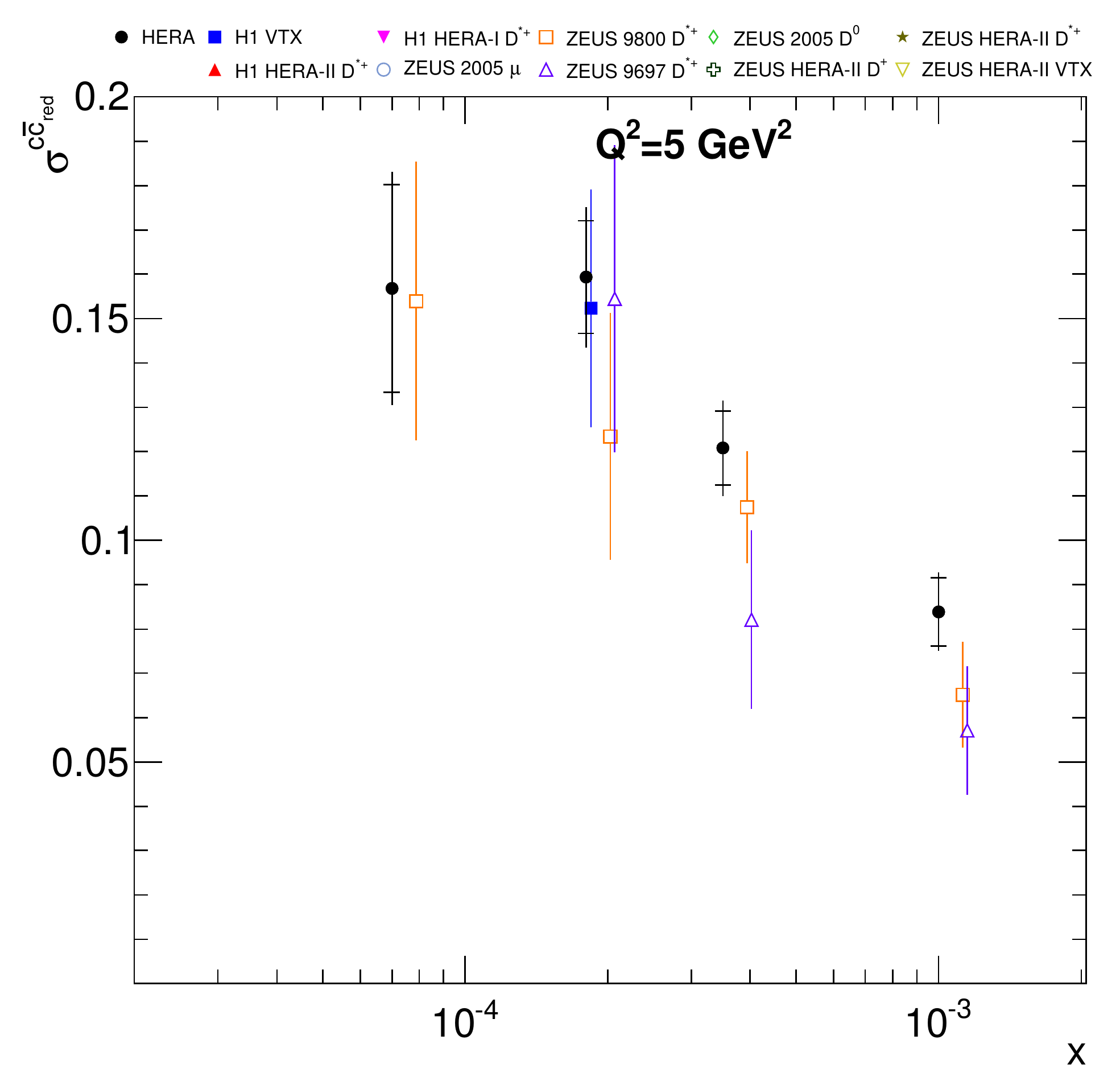}
  \caption[Combined reduced charm cross sections for $Q^2=\SI{5}{GeV^2}$]
  {Combined measurements of $\sigma_{red}^{c\bar{c}}$ (closed circles) shown as a function of $x$ for $Q^2=\SI{5}{GeV^2}$. 
  The input measurements are also shown with different markers. 
	For the combined data, the inner error bars indicate the uncorrelated part of the uncertainties and 
	the outer error bars represent the total uncertainties. 
  For presentation purposes each individual measurement is shifted in $x$.}  
	\label{fig:comb:red:comb_q2_2}
\end{figure}

\begin{figure}[tbp]
  \centering
  \includegraphics[width=1.0\figwidth,trim=0 0 0 0mm,clip=true]{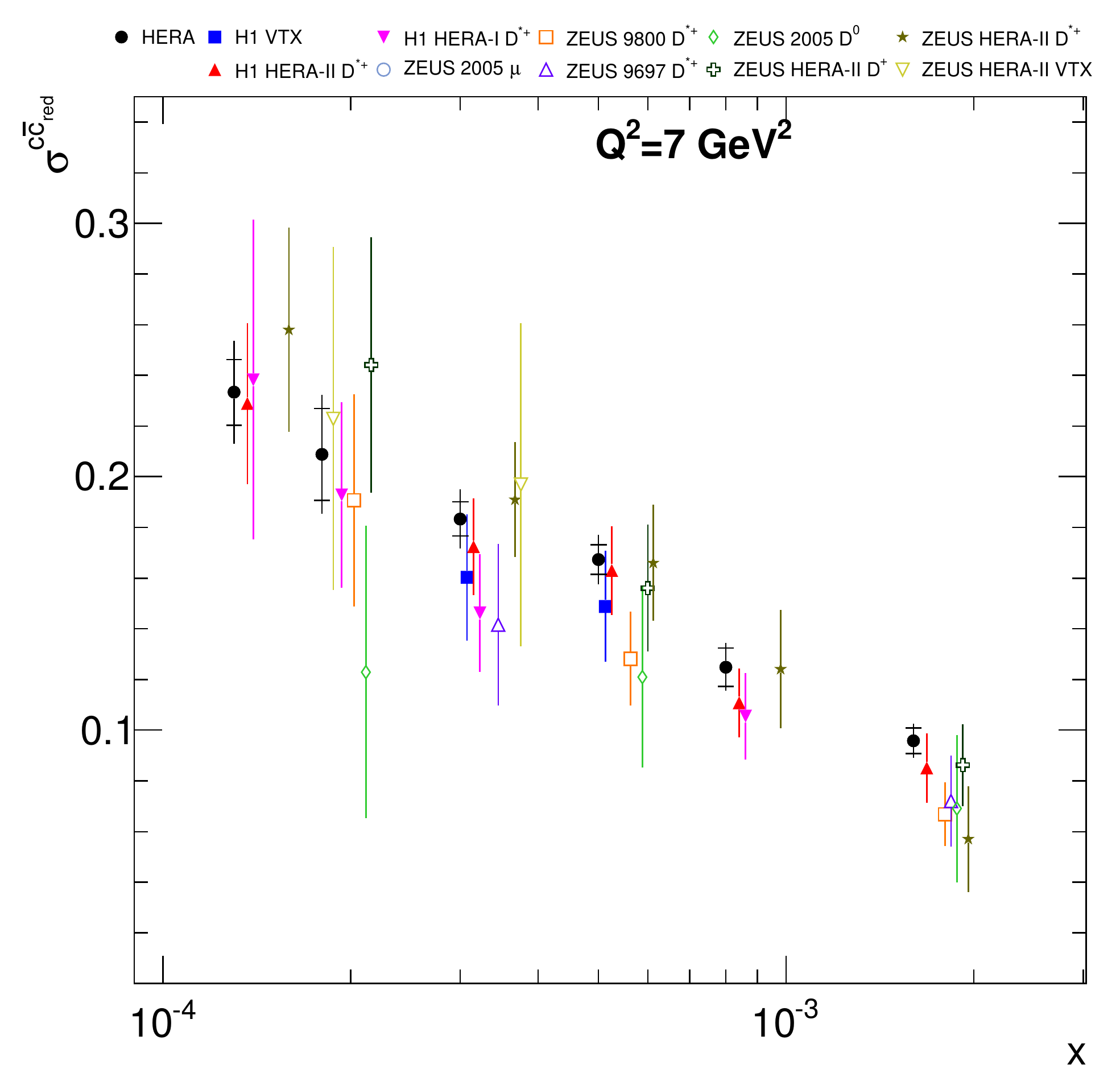}
  \caption[Combined reduced charm cross sections for $Q^2=\SI{7}{GeV^2}$]
  {Combined measurements of $\sigma_{red}^{c\bar{c}}$ (closed circles) shown as a function of $x$ for $Q^2=\SI{7}{GeV^2}$. 
  The input measurements are also shown with different markers. 
	For the combined data, the inner error bars indicate the uncorrelated part of the uncertainties and 
	the outer error bars represent the total uncertainties. 
  For presentation purposes each individual measurement is shifted in $x$.}  
	\label{fig:comb:red:comb_q2_3}
\end{figure}

\begin{figure}[tbp]
  \centering
  \includegraphics[width=1.0\figwidth,trim=0 0 0 0mm,clip=true]{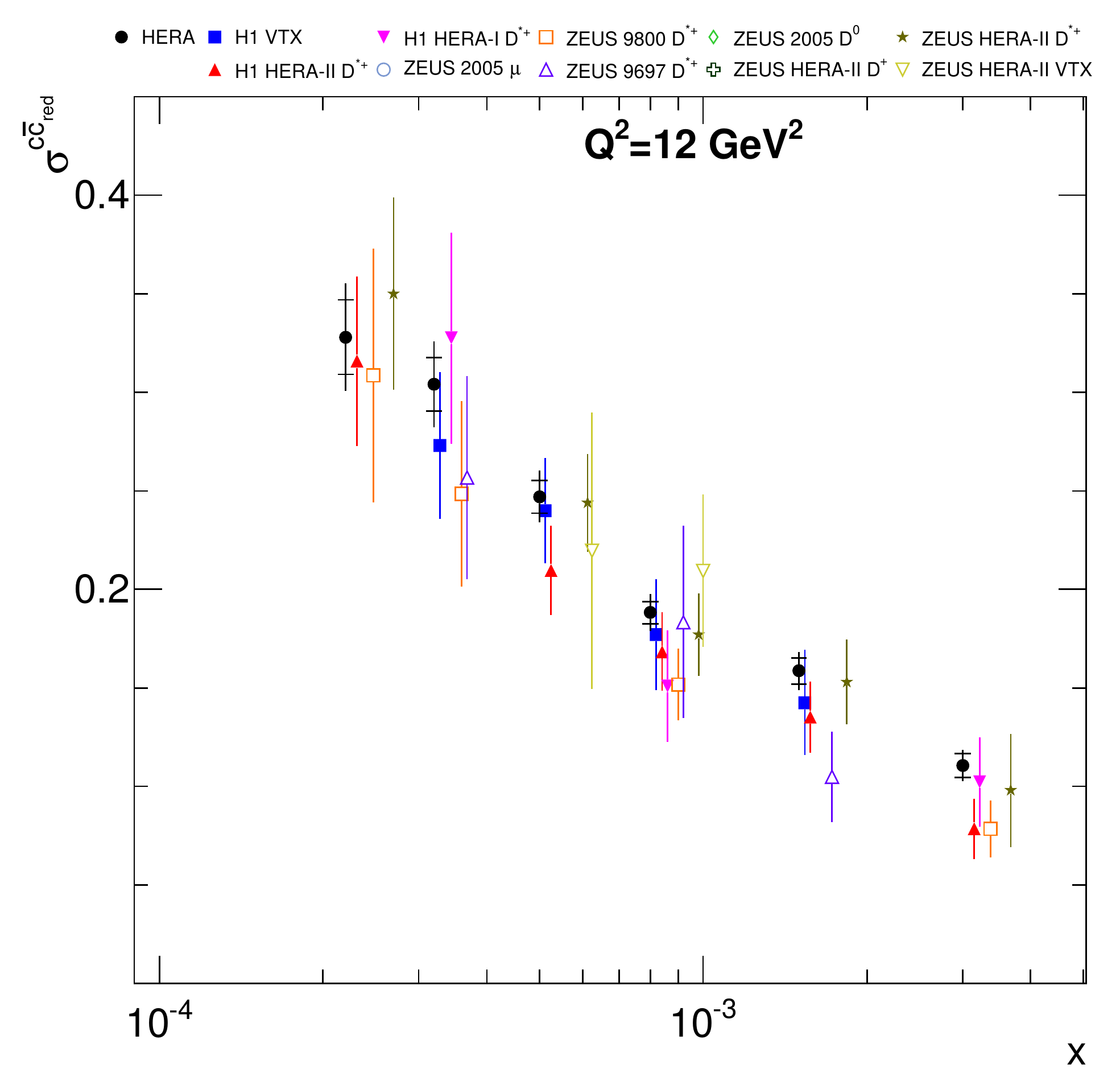}
  \caption[Combined reduced charm cross sections for $Q^2=\SI{12}{GeV^2}$]
  {Combined measurements of $\sigma_{red}^{c\bar{c}}$ (closed circles) shown as a function of $x$ for $Q^2=\SI{12}{GeV^2}$. 
  The input measurements are also shown with different markers. 
	For the combined data, the inner error bars indicate the uncorrelated part of the uncertainties and 
	the outer error bars represent the total uncertainties. 
  For presentation purposes each individual measurement is shifted in $x$.}  
	\label{fig:comb:red:comb_q2_4}
\end{figure}

\begin{figure}[tbp]
  \centering
  \includegraphics[width=1.0\figwidth,trim=0 0 0 0mm,clip=true]{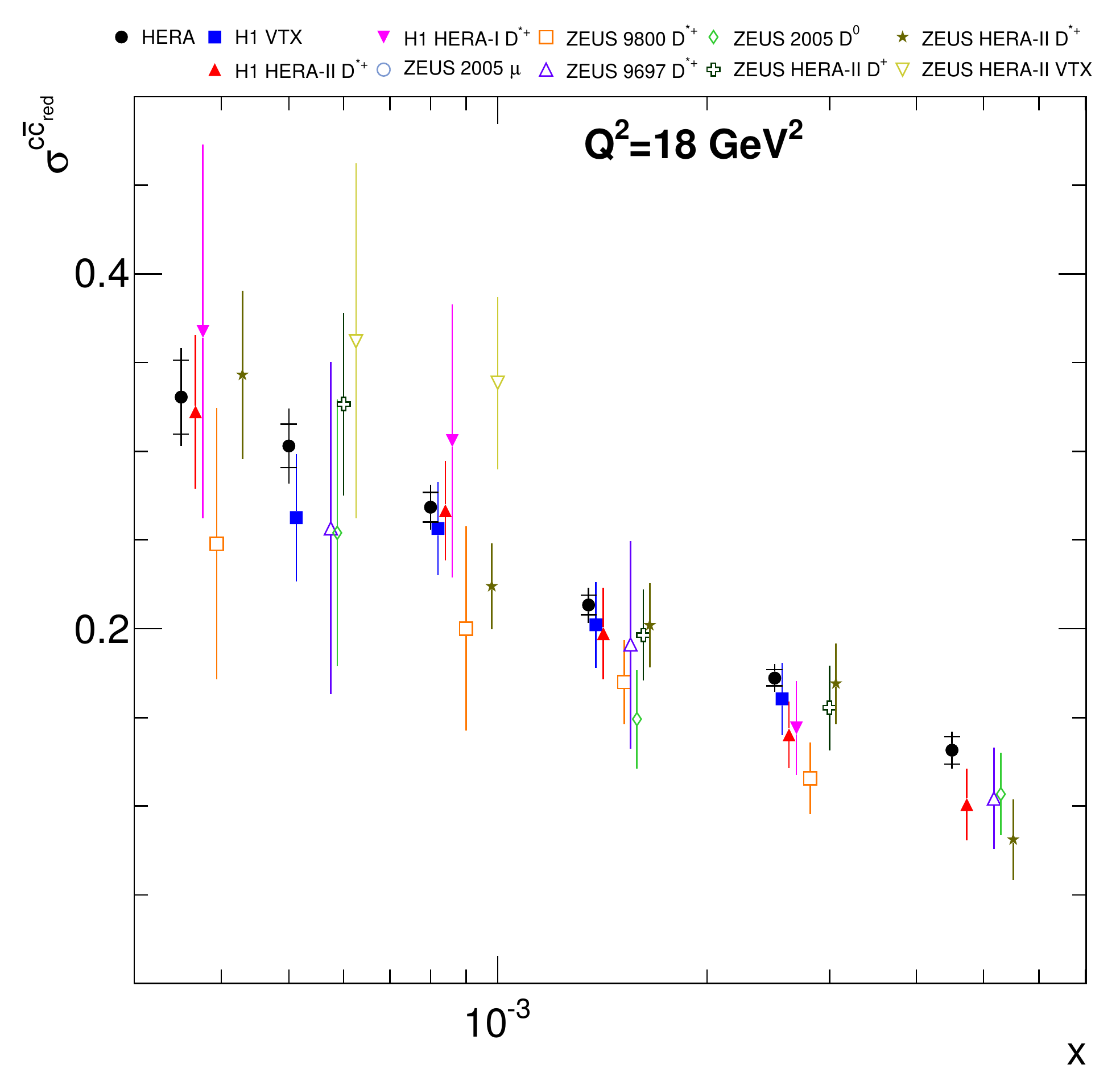}
  \caption[Combined reduced charm cross sections for $Q^2=\SI{18}{GeV^2}$]
  {Combined measurements of $\sigma_{red}^{c\bar{c}}$ (closed circles) shown as a function of $x$ for $Q^2=\SI{18}{GeV^2}$. 
  The input measurements are also shown with different markers. 
	For the combined data, the inner error bars indicate the uncorrelated part of the uncertainties and 
	the outer error bars represent the total uncertainties. 
  For presentation purposes each individual measurement is shifted in $x$.}  
	\label{fig:comb:red:comb_q2_5}
\end{figure}

\begin{figure}[tbp]
  \centering
  \includegraphics[width=1.0\figwidth,trim=0 0 0 0mm,clip=true]{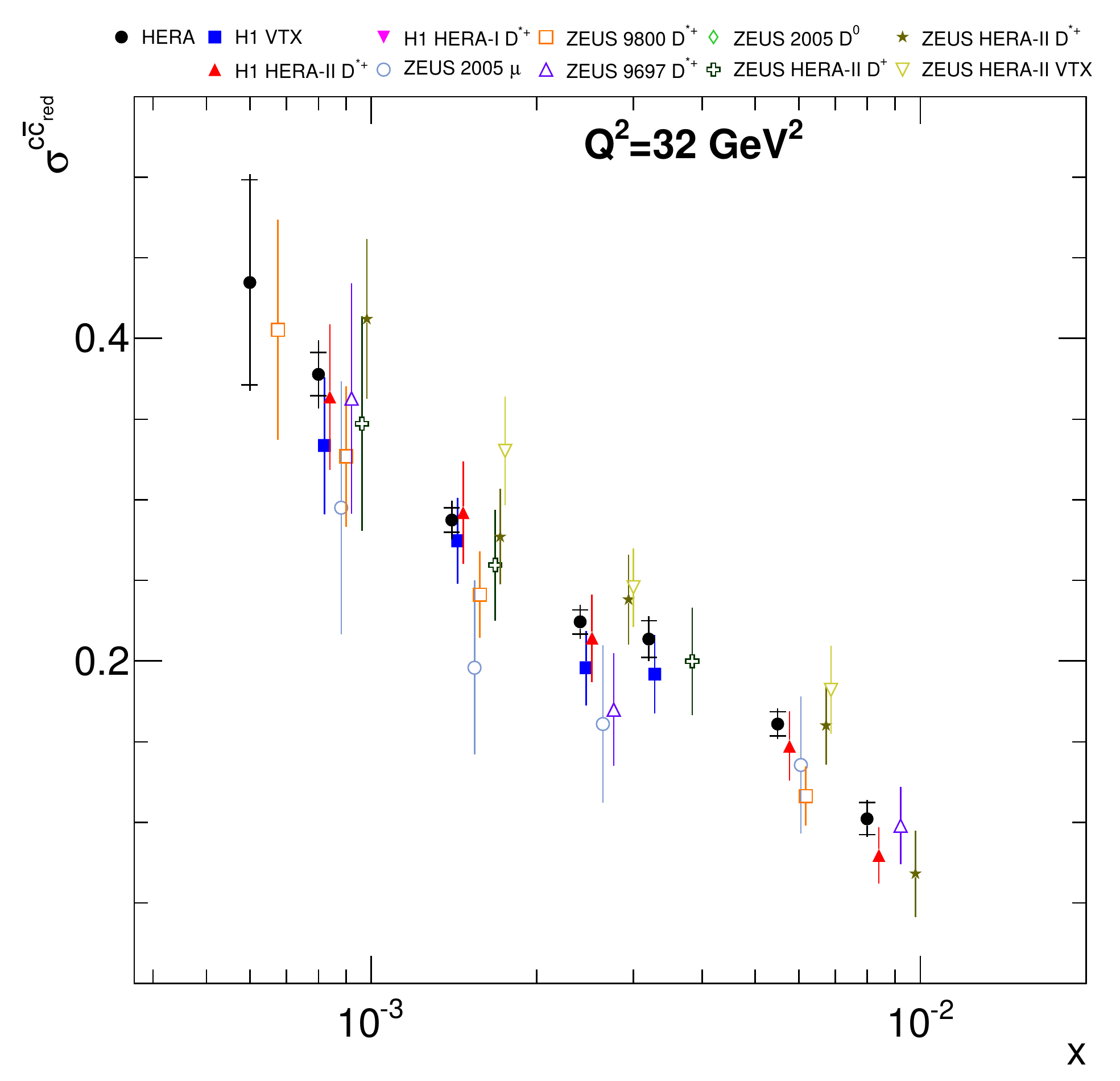}
  \caption[Combined reduced charm cross sections for $Q^2=\SI{32}{GeV^2}$]
  {Combined measurements of $\sigma_{red}^{c\bar{c}}$ (closed circles) shown as a function of $x$ for $Q^2=\SI{32}{GeV^2}$. 
  The input measurements are also shown with different markers. 
	For the combined data, the inner error bars indicate the uncorrelated part of the uncertainties and 
	the outer error bars represent the total uncertainties. 
  For presentation purposes each individual measurement is shifted in $x$.}  
	\label{fig:comb:red:comb_q2_6}
\end{figure}

\begin{figure}[tbp]
  \centering
  \includegraphics[width=1.0\figwidth,trim=0 0 0 0mm,clip=true]{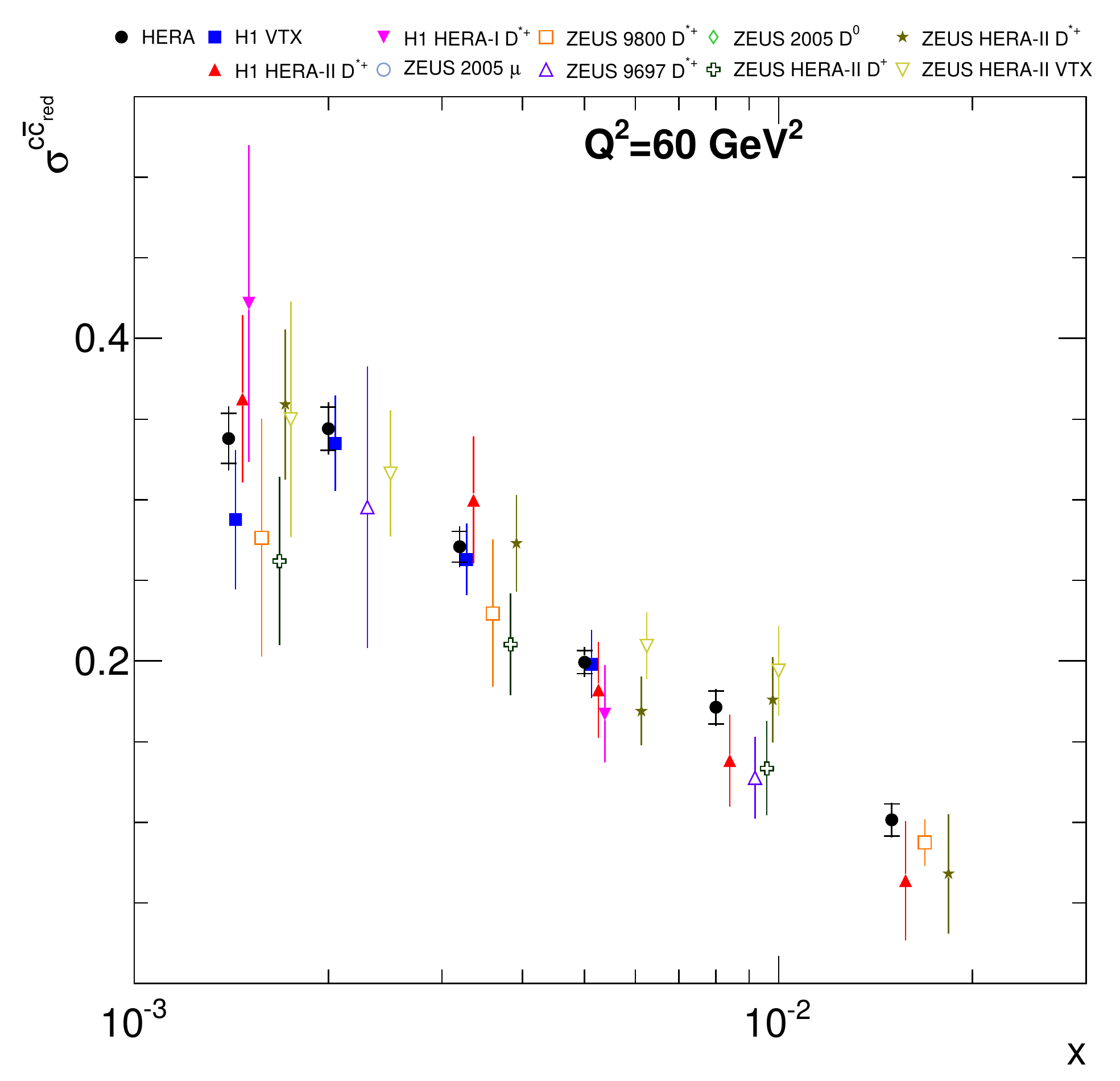}
  \caption[Combined reduced charm cross sections for $Q^2=\SI{60}{GeV^2}$]
  {Combined measurements of $\sigma_{red}^{c\bar{c}}$ (closed circles) shown as a function of $x$ for $Q^2=\SI{60}{GeV^2}$. 
  The input measurements are also shown with different markers. 
	For the combined data, the inner error bars indicate the uncorrelated part of the uncertainties and 
	the outer error bars represent the total uncertainties. 
  For presentation purposes each individual measurement is shifted in $x$.}  
	\label{fig:comb:red:comb_q2_7}
\end{figure}

\begin{figure}[tbp]
  \centering
  \includegraphics[width=1.0\figwidth,trim=0 0 0 0mm,clip=true]{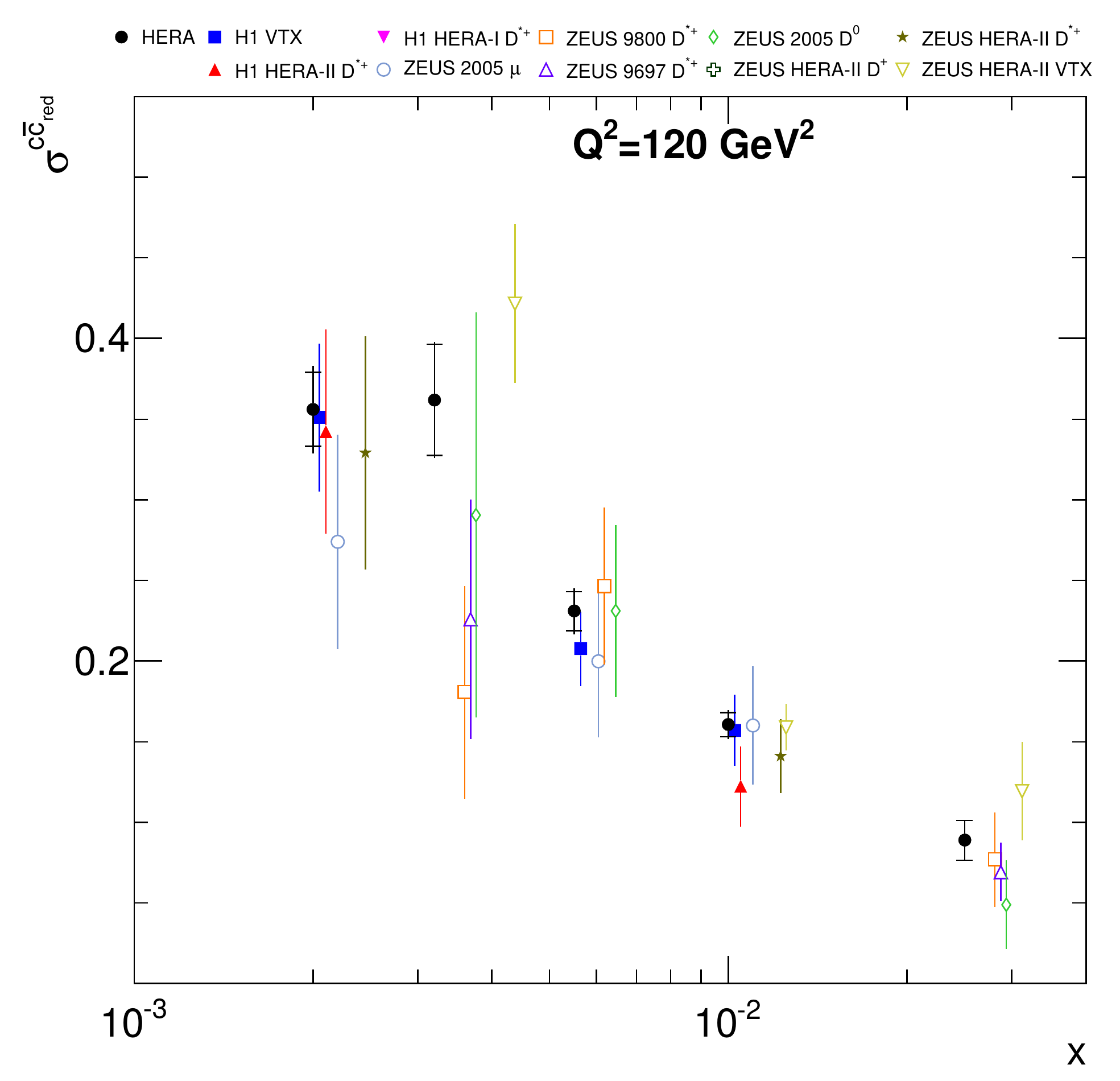}
  \caption[Combined reduced charm cross sections for $Q^2=\SI{120}{GeV^2}$]
  {Combined measurements of $\sigma_{red}^{c\bar{c}}$ (closed circles) shown as a function of $x$ for $Q^2=\SI{120}{GeV^2}$. 
  The input measurements are also shown with different markers. 
	For the combined data, the inner error bars indicate the uncorrelated part of the uncertainties and 
	the outer error bars represent the total uncertainties. 
  For presentation purposes each individual measurement is shifted in $x$.}  
	\label{fig:comb:red:comb_q2_8}
\end{figure}

\begin{figure}[tbp]
  \centering
  \includegraphics[width=1.0\figwidth,trim=0 0 0 0mm,clip=true]{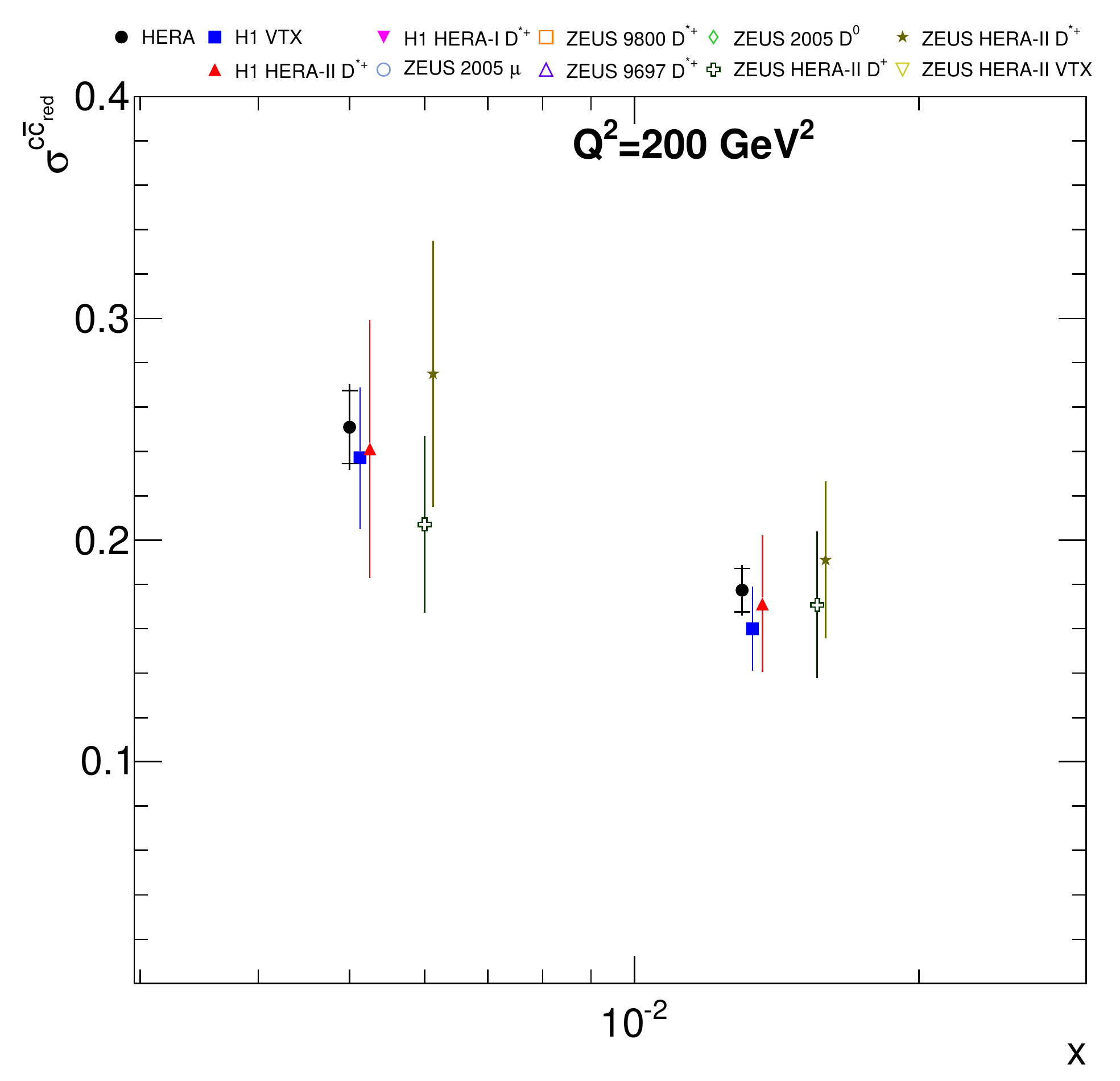}
  \caption[Combined reduced charm cross sections for $Q^2=\SI{200}{GeV^2}$]
  {Combined measurements of $\sigma_{red}^{c\bar{c}}$ (closed circles) shown as a function of $x$ for $Q^2=\SI{200}{GeV^2}$. 
  The input measurements are also shown with different markers. 
	For the combined data, the inner error bars indicate the uncorrelated part of the uncertainties and 
	the outer error bars represent the total uncertainties. 
  For presentation purposes each individual measurement is shifted in $x$.}  
	\label{fig:comb:red:comb_q2_9}
\end{figure}

\begin{figure}[tbp]
  \centering
  \includegraphics[width=1.0\figwidth,trim=0 0 0 0mm,clip=true]{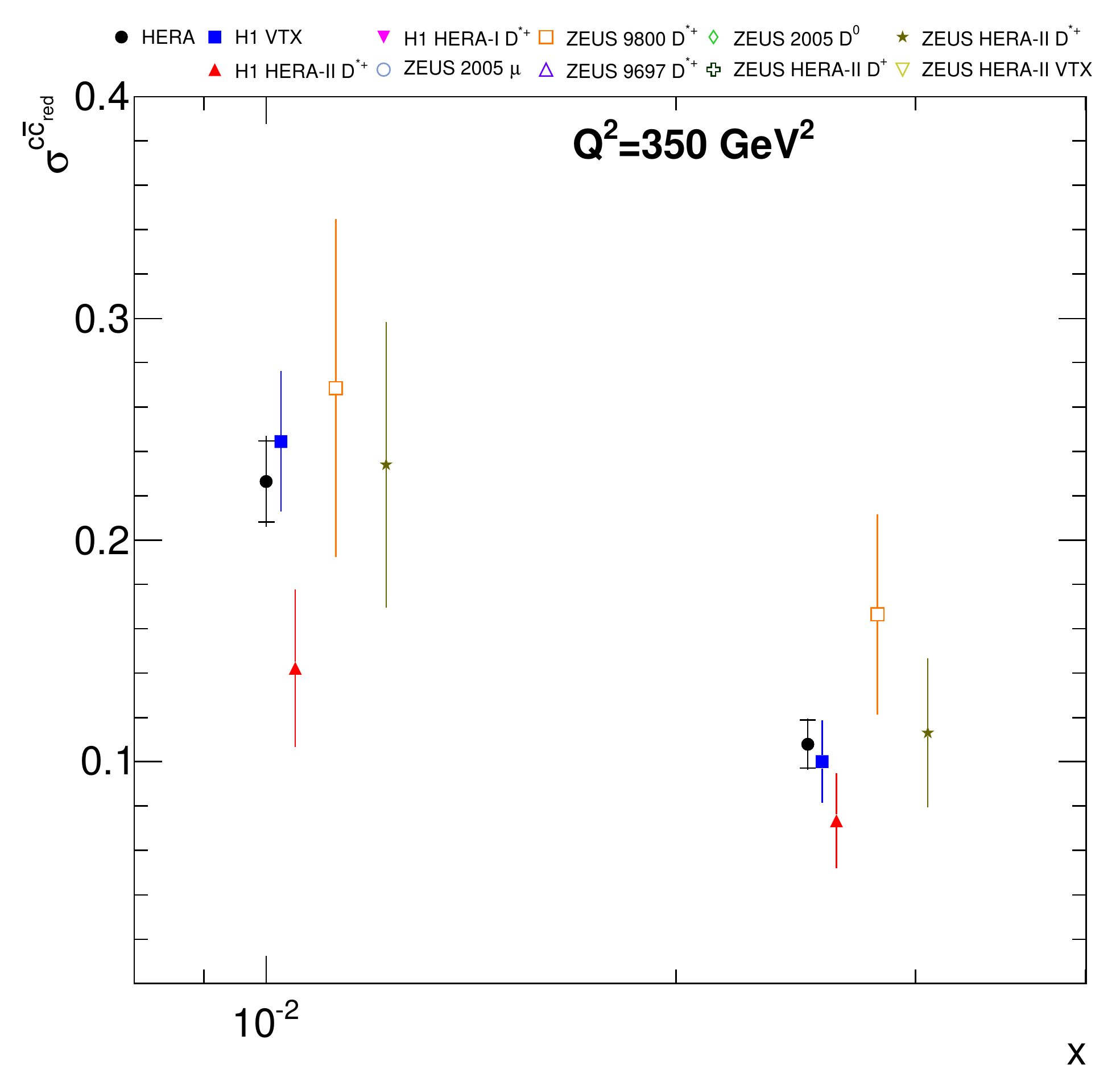}
  \caption[Combined reduced charm cross sections for $Q^2=\SI{350}{GeV^2}$]
  {Combined measurements of $\sigma_{red}^{c\bar{c}}$ (closed circles) shown as a function of $x$ for $Q^2=\SI{350}{GeV^2}$. 
  The input measurements are also shown with different markers. 
	For the combined data, the inner error bars indicate the uncorrelated part of the uncertainties and 
	the outer error bars represent the total uncertainties. 
  For presentation purposes each individual measurement is shifted in $x$.}  
	\label{fig:comb:red:comb_q2_10}
\end{figure}

\begin{figure}[tbp]
  \centering
  \includegraphics[width=1.0\figwidth,trim=0 0 0 0mm,clip=true]{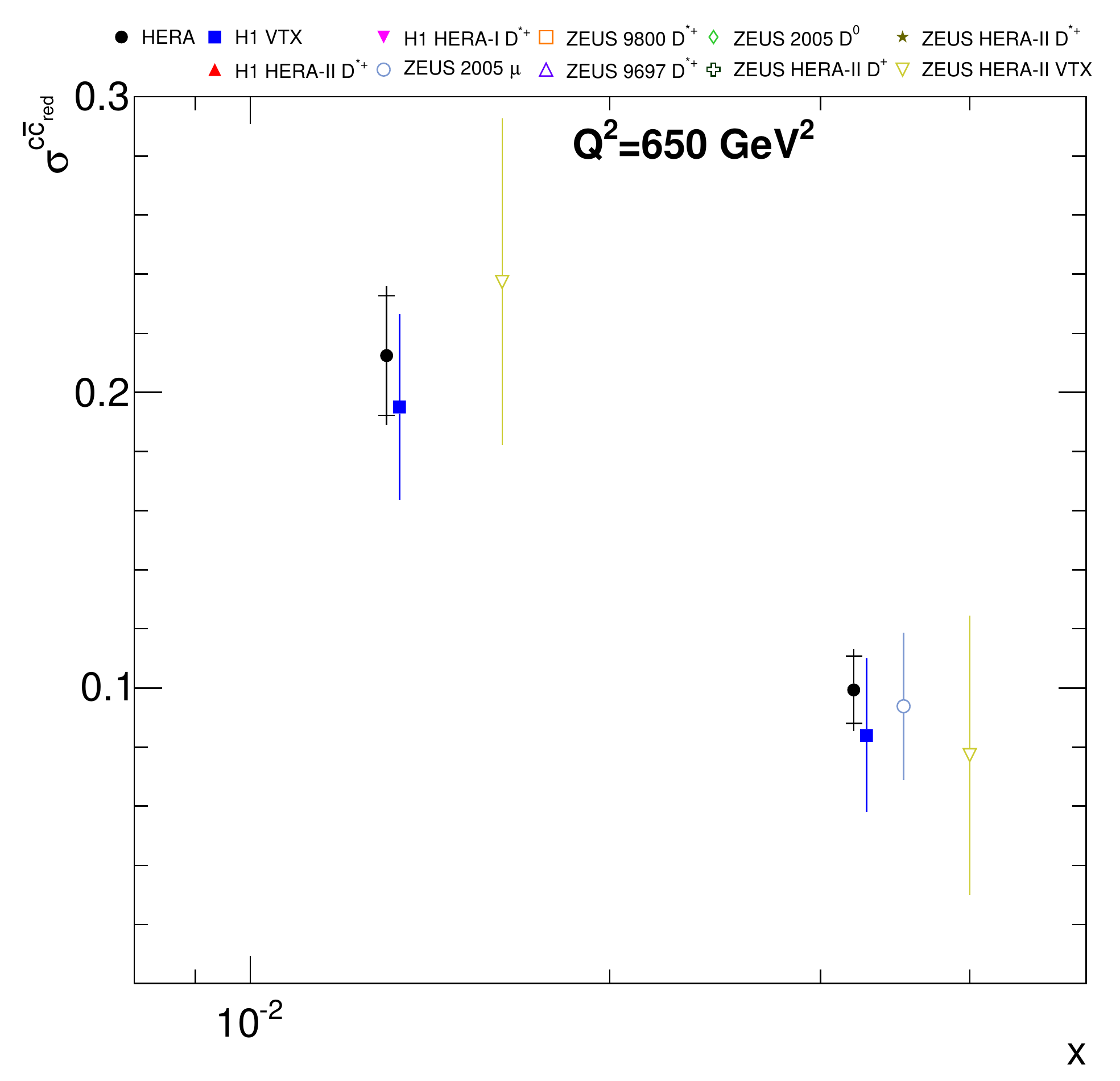}
  \caption[Combined reduced charm cross sections for $Q^2=\SI{650}{GeV^2}$]
  {Combined measurements of $\sigma_{red}^{c\bar{c}}$ (closed circles) shown as a function of $x$ for $Q^2=\SI{650}{GeV^2}$. 
  The input measurements are also shown with different markers. 
	For the combined data, the inner error bars indicate the uncorrelated part of the uncertainties and 
	the outer error bars represent the total uncertainties. 
  For presentation purposes each individual measurement is shifted in $x$.}  
	\label{fig:comb:red:comb_q2_11}
\end{figure}

\begin{figure}[tbp]
  \centering
  \includegraphics[width=1.0\figwidth,trim=0 0 0 0mm,clip=true]{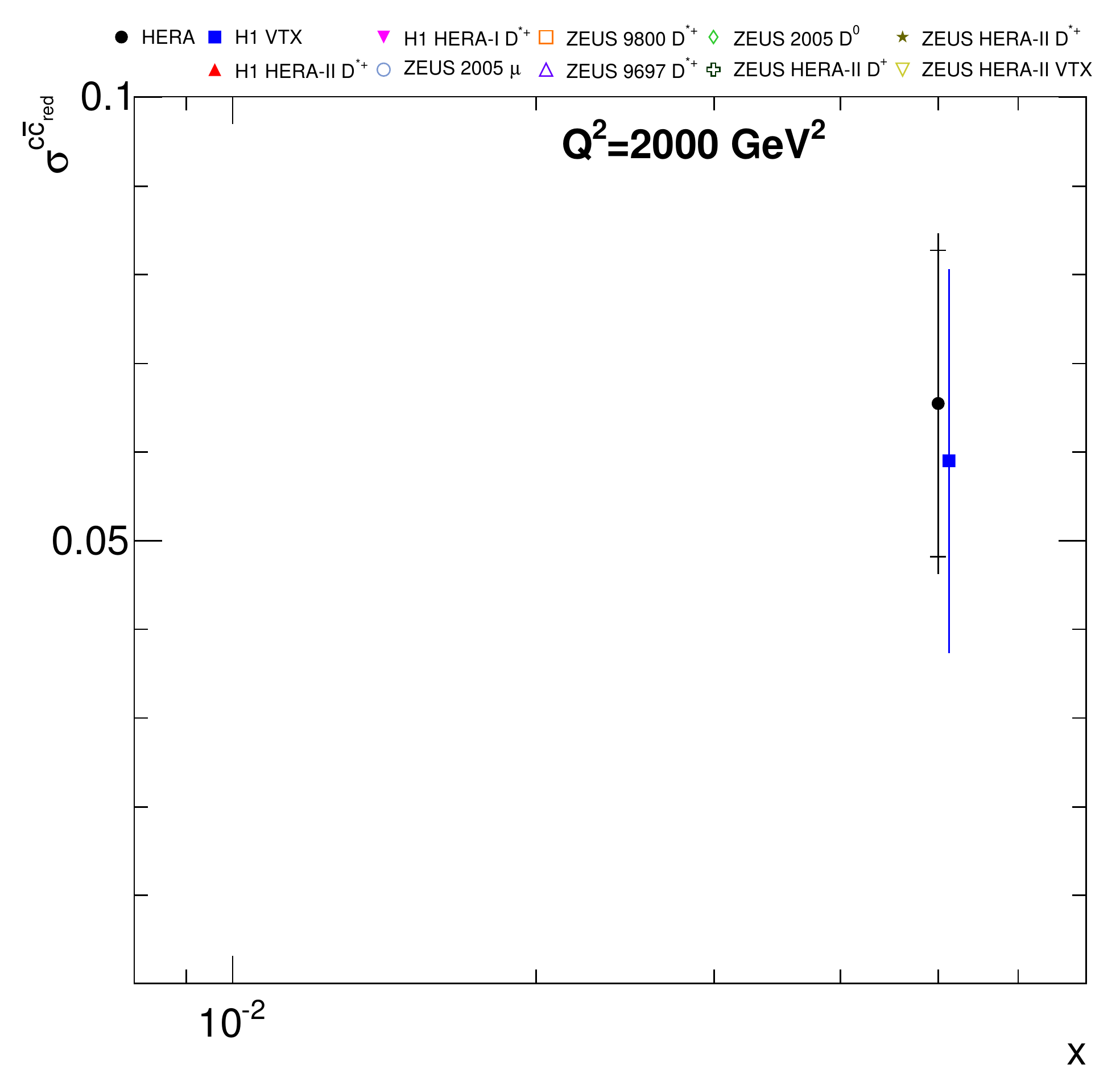}
  \caption[Combined reduced charm cross sections for $Q^2=\SI{2000}{GeV^2}$]
  {Combined measurements of $\sigma_{red}^{c\bar{c}}$ (closed circles) shown as a function of $x$ for $Q^2=\SI{2000}{GeV^2}$. 
  The input measurements are also shown with different markers. 
	For the combined data, the inner error bars indicate the uncorrelated part of the uncertainties and 
	the outer error bars represent the total uncertainties. 
  For presentation purposes each individual measurement is shifted in $x$.}  
	\label{fig:comb:red:comb_q2_12}
\end{figure}

Figs.~\ref{fig:comb:red:comb_q2_1} to~\ref{fig:comb:red:comb_q2_12} show the combined data with the input measurements for individual values of $Q^2$.

\begin{table}[tbp]
\caption[Sources of correlated uncertainties in combination of reduced cross sections]
{Sources of bin-to-bin correlated uncertainties considered in the combination of the reduced charm cross sections. For each source the affected datasets, name, type (see Section~\ref{sec:comb:proc:unc}) and reference to the place, 
  where information can be found, are given, together with the shift (sh) and reduction (red) factor in the combination obtained after the first iteration.}
\label{tab:comb:red:syst}
\scriptsize
\centering
\tabcolsep0.40mm
\renewcommand*{\arraystretch}{0.95}
\begin{tabu} to \columnwidth[t]{|c|l|X[l]|l|l|>{$}r<{$}|>{$}r<{$}|}\hline
Descr. & Datasets & Name & Type & Reference & \text{sh}~[\%] & \text{red}~[\%]\\ \hline
$\delta_{1}$ & 1 & H1 VTX resolution & S & H1 Col. & -0.1 & 0.9 \\
$\delta_{2}$ & 1--4 & H1 CJC efficiency & S &  H1 Col.& 0.2 & 0.8 \\
$\delta_{3}$ & 1 & H1 CST efficiency & S & H1 Col. & 0.1 & 1.0 \\
$\delta_{4}$ & 1 & H1 B multiplicity & S & H1 Col. & -0.2 & 0.9 \\
$\delta_{5}$ & 1--11 & NLO, $c$ longitudinal frag. & T & Theory & -1.3 & 0.7 \\
$\delta_{6}$ & 1,3,4 & H1 PHP background & S & H1 Col. & 0.4 & 0.9 \\
$\delta_{7}$ & 1 & H1 multiplicity $\Dch$ & S & H1 Col. & 0.1 & 1.0 \\
$\delta_{8}$ & 1 & H1 multiplicity $D^{0}$ & S & H1 Col. & -0.1 & 1.0 \\
$\delta_{9}$ & 1 & H1 multiplicity $D_s$ & S & H1 Col. & 0.1 & 1.0 \\
$\delta_{10}$ & 1 & H1 VTX $b$ frag. & S & H1 Col. & -0.1 & 1.0 \\
$\delta_{11}$ & 1 & H1 VTX rew. $x$ & S & H1 Col. & -0.1 & 0.9 \\
$\delta_{12}$ & 1 & H1 VTX rew. $p_T$ & S & H1 Col. & 0.1 & 0.7 \\
$\delta_{13}$ & 1 & H1 VTX rew. $\eta$ & S & H1 Col. & -0.1 & 0.8 \\
$\delta_{14}$ & 1 & H1 VTX $uds$ background & S & H1 Col. & -0.5 & 0.4 \\
$\delta_{15}$ & 1 & H1 VTX $\phi$ of $c$ quark & S & H1 Col. & 0.1 & 0.9 \\
$\delta_{16}$ & 1 & H1 hadronic energy scale & S & H1 Col. & 0.1 & 0.8 \\
$\delta_{17}$ & 1 & H1 VTX $F_2$ normalisation & S & H1 Col. & -0.1 & 1.0 \\
$\delta_{18}$ & 3,4 & H1 primary-vertex fit & S & H1 Col. & 0.2 & 1.0 \\
$\delta_{19}$ & 2--4 & H1 $e$ energy & S & H1 Col. & 0.3 & 0.7 \\
$\delta_{20}$ & 2--4 & H1 $e$ $\theta$ & S & H1 Col. & 0.2 & 0.7 \\
$\delta_{21}$ & 3,4 & H1 luminosity HERA-II & N & H1 Col. & -0.4 & 0.8 \\
$\delta_{22}$ & 3,4 & H1 trigger HERA-II & S & H1 Col. & -0.1 & 1.0 \\
$\delta_{23}$ & 3,4 & H1 MC fragmentation & S & H1 Col. & -0.2 & 0.9 \\
$\delta_{24}$ & 2--7,10 & $br(\Dstar\to K \pi \pi)$ & N & \cite{pdg2012} & 0.4 & 1.0 \\
$\delta_{25}$ & 2--6,10 & $ff(c\to \Dstar)$ & T & \cite{Lohrmann:2011np} & 0.6 & 0.9 \\
$\delta_{26}$ & 2,3 & H1 MC efficiency & S & H1 Col. & 0.3 & 0.7 \\
$\delta_{27}$ & 2--11 & NLO, $m_c$ & T & Theory & 0.6 & 0.6 \\
$\delta_{28}$ & 2--11 & NLO, scale & T & Theory & -0.9 & 0.4 \\
$\delta_{29}$ & 2--11 & NLO, $c$ transverse frag. & T & Theory & 0.1 & 0.7 \\
$\delta_{30}$ & 2--4 & NLO, PDF & T & Theory & 0.9 & 0.9 \\
$\delta_{31}$ & 2--11 & NLO, $\alpha_s(M_Z)$ & T & Theory & -0.2 & 0.7 \\
$\delta_{32}$ & 2 & H1 luminosity 96--97 & N & H1 Col. & -0.0 & 1.0 \\
$\delta_{33}$ & 2 & H1 trigger 96--97 & S & H1 Col. & -0.0 & 0.9 \\
$\delta_{34}$ & 2 & H1 MC alternative frag. & S & H1 Col. & -0.2 & 0.7 \\
$\delta_{35}$ & 8 & ZEUS $\mu$ B/RMUON efficiency & S & \cite{zeus_muon} & -0.3 & 0.9 \\
$\delta_{36}$ & 8 & ZEUS $\mu$ FMUON efficiency & S & \cite{zeus_muon} & 0.2 & 1.0 \\
$\delta_{37}$ & 8 & ZEUS $\mu$ energy scale & S & \cite{zeus_muon} & 0.0 & 0.8 \\
$\delta_{38}$ & 8 & ZEUS $\mu$ $p_T^{\rm miss}$ calibration & S & \cite{zeus_muon} & 0.2 & 0.7 \\
$\delta_{39}$ & 8 & ZEUS $\mu$ hadronic resolution & S & \cite{zeus_muon} & 0.6 & 0.7 \\
$\delta_{40}$ & 8 & ZEUS $\mu$ IP resolution & S & \cite{zeus_muon} & -0.3 & 1.0 \\
$\delta_{41}$ & 8 & ZEUS $\mu$ MC model & S & \cite{zeus_muon} & 0.3 & 0.9 \\
$\delta_{42}$ & 8 & $ff(c\to \mu)$ & T & \cite{pdg2012}  & 0.2 & 1.0 \\
$\delta_{43}$ & 7--11 & ZEUS luminosity HERA-II & N & \cite{zeusdch_hera2,zeusdstar_hera2,zeussecvtx_hera2} & -0.6 & 0.9 \\
$\delta_{44}$ & 5 & ZEUS luminosity 96--97 & N & \cite{zd9697} & 0.7 & 0.9 \\
$\delta_{45}$ & 6 & ZEUS luminosity 98--00 & N & \cite{zd00} & 0.9 & 0.9 \\
$\delta_{46}$ & 7 & ZEUS $D^{0}$ lifetime significance & S & \cite{zd0dp,Nicholass:2008zz} & 0.9 & 0.5 \\
$\delta_{47}$ & 7 & $ff(c\to D^{0})$ & T & \cite{Lohrmann:2011np} & 0.2 & 1.0 \\
$\delta_{48}$ & 9 & $ff(c\to \Dch)$ & T &\cite{Lohrmann:2011np}   & -0.1 & 0.9 \\
$\delta_{49}$ & 9 & ZEUS \Dch electron energy scale & S & \cite{zeusdch_hera2} & 0.3 & 1.0 \\
$\delta_{50}$ & 9 & ZEUS \Dch hadronic energy scale & S & \cite{zeusdch_hera2} & -0.1 & 1.0 \\
$\delta_{51}$ & 9 & ZEUS \Dch trigger inefficiency & S & \cite{zeusdch_hera2} & 0.3 & 0.9 \\
$\delta_{52}$ & 9 & ZEUS \Dch decay length smearing & S & \cite{zeusdch_hera2} & 0.2 & 1.0 \\
$\delta_{53}$ & 9 & ZEUS \Dch MC $b$ normalisation & S & \cite{zeusdch_hera2} & 0.1 & 0.9 \\
$\delta_{54}$ & 9 & ZEUS \Dch MC rew. $p_T\text{--}Q^2$ & S & \cite{zeusdch_hera2} & -0.6 & 0.8 \\
$\delta_{55}$ & 9 & ZEUS \Dch MC rew. $\eta$ & S & \cite{zeusdch_hera2} & 0.5 & 0.7 \\
$\delta_{56}$ & 9 & ZEUS \Dch tracking inefficiency & S & \cite{zeusdch_hera2} & -0.2 & 1.0 \\
$\delta_{57}$ & 9 & ZEUS \Dch MVD hit efficiency & S & \cite{zeusdch_hera2} & -0.0 & 1.0 \\
$\delta_{58}$ & 9 & ZEUS \Dch $\chi^2_{\rm sec.vtx.}$ distribution & S & \cite{zeusdch_hera2} & -0.1 & 1.0 \\
$\delta_{59}$ & 9 & $br(\Dch	\to K \pi \pi)$ & N & \cite{pdg2012} & -0.1 & 1.0 \\
$\delta_{60}$ & 10 & ZEUS \Dstar hadronic energy scale & S & \cite{zeusdstar_hera2} & 0.1 & 0.5 \\
$\delta_{61}$ & 10 & ZEUS \Dstar electron energy scale & S & \cite{zeusdstar_hera2} & -0.3 & 0.6 \\
$\delta_{62}$ & 10 & ZEUS \Dstar $p_T(\pi_s)$ & S & \cite{zeusdstar_hera2} & -1.0 & 0.9 \\
$\delta_{63}$ & 10 & ZEUS \Dstar tracking inefficiency & S & \cite{zeusdstar_hera2} & -0.8 & 0.9 \\
$\delta_{64}$ & 10 & ZEUS \Dstar PHP background & S & \cite{zeusdstar_hera2} & -0.4 & 1.0 \\
$\delta_{65}$ & 10 & ZEUS \Dstar diffractive backgr. & S & \cite{zeusdstar_hera2} & 0.4 & 0.9 \\
$\delta_{66}$ & 10 & ZEUS \Dstar MC rew. $p_T$, $Q^2$ & S & \cite{zeusdstar_hera2} & 0.5 & 0.9 \\
$\delta_{67}$ & 10 & ZEUS \Dstar MC rew. $\eta$ & S & \cite{zeusdstar_hera2} & 0.4 & 0.8 \\
$\delta_{68}$ & 10 & ZEUS \Dstar MC $b$ normalisation & S & \cite{zeusdstar_hera2} & -0.4 & 0.8 \\
$\delta_{69}$ & 11 & ZEUS VTX trigger inefficiency & S & \cite{zeussecvtx_hera2} & -0.3 & 0.9 \\
$\delta_{70}$ & 11 & ZEUS VTX tracking inefficiency & S & \cite{zeussecvtx_hera2} & 0.3 & 1.0 \\
$\delta_{71}$ & 11 & ZEUS VTX jet energy scale & S & \cite{zeussecvtx_hera2} & 0.6 & 0.7 \\
$\delta_{72}$ & 11 & ZEUS VTX electron energy scale & S & \cite{zeussecvtx_hera2} & 0.1 & 1.0 \\
$\delta_{73}$ & 11 & ZEUS VTX $c$ MC rew. $Q^2$ & S & \cite{zeussecvtx_hera2} & 0.3 & 0.9 \\
$\delta_{74}$ & 11 & ZEUS VTX $b$ MC rew. $Q^2$ & S & \cite{zeussecvtx_hera2} & -0.1 & 1.0 \\
$\delta_{75}$ & 11 & ZEUS VTX $c$ MC rew. $\eta$ & S & \cite{zeussecvtx_hera2} & 0.0 & 1.0 \\
$\delta_{76}$ & 11 & ZEUS VTX $b$ MC rew. $\eta$ & S & \cite{zeussecvtx_hera2} & -0.5 & 1.0 \\
$\delta_{77}$ & 11 & ZEUS VTX $c$ MC rew. $E_T$ & S & \cite{zeussecvtx_hera2} & 0.0 & 1.0 \\
$\delta_{78}$ & 11 & ZEUS VTX $b$ MC rew. $E_T$ & S & \cite{zeussecvtx_hera2} & -0.1 & 1.0 \\
\hline
\end{tabu}
\end{table}

\begin{table*}[tbp]
\caption[Combined reduced charm cross section with all correlated sources] 
{The combined reduced charm cross section, with its 
statistical ($\delta_{stat}$),uncorrelated ($\delta_{unc}$) and correlated ($\delta_1$ to $\delta_{78}$) uncertainties.}
{The correlated ($\delta_1$ to $\delta_{78}$) uncertainties
for the combined reduced charm cross section. For the cross section
values and their uncorrelated uncertainties see Table~\ref{tab:comb:red:combined}.} 
\label{tab:comb:red:combinedfull} 
\fontsize{1.87mm}{1em}
\begin{center} 
\tabcolsep0.25mm 
\renewcommand*{\arraystretch}{0.525} 
\begin{tabu} to \textwidth {|>{$}r<{$}|>{$}r<{$}|>{$}r<{$}|>{$}r<{$}|>{$}r<{$}|>{$}r<{$}|>{$}r<{$}|>{$}r<{$}|>{$}r<{$}|>{$}r<{$}|>{$}r<{$}|>{$}r<{$}|>{$}r<{$}|>{$}r<{$}|>{$}r<{$}|>{$}r<{$}|>{$}r<{$}|>{$}r<{$}|>{$}r<{$}|>{$}r<{$}|>{$}r<{$}|>{$}r<{$}|>{$}r<{$}|>{$}r<{$}|>{$}r<{$}|>{$}r<{$}|>{$}r<{$}|>{$}r<{$}|>{$}r<{$}|>{$}r<{$}|>{$}r<{$}|>{$}r<{$}|>{$}r<{$}|>{$}r<{$}|>{$}r<{$}|>{$}r<{$}|>{$}r<{$}|>{$}r<{$}|>{$}r<{$}|>{$}r<{$}|} \hline 
Q^2 & x & \delta_{1} & \delta_{2} & \delta_{3} & \delta_{4} & \delta_{5} & \delta_{6} & \delta_{7} & \delta_{8} & \delta_{9} & \delta_{10} & \delta_{11} & \delta_{12} & \delta_{13} & \delta_{14} & \delta_{15} & \delta_{16} & \delta_{17} & \delta_{18} & \delta_{19} & \delta_{20} & \delta_{21} & \delta_{22} & \delta_{23} & \delta_{24} & \delta_{25} & \delta_{26} & \delta_{27} & \delta_{28} & \delta_{29} & \delta_{30} & \delta_{31} & \delta_{32} & \delta_{33} & \delta_{34} & \delta_{35} & \delta_{36} & \delta_{37} & \delta_{38}\\
$(\SI{}{GeV^2})$ & & [\%] & [\%] & [\%] & [\%] & [\%] & [\%] & [\%] & [\%] & [\%] & [\%] & [\%] & [\%] & [\%] & [\%] & [\%] & [\%] & [\%] & [\%] & [\%] & [\%] & [\%] & [\%] & [\%] & [\%] & [\%] & [\%] & [\%] & [\%] & [\%] & [\%] & [\%] & [\%] & [\%] & [\%] & [\%] & [\%] & [\%] & [\%]\\
\hline 
2.5 & 0.00003  & -0.0 & 0.0 & -0.0 & 0.0 & -0.0 & -0.0 & -0.0 & -0.0 & 0.0 & -0.0 & -0.0 & -0.2 & -0.1 & -0.1 & -0.3 & -0.0 & 0.4 & 0.4 & 0.2 & 0.3 & -0.2 & 0.1 & 0.1 & 0.2 & 0.0 & 0.4 & -0.3 & 0.8 & 0.2 & -0.2 & -0.2 & 0.4 & 0.1 & 0.9 & -1.3 & -0.9 & 0.3 & 0.4 \\ 
2.5 & 0.00007  & -0.0 & 0.0 & -0.0 & 0.0 & -0.0 & -0.0 & -0.0 & -0.0 & 0.0 & -0.0 & -0.1 & -0.2 & -0.1 & -0.1 & -0.3 & 0.0 & 0.3 & 0.4 & 0.2 & 0.3 & -0.2 & 0.0 & 0.0 & 0.2 & -0.1 & 0.4 & -0.3 & 0.8 & 0.2 & -0.1 & -0.2 & 0.6 & 0.2 & 0.9 & -1.6 & -1.0 & 0.4 & 0.3 \\ 
2.5 & 0.00013  & -0.0 & 0.0 & -0.0 & 0.0 & -0.0 & -0.0 & -0.0 & -0.0 & 0.0 & -0.0 & -0.1 & -0.3 & -0.1 & -0.1 & -0.3 & 0.0 & 0.3 & 0.4 & 0.2 & 0.3 & -0.2 & 0.1 & -0.1 & 0.2 & -0.2 & 0.4 & -0.4 & 0.7 & 0.2 & -0.2 & -0.1 & 0.7 & 0.3 & 0.8 & -1.3 & -0.6 & 0.4 & -0.1 \\ 
2.5 & 0.00018  & -0.0 & 0.0 & -0.0 & 0.0 & -0.0 & -0.0 & -0.0 & -0.0 & 0.0 & -0.0 & -0.1 & -0.3 & -0.1 & -0.1 & -0.2 & 0.0 & 0.3 & 0.4 & 0.1 & 0.3 & -0.2 & 0.1 & -0.1 & 0.1 & -0.2 & 0.4 & -0.4 & 0.7 & 0.3 & -0.2 & -0.4 & 0.7 & 0.5 & 0.8 & -1.3 & -0.6 & 0.6 & 0.1 \\ 
2.5 & 0.00035  & -0.0 & 0.0 & -0.0 & 0.0 & -0.0 & -0.0 & -0.0 & -0.0 & 0.0 & -0.0 & -0.1 & -0.3 & -0.1 & -0.1 & -0.2 & 0.0 & 0.3 & 0.4 & 0.1 & 0.4 & -0.2 & 0.2 & -0.1 & 0.0 & -0.3 & 0.3 & -0.4 & 0.7 & 0.2 & -0.3 & -0.3 & 0.6 & 0.6 & 0.6 & -1.0 & -0.5 & 0.6 & -0.1 \\ 
5 & 0.00007 & -0.0 & 0.0 & -0.0 & 0.0 & -0.0 & -0.0 & -0.0 & -0.0 & 0.0 & -0.0 & -0.0 & -0.2 & -0.1 & -0.1 & -0.3 & -0.0 & 0.4 & 0.4 & 0.2 & 0.2 & -0.3 & 0.0 & 0.1 & 0.2 & 0.0 & 0.4 & -0.4 & 0.7 & 0.2 & -0.1 & -0.2 & 0.6 & 0.1 & 0.9 & -1.2 & -0.7 & 0.3 & 0.3 \\ 
5 & 0.00018 & -0.0 & 0.0 & -0.0 & 0.0 & -0.0 & -0.0 & -0.0 & -0.0 & 0.0 & -0.0 & -0.0 & -0.3 & -0.1 & -0.1 & -0.3 & 0.0 & 0.3 & 0.5 & 0.1 & 0.3 & -0.4 & 0.1 & 0.1 & 0.2 & -0.1 & 0.1 & -0.4 & 0.6 & 0.2 & -0.2 & 0.1 & 0.5 & 0.1 & 0.6 & -0.9 & -0.3 & 0.2 & -0.2 \\ 
5 & 0.00035 & -0.0 & 0.0 & -0.0 & 0.0 & -0.0 & -0.0 & -0.0 & -0.0 & 0.0 & -0.0 & -0.1 & -0.3 & -0.1 & -0.1 & -0.3 & 0.0 & 0.3 & 0.4 & 0.2 & 0.3 & -0.3 & 0.1 & 0.0 & 0.1 & -0.1 & 0.3 & -0.4 & 0.7 & 0.1 & -0.2 & -0.2 & 0.5 & 0.3 & 0.5 & -1.1 & -0.6 & 0.3 & 0.1 \\ 
5 & 0.00100 & -0.0 & 0.0 & -0.0 & 0.0 & -0.0 & -0.0 & -0.0 & -0.0 & 0.0 & -0.0 & -0.1 & -0.3 & -0.1 & -0.1 & -0.3 & 0.0 & 0.3 & 0.4 & 0.2 & 0.4 & -0.3 & 0.1 & -0.0 & 0.0 & -0.2 & 0.1 & -0.3 & 0.8 & 0.0 & -0.3 & -0.2 & 0.3 & 0.4 & 0.3 & -0.9 & -0.7 & 0.4 & 0.0 \\ 
7 & 0.00013 & -0.0 & 0.0 & -0.0 & 0.0 & -0.0 & -0.0 & -0.0 & -0.0 & 0.0 & -0.0 & -0.1 & -0.2 & -0.1 & -0.1 & -0.3 & -0.0 & 0.3 & 0.4 & 0.2 & 0.3 & -0.2 & 0.1 & 0.1 & 0.1 & -0.0 & 0.3 & -0.3 & 0.9 & -0.1 & -0.2 & -0.0 & 0.2 & 0.1 & 0.8 & -1.2 & -0.5 & 0.3 & -0.1 \\ 
7 & 0.00018 & -0.0 & 0.0 & -0.0 & 0.0 & -0.0 & -0.0 & -0.0 & -0.0 & 0.0 & -0.0 & -0.0 & -0.3 & -0.1 & -0.0 & -0.3 & -0.0 & 0.3 & 0.4 & 0.2 & 0.3 & -0.2 & 0.1 & 0.0 & 0.2 & -0.1 & 0.4 & -0.2 & 0.6 & 0.1 & -0.2 & -0.0 & 0.8 & 0.3 & 0.8 & -1.2 & -0.4 & -0.0 & -0.3 \\ 
7 & 0.00030 & -0.0 & 0.0 & -0.0 & 0.0 & -0.0 & -0.0 & -0.0 & -0.0 & 0.0 & -0.0 & -0.1 & -0.3 & -0.1 & -0.1 & -0.3 & -0.0 & 0.3 & 0.4 & 0.2 & 0.3 & -0.3 & 0.1 & 0.0 & 0.1 & -0.1 & 0.3 & -0.4 & 0.5 & 0.0 & -0.2 & 0.1 & 0.5 & 0.1 & 0.5 & -0.8 & -0.3 & 0.3 & -0.2 \\ 
7 & 0.00050 & -0.0 & 0.0 & -0.0 & 0.0 & -0.0 & -0.0 & -0.0 & -0.0 & 0.0 & -0.0 & -0.1 & -0.3 & -0.1 & -0.1 & -0.3 & -0.0 & 0.3 & 0.4 & 0.2 & 0.3 & -0.3 & 0.2 & 0.0 & 0.1 & -0.1 & 0.4 & -0.3 & 0.6 & -0.0 & -0.3 & 0.0 & 0.5 & 0.3 & 0.6 & -0.7 & -0.3 & 0.3 & -0.2 \\ 
7 & 0.00080 & -0.0 & 0.0 & -0.0 & 0.0 & -0.0 & -0.0 & -0.0 & -0.0 & 0.0 & -0.0 & -0.1 & -0.3 & -0.1 & -0.1 & -0.3 & 0.0 & 0.3 & 0.4 & 0.2 & 0.3 & -0.3 & 0.1 & 0.0 & 0.1 & 0.0 & 0.2 & -0.4 & 0.5 & 0.0 & -0.3 & -0.0 & 0.4 & 0.1 & 0.3 & -0.5 & -0.1 & 0.2 & -0.2 \\ 
7 & 0.00160 & -0.0 & 0.0 & -0.0 & 0.0 & -0.0 & -0.0 & -0.0 & -0.0 & 0.0 & -0.0 & -0.1 & -0.3 & -0.1 & -0.1 & -0.3 & 0.0 & 0.3 & 0.4 & 0.2 & 0.3 & -0.3 & 0.1 & 0.0 & 0.1 & -0.0 & 0.1 & -0.3 & 0.7 & 0.2 & -0.3 & -0.1 & 0.2 & 0.3 & 0.2 & -0.3 & -0.7 & 0.3 & -0.0 \\ 
12 & 0.00022 & -0.0 & 0.0 & -0.0 & 0.0 & -0.0 & -0.0 & -0.0 & -0.0 & 0.0 & -0.0 & -0.0 & -0.3 & -0.1 & -0.1 & -0.3 & 0.0 & 0.3 & 0.4 & 0.3 & 0.2 & -0.3 & 0.0 & 0.1 & 0.1 & 0.1 & 0.3 & -0.4 & 0.7 & 0.2 & -0.1 & -0.2 & 0.5 & 0.1 & 0.6 & -0.9 & -0.3 & 0.2 & -0.1 \\ 
12 & 0.00032 & -0.0 & 0.0 & 0.0 & 0.0 & -0.0 & 0.0 & -0.0 & -0.0 & 0.0 & -0.0 & -0.0 & -0.3 & -0.1 & -0.1 & -0.3 & -0.0 & 0.3 & 0.4 & 0.3 & 0.3 & -0.3 & 0.1 & 0.0 & 0.3 & 0.1 & 0.2 & -0.4 & 0.3 & -0.1 & -0.4 & -0.0 & 0.5 & 0.3 & 0.5 & -0.7 & -0.1 & 0.3 & -0.0 \\ 
12 & 0.00050 & -0.0 & 0.0 & -0.0 & 0.0 & -0.0 & -0.0 & -0.0 & -0.0 & 0.0 & -0.0 & -0.1 & -0.3 & -0.1 & -0.1 & -0.3 & 0.0 & 0.3 & 0.4 & 0.2 & 0.2 & -0.4 & 0.1 & 0.0 & 0.1 & 0.0 & 0.2 & -0.4 & 0.4 & 0.2 & -0.4 & -0.0 & 0.5 & 0.1 & 0.5 & -0.8 & -0.3 & 0.2 & -0.2 \\ 
12 & 0.00080 & -0.0 & 0.0 & -0.0 & 0.0 & -0.0 & -0.0 & -0.0 & -0.0 & 0.0 & -0.0 & -0.1 & -0.3 & -0.1 & -0.1 & -0.3 & 0.0 & 0.3 & 0.4 & 0.3 & 0.3 & -0.3 & 0.1 & 0.1 & 0.1 & 0.0 & 0.2 & -0.3 & 0.5 & 0.0 & -0.2 & -0.1 & 0.4 & -0.0 & 0.4 & -0.6 & -0.3 & 0.2 & -0.2 \\ 
12 & 0.00150 & -0.0 & 0.0 & -0.0 & 0.0 & -0.0 & -0.0 & -0.0 & -0.0 & 0.0 & -0.0 & -0.1 & -0.3 & -0.1 & -0.1 & -0.3 & -0.0 & 0.3 & 0.4 & 0.2 & 0.3 & -0.3 & 0.1 & 0.1 & 0.2 & -0.0 & -0.1 & -0.3 & 0.6 & -0.1 & -0.3 & -0.0 & 0.1 & 0.2 & 0.1 & -0.6 & -0.2 & 0.1 & 0.0 \\ 
12 & 0.00300 & -0.0 & 0.0 & -0.0 & 0.0 & -0.0 & -0.0 & -0.0 & -0.0 & 0.0 & -0.0 & -0.1 & -0.3 & -0.1 & -0.1 & -0.3 & 0.0 & 0.2 & 0.3 & 0.3 & 0.3 & -0.3 & 0.0 & 0.1 & 0.1 & -0.1 & 0.0 & -0.3 & 0.7 & 0.0 & -0.2 & -0.1 & 0.2 & 0.2 & 0.0 & -0.6 & -0.6 & 0.2 & 0.0 \\ 
18 & 0.00035 & -0.0 & 0.0 & -0.0 & 0.0 & -0.0 & -0.0 & -0.0 & -0.0 & 0.0 & -0.0 & -0.0 & -0.3 & -0.1 & -0.1 & -0.3 & -0.0 & 0.3 & 0.4 & 0.3 & 0.2 & -0.3 & 0.0 & 0.1 & 0.2 & 0.1 & 0.2 & -0.4 & 0.8 & 0.0 & 0.0 & -0.3 & 0.3 & 0.1 & 0.4 & -0.9 & -0.5 & 0.1 & 0.0 \\ 
18 & 0.00050 & -0.0 & 0.0 & -0.0 & 0.0 & -0.0 & -0.0 & -0.0 & -0.0 & 0.0 & -0.0 & -0.1 & -0.3 & -0.1 & -0.1 & -0.3 & -0.0 & 0.4 & 0.3 & 0.3 & 0.2 & -0.2 & -0.0 & 0.1 & 0.1 & -0.0 & 0.3 & -0.2 & 0.5 & -0.1 & -0.1 & 0.1 & 0.7 & 0.2 & 0.5 & -0.6 & -0.1 & 0.4 & 0.1 \\ 
18 & 0.00080 & -0.0 & 0.0 & -0.0 & 0.0 & -0.0 & -0.0 & -0.0 & -0.0 & 0.0 & -0.0 & -0.1 & -0.3 & -0.1 & -0.1 & -0.3 & 0.0 & 0.3 & 0.4 & 0.3 & 0.3 & -0.3 & 0.0 & 0.0 & 0.2 & 0.1 & 0.2 & -0.4 & 0.5 & 0.1 & -0.2 & -0.2 & 0.4 & -0.1 & 0.3 & -0.7 & -0.3 & 0.1 & -0.2 \\ 
18 & 0.00135 & -0.0 & 0.0 & -0.0 & 0.0 & -0.0 & -0.0 & -0.0 & -0.0 & 0.0 & -0.0 & -0.0 & -0.3 & -0.1 & -0.1 & -0.3 & -0.0 & 0.2 & 0.4 & 0.3 & 0.3 & -0.3 & 0.1 & 0.1 & 0.1 & 0.1 & 0.2 & -0.4 & 0.6 & 0.0 & -0.1 & -0.1 & 0.4 & -0.1 & 0.2 & -0.6 & -0.3 & 0.2 & -0.1 \\ 
18 & 0.00250 & -0.0 & 0.0 & -0.0 & 0.0 & -0.0 & -0.0 & -0.0 & -0.0 & 0.0 & -0.0 & -0.0 & -0.3 & -0.0 & -0.1 & -0.4 & -0.0 & 0.3 & 0.3 & 0.3 & 0.3 & -0.4 & 0.0 & 0.2 & 0.1 & 0.1 & 0.0 & -0.3 & 0.7 & -0.1 & -0.2 & -0.2 & 0.3 & -0.0 & 0.1 & -0.5 & -0.1 & 0.2 & 0.1 \\ 
18 & 0.00450 & -0.0 & 0.0 & -0.0 & 0.0 & -0.0 & -0.0 & -0.0 & -0.0 & 0.0 & -0.0 & -0.1 & -0.3 & -0.1 & -0.1 & -0.3 & 0.0 & 0.2 & 0.4 & 0.3 & 0.3 & -0.3 & 0.0 & 0.1 & 0.1 & -0.0 & 0.0 & -0.3 & 0.6 & -0.0 & -0.2 & 0.0 & 0.1 & 0.0 & 0.1 & -0.5 & -0.6 & 0.1 & -0.0 \\ 
32 & 0.00060 & -0.0 & 0.0 & -0.0 & 0.0 & -0.0 & -0.0 & -0.0 & -0.0 & 0.0 & -0.0 & -0.0 & -0.2 & -0.1 & -0.1 & -0.3 & 0.0 & 0.3 & 0.4 & 0.3 & 0.2 & -0.3 & -0.0 & 0.2 & 0.2 & 0.1 & 0.3 & -0.3 & 0.6 & 0.1 & -0.1 & -0.1 & 0.4 & -0.1 & 0.5 & -0.8 & -0.5 & 0.0 & 0.2 \\ 
32 & 0.00080 & -0.0 & 0.0 & -0.0 & 0.0 & -0.0 & -0.0 & -0.0 & -0.0 & 0.0 & -0.0 & -0.0 & -0.3 & -0.1 & -0.1 & -0.4 & 0.0 & 0.3 & 0.4 & 0.3 & 0.2 & -0.3 & -0.0 & 0.1 & 0.2 & 0.1 & 0.1 & -0.4 & 0.6 & 0.1 & -0.2 & -0.1 & 0.4 & -0.3 & 0.3 & -0.7 & -0.3 & 0.0 & -0.1 \\ 
32 & 0.00140 & -0.0 & 0.0 & -0.0 & 0.0 & -0.0 & -0.0 & -0.0 & -0.0 & 0.0 & -0.0 & -0.0 & -0.3 & -0.0 & -0.1 & -0.4 & 0.0 & 0.3 & 0.4 & 0.3 & 0.2 & -0.3 & 0.0 & 0.1 & 0.2 & 0.1 & 0.3 & -0.3 & 0.4 & 0.1 & -0.3 & -0.0 & 0.3 & 0.1 & 0.3 & -0.4 & -0.1 & 0.1 & -0.1 \\ 
32 & 0.00240 & -0.0 & 0.0 & -0.0 & 0.0 & -0.0 & -0.0 & -0.0 & -0.0 & 0.0 & -0.0 & -0.0 & -0.3 & -0.1 & -0.1 & -0.3 & -0.0 & 0.3 & 0.3 & 0.3 & 0.3 & -0.2 & 0.0 & 0.2 & 0.1 & -0.0 & 0.2 & -0.3 & 0.4 & -0.1 & -0.1 & -0.1 & 0.3 & 0.1 & 0.4 & -0.3 & -0.1 & 0.1 & -0.0 \\ 
32 & 0.00320 & -0.0 & 0.0 & -0.0 & 0.0 & -0.0 & -0.0 & -0.0 & -0.0 & 0.0 & 0.0 & -0.1 & -0.3 & -0.0 & -0.1 & -0.5 & -0.0 & 0.2 & 0.3 & 0.4 & 0.4 & -0.6 & -0.1 & 0.2 & 0.2 & 0.3 & -0.0 & -0.4 & 0.0 & -0.5 & -0.5 & -0.3 & 0.0 & 0.2 & 0.2 & -0.2 & -0.3 & 0.3 & 0.1 \\ 
32 & 0.00550 & -0.0 & 0.0 & -0.0 & 0.0 & -0.0 & -0.0 & -0.0 & -0.0 & 0.0 & -0.0 & -0.0 & -0.3 & -0.1 & -0.1 & -0.4 & -0.0 & 0.3 & 0.3 & 0.3 & 0.2 & -0.3 & -0.0 & 0.2 & 0.2 & 0.1 & 0.1 & -0.4 & 0.6 & -0.1 & -0.2 & -0.2 & 0.3 & 0.1 & 0.0 & -0.2 & 0.0 & -0.1 & 0.2 \\ 
32 & 0.00800 & -0.0 & 0.0 & -0.0 & 0.0 & -0.0 & -0.0 & -0.0 & -0.0 & 0.0 & -0.0 & -0.1 & -0.3 & -0.1 & -0.1 & -0.3 & -0.0 & 0.2 & 0.4 & 0.2 & 0.3 & -0.4 & 0.1 & 0.1 & 0.1 & 0.0 & 0.0 & -0.3 & 0.7 & -0.2 & -0.3 & 0.2 & -0.1 & -0.0 & 0.2 & -0.5 & -0.5 & 0.1 & -0.0 \\ 
60 & 0.00140 & -0.0 & 0.0 & -0.0 & 0.0 & -0.0 & -0.0 & -0.0 & -0.0 & 0.0 & -0.0 & -0.0 & -0.3 & -0.0 & -0.1 & -0.4 & -0.0 & 0.3 & 0.4 & 0.3 & 0.2 & -0.4 & 0.0 & 0.1 & 0.2 & 0.3 & 0.1 & -0.4 & 0.4 & -0.0 & -0.1 & -0.1 & 0.4 & -0.1 & 0.1 & -0.4 & -0.1 & 0.0 & 0.1 \\ 
60 & 0.00200 & -0.0 & 0.0 & 0.0 & 0.0 & -0.0 & 0.0 & -0.0 & -0.0 & 0.0 & -0.0 & -0.0 & -0.2 & -0.0 & -0.1 & -0.4 & 0.0 & 0.2 & 0.4 & 0.4 & 0.2 & -0.1 & 0.0 & 0.1 & 0.3 & 0.3 & 0.2 & -0.4 & 0.2 & -0.1 & -0.2 & -0.3 & 0.1 & -0.2 & 0.2 & -0.4 & -0.1 & -0.1 & 0.0 \\ 
60 & 0.00320 & -0.0 & 0.0 & -0.0 & 0.0 & -0.0 & -0.0 & -0.0 & -0.0 & 0.0 & -0.0 & -0.0 & -0.3 & -0.1 & -0.1 & -0.4 & 0.0 & 0.3 & 0.4 & 0.3 & 0.2 & -0.3 & -0.0 & 0.2 & 0.1 & 0.2 & 0.2 & -0.4 & 0.4 & 0.1 & -0.2 & -0.2 & 0.2 & -0.2 & 0.0 & -0.4 & -0.1 & -0.1 & -0.1 \\ 
60 & 0.00500 & -0.0 & 0.0 & -0.0 & 0.0 & -0.0 & -0.0 & -0.0 & -0.0 & 0.0 & -0.0 & -0.0 & -0.2 & -0.0 & -0.1 & -0.4 & 0.0 & 0.2 & 0.4 & 0.3 & 0.2 & -0.3 & 0.1 & 0.1 & 0.2 & 0.2 & -0.0 & -0.2 & 0.6 & -0.1 & -0.0 & -0.1 & 0.4 & -0.1 & 0.3 & -0.1 & -0.1 & 0.2 & 0.1 \\ 
60 & 0.00800 & -0.0 & 0.0 & -0.0 & 0.0 & -0.0 & -0.0 & -0.0 & -0.0 & 0.0 & -0.0 & -0.0 & -0.3 & -0.0 & -0.0 & -0.4 & 0.1 & 0.3 & 0.3 & 0.3 & 0.2 & -0.4 & 0.0 & 0.3 & 0.3 & 0.3 & 0.1 & -0.4 & 0.4 & -0.3 & -0.3 & -0.2 & 0.0 & -0.2 & 0.2 & -0.2 & -0.2 & 0.1 & 0.1 \\ 
60 & 0.01500 & -0.0 & 0.0 & -0.0 & 0.0 & -0.0 & -0.0 & -0.0 & -0.0 & 0.0 & -0.0 & -0.0 & -0.3 & -0.1 & -0.1 & -0.4 & 0.0 & 0.3 & 0.4 & 0.4 & 0.2 & -0.4 & -0.0 & 0.2 & 0.2 & 0.2 & -0.0 & -0.3 & 0.6 & -0.2 & -0.1 & -0.0 & 0.2 & -0.1 & -0.3 & -0.3 & -0.4 & -0.3 & 0.1 \\ 
120 & 0.00200 & -0.0 & 0.0 & -0.0 & 0.0 & -0.0 & -0.0 & -0.0 & -0.0 & 0.0 & 0.0 & -0.0 & -0.2 & -0.0 & -0.2 & -0.5 & 0.0 & 0.3 & 0.4 & 0.1 & 0.3 & -0.2 & 0.4 & 0.2 & 0.3 & 0.3 & 0.5 & 0.1 & 0.5 & -0.3 & -0.6 & -0.2 & 0.2 & 0.0 & -0.2 & -0.1 & 0.2 & -0.1 & 0.3 \\ 
120 & 0.00320 & -0.0 & 0.0 & -0.0 & 0.0 & -0.0 & -0.0 & -0.0 & -0.0 & 0.0 & -0.0 & -0.0 & -0.3 & -0.1 & -0.1 & -0.5 & 0.0 & 0.1 & 0.6 & 0.3 & 0.2 & -0.3 & 0.0 & 0.2 & 0.2 & 0.2 & 0.1 & -0.3 & 0.4 & -0.2 & -0.2 & -0.1 & 0.2 & -0.1 & 0.0 & -0.3 & -0.2 & -0.0 & 0.1 \\ 
120 & 0.00550 & -0.0 & 0.0 & -0.0 & 0.0 & -0.0 & -0.0 & -0.0 & -0.0 & 0.0 & -0.0 & -0.0 & -0.3 & -0.1 & -0.1 & -0.2 & 0.0 & 0.3 & 0.6 & 0.3 & 0.2 & -0.5 & 0.1 & 0.2 & 0.2 & 0.5 & 0.2 & -0.4 & 0.2 & -0.3 & -0.1 & -0.1 & -0.0 & -0.2 & 0.0 & -0.1 & 0.1 & -0.1 & 0.1 \\ 
120 & 0.01000 & -0.0 & 0.0 & -0.0 & 0.0 & -0.0 & -0.0 & -0.0 & -0.0 & 0.0 & -0.0 & -0.1 & -0.2 & -0.1 & -0.1 & -0.5 & 0.1 & 0.3 & 0.5 & 0.1 & 0.3 & -0.4 & 0.1 & 0.1 & 0.5 & 0.1 & 0.2 & -0.1 & 0.1 & -0.2 & -0.1 & -0.2 & 0.1 & -0.0 & 0.2 & -0.2 & -0.1 & 0.2 & 0.1 \\ 
120 & 0.02500 & -0.0 & 0.0 & -0.0 & 0.0 & -0.0 & -0.0 & -0.0 & -0.0 & 0.0 & -0.0 & -0.0 & -0.3 & -0.1 & -0.1 & -0.4 & 0.0 & 0.2 & 0.5 & 0.3 & 0.2 & -0.4 & 0.0 & 0.2 & 0.3 & 0.2 & 0.0 & -0.3 & 0.4 & -0.2 & -0.1 & 0.1 & 0.0 & -0.1 & -0.5 & -0.3 & -0.0 & 0.2 & -0.1 \\ 
200 & 0.00500 & -0.0 & 0.0 & -0.0 & 0.0 & -0.0 & -0.0 & -0.0 & -0.0 & 0.0 & -0.0 & -0.0 & -0.2 & -0.1 & -0.3 & -0.5 & 0.0 & 0.3 & 0.4 & 0.3 & 0.3 & -0.5 & 0.1 & 0.4 & 0.4 & 0.2 & 0.2 & 0.1 & 0.4 & -0.4 & -0.7 & -0.2 & 0.2 & -0.1 & -0.1 & 0.1 & -0.0 & -0.4 & 0.0 \\ 
200 & 0.01300 & -0.0 & 0.0 & -0.0 & 0.0 & -0.0 & -0.0 & -0.0 & -0.0 & 0.0 & -0.0 & -0.0 & -0.2 & -0.1 & -0.1 & -0.5 & 0.1 & 0.3 & 0.4 & 0.1 & 0.4 & -0.2 & 0.1 & 0.3 & 0.4 & -0.0 & 0.4 & 0.2 & 0.6 & -0.1 & -0.7 & -0.5 & 0.1 & -0.2 & 0.1 & -0.2 & -0.2 & -0.1 & -0.1 \\ 
350 & 0.01000 & -0.0 & 0.0 & 0.0 & 0.0 & -0.0 & 0.0 & -0.0 & -0.0 & 0.0 & -0.0 & -0.0 & -0.2 & -0.0 & -0.2 & -0.5 & -0.0 & 0.2 & 0.4 & 0.3 & 0.3 & -0.4 & 0.1 & 0.5 & 0.6 & 0.2 & -0.0 & 0.3 & 0.3 & -0.2 & -0.7 & -0.4 & -0.1 & -0.2 & -0.3 & 0.2 & 0.4 & -0.3 & 0.2 \\ 
350 & 0.02500 & -0.0 & 0.0 & -0.0 & 0.0 & -0.0 & -0.0 & -0.0 & -0.0 & 0.0 & -0.0 & -0.1 & -0.2 & -0.1 & -0.2 & -0.4 & 0.1 & 0.2 & 0.6 & 0.2 & 0.4 & -0.5 & 0.2 & 0.4 & 0.5 & 0.3 & 0.1 & 0.0 & 0.6 & -0.4 & -0.5 & -0.5 & 0.1 & -0.7 & -0.2 & 0.0 & -0.1 & -0.0 & 0.2 \\ 
650 & 0.01300 & -0.0 & 0.0 & -0.0 & 0.0 & -0.0 & -0.0 & -0.0 & -0.0 & 0.1 & 0.0 & -0.0 & -0.0 & -0.1 & -0.3 & -0.6 & 0.1 & 0.3 & 0.3 & -0.3 & 0.1 & -1.2 & 0.4 & 0.5 & 0.3 & 0.1 & -0.2 & 0.3 & 0.6 & -0.1 & -0.4 & -0.1 & -0.1 & -0.6 & -0.7 & 0.2 & 0.3 & -0.8 & -0.3 \\ 
650 & 0.03200 & -0.0 & 0.0 & 0.0 & 0.0 & -0.0 & -0.0 & -0.0 & -0.0 & 0.2 & 0.0 & -0.0 & -0.1 & 0.1 & -0.2 & -1.2 & -0.1 & 0.1 & 0.4 & 0.4 & 0.7 & -0.5 & 0.7 & 0.2 & 0.7 & 0.2 & 0.5 & -0.6 & 0.4 & -0.3 & -0.6 & 0.2 & -0.8 & 0.2 & -0.5 & -0.6 & 0.2 & -1.0 & -0.3 \\ 
2000 & 0.05000 & -0.0 & 0.0 & 0.0 & 0.0 & -0.0 & -0.0 & -0.0 & -0.0 & 0.1 & 0.0 & -0.1 & 0.1 & -0.4 & -0.6 & -1.3 & 0.4 & 1.1 & 0.9 & -1.3 & -0.4 & -0.8 & 0.6 & 0.7 & -0.2 & -0.7 & 0.3 & -0.3 & 0.2 & 0.2 & -0.8 & 0.3 & -1.7 & -0.3 & -1.2 & -0.4 & 0.6 & -2.6 & -1.2 \\ 
\hline
\end{tabu}
\begin{tabu} to \textwidth {|>{$}l<{$}|>{$}l<{$}|>{$}r<{$}|>{$}r<{$}|>{$}r<{$}|>{$}r<{$}|>{$}r<{$}|>{$}r<{$}|>{$}r<{$}|>{$}r<{$}|>{$}r<{$}|>{$}r<{$}|>{$}r<{$}|>{$}r<{$}|>{$}r<{$}|>{$}r<{$}|>{$}r<{$}|>{$}r<{$}|>{$}r<{$}|>{$}r<{$}|>{$}r<{$}|>{$}r<{$}|>{$}r<{$}|>{$}r<{$}|>{$}r<{$}|>{$}r<{$}|>{$}r<{$}|>{$}r<{$}|>{$}r<{$}|>{$}r<{$}|>{$}r<{$}|>{$}r<{$}|>{$}r<{$}|>{$}r<{$}|>{$}r<{$}|>{$}r<{$}|>{$}r<{$}|>{$}r<{$}|>{$}r<{$}|>{$}r<{$}|>{$}r<{$}|>{$}r<{$}|} \hline 
Q^2 & x & \delta_{39} & \delta_{40} & \delta_{41} & \delta_{42} & \delta_{43} & \delta_{44} & \delta_{45} & \delta_{46} & \delta_{47} & \delta_{48} & \delta_{49} & \delta_{50} & \delta_{51} & \delta_{52} & \delta_{53} & \delta_{54} & \delta_{55} & \delta_{56} & \delta_{57} & \delta_{58} & \delta_{59} & \delta_{60} & \delta_{61} & \delta_{62} & \delta_{63} & \delta_{64} & \delta_{65} & \delta_{66} & \delta_{67} & \delta_{68} & \delta_{69} & \delta_{70} & \delta_{71} & \delta_{72} & \delta_{73} & \delta_{74} & \delta_{75} & \delta_{76} & \delta_{77} & \delta_{78}\\
$(\SI{}{GeV^2})$ & & [\%] & [\%] & [\%] & [\%] & [\%] & [\%] & [\%] & [\%] & [\%] & [\%] & [\%] & [\%] & [\%] & [\%] & [\%] & [\%] & [\%] & [\%] & [\%] & [\%] & [\%] & [\%] & [\%] & [\%] & [\%] & [\%] & [\%] & [\%] & [\%] & [\%] & [\%] & [\%] & [\%] & [\%] & [\%] & [\%] & [\%] & [\%] & [\%] & [\%]\\
\hline 
2.5 & 0.00003 & -0.2 & 1.2 & -1.0 & 0.4 & -0.7 & 0.9 & 1.2 & 0.7 & 0.4 & 0.4 & -1.3 & -0.1 & -2.0 & 2.4 & 1.0 & 1.8 & 0.2 & -1.1 & -2.2 & -1.1 & -1.5 & -2.2 & 0.2 & -0.0 & -0.4 & -2.4 & -1.2 & 1.0 & -1.7 & 0.3 & 1.7 & -1.4 & 1.4 & 0.1 & -0.7 & 0.3 & 1.1 & 0.7 & -2.2 & 0.3 \\ 
2.5 & 0.00007 & 0.0 & 1.3 & -0.9 & 0.5 & -0.9 & 0.7 & 1.0 & 0.8 & 0.6 & 0.2 & -0.6 & -0.1 & -2.4 & 2.3 & 1.1 & 2.2 & 0.1 & -1.2 & -2.1 & -1.2 & -1.4 & -2.2 & 0.3 & 0.1 & -0.1 & -2.2 & -0.8 & 1.3 & -2.4 & 1.2 & 0.1 & -1.5 & 1.2 & -0.4 & -0.4 & 0.7 & 0.8 & 0.5 & -1.7 & 0.1 \\ 
2.5 & 0.00013 & 0.4 & 1.4 & -0.8 & 0.7 & -0.8 & 0.8 & 0.9 & 0.5 & 0.5 & 0.2 & -0.5 & 0.3 & -1.6 & 1.5 & 0.7 & 2.4 & 0.3 & -1.2 & -2.2 & -0.5 & -1.4 & -2.6 & 0.4 & 0.4 & -0.0 & -2.1 & -0.5 & 1.3 & -2.4 & 1.6 & -0.6 & -1.7 & 1.3 & -0.5 & -0.4 & 0.7 & 0.6 & 0.3 & -1.2 & -0.1 \\ 
2.5 & 0.00018 & 0.5 & 1.6 & -0.7 & 0.8 & -1.2 & 0.7 & 0.6 & 0.7 & 0.2 & 0.4 & -0.1 & 0.1 & -1.4 & 2.2 & 1.2 & 2.1 & -0.1 & -1.1 & -1.2 & -1.6 & -1.1 & -0.4 & 0.3 & -0.2 & -0.5 & -1.7 & -2.1 & 1.3 & -2.4 & 0.8 & 0.2 & -1.4 & 0.7 & -0.7 & -0.0 & 0.5 & 0.8 & 0.3 & -0.2 & -0.3 \\ 
2.5 & 0.00035 & 0.8 & 1.5 & -0.7 & 1.0 & -1.1 & 1.1 & 0.5 & 0.3 & 0.1 & 0.3 & -0.0 & 0.5 & -0.2 & 1.6 & 0.6 & 2.3 & 0.1 & -1.2 & -1.1 & -1.0 & -0.6 & -1.3 & -0.6 & 0.3 & -0.6 & -1.1 & -1.1 & 1.2 & -2.3 & 1.0 & 0.3 & -1.5 & 1.2 & -0.6 & -0.0 & 0.2 & 0.7 & -0.0 & 0.8 & -0.5 \\ 
5.0 & 0.00007 & -0.0 & 1.3 & -0.8 & 0.4 & -0.8 & 0.7 & 0.9 & 0.9 & 0.5 & 0.3 & -0.5 & -0.5 & -2.5 & 2.5 & 1.0 & 2.1 & -0.2 & -1.1 & -1.6 & -1.8 & -1.2 & -1.5 & -0.2 & -0.2 & -0.5 & -2.1 & -1.0 & 1.0 & -1.9 & 0.5 & 1.1 & -1.2 & 1.1 & -0.2 & -0.5 & 0.4 & 1.0 & 0.6 & -1.8 & 0.2 \\ 
5.0 & 0.00018 & 0.3 & 1.1 & -0.9 & 0.7 & -0.7 & 1.6 & 1.6 & 0.6 & 1.4 & 0.0 & -0.2 & -0.3 & -1.4 & 0.9 & 0.1 & 2.4 & -0.1 & -1.1 & -1.6 & -0.9 & -1.1 & 0.2 & 0.2 & -0.1 & 0.3 & -1.6 & -0.9 & 0.2 & -0.9 & 0.6 & 0.0 & -0.2 & 0.9 & -0.1 & -0.1 & 0.1 & 0.5 & 0.3 & 0.5 & 0.9 \\ 
5.0 & 0.00035 & 0.3 & 1.2 & -0.8 & 0.9 & -0.9 & 0.7 & 0.5 & 0.5 & 0.3 & 0.3 & -0.2 & 0.1 & -1.2 & 1.4 & 0.7 & 2.0 & -0.2 & -0.9 & -1.1 & -1.2 & -0.5 & -0.7 & -0.2 & -0.3 & -0.6 & -1.5 & -0.7 & 0.7 & -1.6 & 0.2 & 0.6 & -1.2 & 0.6 & -0.7 & 0.1 & 0.3 & 0.6 & -0.0 & 0.5 & -0.4 \\ 
5.0 & 0.00100 & 0.5 & 1.2 & -0.8 & 1.1 & -1.0 & 0.9 & 0.2 & 0.1 & -0.1 & 0.2 & 0.2 & 0.9 & -0.1 & 0.9 & 0.7 & 1.3 & -0.0 & -0.6 & -0.7 & -0.6 & 0.0 & -0.2 & -0.3 & -0.2 & -0.8 & -0.5 & -0.3 & 0.7 & -1.6 & 0.0 & 0.8 & -1.1 & 0.8 & -0.7 & 0.3 & -0.1 & 0.4 & -0.5 & 2.3 & -0.7 \\ 
7.0 & 0.00013 & 0.2 & 1.3 & -0.7 & 0.7 & -0.4 & 0.8 & 0.6 & 1.0 & 0.3 & 0.5 & -0.6 & -0.5 & -1.9 & 1.6 & -0.1 & 2.0 & 0.5 & -0.7 & -1.7 & -1.7 & -1.0 & -1.1 & 1.2 & 0.0 & -0.1 & -1.9 & -1.0 & 0.5 & -0.7 & 1.2 & 0.1 & -1.4 & 1.2 & -0.4 & -0.4 & 0.4 & 0.8 & 0.4 & -1.6 & 0.1 \\ 
7.0 & 0.00018 & 0.3 & 1.0 & -0.5 & 0.6 & -0.5 & 0.7 & 0.6 & 1.0 & 0.4 & -0.2 & -0.3 & 0.1 & -1.6 & 1.1 & 0.8 & 2.3 & 0.5 & -0.9 & -2.8 & -0.3 & -1.1 & -2.6 & 0.1 & 0.1 & -0.0 & -2.4 & 0.2 & 0.6 & -1.2 & 0.6 & 0.2 & -1.6 & 1.3 & -0.1 & -0.1 & 0.5 & 0.5 & 0.7 & -1.1 & -0.0 \\ 
7.0 & 0.00030 & 0.4 & 1.1 & -0.7 & 0.9 & -0.5 & 0.8 & 0.6 & 0.4 & 0.4 & 0.3 & -0.0 & -0.0 & -1.0 & 1.1 & 0.2 & 1.8 & 0.3 & -0.7 & -1.4 & -1.1 & -0.8 & -0.3 & 0.7 & -0.3 & -0.2 & -2.0 & -0.7 & 0.3 & -0.8 & 1.2 & -0.3 & -0.8 & 1.2 & -0.3 & -0.1 & 0.3 & 0.5 & -0.0 & 0.3 & 0.2 \\ 
7.0 & 0.00050 & 0.4 & 1.3 & -0.6 & 1.0 & -0.7 & 0.9 & 0.2 & 0.9 & 0.2 & 0.1 & -0.1 & 0.0 & -0.7 & 1.1 & -0.1 & 1.7 & -0.2 & -0.8 & -0.7 & -0.4 & -0.6 & -0.1 & 0.1 & -0.3 & -0.2 & -1.4 & -1.0 & 0.2 & -0.4 & 0.6 & -0.2 & -1.3 & 0.8 & -0.4 & 0.2 & 0.2 & 0.4 & 0.1 & 0.8 & 0.2 \\ 
7.0 & 0.00080 & 0.3 & 1.1 & -0.6 & 1.0 & -1.1 & 0.4 & 0.2 & 0.4 & 0.3 & 0.2 & 0.2 & 0.2 & -0.2 & 0.7 & -0.0 & 0.5 & 0.2 & -0.3 & -0.8 & -0.6 & -1.1 & 0.5 & 1.2 & -0.4 & -0.5 & -1.2 & -0.7 & 0.6 & -0.9 & 0.9 & -0.3 & -1.6 & 0.7 & -0.3 & -0.0 & 0.2 & 0.2 & -0.5 & 0.9 & -0.5 \\ 
7.0 & 0.00160 & 0.4 & 1.1 & -0.7 & 1.0 & -0.9 & 0.5 & 0.1 & 0.1 & -0.0 & -0.1 & 0.3 & 0.8 & 0.1 & 0.9 & 0.9 & 1.7 & -0.3 & -0.9 & 0.1 & -0.3 & 0.4 & -0.5 & -0.2 & 0.4 & -0.2 & -0.3 & -0.7 & 0.2 & -0.5 & 0.4 & 0.3 & -1.3 & 1.4 & 0.1 & -0.0 & -0.2 & -0.1 & -0.4 & 2.0 & -0.7 \\ 
12.0 & 0.00022 & 0.0 & 1.3 & -0.7 & 0.6 & -1.0 & 0.3 & 0.8 & 1.0 & 0.2 & 0.7 & -0.4 & -0.3 & -1.5 & 1.4 & 0.5 & 1.4 & 0.4 & -0.6 & -1.4 & -1.6 & -1.2 & -0.7 & 0.9 & 0.2 & -0.1 & -1.6 & -1.1 & 0.3 & -0.6 & 0.6 & 0.5 & -0.9 & 1.9 & -0.8 & -0.2 & 0.5 & 0.7 & 0.2 & -1.1 & -0.0 \\ 
12.0 & 0.00032 & 0.1 & 1.1 & -1.0 & 0.8 & -0.4 & 0.4 & 0.5 & 0.3 & 0.5 & 0.1 & -0.2 & -0.1 & -0.9 & 1.2 & 0.6 & 2.0 & 0.6 & -0.7 & -1.9 & -1.1 & -0.9 & -0.4 & 0.7 & -0.3 & -0.0 & -2.4 & -1.2 & 0.1 & -1.0 & 0.8 & 0.1 & 0.2 & 1.1 & 0.2 & -0.1 & 0.1 & 0.5 & 0.2 & 0.8 & 0.7 \\ 
12.0 & 0.00050 & 0.3 & 1.2 & -0.7 & 1.0 & -0.3 & 0.6 & 0.2 & 0.8 & 0.2 & 0.3 & -0.1 & -0.1 & -1.0 & 0.5 & -0.0 & 1.3 & 0.0 & -0.6 & -1.0 & -0.6 & -0.7 & -0.0 & 0.4 & -0.1 & 0.0 & -1.4 & -0.8 & 0.0 & -0.4 & 0.4 & -0.3 & -0.4 & 1.0 & -0.3 & 0.0 & 0.1 & 0.5 & -0.0 & 0.5 & 0.3 \\ 
12.0 & 0.00080 & 0.3 & 1.1 & -0.7 & 0.9 & -0.8 & 0.4 & 0.3 & 0.5 & 0.3 & 0.2 & 0.2 & 0.4 & -0.9 & 0.6 & 0.3 & 0.8 & -0.0 & -0.2 & -0.5 & -0.3 & 0.1 & -0.3 & 0.2 & -0.5 & -0.8 & -1.5 & -0.8 & 0.2 & -0.5 & 0.2 & -0.5 & -1.0 & 0.9 & -0.5 & 0.0 & 0.1 & 0.4 & -0.0 & 0.6 & 0.4 \\ 
12.0 & 0.00150 & 0.0 & 1.2 & -0.6 & 1.2 & -0.9 & 0.5 & 0.2 & 0.1 & 0.0 & 0.2 & 0.3 & 0.6 & 0.2 & 0.7 & 0.1 & 1.2 & -0.2 & -0.5 & -0.3 & 0.1 & 0.2 & 0.3 & 0.1 & -0.6 & -0.8 & -0.7 & -1.0 & 0.1 & -0.3 & -0.3 & -0.5 & -1.7 & 1.1 & -0.4 & -0.0 & -0.2 & 0.4 & -0.3 & 1.3 & 0.4 \\ 
12.0 & 0.00300 & 0.3 & 0.9 & -0.6 & 1.2 & -0.9 & 0.5 & -0.2 & -0.0 & -0.0 & 0.1 & 0.5 & 0.9 & 0.1 & 0.2 & 0.8 & 1.2 & -0.2 & -0.5 & -0.2 & -0.0 & 1.0 & 0.2 & -0.4 & 0.1 & -0.5 & 0.7 & -0.4 & 0.5 & -0.9 & 0.4 & -0.5 & -1.0 & 1.0 & 0.2 & -0.1 & -0.4 & 0.0 & -1.1 & 2.5 & -0.7 \\ 
18.0 & 0.00035 & -0.1 & 1.1 & -0.7 & 0.7 & -0.6 & 0.3 & 0.5 & 0.7 & 0.3 & 0.4 & -0.3 & -0.3 & -1.4 & 1.1 & 0.7 & 1.7 & 0.3 & -0.7 & -1.4 & -1.0 & -0.7 & -0.8 & 0.7 & 0.3 & -0.0 & -1.3 & -1.7 & -0.0 & 0.1 & 0.5 & 0.5 & -0.8 & 1.7 & -0.8 & -0.1 & 0.4 & 0.5 & -0.0 & -0.7 & -0.1 \\ 
18.0 & 0.00050 & 0.2 & 0.8 & -0.6 & 0.8 & -0.9 & 0.9 & 0.4 & 0.5 & -0.1 & 0.4 & 0.0 & 0.0 & -1.3 & 1.1 & 0.6 & 1.8 & 0.7 & -0.5 & -1.9 & -1.2 & -0.6 & -0.5 & 0.6 & -0.3 & 0.0 & -2.4 & -1.3 & 0.1 & -0.9 & 1.0 & -0.2 & 0.1 & 1.3 & 0.4 & -0.1 & 0.0 & 0.4 & 0.5 & 0.8 & 0.8 \\ 
18.0 & 0.00080 & 0.3 & 1.0 & -0.6 & 0.9 & -0.2 & 0.5 & 0.0 & 0.5 & 0.2 & 0.2 & -0.0 & 0.0 & -0.8 & 0.5 & -0.0 & 1.2 & 0.0 & -0.5 & -0.8 & -0.3 & -0.4 & -0.0 & 0.3 & -0.2 & 0.1 & -1.1 & -0.9 & -0.0 & -0.3 & 0.3 & -0.3 & -0.3 & 0.8 & -0.4 & 0.1 & 0.1 & 0.4 & -0.0 & 0.6 & 0.3 \\ 
18.0 & 0.00135 & 0.2 & 1.0 & -0.5 & 1.0 & -0.6 & 0.6 & 0.0 & 0.4 & 0.1 & 0.1 & 0.1 & 0.2 & -0.3 & 0.7 & 0.3 & 1.3 & -0.1 & -0.6 & -0.6 & -0.2 & -0.2 & 0.4 & 0.1 & -0.4 & -0.1 & -1.0 & -0.5 & 0.1 & -0.3 & 0.1 & -0.6 & -1.2 & 1.0 & -0.2 & 0.5 & 0.1 & 0.2 & 0.3 & 1.0 & 0.1 \\ 
18.0 & 0.00250 & 0.0 & 0.7 & -0.8 & 0.9 & -0.5 & 0.2 & -0.1 & 0.1 & 0.0 & 0.1 & 0.3 & 0.4 & 0.2 & 1.0 & 0.2 & 1.1 & -0.2 & -0.5 & -0.0 & 0.4 & 0.1 & 0.4 & 0.4 & -0.3 & -0.3 & -0.2 & -0.5 & 0.2 & -0.5 & -0.1 & -0.4 & -1.3 & 1.0 & -0.4 & -0.0 & -0.2 & 0.1 & 0.1 & 1.3 & 0.3 \\ 
18.0 & 0.00450 & 0.3 & 0.8 & -0.6 & 1.2 & -0.6 & 0.7 & -0.2 & -0.3 & -0.2 & 0.1 & 0.1 & 0.8 & 0.5 & 0.1 & 1.3 & 0.8 & -0.1 & -0.3 & -0.2 & -0.4 & 0.7 & 0.5 & 0.3 & 0.2 & -0.3 & 0.4 & 0.7 & 0.0 & -0.1 & -0.0 & -0.8 & -1.7 & 1.4 & 1.5 & 1.2 & -0.3 & 0.5 & -0.8 & 2.4 & -0.7 \\ 
32.0 & 0.00060 & -0.2 & 0.9 & -0.7 & 0.6 & -0.6 & 0.4 & 0.4 & 0.8 & 0.4 & 0.2 & -0.4 & -0.5 & -1.9 & 1.5 & 0.8 & 1.6 & -0.2 & -0.7 & -1.1 & -0.8 & -0.1 & -0.6 & -0.1 & -0.6 & -0.5 & -1.3 & -0.6 & 0.6 & -1.2 & 0.2 & 0.5 & -0.8 & 0.5 & -0.4 & -0.1 & 0.2 & 0.5 & 0.1 & -0.5 & -0.1 \\ 
32.0 & 0.00080 & 0.2 & 0.9 & -0.3 & 0.7 & -0.7 & 0.5 & 0.3 & 0.4 & 0.1 & 0.4 & -0.1 & -0.1 & -1.2 & 0.8 & 0.4 & 1.4 & 0.3 & -0.4 & -1.3 & -0.9 & -0.4 & -0.4 & 0.4 & 0.0 & -0.1 & -1.4 & -1.0 & -0.4 & -0.9 & 0.5 & -0.2 & -0.3 & 1.3 & -0.3 & -0.1 & 0.2 & 0.5 & 0.2 & 0.3 & 0.3 \\ 
32.0 & 0.00140 & 0.1 & 0.8 & -0.5 & 0.8 & -0.4 & 0.4 & -0.2 & 0.4 & 0.3 & 0.3 & 0.0 & -0.1 & -0.6 & 0.3 & -0.0 & 1.3 & -0.1 & -0.3 & -0.8 & -0.2 & -0.1 & 0.3 & 0.1 & -0.3 & -0.3 & -0.9 & -0.7 & 0.1 & -0.3 & 0.3 & -0.2 & -0.6 & 1.0 & -0.3 & -0.1 & 0.1 & 0.3 & 0.4 & 0.4 & 0.0 \\ 
32.0 & 0.00240 & 0.2 & 0.7 & -0.4 & 1.1 & -0.7 & 0.8 & -0.0 & 0.4 & 0.3 & 0.3 & 0.2 & 0.2 & -0.3 & 0.2 & -0.1 & 1.0 & -0.6 & -0.3 & -0.5 & 0.1 & -0.1 & 0.7 & 0.1 & 0.0 & -0.5 & -1.0 & -0.5 & 0.1 & -0.4 & 0.1 & -0.3 & -0.8 & 0.9 & -0.1 & -0.2 & -0.0 & 0.5 & 0.1 & 0.3 & 0.3 \\ 
32.0 & 0.00320 & 0.1 & 0.4 & -1.3 & 0.6 & -0.1 & -0.5 & -0.3 & -0.4 & 0.8 & 0.0 & -0.0 & 0.1 & 0.3 & 0.7 & 0.4 & 0.7 & 0.4 & 0.0 & 0.2 & 0.1 & 0.8 & 0.4 & 0.3 & -0.1 & -0.3 & -0.7 & -0.2 & 0.2 & -0.3 & -0.0 & -0.2 & -1.5 & 1.1 & -0.5 & -0.0 & -0.2 & -0.2 & 0.7 & 1.1 & 0.6 \\ 
32.0 & 0.00550 & 0.4 & 0.7 & -0.5 & 1.0 & -0.7 & 0.5 & -0.1 & 0.1 & 0.2 & 0.1 & 0.4 & -0.1 & 0.2 & -0.3 & 0.1 & 1.2 & -0.7 & -0.6 & -0.3 & -0.1 & 0.5 & 0.8 & -0.2 & -0.2 & -1.0 & -0.6 & -0.7 & 0.5 & -0.4 & -0.4 & -0.3 & -0.7 & 1.0 & 0.6 & -0.6 & -0.3 & 1.0 & -0.2 & 0.6 & -0.3 \\ 
32.0 & 0.00800 & 0.4 & 0.8 & -0.6 & 1.2 & 0.1 & 1.1 & -0.2 & -0.2 & -0.2 & 0.2 & -0.5 & 0.4 & 1.2 & 0.6 & 0.5 & 0.4 & 0.7 & 0.1 & -0.3 & -0.4 & 1.0 & 0.7 & 0.5 & -0.1 & -0.5 & 0.3 & 2.2 & 0.4 & -1.0 & 0.3 & -0.5 & -1.3 & 2.1 & 0.7 & -0.6 & -0.6 & 0.1 & -1.2 & 1.7 & -0.5 \\ 
60.0 & 0.00140 & -0.0 & 0.5 & -0.2 & 1.0 & -0.7 & 0.5 & -0.4 & 0.3 & -0.1 & 0.4 & -0.1 & -0.2 & -1.1 & 0.3 & 0.4 & 0.7 & 0.4 & -0.3 & -0.8 & -0.8 & -0.2 & -0.3 & 0.5 & -0.1 & -0.4 & -1.3 & -1.1 & 0.1 & -0.2 & 0.2 & -0.6 & -0.5 & 1.2 & -0.6 & 0.1 & 0.1 & 0.1 & 0.2 & 0.5 & 0.4 \\ 
60.0 & 0.00200 & 0.0 & 0.4 & -0.2 & 0.9 & -0.4 & 0.4 & -0.5 & 0.3 & 0.1 & 0.4 & -0.0 & -0.1 & -0.6 & -0.2 & 0.1 & 0.6 & 0.1 & -0.2 & -0.1 & -0.3 & -0.0 & -0.2 & 0.1 & -0.2 & 0.4 & -0.8 & -0.5 & 0.1 & -0.3 & 0.1 & -0.2 & -0.5 & 0.4 & -0.2 & 0.0 & 0.1 & 0.1 & 0.1 & 0.4 & 0.7 \\ 
60.0 & 0.00320 & 0.1 & 0.6 & -0.4 & 1.0 & -0.4 & 0.5 & -0.3 & 0.1 & 0.5 & 0.1 & -0.0 & -0.3 & -0.3 & 0.2 & 0.0 & 1.2 & -0.1 & -0.3 & -0.7 & -0.1 & -0.2 & 0.8 & 0.1 & 0.0 & 0.3 & -0.6 & -0.3 & 0.0 & -0.2 & -0.1 & -0.5 & -0.7 & 0.7 & -0.6 & 0.1 & -0.1 & -0.0 & 0.3 & 0.7 & -0.0 \\ 
60.0 & 0.00500 & 0.2 & 0.3 & -0.4 & 0.8 & -0.6 & 0.7 & -0.4 & 0.1 & 0.3 & 0.2 & 0.2 & -0.0 & -0.1 & 0.3 & 0.1 & 1.0 & -0.0 & -0.6 & -0.3 & 0.6 & 0.0 & 0.8 & 0.1 & 0.4 & -0.2 & -0.5 & -0.1 & 0.1 & -0.2 & -0.0 & -0.6 & -0.9 & 0.5 & -0.5 & 0.1 & 0.0 & 0.2 & -0.1 & 0.6 & 0.3 \\ 
60.0 & 0.00800 & 0.0 & 0.6 & -0.7 & 1.0 & -0.3 & 0.5 & -0.1 & -0.6 & 0.5 & 0.6 & 0.1 & 0.1 & -0.2 & 0.3 & -0.0 & 0.2 & -0.4 & 0.0 & -0.2 & 0.1 & 0.1 & 0.4 & 0.2 & -0.4 & -0.0 & -0.2 & -0.3 & 0.1 & -0.1 & -0.4 & -0.5 & -0.8 & 0.7 & 0.1 & -0.1 & -0.6 & -0.1 & 0.2 & 1.1 & -0.4 \\ 
60.0 & 0.01500 & -0.2 & 0.3 & -0.6 & 1.1 & -0.3 & 0.6 & -0.4 & 0.2 & 0.5 & -0.2 & 0.1 & -0.3 & -0.2 & 0.1 & 0.1 & 1.0 & -0.4 & -0.2 & -0.4 & -0.2 & 0.8 & 0.4 & 0.1 & -0.9 & -1.1 & 0.3 & 0.2 & 0.2 & 0.1 & 0.9 & 0.2 & -1.2 & -0.5 & -0.1 & 0.1 & -0.4 & 0.0 & -0.8 & 2.1 & -0.6 \\ 
120.0 & 0.00200 & -0.3 & 0.5 & -0.3 & 1.5 & -0.5 & 0.3 & -0.8 & 0.3 & 0.2 & 0.4 & -0.0 & -0.5 & -0.9 & 0.1 & 0.3 & 1.3 & 0.9 & 0.5 & -0.4 & -0.3 & -0.6 & -0.1 & 0.4 & 0.3 & 0.1 & -1.1 & -0.9 & -0.0 & -0.6 & 0.2 & -0.2 & -0.2 & 1.0 & -0.2 & -0.3 & 0.7 & 0.6 & 0.1 & 0.7 & 0.2 \\ 
120.0 & 0.00320 & 0.2 & 0.3 & -0.7 & 1.0 & -0.5 & 0.6 & 0.3 & -0.4 & -0.4 & -0.4 & 0.2 & -0.5 & -0.2 & -0.6 & -0.2 & 0.7 & -0.0 & -0.7 & 0.1 & -0.4 & 0.1 & -0.1 & -0.0 & -0.7 & -0.0 & -0.5 & -0.2 & 0.1 & -0.2 & 0.0 & -0.3 & -0.2 & -0.0 & -0.1 & 0.3 & 0.2 & 0.1 & 0.2 & -0.1 & -0.1 \\ 
120.0 & 0.00550 & 0.0 & -0.3 & -0.2 & 1.2 & -0.5 & 0.6 & -0.5 & 0.1 & 0.4 & 0.4 & 0.2 & -0.4 & -0.5 & -0.4 & 0.2 & 0.7 & 0.1 & 0.2 & 0.0 & 0.1 & 0.5 & 0.4 & 0.0 & 0.4 & 0.2 & -0.7 & -0.4 & 0.1 & -0.4 & -0.0 & -0.2 & -1.2 & 0.6 & -0.2 & 0.1 & 0.7 & 0.6 & 0.2 & 0.9 & 0.5 \\ 
120.0 & 0.01000 & 0.2 & 0.1 & -0.5 & 1.0 & -0.5 & 0.6 & -0.2 & -0.2 & 0.6 & -0.1 & 0.4 & 0.0 & -0.3 & -0.5 & 0.1 & 0.6 & -0.0 & 0.0 & 0.0 & -0.3 & 0.4 & 0.1 & 0.2 & -0.3 & 0.4 & -0.6 & 0.1 & 0.5 & 0.1 & -0.1 & -0.8 & -0.7 & 0.6 & -0.2 & -0.4 & 0.6 & 0.4 & 0.0 & 0.7 & -0.0 \\ 
120.0 & 0.02500 & 0.2 & 0.3 & -0.9 & 1.0 & -0.4 & 0.5 & 0.5 & -0.6 & -0.7 & -0.5 & -0.3 & -0.4 & 0.0 & -0.8 & -0.4 & 1.2 & -0.7 & 0.1 & -0.2 & -0.0 & 0.1 & 0.4 & -0.2 & -0.4 & -0.5 & -0.4 & -0.2 & 0.1 & -0.3 & -0.2 & -0.6 & -0.6 & -0.1 & -0.4 & 1.0 & 0.2 & -0.0 & -0.3 & 1.4 & -0.5 \\ 
200.0 & 0.00500 & -0.3 & 0.1 & -0.0 & 1.2 & -0.3 & 0.4 & -1.1 & 0.2 & 0.4 & 0.2 & -0.1 & -0.5 & -0.8 & -0.8 & 0.3 & 1.1 & 0.4 & 0.4 & -0.7 & 0.2 & 1.0 & 1.1 & 0.7 & -0.1 & 0.2 & -1.0 & -0.5 & 0.1 & -0.2 & 0.4 & -0.5 & -0.6 & 0.9 & -0.3 & 0.0 & 0.0 & -0.2 & 0.1 & 0.7 & 0.4 \\ 
200.0 & 0.01300 & 0.0 & 0.4 & -0.8 & 0.8 & -0.1 & 0.3 & -0.1 & -0.2 & 1.2 & -0.1 & -0.1 & -0.4 & -0.3 & -0.2 & 0.4 & 0.6 & 0.2 & 0.1 & -0.1 & 0.0 & 0.6 & 0.6 & 0.1 & 0.4 & 0.2 & -0.6 & -0.3 & 0.2 & -0.3 & -0.1 & -0.3 & -1.2 & 0.9 & -0.4 & -0.0 & -0.1 & -0.1 & 0.1 & 0.7 & 0.4 \\ 
350.0 & 0.01000 & -0.5 & -0.1 & -0.0 & 1.3 & -0.5 & 0.9 & -0.8 & 0.3 & 0.9 & 0.1 & 0.0 & -1.2 & -0.7 & -1.0 & 0.4 & 1.0 & -0.4 & 0.0 & -0.3 & -0.0 & 0.1 & 0.7 & 0.1 & 0.8 & 0.8 & -0.8 & -0.7 & 0.1 & -0.1 & -0.0 & -0.4 & -0.4 & 0.7 & -0.3 & 0.1 & 0.2 & -0.2 & -0.5 & 0.5 & 0.2 \\ 
350.0 & 0.02500 & -0.1 & -0.4 & -0.6 & 0.9 & -0.5 & 0.5 & -0.6 & -0.2 & 0.9 & -0.0 & 0.1 & -0.6 & 0.1 & -0.3 & 0.7 & 1.1 & 0.2 & 0.2 & 0.1 & 0.0 & 1.1 & 0.4 & -0.2 & 0.6 & 0.2 & -0.8 & -0.5 & 0.2 & -0.2 & -0.1 & -0.7 & -1.6 & 1.0 & -0.4 & -0.0 & 0.1 & -0.1 & -0.5 & 0.7 & 0.5 \\ 
650.0 & 0.01300 & -0.5 & -0.1 & 0.6 & 1.8 & -0.8 & 1.7 & -0.7 & 0.1 & 1.7 & 0.9 & 0.5 & -1.0 & -1.0 & -1.4 & 0.3 & 0.3 & 0.8 & -0.4 & 0.2 & 0.1 & 0.8 & 0.5 & 0.2 & 1.4 & 0.7 & -1.6 & -0.6 & 0.5 & -0.6 & 0.3 & 0.1 & -0.8 & 0.3 & -0.3 & 0.1 & -0.0 & 0.1 & 0.1 & 0.5 & 0.9 \\ 
650.0 & 0.03200 & -0.9 & 1.0 & 0.0 & 1.9 & -1.4 & 2.2 & -0.1 & -0.6 & 2.3 & 0.3 & 0.9 & -0.3 & -0.4 & -0.6 & 0.8 & 2.2 & 1.1 & 1.8 & 1.5 & 0.9 & 0.8 & 1.5 & -0.3 & 0.6 & 0.2 & -0.7 & -0.3 & 0.4 & -0.5 & -0.5 & -0.3 & -3.1 & 1.0 & -1.3 & -0.2 & 0.9 & 0.9 & 0.4 & 1.4 & 1.4 \\ 
2000.0 & 0.05000 & -1.5 & 1.4 & 2.9 & 2.7 & -1.5 & 4.6 & 2.2 & -0.6 & 4.6 & 0.0 & 0.8 & -2.4 & -2.5 & -3.4 & 0.5 & 0.5 & 0.3 & 1.2 & 1.2 & 0.6 & 2.6 & 1.7 & -0.0 & 2.9 & 2.1 & -2.0 & -1.0 & 1.1 & -1.1 & 0.1 & 0.6 & -2.9 & 0.9 & -1.2 & 0.4 & -0.1 & 0.3 & 0.1 & 1.2 & 2.0 \\ 
\hline
\end{tabu} 
\end{center} 
\end{table*}

\subsection{Comparison to theoretical predictions and QCD analysis in VFNS}
\label{sec:comb:red:vfns}

\begin{figure}[tbp]
  \centering
  \includegraphics[width=1.0\figwidth,trim=4.5mm 3mm 14.5mm 19mm,clip=true]{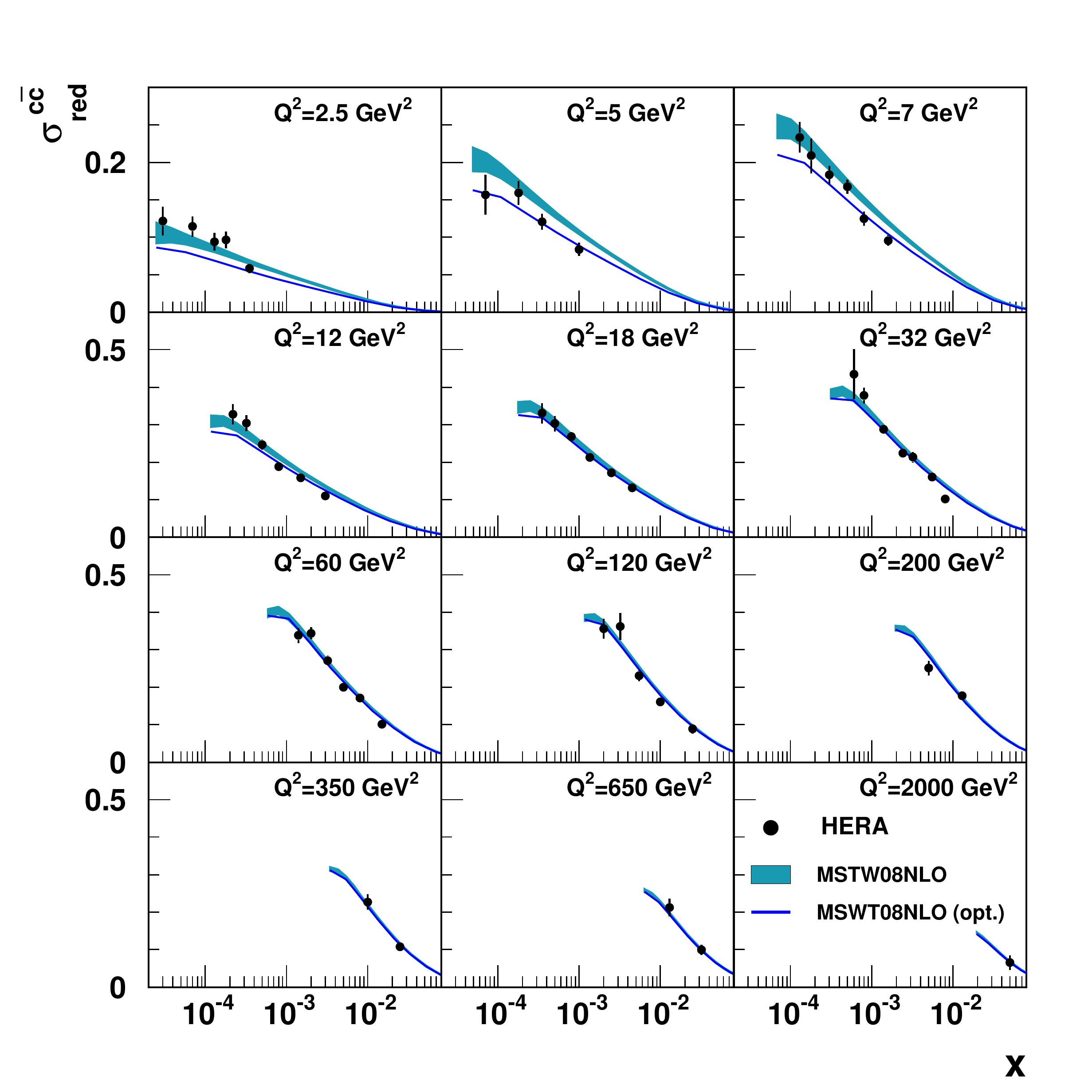}
  \caption[Reduced charm cross sections compared to MSTW NLO predictions]
  {Combined measurements of $\sigma_{red}^{c\bar{c}}$ (closed circles) shown as a function 
	of $x$ for particular $Q^2$, compared to the prediction by MSTW at NLO. 
	The predictions obtained using the standard (optimised) parametrisation are represented by the shaded bands (solid lines). 
	The uncertainties for the optimised parametrisation are not evaluated by the authors of the predictions but are expected to be of same size as those for the standard parametrisation.}
	\label{fig:comb:red:combinedvsmstwnlo}
\end{figure}

\begin{figure}[tbp]
  \centering
  \includegraphics[width=1.0\figwidth,trim=4.5mm 3mm 14.5mm 19mm,clip=true]{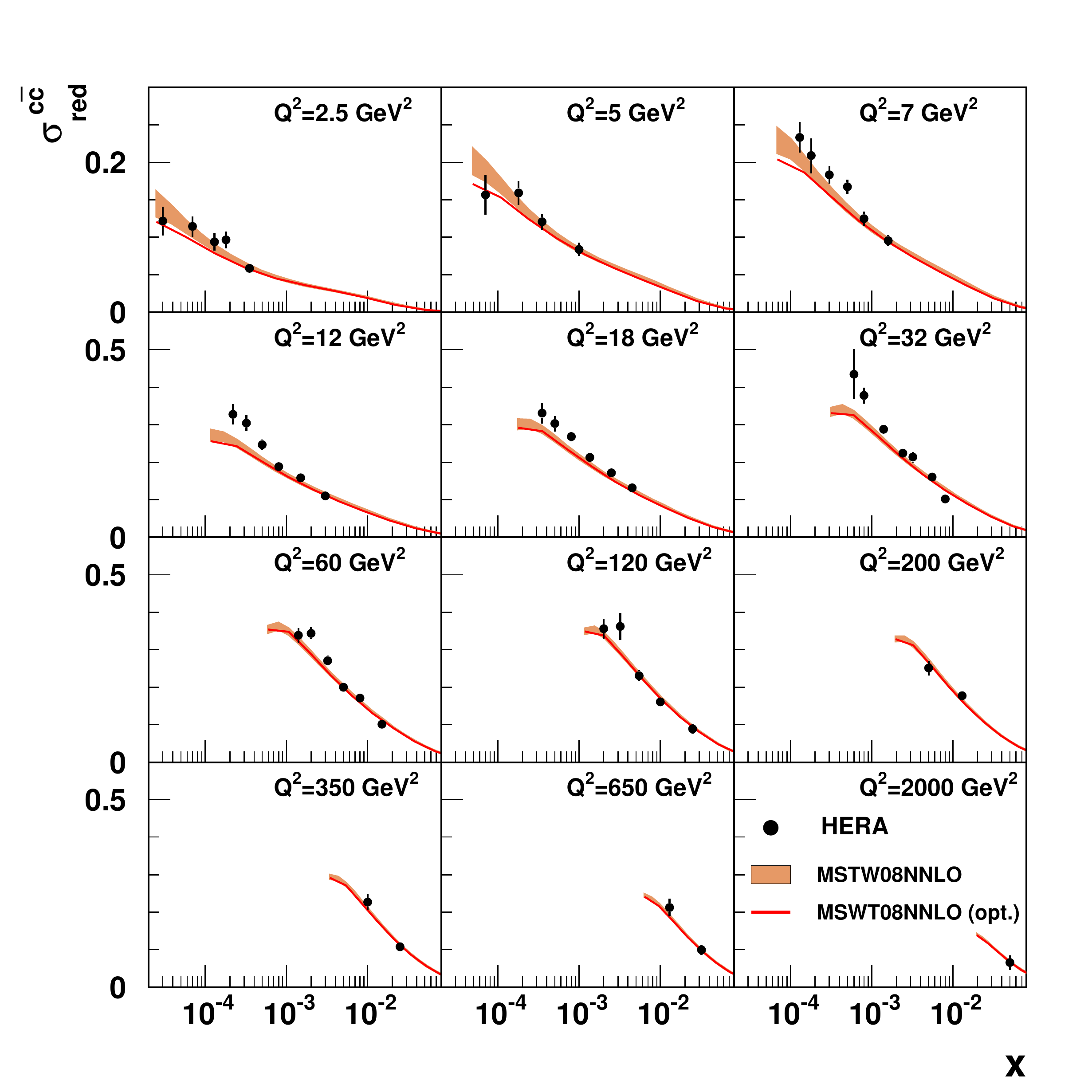}
  \caption[Reduced charm cross sections compared to MSTW NNLO predictions]
  {Combined measurements of $\sigma_{red}^{c\bar{c}}$ (closed circles) shown as a function 
	of $x$ for particular $Q^2$, compared to the prediction by MSTW at NNLO. 
	The predictions obtained using the standard (optimised) parametrisation are represented by the shaded bands (solid lines). 
	The uncertainties for the optimised parametrisation are not evaluated by the authors of the predictions but are expected to be of same size as those for the standard parametrisation.}
	\label{fig:comb:red:combinedvsmstwnnlo}
\end{figure}

In Figs~\ref{fig:comb:red:combinedvsmstwnlo} and~\ref{fig:comb:red:combinedvsmstwnnlo} the combined cross sections are compared with 
predictions of the MSTW group in the GM-VFNS at NLO and NNLO, respectively, 
using the RT-standard~\cite{Thorne:2006qt,Martin:2009iq} and the RT-optimised~\cite{Thorne:2012az} interpolation procedure 
of the cross section at the charm-production threshold. 
At NLO, the optimised prediction tends to describe the data better than 
the standard one at lower $Q^2$. 
The description of the data is improved at NNLO compared to NLO. 

\begin{figure}[tbp]
  \centering
  \includegraphics[width=1.0\figwidth,trim=4.5mm 3mm 14.5mm 19mm,clip=true]{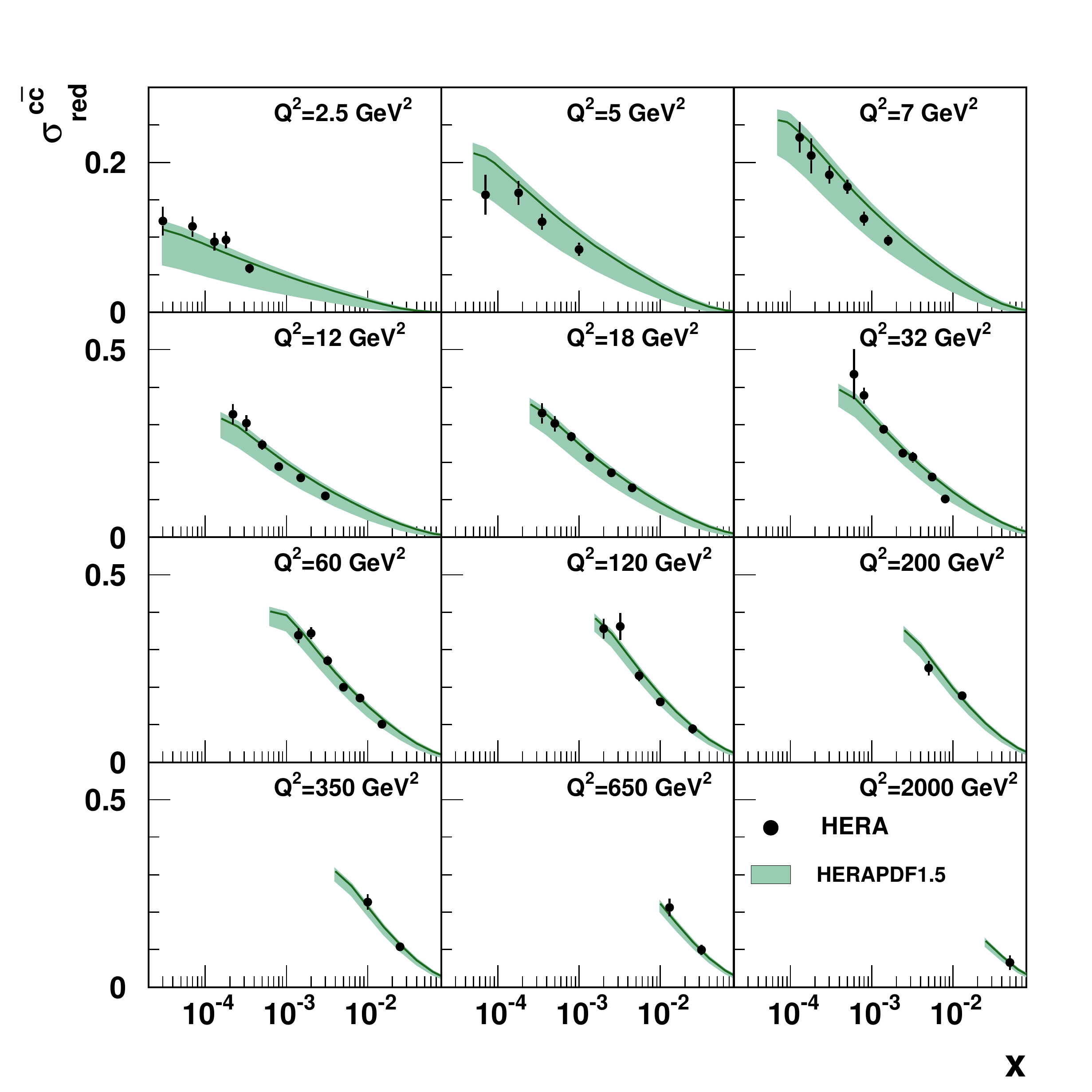}
  \caption[Reduced charm cross sections compared to HERAPDF1.5 predictions]
  {Combined measurements of $\sigma_{red}^{c\bar{c}}$ (closed circles) shown as a function 
	of $x$ for particular $Q^2$, compared to the NLO predictions based on HERAPDF1.5 extracted in the RT standard scheme. 
	The line represents the prediction using $\mct =1.4$~GeV. The uncertainty band shows the full PDF uncertainty which is dominated by the variation of \mct.}
	\label{fig:comb:red:combinedvsherapdf15}
\end{figure}

In Fig.~\ref{fig:comb:red:combinedvsherapdf15} the data are compared to the NLO predictions based on HERAPDF1.5 \cite{Radescu:2010zz} extracted in the RT standard scheme  
using as inputs the published HERA-I \cite{DIScomb} and the preliminary HERA-II combined inclusive DIS data. 
For the central PDF set a $c$-quark mass parameter $\mct=1.4$~GeV is used. 
The uncertainty bands of the predictions reflect the full uncertainties on the HERAPDF1.5 set. 
They are dominated by the uncertainty on \mct~which is varied between $1.35$~GeV and $1.65$~GeV~\cite{DIScomb}. Within these uncertainties the HERAPDF1.5 predictions describe the data well.
The central predictions are very similar to those of the MSTW group for the same scheme. 

\begin{figure}[tbp]
  \centering
  \includegraphics[width=1.0\figwidth,trim=4.5mm 3mm 14.5mm 19mm,clip=true]{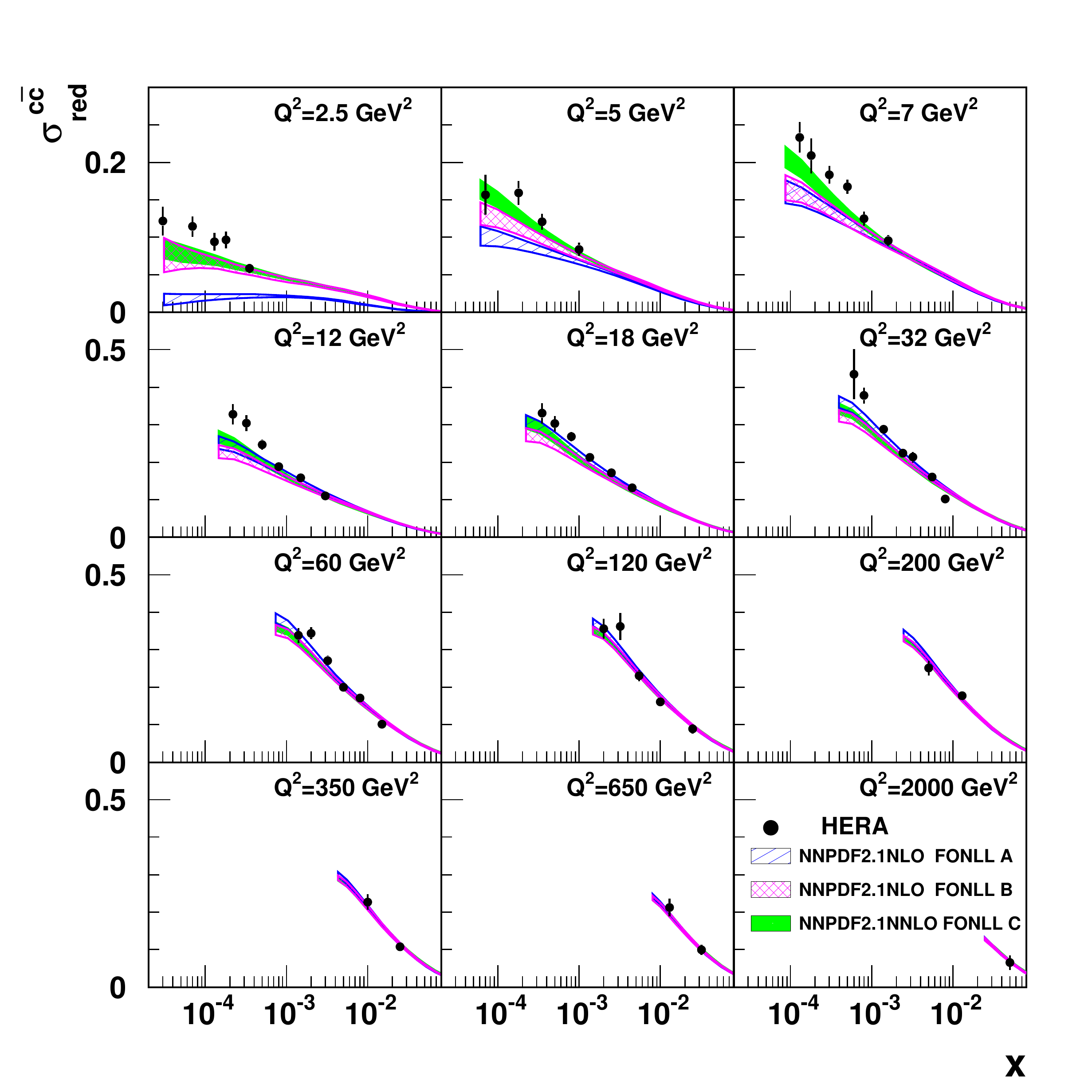}
  \caption[Reduced charm cross sections compared to NNPDF predictions]
  {Combined measurements of $\sigma_{red}^{c\bar{c}}$ (closed circles) shown as a function 
	of $x$ for particular $Q^2$, compared to the predictions by NNPDF. 
	The predictions from NNPDF2.1 in FONNL-A, -B and -C schemes are available with their uncertainties and are represented by bands with different hatch styles.}
	\label{fig:comb:red:combinedvsnnpdf}
\end{figure}

In Fig.~\ref{fig:comb:red:combinedvsnnpdf} the data are compared to the 
predictions in the GM-VFNS by the NNPDF Collaboration. 
Both the NNPDF FONLL-A~\cite{Forte:2010ta} and  FONLL-B~\cite{Ball:2011mu,Ball:2011uy} 
predictions describe the data fairly well at higher $Q^2$, while they fail to describe the 
data at lower $Q^2$. The description of the data at lower $Q^2$ is improved 
in the FONLL-C~\cite{Ball:2011mu,Ball:2011uy} scheme. 

\begin{figure}[tbp]
  \centering
  \includegraphics[width=1.0\figwidth,trim=4.5mm 3mm 14.5mm 19mm,clip=true]{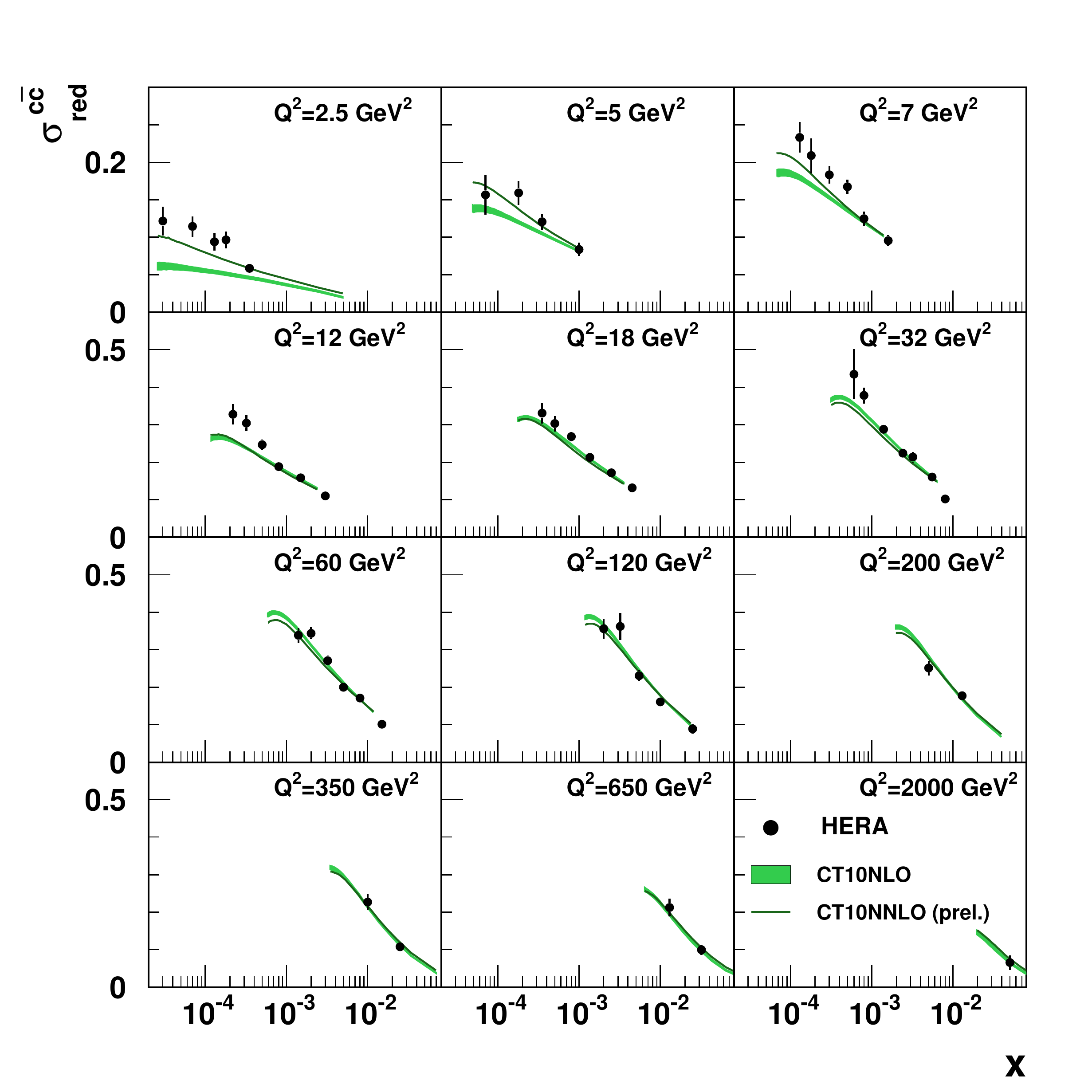}
  \caption[Reduced charm cross sections compared to CTEQ predictions]
  {Combined measurements of $\sigma_{red}^{c\bar{c}}$ (closed circles) shown as a function 
	of $x$ for particular $Q^2$, compared to the predictions by CTEQ. 
	The CT10 NLO prediction with its uncertainties is shown by the shaded bands. 
	The uncertainties on the CT10 NNLO (prel.) predictions are not shown.}
	\label{fig:comb:red:combinedvscteq}
\end{figure}

In Fig.~\ref{fig:comb:red:combinedvscteq} the data are compared to the 
predictions in the GM-VFNS by the CTEQ Collaboration. 
The CT predictions~\cite{ct10f3,ct12nnlo} are based on the S-ACOT-$\chi$ heavy-quark scheme. 
The NLO prediction, which is very similar to the FONLL-A scheme, describes the data well for 
$Q^2>5$ GeV$^2$ but fails to describe the data at lower $Q^2$. 
Similar to the FONNL-C case, the description of the data 
improves significantly at NNLO.

In summary, conclusions similar to~\cite{heracharmcomb} can be drawn. 
The best description of the data is achieved by the predictions 
including partial $O(\alpha_s^3)$ corrections (MSTW NNLO), however 
they do not fully describe the $Q^2$ slope of the data at low $Q^2$ ($2.5<Q^2<5.0~$GeV$^2$). 
The predictions including $O(\alpha_s^2)$ terms in all
parts of the calculation (NNPDF FONLL-C, CT NNLO) as well as the 
MSTW NLO optimal scheme also agree with the data reasonably well. 
The largest deviations are observed for predictions based on $O(\alpha_s)$ 
terms only (NNPDF FONLL-A and CT NLO).
As investigated below, 
further differences can be partially explained 
by the different choices for the value of the respective $c$-quark mass parameter \mct.

Similar to the extraction of the $c$-quark mass in the FFNS, described in Section~\ref{sec:comb:red:ffns}, 
the combined charm data were used to determine the effective parameters of the individual VFNS.

The following implementations of the VFNS were considered: ACOT full~\cite{Aivazis:1993kh,Aivazis:1993pi} as used for the 
CTEQHQ releases of PDFs; S-ACOT-$\chi$~\cite{Collins:1998rz,Kramer:2000hn,Tung:2001mv} as used for the latest CTEQ releases of PDFs, and 
for the FONLL-A scheme \cite{Forte:2010ta} used by NNPDF;  
the RT standard scheme~\cite{Thorne:2006qt,Martin:2009iq} as used for the MRST and MSTW releases of PDFs, as well as the RT optimised 
scheme providing a smoother behaviour across thresholds~\cite{Thorne:2012az}. The ZM-VFNS as implemented by 
the CTEQ group~\cite{Aivazis:1993kh,Aivazis:1993pi} was also used for comparison. In all schemes, the onset of the heavy-quark PDFs is controlled by 
the parameter \mct, in addition to the kinematic constraints. 

The fitting procedure was the same as in the FFNS fit, described in Section~\ref{sec:comb:red:ffns}, except:
\begin{itemize}
	\item since most of the considered VFNS at $O(\alpha_s^2)$ fail to describe the $Q^2$ slope of the data in the range of $2.5<Q^2<\SI{5.0}{GeV^2}$, the first $Q^2$ bin was excluded from the fit;
	\item the strong coupling constant was chosen $\alpha_s^{n_f=5}(M_Z)=0.1176 \pm 0.0020$;
	\item the renormalisation and factorisation scales for the heavy quarks were set to $\mu_f=\mu_r=Q$ and not varied, since it is not technically possible in the framework;
	\item the preferred mass parameters were obtained from the scan, since the implementation of the calculations does not allow for their changes in the PDF fitting procedure. 
	The step size $\SI{0.01}{GeV}$ was used.
\end{itemize}

\begin{figure}[htbp]
  \centering
  \includegraphics[width=1.0\figwidth,trim=4mm 0 4mm 0mm,clip=true]{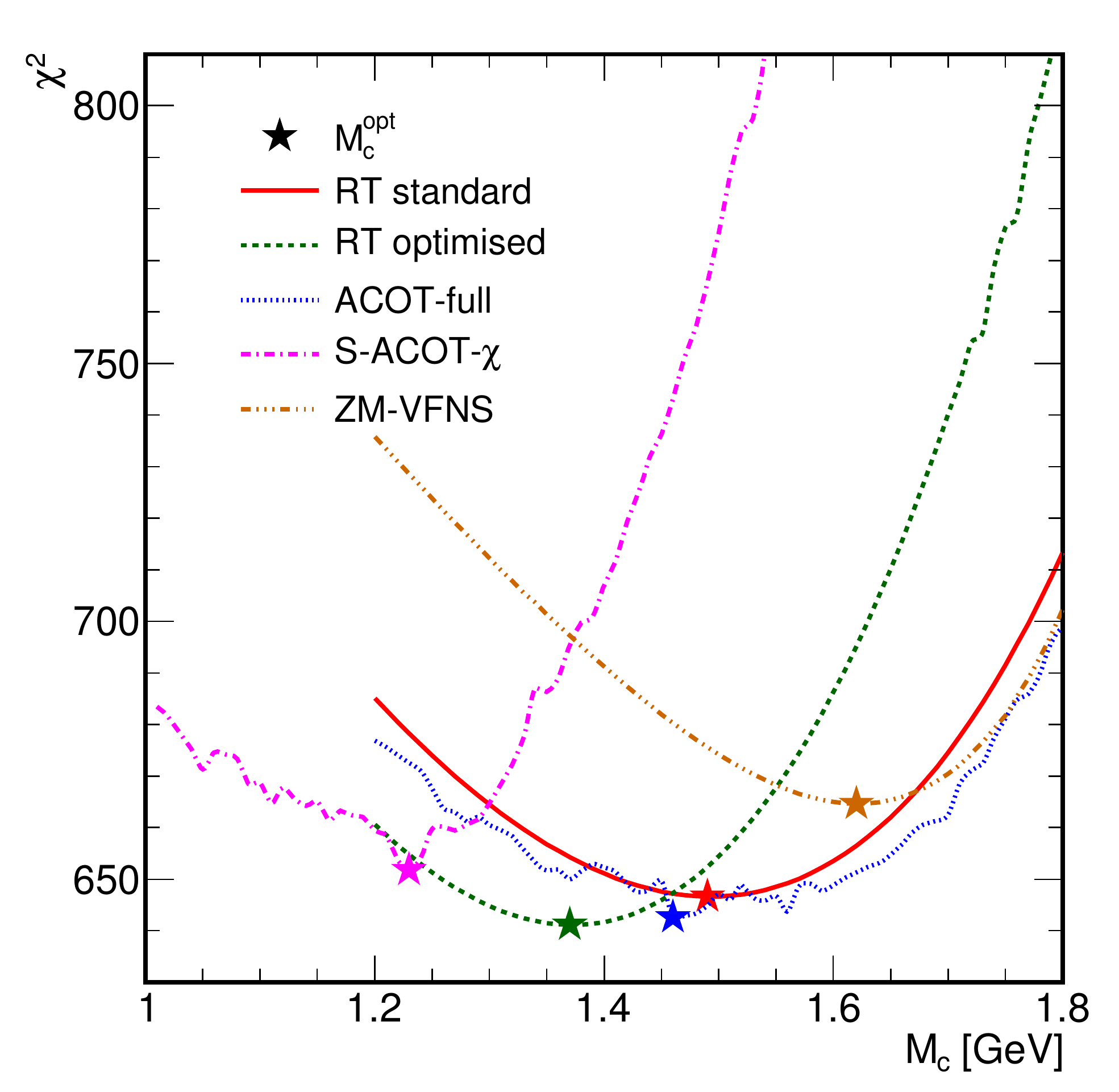}
  \caption[$\chi^2(M_c)$ values for different VFNS]
  {The values of $\chi^2(M_c)$ for the PDF fit to the combined HERA inclusive DIS and charm measurements in different VFNS, 
	 presented by lines with different styles. The values of $\mcto$ for each scheme are indicated by the stars.}
	\label{fig:comb:red:mcopt}
\end{figure}

In Fig.~\ref{fig:comb:red:mcopt} the $\chisq$ values as a function of \mct are shown for all schemes considered. 
Similar minimal $\chi^2$-values are observed for the different schemes, albeit at quite different optimal values of 
the charm-mass parameter, \mcto. 
In the cases of the ACOT full and S-ACOT-$\chi$ schemes the dependence of \chisq on \mct has small discontinuities since 
these schemes are implemented using K-factors.%
\footnote{From a calculation point of view, the theoretical model consists of the numerical integration of an integro-differential equation
and multiple convolution integrals that are evaluated mostly by adaptive algorithms (K-factors). 
Change of a parameter (\mct in this case) results in the appearance of an uncontrolled numerical noise. 
Some details can be found also in~\cite{Pumplin:2000vx}.} 
A smooth curve can be obtained by fitting the points with a parabolic function, although this will not significantly change the preferred \mcto values. 

In Table~\ref{tab:comb:red:mcopt}
the resulting values of \mcto are given together with the uncertainties and the \chisqndof values; 
for comparison the `HERA 2012' results are also given. 
For ACOT full and S-ACOT-$\chi$ schemes the uncertainties are not evaluated, since due to the present discontinuities in the \chisq curves they can be misleading. 
The RT optimised scheme yields the best global $\chi^2$. The fit in the S-ACOT-$\chi$ scheme results in a very low value of \mcto as compared to the other schemes. 
In general the predictions of the different schemes become very similar for $Q^2 \ge \SI{5}{GeV^2}$ and describe the data well, once the charm-mass parameters are set to the preferred values. 
Note, that even the ZM-VFNS, which includes mass effects only indirectly~\cite{Aivazis:1993kh,Aivazis:1993pi}, yields a reasonably good description 
of the combined charm data for $Q^2 \ge 5$~GeV$^2$ (although it predicts a zero cross section in the lowest $Q^2$ bin), however $\sim 20$ units of \chisq worse than the other schemes.

\begin{table*}[tbp]
  \caption[Values of charm-mass parameter \mcto\ in different VFNS]
  {The values of the charm-mass parameter \mcto\ as determined from the \mct scans in different VFNS, with their uncertainties and the corresponding \chisqndof values. 
  The results of the present study (`HERA') are shown together with the previous `HERA 2012' results from~\cite{heracharmcomb}.}
\label{tab:comb:red:mcopt}
\centering
\linespread{1.3}
	\tabcolsep1.15mm
	\renewcommand*{\arraystretch}{1.4}
  \begin{tabularx}{\textwidth}[h]{|X||l|l||l|l|}
    \hline
    \multirow{2}{*}{Scheme} & \multicolumn{2}{c||}{HERA} & \multicolumn{2}{c|}{HERA 2012} \\ \cline{2-5}
                            & \mcto\, [GeV] & $\chisqndof$ & \mcto\, [GeV] & $\chisqndof$ \\
  \hline
  \hline 
   RT standard   & $1.49 ~\pm 0.04{\text{(fit)}} ~\substack{+0.03\\-0.01}{\text{(mod)}} ~\substack{+0.02\\-0.02}{\text{(par)}} ~\substack{+0.01\\-0.00}{(\alpha_s)} $ & $647/626$ & 
                   $1.50 ~\pm 0.06{\text{(fit)}} ~\pm 0.06               {\text{(mod)}} ~\pm 0.01               {\text{(par)}} ~\pm 0.003              {(\alpha_s)} $ & $631/626$ \\
   RT optimised  & $1.37 ~\pm 0.04{\text{(fit)}} ~\substack{+0.02\\-0.00}{\text{(mod)}} ~\substack{+0.02\\-0.00}{\text{(par)}} ~\substack{+0.01\\-0.00}{(\alpha_s)} $ & $641/626$ & 
                   $1.38 ~\pm 0.05{\text{(fit)}} ~\pm 0.03               {\text{(mod)}} ~\pm 0.01               {\text{(par)}} ~\pm 0.01               {(\alpha_s)} $ & $624/626$ \\
   ACOT-full     & $\approx 1.46 $                                                                                                                                            & $643/626$ &
                   $1.52 ~\pm 0.05{\text{(fit)}} ~\pm 0.12               {\text{(mod)}} ~\pm 0.01               {\text{(par)}} ~\pm 0.06               {(\alpha_s)} $ & $607/626$ \\
   S-ACOT-$\chi$ & $\approx 1.23 $                                                                                                                                            & $652/626$ &
                   $1.15 ~\pm 0.04{\text{(fit)}} ~\pm 0.01               {\text{(mod)}} ~\pm 0.01               {\text{(par)}} ~\pm 0.02               {(\alpha_s)} $ & $613/626$ \\
   ZM-VFNS       & $1.62 ~\pm 0.04{\text{(fit)}} ~\substack{+0.04\\-0.00}{\text{(mod)}} ~\substack{+0.02\\-0.02}{\text{(par)}} ~\substack{+0.02\\-0.01}{(\alpha_s)} $ & $665/626$ & 
                   $1.60 ~\pm 0.05{\text{(fit)}} ~\pm 0.03               {\text{(mod)}} ~\pm 0.05               {\text{(par)}} ~\pm 0.01               {(\alpha_s)} $ & $632/626$ \\
  \hline
   \end{tabularx}
\linespread{1.0}
\end{table*}

Similar to the fit in the FFNS, all fitted \mcto values are consistent with those which have been determined in the previous analysis~\cite{heracharmcomb} with the `HERA 2012' combined data. 
Those variants, for which the uncertainties are determined, exhibit improved precision.

Using different charm-mass parameters adjusted to the HERA data allows for a reduction of the theoretical uncertainty 
due to the choice of the heavy-flavour scheme 
for $W^{\pm}$ and $Z$ production at the LHC, as was demonstrated in~\cite{heracharmcomb}.

\clearpage
\section{PDF fit with LHCb heavy-flavour data: additional information}
\label{sec:app:pdffit}

In this Appendix additional information on the PDF fit with the LHCb heavy-flavour data (Section~\ref{sec:pdffit}) is provided. 

\subsection{MNR calculations in HERAFitter: details of implementation}
\label{sec:pdffit:th:impl}

A PDF fit in the framework described in Section~\ref{sec:pdffit:det:framework} typically requires several thousands of iterations to converge. 
In each iteration the theoretical predictions for each dataset must be recomputed. 
Since computation of the NLO predictions is usually very time consuming, this requires a ``smart'' implementation 
of the calculations with separating the bottleneck parts from the iterative procedure. 
Another popular solution is to use ``fast'' techniques, such as K-factors or precomputed perturbative grids 
(see, e.g.~\cite{fastnlo_Kluge:2006xs,fastnlo_Wobisch:2011ij,fastnlo_Britzger:2012bs,applgrid_Carli:2010rw,amcfast_Bertone:2014zva}). 
Although ``fast'' techniques are widely used by modern PDF groups, they usually have shortcomings, since they do not allow changing parameters of the calculations, 
like the factorisation and renormalisation scales or heavy-quark masses.

The MNR calculations (one-particle inclusive variant) as implemented originally in the FORTRAN code~\cite{mnrplace} require about several hours to calculate one set of the predictions 
for one of the considered LHCb datasets.%
\footnote{The timing depends on the number of bins, desired accuracy of the predictions and CPU; the quoted one is for 40 bins from~\cite{LHCbCharm}, 1\% inaccuracy and Intel Core i7-3520M.} 
Numerical multi-dimensional integration over the phase space is done with the MC method using the VEGAS algorithm~\cite{vegas_Lepage:1977sw}. 
The main advantage of the MC integration is that it can be suitably performed for any configuration of the phase space; the only number to be adjusted to reach the desired accuracy is the total number of iterations. 
The disadvantage is that all parts of the calculations have to be repeated in each iteration. 

Therefore in HERAFitter numerical multi-dimensional integration for the MNR calculations was implemented as nested loops using the trapezoidal rule for each one-dimensional integration. 
This allows for separation of the most time consuming parts in the top loop(s). 
The one-particle inclusive variant of the calculations was used. 
All flexibility of the original MNR code was retained: the factorisation and renormalisation scales, heavy-quark mass, strong coupling constant, fragmentation function and PDFs may be changed in each iteration 
(in other words, may be treated as fit parameters). 
The typical timing to calculate one set of the predictions for the considered LHCb datasets is $\sim \SI{1}{s}$ 
and the inaccuracy of the predictions is less than 1\% comparing to the results obtained with the original MNR code. 
This allows a PDF fit with these data to converge typically within a few hours. 
Additionally the results were cross checked with the NLO predictions as calculated by the (semi)independent FONLL program, using the public web interface~\cite{FONLLWeb}%
\footnote{The `NLO' option of the FONLL program was used.}, 
and differences were found to be within $1\text{--}3$\%. 

However note that the integration loops were adjusted for this particular configuration; another phase space and/or binning will need their readjustment.

\subsection{Study of charm fragmentation function}
\label{sec:app:pdffit:frag}

In the `LHCb Abs' fit the following tendency was observed: the LHCb charm data prefer a harder fragmentation function 
than was measured at HERA, since the variation of $\alpha_k$ to upper values results in better \chisq.
This can be seen even from the nominal fit: predictions for the bins $1<p_T<\SI{3}{GeV}$ 
are on average above the data, while the bins with higher $p_T$ are below; this non-perfect description 
of the $p_T$ shape actually explains the somewhat large $\chisq$ values for the LHCb charm datasets in Table~\ref{tab:pdffit:chisqndof}.

In order to investigate this further the fragmentation-function parameter for charm was released 
in the fit. The fit converged to a very large $\alpha_k$ value which corresponds to an almost $z \simeq 1$ parton to hadron 
transition. Another check was performed by using the BCFY fragmentation function~\cite{Braaten:1994bz} with $r=0.1$ extracted 
from \epem colliders within the FONLL approach~\cite{frag06}, which corresponds approximately to $\alpha_k=12$.
A much better description of the charm data was found than with the fragmentation function derived from the HERA data. 
This study qualitatively confirms the recipe for heavy-flavour 
fragmentation provided in~\cite{Baines:2006uw}: since FONLL resummations of NLL provide 
evolution of the perturbative part of the fragmentation function to the scale $\sim m_Q$, with NLO QCD predictions 
for hadro- and electroproduction of heavy flavours for the $p_T$ region close to the threshold it would be more appropriate to use a fragmentation function 
extracted at FONLL (e.g.\ those from \epem at the $Z^0$ resonance), while for the high-$p_T$ region it is 
more appropriate to use a fragmentation function extracted at the NLO approach. 
However such a study was beyond the scope, so the QCD analysis was limited to the usage of the fragmentation function 
measured at HERA. 

For beauty no tendencies were observed: the LHCb data clearly prefer the value 
$\alpha_k \approx 11$ similar to that extracted from the LEP data.

\subsection{Additional tables and plots}
\label{sec:app:pdffit:add}

\begin{figure}[tbp]
  \centering
  \includegraphics[width=0.495\figwidth,trim=2mm 2mm 2mm 8.5mm,clip=true]{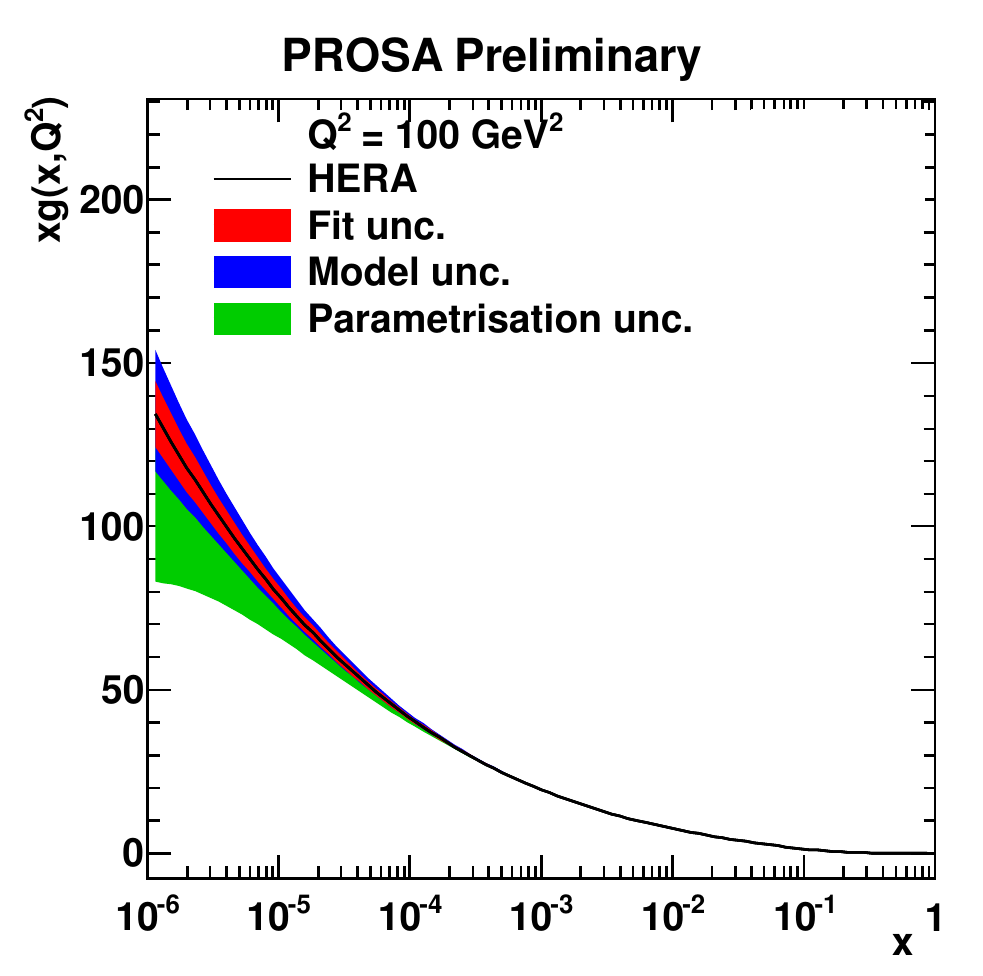}
  \includegraphics[width=0.495\figwidth,trim=2mm 2mm 2mm 8.5mm,clip=true]{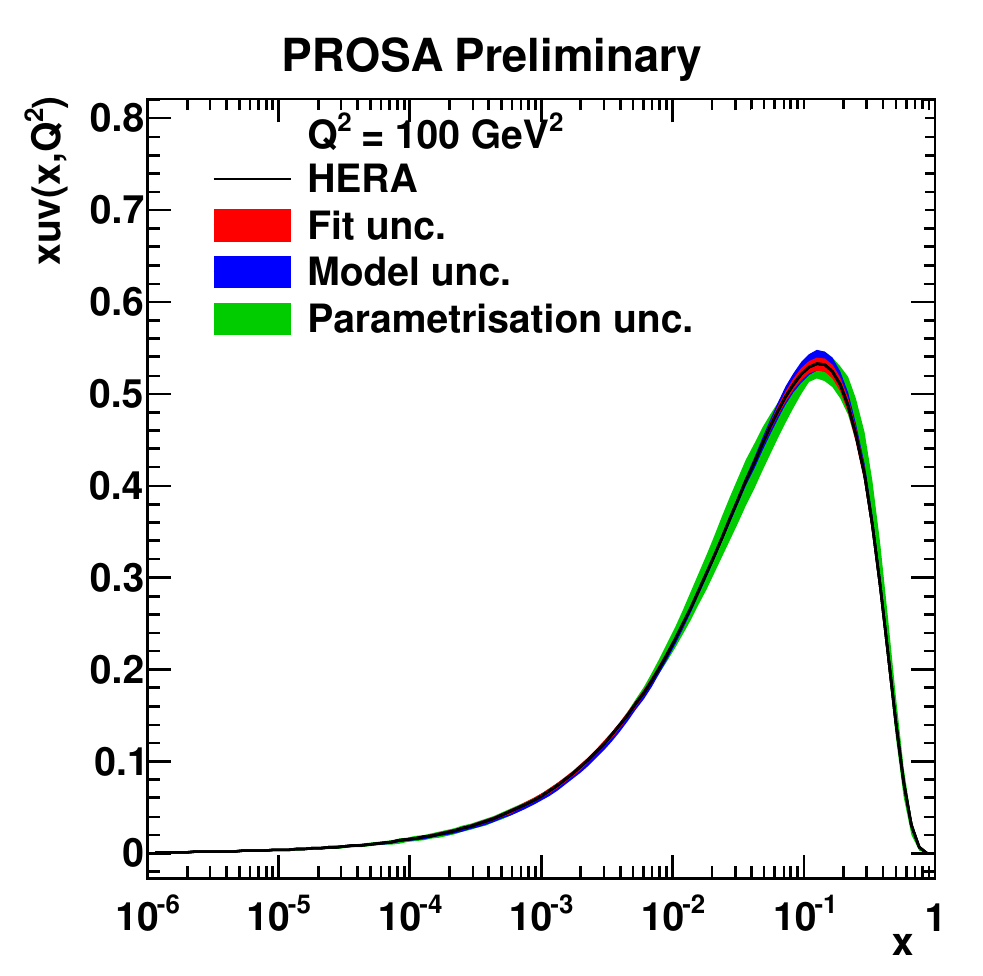}
  \includegraphics[width=0.495\figwidth,trim=2mm 2mm 2mm 8.5mm,clip=true]{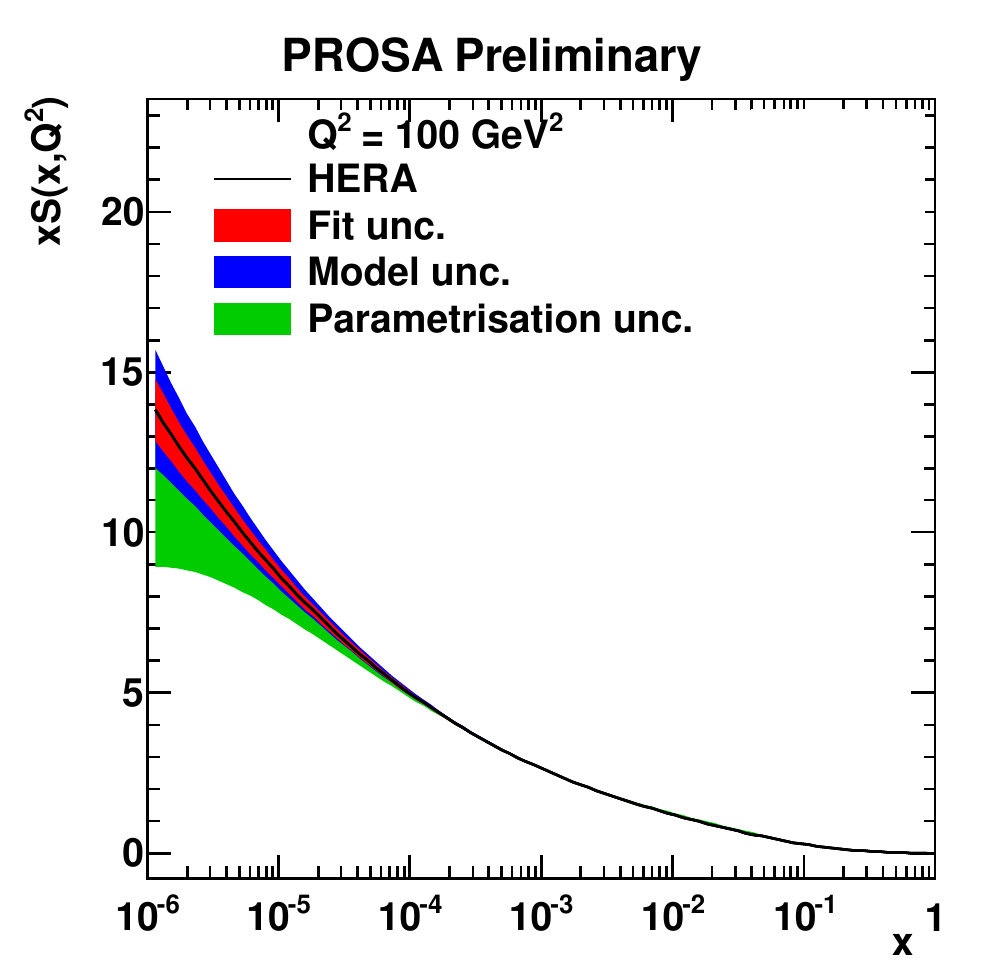}
  \includegraphics[width=0.495\figwidth,trim=2mm 2mm 2mm 8.5mm,clip=true]{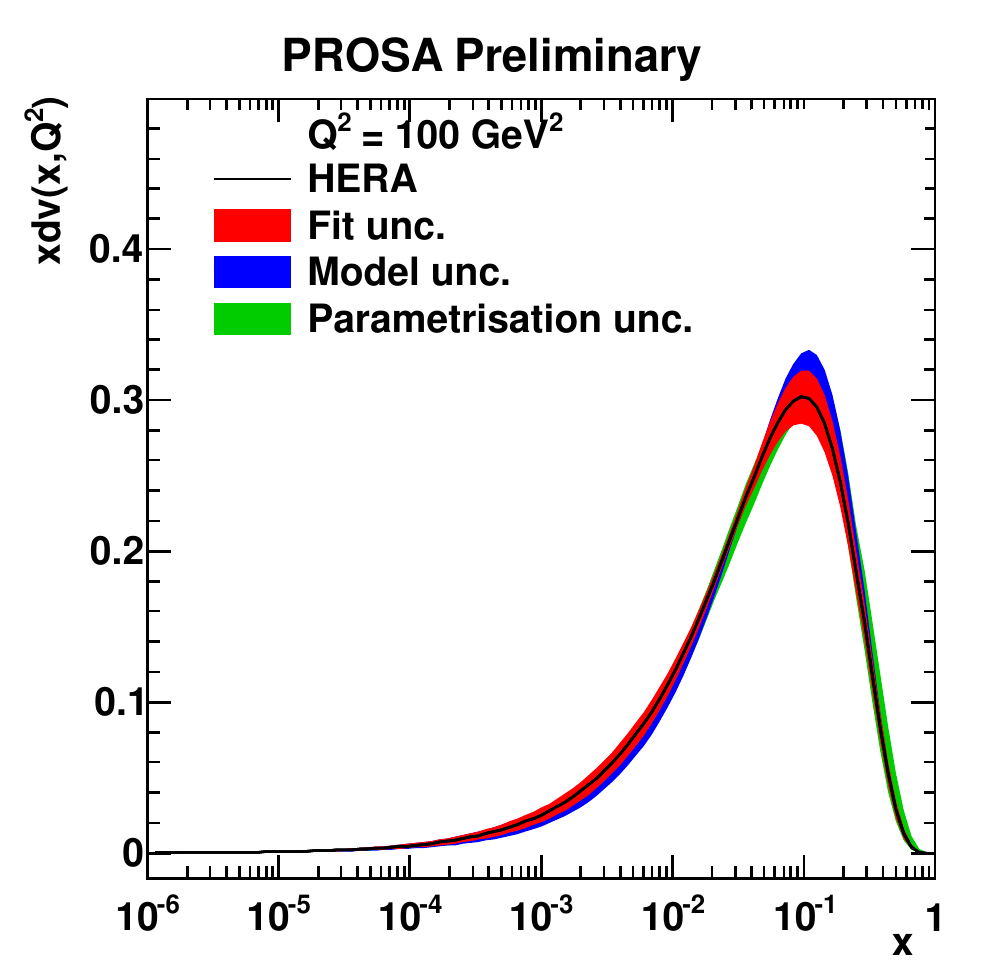}
  \caption[PDF uncertainties at $Q^2=\SI{100}{GeV^2}$ for `HERA only' fit]
  {The individual contributions to the uncertainties of the gluon (top left), $u$-valence (top right), sea (bottom left) and $d$-valence (bottom right) distributions at $Q^2=\SI{100}{GeV^2}$ 
  obtained in the fit with the HERA-only data.}
	\label{fig:pdffit:FittedPDFs_HERAOnly_q2_100}
\end{figure}

\begin{figure}[tbp]
  \centering
  \includegraphics[width=0.495\figwidth,trim=2mm 2mm 2mm 8.5mm,clip=true]{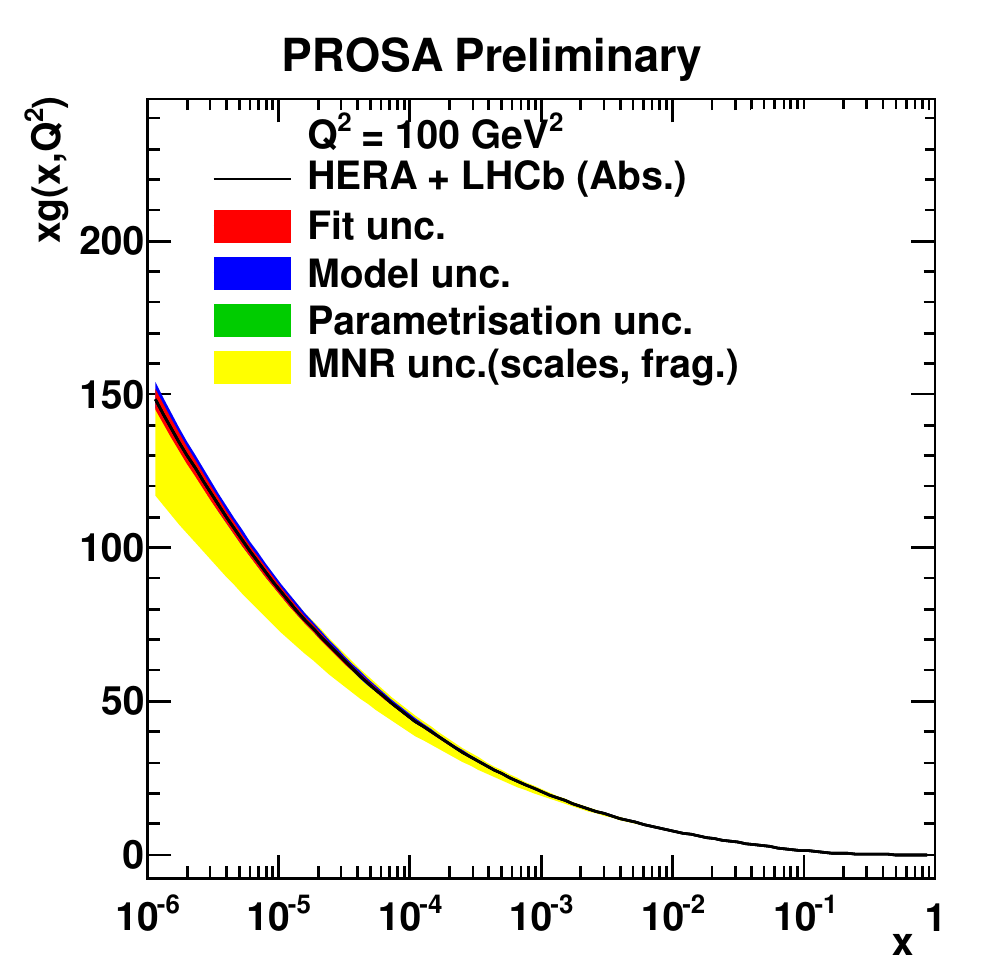}
  \includegraphics[width=0.495\figwidth,trim=2mm 2mm 2mm 8.5mm,clip=true]{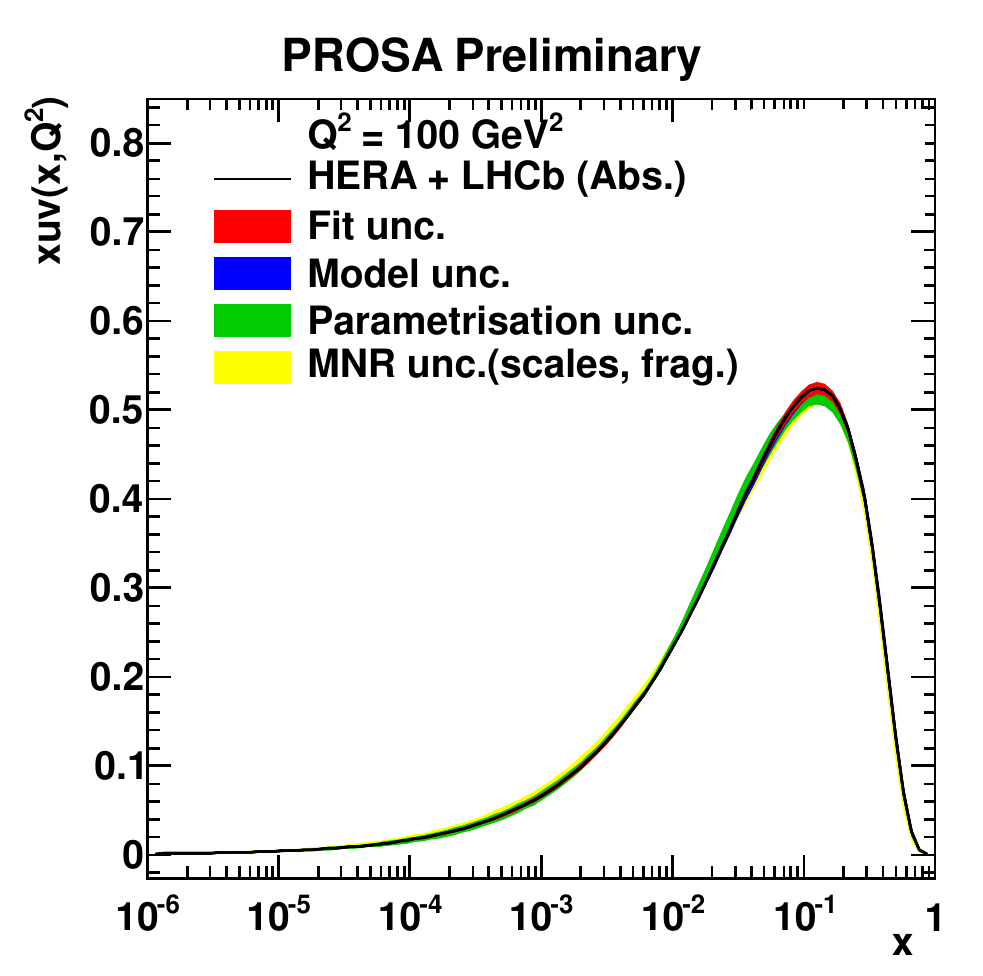}
  \includegraphics[width=0.495\figwidth,trim=2mm 2mm 2mm 8.5mm,clip=true]{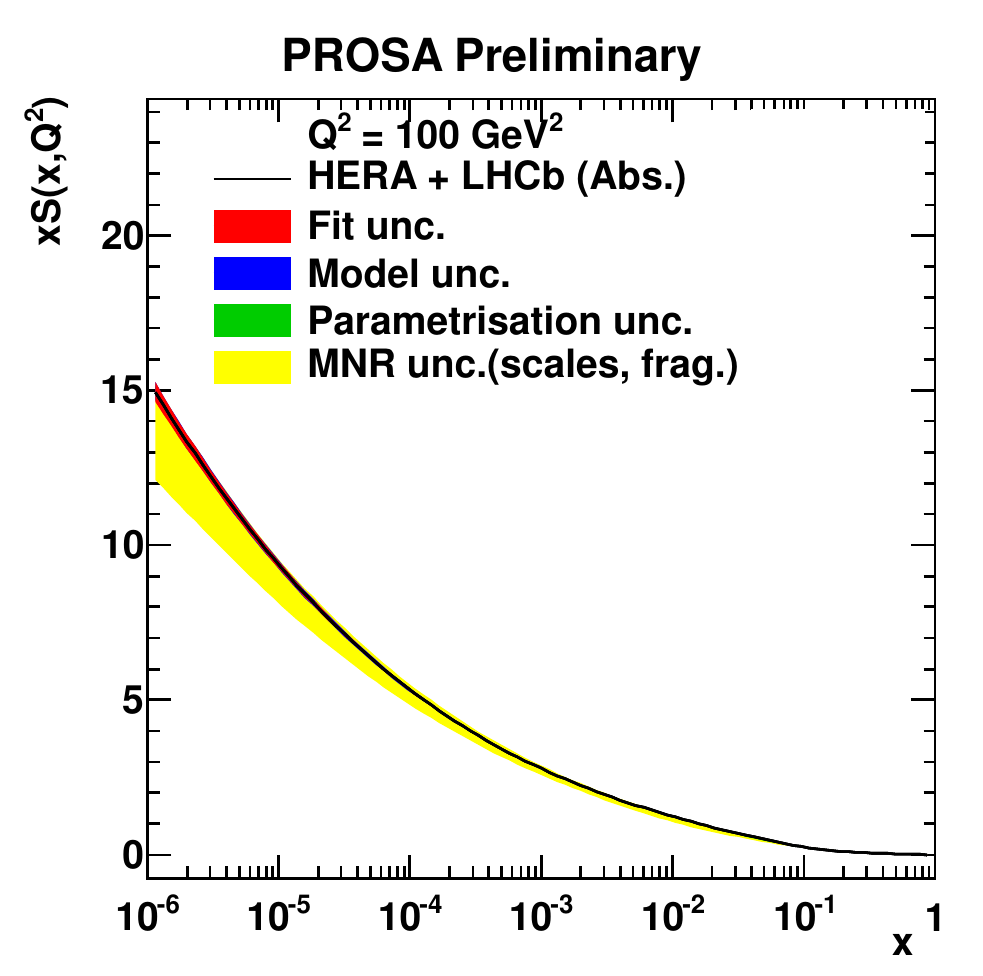}
  \includegraphics[width=0.495\figwidth,trim=2mm 2mm 2mm 8.5mm,clip=true]{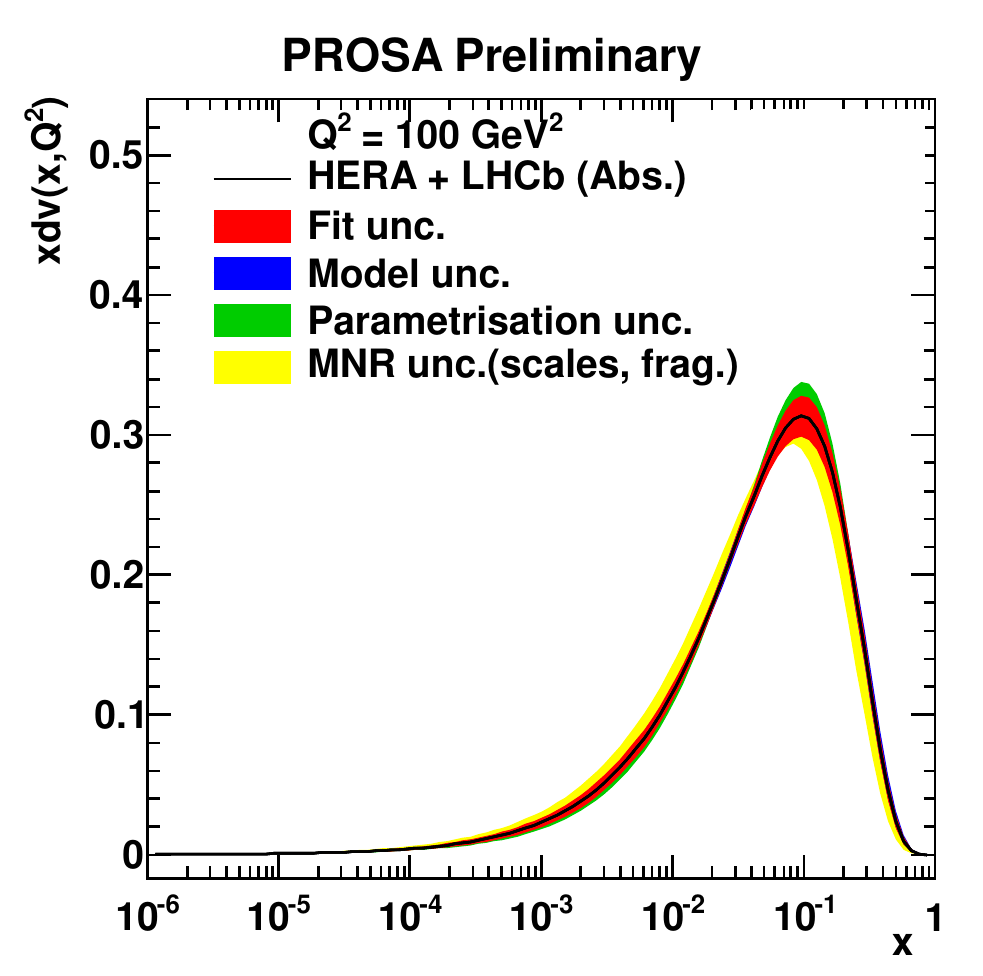}
  \caption[PDF uncertainties at $Q^2=\SI{100}{GeV^2}$ for `LHCb Abs' fit]
  {The individual contributions to the uncertainties of the gluon (top left), $u$-valence (top right), sea (bottom left) and $d$-valence (bottom right) distributions at $Q^2=\SI{100}{GeV^2}$ 
  obtained in the fit with the HERA and LHCb data using the `LHCb Abs' approach.}
	\label{fig:pdffit:FittedPDFs_LHCbAbs_q2_100}
\end{figure}

\begin{figure}[tbp]
  \centering
  \includegraphics[width=0.495\figwidth,trim=2mm 2mm 2mm 8.5mm,clip=true]{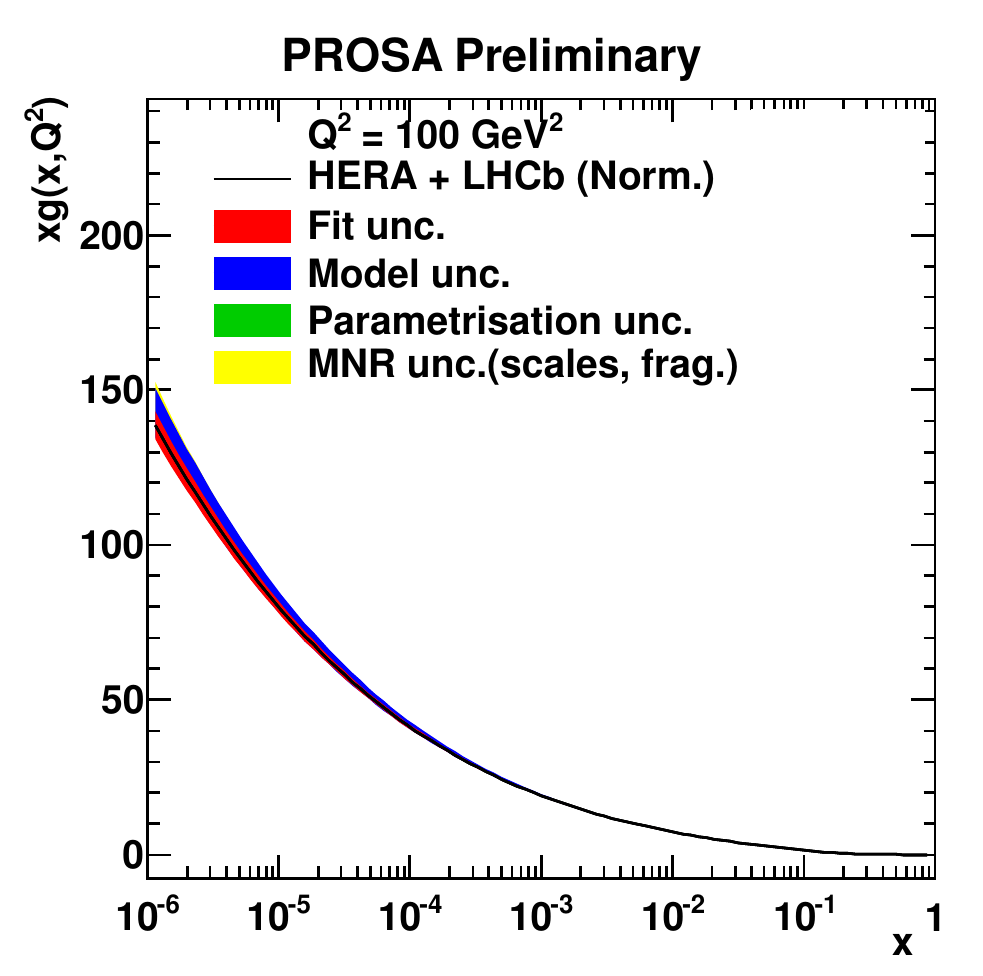}
  \includegraphics[width=0.495\figwidth,trim=2mm 2mm 2mm 8.5mm,clip=true]{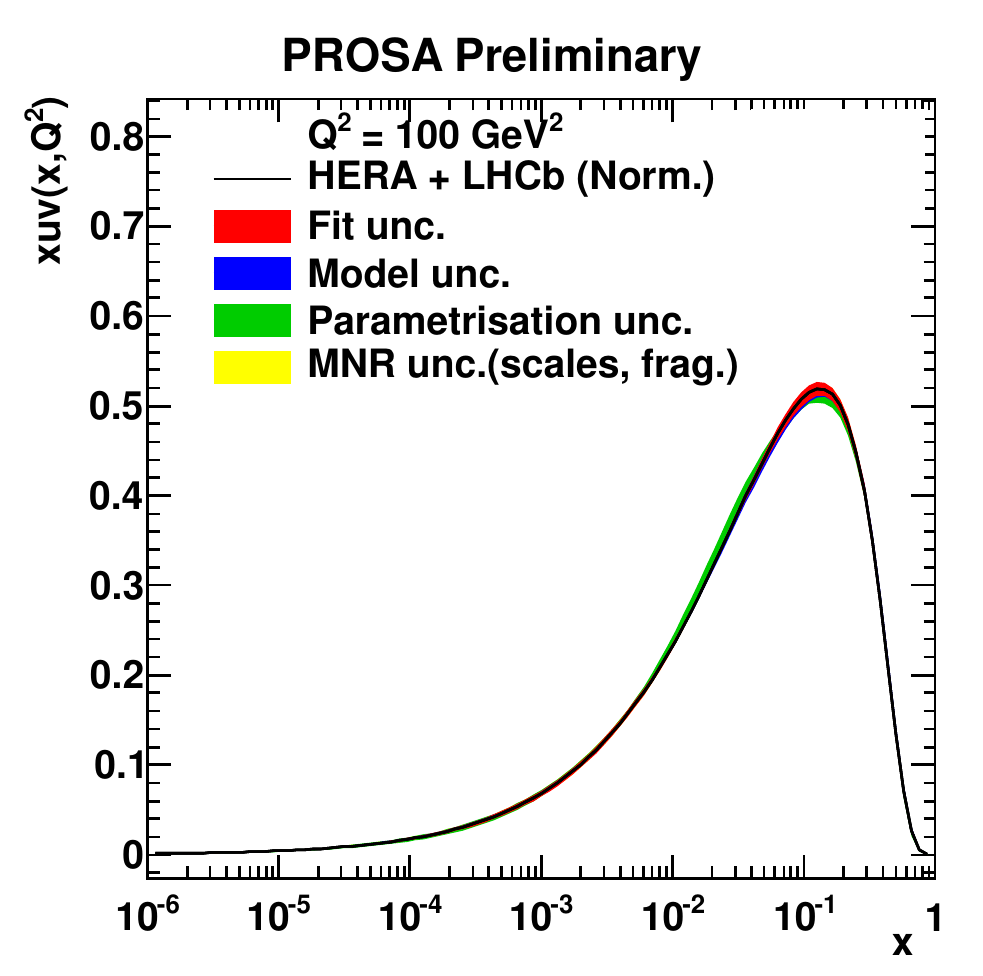}
  \includegraphics[width=0.495\figwidth,trim=2mm 2mm 2mm 8.5mm,clip=true]{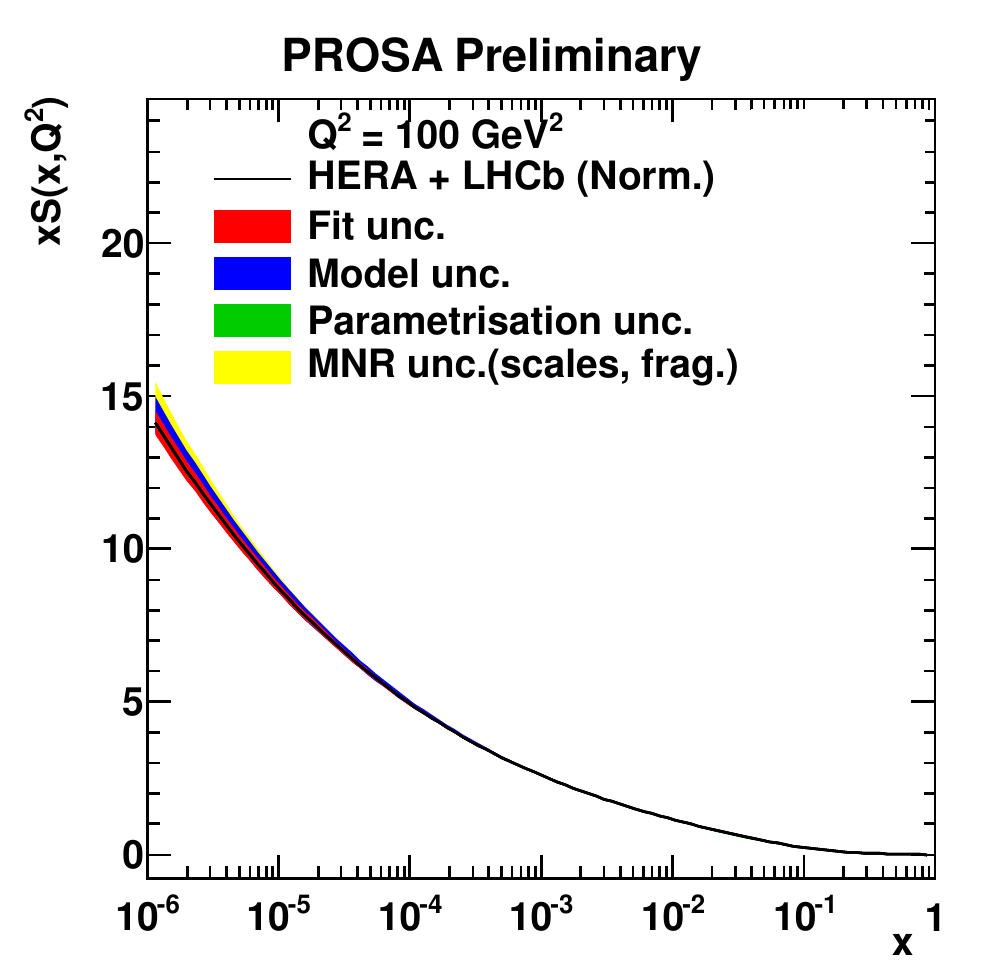}
  \includegraphics[width=0.495\figwidth,trim=2mm 2mm 2mm 8.5mm,clip=true]{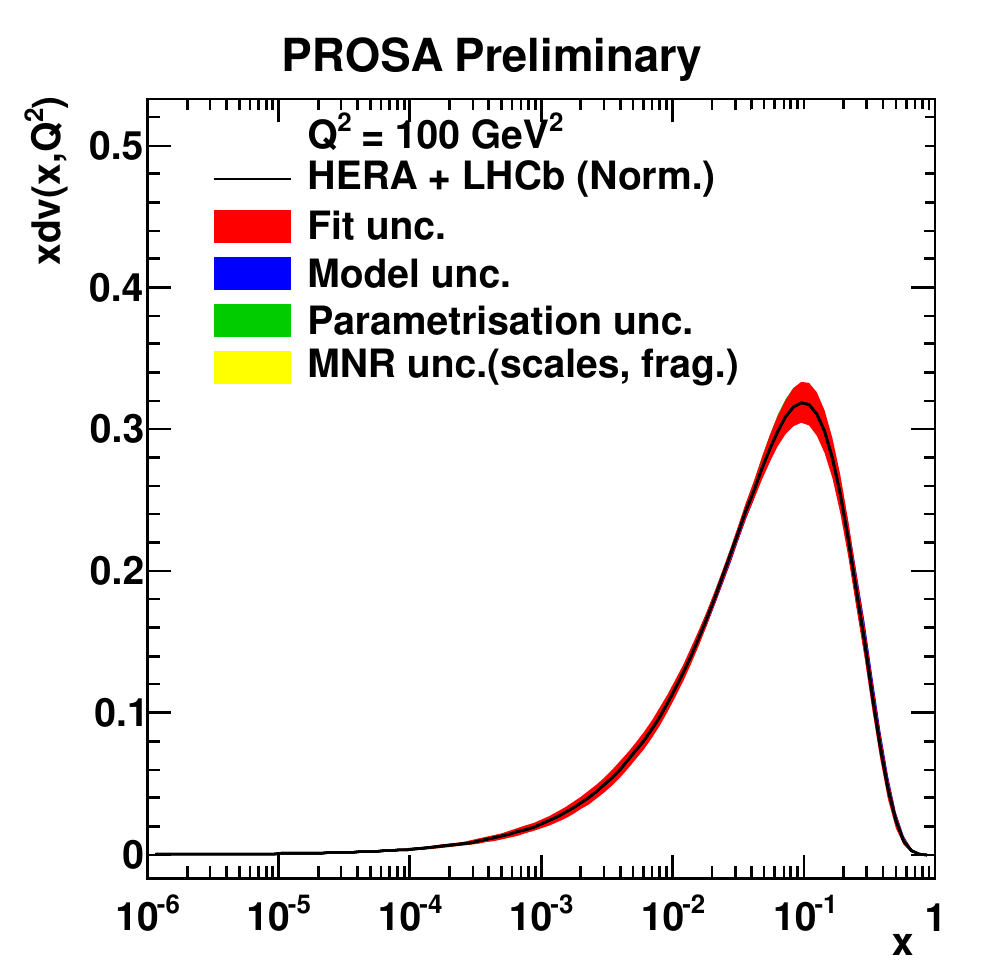}
  \caption[PDF uncertainties at $Q^2=\SI{100}{GeV^2}$ for `LHCb Norm' fit]
  {The individual contributions to the uncertainties of the gluon (top left), $u$-valence (top right), sea (bottom left) and $d$-valence (bottom right) distributions at $Q^2=\SI{100}{GeV^2}$ 
  obtained in the fit with the HERA and LHCb data using the `LHCb Norm' approach.}
	\label{fig:pdffit:FittedPDFs_LHCbNorm_q2_100}
\end{figure}

Figs.~\ref{fig:pdffit:FittedPDFs_HERAOnly_q2_100} to~\ref{fig:pdffit:FittedPDFs_LHCbNorm_q2_100} show individual contributions to the uncertainties 
for the distributions at $Q^2=\SI{100}{GeV^2}$, obtained in the `HERA only', 
`LHCb Abs' and `LHCb Norm' fits.

\begin{figure}[htbp]
  \centering
  \includegraphics[width=0.495\figwidth,trim=2mm 1mm 4mm 8.5mm,clip=true]{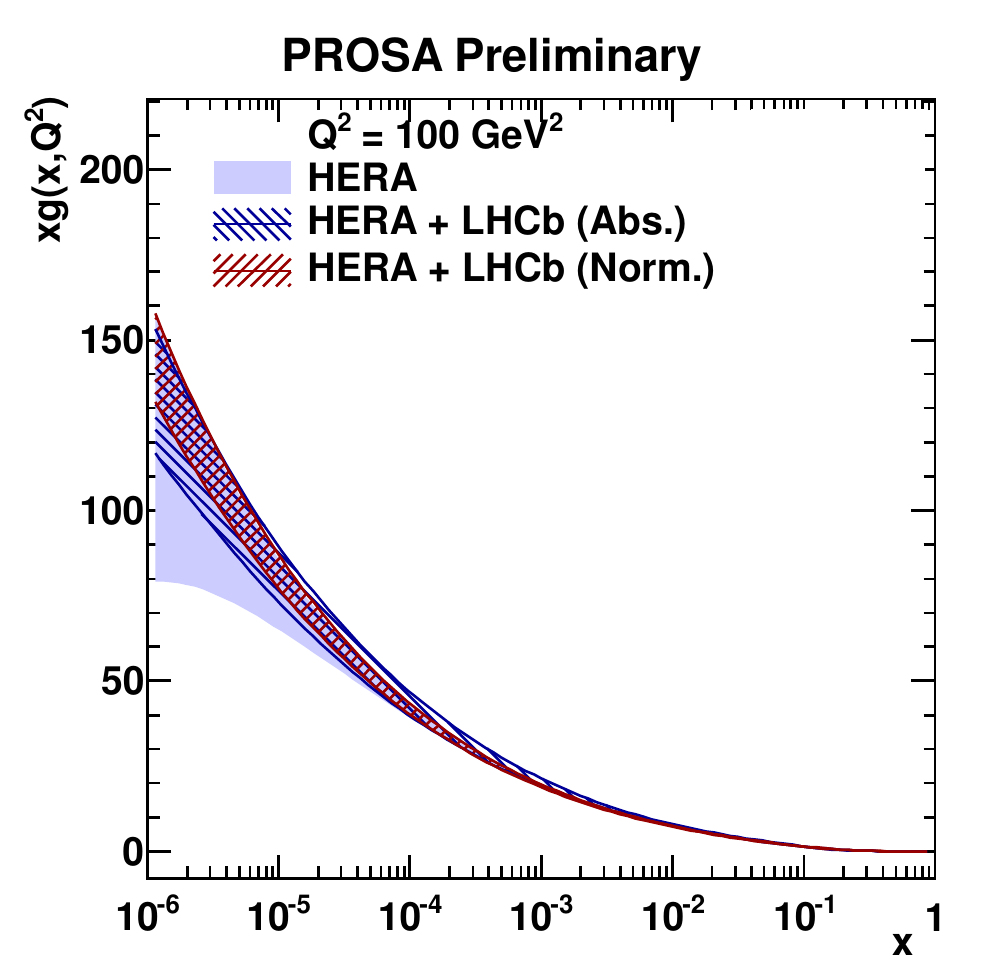}
  \includegraphics[width=0.495\figwidth,trim=2mm 1mm 4mm 8.5mm,clip=true]{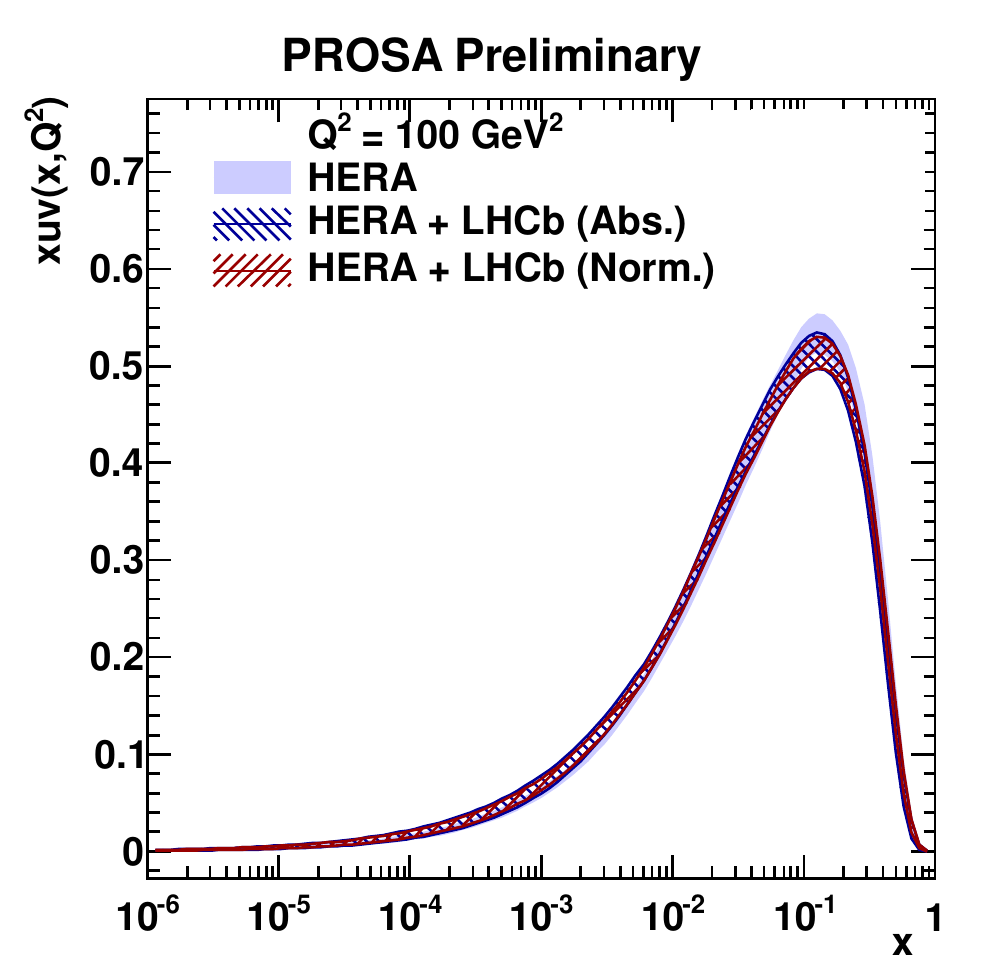}
  \includegraphics[width=0.495\figwidth,trim=2mm 1mm 4mm 8.5mm,clip=true]{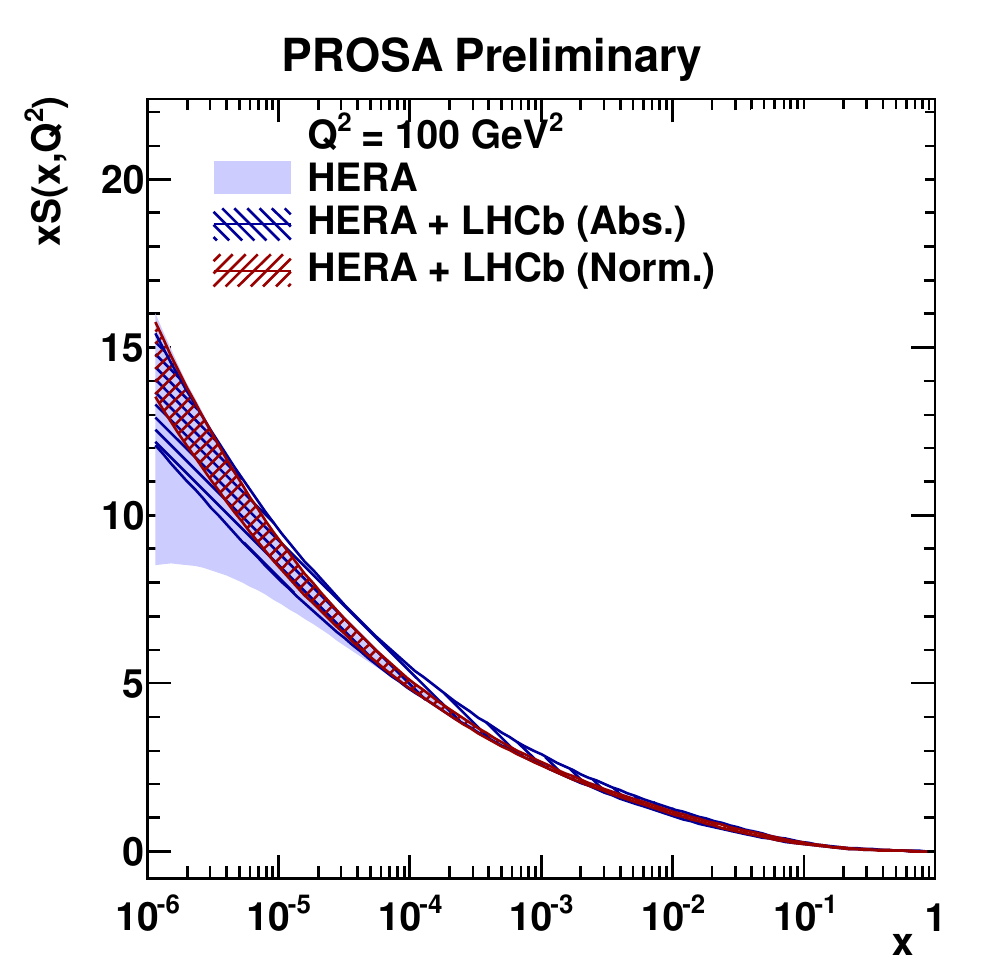}
  \includegraphics[width=0.495\figwidth,trim=2mm 1mm 4mm 8.5mm,clip=true]{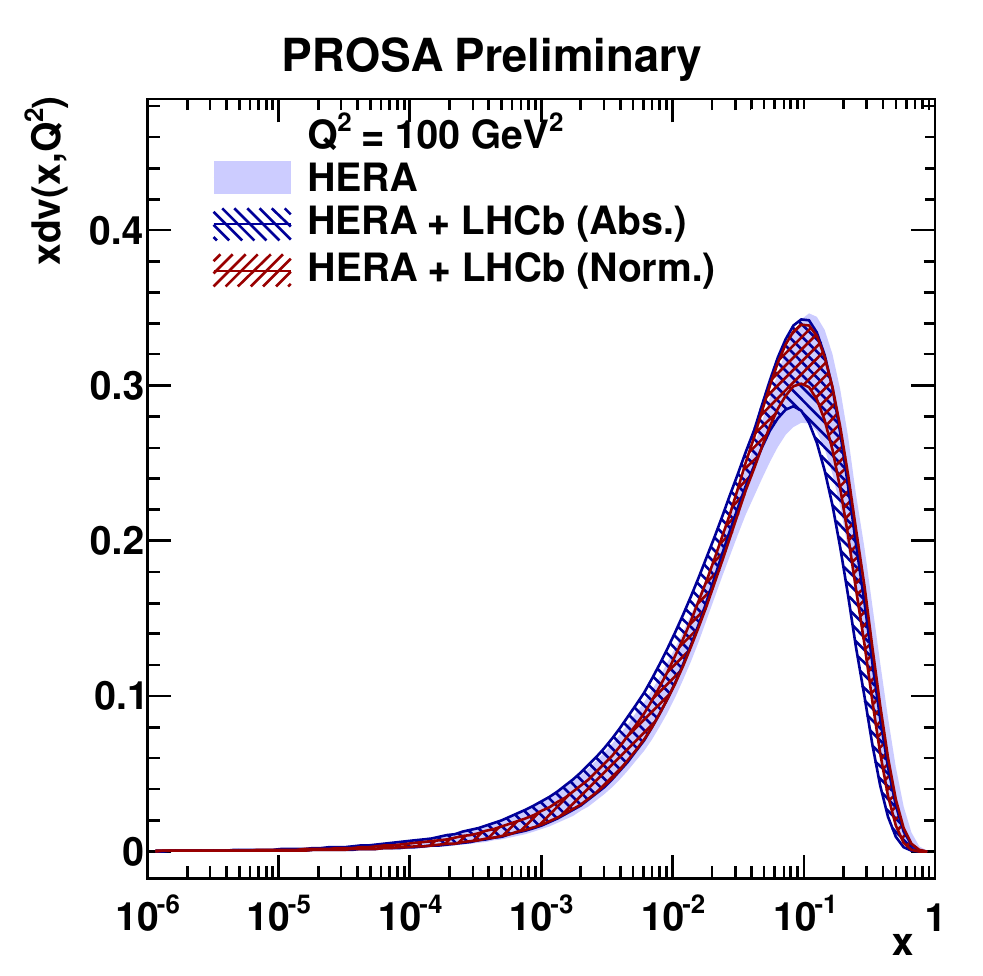}
  \caption[Comparison of PDFs at $Q^2=\SI{100}{GeV^2}$ from three fits]
  {The gluon (top left), $u$-valence (top right), sea (bottom left) and $d$-valence (bottom right) distributions at $Q^2=\SI{100}{GeV^2}$ obtained in the fit with the HERA-only, HERA and LHCb absolute, and HERA and LHCb normalised data.	The widths of the bands represent the total uncertainties.}
	\label{fig:pdffit:FittedPDFs_Comparison_q2_100}
\end{figure}

\begin{figure}[htbp]
  \centering
  \includegraphics[width=0.495\figwidth,trim=2mm 1mm 4mm 8.5mm,clip=true]{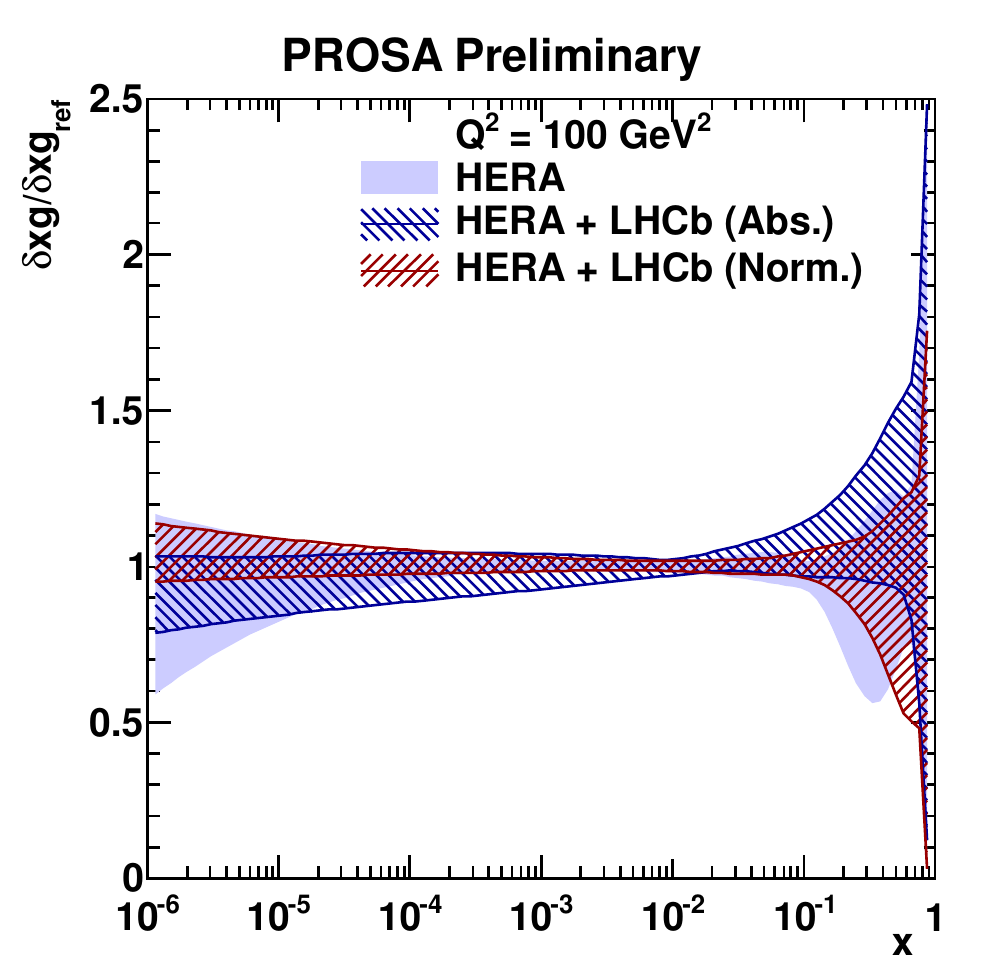}
  \includegraphics[width=0.495\figwidth,trim=2mm 1mm 4mm 8.5mm,clip=true]{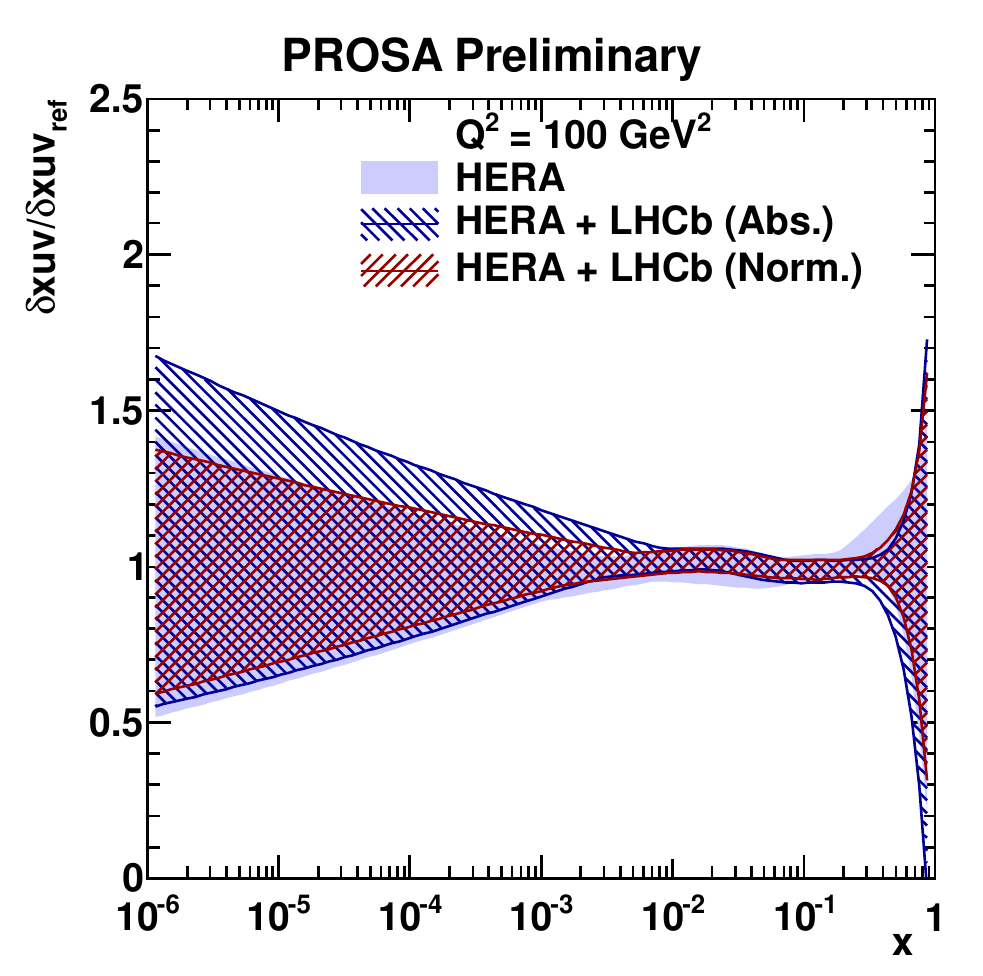}
  \includegraphics[width=0.495\figwidth,trim=2mm 1mm 4mm 8.5mm,clip=true]{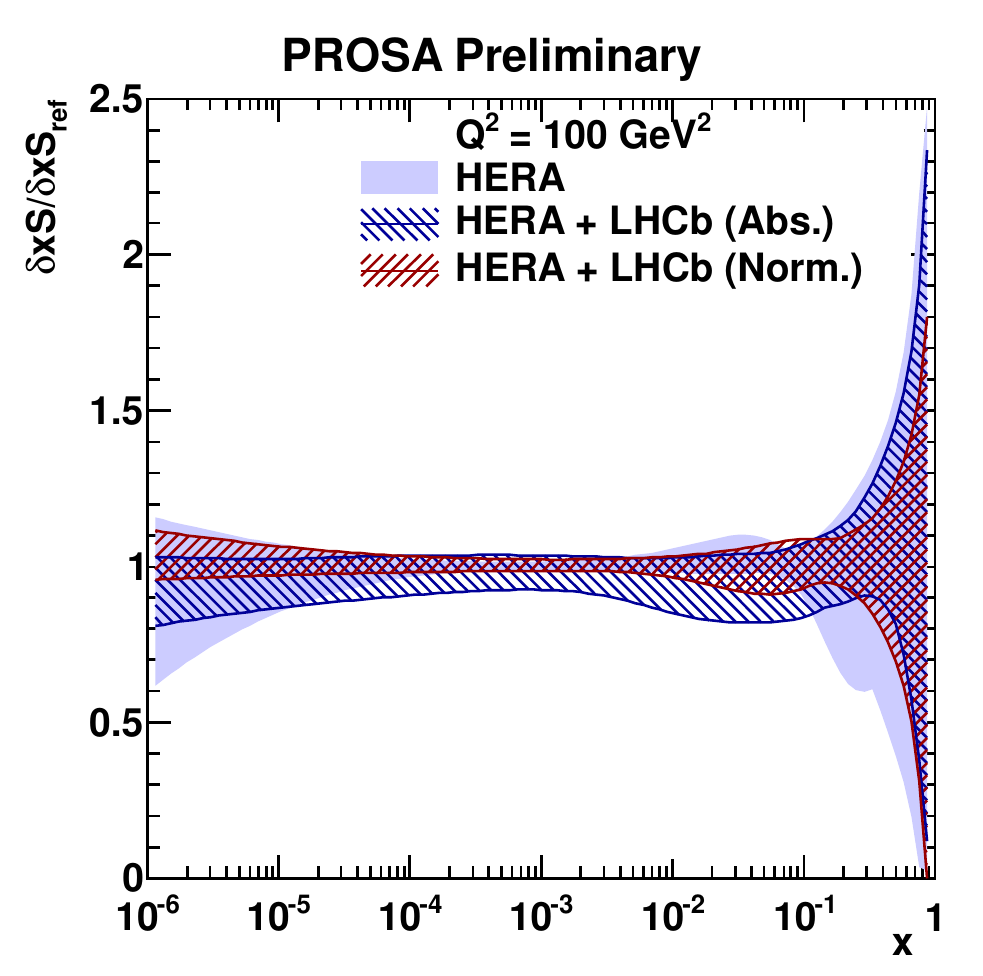}
  \includegraphics[width=0.495\figwidth,trim=2mm 1mm 4mm 8.5mm,clip=true]{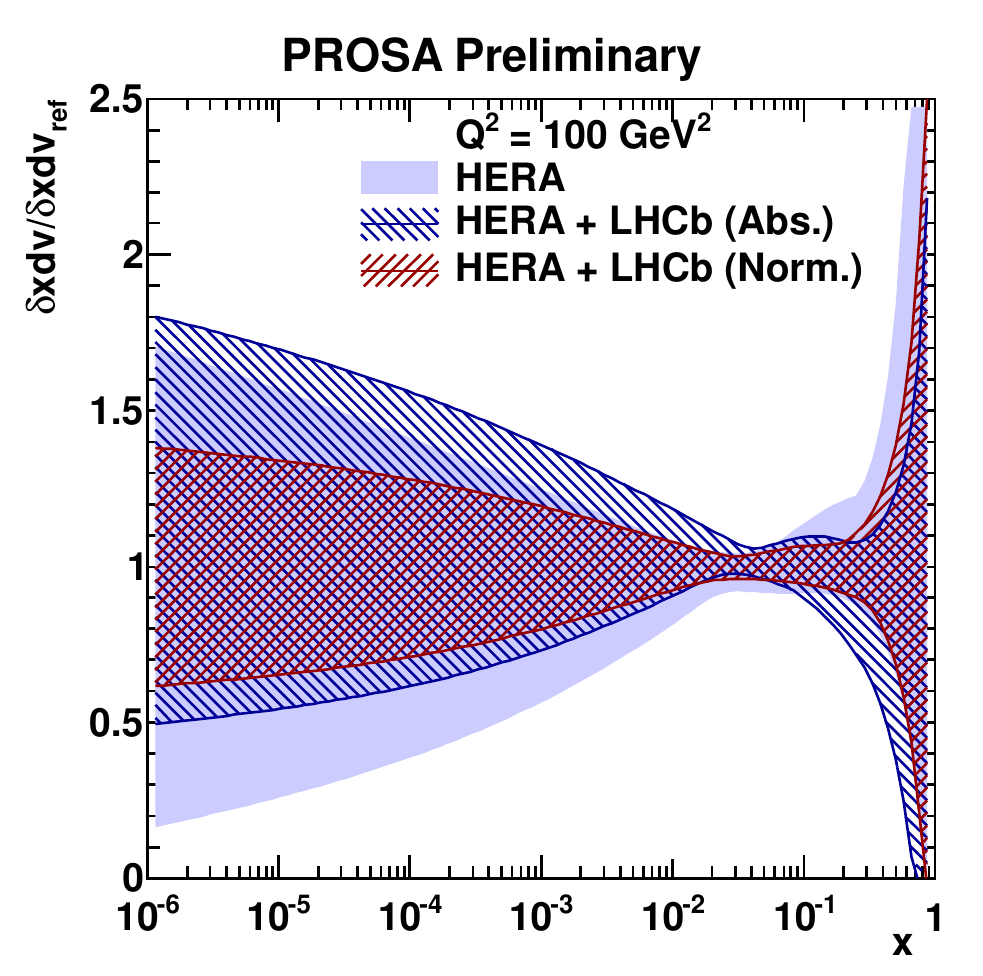}
  \caption[Comparison of relative PDF uncertainties at $Q^2=\SI{100}{GeV^2}$ from three fits]
  {The gluon (top left), $u$-valence (top right), sea (bottom left) and $d$-valence (bottom right) distributions at $Q^2=\SI{100}{GeV^2}$ obtained in the fit with the HERA-only, HERA and LHCb absolute, and HERA and LHCb normalised data, 
	normalised to one for a direct comparison of the uncertainties.
	The widths of the bands represent the total uncertainties.}
	\label{fig:pdffit:FittedPDFs_Comparison_q2_100_ratio}
\end{figure}

Figs.~\ref{fig:pdffit:FittedPDFs_Comparison_q2_100} and~\ref{fig:pdffit:FittedPDFs_Comparison_q2_100_ratio} show 
the PDFs and their relative uncertainties at the scale $Q^2=\SI{100}{GeV^2}$ 
obtained in the `HERA only', `LHCb Abs' and `LHCb Norm' fits.

\end{appendix}
\end{document}